\documentstyle[12pt,a42,twoside,epsfig,cite,axodraw,amstex,amssymb,float,mcite,verbatim,amsxtra,afterpage]{article}
\begin{document}
 \newcommand{\PS}{\hat{P}}
 \newcommand{\SigmaP}{{\sf Sigma}}
 \newcommand\SHU{\,\mbox{$\sqcup \! \sqcup$}\,}
 \newcommand{\D}{\displaystyle}
 \newcommand{\OLR}{\overleftrightarrow}
 \newcommand{\OL}{\overleftarrow}
 \newcommand{\OR}{\overrightarrow}
 \newcommand{\N}{\nonumber}
 \newcommand{\MS}{\overline{{\sf MS}}}
 \newcommand{\MOM}{\tiny{\mbox{MOM}}}
 \newcommand{\ep}{\varepsilon}
 \newcommand{\adag}{/\!\!\!\! }
 \newcommand{\si}{{\rm sign}}
 \newcommand{\adelta}{/\!\!\!\!\Delta }
 \newcommand{\GeV}{{\rm GeV}}
 \newcommand{\MeV}{{\rm MeV}}
 \newcommand{\eV}{{\rm eV}}
 \newcommand{\bra}[1]{\langle#1\mid}
 \newcommand{\ket}[1]{\mid#1\rangle}
 \newcommand{\A}{\mbox{\wedge}}
 \newcommand{\Li}{\mbox{Li}}
 \newcommand{\M}{\mbox{\rm\bf M}}
 \newcommand\ds{\displaystyle}
 \newcommand{\SN}{\mbox{S}}
 \newcommand\SH{\,\mbox{$\sqcup \! \sqcup$}\,}
 \newcommand{\Sf}{\mbox{S}_{1,2}}
 \newtheorem{Cor}{\underline{Corollary}}
 \newtheorem{Lem}{\underline{Lemma}}
 \newcommand{\gsim}{\raisebox{-0.07cm   }
 {$\, \stackrel{>}{{\scriptstyle\sim}}\, $}}
 \newcommand{\PO}{\hat{P}}
 \newcommand{\Mvec}{\mbox{\rm\bf M}}
 \newcommand{\fs}{\footnotesize}
 \newcommand{\eps}{\varepsilon}
 \newcommand{\Ahathat}{\hat{\hspace*{-1mm}\hat{A}}}
 \newcommand{\Atiltil}{\tilde{\hspace*{-1mm}\tilde{A}}}
 \newcommand{\Atil}{\tilde{A}}
 \newcommand{\Ctil}{\tilde{C}}
 \newcommand{\ahathat}{\hat{\hat{a}}}
 \sloppy
%%%%%%%%%%%%%%%%%%%%%%%%%%%%%%%%%%%%%%%%%%%%%%%%%%%%%%%%%%%%%%%%%%%%%%%%%
 \thispagestyle{empty}
\begin{flushleft}
 DESY-THESIS-2009-034
\end{flushleft}

 \setcounter{page}{0}

 \begin{center}
 
  {\Huge\bf\boldmath Mellin Moments of Heavy Flavor Contributions
                     to $F_2(x,Q^2)$ at NNLO \\
   }
  \vspace{3cm}

  \large 
   Dissertation \\
  \vspace{2mm}
   zur Erlangung des wissenschaftlichen Grades \\
  \vspace{2mm}
   Dr. rer. nat. \\
  \vspace{2mm}
   der Fakult{\"at} Physik der Technischen Universit\"at Dortmund

  \vspace{3em}

   {\small vorgelegt von}

  \vspace{2em}
  \large
  Sebastian Werner Gerhard Klein$^{a,b}$ \\
  {\small geboren am 23.01.1980 in Karlsruhe} 

  \vspace{8mm}
  Betreuer: PD Dr. habil. Johannes Bl\"umlein$^{a,b}$ \\
  \normalsize
  \vspace{4em}

   {\it$^a$Technische Universit\"at Dortmund, Fakult\"at Physik}\\
   {\it    Otto-Hahn-Str. 4, D-44227 Dortmund}\\
  \vspace{1mm}
   {\it$^b$Deutsches Elektronen--Synchrotron, DESY}\\
   {\it  Platanenallee 6, D--15738 Zeuthen}\\
 \end{center}
 \vspace*{\fill}
 
 \vspace*{\fill}
 \newpage
%%%%%%%%%%%%%%%%%%%%%%%%%%%%%%%%%%%%%%%%%%%%%%%%%%%%%%%%%%%%%%%%%%%%%%%%%
 \thispagestyle{empty}
 \begin{flushleft}
 \end{flushleft}
 \setcounter{page}{0}

 \vspace{170mm}

 eingereicht am $5$. Juni 2009\\

 \vspace{5mm}

 Gutachter:  
 \begin{itemize}
 \setlength{\leftmargin}{5mm} \item[] PD Dr. habil. Johannes Bl\"umlein, DESY, Zeuthen
  \setlength{\leftmargin}{5mm} \item[] Prof. Dr. Ewald Reya, 
                                       Technische Universit\"at Dortmund 
  \setlength{\leftmargin}{5mm} \item[] Dr. Jos Vermaseren, 
                                       NIKHEF, Holland
 \end{itemize}

 \newpage
%%%%%%%%%%%%%%%%%%%%%%%%%%%%%%%%%%%%%%%%%%%%%%%%%%%%%%%%%%%%%%%%%%%%%%%%
 \tableofcontents
 \listoffigures
 \listoftables
 \newpage
 \thispagestyle{empty}
 \begin{flushleft}
 \end{flushleft}
 \newpage
%%%%%%%%%%%%%%%%%%%%%%%%%%%%%%%%%%%%%%%%%%%%%%%%%%%%%%%%%%%%%%%%%%%%%%%%
%%%%%%%%%%%%%%%%%%%%%%%%%%%%%%%%%%%%%%%%%%%%%%%%%%%%%%%%%%%%%%%%%%%%%%%%
%
% Begin of the main part
%
%%%%%%%%%%%%%%%%%%%%%%%%%%%%%%%%%%%%%%%%%%%%%%%%%%%%%%%%%%%%%%%%%%%%%%%%
%%%%%%%%%%%%%%%%%%%%%%%%%%%%%%%%%%%%%%%%%%%%%%%%%%%%%%%%%%%%%%%%%%%%%%%%
%%%%%%%%%%%%%%%%%%%%%%%%%%%%%%%%%%%%%%%%%%%%%%%%%%%%%%%%%%%%%%%%%%%%%%%%
%%%%%%%%%%%%%%%%%%%%%%%%%%%%%%%%%%%%%%%%%%%%%%%%%%%%%%%%%%%%%%%%%%%%%%%%
%
% 
% Chapter 1
%
% Introduction
%
%%%%%%%%%%%%%%%%%%%%%%%%%%%%%%%%%%%%%%%%%%%%%%%%%%%%%%%%%%%%%%%%%%%%%%%%%
%%%%%%%%%%%%%%%%%%%%%%%%%%%%%%%%%%%%%%%%%%%%%%%%%%%%%%%%%%%%%%%%%%%%%%%%%
\newpage
\section{\bf\boldmath Introduction }
 \label{Sec-Intro} 
 \renewcommand{\theequation}{\thesection.\arabic{equation}}
 \setcounter{equation}{0}
%%%%%%%%%%%%%%%%%%%%%%%%%%%%%%%%%%%%%%%%%%%%%%%%%%%%%%%%%%%%%%%%%%%%%%%%%
  Quantum Chromodynamics (QCD) has been established as the theory of the strong
  interaction and explains the properties of hadrons, such as the proton or the
  neutron, in particular at short distances. Hadrons are composite objects and
  made up of quarks and antiquarks, which are bound together by the exchange of
  gluons, the gauge field of the strong force. The corresponding charge is 
  called color, leading to a $SU(3)_c$ gauge theory. This is analogous to the
  electric charge, which induces the $U(1)$ gauge group of electromagnetism. \\

  The path to the discovery of QCD started in the $1960$ies. By that time, a 
  large amount of hadrons had been observed in cosmic ray and accelerator 
  experiments. Hadrons are strongly interacting particles which occur as mesons
  (spin$~=0,~1$) or baryons (spin$~=1/2,~3/2$). In the early
  $1960$ies investigations were undertaken to classify all hadrons, based on 
  their properties such as flavor-- and spin quantum numbers and masses.
  In 1964, 
  M.~Gell-Mann, \cite{GellMann:1964nj}, and G.~Zweig, \cite{Zweig:1964jf},
  proposed the quark model as a mathematical description for these hadrons.
  Three fractionally charged quark flavors, up ($u$), down ($d$) and strange
  ($s$), known as valence quarks, were sufficient to describe the quantum
  numbers of the hadron spectrum which had been discovered by then.
  Baryons are thus considered as bound states of three quarks and mesons of a
  quark-antiquark pair. Assuming an approximate $SU(3)$ flavor
  symmetry, ``the eightfold way'', \cite{GellMann:1964xy,Kokkedee:1969,Close:1979bt}, 
  mass formulas for hadrons built on the basis of quark states could be 
  derived. A great
  success for the quark model was marked by the prediction of the mass of the 
  $\Omega^-$-baryon before it was finally observed, \cite{Barnes:1964pd}. In 
  the same year, G\"ursey and Radicati, \cite{Gursey:1992dc}, introduced spin 
  into the model and proposed a larger $SU(6)_{spin-flavor}=SU(2)_{spin}\otimes
  SU(3)_{flavor}$ symmetry. This allowed the unification of the mass formulas 
  for the spin--$1/2$ and spin--$3/2$ baryons and provided the tool to 
  calculate the ratio of the magnetic moments of the proton and the neutron to 
  be $\approx-3/2$, which is in agreement with experiment within $3\%$, 
  \cite{Beg:1964nm,Sakita:1964qr}. However, this theory required the quarks 
  that gave the correct low-lying baryons to be in a symmetric state under 
  permutations, which contradicts the spin--statistics theorem,
  \cite{Pauli:1940zz}, since quarks have to be fermions. Greenberg, 
  \cite{Greenberg:1964pe}, resolved this contradiction by introducing a
  ``symmetric quark model''. It allows quarks to have a new hidden 
  three--valued charge, called color, which is expressed in terms of parafermi 
  statistics. Finally, in 1965, Nambu, \cite{Nambu:1966}, and Han and Nambu, 
  \cite{Han:1965pf}, proposed a new symmetry, $SU(3)_{color}$, which makes the 
  hidden three--valued charge degree of freedom explicit and is equivalent to 
  Greenberg's description. Since there was no explicit experimental evidence of
  this new degree of freedom, the assumption was made that all physical 
  bound states must be color-neutral,~\cite{Nambu:1966,Han:1965pf,Fritzsch:1972jv}.  \\

  The possibility to study the substructure of nucleons arose at the end of the
  $1960$ies with the advent of the Stanford Linear Accelerator {\sf SLAC},
  \cite{Mo:1965dv,*Taylor:1967qv}. This facility allowed
  to perform deeply inelastic lepton-nucleon scattering (DIS)
  experiments at much 
  higher resolutions than previously possible. The cross section can be 
  parametrized quite generally in terms of several structure functions $F_i$ of
  the nucleon, \cite{Drell:1963ej,*Derman:1978iz}. 
  These were measured for the proton by the 
  {\sf SLAC-MIT} experiments and depend both on the energy transfer $\nu$
  and the $4$-momentum transfer $q^2=-Q^2$ from the lepton to the nucleon
  in the nucleon's rest frame. In the 
  Bjorken limit, $\{Q^2,~\nu~\rightarrow~\infty$,~$Q^2/\nu=$~fixed$\}$,
  \cite{Bjorken:1968dy}, it was found that the structure functions depend on 
  the ratio of $Q^2$ and $\nu$ only, $F_i(\nu,Q^2)=F_i(Q^2/\nu)$. This 
  phenomenon was called scaling,~\cite{Coward:1967au,*Panofsky:1968pb,*Bloom:1969kc,*Breidenbach:1969kd} cf. also \cite{Kendall:1991np,*Taylor:1991ew,*Friedman:1991nq},
  and had been predicted by Bjorken in his field theoretic analysis based on 
  current algebra,~\cite{Bjorken:1968dy}. As the relevant
  parameter in the deep-inelastic limit he introduced the Bjorken-scaling 
  variable $x=Q^2/2M\nu$, where $M$ is the mass of the nucleon. After scaling 
  was discovered, R. Feynman gave a phenomenological explanation for this 
  behavior of the structure functions within the parton model, 
  \cite{Feynman:1969wa,Feynman:1969ej,Feynman:1973xc}. According to this model,
  the proton consists of several point-like constituents, the partons. His 
  assumption was that during the interaction time - which is very short since 
  high energies are involved - these partons behave as free particles off which
  the electrons scatter elastically. Therefore,
  the total cross section is just 
  the incoherent sum of the individual electron-parton cross-sections, weighted
  by the probability to find the particular parton inside the proton. The 
  latter is described by the parton density $f_i(z)$. 
  It denotes the probability to find parton $i$ in the proton, carrying the 
  fraction $z$ of the total proton momentum $P$. In the limit considered by 
  Feynman, $z$ becomes equal to $x$, giving an explanation for scaling. This is
  a direct consequence of the {\sf rigid} correlation $M\nu=q.P$, as observed 
  in experiment. Even more important for the acceptance of the quark parton 
  model was the observation that the Callan-Gross relation, 
  \cite{Callan:1969uq}, holds, namely that the longitudinal structure function 
  $F_L$ vanishes in the situation of strict scaling. This experimental result 
  favored the idea of the proton containing spin--$1/2$, point-like 
  constituents and ruled out different approaches, such as the algebra of 
  fields, \cite{Lee:1967iu}, or explanations assuming
  vector--meson dominance, \cite{Sakurai:1969,*Sakurai:1969ss,*Tsai:1969yk,*Fraas:1970vj}.
  Finally, Bjorken and Paschos,~\cite{Bjorken:1969ja}, linked the parton model 
  to the group theoretic approach by identifying quarks and partons.  \\ 

  Today QCD forms one part of the Standard Model of elementary particle
  physics, 
  supplementing the electroweak $SU_L(2) \times U_Y(1)$ sector, which had been 
  proposed by S.~Weinberg in 1967,~\cite{Weinberg:1967tq}, extending earlier 
  work by S.~Glashow,~\cite{Glashow:1961tr}, cf. also 
  \cite{Salam:1964ry,*Salam:1968rm}, for the leptonic sector. This theory was 
  proved to be renormalizable by G.~t'Hooft and M.~Veltman in $1972$, 
  \cite{'tHooft:1972ue},~see also \cite{Taylor:1971ff,*Slavnov:1972fg,*Lee:1972fjxLee:1973fn}, if anomalies are canceled,
  \cite{Bell:1969ts,*Adler:1969gk,Bertlmann:1996xk}, requiring 
  an appropriate representation for {\sf all} fermions.
  G.~t'~Hooft also proved renormalization for 
  massless 
  Yang-Mills theories, \cite{'tHooft:1971fh}. These gauge theories had first 
  been studied by C.N.~Yang and R.L.~Mills in $1954$,~\cite{Yang:1954ek}, and 
  have the distinctive property that their gauge group is non-abelian, leading
  to interactions between the gauge--bosons,
  \cite{Fritzsch:1972jv}, contrary
  to the case of Quantum Electrodynamics. In $1972/73$, 
  M.~Gell-Mann, H.~Fritzsch 
  and H.~Leutwyler,~\cite{Fritzsch:1973pi}, cf. also \cite{Nambu:1966}, 
  proposed to gauge color which led to an extension of the Standard Model to 
  $SU_L(2)\times U_Y(1)\times SU_c(3)$, including the strongly interacting 
  sector. The dynamical theory of quarks and gluons, Quantum Chromodynamics, is
  thus a massless Yang-Mills theory which describes the interaction of 
  different quark flavors via massless gluons. Among the semi-simple compact 
  Lie-groups, $SU(3)_c$ turns out to be the only possible gauge group for this 
  theory,~cf.~\cite{Reya:1979zk,Muta:1998vi}. 
  In $1973$, D.~Gross and F.~Wilczek, 
  \cite{Gross:1973id}, and H.~Politzer,~\cite{Politzer:1973fx}, proved by a 
  $1$-loop calculation that Quantum Chromodynamics is an asymptotically free 
  gauge theory, cf. also \cite{tHooft:unpub}, which allows to perform 
  perturbative calculations for processes at large enough scales. There, 
  the strong coupling constant becomes a sufficiently small perturbative 
  parameter.

  In the beginning, QCD was not an experimentally well--established theory, 
  which was mainly due to its non--perturbative nature. The large value of the 
  strong coupling constant over a wide energy range prevents one from using 
  perturbation theory. In the course of performing precision tests of 
  QCD, the operator product expansion near the light--cone,
  the light--cone expansion (LCE),~\cite{Wilson:1969zs,*Zimmermann:1970,*Frishman:1971qn,*Brandt:1970kg},
  proved to be important. By applying it to deep--inelastic processes, one 
  facilitates a separation of hadronic bound state effects and the short 
  distance effects. This is possible, since the cross sections of deeply 
  inelastic processes receive contributions from two different resolution
  scales $\mu^2$.
  One is the short distance region, where perturbative techniques
  can be applied. The other describes the long distance region. Here bound 
  state effects are essential and a perturbative treatment is not possible due 
  to the large coupling involved. By means of the LCE, the two energy scales of
  the process are associated with two different quantities: the Wilson 
  coefficients and the hadronic operator matrix elements or parton densities. 
  The former contain the large scale contributions and can therefore be 
  calculated perturbatively, whereas the latter describe the low scale behavior
  and are quantities which have to be extracted from experimental data or can
  be calculated by applying rigorous non--perturbative methods. 
  Using the LCE, one may derive Feynman's parton model and show the 
  equivalence of the approaches by Feynman and Bjorken in the twist--$2$ 
  approximation,~\cite{Gross:1971wn}. The LCE also allows to go beyond the 
  naive partonic description, which is formulated in the renormalization group
  improved
  parton model. Shortly after the formulation of QCD, logarithmic scaling 
  violations of the deep inelastic cross section where observed,~\cite{Chang:1975sv,*Watanabe:1975su},
  which had to be expected since QCD is not an essentially free field theory, 
  neither is it conformally invariant, \cite{Ferrara:1973eg}.
  The theoretical explanation involves the calculation of higher order 
  corrections to the Wilson coefficients as well as to the anomalous 
  dimensions of the composite operators emerging in the LCE, 
  \cite{Gross:1973juxGross:1974cs,*Georgi:1951sr}, and predicts the correct 
  logarithmic $Q^2$ dependence of the structure functions. In fact, the
  prediction of scaling violations is one of the strongest experimental 
  evidences for QCD.

  Thus deeply inelastic scattering played a crucial role in formulating and 
  testing QCD as the theory governing the dynamics of quark systems. Its two 
  most important properties are the confinement postulate - all physical states
  have to be color singlets - and asymptotic freedom - the strength of the
  interaction becomes weaker at higher scales, i.e. at shorter distances,
  cf. e.g. \cite{PHOHAD:1971,*Politzer:1974fr,*Marciano:1977su,*Ellis:1979kt,Buras:1979yt,Reya:1979zk,Altarelli:1981ax,*Wilczek:1982yx,Jaffe:1985je,Collins:1987pm,Ellis:1988vi,Mueller:1989hs,Roberts:1990ww,Sterman:1994ce,*Ellis:1991qj,*Brock:1993sz,*Blumlein:1993ar,Mulders:1996}.

  An important step toward completing the Standard Model were the observations 
  of the three heavy quarks charm (c), bottom (b) and top (t). In $1974$, two 
  narrow resonances, called $\Psi~$ and $\Psi'$, were observed at 
  ${\sf SLAC}$ in $e^+e^-$ collisions at $3.1~\GeV$ and $3.7~\GeV$, 
  respectively,~\cite{Augustin:1974xw,*Abrams:1974yy}. 
  At the same time another resonance called $J$ was discovered
  in proton-proton collisions at ${\sf BNL}$,~\cite{Aubert:1974js}, 
  which turned out to be the same particle. 
  Its existence could not be explained in terms of the three known 
  quark flavors and was interpreted as a meson consisting of a new quark, the 
  charm quark. This was an important success of the Standard Model since the 
  existence of the charm had been postulated before,~\cite{Maki:1964ux,*Hara:1963gw,*Bjorken:1964gz}. It
  is necessary to cancel anomalies for the $2$nd family as well as for
  the GIM--mechanism, \cite{Glashow:1970gm}, in order to explain the
  absence of flavor changing neutral currents.
  With its mass of 
  $m_c \approx 1.3~\GeV$ it is much heavier than the light quarks, 
  $m_u \approx 2~\MeV~,m_d \approx 5~\MeV~,m_s \approx 104~\MeV$,
  \cite{Amsler:2008zzb}, and heavier than the nucleons. In later experiments, 
  two other heavy quarks were detected. In $1977$, the $\Upsilon$ 
  ($=b\overline{b}$) resonance was observed at ${\sf FERMILAB}$,
  \cite{Herb:1977ek}, and interpreted as a bound state of the even heavier 
  bottom quark, with $m_b \approx 4.2~\GeV$,~\cite{Amsler:2008zzb}. 
  Ultimately, the quark picture was completed in case of three fermionic 
  families by the discovery of the heaviest 
  quark, the top-quark, in $p\overline{p}$ collisions at the ${\sf TEVATRON}$ 
  in $1995$, \cite{Abe:1994xtxAbe:1994stxAbe:1995hr,*Abachi:1995iq}. Its mass
  is given by roughly $m_t \approx 171~\GeV$,~\cite{Amsler:2008zzb}. Due to 
  their large masses, heavy quarks cannot be considered as constituents of 
  hadrons at rest or bound in atomic nuclei. They are rather excited in high 
  energy experiments and may form short-lived hadrons, with the exception of 
  the top-quark, which decays before it can form a bound state.  \\ 

  The theoretical calculation in this thesis relates to the production of heavy
  quarks in unpolarized deeply inelastic scattering via single photon exchange.
  In this case, the double differential scattering cross-section can be 
  expressed in terms of the structure functions $F_2(x,Q^2)$ and $F_L(x,Q^2)$.
  Throughout the last forty years, many DIS experiments have been performed, \cite{Stein:1975yy,*Atwood:1976ys,*Bodek:1979rx,*Mestayer:1982ba,Allkover:1981,*Aubert:1985fx,Bollini:1982ac,*Benvenuti:1984duxBenvenuti:1987zjxBenvenuti:1989rhxBenvenuti:1989fm,Amaudruz:1991nwxAmaudruz:1992bf,*Arneodo:1995cq,Chang:1975sv,*Watanabe:1975su,Anderson:1979mt,Adams:1989emxAdams:1996gu,Jonker:1981dc,*Bergsma:1982ckxBergsma:1984ny,Berge:1989hr,Jones:1994pw,Shaevitz:1995yc,Bosetti:1978kz,*deGroot:1978hr,*Heagy:1980wj,*Morfin:1981kg,*Bosetti:1982yy,*Abramowicz:1982re,*MacFarlane:1983ax,*Allasia:1985hw}.
  The proton was probed to shortest distances at the 
  Hadron-Elektron-Ring-Anlage {\sf HERA} at {\sf DESY} in Hamburg, 
  \cite{:1981uka,Abt:1993wz,Derrick:1992nw,Ackerstaff:1998av,Hartouni:1995cf}. 
  In these experiments, a large amount of data has been acquired, and in the 
  case of {\sf HERA} it is still being processed, especially for those of the 
  last running period, which was also devoted to the measurement of 
  $F_L(x,Q^2)$, \cite{:2008tx,*Collaboration:2009na}. Up to now, the structure 
  function $F_2(x,Q^2)$ is measured in a wide kinematic region, 
  \cite{Amsler:2008zzb}, whereas $F_L(x,Q^2)$ was mainly measured in fixed 
  target experiments, \cite{Whitlow:1990gk,*Dasu:1993vk,*Tao:1995uh,*Arneodo:1996qexArneodo:1996kd,*Liang:2004tk}, and determined in the region of large 
  $\nu$,~\cite{Adloff:1996yz}. In the analysis of DIS data, the contributions 
  of heavy quarks play an important role, cf. e.g. \cite{Klein:2004DISProc,Feltesse:2005aa,Lipka:2006ny,Thompson:2007mx,Jung:2009eq}.
  One finds that the scaling violations of the heavy quark contributions differ
  significantly from those of the light partons in a rather wide range starting
  from lower values of $Q^2$. This demands a detailed description. 
  Additionally, it turns out that the heavy quark contributions to the 
  structure functions may amount up to 25-35\%, especially in the small--$x$ 
  region,~\cite{Lipka:2006ny,Chekanov:2008yd,Blumlein:1996sc,Thompson:2007mx},
  which requires a more precise theoretical evaluation of these terms.
 \\ 

  Due to the kinematic range of {\sf HERA} and the previous DIS experiments, 
  charm is produced much more abundantly and gives a higher contribution to the
  cross section than bottom, \cite{Thompson:2007mx}. Therefore we subsequently
  limit our discussion to one species of a heavy quark.
  Intrinsic heavy quark
  production is not considered, since data from {\sf HERA} show that this 
  production mechanism hardly gives any contribution, cf. \cite{Brodsky:1980pb,*Hoffmann:1983ah,*Derrick:1995sc,*Harris:1995jx,Adloff:1996xq}.
  The need for considering heavy quark production has several aspects. One of 
  them is to obtain a better description of heavy flavor production and its 
  contribution to the structure functions of the nucleon. On the other hand, 
  increasing our knowledge on the perturbative part of deep--inelastic 
  processes allows for a more precise determination of the QCD--scale 
  $\Lambda_{\rm QCD}$ and the strong coupling constant $\alpha_s$,
  as well as of
  the parton--densities from experimental data. For the former, sufficient 
  knowledge of the ${\sf NNLO}$ massive corrections in DIS is required to 
  control the theory--errors on the level of the experimental accuracy and below, \cite{Bethke:2000ai,*Bethke:2004uy,Blumlein:2004ip,Alekhin:2005dxxAlekhin:2005dy,Dittmar:2005ed,Gluck:2006yz,Alekhin:2006zm,Blumlein:2006be,Blumlein:2007dk,Jung:2008tq}. 
  The parton distribution functions are process independent quantities and can
  be used to describe not only deeply inelastic scattering, but also a large
  variety of scattering events at (anti--)proton--proton colliders such 
  as the 
  ${\sf TEVATRON}$ at ${\sf FERMILAB}$, and the Large--Hadron--Collider
  (${\sf LHC}$) at ${\sf CERN}$, \cite{Jung:2009eq}. Heavy quark 
  production is well suited to extract the gluon density since at leading order
  (LO) only the photon--gluon fusion process contributes to the cross section,
  \cite{Witten:1975bh,*Babcock:1977fi,*Shifman:1977yb,*Leveille:1978px,Gluck:1980cp}. 
  Next-to-leading order (NLO) calculations, as performed in 
  Refs.~\cite{Laenen:1992zkxLaenen:1992xs,*Riemersma:1994hv}, showed that this 
  process is still 
  dominant, although now other processes contribute, too. The gluon density 
  plays a special role, since it carries roughly $50~\%$ of the proton
  momentum, as data from ${\sf FERMILAB}$ and ${\sf CERN}$ showed 
  already in the $1970$ies, \cite{Taylor:1976rk}. Improved knowledge on the 
  gluon distribution $G(x,Q^2)$ is also necessary to describe gluon-initiated 
  processes at the ${\sf TEVATRON}$ and at the ${\sf LHC}$. The 
  study of heavy quark production will also help to further understand the 
  small-$x$
  behavior of the structure functions, showing a steep rise, which is mainly 
  attributed to properties of the gluon density. \\

  The perturbatively calculable contributions to the DIS cross section are the 
  Wilson coefficients. In case of light flavors only, these are denoted by 
  $C_{(q,g),(2,L)}(x,Q^2/\mu^2)$~\footnote{$q$=quark, $g$=gluon} and at present
  they are known up to the third order in the strong coupling constant,~\cite{Zee:1974du,Bardeen:1978yd,Furmanski:1981cw,Duke:1981ga,*Devoto:1984wu,*Kazakov:1987jk,*Kazakov:1990fu,*SanchezGuillen:1990iq,*vanNeerven:1991nnxZijlstra:1991qcxZijlstra:1992qd,*Kazakov:1992xj,*Larin:1991fv,Moch:1999eb,Larin:1993vu,Larin:1996wd,Retey:2000nq,Moch:2004xu,Blumlein:2004xt,Vermaseren:2005qc}.
  Including massive quarks into the analysis, the corresponding terms
  are known exactly at ${\sf NLO}$. The ${\sf LO}$ terms have been derived in 
  the late seventies,~\cite{Witten:1975bh,*Babcock:1977fi,*Shifman:1977yb,*Leveille:1978px,Gluck:1980cp},
  and the ${\sf NLO}$ corrections semi--analytically in $z$--space in the 
  mid--90ies, 
  \cite{Laenen:1992zkxLaenen:1992xs,*Riemersma:1994hv}. A fast numerical
  implementation was
  given in \cite{Alekhin:2003ev}. In order to describe DIS at the level of 
  twist $\tau=2$, also the anomalous dimensions of the local composite 
  operators emerging in the LCE are needed. These have to be combined with the 
  Wilson coefficients and describe, e.g.,
  the scaling violations of the structure
  functions and parton densities, \cite{Gross:1973juxGross:1974cs,*Georgi:1951sr}. 
  This description is equivalent to the picture in $z$--space in terms of 
  the splitting functions, \cite{Altarelli:1977zs}. The unpolarized anomalous
  dimensions are known up to ${\sf NNLO}$~\footnote{In Ref.~\cite{Baikov:2006ai}, the $2$nd moment of the $4$--loop ${\sf NS^+}$ anomalous dimension was calculated.}.
  At leading,~\cite{Gross:1973juxGross:1974cs,*Georgi:1951sr}, and at next--to--leading--order level,~\cite{Floratos:1977auxFloratos:1977aue1,Floratos:1978ny,GonzalezArroyo:1979df,GonzalezArroyo:1979he,*Curci:1980uw,*Furmanski:1980cm,Hamberg:1991qt},
  they have been known for a long time and were confirmed several times. The 
  ${\sf NNLO}$ anomalous dimension were calculated by Vermaseren et. al.
  First, the fixed moments were calculated in 
  Refs.~\cite{Larin:1996wd,Retey:2000nq,Blumlein:2004xt} and the complete
  result was obtained in Refs.~\cite{Moch:2004pa,Vogt:2004mw}. \\

  The main parts of this thesis are the extension of the description of the 
  contributions of heavy quark mass--effects to the deep--inelastic Wilson 
  coefficients to ${\sf NNLO}$. In course of that, we also obtain a first 
  independent calculation 
  of fixed moments of the fermionic parts of the ${\sf NNLO}$ anomalous 
  dimensions given in Refs.~\cite{Larin:1996wd,Retey:2000nq} before.

  The calculation of the 3-loop heavy flavor Wilson coefficients in the whole 
  $Q^2$ region is currently not within reach. However, as noticed in Ref.
  \cite{Buza:1995ie}, a very precise description of the heavy flavor Wilson 
  coefficients contributing to the structure function $F_2(x,Q^2)$ at 
  ${\sf NLO}$ is obtained for $Q^2 \gsim 10~m^2_Q$, disregarding the power 
  corrections $\propto (m_Q^2/Q^2)^k, k \geq 1$. If one considers the charm 
  quark, this covers an important
  region for deep--inelastic physics at ${\sf HERA}$.
  In this limit, the massive 
  Wilson coefficients factorize into universal massive operator matrix elements
  (OMEs) $A_{ij}(x,\mu^2/m^2_Q)$ and the light flavor Wilson coefficients 
  $C_{(q,g),(2,L)}(x,Q^2/\mu^2)$. The former are process independent
  quantities 
  and describe all quark mass effects. They are given by matrix 
  elements of the leading twist
  local composite operators $O_i$ between partonic states $j$ ($i,j=q,g$), 
  including quark masses. The process dependence is described by the massless
  Wilson coefficients. This factorization has been applied in 
  Ref.~\cite{Blumlein:2006mh} to obtain the asymptotic limit for 
  $F_L^{c\overline{c}}(x,Q^2)$ at 
  {\sf NNLO}. However, unlike the case for $F_2^{c\overline{c}}$,
  the asymptotic result in 
  this case is only valid for much higher values
  $Q^2 \gsim 800~m^2_Q$, outside the kinematic domain
  at ${\sf HERA}$ for this quantity.
  An analytic result for the ${\sf NLO}$ quarkonic massive 
  operator matrix elements $A_{qj}$ needed for the description of the structure
  functions at this order was derived in Ref.~\cite{Buza:1995ie} and confirmed 
  in Ref.~\cite{Bierenbaum:2007qe}. A related application of the massive OMEs 
  concerns the formulation of a variable flavor number scheme (VFNS) to
  describe parton densities of massive quarks at sufficiently high scales.
  This procedure has been described in 
  detail in Ref.~\cite{Buza:1996wv}, where the remaining gluonic massive OMEs 
  $A_{gj}$ were calculated up to $2$--loop order, thereby giving a full 
  ${\sf NLO}$ description. This calculation was confirmed and extended in 
  \cite{Bierenbaum:2009zt}. \\

  In this work, fixed moments of all contributing massive OMEs at the 
  $3$--loop level are calculated and presented, which is a new result,
  \cite{Bierenbaum:2008dk,Bierenbaum:2008tt,Bierenbaum:2009HERA,Bierenbaum:2009mv}. 
  The OMEs are then matched
  with the corresponding known $O(\alpha_s^3)$ light flavor Wilson 
  coefficients to obtain
  the heavy flavor Wilson coefficients in the limit $Q^2\gg~m^2$, which leads
  to a precise description for $Q^2/m^2\gsim~10$ in case of $F_2(x,Q^2)$.
  It is now possible to calculate all logarithmic contributions
  $\propto\ln(Q^2/m^2)^k$ to the massive Wilson coefficients in the 
  asymptotic region for general values of the Mellin variable $N$.
  This applies as well 
  for a large part of the constant term, where also the $O(\ep)$ contributions 
  at the $2$--loop level occur. 
  The first calculation of the latter for all--$N$ forms a part of this thesis,
  too, \cite{Bierenbaum:2007rg,Bierenbaum:2008dk,Bierenbaum:2008tm,Bierenbaum:2008yu,Bierenbaum:2009zt,Bierenbaum:2009HERA}.
  Thus only the constant terms of the 
  unrenormalized $3$--loop results are at present only known for fixed moments.
  Since the OMEs are given by the twist $\tau=2$ composite operators between 
  on--shell partonic states, also fixed moments of the fermionic contributions 
  to the ${\sf NNLO}$ unpolarized anomalous are obtained, which are thereby 
  confirmed for the first time in an independent calculation,
  \cite{Bierenbaum:2008dk,Bierenbaum:2008tt,Bierenbaum:2009HERA,Bierenbaum:2009mv}.

  A more technical aspect of this thesis is the study of the mathematical 
  structure of single scale quantities in renormalizable quantum field 
  theories, \cite{Blumlein:2007dj,Bierenbaum:2007zu,Blumlein:2009tm,Blumlein:2009tj}.
  One finds that the known results for a large number of different 
  hard scattering processes are most simply expressed in terms
  of nested harmonic sums, cf. \cite{Blumlein:1998if,Vermaseren:1998uu}. This
  holds at least up to 3--loop order for massless Yang--Mills theories, 
cf.~\cite{Blumlein:2004bb,Moch:2004pa,Vogt:2004mw,Vermaseren:2005qc,Dittmar:2005ed,Blumlein:2005im,*Blumlein:2006rr,Blumlein:2007dj},
  including the $3$--loop Wilson coefficients and anomalous dimensions. By 
  studying properties of harmonic sums, one may thus obtain significant
  simplifications, \cite{GonzalezArroyo:1979df}, since they obey
  algebraic, \cite{Blumlein:2003gb}, and structural relations,
  \cite{Blumlein:2009ta,Blumlein:2009fz}. 
  Performing the calculation in Mellin--space 
  one is naturally led to harmonic sums, which is 
  an approach we thoroughly adopt in our
  calculation. In course of this, new types of infinite sums occur if 
  compared to massless calculations. In the latter case, summation algorithms 
  such as presented in Refs. 
  \cite{Vermaseren:1998uu,Weinzierl:2002hv,Moch:2005uc} may be used to 
  calculate the respective sums. The new sums which emerge were calculated 
  using the recent summation 
  package~\SigmaP,~\cite{Refined,Schneider:2007,sigma1,sigma2},
  written in ${\sf MATHEMATICA}$, which opens up completely new possibilities 
  in symbolic summation and has been tested extensively through this work, 
  \cite{Bierenbaum:2007zu}.

  For fixed values of $N$, single scale quantities reduce to zero--scale 
  quantities, which can be expressed by rational numbers and certain special 
  numbers as \emph{multiple zeta values (MZVs)}, \cite{Borwein:1999js,Blumlein:2009Zet},
  and related quantities. 
  Zero scale problems are much easier to calculate than 
  single scale problems. By working in Mellin--space, single scale quantities
  are discrete and one can seek a description in terms of difference equations.
  One may think of an automated reconstruction
  of the all--$N$ relation out of a \emph{finite number} of Mellin moments 
  given in analytic form. This is possible for recurrent quantities.
  At least up to {3-loop order}, presumably to even higher orders, single
  scale quantities belong to this class. In this work, \cite{Blumlein:2009tm,Blumlein:2009tj},
  we report on a general
  algorithm for this purpose, which we applied to a problem being currently
  one of the most sophisticated ones: the determination of the anomalous
  dimensions and Wilson 
  coefficients to 3--loop order for unpolarized deeply-inelastic scattering, 
  \cite{Moch:2004pa,Vogt:2004mw,Vermaseren:2005qc}. \\

  The thesis is based on the publications Refs.~\cite{Bierenbaum:2008yu,Bierenbaum:2009zt,Blumlein:2009tj,Bierenbaum:2009mv}, the conference contributions \cite{Bierenbaum:2007pn,Bierenbaum:2007zz,Bierenbaum:2007rg,Bierenbaum:2007zu,Blumlein:2007dj,Bierenbaum:2008tt,Bierenbaum:2008dk,Bierenbaum:2008tm,Bierenbaum:2009HERA,Blumlein:2009tm} and the papers 
  in preparation \cite{Bierenbaum:prep1,Blumlein:trans}. 
  It is organized as follows. Deeply inelastic scattering within the parton
  model, the LCE and how one obtains improved results using the renormalization
  group are described in Section~\ref{Sec-DIS}. Section~\ref{Sec-HQDIS}
  is devoted to the production mechanisms of heavy quarks and 
  their contributions to the cross section. We also discuss the framework of
  obtaining the heavy flavor Wilson coefficients using massive OMEs in the 
  asymptotic limit $Q^2 \gg m_Q^2$ and comment on the different schemes one may
  apply to treat heavy quark production,
  \cite{Bierenbaum:2009zt,Bierenbaum:2009mv}.
  The massive operator matrix elements are considered in Section~\ref{Sec-REN} 
  and we describe in detail the renormalization of these objects to $3$--loop
  order, cf. \cite{Bierenbaum:2008dk,Bierenbaum:2008tt,Bierenbaum:2008yu,Bierenbaum:2009HERA,Bierenbaum:2009zt,Bierenbaum:2009mv}.
  Section~\ref{Sec-REP} contains transformation formulas between the different 
  renormalization schemes. We clarify an apparent
  inconsistency which we find in the renormalization of the massive 
  contributions to the ${\sf NLO}$ 
  Wilson coefficients given in Refs.~\cite{Laenen:1992zkxLaenen:1992xs,*Riemersma:1994hv} and the massive OMEs 
  as presented in Refs.~\cite{Buza:1995ie,Buza:1996wv}. This is due to the
  renormalization scheme chosen, cf. Ref.~\cite{Bierenbaum:2009zt,Bierenbaum:2009mv}. 
  In Section~\ref{Sec-2L} the calculation and the 
  results for the $2$--loop massive operator matrix elements up to $O(\ep)$ in 
  dimensional regularization are presented. This confirms the results of Ref.
  \cite{Buza:1996wv}, cf. \cite{Bierenbaum:2009zt}. The $O(\ep)$ terms are new
  results and are needed for renormalization at $O(\alpha_s^3)$, cf.
  \cite{Bierenbaum:2007rg,Bierenbaum:2008tm,Bierenbaum:2008dk,Bierenbaum:2008yu,Bierenbaum:2009zt,Bierenbaum:2009HERA}.
  We describe the calculation using hypergeometric functions to set up
  infinite sums containing the parameter $N$ as well. 
  These sums are solved using the summation package \SigmaP, cf.
  \cite{Bierenbaum:2008yu,Bierenbaum:2007zu}. All sums can then be expressed in
  terms of nested harmonic sums. The same structure is expected for the 
  $3$--loop terms, of which we calculate fixed moments ($N=2,...,10(14)$) using
  the programs {\sf QGRAF}, \cite{Nogueira:1991ex}, {\sf FORM}, 
  \cite{Vermaseren:2000nd,vanRitbergen:1998pn}, and {\sf MATAD},
  \cite{Steinhauser:2000ry} in Section~\ref{Sec-3L}, cf. \cite{Bierenbaum:2008dk,Bierenbaum:2008tt,Bierenbaum:2009HERA,Bierenbaum:2009mv}.
  Thus we confirm the corresponding moments of the fermionic contributions to 
  all unpolarized $3$--loop anomalous dimensions which have been calculated 
  before in Refs.~\cite{Larin:1996wd,Retey:2000nq,Blumlein:2004xt,Moch:2004pa,Vogt:2004mw}.
  In Section~\ref{Sec-POL} we calculate the asymptotic heavy flavor Wilson 
  coefficients for the polarized structure function $g_1(x,Q^2)$ to
  $O(\alpha_s^2)$ following Ref.~\cite{Buza:1996xr}
  and compare them with the results
  given there. We newly present the terms of $O(\alpha_s^2\ep)$ which 
  contribute to the polarized massive OMEs at $O(\alpha_s^3)$ through 
  renormalization, \cite{Bierenbaum:2007zz,Bierenbaum:2007pn,Bierenbaum:prep1}.
  One may also consider the local flavor non--singlet 
  tensor operator for transversity, \cite{Barone:2001sp}. This is done in
  Section~\ref{sec-1}. We derive the corresponding massive OMEs for general 
  values of $N$ up to $O(\alpha_s^2\ep)$ and for the 
  fixed moments $N=1\ldots 13$ at $O(\alpha_s^3)$, \cite{Blumlein:trans}. 
  A calculation keeping the full $N$ dependence has not been performed
  yet. In Section~\ref{Sec-FULL3L} we describe several steps which have been
  undertaken in this direction so far. 
  This involves the calculation of several
  non--trivial $3$--loop scalar integrals for all $N$ and the description of a 
  technique to reconstruct the complete result starting from a fixed number of 
  moments, cf. \cite{Blumlein:2009tm,Blumlein:2009tj}. Section~\ref{Sec-CONC} 
  contains the conclusions. Our conventions are summarized in Appendix 
  \ref{App-Con}.
  The set of Feynman--rules used, in particular for the composite
  operators, is given in Appendix \ref{App-FeynRules}. 
  In Appendix \ref{App-SpeFun} we summarize properties of special
  functions which frequently occurred in this work. 
  Appendix \ref{App-Sums} contains examples of different types of infinite sums
  which had to be computed in the present calculation.
  The main results are shown in Appendices \ref{App-AnDim}--\ref{App-Trans}:
  various anomalous
  dimensions and the constant contributions of the different massive OMEs 
  for fixed values of $N$ at $O(\alpha_s^3)$. All Figures in this work 
  have been drawn using ${\sf Axodraw}$, \cite{Vermaseren:1994je}.
%%%%%%%%%%%%%%%%%%%%%%%%%%%%%%%%%%%%%%%%%%%%%%%%%%%%%%%%%%%%%%%%%%%%%%%%%%%%%%%
%%%%%%%%%%%%%%%%%%%%%%%%%%%%%%%%%%%%%%%%%%%%%%%%%%%%%%%%%%%%%%%%%%%%%%%%%%%%%%%
%
% Chapter 2
%
% Deeply Inelastic Scattering
%
%%%%%%%%%%%%%%%%%%%%%%%%%%%%%%%%%%%%%%%%%%%%%%%%%%%%%%%%%%%%%%%%%%%%%%%%%%%%%%%
%%%%%%%%%%%%%%%%%%%%%%%%%%%%%%%%%%%%%%%%%%%%%%%%%%%%%%%%%%%%%%%%%%%%%%%%%%%%%%%
\newpage
 \section{\bf\boldmath Deeply Inelastic Scattering}
  \label{Sec-DIS}
  \renewcommand{\theequation}{\thesection.\arabic{equation}}
  \setcounter{equation}{0}
%%%%%%%%%%%%%%%%%%%%%%%%%%%%%%%%%%%%%%%%%%%%%%%%%%%%%%%%%%%%%%%%%%%%%%%%%%%%%%%
   Deep--inelastic scattering experiments provide one of the cleanest
   possibilities to probe the space--like short distance structure of hadrons
   through the reactions
   \begin{eqnarray}
    l^{\pm} N &\rightarrow& l^{\pm} + X \\
    \nu_l (\overline{\nu}_l) N &\rightarrow& l^{\mp} + X \\
    l^{\mp} N &\rightarrow& \nu_l (\overline{\nu}_l) + X~, 
   \end{eqnarray}
   with $l = e, \mu$, $\nu_l = \nu_{e, \mu, \tau}$, $N = p, d$ or a nucleus,
   and $X$ the inclusive hadronic final state. The $4$-momentum transfers 
   $q^2=-Q^2$ involved are at least of the order of $Q^2 \ge 4~\GeV^2$ and 
   one may resolve spatial scales of approximately $1/\sqrt{Q^2}$. The
   different deep inelastic charged-- and neutral current reactions offer 
   complementary sensitivity to unfold the quark flavor and gluonic structure 
   of the nucleons. Furthermore, polarized lepton scattering off polarized 
   targets is studied in order to investigate the spin structure of the 
   nucleons.

   The electron--proton experiments performed at ${\sf SLAC}$ in $1968$,
   \cite{Mo:1965dv,*Taylor:1967qv,Coward:1967au,*Panofsky:1968pb,*Bloom:1969kc,*Breidenbach:1969kd}, cf. also \cite{Kendall:1991np,*Taylor:1991ew,*Friedman:1991nq},
   and at ${\sf DESY}$, \cite{Albrecht:1969zyxAlbrecht:1969qmxAlbrecht:1969zb},
   found the famous scaling behavior of the structure functions which had been 
   predicted by Bjorken before, \cite{Bjorken:1968dy}. These measurements led 
   to the creation of the parton model, 
   \cite{Feynman:1969wa,Feynman:1969ej,Bjorken:1969ja}. Several years later, 
   after a series of experiments had confirmed its main predictions, the 
   partons 
   were identified with the quarks, anti-quarks and gluons as real quantum 
   fields, which are confined inside hadrons. Being formerly merely
   mathematical objects,~\cite{GellMann:1964nj,Zweig:1964jf},
   they became essential 
   building blocks of the Standard Model of elementary particle physics,
   besides the 
   leptons and the electroweak gauge fields, thereby solving the anomaly-problem,~\cite{Bell:1969ts,*Adler:1969gk,Bertlmann:1996xk}. 
 
   In the following years, more studies were undertaken at higher energies, 
   such as the electron--proton/neutron scattering experiments at 
   ${\sf SLAC}$,~\cite{Stein:1975yy,*Atwood:1976ys,*Bodek:1979rx,*Mestayer:1982ba}. 
   Muons were used as probes of the nucleons by 
   ${\sf EMC}$,~\cite{Allkover:1981,*Aubert:1985fx}, ${\sf BCDMS}$,~\cite{Bollini:1982ac,*Benvenuti:1984duxBenvenuti:1987zjxBenvenuti:1989rhxBenvenuti:1989fm}, 
   and ${\sf NMC}$,~\cite{Amaudruz:1991nwxAmaudruz:1992bf,*Arneodo:1995cq}, at
   the ${\sf SPS}$,~\cite{Clifft:1974zt}, at ${\sf CERN}$, as well as by the 
   ${\sf E26}$--,~\cite{Fox:1974ry,Chang:1975sv,*Watanabe:1975su}, 
   ${\sf~CHIO}$--,~\cite{Anderson:1979mt}, and ${\sf E665}$--,~\cite{Adams:1989emxAdams:1996gu}, collaborations at ${\sf FERMILAB}$. For a general review of
   $\mu^{\pm}~N$--scattering, see \cite{Sloan:1988qj}.
   The latter experiments 
   were augmented by several high energy neutrino scattering experiments by 
   the ${\sf CHARM}$-- and ${\sf CDHSW}$--collaborations, \cite{Jonker:1981dc,*Bergsma:1982ckxBergsma:1984ny,Holder:1977gn,*VonRuden:1982fp,Berge:1989hr},
   and the ${\sf WA21/25}$--experiments, \cite{Harigel:1977,Jones:1994pw}, at 
   the ${\sf SPS}$, and by the ${\sf CCFR}$--collaboration,~\cite{Sakumoto:1990py,*King:1991gs,Shaevitz:1995yc}, 
   at ${\sf FERMILAB}$. Further results on neutrinos were reported in Refs.~\cite{Bosetti:1978kz,*deGroot:1978hr,*Heagy:1980wj,*Morfin:1981kg,*Bosetti:1982yy,*Abramowicz:1982re,*MacFarlane:1983ax,*Allasia:1985hw}, cf. also \cite{Diemoz:1986kt,*Eisele:1986uz,*Mishra:1989jc,*Winter:1991ua,*Schmitz:1997}. The data of these experiments confirmed QCD as the 
   theory describing the strong interactions within hadrons, most notably by 
   the observation of logarithmic scaling violations of the structure 
   functions at higher energies and lower values of $x$, which had been 
   precisely predicted by theoretical calculations,~\cite{Gross:1973juxGross:1974cs,*Georgi:1951sr}.
   
   All these experiments had in common that they were fixed target experiments 
   and therefore could only probe a limited region of phase space, up to 
   $x\ge~10^{-3},~Q^2\le~500\GeV^2$.
   The first 
   electron--proton collider experiments became possible with the advent of 
   the ${\sf HERA}$ facility, which began operating in the beginning of the 
   $1990$ies at {\sf DESY},~\cite{:1981uka}. This allowed measurements at much 
   larger values of $Q^2$ and at far smaller values of $x$ than before, 
   $x\ge~10^{-4},~Q^2\le~20000\GeV^2$. 
   The physics potential for the deep--inelastic experiments at ${\sf HERA}$
   was studied during a series of workshops, see \cite{Peccei:1988pa,*Buchmuller:1992rq,*Blumlein:1992qi,*Faessler:1993ku,*Mathiot:1995ir,Bluemlein:1995uc,Ingelman:1996ge,Blumlein:1997ch,Blumlein:2001je}.
   ${\sf HERA}$ collected a vast amount of data until its shutdown in $2007$, 
   a part of which is still being analyzed, reaching unprecedented experimental
   precisions below the level of $1$ $\%$. 
   Two general 
   purpose experiments to study inclusive and various semi-inclusive 
   unpolarized deep--inelastic reactions, ${\sf H1}$, \cite{Abt:1993wz}, 
   and ${\sf ZEUS}$, \cite{Derrick:1992nw}, were performed. Both experiments
   measured the structure functions $F_{2,L}(x,Q^2)$ as well as the 
   heavy quark contributions to these structure functions to high precision.
   The theoretical calculations in this thesis are important for the analysis
   and understanding of the latter, as will be outlined in 
   Section~\ref{Sec-HQDIS}. The 
   {\sf HERMES}--experiment, \cite{Ackerstaff:1998av}, studied scattering of
   polarized electrons and positrons off polarized gas--targets.
   {\sf HERA-B}, \cite{Hartouni:1995cf},
   was dedicated to the study of {\sf CP}--violations in the $B$--sector.

   In the following, we give a brief introduction into the theory of DIS and 
   the theoretical tools which are used to predict the
   properties of structure functions, such as asymptotic scaling
   and scaling violations. In Section~\ref{SubSec-DISKin},
   we discuss the kinematics of the DIS process and derive
   the cross section for unpolarized electromagnetic 
   electron-proton scattering. In Section~\ref{SubSec-DISParton}, we give a 
   description of the naive parton model, which was employed to explain the 
   results obtained at {\sf SLAC} and gave a first correct qualitative 
   prediction of the observed experimental data. A rigorous treatment of DIS 
   can be obtained by applying the light--cone expansion to the
   forward Compton amplitude,~\cite{Wilson:1969zs,*Zimmermann:1970,*Frishman:1971qn,*Brandt:1970kg},
   which is described in Section~\ref{SubSec-DISComptLCE}. This is equivalent
   to the QCD--improved parton model at the level of twist $\tau=2$,
   cf. e.g. \cite{Kogut:1972di,*Yan:1976np,Reya:1979zk,Roberts:1990ww,Muta:1998vi}.
   One obtains evolution equations for the structure functions and 
   the parton densities with respect to the mass scales considered. The
   evolution is governed by the splitting functions,~\cite{Altarelli:1977zs},
   or the anomalous
   dimensions,~\cite{Gross:1973juxGross:1974cs,*Georgi:1951sr},
   cf. Section~\ref{SubSec-DISEvol}.
%%%%%%%%%%%%%%%%%%%%%%%%%%%%%%%%%%%%%%%%%%%%%%%%%%%%%%%%%%%%%%%%%%%%%%%%%%%%%%%
   \subsection{\bf\boldmath Kinematics and Cross Section} 
    \label{SubSec-DISKin}
\begin{figure}[h]
      \begin{center}
       \includegraphics[angle=0, width=8.0cm]{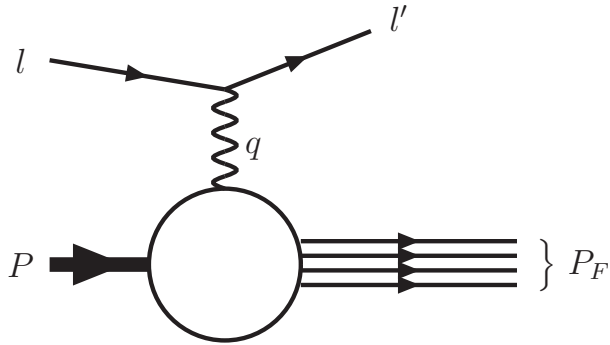}
      \end{center}
      \begin{center} 
       \caption{\sf Schematic graph of deeply inelastic scattering for single 
               boson exchange.}
       \label{DISLO1}
       \noindent
       \small
      \end{center}
      \normalsize
     \end{figure}

%%%%%%%%%%%%%%%%%%%%%%%%%%%%%%%%%%%%%%%%%%%%%%%%%%%%%%%%%%%%%%%%%%%%%%%%%%%%%%%
     The schematic diagram for the Born cross section of DIS is shown in 
     Figure~\ref{DISLO1} for single gauge boson exchange.
%%%%%%%%%%%%%%%%%%%%%%%%%%%%%%%%%%%%%%%%%%%%%%%%%%%%%%%%%%%%%%%%%%%%%%%%%%%%%%%
%%%%%%%%%%%%%%%%%%%%%%%%%%%%%%%%%%%%%%%%%%%%%%%%%%%%%%%%%%%%%%%%%%%%%%%%%%%%%%%
     A lepton with momentum $l$ scatters off a nucleon of mass $M$ and momentum
     $P$ via the exchange of a virtual vector boson with momentum $q$. The 
     momenta of the outgoing lepton and the set of hadrons are given by $l'$ 
     and $P_F$, respectively. Here $F$ can consist of any combination of 
     hadronic final states allowed by quantum number conservation. We consider 
     inclusive final states and thus all the hadronic states contributing to
     $F$ are summed over. The kinematics of the process can be measured from 
     the scattered lepton or the hadronic final states, cf. e.g. \cite{Blumlein:1992we,Blumlein:1994ii,Arbuzov:1995id},
     depending on the respective experiment. The virtual vector boson has 
     space-like momentum with a virtuality $Q^2$
     \begin{eqnarray}
      Q^2&\equiv&-q^2~,\quad  q=l-l'~. \label{virtuality}
     \end{eqnarray}
     There are two additional independent kinematic variables for which we
     choose
     \begin{eqnarray}
      s  &\equiv&(P+l)^2~, \label{sdeep} \\
      W^2&\equiv&(P+q)^2=P_F^2~. \label{pdeepf}
     \end{eqnarray}
     Here, $s$ is the total cms energy squared and $W$ denotes the invariant
     mass 
     of the hadronic final state. In order to describe the process, one 
     usually refers to Bjorken's scaling variable $x$, the inelasticity $y$, 
     and the total energy transfer $\nu$ of the lepton to the nucleon in the 
     nucleon's rest frame,~\cite{Bjorken:1969mm}. They are defined by
     \begin{eqnarray}
      \nu&\equiv&\frac{P.q}{M}~
             \,\,\,
             =~\frac{W^2+Q^2-M^2}{2M}~, 
               \label{nudef} \\
      x  &\equiv&\frac{-q^2}{2P. q}~
             =~\frac{Q^2}{2M\nu}~
             \,\,\,\,\,
             =~\frac{Q^2}{W^2+Q^2-M^2}~, 
               \label{Bjorkenx} \\
      y  &\equiv&\frac{P.q}{P.l}\hspace{3mm}
             =~\frac{2M\nu}{s-M^2}~
             =~\frac{W^2+Q^2-M^2}{s-M^2}~,
               \label{Bjorkeny}
     \end{eqnarray}
     where lepton masses are disregarded. In general, the virtual vector boson
     exchanged can be a $\gamma,~Z$ or $W^{\pm}$--boson with the in--
     and outgoing lepton, respectively, 
     being an electron, muon or neutrino. In the 
     following, we consider only unpolarized neutral current charged 
     lepton--nucleon scattering. In addition, we will disregard weak gauge
     boson effects caused by the exchange of a $Z$--boson. This is justified
     as long as the 
     virtuality is not too large, i.e. $Q^2 < 500~\GeV^2$,
     cf.~\cite{Blumlein:1987xk}. We assume the QED- and electroweak radiative 
    corrections to have been carried out,~\cite{Kwiatkowski:1990es,Blumlein:1994ii,Arbuzov:1995id}. 

     The kinematic region of DIS is limited by a series of 
     conditions. The hadronic mass obeys
     \begin{eqnarray}
      W^2 \ge M^2~. \label{physreg}
     \end{eqnarray}
     Furthermore,
     \begin{eqnarray}
      \nu \ge 0~, \quad 0\le y \le 1~, s\ge M^2~. \label{physreg2}
     \end{eqnarray}
     From (\ref{physreg}) follows the kinematic region for Bjorken-$x$ via
      \begin{eqnarray}
       &&W^2=(P+q)^2=M^2-Q^2\Bigl(1-\frac{1}{x}\Bigr) \ge M^2
% \N\\  &&
       \hspace{6mm}\Longrightarrow~0 \le x \le 1~. \label{xregion}
      \end{eqnarray}
     Note that $x=1$ describes the elastic process, while the inelastic region
     is defined by $x < 1$. Additional kinematic constraints follow from the 
     design parameters of the accelerator,
     \cite{Engelen:1998rf,*Abramowicz:1998ii}. 
     In the case of {\sf HERA}, these were $820(920)~\GeV$ for the proton beam 
     and $27.5~\GeV$ for the electron beam, resulting in a cms--energy 
     $\sqrt{s}$ of $300.3(319)~\GeV$~\footnote{During the final running period 
      of ${\sf HERA}$, low--energy measurements were carried out with 
      $E_p=460~(575)~\GeV$ in order to extract the longitudinal structure 
      function $F_L(x,Q^2)$,~\cite{:2008tx,*Collaboration:2009na}.}. 
     This additionally imposes kinematic constraints which follow from
     \begin{eqnarray} 
      Q^2&=&xy(s-M^2)~, \label{kincon1}
     \end{eqnarray}
     correlating $s$ and $Q^2$.
     For the kinematics at ${\sf HERA}$, this implies
     \begin{eqnarray}
      Q^2 \le sx \approx 10^5 x~. \label{kincon2}
     \end{eqnarray}
%%%%%%%%%%%%%%%%%%%%%%%%%%%%%%%%%%%%%%%%%%%%%%%%%%%%%%%%%%%%%%%%%%%%%%%%%%%%%%%
%
%    Cross Section
%
%%%%%%%%%%%%%%%%%%%%%%%%%%%%%%%%%%%%%%%%%%%%%%%%%%%%%%%%%%%%%%%%%%%%%%%%%%%%%%%

     In order to calculate the cross section of deeply inelastic 
     $ep$--scattering, one considers the tree--level transition matrix element 
     for the electromagnetic current. It is given by, cf. e.g. 
     \cite{Reya:1979zk,Roberts:1990ww,Muta:1998vi}, 
     \begin{eqnarray}
      M_{fi} 
            =e^2\overline{u}(l',\eta')\gamma^{\mu}u(l,\eta)
             \frac{1}{q^2}\bra{P_F}J^{em}_{\mu}(0)\ket{P,\sigma}~.\label{mfi}
     \end{eqnarray}
     Here, the spin of the charged lepton or nucleon is denoted by 
     $\eta (\eta')$ and $\sigma$, respectively. The state vectors of the 
     initial--state nucleons and the hadronic final state are 
     $\ket{P,\sigma}$ and $\ket{P_F}$. 
     The Dirac--matrices are denoted by $\gamma_{\mu}$ and
     bi--spinors by $u$, see Appendix~\ref{App-Con}. 
     Further $e$ is the electric 
     unit charge and $J^{em}_{\mu}(\xi)$ the quarkonic part of the 
     electromagnetic current operator, which is self-adjoint~:
     \begin{eqnarray}
        J_{\mu}^{\dagger}(\xi)=J_{\mu}(\xi)~.\label{jself} 
     \end{eqnarray}
     In QCD, it is given by
     \begin{eqnarray}
      J^{em}_{\mu}(\xi)=\sum_{f,f'} \overline{\Psi}_f(\xi)\gamma_{\mu}
                        \lambda^{em}_{ff'}\Psi_{f'}(\xi)~, \label{current}
     \end{eqnarray}
     where $\Psi_f(\xi)$ denotes the quark field of flavor $f$. For three 
     light flavors, $\lambda^{em}$ is given by the following combination of 
     Gell--Mann matrices of the flavor group $SU(3)_{flavor}$, cf. 
     \cite{Blumlein:1999sc,Yndurain:1999ui},
     \begin{eqnarray}
      \lambda^{em}=\frac{1}{2}\Bigl(\lambda_{flavor}^3
                   +\frac{1}{\sqrt{3}}\lambda_{flavor}^8
                   \Bigr)~. \label{lambdaem} 
     \end{eqnarray}
     According to standard definitions,~\cite{Reya:1979zk,Field:1989uq,Roberts:1990ww,Muta:1998vi}, the differential inclusive cross section is then given by
     \begin{eqnarray}
      l_0'\frac{d\sigma}{d^3l'}=\frac{1}{32(2\pi)^3(l.P)}
                                \sum_{\eta',\eta ,\sigma ,F}
                                (2\pi)^4\delta^4(P_F+l'-P-l)
                                |M_{fi}|^2~. \label{scatcro}
     \end{eqnarray}
     Inserting the transition matrix element (\ref{mfi}) into the relation
     for the scattering cross section (\ref{scatcro}),
     one notices that the trace over the 
     leptonic states forms a separate tensor, $L^{\mu\nu}$. Similarly, the 
     hadronic tensor $W_{\mu\nu}$ is obtained,
     \begin{eqnarray}
      L_{\mu \nu}(l,l')&=&\sum_{\eta',\eta}
                          \Bigl[\overline{u}(l',\eta')\gamma^{\mu}u(l,
                          \eta)\Bigr]^*
                          \Bigl[\overline{u}(l',\eta')\gamma^{\nu}u(l
                          ,\eta)\Bigr]~, \label{leptontens}\\
      W_{\mu\nu}(q,P)&=&\frac{1}{4\pi}\sum_{\sigma ,F}
                        (2\pi)^4\delta^4(P_F-q-P)
                        \bra{P,\sigma}J^{em}_{\mu}(0)\ket{P_F} 
                        \bra{P_F}J^{em}_{\nu}(0)\ket{P,\sigma}~. \N\\ 
                        \label{hadrontens}
     \end{eqnarray}
     Thus one arrives at the following relation for the cross section
     \begin{eqnarray}
      l_0'\frac{d\sigma}{d^3l'}&=&\frac{1}{4 P.l}
                                  \frac{\alpha^2}{Q^4}
                                  L^{\mu\nu}W_{\mu\nu}
                                =\frac{1}{2(s-M^2)}
                                  \frac{\alpha^2}{Q^4}
                                  L^{\mu\nu}W_{\mu\nu}~,\label{crosssec}
     \end{eqnarray}
     where $\alpha$ denotes the fine-structure constant, 
     see Appendix~\ref{App-Con}.
     The leptonic tensor in~(\ref{crosssec}) can be easily computed in the 
     context of the Standard Model,
     \begin{eqnarray}
      L_{\mu \nu}(l,l')&=&Tr[l \hspace*{-1.3mm}/ \gamma^{\mu}
                        l' \hspace*{-1.7mm}/ \gamma^{\nu}]
                        =4\left(l_{\mu}l_{\nu}'+l_{\mu}'l_{\nu}-\frac{Q^2}{2}
                        g_{\mu \nu}\right)~. \label{leptontens2}
     \end{eqnarray}
     This is not the case for the hadronic tensor, which contains
     non--perturbative hadronic contributions due to long-distance effects.
     To calculate these effects a priori, non-perturbative QCD
     calculations have to be performed, as in QCD lattice simulations.
     During the  
     last years these calculations were performed with increasing 
     systematic and numerical accuracy,~cf.~e.g.~\cite{Dolgov:2002zm,Gockeler:2007qs,*Baron:2007ti,*Bietenholz:2008fe,*Syritsyn:2009np}.

     The general structure of the hadronic tensor can be fixed using 
     $S$--matrix theory and the global symmetries of the process. In order to 
     obtain a form suitable for the subsequent calculations, one rewrites
     Eq.~(\ref{hadrontens}) as, cf. \cite{Itzykson:1980rh,Muta:1998vi}, 
     \begin{eqnarray}
      W_{\mu\nu}(q,P)
                     &=&\frac{1}{4\pi}\sum_{\sigma}
                        \int d^4\xi\exp(iq\xi)
                        \bra{P}[J^{em}_{\mu}(\xi),
                        J^{em}_{\nu}(0)]\ket{P} \N\\
                     &=&\frac{1}{2\pi}
                        \int d^4\xi\exp(iq\xi)
                        \bra{P}[J^{em}_{\mu}(\xi),
                        J^{em}_{\nu}(0)]\ket{P}~. \label{hadrontens4}
     \end{eqnarray}
     Here, the following notation for the spin-average is introduced in
     Eq.~(\ref{hadrontens4})
     \begin{eqnarray}
      \frac{1}{2}\sum_{\sigma}\bra{P,\sigma}X\ket{P,\sigma}
      \equiv \bra{P}X\ket{P}~. \label{spinshort}
     \end{eqnarray}
     Further, $[a,b]$ denotes the commutator of $a$ and $b$. Using 
     symmetry and conservation laws, the hadronic tensor can be decomposed into
     different scalar structure functions and thus be stripped of its 
     Lorentz--structure. In the most general case, including polarization, 
     there are $14$ independent 
     structure functions,~\cite{Blumlein:1996vs,Blumlein:1998nv}, which contain
     all information on the structure of the proton. However, in the case
     considered here, only two structure functions contribute. One uses 
     Lorentz-- and time--reversal invariance,~\cite{Wilson:1969zs,*Zimmermann:1970,*Frishman:1971qn,*Brandt:1970kg},
     and additionally the fact that the electromagnetic 
     current is conserved.
     This enforces electromagnetic gauge invariance for the hadronic tensor,
     \begin{eqnarray}
      q_{\mu}W^{\mu\nu}=0~.
     \end{eqnarray}
     The leptonic tensor (\ref{leptontens2}) is symmetric and thus $W_{\mu\nu}$
     can be taken to be symmetric as well, since all antisymmetric parts are 
     canceled in the contraction. By making a general ansatz for the hadronic 
     tensor using these properties, one obtains
     \begin{eqnarray}
      W_{\mu \nu}(q,P)=&&
                       \frac{1}{2x}\left(g_{\mu \nu}+\frac{q_{\mu}q_{\nu}}{Q^2}
                  \right)F_{L}(x,Q^2) \N\\
                  &+&\frac{2x}{Q^2}\left(
                   P_{\mu}P_{\nu}+\frac{q_{\mu}P_{\nu}+q_{\nu}P_{\mu}}{2x}
                   -\frac{Q^2}{4x^2}g_{\mu\nu}\right)F_{2}(x,Q^2)~.
      \label{hadrontens2}
     \end{eqnarray} 
     The dimensionless structure functions $F_2(x,Q^2)$ and $F_L(x,Q^2)$ 
     depend on two variables, Bjorken-$x$ and $Q^2$, contrary to the case of 
     elastic scattering, in which only one variable, e.g. $Q^2$,
     determines the cross section. Due to hermiticity of the hadronic tensor, 
     the structure functions are real. The decomposition (\ref{hadrontens2}) of
     the hadronic tensor leads to the differential cross section 
     of unpolarized DIS in case of single photon exchange
     \begin{eqnarray}
      \frac{d\sigma}{dxdy}=\frac{2\pi\alpha^2}{xyQ^2}
                           \Bigg\{\Bigl[1+(1-y)^2\Bigr]F_2(x,Q^2)
                                  -y^2F_L(x,Q^2)\Biggr\}~.\label{crosssec1}
     \end{eqnarray}
     A third structure function, $F_1(x,Q^2)$, 
     \begin{eqnarray}
      F_1(x,Q^2)=\frac{1}{2x}\Bigl[F_2(x,Q^2)-F_L(x,Q^2)\Bigr]~,\label{F1}
     \end{eqnarray}
     which is often found in the literature, is not independent of the previous
     ones.

     For completeness, we finally give the full Born cross section for the 
     neutral current, including the exchange of $Z$--bosons, 
     cf.~\cite{Arbuzov:1995id}. Not neglecting the lepton mass $m$, it is given
     by 
     \begin{eqnarray}
      \frac {d^2 \sigma_{\mathrm{NC}}} {dx dy}
       &=&
          \frac{2\pi\alpha^2 }{xyQ^2}
            \Biggl\{ 
               \Biggl[
                      2\left(1-y\right)-2xy\frac{M^2}{s}
                     +\left(1-2\frac{m^2}{Q^2}\right)
                      \left(1+4x^2\frac{M^2}{Q^2} \right) 
\N \\ &&
                      \times \frac{y^2}{1+R(x,Q^2)} 
              \Biggr] {\cal F}_{2}(x,Q^2)
              +~x y(2-y){\cal F}_{3}(x,Q^2) \Biggr\}~. \label{born}
     \end{eqnarray}
     Here, $R(x,Q^2)$ denotes the ratio
     \begin{eqnarray}
      R(x,Q^2) = \frac{\sigma_L}{\sigma_T}
      &=& \left(1+4x^2\frac{M^2}{Q^2}\right) \frac{{\cal F}_{2}(x,Q^2)}
          {2x{\cal F}_{1}(x,Q^2)} - 1~, \label{rqcd}
     \end{eqnarray}
     and the {\sf effective} structure functions 
     ${\cal F}_{l}(x,Q^2),~l = 1 ... 3$ are represented by the structure 
     functions $F_l, G_l$ and $H_l$ via
     \begin{eqnarray} 
      {\cal F}_{1,2}(x,Q^2)
        &=& F_{1,2}(x,Q^2) + 2 |Q_{e}| \left( v_{e} + \lambda a_e \right)
            \chi(Q^2) G_{1,2}(x,Q^2)
\N \\ &&
            +~4 \left( v_{e}^{2} + a_{e}^{2} + 2 \lambda v_e a_e \right)
             \chi^2(Q^2) H_{1,2}(x,Q^2)~, \label{f112} \\
      x{\cal F}_3(x,Q^2)
        &=& -2~\mbox{\rm{sign}}(Q_e) 
            \Biggl\{
                    |Q_{e}|\left( a_{e} + \lambda v_e \right) 
                    \chi(Q^2) xG_{3}(x,Q^2)
\N \\ &&
                   +\left[2v_{e} a_e + \lambda \left(v_e^2 + a_e^2 \right) 
                    \right]\chi^2(Q^2) xH_{3}(x,Q^2)
           \Biggr\}~. \label{f123}  
     \end{eqnarray}
     Here, $Q_e=-1,~a_e=1$ in case of electrons and 
     \begin{eqnarray}
       \lambda&=&\xi \, \mbox{\rm{sign}}(Q_e)~, \label{laxi} \\
        v_e&=&1-4 \sin^{2}\theta_{W}^{\mathrm{eff}}~,\\
        \chi (Q^2) &=&
       {G_\mu \over\sqrt{2}}{M_{Z}^{2} \over{8\pi\alpha(Q^2)}}{Q^2 
          \over{Q^2+M_{Z}^{2}}}~, \label{chiq}
     \end{eqnarray}
     with $\xi$ the electron polarization, $\theta_{W}^{\mathrm{eff}}$ the
     effective weak mixing angle, $G_\mu$ the Fermi constant and $M_Z$ the 
     $Z$--boson mass. 
%%%%%%%%%%%%%%%%%%%%%%%%%%%%%%%%%%%%%%%%%%%%%%%%%%%%%%%%%%%%%%%%%%%%%%%%%%%%%%%
   \subsection{\bf\boldmath The Parton Model} 
    \label{SubSec-DISParton}
\begin{figure}[H]
     \begin{center}
       \includegraphics[angle=0, width=8.0cm]{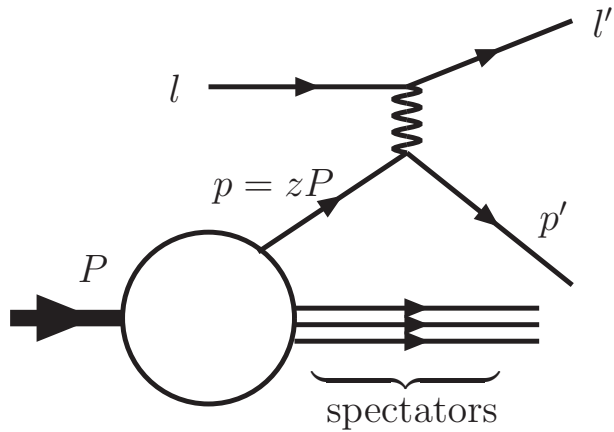}
     \end{center}
     \begin{center} 
      \caption{\sf Deeply inelastic electron-proton scattering
              in the parton model.}
      \label{partmod}
      \noindent
      \small
     \end{center}
     \normalsize
     \end{figure}

%%%%%%%%%%%%%%%%%%%%%%%%%%%%%%%%%%%%%%%%%%%%%%%%%%%%%%%%%%%%%%%%%%%%%%%%%%%%%%%
     The structure functions (\ref{hadrontens2}) depend on two kinematic 
     variables, $x$ and $Q^2$. Based on an analysis using current algebra, 
     Bjorken predicted scaling of the structure functions,
     cf. \cite{Bjorken:1968dy},
     \begin{eqnarray}
      \lim_{\{Q^2,~\nu\}~\rightarrow~\infty,~x=const.} 
            F_{(2,L)}(x,Q^2)=F_{(2,L)}(x)~.
      \label{scaling}
     \end{eqnarray}
     This means that in the Bjorken limit $\{Q^2,~\nu~\}\rightarrow~\infty$, 
     with $x$ fixed, the structure functions depend on the ratio $Q^2/\nu$ 
     only. Soon after this prediction, approximate scaling was observed 
     experimentally in 
     electron-proton collisions at ${\sf SLAC}~(1968)$, 
     \cite{Coward:1967au,*Panofsky:1968pb,*Bloom:1969kc,*Breidenbach:1969kd},
     cf. also \cite{Kendall:1991np,*Taylor:1991ew,*Friedman:1991nq}~\footnote{The results obtained at ${\sf DESY}$, \cite{Albrecht:1969zyxAlbrecht:1969qmxAlbrecht:1969zb}, pointed in the same direction, but were less decisive, because not as large values of $Q^2$ as at ${\sf SLAC}$ could be reached.}. Similar to the 
     $\alpha-$particle scattering experiments by Rutherford in $1911$, 
     \cite{Rutherford:1911zz}, the cross section remained large at high 
     momentum transfer $Q^2$, a behavior which is known from point--like 
     targets. This was found in contradiction to the expectation that the
     cross 
     section should decrease rapidly with increasing $Q^2$, since the size of 
     the proton had been determined to be about $10^{-13}~$cm with a smooth 
     charge distribution, \cite{Mcallister:1956ng,*Schopper:1961,*Hofstadter:1963}. 
     However, only in rare cases
     a single proton was detected in the final state, instead it consisted of a
     large number of hadrons. A proposal by Feynman contained the correct
     ansatz. To 
     account for the observations, he introduced the parton model, 
     \cite{Feynman:1969wa,Feynman:1969ej}, cf. also \cite{Bjorken:1969ja,Feynman:1973xc,Roberts:1990ww,Reya:1979zk,Kogut:1972di,*Yan:1976np}. He assumed the proton as an 
     extended object, consisting of several point-like particles, the partons. 
     They are bound together by their mutual interaction and behave like free 
     particles during the interaction with the highly virtual photon in the 
     Bjorken-limit~\footnote{Asymptotic freedom, which was discovered later, is instrumental for this property.}.
     One arrives at the picture of the proton being ``frozen'' while the 
     scattering takes place. The electron scatters elastically off the partons 
     and this process does not interfere with the other partonic states, 
     the ``spectators''. The DIS cross section is then given by the incoherent 
     sum over the individual virtual electron--parton cross sections. Since no 
     information on the particular proton structure is known, Feynman described
     parton $i$ by the parton distribution function (PDF) $f_i(z)$. It gives 
     the probability to find parton $i$ in the ``frozen'' proton, carrying the 
     fraction $z$ of its momentum. Figure \ref{partmod} shows a schematic 
     picture of the parton model in Born approximation. The in-- and outgoing 
     parton momenta are denoted by $p$ and $p'$, respectively. 
%%%%%%%%%%%%%%%%%%%%%%%%%%%%%%%%%%%%%%%%%%%%%%%%%%%%%%%%%%%%%%%%%%%%%%%%%%
%%%%%%%%%%%%%%%%%%%%%%%%%%%%%%%%%%%%%%%%%%%%%%%%%%%%%%%%%%%%%%%%%%%%%%%%%
     \noindent
     Similar to the scaling variable $x$, one defines the partonic scaling 
     variable $\tau$, 
     \begin{eqnarray}
      \tau\equiv\frac{Q^2}{2 p.q}~. \label{taudef}
     \end{eqnarray}
     It plays the same role as the Bjorken-variable, but for
     the partonic sub-process. 
     In the collinear parton model~\footnote{For other parton models, as the covariant parton model, cf. \cite{Nash:1971aw,*Landshoff:1971xb,Jackson:1989ph,*Roberts:1996ub,Blumlein:1996tp,Blumlein:2003wk}.}, 
     which is applied throughout this thesis, $p=zP$ holds, i.e., the momentum 
     of the partons is taken to be collinear to the proton momentum. 
     From (\ref{taudef}) one obtains
     \begin{eqnarray}
      \tau z=x~.
     \end{eqnarray}
     Feynman's original parton model, referred to as the naive parton model, 
     neglects the mass of the partons and enforces the strict correlation
     \begin{eqnarray}
       \delta\left(\frac{q.p}{M}-\frac{Q^2}{2M}\right)~,\label{feyncor}
     \end{eqnarray}
     due to the {\sf experimentally observed scaling behavior}, which leads to 
     $z=x$. The naive parton model then assumes, in 
     accordance with the quark hypothesis, \cite{Bjorken:1969ja,GellMann:1964nj,Zweig:1964jf}, 
     that the proton is made up of three valence quarks, two up and one down 
     type, cf. e.g. \cite{Close:1979bt}. This conclusion was generally 
     accepted only several years after the introduction of the parton model, 
     when various experiments had verified its predictions.

     Let us consider a simple example, which reproduces the naive parton model 
     at {\sf LO} and incorporates already some aspects of the improved parton 
     model. The latter allows virtual quark states (sea-quarks) and gluons as 
     partons as well. In the QCD--improved parton model, cf. \cite{Roberts:1990ww,Reya:1979zk,Kogut:1972di,*Yan:1976np}, besides the $\delta$-distribution,
     (\ref{feyncor}), a 
     function ${\cal W}^i_{\mu\nu}(\tau,Q^2)$ contributes to the hadronic 
     tensor. It is called partonic tensor and given by the hadronic tensor, Eq.
     (\ref{hadrontens4}), replacing the hadronic states by 
     partonic states $i$. The basic assumption is that the hadronic tensor can 
     be factorized into the PDFs and the partonic tensor, cf. e.g. 
     \cite{Amati:1978wx,*Libby:1978qf,*Libby:1978bx,*Mueller:1978xu,*Collins:1981ta,*Bodwin:1984hc,*Collins:1985ue,Collins:1987pm}.
     The PDFs are non-perturbative quantities and have 
     to be extracted from experiment, whereas the partonic tensors are 
     calculable perturbatively. A more detailed discussion of this using the 
     LCE is given in Section~\ref{SubSec-DISComptLCE}. The hadronic tensor
     reads, cf. \cite{Mulders:1996}, 
     \begin{eqnarray}
      W_{\mu\nu}(x,Q^2)=\frac{1}{4\pi}\sum_i \int_0^1 dz
                        \int_0^1 d\tau \left(f_i(z)+f_{\overline{i}}(z)
                        \right){\cal W}_{\mu\nu}^i(\tau,Q^2)
                        \delta(x-z \tau)~. \label{hadrontens8}
     \end{eqnarray}
     Here, the number of partons and their respective type are not yet
     specified
     and we have included the corresponding PDF of the respective anti-parton, 
     denoted by $f_{\overline{i}}(z)$. Let us assume that the electromagnetic 
     parton current takes the simple form 
     \begin{eqnarray} 
      \bra{i}j^i_{\mu}(\tau)\ket{i}=-ie_i\overline{u}^i\gamma_{\mu}u^i~,
     \end{eqnarray}
     similar to the leptonic current, (\ref{mfi}). Here $e_i$ is the electric 
     charge of parton $i$. At {\sf LO} one finds
     \begin{eqnarray}
      {\cal W}^i_{\mu\nu}(\tau,Q^2)=\frac{2\pi e_i^2}{q. p^i}\delta(1-\tau)
                             \Bigl[2p^i_{\mu}p^i_{\nu}+p^i_{\mu}q_{\nu}
                               +p^i_{\nu}q_{\mu}
                             -g_{\mu\nu}q.p^i\Bigr]~.\label{partontensLO}
     \end{eqnarray}
     The $\delta$-distribution in (\ref{partontensLO}), together with the 
     $\delta$-distribution in (\ref{hadrontens8}), just reproduces Feynman's
     assumption of the naive parton model, $z=x$. Substitution of 
     (\ref{partontensLO}) into the general expression for the hadronic tensor 
     (\ref{hadrontens2}) and projecting onto the structure functions yields
     \begin{eqnarray} 
      F_L(x,Q^2)&=&0~,\N\\
      F_2(x,Q^2)&=&x\sum_ie^2_i\left(f_i(x)+f_{\overline{i}}(x)\right)~.
      \label{resfeynLO}
     \end{eqnarray} 
     This result, at {\sf LO}, is the same as in the naive parton model.
     It predicts
     \begin{itemize}
      \item the Callan-Gross relation, cf. \cite{Callan:1969uq},
           \begin{eqnarray}
            F_L(x,Q^2)=F_2(x,Q^2)-2xF_1(x,Q^2)=0~.
           \end{eqnarray}
      \item the structure functions are scale-independent.
     \end{itemize}
     These findings were a success of the parton model, 
     since they reproduced the general behavior of the data 
     as observed by the ${\sf MIT/SLAC}$ experiments.

     Finally, we present for completeness the remaining structure functions 
     $G_{2,3}$ and  $H_{2,3}$ at the Born level for the complete neutral 
     current,
     cf. Eq.~(\ref{born}),
     \begin{eqnarray}
      G_2(x,Q^2)&=&x\sum_i|e_i|v_i\left(f_i(x)+f_{\overline{i}}(x)\right), \\
      H_2(x,Q^2)&=&x\sum_i\frac{1}{4}\left(v_i^2+a_i^2\right)
                          \left(f_i(x)+f_{\overline{i}}(x)\right), \\
      xG_3(x,Q^2)&=&x\sum_i|e_i|a_i\left(f_i(x)-f_{\overline{i}}(x)\right), \\
      xH_3(x,Q^2)&=&x\sum_i\frac{1}{2}v_i a_i
                     \left(f_i(x)-f_{\overline{i}}(x)\right), 
     \end{eqnarray}
     with $a_i=1$ and
     \begin{eqnarray}
      v_i&=&1-4 |e_i| \sin^{2}\theta_{W}^{\mathrm{eff}}.
     \end{eqnarray}
%%%%%%%%%%%%%%%%%%%%%%%%%%%%%%%%%%%%%%%%%%%%%%%%%%%%%%%%%%%%%%%%%%%%%%%%%%
     \subsubsection{Validity of the Parton Model}
      \label{SubSubSec-DISValpart}
%%%%%%%%%%%%%%%%%%%%%%%%%%%%%%%%%%%%%%%%%%%%%%%%%%%%%%%%%%%%%%%%%%%%%%%%%%
       The validity of the parton picture can be justified by considering an 
       impulse approximation of the scattering process as seen from a certain 
       class of reference frames, in which the proton momentum is taken to be 
       very large  ($P_\infty$-frames). Two things happen to the proton when 
       combining this limit with the Bjorken--limit: The internal interactions 
       of its partons are time dilated, and it is Lorentz contracted in the 
       direction of the collision. As the cms energy increases, the parton 
       lifetimes are lengthened and the time it takes the electron to interact
       with the proton is shortened. Therefore the condition for the validity 
       of the parton model is given by, cf. \cite{Drell:1970yt,Bjorken:1969ja},
       \begin{eqnarray}
        \frac{\tau_{\rm int}}{\tau_{\rm life}} \ll 1~.\label{cond}
       \end{eqnarray}
       Here $\tau_{\rm int}$ denotes the interaction time and 
       $\tau_{\rm life}$ the average life time of a parton. If (\ref{cond}) 
       holds, the proton will be in a single virtual state characterized by a 
       certain number of partons during the entire interaction time. This 
       justifies the assumption that parton $i$ carries a definite momentum 
       fraction $z_i$, $0 \le z_i \le 1$, of the proton in the cms.
       This parton model is also referred to as 
       collinear parton model, since the proton is assumed to consist out 
       of a stream of partons with parallel momenta. Further $\sum_i z_i =1$ 
       holds. In order to derive the fraction of times in (\ref{cond}), one 
       aligns the coordinate system parallel to the proton's momentum. Thus one
       obtains in the limit ${P^2_3} \gg M^2$, \cite{Blumlein:1997},
       \begin{eqnarray}
        P=\left(\sqrt{P_3^2+M^2};0,0,P_3\right) \approx 
          \left(P_3+\frac{M^2}{2\cdot P_3};0,0,P_3\right)~. \label{Pnuc}
       \end{eqnarray}
       The photon momentum can be parametrized by
       \begin{eqnarray}
        q=(q_{0};q_{3},\vec q_{\bot})~,  \label{qinf}
       \end{eqnarray}
       where $\vec q_{\bot}$ denotes its transverse momentum with respect to 
       the proton. By choosing the cms of the initial states as 
       reference and requiring that $\nu M$ and $q^2$ approach a limit 
       independent of ${ P_3}$ as ${P_3} \rightarrow \infty$, one finds for the
       characteristic interaction time scale, using an (approximate) 
       time--energy uncertainty relation,
       \begin{eqnarray}
        \tau_{\rm int}  &\simeq&  \frac{1}{q_0}=\frac{4P_3x}{Q^2(1-x)}~. 
           \label{tauint}
       \end{eqnarray}
       The life time of the individual partons is estimated accordingly to be 
       inversely proportional to the energy fluctuations of the partons around 
       the average energy $E$
       \begin{eqnarray}
        \tau_{\rm life} \simeq \frac{1}{\sum_{i}E_{i}-E} \label{taulife}~.
       \end{eqnarray}
       Here $E_i$ denote the energies of the individual partons. After 
       introducing the two-momentum $\vec{k}_{\bot i}$ of the partons 
       perpendicular to the direction of motion of the proton as given in 
       (\ref{Pnuc}), a simple calculation yields,~cf.~\cite{Blumlein:1997},
       \begin{eqnarray}
        \frac{\tau_{\rm int}}{\tau_{\rm life}}
              &=&
                 \D\frac{2x}{Q^2(1-x)}\left(
                     \sum_{i}\frac{(m_i^2+k_{\bot i}^2)}{z_i}-M^2
                                     \right)~, \label{cond2}
       \end{eqnarray}
       where $m_i$ denotes the mass of the $i$-th parton. This expression is 
       independent of $P_3$. The above procedure allows therefore to estimate 
       the probability of deeply inelastic scattering to occur independently of
       the large momentum of the proton. Accordingly, we consider now the case
       of two 
       partons with momentum fractions $x$~and~$1-x$~ and equal perpendicular 
       momentum, neglecting all masses. One obtains
       \begin{eqnarray}
        \frac{\tau_{\rm int}}{\tau_{\rm life}}\approx 
          \frac{2k_{\bot}^2}{Q^2(1-x)^2}~.
       \end{eqnarray}
       This example leads to the conclusion, that deeply 
       inelastic scattering probes single partons if the virtuality of the 
       photon is much larger than the transverse momenta squared of the
       partons and Bjorken-$x$ is neither close to one nor zero. In the 
       latter case, $xP_3$ would be the large momentum to be considered.
       If one does not neglect the quark masses,
       one has to adjust this picture, as will be described in 
       Section~\ref{SubSec-HQFlav}.
%%%%%%%%%%%%%%%%%%%%%%%%%%%%%%%%%%%%%%%%%%%%%%%%%%%%%%%%%%%%%%%%%%%%%%%%%%%%%%%
   \subsection{\bf\boldmath The Light--Cone Expansion}
    \label{SubSec-DISComptLCE}
\begin{figure}[H]
       \begin{center}
        \includegraphics[angle=0, width=10.0cm]{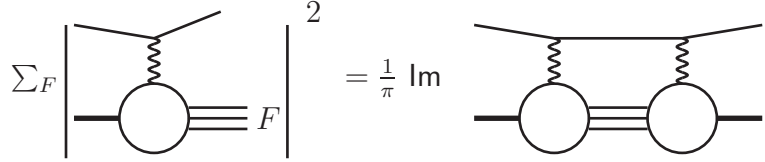}
       \end{center}
       \begin{center} 
        \caption{\sf Schematic picture of the optical theorem.}
       \label{picopt}
       \noindent
       \small
      \end{center}
      \normalsize
     \end{figure}

%%%%%%%%%%%%%%%%%%%%%%%%%%%%%%%%%%%%%%%%%%%%%%%%%%%%%%%%%%%%%%%%%%%%%%%%%%%%%%%
     In quantum field theory one usually considers time-ordered products, 
     denoted by ${\sf T}$, rather than a commutator as it appears in the 
     hadronic tensor in Eq.~(\ref{hadrontens4}). The hadronic 
     tensor can be expressed as the imaginary part of the forward Compton 
     amplitude for virtual gauge boson--nucleon scattering,
     $T_{\mu\nu}(q,P)$. The optical theorem, depicted graphically in Figure 
     \ref{picopt}, yields 
     \begin{eqnarray}
      W_{\mu\nu}(q,P)   &=&\frac{1}{\pi} {\sf Im}~T_{\mu\nu}(q,P)
                            \label{opttheo}~,
     \end{eqnarray}
     where the Compton amplitude is given by, cf.~\cite{Blumlein:1999sc},
     \begin{eqnarray} 
      T_{\mu \nu}(q,P)&=&i\int d^4\xi~\exp(iq\xi)\bra{P}
                            {\sf T} J_{\mu}(\xi) 
                            J_{\nu}(0)\ket{P}~.
                            \label{comptontensor}
     \end{eqnarray}
%%%%%%%%%%%%%%%%%%%%%%%%%%%%%%%%%%%%%%%%%%%%%%%%%%%%%%%%%%%%%%%%%%%%%%%%%%%%%%%
%%%%%%%%%%%%%%%%%%%%%%%%%%%%%%%%%%%%%%%%%%%%%%%%%%%%%%%%%%%%%%%%%%%%%%%%%%%%%%%
     By applying the same invariance and conservation conditions as for the 
     hadronic tensor, the Compton amplitude can be expressed in the unpolarized
     case by two amplitudes $T_L(x,Q^2)$ and $T_2(x,Q^2)$. It is then given by
     \begin{eqnarray}
      T_{\mu\nu}(q,P)=&&\frac{1}{2x}\left( 
                                         g_{\mu\nu}
                                        +\frac{q_{\mu}q_{\nu}}{Q^2}
                                 \right)
                                 T_{L}(x,Q^2)
\N\\ &+&
                     \frac{2x}{Q^2}\left(
                                           P_{\mu}P_{\nu}
                                          +\frac{ q_{\mu}P_{\nu}
                                                 +q_{\nu}P_{\mu}
                                                }{2x}
                                         -\frac{Q^2}{4x^2}g_{\mu\nu}
                                    \right)
                      T_{2}(x,Q^2)~.
                       \label{comptontens}
     \end{eqnarray}
     Using translation invariance, one can show that (\ref{comptontensor}) is 
     crossing symmetric under $q~\rightarrow~-q$,
     cf. \cite{Jackiw:1972ee,Blumlein:1996vs},
     \begin{eqnarray}
      T_{\mu\nu}(q,P)=T_{\mu\nu}(-q,P)~,\label{crossing1}
     \end{eqnarray}
     with $q \rightarrow -q$ being equivalent to 
     $\nu,x \rightarrow (-\nu),(-x)$. The corresponding relations for the 
     amplitudes are then obtained by considering (\ref{comptontens})
     \begin{eqnarray}
      T_{(2,L)}(x,Q^2)&=&T_{(2,L)}(-x,Q^2)~.\label{crossing2}
     \end{eqnarray}
     By (\ref{opttheo}) these amplitudes relate to the structure functions 
     $F_L$ and $F_2$ as
     \begin{eqnarray}
      F_{(2,L)}(x,Q^2)&=&\frac{1}{\pi}{\sf Im}~T_{(2,L)}
                       (x,Q^2)~.\label{structamp}
     \end{eqnarray}
     Another general property of the Compton amplitude is that $T_L$ and $T_2$ 
     are real analytic functions of $x$ at fixed $Q^2$,
     cf.~\cite{Jaffe:1985je}, i.e.
     \begin{eqnarray}
      T_{(2,L)}(x^*,Q^2)&=&T^*_{(2,L)}(x,Q^2)~.\label{realan}
     \end{eqnarray}
     Using this description one can perform the LCE,
     \cite{Wilson:1969zs,*Zimmermann:1970,*Frishman:1971qn,*Brandt:1970kg},
     or the cut--vertex method in the time--like case, 
     \cite{Frishman:1973pp,Geyer:1977gv,Mueller:1981sg},
     respectively, and derive 
     general properties of the moments of the structure functions as will be 
     shown in the subsequent Section. A technical aspect which has been proved 
     very useful is to work in Mellin space rather than in $x$--space. The 
     $N$th Mellin moment of a function $f$ is defined through the integral
     \begin{eqnarray}
      \M[f](N)\equiv\int_0^1 dz~z^{N-1}f(z)~. \label{Mellintrans} 
     \end{eqnarray} 
     This transform diagonalizes the Mellin--convolution $f\otimes~g$ of two 
     functions $f,~g$
     \begin{eqnarray}
      [f \otimes g](z) = \int_0^1 dz_1 \int_0^1 dz_2~~ \delta(z - z_1 z_2) 
      ~f(z_1) g(z_2)~. \label{Mellinconz}
     \end{eqnarray}
     The convolution (\ref{Mellinconz}) 
     decomposes into a simple product of the Mellin-transforms 
     of the two functions,
     \begin{eqnarray}
      \M[f \otimes g](N) = \M[f](N)\M[g](N)~. \label{MellinconN}
     \end{eqnarray}
     In Eqs. (\ref{Mellintrans}, \ref{MellinconN}), $N$ is taken to be an 
     integer. However, later on one may perform an analytic continuation to 
     arbitrary complex values of $N$,~\cite{Blumlein:2000hw,*Blumlein:2005jg}.
     Note that it is enough to know all even {\sf or} odd integer moments 
     -- as is the case for inclusive DIS -- of the functions $f,~g$ to perform 
     an analytic continuation to arbitrary complex values $N\in\mathbb{C}$,
     \cite{Carlson:thesis,*Titchmarsh:1939}. 
     Then Eq.~(\ref{Mellinconz}) can be 
     obtained from the relation for the moments, (\ref{MellinconN}), by an 
     inverse Mellin--transform. Hence in this case the $z$-- and $N$--space 
     description are equivalent, which we will frequently use later on.
%%%%%%%%%%%%%%%%%%%%%%%%%%%%%%%%%%%%%%%%%%%%%%%%%%%%%%%%%%%%%%%%%%%%%%%%%%%%%%
     \subsubsection{Light--Cone Dominance}
      \label{SubSubSec-LCEdomLCE}
%%%%%%%%%%%%%%%%%%%%%%%%%%%%%%%%%%%%%%%%%%%%%%%%%%%%%%%%%%%%%%%%%%%%%%%%%%%%%%
       It can be shown that in the Bjorken limit,  
       $Q^2\rightarrow \infty,~ \nu \rightarrow \infty$, $x$ fixed, the 
       hadronic tensor is dominated by its contribution near the light--cone, 
       i.e. by the values of the integrand in (\ref{hadrontens4}) at 
       $\xi^2 \approx 0$,~cf.~\cite{Wilson:1969zs,*Zimmermann:1970,*Frishman:1971qn,*Brandt:1970kg}. 
       This can be understood by considering the infinite momentum frame, see 
       Section~\ref{SubSubSec-DISValpart},
       \begin{eqnarray}
        P&=&(P_3;0,0,P_3)~,\\
        q&=&\Bigl(\frac{\nu}{2 P_3};\sqrt{Q^2},0,\frac{-\nu}{2 P_3}\Bigr)~,\\
        P_3 &\approx& \sqrt{\nu} \rightarrow \infty~.
       \end{eqnarray}
       According to the Riemann--Lebesgue theorem, the integral in 
       (\ref{hadrontens4}) is dominated by the region where $q.\xi \approx 0$ 
       due to the rapidly oscillating exponential $\exp(iq.\xi)$, 
       \cite{Reya:1979zk}. One can now 
       rewrite the dot product as, cf. \cite{Yndurain:1999ui},
       \begin{eqnarray}
        q. \xi=\frac{1}{2}(q^0-q^3)(\xi^0+\xi^3)
                    +\frac{1}{2}(q^0+q^3)(\xi^0-\xi^3)
                    -q^1\xi^1 \label{qdotz}~,
       \end{eqnarray}
       and infer that the condition $q.\xi \approx 0$ in the Bjorken-limit is 
       equivalent to 
       \begin{eqnarray}
        \xi^0 \pm \xi^3 \propto \frac{1}{\sqrt{\nu}}~,\quad 
        \xi^1  \propto \frac{1}{\sqrt{\nu}}~, \label{limit} 
       \end{eqnarray}
       which results in
       \begin{eqnarray}
        \xi^2~\approx~0~,
       \end{eqnarray}
       called {\sf light--cone dominance}: for DIS in the 
       Bjorken-limit the dominant contribution to the hadronic tensor 
       $W_{\mu\nu}(q,P)$ and the Compton Amplitude comes from the region where 
       $\xi^2 \approx 0$. 
       
       This property allows to apply the LCE of the current--current 
       correlation in Eq.~(\ref{hadrontens4}) and for the time ordered product 
       in Eq.~(\ref{comptontensor}), respectively. In the latter case it reads 
       for scalar currents, cf. \cite{Wilson:1969zs,*Zimmermann:1970,*Frishman:1971qn,*Brandt:1970kg},
       \begin{eqnarray}
        \lim_{\xi^2 \rightarrow 0} {\sf T} J(\xi),J(0)
        \propto \sum_{i,N,\tau} \overline{C}^N_{i,\tau}(\xi^2,\mu^2) \xi_{\mu_1}... 
        \xi_{\mu_N} O_{i,\tau}^{\mu_1...\mu_N}(0,\mu^2)~.\label{lighex}
       \end{eqnarray}
       The $O_{i,\tau}(\xi,\mu^2)$ are local operators which are finite as 
       $\xi^2 \rightarrow 0$. The singularities which appear for the product of
       two operators as their arguments become equal are shifted to the 
       $c$-number coefficients $\overline{C}^N_{i,\tau}(\xi^2,\mu^2)$, the 
       Wilson coefficients, and can therefore be treated separately. 
       In Eq.~(\ref{lighex}), $\mu^2$ is the factorization scale describing 
       at which point the separation between the perturbative and 
       non--perturbative contributions takes place. The summation index
       $i$ runs over the set of allowed operators in the model, while the sum 
       over $N$ extends to infinity. Dimensional analysis shows that the degree
       of divergence of the functions $\overline{C}_{i,\tau}^N$ as 
       $\xi^2 \rightarrow 0$ is given by 
       \begin{eqnarray}
        \overline{C}^{N}_{i,\tau}(\xi^2,\mu^2) \propto
                \Biggl(\frac{1}{\xi^2}\Biggr)^{-\tau/2+d_J}~. \label{Cdiv}
       \end{eqnarray}
       Here, $d_J$ denotes the canonical dimension of the current $J(\xi)$ and 
       $\tau$ is the twist of the local operator $O_{i,\tau}^{\mu_1..\mu_N}(\xi,\mu^2)$, which is defined by,~cf.~\cite{Gross:1971wn},
       \begin{eqnarray}
         \tau\equiv D_O-N~. \label{twist}
       \end{eqnarray}
       $D_O$ is the canonical (mass) dimension of 
       $O_{i,\tau}^{\mu_1..\mu_N}(\xi,\mu^2)$ and $N$ is called its spin. From 
       (\ref{Cdiv}) one can infer that the most singular coefficients are those
       related to the operators of lowest twist, 
       i.e. in the case of the LCE of the 
       electromagnetic current (\ref{current}), twist $\tau=2$. 
       The contributions due to higher twist operators are
       suppressed by factors of $(\overline{\mu}^2/Q^2)^k$, with 
       $\overline{\mu}$ a typical hadronic mass scale of $O(1~\GeV)$. In a wide
       range of phase--space it is thus sufficient to consider the leading 
       twist contributions only, which we will do in the following and omit the
       index $\tau$. 
%%%%%%%%%%%%%%%%%%%%%%%%%%%%%%%%%%%%%%%%%%%%%%%%%%%%%%%%%%%%%%%%%%%%%%%%%%%%%%
     \subsubsection{A Simple Example}
      \label{SubSubSec-DISex}
\begin{figure}[H]
        \begin{center}
         \includegraphics[angle=0, width=4.0cm]{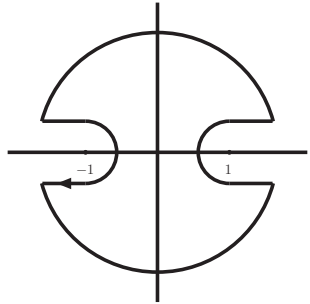}
        \end{center}
        \begin{center} 
         \caption{\sf Integration contour in the complex $x'$-plane.}
         \label{CONTOUR}
         \noindent
         \small
        \end{center}
        \normalsize
       \end{figure}

%%%%%%%%%%%%%%%%%%%%%%%%%%%%%%%%%%%%%%%%%%%%%%%%%%%%%%%%%%%%%%%%%%%%%%%%%%%%%%
       In this Section, we consider a simple example of the LCE applied to the 
       Compton amplitude and its relation to the hadronic tensor, neglecting 
       all Lorentz--indices and model dependence,
       cf. Ref.~\cite{Muta:1998vi,Bardeen:1978yd}.
       The generalization to the 
       case of QCD is straightforward and hence we will already make some 
       physical arguments which apply in both cases. The scalar expressions 
       corresponding to the hadronic tensor and the Compton amplitude are given
       by
       \begin{eqnarray}
        W(x,Q^2)&=&       \frac{1}{2\pi}
                          \int d^4\xi\exp(iq\xi)
                          \bra{P}[J(\xi),
                          J(0)]\ket{P}~, \label{hadrontensEx} \\
        T(x,Q^2)&=&i\int d^4\xi~\exp(iq\xi)\bra{P}
                              {\sf T} J(\xi) 
                              J(0)\ket{P}~.
                              \label{comptontensorEx}
       \end{eqnarray}
       Eq.~(\ref{comptontensorEx}) can be evaluated in the limit 
       $\xi^2 \rightarrow 0$ for twist $\tau=2$ by using the LCE given in 
       Eq.~(\ref{lighex}), where for brevity only one local operator is 
       considered. The coefficient functions in momentum space are defined as
       \begin{eqnarray}
        \int \exp(iq.\xi) \xi_{\mu_1}..\xi_{\mu_N} 
             \overline{C}^{N}(\xi^2,\mu^2) 
          &\equiv&
           -i\Bigl(\frac{2}{-q^2}\Bigr)^{N} q_{\mu_1}...q_{\mu_N}
           C^{N}\Bigl(\frac{Q^2}{\mu^2}\Bigr)  ~. \label{CLExFour}
       \end{eqnarray} 
       The nucleon states act on the composite operators only and the 
       corresponding matrix elements can be expressed as
       \begin{eqnarray}
        \bra{P} O^{\mu_1...\mu_N}(0,\mu^2) \ket{P}
         &=& A^{N}\Bigl(\frac{P^2}{\mu^2}\Bigr) P^{\mu_1}...P^{\mu_N} 
                + \mbox{\rm trace terms}. 
          \label{HadOMEs}
       \end{eqnarray}
       The trace terms in the above equation can be neglected, because due to
       dimensional counting they would give contributions of the order 
       $1/Q^2,~1/\nu$ and hence are irrelevant in the Bjorken--limit. Thus the 
       Compton amplitude reads, cf. e.g. \cite{Roberts:1990ww,Muta:1998vi}, 
       \begin{eqnarray}
        T(x',Q^2)&=&2\sum_{N=0,2,4,..} C^{N}\Bigl(\frac{Q^2}{\mu^2}\Bigr)
                                 A^{N}\Bigl(\frac{P^2}{\mu^2}\Bigr) 
                                 {x'}^N~,~x'=\frac{1}{x} \label{CompAmplEx1}
       \end{eqnarray}
       In (\ref{CompAmplEx1}) only the even moments contribute. This is a 
       consequence of crossing symmetry, Eq.~(\ref{crossing2}), and holds as 
       well in the general case of unpolarized DIS for single photon exchange. 
       In other cases the projection is onto the odd moments. Depending on the 
       type of the observable the series may start at different initial values,
       cf. e.g.~\cite{Blumlein:1996vs,Blumlein:1998nv}. The sum in Eq.
       (\ref{CompAmplEx1}) is convergent in the unphysical region $x\ge~1$ and 
       an analytic continuation to the physical region $0 \leq x \leq 1$ has to
       be performed. Here, one of the assumptions is that scattering 
       amplitudes are analytic in the complex plane except at values of 
       kinematic variables allowing intermediate states to be on mass--shell. 
       This general feature has been proved to all orders in perturbation
       theory, \cite{Landau:1959fi,Bjorken:1959fd}.
       In QCD, it is justified on grounds of the parton model.
       When $\nu \ge Q^2/2M$, i.e. 
       $0\le~x\le~1$, the virtual photon-proton system can produce a physical 
       hadronic intermediate state, so the $T_{(2,L)}(x,Q^2)$ and $T(x,Q^2)$, 
       respectively, have cuts along the positive (negative) real $x$-axis 
       starting from $1$($-1$) and poles at $\nu=Q^2/2M$
       $(x = 1,-1)$.
       The 
       discontinuity along the cut is then just given by (\ref{opttheo}). The 
       Compton amplitude can be further analyzed by applying (subtracted) 
       dispersion relations, cf. \cite{Blumlein:1996vs,Blumlein:1998nv}. 
       Equivalently, one can divide both sides of 
       Eq.~(\ref{CompAmplEx1}) by ${x'}^{m}$ and integrate along the path shown
       in Figure \ref{CONTOUR}, cf. \cite{Muta:1998vi,Mulders:1996}.
%%%%%%%%%%%%%%%%%%%%%%%%%%%%%%%%%%%%%%%%%%%%%%%%%%%%%%%%%%%%%%%%%%%%%%%%%%%%%%%
%%%%%%%%%%%%%%%%%%%%%%%%%%%%%%%%%%%%%%%%%%%%%%%%%%%%%%%%%%%%%%%%%%%%%%%%%%%%%%%
       \noindent
       For the left--hand side of (\ref{CompAmplEx1}) one obtains
       \begin{eqnarray}
        \frac{1}{2\pi i}\oint  dx' \frac{T(x',Q^2)}{{x'}^m} 
        = \frac{2}{\pi} \int_1^{\infty} \frac{dx'}{{x'}^m} {\sf Im} T(x',Q^2)
        = 2\int_0^1 dx~x^{m-2}  W(x,Q^2)~, \label{ContInt1}
       \end{eqnarray}
       where the optical theorem, (\ref{opttheo}), and crossing symmetry, 
       (\ref{crossing2}) have been used.
       The right--hand side of (\ref{CompAmplEx1}) yields
       \begin{eqnarray}
         \frac{1}{\pi i}\sum_{N=0,2,4,..} C^{N}\Bigl(\frac{Q^2}{\mu^2}\Bigr)
                                    A^{N}\Bigl(\frac{P^2}{\mu^2}\Bigr) 
         \oint  dx'~{x'}^{N-m}=2 C^{m-1}\Bigl(\frac{Q^2}{\mu^2}\Bigr) 
                                        A^{m-1}\Bigl(\frac{P^2}{\mu^2}\Bigr)~.
         \label{ContInt2}
       \end{eqnarray}
       Thus from Eqs. (\ref{ContInt1}) and (\ref{ContInt2}) one obtains for the
       moments of the scalar hadronic tensor defined in 
       Eq.~(\ref{hadrontensEx})
       \begin{eqnarray}
         \int_0^1 dx~x^{N-1} W(x,Q^2)&=&
                C^{N}\Bigl(\frac{Q^2}{\mu^2}\Bigr)
                A^{N}\Bigl(\frac{P^2}{\mu^2}\Bigr)~. \label{MomHadEx}
       \end{eqnarray}
%%%%%%%%%%%%%%%%%%%%%%%%%%%%%%%%%%%%%%%%%%%%%%%%%%%%%%%%%%%%%%%%%%%%%%%%%%%%%%
     \subsubsection{The Light--Cone Expansion applied to DIS}
      \label{SubSec-LCE}
%%%%%%%%%%%%%%%%%%%%%%%%%%%%%%%%%%%%%%%%%%%%%%%%%%%%%%%%%%%%%%%%%%%%%%%%%%%%%%
       In order to derive the moment--decomposition of the structure functions
       one essentially has to go through the same steps as in the previous 
       Section. The LCE of the physical forward Compton amplitude 
       (\ref{comptontensor}) at the level of twist $\tau=2$ in the 
       Bjorken--limit is given by, cf. \cite{Bardeen:1978yd,Buras:1979yt},
       \begin{eqnarray}
           T_{\mu\nu}(q,P)\!\!&\rightarrow&\!\!\sum_{i,N}\Biggl\{
\hspace{1mm}
                            \Bigl[
                                    Q^2g_{\mu\mu_1}g_{\nu\mu_2}
                                   +g_{\mu\mu_1}q_{\nu}q_{\mu_2}
                                   +g_{\nu\mu_2}q_{\mu}q_{\mu_1}
                                   -g_{\mu\nu}q_{\mu_1}q_{\mu_2}
                            \Bigr] C_{i,2}\Bigl(N,\frac{Q^2}{\mu^2}\Bigr)
\N\\ && \hspace{-4mm}
                           +\Bigl[
                                    g_{\mu\nu}
                                   +\frac{q_{\mu}q_{\mu}}{Q^2}
                            \Bigr] q_{\mu_1}q_{\mu_2} C_{i,L}
                                     \Bigl(N,\frac{Q^2}{\mu^2}\Bigr)
                                    \Biggr\}
                                  q_{\mu_3}...q_{\mu_N}
                                  \Bigl(\frac{2}{Q^2}\Bigr)^N
                                  \bra{P}O_i^{\mu_1...\mu_N}(\mu^2)\ket{P}~.
            \N\\
                            \label{LCETmunu}
       \end{eqnarray}
       Additionally to Section~\ref{SubSubSec-DISex}, the index $i$ runs over 
       the allowed operators which emerge from the expansion of the 
       product of two electromagnetic currents, Eq.~(\ref{current}). The 
       possible twist--2 operators are given by~\footnote{Here we consider only the spin--averaged case for single photon exchange. Other operators contribute for parity--violating processes, in the polarized case and for transversity, cf. Sections \ref{Sec-POL} and \ref{sec-1}.},~\cite{Geyer:1977gv}, 
       \begin{eqnarray}
        \label{COMP1}
         O^{\sf NS}_{q,r;\mu_1, \ldots, \mu_N} &=& i^{N-1} {\bf S} 
                 [\overline{\psi}\gamma_{\mu_1} D_{\mu_2} \ldots D_{\mu_N} 
                  \frac{\lambda_r}{2}\psi] - {\rm trace~terms}~, \\
        \label{COMP2}
         O^{\sf S}_{q;\mu_1, \ldots, \mu_N} &=& i^{N-1} {\bf S} 
                 [\overline{\psi}\gamma_{\mu_1} D_{\mu_2} \ldots D_{\mu_N}
                  \psi] - {\rm trace~terms}~, \\
        \label{COMP3}
         O^{\sf S}_{g;\mu_1, \ldots, \mu_N} &=& 2 i^{N-2} {\bf S} 
           {\rm \bf Sp}[F_{\mu_1 \alpha}^a D_{\mu_2} \ldots D_{\mu_{N-1}} 
            F_{\mu_N}^{\alpha,a}] - {\rm trace~terms}~.
       \end{eqnarray}
       Here, $\bf S$ denotes the symmetrization operator of the Lorentz indices
       $\mu_1, \ldots, \mu_N$. $\lambda_r$ is the flavor matrix of $SU(n_f)$ 
       with $n_f$ light flavors, $\psi$ denotes the quark field, 
       $F_{\mu\nu}^a$ the gluon field--strength tensor, and 
       $D_{\mu}$ the covariant derivative. The indices $q,~g$ represent the 
       quark-- and gluon--operator, respectively. ${\bf Sp}$ in (\ref{COMP3}) 
       is the color--trace and $a$ the color index in the adjoint 
       representation, cf. Appendix \ref{App-Con}.
       The quark--fields carry color indices in the fundamental
       representation, which have been suppressed. The classification of the 
       composite operators (\ref{COMP1}--\ref{COMP3}) in terms of flavor 
       singlet (${\sf S}$) and non-singlet (${\sf NS}$) refers to their 
       symmetry properties with respect to the flavor group $SU(n_f)$. The 
       operator in Eq.~(\ref{COMP1}) belongs to the adjoint representation of 
       $SU(n_f)$, whereas the operators in Eqs. (\ref{COMP2}, \ref{COMP3}) are 
       singlets under $SU(n_f)$. Neglecting the trace terms, one rewrites the 
       matrix element of the composite operators in terms of its Lorentz 
       structure and the scalar operator matrix elements, cf. \cite{Yndurain:1999ui,Roberts:1990ww},
       \begin{eqnarray}
        \bra{P}O_i^{\mu_1...\mu_N}\ket{P}&=&
                                            A_i\Bigl(N,\frac{P^2}{\mu^2}\Bigr)
                                            P^{\mu_1}...P^{\mu_N}~. 
          \label{NucOMEs}
       \end{eqnarray}
       Eq.~(\ref{LCETmunu}) then becomes
       \begin{eqnarray}
        T_{\mu\nu}(q,P)&=&2\sum_{i,N}\Biggl\{
\hspace{3mm}
                            \frac{2x}{Q^2}\Bigl[
                                    P_{\mu}P_{\nu}
                                   +\frac{P_{\mu}q_{\nu}+P_{\nu}q_{\mu}}{2x}
                                   -\frac{Q^2}{4x^2}g_{\mu\nu}
                            \Bigr] C_{i,2}\Bigl(N,\frac{Q^2}{\mu^2}\Bigr)
\N \\ && \hspace{12mm}
                           +\frac{1}{2x}\Bigl[
                                    g_{\mu\nu}
                                   +\frac{q_{\mu}q_{\mu}}{Q^2}
                            \Bigr] C_{i,L}
                                     \Bigl(N,\frac{Q^2}{\mu^2}\Bigr)
                                    \Biggr\}
                                  \frac{1}{x^{N-1}}
                                  A_i\Bigl(N,\frac{P^2}{\mu^2}\Bigr)~.
         \label{LCETmunu2}
       \end{eqnarray}
       Comparing Eq.~(\ref{LCETmunu2}) with the general Lorentz structure 
       expected for the Compton amplitude, Eq.~(\ref{comptontens}), the 
       relations of the scalar forward amplitudes to the Wilson coefficients 
       and nucleon matrix elements can be read off
       \begin{eqnarray}
       \label{eqT2l}
        T_{(2,L)}(x,Q^2)&=&2\sum_{i,N}\frac{1}{x^{N-1}}
                           C_{i,(2,L)}\Bigl(N,\frac{Q^2}{\mu^2}\Bigr)
                           A_i\Bigl(N,\frac{P^2}{\mu^2}\Bigr)~. \label{T2LMOM}
       \end{eqnarray}
       Eq.~(\ref{eqT2l}) is of the same type as Eq.~(\ref{CompAmplEx1}) and one
       thus obtains for the moments of the structure functions
       \begin{eqnarray}
         F_{(2,L)}(N,Q^2)&=&\M[F_{(2,L)}(x,Q^2)](N) \label{eqMOM} \\
                         &=&\sum_{i}
                           C_{i,(2,L)}\Bigl(\frac{Q^2}{\mu^2},N\Bigr)
                           A_i\Bigl(\frac{P^2}{\mu^2},N\Bigr)~.
                            \label{F2LMOM}
       \end{eqnarray}
       The above equations have already been written in Mellin space, which we 
       will always do from now on, if not indicated otherwise. Eqs. 
       (\ref{T2LMOM}, \ref{F2LMOM}), together with the general structure of the
       Compton amplitude, Eqs. (\ref{comptontens}, \ref{LCETmunu2}), and the 
       hadronic tensor, Eq.~(\ref{hadrontens2}), are the basic equations for 
       theoretical or phenomenological analysis of DIS in the kinematic regions
       where higher twist effects can be safely disregarded. 
       Note that the generalization of these equations to
       electroweak or polarized interactions is straightforward by including 
       additional operators and Wilson coefficients. In order to interpret    
       Eqs. (\ref{T2LMOM}, \ref{F2LMOM}), one uses the fact that the Wilson 
       coefficients $C_{i,(2,L)}$ are independent of the proton state. This is 
       obvious since the wave function of the proton only enters into the 
       definition of the operator matrix elements, cf. Eq.~(\ref{NucOMEs}). In
       order to calculate the Wilson coefficients, the proton state has 
       therefore to be replaced by a suitably chosen quark or gluon state 
       $i$ with momentum $p$. The corresponding partonic tensor is denoted by
       ${\cal W}^i_{\mu\nu}(q,p)$, cf. below Eq.~(\ref{feyncor}),
       with scalar amplitudes
       ${\cal F}^i_{(2,L)}(\tau,Q^2)$. Here $\tau$ is the partonic scaling
       variable defined in Eq.~(\ref{taudef}). The LCE of the electromagnetic 
       current does not change and the replacement only affects the operator 
       matrix elements. The forward Compton amplitude for
       photon--quark (gluon) 
       scattering corresponding to ${\cal W}^i_{\mu\nu}(q,p)$ can be calculated
       order by order in perturbation theory, provided the scale $Q^2$ is large
       enough for the strong coupling constant to be small. In the same manner,
       the contributing operator matrix elements with external partons may be 
       evaluated. Finally, one can read off the Wilson coefficients from the 
       partonic equivalent of Eq.~(\ref{T2LMOM})~\footnote{Due to the optical theorem, one may also obtain the Wilson coefficients by calculating the inclusive hard scattering cross sections of a virtual photon with a quark(gluon) using the standard Feynman--rules and phase--space kinematics.}. By identifying the 
       nucleon OMEs (\ref{NucOMEs}) with the PDFs, one obtains the QCD improved
       parton model. At ${\sf LO}$ it coincides with the naive parton model, 
       which we described in Section~\ref{SubSec-DISParton}, as can be inferred
       from the discussion below Eq.~(\ref{hadrontens8}). The improved parton 
       model states that in the Bjorken limit at the level of twist $\tau=2$ 
       the unpolarized nucleon structure functions $F_i(x,Q^2)$ are obtained in
       Mellin space as products of the universal parton densities 
       $f_i(N,\mu^2)$ with process--dependent Wilson coefficients 
       $C_{i,(2,L)}(N,Q^2/\mu^2)$
       \begin{eqnarray}
        F_{(2,L)}(N,Q^2) = 
                         \sum_i C_{i,(2,L)}\Bigl(N,\frac{Q^2}{\mu^2}\Bigr) 
                         f_i(N,\mu^2)     \label{STR}
       \end{eqnarray}
       to all orders in perturbation theory. This property is also formulated 
       in the factorization theorems,~\cite{Amati:1978wx,*Libby:1978qf,*Libby:1978bx,*Mueller:1978xu,*Collins:1981ta,*Bodwin:1984hc,*Collins:1985ue},
       cf. also \cite{Collins:1987pm},
       where it is essential that an inclusive, 
       infrared--safe cross section is considered, \cite{Bassetto:1984ik}.
       We have not yet dealt with the question of 
       how renormalization is being performed. However, we have already 
       introduced the scale $\mu^2$ into the right--hand side of Eq. 
       (\ref{STR}). This scale is called factorization scale. It describes a
       mass scale at which the separation of the structure functions into the 
       perturbative hard scattering coefficients $C_{i,(2,L)}$ and the 
       non--perturbative parton densities $f_i$ can be performed. This choice 
       is arbitrary at large enough scales and  the physical structure 
       functions do not depend on it. This {\sf independence} is used in turn 
       to establish the corresponding renormalization group equation, 
       \cite{Stuckelberg:1951gg,*GellMann:1954fq,*Bogolyubov:1980nc,Symanzik:1970rt,*Callan:1970yg}, which describes the 
       scale--evolution of the Wilson coefficients, parton densities and 
       structure functions w.r.t. to $\mu^2$ and $Q^2$, 
       cf. Refs.~\cite{Buras:1979yt,Reya:1979zk,Altarelli:1989ue,Owens:1992hd,Roberts:1990ww,Blumlein:1995cm}
       and Section~\ref{SubSec-DISEvol}.

       These evolution equations then predict scaling violations and are used 
       to analyze experimental data in order to unfold the twist--2 parton 
       distributions at some scale $Q^2_0$, together with the
       QCD--scale $\Lambda_{\rm QCD}$, cf. \cite{Duke:1984ge,Altarelli:1989ue,Bethke:1992gh}. 

       Before finishing this Section, we describe the quantities appearing in 
       Eq.~(\ref{STR}) in detail. Starting from the operators defined in 
       Eqs. (\ref{COMP1})--(\ref{COMP3}), three types of parton densities are 
       expected. Since the question how heavy quarks are treated in this 
       framework will be discussed in Section~\ref{Sec-HQDIS}, we write the 
       following equations for $n_f$ light flavors in massless QCD. The gluon 
       density is denoted by $G(n_f,N,\mu^2)$ and multiplies the gluonic Wilson
       coefficients $C_{g,(2,L)}(n_f,N,Q^2/\mu^2)$, which describe the 
       interaction of a gluon with a photon and emerge for the first time at 
       $O(\alpha_s)$.
       Each quark and its anti--quark have a parton density, denoted 
       by $f_{k(\overline{k})}(n_f,N,\mu^2)$. These are grouped together into 
       the flavor singlet combination $\Sigma(n_f,N,\mu^2)$ and a non--singlet 
       combination $\Delta_k(n_f,N,\mu^2)$ as follows 
       \begin{eqnarray}
           \Sigma(n_f,N,\mu^2) &=& \sum_{l=1}^{n_f} \Big[
           f_l(n_f,N,\mu^2) +  f_{\bar l}(n_f,N,\mu^2) \Big]~, 
             \label{SIGMAPDF}\\
           \Delta_k(n_f,N,\mu^2) &=& 
             f_k(n_f,N,\mu^2) +  f_{\bar k}(n_f,N,\mu^2)
            -\frac{1}{n_f}\Sigma(n_f,N,\mu^2)~. \label{DELTAPDF}
       \end{eqnarray}
       The distributions multiply the quarkonic Wilson coefficients 
       $C^{\sf S,NS}_{q,(2,L)}(n_f,N,Q^2/\mu^2)$, which describe the hard 
       scattering of a photon with a light quark. The complete factorization 
       formula for the structure functions is then given by
       \begin{eqnarray}
         F_{(2,L)}(n_f,N,Q^2)=\frac{1}{n_f} \sum_{k=1}^{n_f} e_k^2 
         \Biggl[&& \Sigma(n_f,N,\mu^2)
                  C_{q,(2,L)}^{\sf S}\Big(n_f,N,\frac{Q^2}{\mu^2}\Big)
\N \\ 
                &+&G(n_f,N,\mu^2)
                 C_{g,(2,L)}^{\sf S}\Big(n_f,N,\frac{Q^2}{\mu^2}\Big) 
\N \\ 
                &+&n_f \Delta_k(n_f,N,\mu^2)
                 C_{q,(2,L)}^{\sf NS}\Big(n_f,N,\frac{Q^2}{\mu^2}\Big)\Biggr]~.
          \N\\ \label{FACT2}
       \end{eqnarray}
       Note, that one usually splits the quarkonic ${\sf S}$ contributions 
       into a ${\sf NS}$ and pure--singlet (${\sf PS}$) part via 
       ${\sf S~=PS+NS}$. The perturbative expansions of the Wilson coefficients
       read
       \begin{eqnarray}
        C_{g,(2,L)}^{\sf S}\Big(n_f,N,\frac{Q^2}{\mu^2}\Big)&=&
         \sum_{i=1}^{\infty}a_s^i
            C_{g,(2,L)}^{(i),{\sf S}}\Big(n_f,N,\frac{Q^2}{\mu^2}\Big)~,
\label{Cg2Lpert}
 \\
        C_{q,(2,L)}^{\sf PS}\Big(n_f,N,\frac{Q^2}{\mu^2}\Big)&=&
          \sum_{i=2}^{\infty}a_s^i
            C_{q,(2,L)}^{(i),{\sf PS}}\Big(n_f,N,\frac{Q^2}{\mu^2}\Big)~,
\label{Cq2LPSpert} \\
        C_{q,(2,L)}^{\sf NS}\Big(n_f,N,\frac{Q^2}{\mu^2}\Big)&=&
         \delta_2+\sum_{i=1}^{\infty}a_s^i
            C_{q,(2,L)}^{(i),{\sf NS}}\Big(n_f,N,\frac{Q^2}{\mu^2}\Big)~,
\label{Cq2LNSpert}
      \end{eqnarray}
      where $a_s\equiv\alpha_s/(4\pi)$ and
      \begin{eqnarray}
       \delta_2&=&1\mbox{ for }F_2\mbox{ and }\delta_2=0\mbox{ for }F_L~.
       \label{delta2def}
      \end{eqnarray}
      These terms are at present known up to $O(a_s^3)$. 
      The $O(a_s)$ terms have been calculated in Refs.~\cite{Zee:1974du,Bardeen:1978yd,Furmanski:1981cw} and the $O(a_s^2)$ contributions by 
      various groups in Refs.~\cite{Duke:1981ga,*Devoto:1984wu,*Kazakov:1987jk,*Kazakov:1990fu,*SanchezGuillen:1990iq,*vanNeerven:1991nnxZijlstra:1991qcxZijlstra:1992qd,*Kazakov:1992xj,*Larin:1991fv,Moch:1999eb}. 
      The $O(a_s^3)$ terms have first been calculated for fixed moments in 
      Refs.~\cite{Larin:1993vu,Larin:1996wd,Retey:2000nq,Blumlein:2004xt} and 
      the complete result for all $N$ has been obtained in Refs.~\cite{Vermaseren:2005qc}~\footnote{Recently, the $O(a_s^3)$ Wilson
      coefficient for the structure function $xF_3(x,Q^2)$ was calculated in
      Ref.~\cite{Moch:2008fj}.}.
%%%%%%%%%%%%%%%%%%%%%%%%%%%%%%%%%%%%%%%%%%%%%%%%%%%%%%%%%%%%%%%%%%%%%%%%%%%%%%%
   \subsection{\bf\boldmath RGE--improved Parton Model and Anomalous Dimensions} 
    \label{SubSec-DISEvol}
%%%%%%%%%%%%%%%%%%%%%%%%%%%%%%%%%%%%%%%%%%%%%%%%%%%%%%%%%%%%%%%%%%%%%%%%%%%%%%%
      In the following, we present a derivation of the RGEs 
      for the Wilson coefficients, and subsequently, the evolution 
      equations for the parton densities. When calculating scattering
      cross sections in quantum field theories, they usually
      contain divergences of different origin. The 
      infrared and collinear singularities are connected to the limit of 
      soft-- and collinear radiation, respectively. 
      Due to the Bloch--Nordsieck 
      theorem,~\cite{Bloch:1937pw,*Yennie:1961ad}, 
      it is known that the infrared divergences cancel between virtual 
      and bremsstrahlung contributions. The structure functions are inclusive 
      quantities. Therefore, all final state collinear (mass) singularities
      cancel as well, which is formulated in the Lee--Kinoshita--Nauenberg
      theorem, \cite{Kinoshita:1962ur,*Lee:1964is}. Thus in case 
      of the Wilson coefficients, only the initial state collinear 
      divergences of the external light partons and the ultraviolet 
      divergences remain. The latter are connected to the large 
      scale behavior and are renormalized by a redefinition of the 
      parameters of the theory, as the coupling constant, the masses, the 
      fields, and the composite operators, \cite{Peterman:1978tb,Collins:1984xc}.
      This introduces a renormalization scale $\mu_r$, which forms the 
      subtraction point for renormalization. The scale 
      which appears in the factorization formulas (\ref{STR}, \ref{FACT2})
      is denoted by $\mu_f$ and called factorization scale, cf. 
\cite{Amati:1978wx,*Libby:1978qf,*Libby:1978bx,*Mueller:1978xu,*Collins:1981ta,*Bodwin:1984hc,*Collins:1985ue,Collins:1987pm}.
       Its origin lies in the arbitrariness of the point at which 
      short-- and long--distance effects are separated and is connected 
      to the redefinition of the bare parton densities by absorbing 
      the initial state collinear singularities of the Wilson coefficients 
      into them. Note, that
      one usually adopts dimensional regularization to regularize 
      the infinities in perturbative calculations, cf. Section~\ref{Sec-REN}, 
      which causes another scale $\mu$ to appear. It is associated to the
      mass dimension of the coupling constant in $D\neq 4$ dimensions. In 
      principle all these three scales have to be treated 
      separately, but we will set them equal in the subsequent analysis, 
      $\mu = \mu_r = \mu_f$. 

      The renormalization group equations are obtained using the 
      argument that all these scales are arbitrary and therefore 
      physical quantities do not alter when changing these scales, 
      \cite{Stuckelberg:1951gg,*GellMann:1954fq,*Bogolyubov:1980nc,Symanzik:1970rt,*Callan:1970yg,Collins:1984xc,Peterman:1978tb}.
      One therefore defines the total derivative w.r.t. to $\mu^2$
       \begin{eqnarray}
        {\cal D}(\mu^2)&\equiv& 
                  \mu^2\frac{\partial}{\partial\mu^2}
                 +\beta(a_s(\mu^2))\frac{\partial}{\partial a_s(\mu^2)}
             -\gamma_m(a_s(\mu^2)) m(\mu^2)\frac{\partial}{\partial m(\mu^2)}.
         \label{totdiff}
       \end{eqnarray}
       Here the $\beta$--function and the anomalous dimension of the mass,
       $\gamma_m$, are given by
       \begin{eqnarray}
        \beta(a_s(\mu^2))&\equiv&
                 \mu^2\frac{\partial a_s(\mu^2)}{\partial \mu^2}~,
                   \label{betdef1} \\
        \gamma_m(a_s(\mu^2)))&\equiv&
                 -\frac{\mu^2}{m(\mu^2)}
                  \frac{\partial m(\mu^2)}{\partial \mu^2}~,
       \end{eqnarray}
       cf. Sections \ref{SubSec-RENMa}, \ref{SubSec-RENCo}.
       The derivatives have to be performed keeping the bare
       quantities $\hat{a}_s$, $\hat{m}$ fixed. Additionally, 
       we work in Feynman--gauge and therefore the gauge--parameter 
       is not present in Eq.~(\ref{totdiff}). In the following we will 
       consider only 
       one mass $m$. The composite operators  
       (\ref{COMP1})--(\ref{COMP3}) are renormalized introducing
       operator $Z$--factors
       \begin{eqnarray}
        O^{\sf NS}_{q,r;\mu_1,...,\mu_N}&=&
                    Z^{\sf NS}(\mu^2)\hat{O}^{\sf NS}_{q,r;\mu_1,...,\mu_N}~,
                    \label{ZNSdef}\\
        O^{\sf S}_{i;\mu_1,...,\mu_N}&=&
                  Z^{\sf S}_{ij}(\mu^2)
                  \hat{O}^{\sf S}_{j;\mu_1,...,\mu_N}~,\quad~i=q,g~,
                  \label{ZSijdef}
       \end{eqnarray}
       where in the singlet case mixing occurs since these operators carry 
       the same quantum numbers. The anomalous dimensions of the operators 
       are defined by 
       \begin{eqnarray}
        \gamma_{qq}^{\sf NS}&=&
                  \mu Z^{-1, {\sf NS}}(\mu^2)
                  \frac{\partial}{\partial \mu} Z^{\sf NS}(\mu^2)~,
                                               \label{gammazetNS}\\
        \gamma_{ij}^{\sf S}&=&
                           \mu Z^{-1, {\sf S}}_{il}(\mu^2)
                         \frac{\partial}{\partial \mu} Z_{lj}^{\sf S}(\mu^2)~.
                                               \label{gammazetS}
       \end{eqnarray}
       We begin by considering the partonic structure functions calculated
       with external fields $l$. Here we would like to point out that 
       we calculate matrix elements of currents, operators, etc. and not 
       vacuum expectation values of time--ordered products with the external 
       fields included. The anomalous dimensions of the latter therefore 
       do not contribute, \cite{Yndurain:1999ui}, and they 
       are parts of the anomalous dimensions of the composite operators,
       respectively.
       The RGE reads
       \begin{eqnarray}
         {\cal D}(\mu^2) {\cal F}^l_{(2,L)}(N,Q^2) =0~.
        \label{RENFi2L}
       \end{eqnarray}
       On the partonic level, Eq.~(\ref{F2LMOM}) takes the form 
       \begin{eqnarray}
        {\cal F}^l_{(2,L)}(N,Q^2)=\sum_j C_{j,(2,L)}
                      \Bigl(N,\frac{Q^2}{\mu^2}\Bigr)\bra{l}O_j(\mu^2)\ket{l}~.
        \label{Fi2Lfac}
       \end{eqnarray}
       From the operator renormalization constants of the $O_i$, 
       Eqs. (\ref{ZNSdef}, \ref{ZSijdef}), the following RGE is derived 
       for the matrix elements, \cite{Buras:1979yt},
       \begin{eqnarray}
        \label{EVOL1}
        \sum_j \Bigl({\cal D}(\mu^2)~\delta_{ij}
               +\frac{1}{2}\gamma_{ij}^{\sf S, NS}\Bigr)
                \bra{l}O_j(\mu^2)\ket{l}&=&0~, 
       \end{eqnarray}
       where we write the ${\sf S}$ and ${\sf NS}$ case in one equation 
       for brevity and we remind the reader that in the latter case,
       $i,j,l=q$ only. 
       Combining Eqs. (\ref{RENFi2L}, \ref{Fi2Lfac}, \ref{EVOL1}), 
       one can determine the RGE for the Wilson coefficients. It reads
       \begin{eqnarray}
        \label{EVOL2}
        \sum_i \Bigl({\cal D}(\mu^2)~\delta_{ij}
              -\frac{1}{2}\gamma_{ij}^{\sf S, NS}\Bigr)
        C_{i,(2,L)}\Bigl(N,\frac{Q^2}{\mu^2}\Bigr)&=&0~.
       \end{eqnarray}
       The structure functions, which are observables, obey the same
       RGE as on the partonic level
       \begin{eqnarray}
           {\cal D}(\mu^2)F_{(2,L)}(N,Q^2) 
         = \mu^2 \frac{d}{d\mu^2}F_{(2,L)}(N,Q^2) 
         = 0~.
       \end{eqnarray}
       Using the factorization of the structure functions into Wilson 
       coefficients and parton densities, Eqs. (\ref{STR}, \ref{FACT2}), 
       together with the RGE derived for the Wilson coefficients in Eq. 
       (\ref{EVOL2}), one obtains from the above formula the QCD evolution 
       equations for the parton densities, cf. e.g. \cite{Buras:1979yt,Reya:1979zk,Altarelli:1989ue,Owens:1992hd,Roberts:1990ww,Blumlein:1995cm},
       \begin{eqnarray}
        \label{EVOL3}
         \frac{d}{d\ln \mu^2} f^{\sf S, NS}_i(n_f,N,\mu^2)
              &=&- \frac{1}{2} \sum_j \gamma_{ij}^{\sf S, NS}
                f_j^{\sf S, NS}(n_f,N,\mu^2)~.
       \end{eqnarray}
       Eq.~(\ref{EVOL3}) describes the change of the parton densities
       w.r.t. the scale $\mu$. In the more familiar matrix notation, 
       these equations read
       \begin{eqnarray}
        \frac{d}{d\ln \mu^2}
         \begin{pmatrix} 
             \Sigma(n_f,N,\mu^2) \\
             G(n_f,N,\mu^2)        
         \end{pmatrix} &=&-\frac{1}{2}
         \begin{pmatrix} 
           \gamma_{qq} & \gamma_{qg} \\ 
           \gamma_{gq} & \gamma_{gg} 
         \end{pmatrix}
         \begin{pmatrix} 
           \Sigma(n_f,N,\mu^2) \\
           G(n_f,N,\mu^2)
         \end{pmatrix}~,         \label{EVOL4} \\
        \frac{d}{d\ln \mu^2}
         \Delta_k(n_f,N,\mu^2)
         &=&
            -\frac{1}{2}\gamma_{qq}^{\sf NS}
             \Delta_k(n_f,N,\mu^2)~,\label{EVOL5}
       \end{eqnarray}
       where we have used the definition for the parton densities in 
       Eqs.~(\ref{SIGMAPDF}, \ref{DELTAPDF}).
       The anomalous dimensions in the above equations can be calculated 
       order by order in perturbation theory. At ${\sf LO}$, \cite{Gross:1973juxGross:1974cs,*Georgi:1951sr}, and ${\sf NLO}$, \cite{Floratos:1977auxFloratos:1977aue1,Floratos:1978ny,GonzalezArroyo:1979df,GonzalezArroyo:1979he,*Curci:1980uw,*Furmanski:1980cm,Hamberg:1991qt},
       they have been known for a long time. The ${\sf NNLO}$
       anomalous dimension were calculated first for fixed moments in 
       Refs.~\cite{Larin:1996wd,Retey:2000nq,Blumlein:2004xt}
       and the complete result
       for all moments has been obtained in Refs.~\cite{Moch:2004pa,Vogt:2004mw}~\footnote{Note that from our convention in Eqs. (\ref{gammazetNS}, \ref{gammazetS}) follows a relative factor $2$ between the anomalous dimensions considered in this work compared to Refs.~\cite{Moch:2004pa,Vogt:2004mw}.}.
       As described, the PDFs are 
       non--perturbative quantities and have to be extracted at a certain 
       scale from experimental data using the factorization relation 
       (\ref{STR}). If the scale $\mu^2$ is large enough to apply perturbation 
       theory, the evolution equations can be used to calculate the PDFs 
       at another perturbative scale, which provides a detailed QCD test 
       comparing to precision data.
       There are similar evolution equations for the structure functions and 
       Wilson coefficients, cf. e.g. \cite{Buras:1979yt,Reya:1979zk,Altarelli:1989ue,Owens:1992hd,Roberts:1990ww,Blumlein:1995cm}.
       Different groups analyze the evolution of the parton distribution 
       functions based on precision data from deep--inelastic scattering 
       experiments and other hard scattering cross sections. Analyzes were 
       performed by the Dortmund group, \cite{Gluck:1980cp,Gluck:1988xx,Gluck:1989ze,Gluck:1991ng,Gluck:1994uf,Gluck:1998xa,Gluck:2006yz,Gluck:2007ck,JimenezDelgado:2008hf}, by Alekhin et. al., \cite{Alekhin:2005gq,Alekhin:2006zm}, Bl{\"u}mlein
       et. al., \cite{Blumlein:2004ip,Blumlein:2006be},
       the ${\sf MSTW}$--, \cite{Martin:2009iq}, ${\sf QTEQ}$--,
       \cite{Lai:1999wy},
       and the ${\sf NNPDF}$--collaborations, \cite{Ball:2008by}.
       The PDFs determined in this way can e.g. be used as input data for the 
       $pp$ collisions at the LHC, since they are universal quantities and 
       only relate to the structure of the proton and not to the particular
       kind of scattering events considered. 
       Apart from performing precision 
       analyzes of the PDFs, one can also use the evolution equations 
       to determine $a_s$ more precisely, \cite{Gluck:1980cp,Blumlein:2004ip,Alekhin:2005gq,Alekhin:2006zm,Blumlein:2006be,Gluck:2006yz,Gluck:2007ck,JimenezDelgado:2008hf,Martin:2009iq}.
       
       The evolution equations (\ref{EVOL3}, \ref{EVOL4}, \ref{EVOL5})
       are written for moments only. The representation in $x$--space is 
       obtained by using~(\ref{Mellintrans},~\ref{Mellinconz},~\ref{MellinconN})
       and is usually expressed in terms of the splitting functions 
       $P_{ij}(x)$, \cite{Altarelli:1977zs}. 
       At the level of twist--$2$ the latter are connected to the anomalous 
       dimensions by the Mellin--transform
       \begin{eqnarray}
        \gamma_{ij}(N)=-\M[P_{ij}](N)~. \label{splitan}
       \end{eqnarray}
       The behavior of parton distribution functions in the small $x$ region
       attracted special interest due to possibly new dynamical
       contributions, such as Glauber--model based screening corrections,
       \cite{Gribov:1981ac,*Mueller:1985wy,*Collins:1990cw,*Bartels:1990zk,*Altmann:1992vm,*DelDuca:1995hf,*Lipatov:1996ts},
       and the so-called BFKL contributions, 
       a `leading singularity' resummation in the anomalous dimensions for all
       orders in the strong coupling constant, \cite{Fadin:1975cb,*Balitsky:1978ic,*Kirschner:1983di,*Bartels:1996wc,*Fadin:1998py}. 
       For both
       effects there is no evidence yet in the data both for $F_2(x,Q^2)$ and 
       $F_L(x,Q^2)$, beyond the known perturbative contributions to $O(a_s^3)$.
       This does not exclude that at even smaller values of $x$ contributions 
       of this kind will be found. The BFKL contributions were investigated on
       the basis of a consistent renormalization group treatment, together with
       the fixed order contributions in Refs.~\cite{Blumlein:1995jpxBlumlein:1996ddxBlumlein:1996hbxBlumlein:1997em,Blumlein:1998pp}.
       One main characteristic, comparing with the fixed order case, is that
       several sub-leading series, which are unknown, are required to stabilize
       the results, see also~\cite{Blumlein:1998mg}. 
       This aspect also has to be studied within 
       the framework of recent approaches,
      \cite{Altarelli:2008aj,*Ciafaloni:2007gf}.
%%%%%%%%%%%%%%%%%%%%%%%%%%%%%%%%%%%%%%%%%%%%%%%%%%%%%%%%%%%%%%%%%%%%%%%%%%%%%%%
%
% Chapter 3
%
% Heavy Quark Production in Deeply Inelastic Scattering
%
%%%%%%%%%%%%%%%%%%%%%%%%%%%%%%%%%%%%%%%%%%%%%%%%%%%%%%%%%%%%%%%%%%%%%%%%%%%%%%%
\newpage
 \section{\bf\boldmath Heavy Quark Production in DIS}
  \label{Sec-HQDIS}
  \renewcommand{\theequation}{\thesection.\arabic{equation}}
  \setcounter{equation}{0}
%%%%%%%%%%%%%%%%%%%%%%%%%%%%%%%%%%%%%%%%%%%%%%%%%%%%%%%%%%%%%%%%%%%%%%%%%%%%%%%
   In the Standard Model, the charm, bottom and top quark are treated as 
   heavy quarks, all of which have a mass larger than the QCD--scale 
   $\Lambda_{\sf QCD}(n_f=4)\approx~240~\MeV$,~\cite{Alekhin:2002fv,Blumlein:2004ip,Alekhin:2006zm,Blumlein:2006be,Gluck:2006yz,Gluck:2007ck,JimenezDelgado:2008hf}. The up, 
   down and strange quark are usually treated as massless. Because of 
   confinement, the quarks can only be observed via the asymptotic states 
   baryons and mesons, in which they are contained. In the following, we 
   concentrate on the inclusive production of one species of a heavy quark,
   denoted by 
   $Q(\overline{Q})$, with mass $m$. In the case of ${\sf HERA}$ kinematics, 
   $Q=c$. This is justified to a certain 
   extent by the observation that bottom quark contributions to DIS structure 
   functions are much smaller, cf. \cite{Thompson:2007mx}~\footnote{Likewise, for even higher scales the $b$--quark could be considered as the heavy quark with $u,d,s,c$ being effectively massless, cf. e.g. \cite{Chuvakin:2000jm}.}.
   Since the ratio $m_c^2/m_b^2\approx~1/10$ is not small, there are regions
   in which both masses are potentially important. The description of these 
   effects is beyond the scope of the formalism outlined below and of 
   comparable order as the $m_c^2/Q^2$ corrections. Top--quark production in
   $l^{\pm}N$ scattering is usually treated as a semi--inclusive process, 
   cf. \cite{Schuler:1987wj,*Baur:1987ai,Gluck:1987ukxGluck:1987uke1}.
   
   Charmed mesons are more abundantly produced at ${\sf HERA}$ than baryons.
   $D$--mesons are bound states of charm and 
   lighter quarks, e.g. $D_u=\overline{u}c$, $D_d=\overline{d}c$ etc. 
   Furthermore also $c\overline{c}$ resonances contribute, such as $J/\Psi$,
   by the observation of which charm was discovered,~\cite{Augustin:1974xw,*Abrams:1974yy,Aubert:1974js}. The charm contributions to the structure functions 
   are determined in experiment by tagging charm quarks in the final state,
   e.g. through the $D$--meson decay channel 
   $D^*\rightarrow~D^0\pi_s\rightarrow~K\pi\pi_s$. In the case of DIS, the 
   measured visible cross section is then extrapolated to the full inclusive 
   phase space using theoretical models if structure functions are considered,~\cite{Adloff:1996xq,Breitweg:1999ad,*Adloff:2001zj,Chekanov:2003rb,*Aktas:2004az,Aktas:2005iw,Thompson:2007mx}.

   Within the approach of this thesis, the main objective for studying heavy
   quark production in DIS is to provide a framework allowing for more
   precise measurements of $a_s$ and of the parton densities
   and for a better description of the structure functions
   $F_2^{c\overline{c}}$, $F_2^{b\overline{b}}$.
   The current world data for the nucleon structure functions 
   $F_2^{p,d}(x,Q^2)$ reached the precision of a few per cent over 
   a wide kinematic region. 
   Both the measurements of the heavy flavor contributions to the deep-inelastic structure functions, cf.~\cite{Chekanov:2003rb,*Aktas:2004az,Lipka:2006ny,Thompson:2007mx}, 
   and numerical studies,~\cite{Eichten:1984eu,Gluck:1993dpa,Blumlein:1996sc}, based on the leading, \cite{Witten:1975bh,*Babcock:1977fi,*Shifman:1977yb,*Leveille:1978px,Gluck:1980cp}, 
   and next-to-leading order heavy flavor Wilson coefficients, 
   \cite{Laenen:1992zkxLaenen:1992xs,*Riemersma:1994hv}, show that the scaling violations of the light and the heavy contributions to (\ref{FACT2}) 
   exhibit a different behavior over 
   a wide range of $Q^2$. This is both due to the logarithmic 
   contributions $\ln^k(Q^2/m^2)$ and power corrections 
   $\propto (m^2/Q^2)^k,~k \geq 1$. Moreover, in the region of smaller 
   values of $x$ the heavy flavor contributions amount to 20--40\%. 
   Therefore, the precision measurement of the QCD parameter 
   $\Lambda_{\rm QCD}$,~\cite{Gluck:1980cp,Alekhin:2002fv,Bethke:2000ai,*Bethke:2004uy,Blumlein:2004ip,Blumlein:2006be,Gluck:2006yz,Alekhin:2006zm,Gluck:2007ck,Blumlein:2007dk,JimenezDelgado:2008hf}, and of the parton distribution functions in 
   deeply inelastic scattering requires the analysis
   at the level of the $O(a_s^3)$ corrections to control the
   theory-errors at the level of the experimental accuracy 
   and below, \cite{Bethke:2000ai,*Bethke:2004uy,Gluck:2006yz,Alekhin:2006zm,Blumlein:2006be,Blumlein:2007dk}. 

   The precise value of $\Lambda_{\rm QCD}$, a fundamental parameter of 
   the Standard Model, is of central importance for the quantitative 
   understanding of all strongly interacting processes. Moreover, the 
   possible unification of the gauge forces, \cite{Georgi:1974sy,*Fritzsch:1974nn},
   depends crucially on its value. In recent non--singlet 
   analyzes, \cite{Blumlein:2004ip,Blumlein:2006be,Gluck:2006yz}, errors for $a_s(M_Z^2)$ 
   of $O(1.5~\%)$ were obtained, partially extending the analysis effectively 
   to ${\sf N^3LO}$. In the flavor singlet case the so far unknown 3--loop heavy 
   flavor Wilson coefficients do yet
   prevent a consistent 3--loop analysis,\cite{Alekhin:2005dxxAlekhin:2005dy,Dittmar:2005ed,Jung:2008tq}. Due to the large statistics in the 
   lower $x$ region, one may hope to eventually improve the accuracy of 
   $a_s(M_Z^2)$ beyond the above value.

   Of similar importance is the detailed 
   knowledge of the PDFs for all hadron-induced processes,
   notably for the interpretation of all scattering cross 
   sections measured at the ${\sf TEVATRON}$ and the LHC. 
   For example, the process 
   of Higgs-boson production at the LHC, cf. e.g. \cite{Djouadi:2005gixDjouadi:2005gj},
   depends on the gluon density and its accuracy is widely determined 
   by this distribution.

   In Section~\ref{SubSec-HQElProd}, we describe the general 
   framework of electroproduction of heavy quarks in DIS within 
   the fixed--flavor--number--scheme (FFNS), treating only the light 
   quarks and the gluon as constituents of the nucleon. In the following
   Section,~\ref{SubSec-HQAsym}, we outline the method, which 
   we use to extract all but the power suppressed contributions 
   $\propto~(m^2/Q^2)^k,k\ge~1$ of the heavy flavor Wilson 
   coefficients, \cite{Buza:1995ie}. 
   The latter are equivalent to the Wilson coefficients 
   introduced in Section~\ref{SubSec-LCE}, including heavy quarks. 
   Finally, in Section~\ref{SubSec-HQFlav}
   we comment on the possibility to define heavy quark parton 
   densities within a variable--flavor--number--scheme~(VFNS),
   \cite{Buza:1996wv}. 
%%%%%%%%%%%%%%%%%%%%%%%%%%%%%%%%%%%%%%%%%%%%%%%%%%%%%%%%%%%%%%%%%%%%%%%%%%%%%%%
%
%	Electroproduction of Heavy Quarks
%
%%%%%%%%%%%%%%%%%%%%%%%%%%%%%%%%%%%%%%%%%%%%%%%%%%%%%%%%%%%%%%%%%%%%%%%%%%%%%%%
  \subsection{\bf\boldmath Electroproduction of Heavy Quarks}
   \label{SubSec-HQElProd}
\begin{figure}[htb]
      \begin{center}
       \includegraphics[angle=0, width=7.0cm]{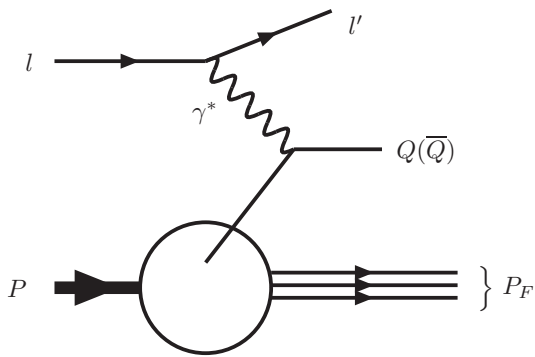}
      \end{center}
      \begin{center} 
       \caption{\sf ${\sf LO}$ intrinsic heavy quark production.}
       \label{CintLO}
       \noindent
       \small
      \end{center}
      \normalsize
     \end{figure}

%%%%%%%%%%%%%%%%%%%%%%%%%%%%%%%%%%%%%%%%%%%%%%%%%%%%%%%%%%%%%%%%%%%%%%%%%%%%%%%
    We study electroproduction of heavy quarks in unpolarized DIS
    via single photon exchange, cf. \cite{Witten:1975bh,*Babcock:1977fi,*Shifman:1977yb,*Leveille:1978px,Gluck:1979aw,Gluck:1980cp}, at sufficiently large 
    virtualities $Q^2$, $Q^2\ge~5\GeV$~\footnote{One may however, also consider photoproduction of heavy quarks in $ep$ collisions where $Q^2\approx~0$, which is a widely hadronic process, cf. \cite{Gluck:1978bf,Berger:1980ni}, and especially important for the production of heavy quark resonances, as e.g. the $J/\Psi$.}.  
    Here, one can distinguish two possible production 
    mechanisms for heavy quarks: extrinsic production and intrinsic heavy quark
    excitation. In the latter case, one introduces a heavy quark state in the 
    nucleon wave function, i.e. the heavy quark is treated at the same level as
    the light quarks in the factorization of the structure functions, 
    cf. Eqs.~(\ref{FACT2})--(\ref{Cq2LNSpert}).
    The ${\sf LO}$ contribution is then given by the 
    flavor excitation process shown in Figure \ref{CintLO},
    \begin{eqnarray}
     \gamma^*+Q(\overline{Q})\rightarrow~Q(\overline{Q})~.  \label{CintLOa}
    \end{eqnarray}
%%%%%%%%%%%%%%%%%%%%%%%%%%%%%%%%%%%%%%%%%%%%%%%%%%%%%%%%%%%%%%%%%%%%%%%%%%%%%%%
%%%%%%%%%%%%%%%%%%%%%%%%%%%%%%%%%%%%%%%%%%%%%%%%%%%%%%%%%%%%%%%%%%%%%%%%%%%%%%%

    Several experimental and theoretical studies suggest that the intrinsic 
    contribution to the heavy flavor cross section is of the order of $1\%$
    or smaller, \cite{Brodsky:1980pb,*Hoffmann:1983ah,*Derrick:1995sc,*Harris:1995jx,Adloff:1996xq}, and we will not consider it any further. 

    In extrinsic heavy flavor production, the heavy quarks are produced 
    as final states in virtual gauge boson scattering off 
    massless partons. This description is also referred to as the fixed flavor
    number 
    scheme. At higher orders, one has to make the distinction between whether 
    one considers the complete inclusive structure functions or only those
    heavy quark contributions, which can be determined in experiments 
    by tagging the final state heavy quarks. In the former case, virtual 
    corrections containing heavy quark loops have to be included into the 
    theoretical calculation as well, cf. also
    Section~\ref{SubSec-HQElProdWave}. 

    We consider only twist-2 parton densities in the Bjorken limit.
    Therefore no transverse momentum effects in the initial 
    parton distributions will be allowed, since these contributions are
    related, in the kinematic sense, to higher twist operators.
    From the conditions for the validity of the parton model, Eqs. 
    (\ref{cond}, \ref{cond2}), it follows that in 
    the region of not too small nor too large values of the Bjorken variable 
    $x$, the partonic description holds for massless partons. Evidently, 
    iff $Q^2 (1-x)^2/m^2 \gg \hspace*{-5mm}/ \hspace*{2mm}1$ {\sf no} partonic 
    description for a potential heavy quark distribution can be obtained. The 
    question under which circumstances one may introduce a heavy flavor 
    parton density will be further 
    discussed in Section~\ref{SubSec-HQFlav}. In a
    general kinematic region the parton densities in Eq.~(\ref{FACT2}) are
    enforced to be massless and the heavy quark mass effects are contained in 
    the inclusive Wilson coefficients. These are calculable perturbatively and 
    denoted by
    \begin{eqnarray}
      {\cal C}^{\sf S,PS,NS}_{i, (2,L)}
          \Bigl(\tau,n_f+1,\frac{Q^2}{\mu^2},\frac{m^2}{\mu^2}\Bigr)~. 
      \label{Calldef}
     \end{eqnarray}
    The argument $n_f+1$ denotes the presence of $n_f$ light and one heavy 
    flavor. $\tau$ is the partonic scaling variable defined in 
    Eq.~(\ref{taudef}) and we will present some of the following equations in 
    $x$--space rather than in Mellin space. 

\begin{figure}[h]
      \begin{center}
       \includegraphics[angle=0, width=7.0cm]{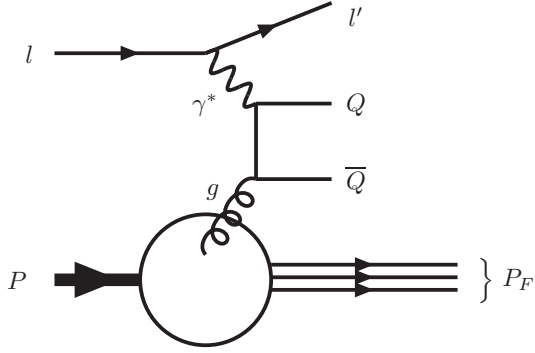}
      \end{center}
      \begin{center} 
       \caption{\sf ${\sf LO}$ extrinsic heavy quark production.}
       \label{CCbarLO}
       \noindent
       \small
      \end{center}
      \normalsize
     \end{figure}

    One may identify the massless flavor contributions in Eq.~(\ref{Calldef}) 
    and separate the Wilson coefficients into a purely light part 
    $C_{i,(2,L)}$, cf. Eq.~(\ref{FACT2}), and a heavy part
    \begin{eqnarray}
      \label{eqLH}
      {\cal C}^{\sf S,PS,NS}_{i,(2,L)}
               \left(\tau,n_f+1,\frac{Q^2}{\mu^2},\frac{m^2}{\mu^2}\right) 
      = 
       && C_{i,(2,L)}^{\sf S,PS,NS}\left(\tau,n_f,\frac{Q^2}{\mu^2}\right)
\N\\ &&\hspace{-25mm}
      + H_{i,(2,L)}^{\sf S,PS}
               \left(\tau,n_f+1,\frac{Q^2}{\mu^2},\frac{m^2}{\mu^2}\right)
      + L_{i,(2,L)}^{\sf S,PS,NS}
             \left(\tau,n_f+1,\frac{Q^2}{\mu^2},\frac{m^2}{\mu^2}\right)~.\N\\
       \label{Callsplit}
     \end{eqnarray}
     Here, we denote the heavy flavor Wilson coefficients by $L_{i,j}$ and 
     $H_{i,j}$, respectively, depending on whether the photon couples to a 
     light $(L)$ or heavy $(H)$ quark line. From this it follows that the light
     flavor Wilson coefficients $C_{i,j}$ depend on $n_f$ light flavors only,
     whereas 
     $H_{i,j}$ and $L_{i,j}$ may contain light flavors in addition to the heavy
     quark, indicated by the argument $n_f+1$. The perturbative series of the 
     heavy flavor Wilson coefficients read
     \begin{eqnarray}
      H_{g,(2,L)}^{\sf S}
          \left(\tau,n_f+1,\frac{Q^2}{\mu^2},\frac{m^2}{\mu^2}\right)&=&
           \sum_{i=1}^{\infty}a_s^i
             H_{g,(2,L)}^{(i), \sf S}
          \left(\tau,n_f+1,\frac{Q^2}{\mu^2},\frac{m^2}{\mu^2}\right)~,
\label{Hg2Lpert}
\\
      H_{q,(2,L)}^{\sf PS}
          \left(\tau,n_f+1,\frac{Q^2}{\mu^2},\frac{m^2}{\mu^2}\right)&=&
           \sum_{i=2}^{\infty}a_s^i
             H_{q,(2,L)}^{(i), \sf PS}
          \left(\tau,n_f+1,\frac{Q^2}{\mu^2},\frac{m^2}{\mu^2}\right)~,
\label{Hq2LPSpert}
\\
      L_{g,(2,L)}^{\sf S}
          \left(\tau,n_f+1,\frac{Q^2}{\mu^2},\frac{m^2}{\mu^2}\right)&=&
           \sum_{i=2}^{\infty}a_s^i
             L_{g,(2,L)}^{(i), \sf S}
          \left(\tau,n_f+1,\frac{Q^2}{\mu^2},\frac{m^2}{\mu^2}\right)~,
\label{Lg2Lpert}
     \end{eqnarray}
     \begin{eqnarray}
%\\
      L_{q,(2,L)}^{\sf S}
          \left(\tau,n_f+1,\frac{Q^2}{\mu^2},\frac{m^2}{\mu^2}\right)&=&
           \sum_{i=2}^{\infty}a_s^i
             L_{q,(2,L)}^{(i), \sf S}
          \left(\tau,n_f+1,\frac{Q^2}{\mu^2},\frac{m^2}{\mu^2}\right)~.
\label{Lq2LSpert}
     \end{eqnarray}
    Note that we have not yet specified a scheme for treating $a_s$, but
    one has to use the same scheme when 
    combining the above terms with the light flavor Wilson coefficients. At 
    ${\sf LO}$, only the term $H_{g,(2,L)}$ contributes via the 
    photon--gluon fusion process shown in Figure \ref{CCbarLO},
    \begin{eqnarray}
     \gamma^*+g\rightarrow~Q+\overline{Q}~. \label{VBfusion}
    \end{eqnarray}
%%%%%%%%%%%%%%%%%%%%%%%%%%%%%%%%%%%%%%%%%%%%%%%%%%%%%%%%%%%%%%%%%%%%%%%%%%%%%%%
%%%%%%%%%%%%%%%%%%%%%%%%%%%%%%%%%%%%%%%%%%%%%%%%%%%%%%%%%%%%%%%%%%%%%%%%%%%%%%%
    The ${\sf LO}$ Wilson coefficients corresponding to this 
    process are given by, \cite{Witten:1975bh,*Babcock:1977fi,*Shifman:1977yb,*Leveille:1978px,Gluck:1979aw,Gluck:1980cp}~\footnote{Eqs. (16), (17) in Ref.~\cite{Bierenbaum:2009zt} contain misprints}, 
     \begin{eqnarray}
      H_{g,2}^{(1)}\left(\tau,\frac{m^2}{Q^2}\right) 
          &=& 
             8T_F\Bigl\{
                 v\Bigl[
                        -\frac{1}{2}+4\tau(1-\tau)+2\frac{m^2}{Q^2}\tau(\tau-1)
                  \Bigr]
\N\\ 
          &+&     \Bigl[
                        -\frac{1}{2}+\tau-\tau^2+2\frac{m^2}{Q^2}\tau(3\tau-1)
                        +4\frac{m^4}{Q^4}\tau^2 
                  \Bigr]\ln\left(\frac{1-v}{1+v}\right)
                 \Bigr\}                \label{Hg2LO}  ~,\\
      H_{g,L}^{(1)}\left(\tau,\frac{m^2}{Q^2}\right) &=& 
           16T_F\Bigl[
                       \tau(1-\tau)v
                      +2\frac{m^2}{Q^2}\tau^2\ln\left(\frac{1-v}{1+v}\right)
                \Bigr]                \label{HgLLO}~.
     \end{eqnarray}
     The cms velocity $v$ of the produced heavy quark pair is given by
     \begin{eqnarray}
      v &=& \sqrt{1 - \frac{4 m^2 \tau}{Q^2(1-\tau)}}~. 
     \end{eqnarray}
     The ${\sf LO}$ heavy flavor contributions to the structure functions are 
     then
     \begin{eqnarray}
      F^{Q\overline{Q}}_{(2,L)}(x,Q^2,m^2) = 
            e_Q^2 a_s \int_{ax}^1 \frac{dz}{z} 
           H^{(1)}_{g,(2,L)}\left(\frac{x}{z},
          \frac{m^2}{Q^2}\right) G(n_f,z,Q^2)~,~~a=1+4 m^2/Q^2~, 
          \label{FcLO} 
     \end{eqnarray}
     where the integration boundaries follow from the kinematics of the 
     process. Here $e_Q$ denotes the electric charge of the heavy quark.

     At $O(a_s^2)$, the terms $H_{q,(2,L)}^{\sf PS}$ and $L_{q,(2,L)}^{\sf S}$ 
     contribute as well. They result from the process 
     \begin{eqnarray}
      \gamma^*+q(\overline{q})\rightarrow~q(\overline{q})+X~,
     \end{eqnarray}
     where $X=Q+\overline{Q}$ in case of extrinsic heavy flavor production.
     The latter is of phenomenological relevance if the heavy quarks are 
     detected 
     in the final states, e.g. via the produced $D_c$--mesons in case $Q=c$.
     For
     a complete inclusive analysis summing over all final states, both light 
     and heavy, one has to include radiative corrections 
     containing virtual heavy quark contributions as well. 
     The term $L_{q,(2,L)}^{\sf S}$ can be split into a ${\sf NS}$ and a
     ${\sf PS}$ piece via 
     \begin{eqnarray}
      L_{q,(2,L)}^{\sf S}=L_{q,(2,L)}^{\sf NS}+L_{q,(2,L)}^{\sf PS}, 
     \end{eqnarray}
     where the ${\sf PS}$--term emerges for the first time at $O(a_s^3)$ and 
     the ${\sf NS}$--term at $O(a_s^2)$, respectively. 
     Finally, $L_{g,(2,L)}^{\sf S}$ 
     contributes for the first time at $O(a_s^3)$ in case of heavy quarks in 
     the final state but there is a $O(a_s^2)$ term involving radiative 
     corrections, which will be commented on in 
     Section~\ref{SubSec-HQElProdWave}. 
     The terms $H_{g,(2,L)}^{(2)}$, $H_{q,(2,L)}^{(2), {\sf PS}}$ and 
     $L_{q,(2,L)}^{(2), {\sf NS}}$
     have been calculated in $x$--space in the 
     complete kinematic range in semi-analytic form in 
     Refs.~\cite{Laenen:1992zkxLaenen:1992xs,*Riemersma:1994hv}~\footnote{A precise representation in Mellin space was given in \cite{Alekhin:2003ev}.},
     considering heavy quarks in the final states only. 

     The heavy quark contribution to the structure functions $F_{(2,L)}(x,Q^2)$
     for one heavy quark of mass $m$ and $n_f$ light flavors is then given by,
     cf. \cite{Buza:1996wv} and Eq.~(\ref{FACT2}),
     \begin{eqnarray}
      \label{eqF2}
       F_{(2,L)}^{Q\overline{Q}}(x,n_f\!\!\!&+&\!\!\!1,Q^2,m^2) =\N\\
       &&\sum_{k=1}^{n_f}e_k^2\Biggl\{ 
                   L_{q,(2,L)}^{\sf NS}\left(x,n_f+1,\frac{Q^2}{m^2}
                                                ,\frac{m^2}{\mu^2}\right)
                \otimes 
                   \Bigl[f_k(x,\mu^2,n_f)+f_{\overline{k}}(x,\mu^2,n_f)\Bigr]
\N\\ &&\hspace{14mm}
               +\frac{1}{n_f}L_{q,(2,L)}^{\sf PS}\left(x,n_f+1,\frac{Q^2}{m^2}
                                                ,\frac{m^2}{\mu^2}\right) 
                \otimes 
                   \Sigma(x,\mu^2,n_f)
\N\\ &&\hspace{14mm}
               +\frac{1}{n_f}L_{g,(2,L)}^{\sf S}\left(x,n_f+1,\frac{Q^2}{m^2}
                                                 ,\frac{m^2}{\mu^2}\right)
                \otimes 
                   G(x,\mu^2,n_f) 
                             \Biggr\}
\N\\
       &+&e_Q^2\Biggl[
                   H_{q,(2,L)}^{\sf PS}\left(x,n_f+1,\frac{Q^2}{m^2}
                                        ,\frac{m^2}{\mu^2}\right) 
                \otimes 
                   \Sigma(x,\mu^2,n_f)
\N\\ &&\hspace{7mm}
                  +H_{g,(2,L)}^{\sf S}\left(x,n_f+1,\frac{Q^2}{m^2}
                                           ,\frac{m^2}{\mu^2}\right)
                \otimes 
                   G(x,\mu^2,n_f)
                                  \Biggr]~,
    \end{eqnarray}
    where the integration boundaries of the Mellin--convolutions follow from 
    phase space kinematics, cf. Eq.~(\ref{FcLO}).
%%%%%%%%%%%%%%%%%%%%%%%%%%%%%%%%%%%%%%%%%%%%%%%%%%%%%%%%%%%%%%%%%%%%%%%%%%%%%%%
  \subsection{\bf\boldmath Asymptotic Heavy Quark Coefficient Functions} 
   \label{SubSec-HQAsym}
%%%%%%%%%%%%%%%%%%%%%%%%%%%%%%%%%%%%%%%%%%%%%%%%%%%%%%%%%%%%%%%%%%%%%%%%%%%%%%%
    An important part of the kinematic region in case of heavy flavor 
    production in DIS is located at larger values of $Q^2$, cf. e.g. 
    \cite{Gluck:1987ukxGluck:1987uke1,Ingelman:1988qn}. As has been shown in 
    Ref.~\cite{Buza:1995ie}, cf. also ~\cite{Berends:1987abxBerends:1987abe1,*vanNeerven:1997gf,Buza:1996wv}, the heavy flavor Wilson coefficients 
    $H_{i,j},~L_{i,j}$ factorize in the limit $Q^2\gg~m^2$
    into massive operator matrix elements $A_{ki}$ and the massless 
    Wilson coefficients $C_{i,j}$, if one heavy quark flavor
    and $n_f$ light flavors are considered.
    The massive OMEs are process independent quantities and contain all the 
    mass dependence except for the power corrections 
    $\propto~(m^2/Q^2)^k,~k\ge~1$. 
    The process dependence is given by the light flavor Wilson coefficients
    only. This allows the analytic calculation of the ${\sf NLO}$ heavy 
    flavor Wilson coefficients, \cite{Buza:1995ie,Bierenbaum:2007qe}. 
    Comparing these asymptotic expressions with the exact ${\sf LO}$ and 
    ${\sf NLO}$ results obtained in Refs.~\cite{Witten:1975bh,*Babcock:1977fi,*Shifman:1977yb,*Leveille:1978px,Gluck:1980cp} and \cite{Laenen:1992zkxLaenen:1992xs,*Riemersma:1994hv}, respectively, one finds that this approximation becomes 
    valid in case of $F_2^{Q\overline{Q}}$ 
    for $Q^2/m^2 \gsim 10$. These scales are 
    sufficiently low and match with the region analyzed in 
    deeply inelastic scattering for precision measurements. 
    In case of $F_L^{Q\overline{Q}}$, this approximation is only 
    valid for $Q^2/m^2 \gsim 800$, \cite{Buza:1995ie}. For the latter 
    case, the 3--loop corrections were calculated in 
    Ref.~\cite{Blumlein:2006mh}. This difference is due to the emergence
    of terms $\propto (m^2/Q^2) \ln(m^2/Q^2)$, which vanish 
    only slowly in the limit $Q^2/m^2 \rightarrow \infty$. 

    In order to derive the factorization formula, one considers 
    the inclusive Wilson coefficients ${\cal C}^{\sf S,PS,NS}_{i,j}$, 
    which have been 
    defined in Eq.~(\ref{Calldef}). After applying the LCE to the partonic 
    tensor, or the forward Compton amplitude, corresponding to 
    the respective
    Wilson coefficients, one arrives at the factorization relation, cf. 
    Eq.~(\ref{F2LMOM}), 
    \begin{eqnarray}
     {\cal C}^{{\sf S,PS,NS}, \small{{\sf \sf asymp}}}_{j,(2,L)}
          \Bigl(N,n_f+1,\frac{Q^2}{\mu^2},\frac{m^2}{\mu^2}\Bigr) 
        &=& \N\\ && \hspace{-55mm}
           \sum_{i} A^{\sf S,PS,NS}_{ij}\Bigl(N,n_f+1,\frac{m^2}{\mu^2}\Bigr)
                    C^{\sf S,PS,NS}_{i,(2,L)}
                      \Bigl(N,n_f+1,\frac{Q^2}{\mu^2}\Bigr)
           +O\Bigl(\frac{m^2}{Q^2}\Bigr)~. \label{CallFAC}
    \end{eqnarray}
    Here $\mu$ refers to the factorization scale
    between the heavy and light contributions in ${\cal {C}}_{j,i}$ and 
    {\sf 'asymp'} denotes the limit $Q^2\gg~m^2$.
    The $C_{i,j}$ are precisely the light Wilson coefficients, 
    cf. Eqs. (\ref{FACT2})--(\ref{Cq2LPSpert}), taken at $n_f+1$
    flavors. This can be inferred from the fact that in the LCE, 
    Eq.~(\ref{lighex}), the Wilson coefficients describe the singularities 
    for very large values of $Q^2$, which can not depend 
    on the presence of a quark mass. The mass dependence is given by the 
    OMEs $A_{ij}$, cf. Eqs. (\ref{HadOMEs},\ref{NucOMEs}), between partonic 
    states. Eq.~(\ref{CallFAC}) accounts for all mass effects 
    but corrections
    which are power suppressed, $(m^2/Q^2)^k, k\ge~1$. This factorization 
    is only valid if the heavy quark coefficient functions are defined in such
    a way that all radiative corrections containing heavy quark loops 
    are included. Otherwise, (\ref{CallFAC}), would not show the correct 
    asymptotic $Q^2$--behavior, \cite{Buza:1996wv}. 

    An equivalent way of describing Eq.~(\ref{CallFAC}) is obtained by 
    considering the calculation of the massless Wilson coefficients.
    Here, the initial state collinear singularities are given 
    by evaluating the massless OMEs between off--shell partons, 
    leading to transition functions $\Gamma_{ij}$. 
    The $\Gamma_{ij}$ are given in terms
    of the anomalous dimensions of the twist--$2$ operators and transfer the
    initial state singularities to the bare parton--densities due to 
    mass factorization, cf. e.g. \cite{Buza:1995ie,Buza:1996wv}.
    In the case at hand, something similar happens: The initial 
    state collinear singularities are transferred to the parton densities
    except for those which are regulated by the quark mass and described by 
    the OMEs. Instead of absorbing these terms
    into the parton densities as well, they are used to reconstruct the 
    asymptotic behavior of the heavy flavor Wilson coefficients. 
    Here,
    \begin{eqnarray}
     \label{eqAij}
      A_{ij}^{\sf S,NS}\Bigl(N,n_f+1,\frac{m^2}{\mu^2}\Bigr)
               = \langle j| O_i^{\sf S,NS}|j \rangle 
               =\delta_{ij}+\sum_{i=1}^{\infty}a_s^i A_{ij}^{(i),{\sf S,NS}}
                \label{pertomeren}
    \end{eqnarray}
    are the operator matrix elements of the local twist--2 
    operators being defined in Eqs. (\ref{COMP1})--(\ref{COMP3})
    between on--shell partonic states $|j\rangle,~~j = q, g$. As usual, 
    the ${\sf S}$ contribution can be split into a ${\sf NS}$
    and ${\sf PS}$ part via
    \begin{eqnarray}
     A_{qq}^{\sf S} = A_{qq}^{\sf NS} + A_{qq}^{\sf PS}~. \label{splitS}
    \end{eqnarray}
    Due to the on--shell condition, all contributions but 
    the $O(a_s^0)$ terms vanish~\footnote{In Ref.~\cite{Larin:1996wd} use was made of this fact to calculate the massless Wilson coefficients without having to calculate the massless OMEs.} if no heavy quark is present 
    in the virtual loops. This is due to the fact that 
    integrals without scale vanish in dimensional
    regularization, cf. Section~\ref{SubSec-RENReg}. 
    Hence only those terms with a mass remain and these are referred to 
    as massive OMEs. The calculation of these massive OMEs is the main
    objective
    of this thesis. In case of the gluon operator, (\ref{COMP3}), 
    the contributing terms are denoted by 
    $A_{gq,Q}$ and $A_{gg,Q}$, where the perturbative series of the  
    former starts at $O(a_s^2)$ and the one of the latter at $O(a_s^0)$~\footnote{The $O(a_s^0)$ term of $A_{gg}$ does not contain a heavy quark, but still remains in Eq.~(\ref{CallFAC}) because no loops have to be calculated.}. For the quark operator, 
    one distinguishes whether the operator couples to a heavy or light
    quark. In the ${\sf NS}$--case, the operator by definition couples to the 
    light quark. Thus there is only one term, $A_{qq,Q}^{\sf NS}$, 
    which contributes at $O(a_s^0)$. In the ${\sf S}$ and ${\sf PS}$--case, 
    two OMEs can be distinguished,
    $\D{\{A_{qq,Q}^{\sf PS},~A_{qg,Q}^{\sf S}\}}$
    and $\D{\{A_{Qq}^{\sf PS},~A_{Qg}^{\sf S}\}}$, where in the former case 
    the operator couples to a light quark and in the latter case to a heavy 
    quark. The terms $A_{qi,Q}$ emerge for the first time at $O(a_s^3)$, 
    $A_{Qq}^{\sf PS}$ at $O(a_s^2)$ and $A_{Qg}^{\sf S}$ at $O(a_s)$. 

    In this work we refer only to the even moments,
    cf. Section~\ref{SubSec-DISComptLCE}. In the non--singlet case we will
    obtain, however, 
    besides the ${\sf NS^+}$ contributions for the even moments also the ${\sf NS^-}$ terms,
    which correspond to the odd moments. 

    Eq.~(\ref{CallFAC}) can now be split into its parts
    by considering the different $n_f$--terms. 
    We adopt the following notation for a function $f(n_f)$
    \begin{eqnarray}
       {\tilde{f}}(n_f)&\equiv&\frac{f(n_f)}{n_f}~. \label{gammapres2}
    \end{eqnarray}
    This is necessary in order to separate the different types of contributions
    in Eq.~(\ref{eqF2}), weighted by the electric charges of the light 
    and heavy flavors, respectively. Since we concentrate on only the heavy 
    flavor part, we define as well for later use     
    \begin{eqnarray}
       \hat{f}(n_f)&\equiv&f(n_f+1)-f(n_f)~, \label{gammapres1}
    \end{eqnarray}
    with $\hat{\hspace*{-1mm}{\tilde{f}}}(n_f) \equiv 
    \widehat{[{\tilde{f}}(n_f)]}$.
    The following Eqs. (\ref{LNSFAC})--(\ref{HgFAC}) are the same as 
    Eqs.~(2.31)--(2.35) in Ref.~\cite{Buza:1996wv}. 
    We present these terms here again, however, since 
    Ref.~\cite{Buza:1996wv} contains a few inconsistencies regarding 
    the $\tilde{f}$--description.
    Contrary to the latter reference, the argument corresponding 
    to the number of flavors stands for all flavors, light or heavy. 
    The separation for the ${\sf NS}$--term is given by
    \begin{eqnarray}
   C_{q,(2,L)}^{\sf NS}\Bigl(N,n_f,\frac{Q^2}{\mu^2}\Bigr)
     + L_{q,(2,L)}^{\sf NS}
          \Bigl(N,n_f+1,\frac{Q^2}{\mu^2},\frac{m^2}{\mu^2}\Bigr)
     &=&
\N\\ &&
       \hspace{-50mm}
        A_{qq,Q}^{\sf NS}\Bigl(N,n_f+1,\frac{m^2}{\mu^2}\Bigr)
        C_{q,(2,L)}^{\sf NS}\Bigl(N,n_f+1,\frac{Q^2}{\mu^2}\Bigr)~.
        \label{LNSFAC}
    \end{eqnarray}
    Here and in the following, we omit the index $"{\sf asymp}"$
    to denote the asymptotic heavy flavor Wilson coefficients, since 
    no confusion is to be expected.
    For the remaining terms, we suppress for brevity the arguments 
    $N$, $Q^2/\mu^2$ and $m^2/\mu^2$, all of which can be inferred from 
    Eqs. (\ref{Callsplit}, \ref{CallFAC}). Additionally, we will suppress 
    from now on the index ${\sf S}$ and label only the ${\sf NS}$
    and ${\sf PS}$ terms explicitly. 
    The contributions to $L_{i,j}$ read
    \begin{eqnarray}
     C_{q,(2,L)}^{\sf PS}(n_f)
       +L_{q,(2,L)}^{\sf PS}
            (n_f+1)
     &=&
        \Bigl[
               A_{qq,Q}^{\sf NS}(n_f+1)
              +A_{qq,Q}^{\sf PS}(n_f+1)
              +A_{Qq}^{\sf PS}(n_f+1)
         \Bigr]
\N\\ &&
         \times
         n_f \tilde{C}_{q,(2,L)}^{\sf PS}(n_f+1)
        +A_{qq,Q}^{\sf PS}(n_f+1)
         C_{q,(2,L)}^{\sf NS}(n_f+1)
\N\\ &&        
+A_{gq,Q}(n_f+1)
         n_f \tilde{C}_{g,(2,L)}(n_f+1)~, \N\\ 
                 \label{LPSFAC} \\
      C_{g,(2,L)}(n_f)
     +L_{g,(2,L)}(n_f+1)
    &=&
           A_{gg,Q}(n_f+1)
           n_f \tilde{C}_{g,(2,L)}(n_f+1)
\N\\ &&
         + A_{qg,Q}(n_f+1)
           C_{q,(2,L)}^{\sf NS}(n_f+1)
\N\\ &&         
+\Bigl[
                A_{qg,Q}(n_f+1)
               +A_{Qg}(n_f+1)
          \Bigr]
        n_f\tilde{C}_{q,(2,L)}^{\sf PS}(n_f+1)~.\N\\
        \label{LgFAC}
    \end{eqnarray}
    The terms $H_{i,j}$ are given by
    \begin{eqnarray}
     H_{q,(2,L)}^{\sf PS}
          (n_f+1)
     &=&  
        A_{Qq}^{\sf PS}(n_f+1)
           \Bigl[ 
                 C_{q,(2,L)}^{\sf NS}(n_f+1)
                +\tilde C_{q,(2,L)}^{\sf PS}
                         (n_f+1)
          \Bigr]
\N\\ &&
          +\Bigl[ 
                A_{qq,Q}^{\sf NS}(n_f+1)
               +A_{qq,Q}^{\sf PS}(n_f+1)
         \Bigr]
        \tilde{C}_{q,(2,L)}^{\sf PS}(n_f+1)
\N\\ &&
       +A_{gq,Q}(n_f+1)
       \tilde{C}_{g,(2,L)}(n_f+1)~,         \label{HPSFAC} \\
     H_{g,(2,L)}(n_f+1)
      &=&
         A_{gg,Q}(n_f+1)
         \tilde{C}_{g,(2,L)}(n_f+1)
        +A_{qg,Q}(n_f+1)
          \tilde{C}_{q,(2,L)}^{\sf PS}(n_f+1)
\N\\ &&
        + A_{Qg}(n_f+1)
          \Bigl[ C_{q,(2,L)}^{\sf NS}(n_f+1)
            +\tilde{C}_{q,(2,L)}^{\sf PS}(n_f+1)
             \Bigr]~.         \label{HgFAC}
    \end{eqnarray}
    Expanding the above equations up to $O(a_s^3)$, we obtain, using
    Eqs. (\ref{gammapres2}, \ref{gammapres1}), the heavy flavor
    Wilson coefficients in the asymptotic limit~: 
    \begin{eqnarray}
     \label{eqWIL1}
     L_{q,(2,L)}^{\sf NS}(n_f+1) &=& 
     a_s^2 \Bigl[A_{qq,Q}^{(2), {\sf NS}}(n_f+1)~\delta_2 +
     \hat{C}^{(2), {\sf NS}}_{q,(2,L)}(n_f)\Bigr]
     \N\\
     &+&
     a_s^3 \Bigl[A_{qq,Q}^{(3), {\sf NS}}(n_f+1)~\delta_2
     +  A_{qq,Q}^{(2), {\sf NS}}(n_f+1) C_{q,(2,L)}^{(1), {\sf NS}}(n_f+1)
       \N \\
     && \hspace*{5mm}
     + \hat{C}^{(3), {\sf NS}}_{q,(2,L)}(n_f)\Bigr]~,  \\
      \label{eqWIL2}
      L_{q,(2,L)}^{\sf PS}(n_f+1) &=& 
     a_s^3 \Bigl[~A_{qq,Q}^{(3), {\sf PS}}(n_f+1)~\delta_2
     +  A_{gq,Q}^{(2)}(n_f)~~n_f\Ctil_{g,(2,L)}^{(1)}(n_f+1) \N \\
     && \hspace*{5mm}
     + n_f \hat{\Ctil}^{(3), {\sf PS}}_{q,(2,L)}(n_f)\Bigr]~,
     \\
     \label{eqWIL3}
      L_{g,(2,L)}^{\sf S}(n_f+1) &=& 
     a_s^2 A_{gg,Q}^{(1)}(n_f+1)n_f \Ctil_{g,(2,L)}^{(1)}(n_f+1)
     \N\\ &+&
      a_s^3 \Bigl[~A_{qg,Q}^{(3)}(n_f+1)~\delta_2 
     +  A_{gg,Q}^{(1)}(n_f+1)~~n_f\Ctil_{g,(2,L)}^{(2)}(n_f+1)
     \N\\ && \hspace*{5mm}
     +  A_{gg,Q}^{(2)}(n_f+1)~~n_f\Ctil_{g,(2,L)}^{(1)}(n_f+1)
     \N\\ && \hspace*{5mm}
     +  ~A^{(1)}_{Qg}(n_f+1)~~n_f\Ctil_{q,(2,L)}^{(2), {\sf PS}}(n_f+1)
     + n_f \hat{\Ctil}^{(3)}_{g,(2,L)}(n_f)\Bigr]~,
 \\ \N \\
     H_{q,(2,L)}^{\sf PS}(n_f+1)
     &=& a_s^2 \Bigl[~A_{Qq}^{(2), {\sf PS}}(n_f+1)~\delta_2
     +~\Ctil_{q,(2,L)}^{(2), {\sf PS}}(n_f+1)\Bigr]
     \N\\
     &+& a_s^3 \Bigl[~A_{Qq}^{(3), {\sf PS}}(n_f+1)~\delta_2
     +~\Ctil_{q,(2,L)}^{(3), {\sf PS}}(n_f+1) \N
    \end{eqnarray}
    \begin{eqnarray}
%     \N \\
 && \hspace*{-20mm}
     + A_{gq,Q}^{(2)}(n_f+1)~\Ctil_{g,(2,L)}^{(1)}(n_f+1) 
     + A_{Qq}^{(2), {\sf PS}}(n_f+1)~C_{q,(2,L)}^{(1), {\sf NS}}(n_f+1) 
        \Bigr]~,       \label{eqWIL4}
        \\ \N\\ 
     H_{g,(2,L)}^{\sf S}(n_f+1) &=& a_s \Bigl[~A_{Qg}^{(1)}(n_f+1)~\delta_2
     +~\Ctil^{(1)}_{g,(2,L)}(n_f+1) \Bigr] \N\\
     &+& a_s^2 \Bigl[~A_{Qg}^{(2)}(n_f+1)~\delta_2
     +~A_{Qg}^{(1)}(n_f+1)~C^{(1), {\sf NS}}_{q,(2,L)}(n_f+1)\N\\ && 
     \hspace*{5mm}
     +~A_{gg,Q}^{(1)}(n_f+1)~\Ctil^{(1)}_{g,(2,L)}(n_f+1) 
     +~\Ctil^{(2)}_{g,(2,L)}(n_f+1) \Bigr]
     \N\\ &+&
     a_s^3 \Bigl[~A_{Qg}^{(3)}(n_f+1)~\delta_2
     +~A_{Qg}^{(2)}(n_f+1)~C^{(1), {\sf NS}}_{q,(2,L)}(n_f+1)
     \N\\ &&
     \hspace*{5mm}
     +~A_{gg,Q}^{(2)}(n_f+1)~\Ctil^{(1)}_{g,(2,L)}(n_f+1)
     \N\\ && \hspace*{5mm}
     +~A_{Qg}^{(1)}(n_f+1)\Bigl\{
     C^{(2), {\sf NS}}_{q,(2,L)}(n_f+1)
     +~\Ctil^{(2), {\sf PS}}_{q,(2,L)}(n_f+1)\Bigr\}
     \N\\ && \hspace*{5mm}
     +~A_{gg,Q}^{(1)}(n_f+1)~\Ctil^{(2)}_{g,(2,L)}(n_f+1)
     +~\Ctil^{(3)}_{g,(2,L)}(n_f+1) \Bigr]~. \label{eqWIL5}
    \end{eqnarray}
    Note that $\delta_2$ has been defined in Eq.~(\ref{delta2def}).
    The above equations include radiative corrections due to heavy quark loops
    to the Wilson coefficients. Therefore, in order to compare e.g. with 
    the calculation in 
    Refs.~\cite{Laenen:1992zkxLaenen:1992xs,*Riemersma:1994hv},
    these terms still have to be subtracted. 
    Since the light flavor Wilson coefficients were calculated in the 
    $\overline{\sf MS}$--scheme, the {\sf same} scheme has to be used for the 
    massive OMEs. It should also be thoroughly used for renormalization
    to derive consistent results in QCD 
    analyzes of deep-inelastic scattering data and to be able to compare to 
    other analyzes directly. This means that one has to 
    take special attendance of which scheme for the definition of 
    $a_s$ was used. In Section~\ref{SubSec-RENCo} we will describe a scheme
    for $a_s$, to which one is naturally led in the course of renormalization.
    We refer to this scheme as ${\sf MOM}$--scheme and present the 
    transformation formula to the $\overline{\sf MS}$ as well. How this 
    affects the asymptotic heavy flavor Wilson coefficients 
    is described in Section~\ref{SubSec-HQElProdWave}, where we compare 
    Eqs. (\ref{eqWIL1})--(\ref{eqWIL5}) to those presented in 
    Ref.~\cite{Buza:1995ie}.
%%%%%%%%%%%%%%%%%%%%%%%%%%%%%%%%%%%%%%%%%%%%%%%%%%%%%%%%%%%%%%%%%%%%%%%%%%%%%%%
%
%	Heavy Quark Parton densities
%
%%%%%%%%%%%%%%%%%%%%%%%%%%%%%%%%%%%%%%%%%%%%%%%%%%%%%%%%%%%%%%%%%%%%%%%%%%%%%%%
  \subsection{\bf\boldmath Heavy Quark Parton Densities}
   \label{SubSec-HQFlav}
%%%%%%%%%%%%%%%%%%%%%%%%%%%%%%%%%%%%%%%%%%%%%%%%%%%%%%%%%%%%%%%%%%%%%%%%%%%%%%%
    The FFNS forms a general starting point to describe and to calculate the
    heavy flavor contributions to the DIS structure functions.
    Approaching higher values of $Q^2$, one may think of the heavy quark 
    becoming effectively light and thus acquiring an own parton density.
    Different variable flavor scheme treatments were considered in the past, 
    cf. e.g. \cite{Aivazis:1993pi,*Thorne:2008xf}. Here we follow 
    \cite{Buza:1996wv} to obtain a description in complete accordance with the
    renormalization group in the ${\sf \overline{MS}}$--scheme.
    In the kinematic region in which the factorization relation (\ref{CallFAC})
    holds, one may redefine the results obtained in the FFNS,
    which allows for a partonic description at the level of $(n_f+1)$ flavors. 

    In the strict sense, only massless particles can be interpreted as partons
    in hard scattering processes since the lifetime of these 
    quantum-fluctuations off the hadronic background 
    $\tau_{\sf life}\propto 1/(k_\perp^2+m_Q^2)$ 
    has to be large against the interaction 
    time $\tau_{\sf int} \propto 1/Q^2$ in the 
    infinite momentum frame, \cite{Drell:1970yt},
    cf. also Section~\ref{SubSubSec-DISValpart}.
    In the massive case, $\tau_{\sf life}$ is necessarily finite and there
    exists a larger scale $Q^2_0$ below which any partonic 
    description fails. From this it follows, that
    the heavy quark effects are genuinely described by the 
    {\sf process dependent} Wilson coefficients. Since parton-densities 
    are {\sf process independent} quantities, only process independent 
    pieces out of the 
    Wilson coefficients can be used to define them for heavy quarks at all. 
    Clearly this is impossible in 
    the region close to threshold but requires $Q^2/m_Q^2 = r \gg 1$, with $ r 
    \gsim 
    10$ in case of $F_2(x,Q^2)$. For $F_L(x,Q^2)$ the corresponding ratio 
    even turns out to be 
    $r \gsim 800$,~\cite{Buza:1995ie,Blumlein:2006mh,Gluck:1993dpa}. 
    Heavy flavor parton distributions can thus be constructed only for scales 
    $\mu^2 \gg m_Q^2$. 
    This is done under the further assumption that for the other
    heavy flavors the masses $m_{Q_i}$ form a hierarchy 
    $m_{Q_1}^2 \ll m_{Q_2}^2 \ll~~{\sf etc.}$ 
    Their
    use in observables is restricted to a region, in which the power 
    corrections can be safely neglected. This range may strongly depend 
    on the observable considered as the  examples of $F_2$ and $F_L$ show.
    Also in case of the structure functions associated to 
    transverse virtual gauge boson polarizations, like $F_2(x,Q^2)$, 
    the factorization (\ref{CallFAC}) only 
    occurs far above threshold, $Q_{{\sf thr}}^2 \sim 4 m_Q^2 x/(1-x)$,
    and at even larger scales for $F_L(x,Q^2)$.  

    In order to maintain the process independence of the parton 
    distributions, we define them for $(n_f+1)$ flavors 
    from the light flavor parton distribution functions for $n_f$
    flavors together with the massive operator matrix elements.
    The following set of parton densities is obtained in 
    Mellin--space,~\cite{Buza:1996wv}~:
    \begin{eqnarray}
     && f_k(n_f+1,N,\mu^2,m^2) + f_{\overline{k}}(n_f+1,N,\mu^2,m^2)= 
\N\\ &&  \hspace{10mm}
        A_{qq,Q}^{\sf NS}\left(N,n_f+1,\frac{\mu^2}{m^2}\right)
          \cdot\bigl[f_k(n_f,N,\mu^2)+f_{\overline{k}}(n_f,N,\mu^2)\bigr]
\N\\ &&  \hspace{10mm}
       +\frac{1}{n_f}A_{qq,Q}^{\sf PS}\left(N,n_f+1,\frac{\mu^2}{m^2}\right)
          \cdot\Sigma(n_f,N,\mu^2)
\N\\ &&  \hspace{10mm}
       +\frac{1}{n_f}A_{qg,Q}\left(N,n_f+1,\frac{\mu^2}{m^2}\right)
          \cdot G(n_f,N,\mu^2), \label{HPDF1} \\
     && 
        f_Q(n_f+1,N,\mu^2,m^2) + f_{\overline{Q}}(n_f+1,N,\mu^2,m^2)=
\N\\ &&  \hspace{10mm}
        A_{Qq}^{\sf PS}\left(N,n_f+1,\frac{\mu^2}{m^2}\right)
          \cdot \Sigma(n_f,N,\mu^2)
\N\\ &&  \hspace{10mm}
       +A_{Qg}\left(N,n_f+1,\frac{\mu^2}{m^2}\right)
          \cdot G(n_f,N,\mu^2)~.   \label{fQQB}
    \end{eqnarray}
    The flavor singlet, non--singlet and gluon 
    densities for $(n_f+1)$ flavors are given by
    \begin{eqnarray}
     \Sigma(n_f+1,N,\mu^2,m^2) 
      &=& \Biggl[
             A_{qq,Q}^{\sf NS}\left(N,n_f+1,\frac{\mu^2}{m^2}\right)
            +A_{qq,Q}^{\sf PS}\left(N,n_f+1,\frac{\mu^2}{m^2}\right)
\N \\ && \hspace*{-20mm}
            +A_{Qq}^{\sf PS}\left(N,n_f+1,\frac{\mu^2}{m^2}\right)
          \Biggr]  \cdot \Sigma(n_f,N,\mu^2)
\N \\ && \hspace*{-23mm} 
          +\left[
                A_{qg,Q}\left(N,n_f+1,\frac{\mu^2}{m^2}\right)
               +A_{Qg}\left(N,n_f+1,\frac{\mu^2}{m^2}\right)
          \right]   \cdot G(n_f,N,\mu^2)~,
\N\\ \\
     \Delta_k(n_f+1,N,\mu^2,m^2)
      &=& f_k(n_f+1,N,\mu^2,m^2)+f_{\overline{k}}(n_f+1,N,\mu^2,m^2)
\N\\ &&
         -\frac{1}{n_f+1}\Sigma(n_f+1,N,\mu^2,m^2)~, \\
     \label{HPDF2}
     G(n_f+1,N,\mu^2,m^2) 
      &=& A_{gq,Q}\left(N,n_f+1,\frac{\mu^2}{m^2}\right) 
                    \cdot \Sigma(n_f,N,\mu^2)
\N\\ && 
         +A_{gg,Q}\left(N,n_f+1,\frac{\mu^2}{m^2}\right) 
                    \cdot G(n_f,N,\mu^2)~.
    \end{eqnarray}
    Note, that the {\sf new} parton densities depend on the renormalized heavy 
    quark mass $m^2=m^2(a_s^2(\mu^2))$. As will be outlined 
    in Sections \ref{Sec-REN}, \ref{Sec-REP}, 
    the corresponding relations for the 
    operator matrix elements depend on the mass--renormalization scheme. 
    This has to be taken into account in 
    QCD-analyzes, in particular, $m^2$ cannot be chosen constant. 
    The quarkonic and gluonic operators obtained 
    in the light--cone expansion can be normalized arbitrarily. 
    It is, however, convenient to chose 
    the relative factor such, that the non-perturbative nucleon-state 
    expectation values, 
    $\Sigma(n_f,N,\mu^2)$ and $G(n_f,N,\mu^2)$, obey
    \begin{eqnarray}
     \Sigma(n_f,N=2,\mu^2)+G(n_f,N=2,\mu^2) = 1
    \end{eqnarray}
    due to 4-momentum conservation. As a consequence, the OMEs fulfill the 
    relations, \cite{Buza:1996wv},
    \begin{eqnarray}
     A_{qq,Q}^{\sf NS}(N=2)
      +A_{qq,Q}^{\sf PS}(N=2)
      +A_{Qq}^{\sf PS}(N=2)
      +A_{gq,Q}(N=2) &=& 1~, \label{sumrule1}    \\
      A_{qg,Q}(N=2)
     +A_{Qg}(N=2)
     + A_{gg,Q}(N=2) &=& 1~.  \label{sumrule2}
    \end{eqnarray}
    The above scenario can be easily followed up to 2-loop order. Also here 
    diagrams contribute which carry two different heavy quark flavors. At this 
    level, the additional heavy degree of freedom may be absorbed into the 
    coupling constant and thus decoupled temporarily. 
    Beginning with 3-loop order the situation becomes more 
    involved since there are graphs in which two different 
    heavy quark flavors occur in nested topologies, i.e., the 
    corresponding diagrams depend on the ratio $\rho = m_c^2/m_b^2$ yielding 
    power corrections in $\rho$. There is no strong hierarchy between these 
    two masses. The above picture, leading to heavy flavor parton 
    distributions whenever $Q^2 \gg m^2$ will not hold 
    anymore, since, in case of the two-flavor graphs, one cannot decide immediately 
    whether they belong to the $c$-- or the $b$--quark distribution. 
    Hence, the partonic description can only be maintained within a certain 
    approximation by {\sf assuming} $\rho \ll 1$.

    Conversely, one may extend the kinematic regime for deep-inelastic 
    scattering to define the distribution functions  
    (\ref{HPDF1})--(\ref{HPDF2}) upon knowing the power corrections which 
    occur in the heavy flavor Wilson coefficients 
    ${\sf H}_{i,j}=H_{i,j},~L_{i,j}$. This is the case for 2-loop order. 
    We separate   
    \begin{eqnarray}
     {\sf H}_{i,j}\left(x,\frac{Q^2}{m^2},\frac{m^2}{\mu^2}\right) = 
     {\sf H}_{i,j}^{{\sf asymp}}\left(x,\frac{Q^2}{m^2},
                                             \frac{m^2}{\mu^2}\right)
     + {\sf H}_{i,j}^{{\sf power}}\left(x,\frac{Q^2}{m^2}
                   ,\frac{m^2}{\mu^2}\right)~, \label{splitpower}
    \end{eqnarray}
    where ${\sf H}_{i,j}^{{\sf asymp}}(x,Q^2/m^2,m^2/\mu^2)$
    denotes the 
    part of the Wilson coefficient given in Eq.~(\ref{CallFAC}). 
    If one accounts
    for ${\sf H}_{i,j}^{{\sf power}}(x,Q^2/m^2,m^2/\mu^2)$ in the fixed 
    flavor number scheme, Eqs.~(\ref{HPDF1})--(\ref{HPDF2}) are still valid, 
    but they do not necessarily 
    yield the dominant contributions in the region closer to threshold. There,
    the kinematics of heavy quarks is by far not collinear, which is the main 
    reason that a partonic description has to fail. Moreover, relation 
    Eq.~(\ref{cond}) may be violated. In any case, it is not possible 
    to use the partonic description (\ref{HPDF1})--(\ref{HPDF2}) alone for 
    other hard processes in a kinematic domain with significant power
    corrections.

    For processes in the high $p_\perp$ region at the LHC, in which
    condition (\ref{cond}) is fulfilled and the characteristic scale 
    $\mu^2$ obeys $\mu^2\gg~m^2$,
    one may use heavy flavor parton distributions by proceeding
    as follows. In the region $Q^2 \gsim 10~m^2$ the heavy flavor 
    contributions to the $F_2(x,Q^2)$--world data are very well described 
    by the asymptotic representation in the FFNS. For large scales one can 
    then form a variable flavor representation including one heavy flavor 
    distribution, \cite{Buza:1996wv}.
    This process can be iterated towards the next heavier flavor, provided the 
    {\sf universal} representation holds  and all power corrections can be 
    safely neglected. One has to take special care of the fact, that
    the matching scale in the coupling constant, at which the 
    transition $n_f \rightarrow n_f+1$ is to be performed, often differs rather
    significantly from $m$, cf. \cite{Blumlein:1998sh}.
%%%%%%%%%%%%%%%%%%%%%%%%%%%%%%%%%%%%%%%%%%%%%%%%%%%%%%%%%%%%%%%%%%%%%%%%%%%%%%%
%
% Chapter 4
%
% Renormalization of Composite Operators
%
%%%%%%%%%%%%%%%%%%%%%%%%%%%%%%%%%%%%%%%%%%%%%%%%%%%%%%%%%%%%%%%%%%%%%%%%%%%%%%%
\newpage
 \section{\bf\boldmath Renormalization of Composite Operator Matrix Elements}
  \label{Sec-REN}
  \renewcommand{\theequation}{\thesection.\arabic{equation}}
  \setcounter{equation}{0}
%%%%%%%%%%%%%%%%%%%%%%%%%%%%%%%%%%%%%%%%%%%%%%%%%%%%%%%%%%%%%%%%%%%%%%%%%%%%%%%
   Before renormalizing the massive OMEs, they have to be calculated applying 
   a suitable regularization scheme, for which we apply dimensional
   regularization in 
   $D=4+\ep$ dimensions, see Section~\ref{SubSec-RENReg}. The unrenormalized 
   massive OMEs are then denoted by a double--hat and are expanded into a 
   perturbative series in the bare coupling constant $\hat{a}_s$~\footnote{We would like to remind the reader of the definition of the hat--symbol for a function f, Eq. (\ref{gammapres1}), which is not to be confused with the hat--symbol denoting unrenormalized quantities} via
   \begin{eqnarray}
    \Ahathat_{ij}\Bigl(\frac{\hat{m}^2}{\mu^2},\ep,N\Bigr)&=&
                         \delta_{ij}+
                         \sum_{l=1}^{\infty}
                         \hat{a}_s^l~
                               \Ahathat_{ij}^{(l)}
                               \Bigl(\frac{\hat{m}^2}{\mu^2},\ep,N\Bigr) 
\N\\ 
                      &=&\delta_{ij}+
                        \sum_{l=1}^{\infty}
                        \hat{a}_s^l\Bigl(\frac{\hat{m}^2}{\mu^2}\Bigr)^{l\ep/2}
                             ~\Ahathat_{ij}^{(l)}
                              \Bigl(\hat{m}^2=\mu^2,\ep,N\Bigr)~.
                         \label{pertome1}
   \end{eqnarray}
   The OMEs in Eq.~(\ref{pertome1}) depend on $\ep$, the Mellin--Parameter $N$,
   the bare mass $\hat{m}$ and the renormalization scale $\mu=\mu_R$. Also the
   factorization scale $\mu_F$ will be identified with $\mu$ in the following.
   Note that in the last line of (\ref{pertome1}), the dependence on the ratio
   of the mass and the renormalization scale was made explicit for each order
   in $\hat{a}_s$. 
   The possible values of the indices $ij$ have been described in 
   Section~\ref{SubSec-HQAsym}, below Eq.~(\ref{pertomeren}).

   The factorization between the massive OMEs and the massless Wilson 
   coefficients (\ref{CallFAC}) requires the external legs of the operator 
   matrix elements to be on--shell, 
   \begin{eqnarray}
    \label{OS}
                      p^2 = 0~,
   \end{eqnarray}
   where $p$ denotes the external momentum. Unlike in the massless case, where
   the scale of the OMEs is set by an off--shell momentum $-p^2 < 0$, in our 
   framework the internal heavy quark mass yields the scale.
   In the former case, 
   one observes a mixing of the physical OMEs with non--gauge invariant (NGI) 
   operators, cf. \cite{Hamberg:1991qt,Collins:1994ee,*Harris:1994tp}, and 
   contributions originating in the violation of the equations of motion (EOM).
   Terms of this kind do not contribute in the present case, as will be 
   discussed in Section~\ref{SubSec-RENProj}.

   Renormalizing the OMEs then consists of four steps. First, mass and charge 
   renormalization have to performed. The former is done in the 
   on--mass--shell--scheme 
   and described in Section~\ref{SubSec-RENMa}. For the latter, we 
   present the final result in the $\overline{\sf MS}$--scheme, but in an 
   intermediate step, we adopt an on--shell subtraction scheme 
   (${\sf MOM}$--scheme) for the gluon 
   propagator, cf. Section~\ref{SubSec-RENCo}. This is necessary to maintain 
   condition (\ref{OS}), i.e., to keep the 
   external massless partons on--shell. Note, that there are other, differing 
   ${\sf MOM}$--schemes used in the literature,
   cf. e.g.~\cite{Chetyrkin:2008jk}. 

   After mass and coupling constant renormalization,
   we denote the OMEs with a single 
   hat, $\hat{A}_{ij}$. The remaining singularities are then connected to the 
   composite operators and the particle kinematics of the corresponding 
   Feynman--diagrams. One can distinguish between ultraviolet (UV) and 
   collinear (C) divergences. In Section~\ref{SubSec-RENOp}, we describe 
   how the former are renormalized via the operator 
   $Z$--factors. The UV--finite OMEs are denoted by a bar, $\bar{A}_{ij}$.
   Finally, the C--divergences are removed via mass factorization, 
   cf. Section~\ref{SubSec-RENMassFac}. The renormalized OMEs are then denoted
   by $A_{ij}$. Section~\ref{SubSec-RENPred} contains the general structure 
   of the massive OMEs up to $O(a_s^3)$ in terms of renormalization constants
   and lower order contributions.
%%%%%%%%%%%%%%%%%%%%%%%%%%%%%%%%%%%%%%%%%%%%%%%%%%%%%%%%%%%%%%%%%%%%%%%%%%%%%%%
  \subsection{\bf\boldmath Regularization Scheme}
   \label{SubSec-RENReg}
%%%%%%%%%%%%%%%%%%%%%%%%%%%%%%%%%%%%%%%%%%%%%%%%%%%%%%%%%%%%%%%%%%%%%%%%%%%%%%%
   When evaluating momentum integrals of 
   Feynman diagrams in $D=4$ dimensions, one encounters singularities,
   which have to be regularized. A convenient method  
   is to apply $D$-dimensional regularization, 
   \cite{'tHooft:1972fi,Ashmore:1972uj,*Cicuta:1972jf,*Bollini:1972ui}.
   The dimensionality of space--time is analytically continued to 
   values $D\neq 4$, for which the corresponding integrals converge. 
   After performing a Wick rotation, integrals in Euclidean space
   of the form
   \begin{eqnarray}
    \int \frac{d^Dk}{(2\pi)^D}\frac{(k^2)^r}{(k^2+R^2)^m}=
     \frac{1}{(4\pi)^{D/2}}\frac{\Gamma(r+D/2)\Gamma(m-r-D/2)}{\Gamma(D/2)
     \Gamma(m)}(R^2)^{r+D/2-m}~ \label{Dint}
   \end{eqnarray}
   are obtained. Note that within dimensional regularization, this integral 
   vanishes if $R=0$, i.e. if it 
   does not contain a scale, \cite{Collins:1984xc}.
   The properties of the $\Gamma$--function in the complex plane are well 
   known, 
   see Appendix~\ref{App-SpeFun}. Therefore one can analytically 
   continue the right-hand side of Eq.~(\ref{Dint}) from integer values of 
   $D$ to arbitrary complex values.
   In order to recover the physical space-time dimension, we set $D=4+\ep$.
   The singularities can now be isolated by expanding the $\Gamma$--functions
   into Laurent-series around $\ep=0$.
   Note that this method regularizes both UV- and C- singularities and 
   one could in principle distinguish their  
   origins by a label, $\ep_{UV},~\ep_{C}$, but 
   we treat all singularities by a common parameter $\ep$ in the following.
   Additionally, all other quantities have to be considered in $D$ dimensions.
   This applies for the metric tensor $g_{\mu\nu}$ 
    and the Clifford-Algebra of $\gamma$--matrices, 
   see Appendix~\ref{App-Con}.
   Also the 
   bare coupling constant $\hat{g}_s$, which is dimensionless in
   $D=4$, has to be continued to $D$ dimensions.
   Due to this it acquires the dimension of mass,
   \begin{eqnarray}
    \hat{g}_{s,D}=\mu^{-\ep/2}\hat{g}_s \label{mudef}~,
   \end{eqnarray}
   which is described by a scale $\mu$ corresponding 
   to the renormalization scale in Eq.~(\ref{pertome1}).
   From now on, Eq.~(\ref{mudef}) is understood to have been applied and 
   we set
   \begin{eqnarray}
    \frac{\hat{g}^2_s}{(4\pi)^2}=\hat{a}_s~. \label{asdef} 
   \end{eqnarray}
   Dimensional regularization has the advantage, unlike 
   the Pauli--Villars regularization, \cite{Pauli:1949zm}, 
   that it obeys all physical requirements such as 
   Lorentz-invariance, gauge invariance and unitarity,
   \cite{'tHooft:1972fi,Speer:1974cz}. 
   Hence it is suitable
   to be applied in perturbative calculations in quantum field theory including
   Yang--Mills fields. 

   Using dimensional regularization, the poles of the unrenormalized 
   results appear as terms $1/\ep^i$, where in the calculations
   in this thesis $i$ can run from $1$ to the number of loops. 
   In order to remove 
   these singularities, one has to perform renormalization and mass 
   factorization. To do this, a
   suitable scheme has to be chosen. The most commonly used
   schemes in perturbation theory are the ${\rm MS}$-scheme, 
   \cite{'tHooft:1973mm}, and the 
   $\overline{\sf MS}$-scheme, \cite{Bardeen:1978yd}, to which we will refer in 
   the following.   \\
   In the ${\rm MS}$-scheme
   only the pole terms in $\ep$ are subtracted. More generally, the 
   $\overline{\sf MS}$-scheme 
   makes use of the observation that $1/\ep$--poles always appear in 
   combination with the spherical factor 
   \begin{eqnarray}
     S_{\ep} \equiv \exp \Bigl[\frac{\ep}{2}(\gamma_E-\ln(4\pi))\Bigr]
     \label{Sep}~,
   \end{eqnarray}
   which may be bracketed out for each loop order.
   Here $\gamma_E$ denotes the Euler-Mascheroni constant
    \begin{eqnarray}
     \gamma_E\equiv\lim_{N\rightarrow \infty} \Bigl(\sum_{k=1}^{N}\frac{1}{k}
              -\ln(N)\Bigr)
             \approx 0.577215664901\ldots~. \label{gammaesum}
    \end{eqnarray}
   By subtracting the poles in the form $S_{\ep}/\ep$ in the 
   $\overline{\sf MS}$-scheme, no terms containing $\ln^k(4\pi ),~\gamma^k_E$
   will appear in the renormalized result, simplifying the expression.
   This is due to the fact that for a $k$--loop calculation, 
   one will always obtain the overall term
   \begin{eqnarray}
    \frac{\Gamma(1-k\frac{\ep}{2})}{(4\pi)^{\frac{k\ep}{2}}}&=&S_{\ep}^k
     \exp \Bigl(\sum_{i=2}^{\infty}\frac{\zeta_i}{i}
                        \Bigl(\frac{k\ep}{2}\Bigr)^{i}\Bigr)~, \label{SepA}
   \end{eqnarray}
   with $\zeta_i$ being Riemann's $\zeta$--values, 
   cf. Appendix~\ref{App-SpeFun}.
   In the following, we will always assume that the 
   $\overline{\sf MS}$-scheme is applied and set $S_{\ep}\equiv 1$.
%%%%%%%%%%%%%%%%%%%%%%%%%%%%%%%%%%%%%%%%%%%%%%%%%%%%%%%%%%%%%%%%%%%%%%%%%%%%%%%
  \subsection{\bf\boldmath Projectors }
   \label{SubSec-RENProj}
%%%%%%%%%%%%%%%%%%%%%%%%%%%%%%%%%%%%%%%%%%%%%%%%%%%%%%%%%%%%%%%%%%%%%%%%%%%%%%%
   We consider the expectation values of the local operators 
   (\ref{COMP1})--(\ref{COMP3}) between partonic states $j$
   \begin{eqnarray}
    G_{ij,Q}=\bra{j}O_i\ket{j}_Q~. \label{Greens}
   \end{eqnarray}
   Here, $i,j=q,g$ and the subscript $Q$ denotes the presence of one 
   heavy quark. 
   In case of massless QCD, one has to take
   the external parton $j$ of momentum $p$ off--shell, $p^2 < 0$, which 
   implies that the OMEs derived from Eq.~(\ref{Greens}) are not 
   gauge invariant. As has been outlined in 
   Ref.~\cite{Matiounine:1998ky}, they acquire unphysical parts 
   which are due to the breakdown of the equations of motion 
   (EOM) and the mixing with additional non--gauge--invariant (NGI)
   operators. The EOM terms 
   may be dealt with by applying a suitable projection 
   operator to eliminate them, \cite{Matiounine:1998ky}. The NGI 
   terms are more difficult to deal with, since they 
   affect the renormalization constants and one has to consider 
   additional ghost-- and alien-- OMEs, see 
   \cite{Hamberg:thesis,Hamberg:1991qt,Collins:1994ee,*Harris:1994tp,Matiounine:1998ky} for details. 

   In the case of massive OMEs, these difficulties do not occur. The 
   external particles are massless and taken to be on--shell. Hence 
   the equations of motion are not violated. Additionally, 
   the OMEs remain gauge invariant quantities, since the external states 
   are physical and therefore no mixing with NGI--operators occurs, 
   \cite{Hamberg:thesis,Hamberg:1991qt,Matiounine:1998ky,Collins:1984xc}. \\
   \noindent
   The computation of the Green's functions will reveal trace terms which 
   do not contribute since the local operators are traceless and symmetric 
   under the Lorentz group. It is convenient to project these 
   terms out from the beginning 
   by contracting with an external source term
   \begin{eqnarray}
    J_N \equiv \Delta_{\mu_1}...\Delta_{\mu_N}~. \label{Jsource}
   \end{eqnarray}
   Here $\Delta_{\mu}$ is a light-like vector, $\Delta^2 = 0$. 
   In this way, the 
   Feynman--rules for composite operators can be derived, cf.
   Appendix \ref{App-FeynRules}. In 
   addition, one has to amputate the external field. Note that we 
   nonetheless choose to renormalize the mass and the coupling
   multiplicative and include self--energy insertions 
   containing massive lines on external legs into our calculation.
   The Green's functions in momentum space corresponding to the OMEs 
   with external gluons are then given by 
   \begin{eqnarray} 
    \epsilon^\mu(p) G^{ab}_{Q,\mu\nu} \epsilon^\nu(p)&=&
    \epsilon^\mu(p)
    J_N \bra{A^a_{\mu}(p)} O_{Q;\mu_1 ... \mu_N} \ket{A^b_{\nu}(p)}
    \epsilon^\nu(p)~, 
        \label{GabmnQgdef} \\
    \epsilon^\mu(p) G^{ab}_{q,Q,\mu\nu} \epsilon^\nu(p)&=&
    \epsilon^\mu(p)
    J_N \bra{A^a_{\mu}(p)} O_{q;\mu_1 ... \mu_N} \ket{A^b_{\nu}(p)}_Q
    \epsilon^\nu(p)~,
     \label{GabmnqgQdef} \\
    \epsilon^\mu(p) G^{ab}_{g,Q,\mu\nu} \epsilon^\nu(p)&=&
    \epsilon^\mu(p)
    J_N \bra{A^a_{\mu}(p)} O_{g;\mu_1 ... \mu_N} \ket{A^b_{\nu}(p)}_Q
    \epsilon^\nu(p)~.
        \label{GabmnggQdef}
   \end{eqnarray}
   In Eqs. (\ref{GabmnQgdef}-\ref{GabmnggQdef}), $A^{a}_{\mu}$
   denote the external gluon fields with color index $a$, Lorentz 
   index $\mu$ and momentum $p$. 
   The polarization vector of the external gluon is given by 
   $\epsilon^{\mu}(p)$. Note that 
   in Eq.~(\ref{GabmnQgdef}), the operator couples to the heavy 
   quark. In Eqs. (\ref{GabmnqgQdef},~\ref{GabmnggQdef}) it couples 
   to a light quark or gluon, respectively, with the heavy quark still 
   being present in virtual loops. \\
   \noindent
   In the flavor non--singlet case, there is only one term which 
   reads 
   \begin{eqnarray} 
    \overline{u}(p,s) G^{ij, {\sf NS}}_{q,Q} \lambda_r u(p,s)&=&
    J_N
    \bra{\overline{\Psi}_i(p)}O_{q,r;\mu_1...\mu_N}^{\sf NS}\ket{\Psi^j(p)}_Q~
    \label{GijNS}~,
   \end{eqnarray}
   with $u(p,s),~\overline{u}(p,s)$ being the bi--spinors of the 
   external massless quark and anti--quark, respectively. 
   The remaining Green's functions with an outer quark are given by
   \begin{eqnarray} 
    \overline{u}(p,s) G^{ij}_{Q}  u(p,s)&=&
    J_N\bra{\overline{\Psi}_i(p)} O_{Q,\mu_1 ... \mu_N}  \ket{\Psi^j(p)}~, 
      \label{GijQqPS} \\ 
    \overline{u}(p,s) G^{ij}_{q,Q}  u(p,s)&=&
    J_N\bra{\overline{\Psi}_i(p)}O_{q,\mu_1...\mu_N}  \ket{\Psi^j(p)}_Q
       \label{GijqqQPS} ~, \\
    \overline{u}(p,s) G^{ij}_{g,Q}  u(p,s)&=&
    J_N\bra{\overline{\Psi}_i(p)}O_{g,\mu_1...\mu_N} \ket{\Psi^j(p)}_Q
      \label{GijgqQ}~. 
   \end{eqnarray}
   Note that in the quarkonic case the fields $\overline{\Psi},~\Psi$ 
   with color indices $i,j$ stand for the external light quarks only. Further, 
   we remind that the ${\sf S}$-- contributions are split up 
   according to Eq.~(\ref{splitS}), which is of relevance
   for Eq.~(\ref{GijqqQPS}).

   The above tensors have the general form, cf. 
   \cite{Buza:1995ie,Matiounine:1998ky}, 
   \begin{eqnarray} 
    \hat{G}^{ab}_{Q,\mu\nu}&=&
                       \Ahathat_{Qg}
                          \Bigl(\frac{\hat{m}^2}{\mu^2},\ep,N\Bigr)
                           \delta^{ab}
                          (\Delta \cdot p)^N 
                      \Big [- g_{\mu\nu}
                            +\frac{p_{\mu}\Delta_{\nu}+\Delta_{\mu}p_{\nu}}
                                  {\Delta \cdot p}
                      \Big ] ~, \label{omeGluOpQ} \\
    \hat{G}^{ab}_{l,Q,\mu\nu}&=&
                       \Ahathat_{lg,Q}
                          \Bigl(\frac{\hat{m}^2}{\mu^2},\ep,N\Bigr)
                           \delta^{ab}
                          (\Delta \cdot p)^N 
                      \Big [- g_{\mu\nu}
                            +\frac{p_{\mu}\Delta_{\nu}+\Delta_{\mu}p_{\nu}}
                                  {\Delta \cdot p}
                      \Big ] ~,~~ l=g,q~, \label{omeGluOpgq} \\
   \hat{G}_{Q}^{ij}     &=&
                    \Ahathat_{Qq}^{\sf PS}
                    \Bigl(\frac{\hat{m}^2}{\mu^2},\ep,N\Bigr)
                    \delta^{ij}  
                          (\Delta \cdot p)^{N-1}
                           \adag \Delta ~,\label{omeQuaPS} \\
   \hat{G}_{l,Q}^{ij,r}     &=&
                    \Ahathat_{lq,Q}^{r}
                    \Bigl(\frac{\hat{m}^2}{\mu^2},\ep,N\Bigr)
                    \delta^{ij}  
                          (\Delta \cdot p)^{N-1}
                           \adag \Delta ~, \quad l=g,q~,~
                           \quad r={\sf S,~NS,~PS}~. \label{omelqproj}
   \end{eqnarray}
   Here, we have denoted the Green's function with a hat to signify that 
   the above equations are written on the unrenormalized level.
   In order to simplify the evaluation, it is useful to define 
   projection operators which, applied to the Green's function, yield 
   the corresponding OME. For outer gluons, one defines
   \begin{eqnarray}
    P^{(1)}_g \hat{G}^{ab}_{l,(Q),\mu\nu}
               &\equiv& 
            - \frac{\delta_{ab}}{N_c^2-1} \frac{g^{\mu\nu}}{D-2}
               (\Delta\cdot p)^{-N} \hat{G}^{ab}_{l,(Q),\mu\nu} ~, 
               \label{projG1} \\
    P^{(2)}_g \hat{G}^{ab}_{l,(Q),\mu\nu} &\equiv& 
              \frac{\delta_{ab} }{N_c^2-1} \frac{1}{D-2}
              (\Delta\cdot p)^{-N}   
              \Bigl(-g^{\mu\nu}
                 +\frac{p^{\mu}\Delta^{\nu}+p^{\nu}\Delta^{\mu}}{\Delta\cdot p}
              \Bigr)\hat{G}^{ab}_{l,(Q),\mu\nu}
              ~. \label{projG2}
   \end{eqnarray}
   The difference between the gluonic projectors, Eq.  
   (\ref{projG1}) and Eq.~(\ref{projG2}), can be traced back to the fact 
   that in the former case, the summation over indices 
   $\mu,\nu$ includes unphysical transverse gluon states. These 
   have to be compensated by adding diagrams with external ghost 
   lines, which is not the case when using the physical projector in Eq.
   (\ref{projG2}).

   In the case of external quarks there is only one projector which reads
   \begin{eqnarray} 
    P_q \hat{G}^{ij}_{l,(Q)} &\equiv& 
              \frac{\delta^{ij}}{N_c} ( \Delta\cdot p)^{-N} 
               \frac{1}{4} {\sf Tr}[~\adag p~\hat{G}^{ij}_{l,(Q)}]~. 
               \label{projQ}
   \end{eqnarray}
   In Eqs. (\ref{projG1})--(\ref{projQ}), $N_c$ denotes the number of colors,
   cf. Appendix \ref{App-Con}.
   The unrenormalized OMEs are then obtained by
   \begin{eqnarray}
    \Ahathat_{lg}\Bigl(\frac{\hat{m}^2}{\mu^2},\ep,N\Bigr)
                      &=&P_g^{(1,2)}\hat{G}^{ab}_{l,(Q),\mu\nu}~,
                         \label{proGa}\\
    \Ahathat_{lq}\Bigl(\frac{\hat{m}^2}{\mu^2},\ep,N\Bigr)
                      &=&P_q \hat{G}^{ij}_{l,(Q)}~. \label{proQa}
   \end{eqnarray}
   The advantage of these projection operators is that 
   one does not have to resort to complicated tensorial reduction.
   In perturbation theory, the expressions in Eqs. 
   (\ref{proGa}, \ref{proQa}) can then be evaluated order by 
   order in the coupling constant by applying the
   Feynman-rules given in Appendix~\ref{App-FeynRules}.
%%%%%%%%%%%%%%%%%%%%%%%%%%%%%%%%%%%%%%%%%%%%%%%%%%%%%%%%%%%%%%%%%%%%%%%%%%%%%%%
  \subsection{\bf\boldmath Renormalization of the Mass}
   \label{SubSec-RENMa}
%%%%%%%%%%%%%%%%%%%%%%%%%%%%%%%%%%%%%%%%%%%%%%%%%%%%%%%%%%%%%%%%%%%%%%%%%%%%%%%
   In a first step, we perform mass renormalization.
   There are two traditional schemes for mass renormalization:
   the on--shell--scheme and the $\overline{\sf MS}$--scheme. In the following,
   we will 
   apply the on--shell--scheme, defining the renormalized mass $m$ as the pole 
   of the quark propagator. The differences to the $\overline{\sf MS}$--scheme
   will be discussed
   in Section~\ref{Sec-REP}. The bare mass in Eq.~(\ref{pertome1})
   is replaced by the renormalized on--shell mass $m$ via 
   \begin{eqnarray}
    \hat{m}&=&Z_m m
            =  m \Bigl[ 1 
                       + \hat{a}_s \Bigl(\frac{m^2}{\mu^2}\Bigr)^{\ep/2}
                                   \delta m_1 
                       + \hat{a}_s^2 \Bigl(\frac{m^2}{\mu^2}\Bigr)^{\ep}
                                     \delta m_2
                 \Bigr] + O(\hat{a}_s^3)~.
            \label{mren1}
   \end{eqnarray}
   The constants in the above equation are given by~\footnote{Note that there is a misprint in the double--pole term of Eq.~(28) in Ref.~\cite{Bierenbaum:2008yu}.}
   \begin{eqnarray}
    \delta m_1 &=&C_F
                  \Bigl[\frac{6}{\ep}-4+\Bigl(4+\frac{3}{4}\zeta_2\Bigr)\ep
                  \Bigr] \label{delm1}  \\
               &\equiv&  \frac{\delta m_1^{(-1)}}{\ep}
                        +\delta m_1^{(0)}
                        +\delta m_1^{(1)}\ep~, \label{delm1exp} \\
    \delta m_2 &=& C_F
                   \Biggl\{\frac{1}{\ep^2}\Bigl(18 C_F-22 C_A+8T_F(n_f+N_h)
                    \Bigr)
                  +\frac{1}{\ep}\Bigl(-\frac{45}{2}C_F+\frac{91}{2}C_A
 \N\\ &&
                   -14T_F(n_f+N_h)\Bigr)
                  +C_F\Bigl(\frac{199}{8}-\frac{51}{2}\zeta_2+48\ln(2)\zeta_2
                   -12\zeta_3\Bigr)
 \N\\ &&
                  +C_A\Bigl(-\frac{605}{8}
                  +\frac{5}{2}\zeta_2-24\ln(2)\zeta_2+6\zeta_3\Bigr)
 \N\\ &&
                  +T_F\Bigl[n_f\Bigl(\frac{45}{2}+10\zeta_2\Bigr)+N_h
                  \Bigl(\frac{69}{2}-14\zeta_2\Bigr)\Bigr]\Biggr\}
                  \label{delm2}  \\
               &\equiv&  \frac{\delta m_2^{(-2)}}{\ep^2}
                        +\frac{\delta m_2^{(-1)}}{\ep}
                        +\delta m_2^{(0)}~. \label{delm2exp}
   \end{eqnarray}
   Eq.~(\ref{delm1}) is easily obtained. 
   In Eq.~(\ref{delm2}), $n_f$ denotes the number of light flavors
   and $N_h$ the number of heavy flavors,
   which we will set equal to $N_h=1$ from now on.
   The pole contributions were given in 
   Refs.~\cite{Tarrach:1980up,Nachtmann:1981zg},
   and the constant term was derived 
   in Refs.~\cite{Gray:1990yh,*Broadhurst:1991fy}, 
   cf. also \cite{Fleischer:1998dw}.
   In Eqs. (\ref{delm1exp}, \ref{delm2exp}), we have defined 
   the expansion coefficients in $\ep$ of the corresponding quantities.
   After mass renormalization, the OMEs read up to $O(\hat{a}_s^3)$
   \begin{eqnarray}
    \Ahathat_{ij}\Bigl(\frac{m^2}{\mu^2},\ep,N\Bigr) 
                &=&\delta_{ij}+
                  \hat{a}_s~ 
                   \Ahathat_{ij}^{(1)}\Bigl(\frac{m^2}{\mu^2},\ep,N\Bigr) 
\N\\ 
    && \hspace{-25mm}
                         + \hat{a}_s^2 \left[~
                                        \Ahathat^{(2)}_{ij}
                                        \Bigl(\frac{m^2}{\mu^2},\ep,N\Bigr) 
                                      + \delta m_1 
                                        \Bigl(\frac{m^2}{\mu^2}\Bigr)^{\ep/2}
                                        \frac{md}{dm}~
                                                   \Ahathat_{ij}^{(1)}
                                           \Bigl(\frac{m^2}{\mu^2},\ep,N\Bigr) 
                                \right]
\N\\ && \hspace{-25mm}
                         + \hat{a}_s^3 \Biggl[~ 
                                         \Ahathat^{(3)}_{ij}
                                           \Bigl(\frac{m^2}{\mu^2},\ep,N\Bigr) 
                                        +\delta m_1 
                                         \Bigl(\frac{m^2}{\mu^2}\Bigr)^{\ep/2}
                                         \frac{md}{dm}~ 
                                                    \Ahathat_{ij}^{(2)}
                                           \Bigl(\frac{m^2}{\mu^2},\ep,N\Bigr)
\N\\ && \hspace{-15mm}
                                        + \delta m_2 
                                          \Bigl(\frac{m^2}{\mu^2}\Bigr)^{\ep}
                                          \frac{md}{dm}~ 
                                                    \Ahathat_{ij}^{(1)}
                                           \Bigl(\frac{m^2}{\mu^2},\ep,N\Bigr) 
                                        + \frac{\delta m_1^2}{2}
                                          \Bigl(\frac{m^2}{\mu^2}\Bigr)^{\ep}
                                                     \frac{m^2d^2}{dm^2}~
                                                     \Ahathat_{ij}^{(1)}
                                           \Bigl(\frac{m^2}{\mu^2},\ep,N\Bigr) 
    \Biggr]~. \N \\ \label{maren}
  \end{eqnarray}
%%%%%%%%%%%%%%%%%%%%%%%%%%%%%%%%%%%%%%%%%%%%%%%%%%%%%%%%%%%%%%%%%%%%%%%%%%%%%%%
  \subsection{\bf\boldmath Renormalization of the Coupling}
   \label{SubSec-RENCo}
%%%%%%%%%%%%%%%%%%%%%%%%%%%%%%%%%%%%%%%%%%%%%%%%%%%%%%%%%%%%%%%%%%%%%%%%%%%%%%%
  Next, we consider charge renormalization. 
  At this point it becomes important to define in which scheme 
  the strong coupling constant is renormalized,
  cf. Section~\ref{SubSec-HQAsym}. We briefly 
  summarize the main steps in the massless case for $n_f$ flavors in the 
  $\overline{\sf MS}$--scheme. 
  The bare coupling constant $\hat{a}_s$ is expressed by 
  the renormalized coupling $a_s^{\MS}$ via
  \begin{eqnarray}
   \hat{a}_s             &=& {Z_g^{\MS}}^2(\ep,n_f) 
                             a^{\MS}_s(\mu^2) \N\\
                         &=& a^{\MS}_s(\mu^2)\left[
                                   1 
                                 + \delta a^{\MS}_{s, 1}(n_f) 
                                      a^{\MS}_s(\mu^2)
                                 + \delta a^{\MS}_{s, 2}(n_f) 
                                      {a^{\MS}_s}^2(\mu^2)    
                                     \right] + O({a^{\MS}_s}^3)~. 
                            \label{asrenMSb}
  \end{eqnarray}
  The coefficients in Eq.~(\ref{asrenMSb}) are, 
  \cite{Khriplovich:1969aa,tHooft:unpub,Politzer:1973fx,Gross:1973id} and 
  \cite{Caswell:1974gg,*Jones:1974mm}, 
  \begin{eqnarray}
    \delta a^{\MS}_{s, 1}(n_f) &=& \frac{2}{\ep} \beta_0(n_f)~,
                             \label{deltasMSb1} \\
    \delta a^{\MS}_{s, 2}(n_f) &=& \frac{4}{\ep^2} \beta_0^2(n_f)
                           + \frac{1}{\ep} \beta_1(n_f)~,
                             \label{deltasMSb2}
  \end{eqnarray}
  with 
  \begin{eqnarray}
   \beta_0(n_f)
                 &=& \frac{11}{3} C_A - \frac{4}{3} T_F n_f \label{beta0}~, \\
   \beta_1(n_f)
                 &=& \frac{34}{3} C_A^2 
               - 4 \left(\frac{5}{3} C_A + C_F\right) T_F n_f \label{beta1}~.
  \end{eqnarray}
  From the above equations, one can determine the $\beta$--function,
  Eq.~(\ref{betdef1}), 
  which describes the running of the strong coupling constant and
  leads to asymptotic freedom in case of QCD,
  \cite{Politzer:1973fx,Gross:1973id}. It can be calculated using the fact that
  the bare strong coupling constant does not depend on the 
  renormalization scale $\mu$. Using Eq.~(\ref{mudef}), one obtains
  \begin{eqnarray}
   0&=&\frac{d\hat{a}_{s,D}}{d \ln \mu^2}
     =\frac{d}{d \ln \mu^2} \hat{a}_s \mu^{-\ep}
     =\frac{d}{d \ln \mu^2} a_s(\mu^2)Z_g^2(\ep,n_f,\mu^2) \mu^{-\ep}~, \\
   \Longrightarrow \beta &=&
                             \frac{\ep}{2}a_s(\mu^2)
                            -2a_s(\mu^2)
                            \frac{d}{d \ln \mu^2}\ln Z_g(\ep,n_f,\mu^2)~.
                                \label{betdef2}
  \end{eqnarray}
  Note that in Eq.~(\ref{betdef2}) we have not specified a scheme yet and kept 
  a possible $\mu$--dependence for $Z_g$, which is not present in case of the 
  $\overline{\sf MS}$--scheme. From (\ref{betdef2}), one can calculate the
  expansion coefficients of the $\beta$--function. Combining it with the result
  for $Z_g^{\MS}$ in Eqs. (\ref{deltasMSb1}, \ref{deltasMSb2}), one 
  obtains in the $\overline{\sf MS}$-scheme 
  for $n_f$ light flavors, cf. \cite{Khriplovich:1969aa,Gross:1973id,Politzer:1973fx,tHooft:unpub,Caswell:1974gg,*Jones:1974mm},
  \begin{eqnarray}
   \beta^{\MS}(n_f)&=&-\beta_0(n_f){a^{\MS}_s}^2-\beta_1(n_f){a^{\MS}_s}^3
                     +O({a^{\MS}_s}^4)~.
  \end{eqnarray}
  Additionally, it follows
  \begin{eqnarray}
   \frac{d a_s(\mu^2)}{d \ln(\mu^2)} 
     &=& \frac{1}{2} \ep a_s(\mu^2)-\sum_{k=0}^\infty
               \beta_k a_s^{k+2}(\mu^2)~.
         \label{runningas}
  \end{eqnarray}
  The factorization relation (\ref{CallFAC}) strictly requires that the 
  external massless particles are on--shell. Massive loop corrections to the 
  gluon propagator violate this condition, which has to be 
  enforced subtracting the corresponding corrections. These can be uniquely 
  absorbed into the strong coupling constant applying the background field 
  method,~\cite{Abbott:1980hw,*Rebhan:1985yf,*Jegerlehner:1998zg}, to maintain 
  the Slavnov-Taylor identities of QCD. We thus determine the coupling constant
  renormalization in the $\overline{\sf MS}$-scheme as far as the light flavors and the gluon
  are concerned. In addition, we make the choice that the heavy quark decouples
  in the running coupling constant $a_s(\mu^2)$ for $\mu^2 < m^2$ and thus from
  the renormalized OMEs. This implies the requirement that $\Pi_H(0, m^2) = 0$,
  where $\Pi_H(p^2, m^2)$ is the contribution to the gluon self-energy due to 
  the heavy quark loops, \cite{Buza:1995ie}. 
  Since this condition introduces higher order terms in 
  $\ep$ into $Z_g$, we left the $\overline{\sf MS}$--scheme. This new scheme
  is a ${\sf MOM}$--scheme.
  After mass renormalization in the on--shell--scheme via Eq.~(\ref{mren1}), we 
  obtain for the heavy quark contributions to the gluon self--energy in the 
  background field formalism
  \begin{eqnarray}
   \hat{\Pi}^{\mu\nu}_{H,ab,\mbox{\tiny{BF}}}(p^2,m^2,\mu^2,\ep,\hat{a}_s)&=&
                     i(-p^2g^{\mu\nu}+p^{\mu}p^{\nu})\delta_{ab}
          \hat{\Pi}_{H,\mbox{\tiny{BF}}}(p^2,m^2,\mu^2,\ep,\hat{a}_s)~, \N\\
   \hat{\Pi}_{H,\mbox{\tiny{BF}}}(0,m^2,\mu^2,\ep,\hat{a}_s)&=&
                    \hat{a}_s   \frac{2\beta_{0,Q}}{\ep}
                         \Bigl(\frac{m^2}{\mu^2}\Bigr)^{\ep/2}
                          \exp \Bigl(\sum_{i=2}^{\infty}\frac{\zeta_i}{i}
                          \Bigl(\frac{\ep}{2}\Bigr)^{i}\Bigr)
\N\\ && \hspace{-18mm}
                   +\hat{a}_s^2 \Bigl(\frac{m^2}{\mu^2}\Bigr)^{\ep}
                        \Biggl[
                       \frac{1}{\ep}\Bigl(
                                          -\frac{20}{3}T_FC_A
                                          -4T_FC_F
                                    \Bigr)
                      -\frac{32}{9}T_FC_A
                      +15T_FC_F
\N\\ &&  \hspace{-18mm}
                     +\ep            \Bigl(
                                          -\frac{86}{27}T_FC_A
                                          -\frac{31}{4}T_FC_F
                                          -\frac{5}{3}\zeta_2T_FC_A
                                          -\zeta_2T_FC_F
                                   \Bigr)
                         \Biggl]~, \label{GluSelfBack}
  \end{eqnarray}
  with 
  \begin{eqnarray}
   \beta_{0,Q} &=&\hat{\beta}_0(n_f)=-\frac{4}{3}T_F~. \label{b0Q}
  \end{eqnarray}
  Note that Eq.~(\ref{GluSelfBack}) holds only up to order $O(\ep)$, although 
  we have partially included higher orders in $\ep$ in order to keep the 
  expressions shorter. We have used the Feynman--rules of the background 
  field formalism as given in Ref.~\cite{Yndurain:1999ui}.
  In the following, we define 
  \begin{eqnarray}
   f(\ep)&\equiv&
                 \Bigl(\frac{m^2}{\mu^2}\Bigr)^{\ep/2}
    \exp \Bigl(\sum_{i=2}^{\infty}\frac{\zeta_i}{i}
                       \Bigl(\frac{\ep}{2}\Bigr)^{i}\Bigr)~. \label{fep}
  \end{eqnarray}
  The renormalization constant of the background field $Z_A$ 
  is related to $Z_g$ via 
  \begin{eqnarray} 
   Z_A=Z_g^{-2}~. \label{ZAZg}
  \end{eqnarray}
  The light flavor contributions to $Z_A$, $Z_{A,l}$, can thus be determined 
  by combining Eqs.~(\ref{deltasMSb1},~\ref{deltasMSb2},~\ref{ZAZg}). 
  The heavy flavor part follows from the condition 
  \begin{eqnarray}
   \Pi_{H,\mbox{\tiny{BF}}}(0,m^2)+Z_{A,H}\equiv 0~, \label{ZAcond}
  \end{eqnarray}
  which ensures that the on--shell gluon remains strictly massless. 
  Thus we newly define the renormalization constant 
  of the strong coupling with $n_f$ light and one heavy flavor as 
  \begin{eqnarray}
   Z^{\MOM}_g(\ep,n_f+1,\mu^2,m^2)
         \equiv \frac{1}{(Z_{A,l}+Z_{A,H})^{1/2}}~ \label{Zgnfp1}
  \end{eqnarray}
  and obtain
  \begin{eqnarray}
   {Z_g^{\MOM}}^2(\ep,n_f+1,\mu^2,m^2)&=&
                  1+a^{\MOM}_s(\mu^2) \Bigl[
                              \frac{2}{\ep} (\beta_0(n_f)+\beta_{0,Q}f(\ep))
                        \Bigr]
\N\\ && \hspace{-18mm}
                  +{a^{\MOM}_s}^2(\mu^2) \Bigl[
                                \frac{\beta_1(n_f)}{\ep}
                         +\frac{4}{\ep^2} (\beta_0(n_f)+\beta_{0,Q}f(\ep))^2
\N\\ && \hspace{-18mm}
                          +\frac{1}{\ep}\Bigl(\frac{m^2}{\mu^2}\Bigr)^{\ep}
                           \Bigl(\beta_{1,Q}+\ep\beta_{1,Q}^{(1)}
                                            +\ep^2\beta_{1,Q}^{(2)}
                           \Bigr)
                          \Bigr]+O({a^{\MOM}_s}^3)~, \label{Zgheavy2}
  \end{eqnarray}
  with 
  \begin{eqnarray}
   \beta_{1,Q} &=&\hat{\beta_1}(n_f)=
                  - 4 \left(\frac{5}{3} C_A + C_F \right) T_F~, \label{b1Q} \\
   \beta_{1,Q}^{(1)}&=&
                           -\frac{32}{9}T_FC_A
                           +15T_FC_F~, \label{b1Q1} \\
   \beta_{1,Q}^{(2)}&=&
                               -\frac{86}{27}T_FC_A
                               -\frac{31}{4}T_FC_F
                               -\zeta_2\left(\frac{5}{3}T_FC_A
                                        +T_FC_F\right)~. \label{b1Q2}
  \end{eqnarray}
  The coefficients corresponding to Eq.~(\ref{asrenMSb}) then read
  in the ${\sf MOM}$--scheme
  \begin{eqnarray}
   \delta a_{s,1}^{\MOM}&=&\Bigl[\frac{2\beta_0(n_f)}{\ep}
                           +\frac{2\beta_{0,Q}}{\ep}f(\ep)
                            \Bigr]~,\label{dela1} \\
   \delta a_{s,2}^{\MOM}&=&\Bigl[\frac{\beta_1(n_f)}{\ep}+
                            \Bigl(\frac{2\beta_0(n_f)}{\ep}
                              +\frac{2\beta_{0,Q}}{\ep}f(\ep)\Bigr)^2
\N\\ &&
                          +\frac{1}{\ep}\Bigl(\frac{m^2}{\mu^2}\Bigr)^{\ep}
                           \Bigl(\beta_{1,Q}+\ep\beta_{1,Q}^{(1)}
                                            +\ep^2\beta_{1,Q}^{(2)}
                           \Bigr)\Bigr] +O(\ep^2)~.\label{dela2}
  \end{eqnarray}

  Since the $\overline{\sf MS}$--scheme is commonly used, we transform our
  results back from
  the {\sf MOM}--description into the $\overline{\sf MS}$--scheme, in order to be able to
  compare to other analyzes. This is achieved by observing that the bare
  coupling does not change under this transformation and one obtains the
  condition
  \begin{eqnarray}
      {Z_g^{\MS}}^2(\ep,n_f+1) a^{\MS}_s(\mu^2) = 
      {Z_g^{\MOM}}^2(\ep,n_f+1,\mu^2,m^2) a^{\MOM}_s(\mu^2) \label{condas1}~.
  \end{eqnarray}
  The following relations hold~:
  \begin{eqnarray}
   a_s^{\MOM}&=& a_s^{\MS}
                -\beta_{0,Q}\ln \Bigl(\frac{m^2}{\mu^2}\Bigr) {a_s^{\MS}}^2
\N \\ &&
                +\Biggl[ \beta^2_{0,Q}\ln^2 \Bigl(\frac{m^2}{\mu^2}\Bigr) 
                        -\beta_{1,Q}\ln \Bigl(\frac{m^2}{\mu^2}\Bigr) 
                        -\beta_{1,Q}^{(1)}
                 \Biggr] {a_s^{\MS}}^3
                         +O({a_s^{\MS}}^4)~, \label{asmoma}
  \end{eqnarray}
  or, 
  \begin{eqnarray}
   a_s^{\MS}&=&
               a_s^{\MOM}
              +{a_s^{\MOM}}^2\Biggl(
                          \delta a^{\MOM}_{s, 1}
                         -\delta a^{\MS}_{s, 1}(n_f+1)
                             \Biggr)
              +{a_s^{\MOM}}^{3}\Biggl(
                          \delta a^{\MOM}_{s, 2}
                         -\delta a^{\MS}_{s, 2}(n_f+1)
     \N\\ &&
                        -2\delta a^{\MS}_{s, 1}(n_f+1)\Bigl[
                             \delta a^{\MOM}_{s, 1}
                            -\delta a^{\MS}_{s, 1}(n_f+1)
                                                      \Bigr]
                             \Biggr)+O({a_s^{\MOM}}^4)~, \label{asmsa}
  \end{eqnarray}
  vice versa. 
  Eq.~(\ref{asmsa}) is valid to all orders in $\ep$. 
  Here, $a_s^{\sf \MS} = a_s^{\sf \MS}(n_f + 1)$. Applying the on--shell--scheme
  for mass renormalization and the described {\sf MOM}--scheme for the 
  renormalization of the coupling, one obtains as general formula for mass and 
  coupling constant renormalization up to $O({a^{\MOM}_s}^3)$
  \begin{eqnarray}
   {\hat{A}}_{ij} &=&  \delta_{ij} 
                     + a^{\MOM}_s~\Ahathat_{ij}^{(1)}
                     + {a^{\MOM}_s}^2 \left[~\Ahathat^{(2)}_{ij}
                     + \delta m_1 \Bigl(\frac{m^2}{\mu^2}\Bigr)^{\ep/2} 
                                 m  \frac{d}{dm}~\Ahathat_{ij}^{(1)}
                     + \delta a^{\MOM}_{s,1}~\Ahathat_{ij}^{(1)}\right]
\N\\ &&
   + {a^{\MOM}_s}^3 \Biggl[~\Ahathat^{(3)}_{ij}
   + \delta a^{\MOM}_{s,2}~\Ahathat_{ij}^{(1)}
   + 2 \delta a^{\MOM}_{s,1} \left(~\Ahathat^{(2)}_{ij} 
   +  \delta m_1 \Bigl(\frac{m^2}{\mu^2}\Bigr)^{\ep/2}
      m \frac{d}{dm}~\Ahathat_{ij}^{(1)} \right)
\N\\ &&
                + \delta m_1 \Bigl(\frac{m^2}{\mu^2}\Bigr)^{\ep/2}
                           m  \frac{d}{dm}~\Ahathat_{ij}^{(2)}
                + \delta m_2 \Bigl(\frac{m^2}{\mu^2}\Bigr)^{\ep}
                          m  \frac{d}{dm}~\Ahathat_{ij}^{(1)}
\N\\ &&
                + \frac{\delta m_1^2}{2} \Bigl(\frac{m^2}{\mu^2}\Bigr)^{\ep}
                m^2  \frac{d^2}{{dm}^2}~\Ahathat_{ij}^{(1)}
    \Biggr]~,\label{macoren}
  \end{eqnarray}
  where we have suppressed the dependence on $m,~\ep$ and $N$ in the arguments~\footnote{Here we corrected a typographical error in \cite{Bierenbaum:2008yu}, Eq.~(48).}.  
%%%%%%%%%%%%%%%%%%%%%%%%%%%%%%%%%%%%%%%%%%%%%%%%%%%%%%%%%%%%%%%%%%%%%%%%%%%%%%%
  \subsection{\bf\boldmath Operator Renormalization}
   \label{SubSec-RENOp}
%%%%%%%%%%%%%%%%%%%%%%%%%%%%%%%%%%%%%%%%%%%%%%%%%%%%%%%%%%%%%%%%%%%%%%%%%%%%%%%
    The renormalization of the UV singularities of the composite operators is 
    being performed introducing the corresponding $Z_{ij}$-factors, which have
    been 
    defined in Eqs. (\ref{ZNSdef}, \ref{ZSijdef}). We consider first only
    $n_f$ massless flavors, cf.~\cite{Matiounine:1998ky}, and do then 
    include subsequently one heavy quark. In the former case, 
    renormalization proceeds in the $\overline{\sf MS}$--scheme via
    \begin{eqnarray}
     A_{qq}^{\sf NS}\Bigl(\frac{-p^2}{\mu^2},a_s^{\MS},n_f,N\Bigr)
            &=&Z^{-1,{\sf NS}}_{qq}(a_s^{\MS},n_f,\ep,N)
               \hat{A}_{qq}^{\sf NS}\Bigl(\frac{-p^2}{\mu^2},a_s^{\MS},n_f
                 ,\ep,N\Bigr)  \label{renAqqnf}~, \\
     A_{ij}\Bigl(\frac{-p^2}{\mu^2},a_s^{\MS},n_f,N\Bigr)
           &=&Z^{-1}_{il}(a_s^{\MS},n_f,\ep,N)
              \hat{A}_{lj}\Bigl(\frac{-p^2}{\mu^2},a_s^{\MS},n_f,\ep,N\Bigr)
                   ~,~i,j=q,g, \N\\
              \label{renAijnf}
   \end{eqnarray}
   with $p$ a space--like momentum. As is well known, 
   operator mixing occurs in the singlet case, Eq.~(\ref{renAijnf}).
   As mentioned before, we neglected all terms being associated to EOM
   and NGI parts, since they do not contribute in the renormalization of the
   massive on--shell operator matrix elements. The ${\sf NS}$ and ${\sf PS}$
   contributions are separated via
   \begin{eqnarray}
    Z_{qq}^{-1}&=&Z_{qq}^{-1, {\sf PS}}+Z_{qq}^{-1, {\sf NS}}~,\\
    A_{qq}     &=&A_{qq}^{\sf PS}+A_{qq}^{\sf NS} \label{ZPSNS}~. 
   \end{eqnarray}
   The anomalous dimensions $\gamma_{ij}$ of the operators are defined in Eqs. 
   (\ref{gammazetNS}, \ref{gammazetS})
   and can be expanded in a perturbative series as follows
   \begin{eqnarray}
    \gamma_{ij}^{{\sf S,PS,NS}}(a_s^{\MS},n_f,N)
        &=&\sum_{l=1}^{\infty}{a^{\MS}_s}^l 
                      \gamma_{ij}^{(l), {\sf S,PS,NS}}(n_f,N)~.
            \label{pertgamma}
   \end{eqnarray}
   Here, the ${\sf PS}$ contribution starts at $O(a_s^2)$. 
   In the following, we do not write the dependence
   on the Mellin--variable $N$ for the OMEs, the operator $Z$--factors 
   and the anomalous dimensions explicitly. Further, we 
   will suppress the dependence on $\ep$ for unrenormalized quantities 
   and $Z$--factors. 
   From Eqs. (\ref{gammazetNS}, \ref{gammazetS}), one can determine 
   the relation between the anomalous dimensions and the $Z$--factors 
   order by order in perturbation theory. In the general case, one finds
   up to $O({a_s^{\MS}}^3)$
   \begin{eqnarray}
    Z_{ij}(a^{\MS}_s,n_f) &=&
                            \delta_{ij}
                           +a^{\MS}_s \frac{\gamma_{ij}^{(0)}}{\ep}
                           +{a^{\MS}_s}^2 \Biggl\{
                                 \frac{1}{\ep^2} \Bigl(
                                     \frac{1}{2} \gamma_{il}^{(0)}
                                                 \gamma_{lj}^{(0)}
                                   + \beta_0 \gamma_{ij}^{(0)}
                                                 \Bigr)
                               + \frac{1}{2 \ep} \gamma_{ij}^{(1)}
                                   \Biggr\}
 \N \\ &&
                           + {a^{\MS}_s}^3 \Biggl\{
                                 \frac{1}{\ep^3} \Bigl(
                                     \frac{1}{6}\gamma_{il}^{(0)}
                                                \gamma_{lk}^{(0)}
                                                \gamma_{kj}^{(0)}
                                   + \beta_0 \gamma_{il}^{(0)} 
                                             \gamma_{lj}^{(0)}
                                   + \frac{4}{3} \beta_0^2 \gamma_{ij}^{(0)}
                                                  \Bigr)
\N\\ &&
                               + \frac{1}{\ep^2}  \Bigl(
                                     \frac{1}{6} \gamma_{il}^{(1)} 
                                                 \gamma_{lj}^{(0)}
                                   + \frac{1}{3} \gamma_{il}^{(0)} 
                                                 \gamma_{lj}^{(1)}
                                   + \frac{2}{3} \beta_0 \gamma_{ij}^{(1)} 
                                   + \frac{2}{3} \beta_1 \gamma_{ij}^{(0)}
                                                  \Bigr)
                              + \frac{\gamma_{ij}^{(2)}}{3 \ep}
                                   \Biggr\}~. \label{Zijnf}
  \end{eqnarray}
  The ${\sf NS}$ and ${\sf PS}$ $Z$--factors are given by~\footnote{In Eq.~(\ref{ZqqPSnf}) we corrected typographical errors contained in Eq.~(34), \cite{Bierenbaum:2008yu}.}
  \begin{eqnarray}
   Z_{qq}^{\sf NS}(a^{\MS}_s,n_f) &=& 
                             1 
                           +a^{\MS}_s \frac{\gamma_{qq}^{(0),{\sf NS}}}{\ep}
                           +{a^{\MS}_s}^2 \Biggl\{
                                 \frac{1}{\ep^2} \Bigl(
                                     \frac{1}{2}{\gamma_{qq}^{(0),{\sf NS}}}^2 
                                   + \beta_0 \gamma_{qq}^{(0),{\sf NS}}
                                                 \Bigr)
                              + \frac{1}{2 \ep} \gamma_{qq}^{(1),{\sf NS}} 
                                        \Biggr\}
\N\\ &&
                           +{a^{\MS}_s}^3 \Biggl\{
                                 \frac{1}{\ep^3} \Bigl(
                                     \frac{1}{6} {\gamma_{qq}^{(0),{\sf NS}}}^3
                                   + \beta_0 {\gamma_{qq}^{(0),{\sf NS}}}^2 
                                   + \frac{4}{3} \beta_0^2 
                                                 \gamma_{qq}^{(0),{\sf NS}}
                                                 \Bigr)
\N\\ &&
                               + \frac{1}{\ep^2} \Bigl(
                                     \frac{1}{2} \gamma_{qq}^{(0),{\sf NS}} 
                                                 \gamma_{qq}^{(1),{\sf NS}}
                                    +\frac{2}{3} \beta_0 
                                                 \gamma_{qq}^{(1),{\sf NS}}
                                    +\frac{2}{3} \beta_1 
                                                 \gamma_{qq}^{(0),{\sf NS}} 
                                                 \Bigr)
                               + \frac{1}{3 \ep} \gamma_{qq}^{(2), {\sf NS}} 
                                        \Biggr\}~, \N \\ \label{ZqqNSnf}\\
   Z_{qq}^{\sf PS}(a^{\MS}_s,n_f) &=&
                            {a^{\MS}_s}^2 \Biggl\{
                                 \frac{1}{2\ep^2} \gamma_{qg}^{(0)}
                                                  \gamma_{gq}^{(0)}   
                               + \frac{1}{2\ep}   \gamma_{qq}^{(1), {\sf PS}}
                                        \Biggr\}
                           +{a^{\MS}_s}^3 \Biggl\{
                                 \frac{1}{\ep^3} \Bigl(
                                     \frac{1}{3}\gamma_{qq}^{(0)} 
                                      \gamma_{qg}^{(0)}
                                      \gamma_{gq}^{(0)}
\N\\ &&
                                    +\frac{1}{6}\gamma_{qg}^{(0)}
                                     \gamma_{gg}^{(0)} \gamma_{gq}^{(0)}
                                    +\beta_0 \gamma_{qg}^{(0)}
                                             \gamma_{gq}^{(0)}
                                                 \Bigr)
                               + \frac{1}{\ep^2} \Bigl(
                                     \frac{1}{3}\gamma_{qg}^{(0)} 
                                                \gamma_{gq}^{(1)}
\N\\ &&
                                    +\frac{1}{6}\gamma_{qg}^{(1)} 
                                                \gamma_{gq}^{(0)}
                                    +\frac{1}{2} \gamma_{qq}^{(0)}
                                                 \gamma_{qq}^{(1), {\sf PS}}
                                    +\frac{2}{3} \beta_0
                                                 \gamma_{qq}^{(1), {\sf PS}}
                                                 \Bigr)
                                    +\frac{\gamma_{qq}^{(2), {\sf PS}}}{3\ep} 
                                        \Biggr\}~. \label{ZqqPSnf}
  \end{eqnarray}
  All quantities in Eqs. (\ref{Zijnf})--(\ref{ZqqPSnf}) refer to 
  $n_f$ light flavors and renormalize 
  the massless off--shell OMEs given in Eqs. 
  (\ref{renAqqnf}, \ref{renAijnf}). 
  
  In the next step, we consider an additional heavy quark with mass $m$. 
  We keep the external momentum artificially
  off--shell for the moment, in order to deal with the 
  UV--singularities only.
  For the additional massive quark, one has to account for
  the renormalization of the coupling constant we defined in 
  Eqs.~(\ref{dela1}, \ref{dela2}). The $Z$--factors including one massive 
  quark are then obtained
  by taking Eqs. (\ref{Zijnf})--(\ref{ZqqPSnf}) at $(n_f+1)$
  flavors and performing
  the scheme transformation given in (\ref{asmsa}). The
  emergence of $\delta a_{s,k}^{\sf MOM}$ in $Z_{ij}$ is due to the finite 
  mass effects and cancels singularities which emerge for real radiation and 
  virtual processes at $p^2 \rightarrow 0$.  Thus one obtains up to 
  $O({a_s^{\MOM}}^3)$
%%%%%%%%%%%%%%%%%%%%%%%%%%%%%%%%%
%
%
%
%%%%%%%%%%%%%%%%%%%%%%%%%%%%%%%%%
  \begin{eqnarray}
   Z_{ij}^{-1}(a_s^{\MOM},n_f+1,\mu^2)&=&
      \delta_{ij}
     -a_s^{\MOM}\frac{\gamma_{ij}^{(0)}}{\ep}
     +{a^{\MOM}_s}^2\Biggl[
          \frac{1}{\ep}\Bigl(
                       -\frac{1}{2}\gamma_{ij}^{(1)}
                       -\delta a^{\MOM}_{s,1}\gamma_{ij}^{(0)}
                       \Bigr)
\N\\ &&
         +\frac{1}{\ep^2}\Bigl(
                        \frac{1}{2}\gamma_{il}^{(0)}\gamma_{lj}^{(0)}
                       +\beta_0\gamma_{ij}^{(0)}
                        \Bigr)
         \Biggr]
     +{a^{\MOM}_s}^3\Biggl[
          \frac{1}{\ep}\Bigl(
                       -\frac{1}{3}\gamma_{ij}^{(2)}
                       -\delta a^{\MOM}_{s,1}\gamma_{ij}^{(1)}
\N\\ &&
                       -\delta a^{\MOM}_{s,2}\gamma_{ij}^{(0)}
                       \Bigr)
         +\frac{1}{\ep^2}\Bigl(
                        \frac{4}{3}\beta_0\gamma_{ij}^{(1)}
                       +2\delta a^{\MOM}_{s,1}\beta_0\gamma_{ij}^{(0)}
                       +\frac{1}{3}\beta_1\gamma_{ij}^{(0)}
\N\\ &&
                       +\delta a^{\MOM}_{s,1}\gamma_{il}^{(0)}\gamma_{lj}^{(0)}
                       +\frac{1}{3}\gamma_{il}^{(1)}\gamma_{lj}^{(0)}
                       +\frac{1}{6}\gamma_{il}^{(0)}\gamma_{lj}^{(1)}
                        \Bigr)
         +\frac{1}{\ep^3}\Bigl(
                       -\frac{4}{3}\beta_0^{2}\gamma_{ij}^{(0)}
\N\\ &&
                       -\beta_0\gamma_{il}^{(0)}\gamma_{lj}^{(0)}
                       -\frac{1}{6}\gamma_{il}^{(0)}\gamma_{lk}^{(0)}
                                    \gamma_{kj}^{(0)} 
                     \Bigr)
       \Biggr]~, \label{ZijInfp1}
  \end{eqnarray}
%%%%%%%%%%%%%%%%%%%%%%%%%%%%%%%
%
%
%
%%%%%%%%%%%%%%%%%%%%%%%%%%%%%%%
  and
  \begin{eqnarray}
   Z_{qq}^{-1,{\sf NS}}(a_s^{\MOM},n_f+1,\mu^2)&=&
      1
     -a^{\MOM}_s\frac{\gamma_{qq}^{(0),{\sf NS}}}{\ep}
     +{a^{\MOM}_s}^2\Biggl[
          \frac{1}{\ep}\Bigl(
                       -\frac{1}{2}\gamma_{qq}^{(1),{\sf NS}}
                       -\delta a^{\MOM}_{s,1}\gamma_{qq}^{(0),{\sf NS}}
                       \Bigr)
\N\\ &&
         +\frac{1}{\ep^2}\Bigl(
                        \beta_0\gamma_{qq}^{(0),{\sf NS}}
                       +\frac{1}{2}{\gamma_{qq}^{(0),{\sf NS}}}^{2}
                        \Bigr)
         \Biggr]
     +{a^{\MOM}_s}^3\Biggl[
          \frac{1}{\ep}\Bigl(
                       -\frac{1}{3}\gamma_{qq}^{(2),{\sf NS}}
\N\\  &&
                       -\delta a^{\MOM}_{s,1}\gamma_{qq}^{(1),{\sf NS}}
                       -\delta a^{\MOM}_{s,2}\gamma_{qq}^{(0),{\sf NS}}
                       \Bigr)
         +\frac{1}{\ep^2}\Bigl(
                       \frac{4}{3}\beta_0\gamma_{qq}^{(1),{\sf NS}}
\N\\  &&
                      +2\delta a^{\MOM}_{s,1}\beta_0\gamma_{qq}^{(0),{\sf NS}}
                       +\frac{1}{3}\beta_1\gamma_{qq}^{(0),{\sf NS}}
                       +\frac{1}{2}\gamma_{qq}^{(0),{\sf NS}}
                                   \gamma_{qq}^{(1),{\sf NS}}
\N\\ &&
                       +\delta a^{\MOM}_{s,1}{\gamma_{qq}^{(0),{\sf NS}}}^{2}
                        \Bigr)
         +\frac{1}{\ep^3}\Bigl(
                       -\frac{4}{3}\beta_0^{2}\gamma_{qq}^{(0),{\sf NS}}
\N\\ &&
                       -\beta_0{\gamma_{qq}^{(0),{\sf NS}}}^{2}
                       -\frac{1}{6}{\gamma_{qq}^{(0),{\sf NS}}}^{3}
                     \Bigr)
       \Biggr]~, \label{ZNSInfp1} \\
%%%%%%%%%%%%%%%%%%%%%%%%%%%%%%%%
%
%
%
%%%%%%%%%%%%%%%%%%%%%%%%%%%%%%%%
   Z_{qq}^{-1,{\sf PS}}(a_s^{\MOM},n_f+1,\mu^2)&=&
      {a^{\MOM}_s}^2\Biggl[
          \frac{1}{\ep}\Bigl(
                       -\frac{1}{2}\gamma_{qq}^{(1), {\sf PS}}
                       \Bigr)
         +\frac{1}{\ep^2}\Bigl(
                        \frac{1}{2}\gamma_{qg}^{(0)}\gamma_{gq}^{(0)}
                        \Bigr)
         \Biggr]
 \N\\  &&
     +{a^{\MOM}_s}^3\Biggl[
          \frac{1}{\ep}\Bigl(
                       -\frac{1}{3}\gamma_{qq}^{(2), {\sf PS}}
                       -\delta a^{\MOM}_{s,1}\gamma_{qq}^{(1), {\sf PS}}
                       \Bigr)
         +\frac{1}{\ep^2}\Bigl(
                        \frac{1}{6}\gamma_{qg}^{(0)}\gamma_{gq}^{(1)}
 \N\\  &&
                       +\frac{1}{3}\gamma_{gq}^{(0)}\gamma_{qg}^{(1)}
                       +\frac{1}{2}\gamma_{qq}^{(0)}\gamma_{qq}^{(1), {\sf PS}}
                       +\frac{4}{3}\beta_0\gamma_{qq}^{(1), {\sf PS}}
                       +\delta a^{\MOM}_{s,1}\gamma_{qg}^{(0)}\gamma_{gq}^{(0)}
                        \Bigr)
  \N\\  &&
         +\frac{1}{\ep^3}\Bigl(
                       -\frac{1}{3}\gamma_{qg}^{(0)}\gamma_{gq}^{(0)}
                                   \gamma_{qq}^{(0)}
                       -\frac{1}{6}\gamma_{gq}^{(0)}\gamma_{qg}^{(0)}
                                   \gamma_{gg}^{(0)}
                       -\beta_0\gamma_{qg}^{(0)}\gamma_{gq}^{(0)}
                     \Bigr)
       \Biggr]~.\N\\ \label{ZPSInfp1}
  \end{eqnarray} 
%%%%%%%%%%%%%%%%%%%%%%%%%%%%%%%%
  The above equations are given for $n_f+1$ flavors.
  One re-derives the expressions for $n_f$ light flavors 
  by setting $(n_f+1) =: n_f$ and $\delta a^{\MOM}_s=\delta a^{\MS}_s$.
  As a next step, we split the OMEs into
  a part involving only light flavors and the heavy flavor part 
  \begin{eqnarray}
    {\hat{A}}_{ij}(p^2,m^2,\mu^2,a_s^{\MOM},n_f+1)&=&
                {\hat{A}}_{ij}\Bigl(\frac{-p^2}{\mu^2},a_s^{\MS},n_f\Bigr)
\N\\ &&
             + {\hat{A}}^Q_{ij}(p^2,m^2,\mu^2,a_s^{\MOM},n_f+1)~.
   \label{splitNSHL1}
  \end{eqnarray}
  In (\ref{splitNSHL1}, \ref{eqXX}), the light flavor part depends on 
  $a_s^{\MS}$, since
  the prescription adopted for coupling constant renormalization only applies
  to the massive part. ${\hat{A}}^Q_{ij}$ denotes any massive OME we consider.
  The correct UV--renormalization prescription for the massive contribution 
  is obtained
  by subtracting from Eq.~(\ref{splitNSHL1}) the terms applying to the light
  part only~:
  \begin{eqnarray}
   \bar{A}^Q_{ij}(p^2,m^2,\mu^2,a_s^{\MOM},n_f+1)&=&
               Z^{-1}_{il}(a_s^{\MOM},n_f+1,\mu^2) 
                  \hat{A}^Q_{ij}(p^2,m^2,\mu^2,a_s^{\MOM},n_f+1)
\N\\ &&
              +Z^{-1}_{il}(a_s^{\MOM},n_f+1,\mu^2) 
                   \hat{A}_{ij}\Bigl(\frac{-p^2}{\mu^2},a_s^{\MS},n_f\Bigr)
\N\\ &&
              -Z^{-1}_{il}(a_s^{\MS},n_f,\mu^2)
                   \hat{A}_{ij}\Bigl(\frac{-p^2}{\mu^2},a_s^{\MS},n_f\Bigr)~,
  \label{eqXX} 
  \end{eqnarray}
  where
  \begin{eqnarray}
   Z_{ij}^{-1} = \delta_{ij} + \sum_{k=1}^\infty a_s^k Z_{ij}^{-1, (k)}~.
  \end{eqnarray}
  In the limit $p^2=0$, integrals without a scale vanish within dimensional
  regularization. Hence for the light flavor OMEs only the term 
  $\delta_{ij}$ remains and
  one obtains the UV--finite massive OMEs after expanding in $a_s$
  \begin{eqnarray}
  \bar{A}^Q_{ij}\Bigl(\frac{m^2}{\mu^2},a_s^{\MOM},n_f+1\Bigr) &=& 
             a_s^{\MOM}\Biggl( \hat{A}_{ij}^{(1),Q}
                              \Bigl(\frac{m^2}{\mu^2}\Bigr)
                     +Z^{-1,(1)}_{ij}(n_f+1,\mu^2)
                     -Z^{-1,(1)}_{ij}(n_f)
               \Biggr)
\N\\ && \hspace{-45mm} 
            + {a_s^{\MOM}}^2\Biggl( \hat{A}_{ij}^{(2),Q}
                                    \Bigl(\frac{m^2}{\mu^2}\Bigr)
                       +Z^{-1,(2)}_{ij}(n_f+1,\mu^2)
                       -Z^{-1,(2)}_{ij}(n_f)
\N\\ && \hspace{-45mm}   \phantom{{a_s^{\MOM}}^2\Biggl(}
     +Z^{-1,(1)}_{ik}(n_f+1,\mu^2)
                        \hat{A}_{kj}^{(1),Q}\Bigl(\frac{m^2}{\mu^2}\Bigr)
               \Biggr)
\N\\ && \hspace{-45mm} 
           +{a_s^{\MOM}}^3\Biggl( \hat{A}_{ij}^{(3),Q}
                                   \Bigl(\frac{m^2}{\mu^2}\Bigr)
                       +Z^{-1,(3)}_{ij}(n_f+1,\mu^2)
                       -Z^{-1,(3)}_{ij}(n_f)
  \N\\ && \hspace{-45mm}  \phantom{{a_s^{\MOM}}^3\Biggl(}
                       +Z^{-1,(1)}_{ik}(n_f+1,\mu^2)
                        \hat{A}_{kj}^{(2),Q}\Bigl(\frac{m^2}{\mu^2}\Bigr)
                       +Z^{-1,(2)}_{ik}(n_f+1,\mu^2)
                        \hat{A}_{kj}^{(1),Q}\Bigl(\frac{m^2}{\mu^2}\Bigr)
                        \Biggr)~. \label{GenRen1}
  \end{eqnarray}
  The $Z$--factors at $n_f+1$ flavors refer
  to Eqs. (\ref{ZijInfp1})--(\ref{ZPSInfp1}), whereas those 
  at $n_f$ flavors correspond to the massless case. 
%%%%%%%%%%%%%%%%%%%%%%%%%%%%%%%%%%%%%%%%%%%%%%%%%%%%%%%%%%%%%%%%%%%%%%%%%%%%%%%
  \subsection{\bf\boldmath Mass Factorization}
   \label{SubSec-RENMassFac}
%%%%%%%%%%%%%%%%%%%%%%%%%%%%%%%%%%%%%%%%%%%%%%%%%%%%%%%%%%%%%%%%%%%%%%%%%%%%%%%
  Finally, we have to remove the collinear singularities contained in 
  $\bar{A}_{ij}$, which emerge in the limit $p^2 =  0$. They  
  are absorbed into the parton distribution functions and are not present in 
  case of the off--shell massless OMEs.
  As a generic renormalization formula, generalizing Eqs. (\ref{renAqqnf}, 
  \ref{renAijnf}), one finds
  \begin{eqnarray}
   A_{ij}&=&Z^{-1}_{il} \hat{A}_{lk} \Gamma_{kj}^{-1}~. \label{genren}
  \end{eqnarray}
  The renormalized operator matrix elements are obtained by 
  \begin{eqnarray}
   A^Q_{ij}\Bigl(\frac{m^2}{\mu^2},a_s^{\MOM},n_f+1\Bigr)&=&
     \bar{A}^Q_{il}\Bigl(\frac{m^2}{\mu^2},a_s^{\MOM},n_f+1\Bigr)
     \Gamma_{lj}^{-1}~. \label{genren1}
  \end{eqnarray}
  If all quarks were  massless, the identity, \cite{Buza:1995ie},  
  \begin{eqnarray}
    \Gamma_{ij} = Z^{-1}_{ij}~. \label{GammaZ}
  \end{eqnarray}
  would hold.  However, due to the presence of a heavy quark $Q$, the
  transition functions $\Gamma(n_f)$ refer only to massless sub-graphs.  Hence
  the $\Gamma$--factors contribute up to $O(a_s^2)$ only and do not involve
  the special scheme adopted for the renormalization of the coupling. Due to
  Eq.~(\ref{GammaZ}), they can be read off from
  Eqs. (\ref{Zijnf})--(\ref{ZqqPSnf}).
  The renormalized operator matrix elements are then given by:
  \begin{eqnarray}
   && A^Q_{ij}\Bigl(\frac{m^2}{\mu^2},a_s^{\MOM},n_f+1\Bigr)=
\N\\&&\phantom{+}
                a^{\MOM}_s~\Biggl(
                      \hat{A}_{ij}^{(1),Q}\Bigl(\frac{m^2}{\mu^2}\Bigr)
                     +Z^{-1,(1)}_{ij}(n_f+1)
                     -Z^{-1,(1)}_{ij}(n_f)
                           \Biggr)
\N\\&&
           +{a^{\MOM}_s}^2\Biggl( 
                        \hat{A}_{ij}^{(2),Q}\Bigl(\frac{m^2}{\mu^2}\Bigr)
                       +Z^{-1,(2)}_{ij}(n_f+1)
                       -Z^{-1,(2)}_{ij}(n_f)
                       +Z^{-1,(1)}_{ik}(n_f+1)\hat{A}_{kj}^{(1),Q}
                                              \Bigl(\frac{m^2}{\mu^2}\Bigr)
\N\\
&&\phantom{+{a^{\MOM}_s}^2\Biggl(}
                       +\Bigl[ \hat{A}_{il}^{(1),Q}
                               \Bigl(\frac{m^2}{\mu^2}\Bigr)
                              +Z^{-1,(1)}_{il}(n_f+1)
                              -Z^{-1,(1)}_{il}(n_f)
                        \Bigr] 
                             \Gamma^{-1,(1)}_{lj}(n_f)
                         \Biggr)
\N\\ 
&&
          +{a^{\MOM}_s}^3\Biggl( 
                        \hat{A}_{ij}^{(3),Q}\Bigl(\frac{m^2}{\mu^2}\Bigr)
                       +Z^{-1,(3)}_{ij}(n_f+1)
                       -Z^{-1,(3)}_{ij}(n_f)
                       +Z^{-1,(1)}_{ik}(n_f+1)\hat{A}_{kj}^{(2),Q}
                                              \Bigl(\frac{m^2}{\mu^2}\Bigr)
\N\N\\ 
&&\phantom{+{a^{\MOM}_s}^3\Biggl(}
                       +Z^{-1,(2)}_{ik}(n_f+1)\hat{A}_{kj}^{(1),Q}
                                              \Bigl(\frac{m^2}{\mu^2}\Bigr)
                       +\Bigl[ 
                               \hat{A}_{il}^{(1),Q}
                                 \Bigl(\frac{m^2}{\mu^2}\Bigr)
                              +Z^{-1,(1)}_{il}(n_f+1)
 \N\\ &&\phantom{+{a^{\MOM}_s}^3\Biggl(}
                              -Z^{-1,(1)}_{il}(n_f)
                        \Bigr]
                              \Gamma^{-1,(2)}_{lj}(n_f)
                       +\Bigl[ 
                               \hat{A}_{il}^{(2),Q}
                                 \Bigl(\frac{m^2}{\mu^2}\Bigr)
                              +Z^{-1,(2)}_{il}(n_f+1)
                              -Z^{-1,(2)}_{il}(n_f)
 \N\\ 
&&\phantom{+{a^{\MOM}_s}^3\Biggl(}
                              +Z^{-1,(1)}_{ik}(n_f+1)\hat{A}_{kl}^{(1),Q}
                                              \Bigl(\frac{m^2}{\mu^2}\Bigr)
                        \Bigr]
                              \Gamma^{-1,(1)}_{lj}(n_f)
                        \Biggr)+O({a_s^{\MOM}}^4)~. \label{GenRen3}
  \end{eqnarray}
%%%%%%%%%%%%%%%%%%%%%%%%%%%%%%%%%%%%%%%%%%%%
  From (\ref{GenRen3}) it is obvious that the renormalization of $A^Q_{ij}$ to
  $O(a_s^3)$ requires the $1$--loop terms up to $O(\ep^2)$ and the $2$--loop
  terms up to $O(\ep)$, cf.~\cite{Buza:1995ie,Buza:1996wv,Bierenbaum:2007qe,
    Bierenbaum:2008yu,Bierenbaum:2009zt}. These terms are calculated in 
  Section~\ref{Sec-2L}. 
  Finally, we transform the coupling
  constant back into the $\overline{\sf MS}$--scheme by
  using Eq.~(\ref{asmoma}). We do not
  give the explicit formula here, but present the individual renormalized OMEs
  after this transformation in the next Section as perturbative series in
  $a_s^{\MS}$,
  \begin{eqnarray}
   A_{ij}^Q\Bigl(\frac{m^2}{\mu^2},a_s^{\MS},n_f+1\Bigr)&=&
    \delta_{ij}+
     a_s^{\MS}     A_{ij}^{Q, (1)}\Bigl(\frac{m^2}{\mu^2},n_f+1\Bigr)
   +{a_s^{\MS}}^2  A_{ij}^{Q, (2)}\Bigl(\frac{m^2}{\mu^2},n_f+1\Bigr)
\N \\ \phantom{A_{ij}^Q\Bigl(\frac{m^2}{\mu^2},a_s^{\MS},n_f+1\Bigr)}&& \!\!
   +{a_s^{\MS}}^3  A_{ij}^{Q, (3)}\Bigl(\frac{m^2}{\mu^2},n_f+1\Bigr)
    +O({a_s^{\MS}}^4)~.
    \label{PertOmeren} 
  \end{eqnarray}
  As stated in Section~\ref{Sec-HQDIS}, one has to use the same scheme when 
  combining the massive OMEs with the massless Wilson coefficients in the 
  factorization formula (\ref{CallFAC}). The effects of the transformation 
  between the ${\sf MOM}$-- and $\overline{\sf MS}$--scheme are discussed 
  in Section~\ref{Sec-REP}.
  The subscript $Q$ was introduced in this Section to make the distinction 
  between the massless and massive OMEs explicit and will be dropped from now 
  on, since no confusion is expected. Comparing Eqs. (\ref{GenRen3})
  and (\ref{PertOmeren}), one notices that the term $\delta_{ij}$ is not 
  present in the former because it was subtracted together with the 
  light flavor contributions. However, as one infers from
  Eq.~(\ref{CallFAC}) and the discussion below, this term is necessary when 
  calculating the massive Wilson coefficients in the asymptotic limit and we
  therefore have re--introduced it into Eq.~(\ref{PertOmeren}).
%%%%%%%%%%%%%%%%%%%%%%%%%%%%%%%%%%%%%%%%%%%%%%%%%%%%%%%%%%%%%%%%%%%%%%%%%%%%
  \subsection{\bf\boldmath General Structure of the Massive Operator Matrix
                           Elements}
   \label{SubSec-RENPred}
%%%%%%%%%%%%%%%%%%%%%%%%%%%%%%%%%%%%%%%%%%%%%%%%%%%%%%%%%%%%%%%%%%%%%%%%%%%%
   In the following, we present the general structure of the 
   unrenormalized and renormalized massive 
   operator matrix elements for the specific partonic channels. 
   The former are expressed as a Laurent--series in $\ep$ via
   \begin{eqnarray}
    \Ahathat_{ij}^{(l)}\Bigl(\frac{\hat{m}^2}{\mu^2},\ep,N\Bigr) &=&
      \Bigl(\frac{\hat{m}^2}{\mu^2}\Bigr)^{l\ep/2}
        \sum_{k=0}^{\infty} \frac{a^{(l,k)}_{ij}}{\ep^{l-k}}~. 
   \end{eqnarray}
   Additionally, we set 
   \begin{eqnarray}
           a^{(l,l)}_{ij}\equiv a^{(l)}_{ij}~,\quad
           a^{(l,l+1)}_{ij}\equiv \overline{a}^{(l)}_{ij}~, \mbox{etc.}~.
   \end{eqnarray}
   The pole terms can all be expressed by known
   renormalization constants and lower order contributions to the massive
   OMEs, which provides us with a strong check on our
   calculation. In particular, the complete ${\sf NLO}$ anomalous 
   dimensions, as well as their $T_F$--terms at ${\sf NNLO}$, 
   contribute at $O(a_s^3)$.
   The moments of the $O(\ep^0)$--terms of the unrenormalized OMEs at the 
   $3$--loop level, $a_{ij}^{(3)}$, are a new result of this 
   thesis and will be calculated in Section~\ref{Sec-3L},
   cf. \cite{Bierenbaum:2009mv}. 
   The $O(\ep)$ terms at the $2$--loop level, 
   $\overline{a}_{ij}^{(2)}$, contribute to the non--logarithmic part of the 
   renormalized $3$--loop OMEs and are calculated for general values of 
   $N$ in Section~\ref{Sec-2L}, cf. \cite{Bierenbaum:2008yu,Bierenbaum:2009zt}.
   The pole terms and the $O(\ep^0)$ terms, $a_{ij}^{(2)}$, at $2$--loop 
   have been calculated for the first time in 
   Refs.~\cite{Buza:1995ie,Buza:1996wv}. The terms involving the quark 
   operator, (\ref{COMP1}, \ref{COMP2}), were confirmed in 
   \cite{Bierenbaum:2007qe} and the terms involving the gluon 
   operator (\ref{COMP3}) by the present work, cf. 
   \cite{Bierenbaum:2009zt}. 
   In order to keep up with the notation used in
   \cite{Buza:1995ie,Buza:1996wv}, we define the 2--loop terms
   $a_{ij}^{(2)},~\overline{a}_{ij}^{(2)}$ {\sf after} performing mass
   renormalization in the on--shell--scheme. This we {\sf do not} apply
   for the $3$--loop terms.  
   We choose to calculate one--particle reducible 
   diagrams and therefore have to include external self--energies containing 
   massive quarks into our calculation. 
   Before presenting the operator matrix elements up to three loops, we first 
   summarize the necessary self--energy contributions in the next Section.
   The remaining Sections, (\ref{Sec-NS})--(\ref{SubSec-AggQ}), 
   contain 
   the general structure of the unrenormalized and renormalized massive OMEs
   up to $3$--loops.
   In these Sections, we always proceed as follows: 
   From Eqs. (\ref{macoren},~\ref{GenRen3}), one predicts the pole terms 
   of the respective unrenormalized OMEs by demanding that these terms have 
   to cancel through renormalization. The unrenormalized expressions are then
   renormalized in the ${\sf MOM}$--scheme. Finally, Eq.~(\ref{asmoma}) is
   applied and the renormalized massive OMEs are presented in the 
   $\overline{\sf MS}$--scheme.
%%%%%%%%%%%%%%%%%%%%%%%%%%%%%%%%%%%%%%%%%%%%%%%%%%%%%%%%%%%%%%%%%%%%%%%%%
 \subsubsection{Self--energy contributions}
  \label{Sec-elf}
%%%%%%%%%%%%%%%%%%%%%%%%%%%%%%%%%%%%%%%%%%%%%%%%%%%%%%%%%%%%%%%%%%%%%%%%%
  The gluon and quark self-energy contributions due to heavy quark lines
  are given by
   \begin{eqnarray}
    \hat{\Pi}_{\mu\nu}^{ab}(p^2,\hat{m}^2,\mu^2,\hat{a}_s) &=& i\delta^{ab}
                            \left[-g_{\mu\nu}p^2 +p_\mu p_\nu\right] 
                            \hat{\Pi}(p^2,\hat{m}^2,\mu^2,\hat{a}_s)~,  \\
   \hat{\Pi}(p^2,\hat{m}^2,\mu^2,\hat{a}_s)&=&
        \sum_{k=1}^{\infty}\hat{a}_s^k\hat{\Pi}^{(k)}(p^2,\hat{m}^2,\mu^2). \\
        \label{pertPiGlu}
   \hat{\Sigma}_{ij}(p^2,\hat{m}^2,\mu^2,\hat{a}_s)&=&
                      i~~\delta_{ij}~\adag p~~ 
                      \hat{\Sigma}(p^2,\hat{m}^2,\mu^2,\hat{a_s})~, \\
   \hat{\Sigma}(p^2,\hat{m}^2,\mu^2,\hat{a}_s)&=&
   \sum_{k=2}^{\infty}\hat{a}_s^k\hat{\Sigma}^{(k)}(p^2,\hat{m}^2,\mu^2)~.
        \label{pertSiQu}
  \end{eqnarray}
  Note, that the quark self--energy contributions start at 2--loop order. 
  These self--energies are easily calculated using {\sf MATAD}, 
  \cite{Steinhauser:2000ry}, cf. Section~\ref{Sec-3L}. 
  The expansion coefficients for $p^2=0$ of Eqs.~(\ref{pertPiGlu},~\ref{pertSiQu})
  are needed for the calculation of the gluonic and quarkonic 
  OMEs, respectively.
  The contributions to the gluon vacuum polarization for general gauge 
  parameter $\xi$ are
  \begin{eqnarray}
  \label{eqPI1}
   \hat{\Pi}^{(1)}\Bigl(0,\frac{\hat{m}^2}{\mu^2}\Bigr)&=&
            T_F\Bigl(\frac{\hat{m}^2}{\mu^2}\Bigr)^{\ep/2} 
            \Biggl(
             -\frac{8}{3\ep}
              \exp \Bigl(\sum_{i=2}^{\infty}\frac{\zeta_i}{i}
                       \Bigl(\frac{\ep}{2}\Bigr)^{i}\Bigr)
             \Biggr)~,
               ~\label{GluSelf1} \\
   \hat{\Pi}^{(2)}\Bigl(0,\frac{\hat{m}^2}{\mu^2}\Bigr)&=&
      T_F\Bigl(\frac{\hat{m}^2}{\mu^2}\Bigr)^{\ep}\Biggl(
      -\frac{4}{\ep^2} C_A + \frac{1}{\ep} \Bigl\{-12 C_F + 5 C_A\Bigr\} +
      C_A \Bigl(\frac{13}{12} -\zeta_2\Bigr) - \frac{13}{3} C_F
   \N\\ &&\hspace{-10mm}
      + \ep \left\{C_A \Bigl(\frac{169}{144} + \frac{5}{4} \zeta_2 - 
      \frac{\zeta_3}{3} \Bigr) + C_F \Bigl( - \frac{35}{12} -3 \zeta_2 
       \Bigr) \right\}\Biggr) + O(\ep^2)~,  \label{GluSelf2} \\
   \hat{\Pi}^{(3)}\Bigl(0,\frac{\hat{m}^2}{\mu^2}\Bigr)&=&
       T_F\Bigl(\frac{\hat{m}^2}{\mu^2}\Bigr)^{3\ep/2}\Biggl(
                        \frac{1}{\ep^3}\Biggl\{
                                 -\frac{32}{9}T_FC_A\Bigl(2n_f+1\Bigr)
                                 +C_A^2\Bigl( 
                                              \frac{164}{9}
                                             +\frac{4}{3}\xi
                                       \Bigr)
                                       \Biggr\}
\N\\ 
&&\hspace{-10mm}
                       +\frac{1}{\ep^2}\Biggl\{
                               \frac{80}{27}\Bigl(
                                                  C_A-6C_F
                                                \Bigr)n_fT_F
                              +\frac{8}{27}   \Bigl(
                                                  35C_A-48C_F
                                                \Bigr)T_F
                              +\frac{C_A^2}{27} \Bigl(
                                                  -781
\N\\ &&\hspace{-10mm}
                                                  +63\xi
                                                \Bigr)
                              +\frac{712}{9}C_AC_F
                                       \Biggr\}
                       +\frac{1}{\ep}\Biggl\{
                                \frac{4}{27}\Bigl(
                                                       C_A(-101-18\zeta_2)
                                                      -62C_F
                                                \Bigr)n_fT_F
\N\\ &&\hspace{-10mm}
                              +\frac{2}{27}   \Bigl(
                                                       C_A(-37-18\zeta_2)
                                                       -80C_F
                                                \Bigr)T_F
                              +C_A^2            \Bigl(
                                                  -12\zeta_3
                                                  +\frac{41}{6}\zeta_2
                                                  +\frac{3181}{108}
                                                  +\frac{\zeta_2}{2}\xi
\N\\ &&\hspace{-10mm}
                                                  +\frac{137}{36}\xi
                                                \Bigr)
                              +C_AC_F           \Bigl(
                                                   16\zeta_3
                                                  -\frac{1570}{27}
                                                \Bigr)
                              +\frac{272}{3}C_F^2
                                       \Biggr\}
                       +n_fT_F    \Biggl\{
                                       C_A\Bigl(
                                             \frac{56}{9}\zeta_3
                                            +\frac{10}{9}\zeta_2
\N\\ &&\hspace{-10mm}
                                            -\frac{3203}{243}
                                          \Bigr)
                                      +C_F\Bigl(
                                            -\frac{20}{3}\zeta_2
                                            -\frac{1942}{81}
                                          \Bigr)
                                       \Biggr\}
                       +T_F      \Biggl\{
                                       C_A\Bigl(
                                            -\frac{295}{18}\zeta_3
                                            +\frac{35}{9}\zeta_2
                                            +\frac{6361}{486}
                                          \Bigr)
\N\\ &&\hspace{-10mm}
                                      +C_F\Bigl(
                                            -7\zeta_3
                                            -\frac{16}{3}\zeta_2
                                            -\frac{218}{81}
                                          \Bigr)
                                   \Biggr\}
                       +C_A^2      \Biggl\{
                                       4{\sf B_4}
                                      -27\zeta_4
                                      +\frac{1969}{72}\zeta_3
                                      -\frac{781}{72}\zeta_2
\N\\ &&\hspace{-10mm}
                                      +\frac{42799}{3888}
                                      -\frac{7}{6}\zeta_3\xi
                                      +\frac{7}{8}\zeta_2\xi
                                      +\frac{3577}{432}\xi
                                   \Biggr\}
                       +C_AC_F      \Biggl\{
                                      -8{\sf B_4}
                                      +36\zeta_4
                                      -\frac{1957}{12}\zeta_3\N
   \end{eqnarray}
   \begin{eqnarray}
%\N\\
&&
                                      +\frac{89}{3}\zeta_2
                                      +\frac{10633}{81}
                                   \Biggr\}
                       +C_F^2      \Biggl\{
                                      \frac{95}{3}\zeta_3
                                      +\frac{274}{9}
                                   \Biggr\}
                                                   \Biggr) + O(\ep)~, 
                                          \label{GluSelf3}
   \end{eqnarray}
   and for the quark self--energy,
   \begin{eqnarray}
    \hat{\Sigma}^{(2)}(0,\frac{\hat{m}^2}{\mu^2}) &=&
        T_F C_F \Bigl(\frac{\hat{m}^2}{\mu^2}\Bigr)^{\ep} \left\{\frac{2}{\ep} 
        +\frac{5}{6} + \left[\frac{89}{72} + \frac{\zeta_2}{2} \right] 
             \ep \right\} + O(\ep^2)~, \label{QuSelf2} \\
    \hat{\Sigma}^{(3)}(0,\frac{\hat{m}^2}{\mu^2}) &=&
        T_F C_F \Bigl(\frac{\hat{m}^2}{\mu^2}\Bigr)^{3\ep/2}
        \Biggl(
                        \frac{8}{3\ep^3}C_A \{1-\xi\}
                       +\frac{1}{\ep^2} \Bigl\{
                                    \frac{32}{9}T_F(n_f+2)
                                   -C_A\Bigl(\frac{40}{9}+4\xi\Bigr)
\N\\
&&
                                   -\frac{8}{3}C_F
                                        \Bigl\}
                       +\frac{1}{\ep} \Biggl\{
                                       \frac{40}{27}T_F(n_f+2)
                                      +C_A\Bigl\{
                                             \zeta_2
                                            +\frac{454}{27}
                                            -\zeta_2\xi
                                            -\frac{70}{9}\xi
                                          \Bigr\}
                                       -26C_F
                                        \Biggl\}
\N\\
&&
                       +n_fT_F\Bigl\{
                                \frac{4}{3}\zeta_2
                               +\frac{674}{81}
                              \Bigr\}
                       +  T_F\Bigl\{
                                \frac{8}{3}\zeta_2
                               +\frac{604}{81}
                              \Bigr\}
                       +  C_A\Bigl\{
                                \frac{17}{3}\zeta_3
                               -\frac{5}{3}\zeta_2
                               +\frac{1879}{162}
\N\\
&&
                               +\frac{7}{3}\zeta_3\xi
                               -\frac{3}{2}\zeta_2\xi
                               -\frac{407}{27}\xi
                              \Bigr\}
                       +  C_F\Bigl\{
                               -8\zeta_3
                               -\zeta_2
                               -\frac{335}{18}
                              \Bigr\}
        \Biggr) + O(\ep)~,  \label{QuSelf3}
   \end{eqnarray}
   see also~\cite{Chetyrkin:1999ysxChetyrkin:1999qi,Chetyrkin:2008jk}.
   In Eq.~(\ref{GluSelf3}) the constant
   \begin{eqnarray}
          {\sf B_4}&=&-4\zeta_2\ln^2(2) +\frac{2}{3}\ln^4(2) 
           -\frac{13}{2}\zeta_4
                  +16 {\sf Li}_4\Bigl(\frac{1}{2}\Bigr)
                 ~\approx~  -1.762800093...~  \label{B4}
   \end{eqnarray}
   appears due to genuine massive effects, cf. \cite{Broadhurst:1991fi,Avdeev:1994db,*Laporta:1996mq,Broadhurst:1998rz,Boughezal:2004ef}.
%%%%%%%%%%%%%%%%%%%%%%%%%%%%%%%%%%%%%%%%%%%%%%%%%%%%%%%%%%%%%%%%%%%%%%%%%
%
% Subsubsection
%
% Non--Singlet
%
%%%%%%%%%%%%%%%%%%%%%%%%%%%%%%%%%%%%%%%%%%%%%%%%%%%%%%%%%%%%%%%%%%%%%%%%%
 \subsubsection{$A_{qq,Q}^{\sf NS}$}
  \label{Sec-NS}
%%%%%%%%%%%%%%%%%%%%%%%%%%%%%%%%%%%%%%%%%%%%%%%%%%%%%%%%%%%%%%%%%%%%%%%%%
  The lowest non--trivial ${\sf NS}$--contribution is of $O(a_s^2)$,
  \begin{eqnarray}
   A_{qq,Q}^{\sf NS}&=&1
                       +a_s^2A_{qq,Q}^{(2), {\sf NS}}
                       +a_s^3A_{qq,Q}^{(3), {\sf NS}}
                       +O(a_s^4)~. \label{NSpert}
  \end{eqnarray}
  The expansion coefficients are  obtained in the ${\sf MOM}$--scheme
  from the bare quantities, using Eqs.~(\ref{macoren},~\ref{GenRen3}). 
  After mass-- and coupling constant renormalization, the OMEs are given by
  \begin{eqnarray}
    A_{qq,Q}^{(2), \sf NS, \MOM}&=&
                    \hat{A}_{qq,Q}^{(2),{\sf NS},\MOM}
                   +Z^{-1,(2), {\sf NS}}_{qq}(n_f+1)
                   -Z^{-1,(2), {\sf NS}}_{qq}(n_f)
                ~, \label{2LNSRen1} \\
    A_{qq,Q}^{(3), \sf NS, \MOM}&=&
                     \hat{A}_{qq,Q}^{(3), {\sf NS},\MOM}
                    +Z^{-1,(3), {\sf NS}}_{qq}(n_f+1)
                    -Z^{-1,(3), {\sf NS}}_{qq}(n_f)
\N\\ &&
                    +Z^{-1,(1), {\sf NS}}_{qq}(n_f+1)
                     \hat{A}_{qq,Q}^{(2), {\sf NS},\MOM}
                    +\Bigl[ \hat{A}_{qq,Q}^{(2), {\sf NS},\MOM}
\N\\ &&
                           +Z^{-1,(2), {\sf NS}}_{qq}(n_f+1)
                           -Z^{-1,(2), {\sf NS}}_{qq}(n_f)
                     \Bigr]\Gamma^{-1,(1)}_{qq}(n_f)
                ~. \label{3LNSRen1}
   \end{eqnarray}
   From (\ref{macoren}, \ref{GenRen3}, \ref{2LNSRen1}, \ref{3LNSRen1}), one 
   predicts the pole terms of the unrenormalized OME. 
   At second and third order they read
   \begin{eqnarray}
   \Ahathat_{qq,Q}^{(2),\sf NS}&=&
           \Bigl(\frac{\hat{m}^2}{\mu^2}\Bigr)^{\ep}\Biggl(
                    \frac{\beta_{0,Q}\gamma_{qq}^{(0)}}{\ep^2}
                   +\frac{\hat{\gamma}_{qq}^{(1), {\sf NS}}}{2\ep}
                   +a_{qq,Q}^{(2),{\sf NS}}
                   +\overline{a}_{qq,Q}^{(2),{\sf NS}}\ep
             \Biggr)~, \label{Ahhhqq2NSQ} \\
   \Ahathat_{qq,Q}^{(3),{\sf NS}}&=&
     \Bigl(\frac{\hat{m}^2}{\mu^2}\Bigr)^{3\ep/2}\Biggl\{
            -\frac{4\gamma_{qq}^{(0)}\beta_{0,Q}}{3\ep^3}
                   \Bigl(\beta_0+2\beta_{0,Q}\Bigr)
            +\frac{1}{\ep^2}
              \Biggl(
                      \frac{2\gamma_{qq}^{(1),{\sf NS}}\beta_{0,Q}}{3}
                     -\frac{4\hat{\gamma}_{qq}^{(1),{\sf NS}}}{3}
                             \Bigl[\beta_0+\beta_{0,Q}\Bigr]
\N\\ 
&&
                     +\frac{2\beta_{1,Q}\gamma_{qq}^{(0)}}{3}
                     -2\delta m_1^{(-1)}\beta_{0,Q}\gamma_{qq}^{(0)}
              \Biggr)
            +\frac{1}{\ep} 
              \Biggl(
                      \frac{\hat{\gamma}_{qq}^{(2), {\sf NS}}}{3}
                     -4a_{qq,Q}^{(2),{\sf NS}}\Bigl[\beta_0+\beta_{0,Q}\Bigr]
                     +\beta_{1,Q}^{(1)}\gamma_{qq}^{(0)}\N
\end{eqnarray}
\begin{eqnarray}
%\N\\ 
&&
                     +\frac{\gamma_{qq}^{(0)}\beta_0\beta_{0,Q}\zeta_2}{2}
                     -2 \delta m_1^{(0)} \beta_{0,Q} \gamma_{qq}^{(0)} 
                     -\delta m_1^{(-1)}\hat{\gamma}_{qq}^{(1),{\sf NS}}
                  \Biggr)
         +a_{qq,Q}^{(3), {\sf NS}}
                              \Biggr\}~. \label{Ahhhqq3NSQ}
   \end{eqnarray}
   Note, that we have already used the general structure of the 
   unrenormalized lower order OME in the evaluation of the $O(\hat{a}_s^3)$
   term, as we will always do in the following. 
   Using Eqs.~(\ref{macoren}, \ref{2LNSRen1}, \ref{3LNSRen1}), one can 
   renormalize the above expressions. In addition, we finally transform back 
   to the $\overline{\sf MS}$--scheme using Eq.~(\ref{asmoma}). 
   Thus one obtains the renormalized expansion 
   coefficients of Eq.~(\ref{NSpert}) 
   \begin{eqnarray}
     A_{qq,Q}^{(2),\sf NS, \MS}&=&
                  \frac{\beta_{0,Q}\gamma_{qq}^{(0)}}{4}
                    \ln^2 \Bigl(\frac{m^2}{\mu^2}\Bigr)
                 +\frac{\hat{\gamma}_{qq}^{(1), {\sf NS}}}{2}
                    \ln \Bigl(\frac{m^2}{\mu^2}\Bigr)
                 +a_{qq,Q}^{(2),{\sf NS}}
                 -\frac{\beta_{0,Q}\gamma_{qq}^{(0)}}{4}\zeta_2~,
                  \label{Aqq2NSQMSren} \\
    A_{qq,Q}^{(3),{\sf NS}, \MS}&=&
     -\frac{\gamma_{qq}^{(0)}\beta_{0,Q}}{6}
          \Bigl(
                 \beta_0
                +2\beta_{0,Q}
          \Bigr)
             \ln^3 \Bigl(\frac{m^2}{\mu^2}\Bigr)
         +\frac{1}{4}
          \Biggl\{
                   2\gamma_{qq}^{(1),{\sf NS}}\beta_{0,Q}
                  -2\hat{\gamma}_{qq}^{(1),{\sf NS}}
                             \Bigl(
                                    \beta_0
                                   +\beta_{0,Q}
                             \Bigr)
\N\\ &&
                  +\beta_{1,Q}\gamma_{qq}^{(0)}
          \Biggr\}
             \ln^2 \Bigl(\frac{m^2}{\mu^2}\Bigr)
         +\frac{1}{2}
          \Biggl\{
                   \hat{\gamma}_{qq}^{(2),{\sf NS}}
                  -\Bigl(
                           4a_{qq,Q}^{(2),{\sf NS}}
                          -\zeta_2\beta_{0,Q}\gamma_{qq}^{(0)}
                                    \Bigr)(\beta_0+\beta_{0,Q})
\N\\ &&
                  +\gamma_{qq}^{(0)}\beta_{1,Q}^{(1)}
          \Biggr\}
             \ln \Bigl(\frac{m^2}{\mu^2}\Bigr)
         +4\overline{a}_{qq,Q}^{(2),{\sf NS}}(\beta_0+\beta_{0,Q})
         -\gamma_{qq}^{(0)}\beta_{1,Q}^{(2)}
         -\frac{\gamma_{qq}^{(0)}\beta_0\beta_{0,Q}\zeta_3}{6}
\N\\ &&
         -\frac{\gamma_{qq}^{(1),{\sf NS}}\beta_{0,Q}\zeta_2}{4}
         +2 \delta m_1^{(1)} \beta_{0,Q} \gamma_{qq}^{(0)}
         +\delta m_1^{(0)} \hat{\gamma}_{qq}^{(1),{\sf NS}}
         +2 \delta m_1^{(-1)} a_{qq,Q}^{(2),{\sf NS}}
\N\\ &&
         +a_{qq,Q}^{(3),{\sf NS}}
        ~. \label{Aqq3NSQMSren}
   \end{eqnarray}
   Note that in the ${\sf NS}$--case, one is generically provided 
   with even and odd moments due to a Ward--identity relating the 
   results in the polarized 
   and unpolarized case. The former refer 
   to the anomalous dimensions $\gamma_{qq}^{{\sf NS},+}$ 
   and the latter to $\gamma_{qq}^{{\sf NS},-}$, respectively, 
   as given in Eqs. (3.5, 3.7) and Eqs. (3.6, 3.8) in Ref.~\cite{Moch:2004pa}.
   The relations above also apply to other twist--2 non--singlet massive 
   OMEs, as to transversity, for which  the 2- and 3--loop heavy flavor 
   corrections are given in Section~\ref{sec-1}, cf. also 
   \cite{Blumlein:trans}.
%%%%%%%%%%%%%%%%%%%%%%%%%%%%%%%%%%%%%%%%%%%%%%%%%%%%%%%%%%%%%%%%%%%%%%%%%
%
% Subsubsection
%
% The Pure--Singlet Case
%
%%%%%%%%%%%%%%%%%%%%%%%%%%%%%%%%%%%%%%%%%%%%%%%%%%%%%%%%%%%%%%%%%%%%%%%%%
 \subsubsection{$A_{Qq}^{\sf PS}$ and $A_{qq,Q}^{\sf PS}$}
  \label{SubSec-PS}
%%%%%%%%%%%%%%%%%%%%%%%%%%%%%%%%%%%%%%%%%%%%%%%%%%%%%%%%%%%%%%%%%%%%%%%%%
  There are two different ${\sf PS}$--contributions, cf. the discussion below 
  Eq. \ref{splitS}, 
  \begin{eqnarray}
   A_{Qq}^{\sf PS}&=&
                       a_s^2A_{Qq}^{(2), {\sf PS}}
                       +a_s^3A_{Qq}^{(3), {\sf PS}}
                       +O(a_s^4)~, \label{PSQqpert}\\
   A_{qq,Q}^{\sf PS}&=&
                        a_s^3A_{qq,Q}^{(3), {\sf PS}}
                       +O(a_s^4)~. \label{PSqqQpert}
  \end{eqnarray}
  Separating these contributions is not straightforward, since 
  the generic renormalization formula for operator 
  renormalization and mass factorization, Eq.~(\ref{GenRen3}),
  applies to the sum of these terms only.
  At $O(a_s^2)$, this problem does not occur and renormalization proceeds 
  in the {\sf MOM}--scheme via
  \begin{eqnarray}
    A_{Qq}^{(2), \sf PS, \MOM}&=&
                    \hat{A}_{Qq}^{(2),{\sf PS}, \MOM}
                   +Z^{-1,(2), {\sf PS}}_{qq}(n_f+1)
                   -Z^{-1,(2), {\sf PS}}_{qq}(n_f) \N\\ &&
                   +\Bigl[ \hat{A}_{Qg}^{(1), \MOM}
                          +Z_{qg}^{-1,(1)}(n_f+1)
                          -Z_{qg}^{-1,(1)}(n_f)
                    \Bigr]\Gamma^{-1,(1)}_{gq}(n_f)~.
  \end{eqnarray}
  The unrenormalized expression is given by
  \begin{eqnarray}
   \Ahathat_{Qq}^{(2),\sf PS}&=&
          \Bigl(\frac{\hat{m}^2}{\mu^2}\Bigr)^{\ep}\Biggl(
                  -\frac{\hat{\gamma}_{qg}^{(0)}
                         \gamma_{gq}^{(0)}}{2\ep^2}
                  +\frac{\hat{\gamma}_{qq}^{(1), {\sf PS}}}{2\ep}
                  +a_{Qq}^{(2),{\sf PS}}
                  +\overline{a}_{Qq}^{(2),{\sf PS}}\ep
            \Biggr)~.\label{AhhhQq2PS}
  \end{eqnarray}
  The renormalized result in the ${\sf \MS}$--scheme reads
  \begin{eqnarray}
    A_{Qq}^{(2),\sf PS, \MS}&=&
                -\frac{\hat{\gamma}_{qg}^{(0)}
                             \gamma_{gq}^{(0)}}{8}
                   \ln^2 \Bigl(\frac{m^2}{\mu^2}\Bigr)
                +\frac{\hat{\gamma}_{qq}^{(1), {\sf PS}}}{2}
                   \ln \Bigl(\frac{m^2}{\mu^2}\Bigr)
                +a_{Qq}^{(2),{\sf PS}}
                +\frac{\hat{\gamma}_{qg}^{(0)}
                             \gamma_{gq}^{(0)}}{8}\zeta_2~.
       \label{AQq2PSMSON}
  \end{eqnarray}
  The corresponding renormalization relation at third order is given by
  \begin{eqnarray}
   &&A_{Qq}^{(3), \sf PS, \MOM}+
     A_{qq,Q}^{(3), \sf PS, \MOM}=
                     \hat{A}_{Qq}^{(3), {\sf PS}, \MOM} 
                    +\hat{A}_{qq,Q}^{(3), {\sf PS}, \MOM}
                    +Z^{-1,(3), {\sf PS}}_{qq}(n_f+1)
\N\\ && \phantom{abc}
                    -Z^{-1,(3), {\sf PS}}_{qq}(n_f)
                    +Z^{-1,(1)}_{qq}(n_f+1)\hat{A}_{Qq}^{(2), {\sf PS}, \MOM}
                    +Z^{-1,(1)}_{qg}(n_f+1)\hat{A}_{gq,Q}^{(2), \MOM}
\N\\ && \phantom{abc}
                    +\Bigl[
                            \hat{A}_{Qg}^{(1), \MOM}
                           +Z^{-1,(1)}_{qg}(n_f+1)
                           -Z^{-1,(1)}_{qg}(n_f)
                     \Bigr]\Gamma^{-1,(2)}_{gq}(n_f)
                    +\Bigl[ \hat{A}_{Qq}^{(2), {\sf PS}, \MOM}
\N\\ && \phantom{abc}
                           +Z^{-1,(2), {\sf PS}}_{qq}(n_f+1)
                           -Z^{-1,(2), {\sf PS}}_{qq}(n_f)
                     \Bigr]\Gamma^{-1,(1)}_{qq}(n_f)
                    +\Bigl[ \hat{A}_{Qg}^{(2), \MOM}
                           +Z^{-1,(2)}_{qg}(n_f+1)
\N\\ && \phantom{abc}
                           -Z^{-1,(2)}_{qg}(n_f)
                           +Z^{-1,(1)}_{qq}(n_f+1)A_{Qg}^{(1), \MOM}
                           +Z^{-1,(1)}_{qg}(n_f+1)A_{gg,Q}^{(1), \MOM}
                     \Bigr]\Gamma^{-1,(1)}_{gq}(n_f)~.\N\\ 
                  \label{AQqq3PSRen}
  \end{eqnarray}
  Taking into account the structure of the UV-- and collinear singularities
  of the contributing Feynman--diagrams, these two
  contributions can be separated.  
  For the bare quantities we obtain 
   \begin{eqnarray}
    \Ahathat_{Qq}^{(3),{\sf PS}}&=&
     \Bigl(\frac{\hat{m}^2}{\mu^2}\Bigr)^{3\ep/2}\Biggl[
     \frac{\hat{\gamma}_{qg}^{(0)}\gamma_{gq}^{(0)}}{6\ep^3}
                  \Biggl(
                         \gamma_{gg}^{(0)}
                        -\gamma_{qq}^{(0)}
                        +6\beta_0
                        +16\beta_{0,Q}
                  \Biggr)
   +\frac{1}{\ep^2}\Biggl(     
                        -\frac{4\hat{\gamma}_{qq}^{(1),{\sf PS}}}{3}
                                 \Bigl[
                                        \beta_0
                                       +\beta_{0,Q}
                                 \Bigr]
 \N\\ &&
                        -\frac{\gamma_{gq}^{(0)}\hat{\gamma}_{qg}^{(1)}}{3}
                        +\frac{\hat{\gamma}_{qg}^{(0)}}{6}
                                 \Bigl[
                                        2\hat{\gamma}_{gq}^{(1)}
                                       -\gamma_{gq}^{(1)}
                                 \Bigr]
                        +\delta m_1^{(-1)} \hat{\gamma}_{qg}^{(0)}
                                           \gamma_{gq}^{(0)}
                 \Biggr)
   +\frac{1}{\ep}\Biggl(     
                           \frac{\hat{\gamma}_{qq}^{(2),{\sf PS}}}{3}
                        -n_f\frac{\hat{\tilde{\gamma}}_{qq}^{(2),{\sf PS}}}{3}
         \N\\ &&
                          +\hat{\gamma}_{qg}^{(0)}a_{gq,Q}^{(2)}
                          -\gamma_{gq}^{(0)}a_{Qg}^{(2)}
                          -4(\beta_0+\beta_{0,Q})a_{Qq}^{(2),{\sf PS}}
                   -\frac{\hat{\gamma}_{qg}^{(0)}\gamma_{gq}^{(0)}\zeta_2}{16}
                            \Bigl[
                               \gamma_{gg}^{(0)}
                              -\gamma_{qq}^{(0)}
                              +6\beta_0
                            \Bigr]
         \N\\ &&
                   +\delta m_1^{(0)} \hat{\gamma}_{qg}^{(0)}
                                           \gamma_{gq}^{(0)}
                   -\delta m_1^{(-1)} \hat{\gamma}_{qq}^{(1),{\sf PS}}
                 \Biggr)
   +a_{Qq}^{(3),{\sf PS}}
                              \Biggr]~, \label{AhhhQq3PS} \\
    \Ahathat_{qq,Q}^{(3),{\sf PS}}&=&
           n_f\Bigl(\frac{\hat{m}^2}{\mu^2}\Bigr)^{3\ep/2}\Biggl[
            \frac{2\hat{\gamma}_{qg}^{(0)}\gamma_{gq}^{(0)}\beta_{0,Q}}{3\ep^3}
                +\frac{1}{3\ep^2} \Biggl(
                    2\hat{\gamma}_{qq}^{(1),{\sf PS}}\beta_{0,Q}
                   +\hat{\gamma}_{qg}^{(0)}\hat{\gamma}_{gq}^{(1)}
                                  \Biggr)
\N\\ &&
                +\frac{1}{\ep} \Biggl(
                       \frac{\hat{\tilde{\gamma}}_{qq}^{(2),{\sf PS}}}{3}
                      +\hat{\gamma}_{qg}^{(0)}a_{gq,Q}^{(2)}
         -\frac{\hat{\gamma}_{qg}^{(0)}\gamma_{gq}^{(0)}\beta_{0,Q}\zeta_2}{4}
                                \Biggr)
                +\frac{a_{qq,Q}^{(3), {\sf PS}}}{n_f}
                                                     \Biggr]~. 
                   \label{Ahhhqq3PSQ}
   \end{eqnarray}
   The renormalized terms in the ${\MS}$--scheme are given by
   \begin{eqnarray}
    A_{Qq}^{(3),{\sf PS}, \MS}&=&
      \frac{\hat{\gamma}_{qg}^{(0)}\gamma_{gq}^{(0)}}{48}
                  \Biggl\{
                         \gamma_{gg}^{(0)}
                        -\gamma_{qq}^{(0)}
                        +6\beta_0
                        +16\beta_{0,Q}
                  \Biggr\}
              \ln^3 \Bigl(\frac{m^2}{\mu^2}\Bigr)
  +    \frac{1}{8}\Biggl\{
                         -4\hat{\gamma}_{qq}^{(1),{\sf PS}}
                               \Bigl(
                                 \beta_0
                                +\beta_{0,Q}
                               \Bigr)
\N\\ &&
                        +\hat{\gamma}_{qg}^{(0)}
                               \Bigl(
                                 \hat{\gamma}_{gq}^{(1)}
                                -\gamma_{gq}^{(1)}
                               \Bigr)
                        -\gamma_{gq}^{(0)}\hat{\gamma}_{qg}^{(1)}
                  \Biggr\}
              \ln^2 \Bigl(\frac{m^2}{\mu^2}\Bigr)
  +   \frac{1}{16}\Biggl\{
                         8\hat{\gamma}_{qq}^{(2),{\sf PS}}
                        -8n_f\hat{\tilde{\gamma}}_{qq}^{(2),{\sf PS}}
\N\\ &&
                        -32a_{Qq}^{(2),{\sf PS}}(\beta_0+\beta_{0,Q})
                        +8\hat{\gamma}_{qg}^{(0)}a_{gq,Q}^{(2)}
                        -8\gamma_{gq}^{(0)}a_{Qg}^{(2)}
                        -\hat{\gamma}_{qg}^{(0)}\gamma_{gq}^{(0)}\zeta_2\
                          \Bigl(
                                 \gamma_{gg}^{(0)}
                                -\gamma_{qq}^{(0)}
\N\\ &&
                                +6\beta_0
                                +8\beta_{0,Q}
                          \Bigr)
                  \Biggr\}
              \ln \Bigl(\frac{m^2}{\mu^2}\Bigr)
    +4(\beta_0+\beta_{0,Q})\overline{a}_{Qq}^{(2),{\sf PS}}
    +\gamma_{gq}^{(0)}\overline{a}_{Qg}^{(2)}
    -\hat{\gamma}_{qg}^{(0)}\overline{a}_{gq,Q}^{(2)}
\N\\ &&
    +\frac{\gamma_{gq}^{(0)}\hat{\gamma}_{qg}^{(0)}\zeta_3}{48}
                  \Bigl(
                         \gamma_{gg}^{(0)}
                        -\gamma_{qq}^{(0)}
                        +6\beta_0
                  \Bigr)
    +\frac{\hat{\gamma}_{qg}^{(0)}\gamma_{gq}^{(1)}\zeta_2}{16}
    -\delta m_1^{(1)} \hat{\gamma}_{qg}^{(0)}
                        \gamma_{gq}^{(0)}
    +\delta m_1^{(0)} \hat{\gamma}_{qq}^{(1),{\sf PS}} \N
\end{eqnarray}
\begin{eqnarray}
%\N\\ 
&&
    +2 \delta m_1^{(-1)} a_{Qq}^{(2),{\sf PS}}
    +a_{Qq}^{(3),{\sf PS}}~.     \label{AQq3PSMSren} \\
    A_{qq,Q}^{(3),{\sf PS}, \MS}&=&n_f\Biggl\{
              \frac{\gamma_{gq}^{(0)}\hat{\gamma}_{qg}^{(0)}\beta_{0,Q}}{12}
                         \ln^3 \Bigl(\frac{m^2}{\mu^2}\Bigr)
             +\frac{1}{8}\Bigl(
                     4\hat{\gamma}_{qq}^{(1), {\sf PS}}\beta_{0,Q}
                    +\hat{\gamma}_{qg}^{(0)}\hat{\gamma}_{gq}^{(1)}
              \Bigr)\ln^2 \Bigl(\frac{m^2}{\mu^2}\Bigr)
\N\\
&&
             +\frac{1}{4}\Bigl(
                    2\hat{\tilde{\gamma}}_{qq}^{(2), {\sf PS}}
                   +\hat{\gamma}_{qg}^{(0)}\Bigl\{
                                   2a_{gq,Q}^{(2)}
                                  -\gamma_{gq}^{(0)}\beta_{0,Q}\zeta_2
                                                     \Bigr\}
              \Bigr)\ln \Bigl(\frac{m^2}{\mu^2}\Bigr)
\N\\ &&
              -\hat{\gamma}_{qg}^{(0)}\overline{a}_{gq,Q}^{(2)}
              +\frac{\gamma_{gq}^{(0)}\hat{\gamma}_{qg}^{(0)}
                         \beta_{0,Q}\zeta_3}{12}
              -\frac{\hat{\gamma}_{qq}^{(1), {\sf PS}}\beta_{0,Q}\zeta_2}{4}
                                             \Biggr\}
              +a_{qq,Q}^{(3), {\sf PS}}~. 
              \label{Aqq3PSQMSren}
   \end{eqnarray}
%%%%%%%%%%%%%%%%%%%%%%%%%%%%%%%%%%%%%%%%%%%%%%%%%%%%%%%%%%%%%%%%%%%%%%%%%
%
% Subsubsection 
%
% AQg3
%
%%%%%%%%%%%%%%%%%%%%%%%%%%%%%%%%%%%%%%%%%%%%%%%%%%%%%%%%%%%%%%%%%%%%%%%%%
 \subsubsection{$A_{Qg}$ and $A_{qg,Q}$}
  \label{SubSec-AQqg}
%%%%%%%%%%%%%%%%%%%%%%%%%%%%%%%%%%%%%%%%%%%%%%%%%%%%%%%%%%%%%%%%%%%%%%%%%
  The OME $A_{Qg}$ is the most complex expression. As in the 
${\sf PS}$--case, there are two different contributions
  \begin{eqnarray}
   A_{Qg}&=&
                        a_s  A_{Qg}^{(1)}
                       +a_s^2A_{Qg}^{(2)}
                       +a_s^3A_{Qg}^{(3)}
                       +O(a_s^4)~. \label{AQgpert}\\
   A_{qg,Q}&=&
                        a_s^3A_{qg,Q}^{(3)}
                       +O(a_s^4)~. \label{AqgQpert}
  \end{eqnarray}
  In the {\sf MOM}--scheme
  the $1$-- and $2$--loop contributions obey the following relations 
  \begin{eqnarray} 
    A_{Qg}^{(1), \MOM}&=&
                    \hat{A}_{Qg}^{(1), \MOM}
                   +Z^{-1,(1)}_{qg}(n_f+1)
                   -Z^{-1,(1)}_{qg}(n_f) ~, \\
    A_{Qg}^{(2), \MOM}&=&
                    \hat{A}_{Qg}^{(2), \MOM}
                   +Z^{-1,(2)}_{qg}(n_f+1)
                   -Z^{-1,(2)}_{qg}(n_f)
                   +Z^{-1,(1)}_{qg}(n_f+1)\hat{A}_{gg,Q}^{(1), \MOM}
 \N\\ &&
                   +Z^{-1,(1)}_{qq}(n_f+1)\hat{A}_{Qg}^{(1), \MOM}
                   +\Bigl[ \hat{A}_{Qg}^{(1), \MOM}
                          +Z_{qg}^{-1,(1)}(n_f+1)
\N\\ &&
                          -Z_{qg}^{-1,(1)}(n_f)
                    \Bigr]\Gamma^{-1,(1)}_{gg}(n_f)~. \label{RenAQg2MOM}
  \end{eqnarray}
  The unrenormalized terms are given by 
  \begin{eqnarray}
   \Ahathat_{Qg}^{(1)}&=&
                        \Bigl(\frac{\hat{m}^2}{\mu^2}\Bigr)^{\ep/2}
                        \frac{\hat{\gamma}_{qg}^{(0)}}{\ep}
                          \exp \Bigl(\sum_{i=2}^{\infty}\frac{\zeta_i}{i}
                          \Bigl(\frac{\ep}{2}\Bigr)^{i}\Bigr)~, 
                          \label{AhhhQg1} \\
%\end{eqnarray}\begin{eqnarray}
   \Ahathat_{Qg}^{(2)}&=&
                  \Bigl(\frac{\hat{m}^2}{\mu^2}\Bigr)^{\ep}
                     \Biggl[
                            -\frac{\hat{\gamma}_{qg}^{(0)}}{2\ep^2}
                                \Bigl(
                                       \gamma_{gg}^{(0)}
                                      -\gamma_{qq}^{(0)}
                                      +2\beta_{0}
                                      +4\beta_{0,Q}
                                \Bigr)
                            +\frac{ \hat{\gamma}_{qg}^{(1)}
                                   -2\delta m_1^{(-1)} \hat{\gamma}_{qg}^{(0)}}
                                  {2\ep}
                            +a_{Qg}^{(2)}
\N\\ &&
                           -\delta m_1^{(0)} \hat{\gamma}_{qg}^{(0)}
                           -\frac{\hat{\gamma}_{qg}^{(0)}\beta_{0,Q}\zeta_2}{2}
                            +\ep\Bigl( 
                               \overline{a}_{Qg}^{(2)}
                        -\delta m_1^{(1)} \hat{\gamma}_{qg}^{(0)}
                        -\frac{\hat{\gamma}_{qg}^{(0)}\beta_{0,Q}\zeta_2}{12}
                                \Bigr)
                         \Biggr]
                         ~.\label{AhhhQg2}
  \end{eqnarray}
   Note that we have already made the
   one--particle reducible contributions to Eq.~(\ref{AhhhQg2}) explicit, which
   are given by the ${\sf LO}$--term multiplied
   with the 1--loop gluon--self energy, cf. Eq.~(\ref{GluSelf1}).
   Furthermore, Eq.~(\ref{AhhhQg2}) already contains terms 
   in the $O(\ep^0)$ and $O(\ep)$ expressions which result from
   mass renormalization. At this
   stage of the renormalization procedure they should not be present,
   however, we have included them here in order to have the same notation as
   in Refs.~\cite{Buza:1995ie,Buza:1996wv} at the $2$--loop level. The
   renormalized terms then become in the $\overline{\sf MS}$--scheme
   \begin{eqnarray}
    A_{Qg}^{(1), \MS}&=&
                  \frac{\hat{\gamma}_{qg}^{(0)}}{2}
                  \ln \Bigl(\frac{m^2}{\mu^2}\Bigr)
                  ~, \label{AQg1MSren} \\
    A_{Qg}^{(2), \MS}&=&
                  -\frac{\hat{\gamma}_{qg}^{(0)}}{8}
                     \Biggl[
                             \gamma_{gg}^{(0)}
                            -\gamma_{qq}^{(0)}
                            +2\beta_{0}
                            +4\beta_{0,Q}
                     \Biggr]
                     \ln^2 \Bigl(\frac{m^2}{\mu^2}\Bigr)
                  +\frac{\hat{\gamma}_{qg}^{(1)}}{2}
                      \ln \Bigl(\frac{m^2}{\mu^2}\Bigr)\N
  \end{eqnarray}
  \begin{eqnarray}
% \N\\ 
&&
                  +a_{Qg}^{(2)}
                  +\frac{\hat{\gamma}_{qg}^{(0)}\zeta_2}{8}
                              \Bigl(
                                     \gamma_{gg}^{(0)}
                                    -\gamma_{qq}^{(0)}
                                    +2\beta_{0}
                              \Bigr)~. \label{AQg2MSren}
   \end{eqnarray}
   The generic renormalization relation at the $3$--loop level is given by
   \begin{eqnarray}
    && A_{Qg}^{(3), \MOM}+A_{qg,Q}^{(3), \MOM}
            =
                     \hat{A}_{Qg}^{(3), \MOM}
                    +\hat{A}_{qg,Q}^{(3), \MOM}
                    +Z^{-1,(3)}_{qg}(n_f+1)
                    -Z^{-1,(3)}_{qg}(n_f)
\N\\ && \phantom{abc} 
                    +Z^{-1,(2)}_{qg}(n_f+1)\hat{A}_{gg,Q}^{(1), \MOM}
                    +Z^{-1,(1)}_{qg}(n_f+1)\hat{A}_{gg,Q}^{(2), \MOM}
                    +Z^{-1,(2)}_{qq}(n_f+1)\hat{A}_{Qg}^{(1), \MOM}
\N\\ && \phantom{abc} 
                    +Z^{-1,(1)}_{qq}(n_f+1)\hat{A}_{Qg}^{(2), \MOM}
                    +\Bigl[
                            \hat{A}_{Qg}^{(1), \MOM}
                           +Z^{-1,(1)}_{qg}(n_f+1)
\N\\ && \phantom{abc} 
                           -Z^{-1,(1)}_{qg}(n_f)
                     \Bigr]\Gamma^{-1,(2)}_{gg}(n_f)
                    +\Bigl[ \hat{A}_{Qg}^{(2), \MOM}
                           +Z^{-1,(2)}_{qg}(n_f+1) 
                           -Z^{-1,(2)}_{qg}(n_f)
\N\\ && \phantom{abc} 
                           +Z^{-1,(1)}_{qq}(n_f+1)A_{Qg}^{(1), \MOM}
                           +Z^{-1,(1)}_{qg}(n_f+1)A_{gg,Q}^{(1), \MOM}
                     \Bigr]\Gamma^{-1,(1)}_{gg}(n_f) 
\N\\ && \phantom{abc} 
                    +\Bigl[ \hat{A}_{Qq}^{(2), {\sf PS}, \MOM}
                           +Z^{-1,(2), {\sf PS}}_{qq}(n_f+1)
                           -Z^{-1,(2), {\sf PS}}_{qq}(n_f)
                     \Bigr]\Gamma^{-1,(1)}_{qg}(n_f)
\N\\ && \phantom{abc} 
                    +\Bigl[ \hat{A}_{qq,Q}^{(2), {\sf NS}, \MOM}
                           +Z^{-1,(2), {\sf NS}}_{qq}(n_f+1)
                           -Z^{-1,(2), {\sf NS}}_{qq}(n_f)
                     \Bigr]\Gamma^{-1,(1)}_{qg}(n_f)~.
  \end{eqnarray}
  Similar to the ${\sf PS}$--case, the different contributions can be separated
  and one obtains the following unrenormalized results
   \begin{eqnarray}
   \Ahathat_{Qg}^{(3)}&=&
                  \Bigl(\frac{\hat{m}^2}{\mu^2}\Bigr)^{3\ep/2}
                     \Biggl[
           \frac{\hat{\gamma}_{qg}^{(0)}}{6\ep^3}
             \Biggl(
                   (n_f+1)\gamma_{gq}^{(0)}\hat{\gamma}_{qg}^{(0)}
                 +\gamma_{qq}^{(0)} 
                                \Bigl[
                                        \gamma_{qq}^{(0)}
                                      -2\gamma_{gg}^{(0)}
                                      -6\beta_0
                                      -8\beta_{0,Q}
                                \Bigr]
                 +8\beta_0^2
\N\\ &&
                 +28\beta_{0,Q}\beta_0
                 +24\beta_{0,Q}^2 
                  +\gamma_{gg}^{(0)} 
                                \Bigl[
                                        \gamma_{gg}^{(0)}
                                       +6\beta_0
                                       +14\beta_{0,Q}
                                \Bigr]
             \Biggr)
          +\frac{1}{6\ep^2}
             \Biggl(
                   \hat{\gamma}_{qg}^{(1)}
                      \Bigl[
                              2\gamma_{qq}^{(0)}
                             -2\gamma_{gg}^{(0)}
\N\\ &&
                             -8\beta_0
                             -10\beta_{0,Q} 
                      \Bigr]
                  +\hat{\gamma}_{qg}^{(0)}
                      \Bigl[
                              \hat{\gamma}_{qq}^{(1), {\sf PS}}\{1-2n_f\}
                             +\gamma_{qq}^{(1), {\sf NS}}
                             +\hat{\gamma}_{qq}^{(1), {\sf NS}}
                             +2\hat{\gamma}_{gg}^{(1)}
                             -\gamma_{gg}^{(1)}
                             -2\beta_1
\N\\ &&
                             -2\beta_{1,Q}
                      \Bigr]
                  + 6 \delta m_1^{(-1)} \hat{\gamma}_{qg}^{(0)} 
                      \Bigl[
                              \gamma_{gg}^{(0)}
                             -\gamma_{qq}^{(0)}
                             +3\beta_0
                             +5\beta_{0,Q}
                      \Bigr]
             \Biggr)
          +\frac{1}{\ep}
             \Biggl(
                   \frac{\hat{\gamma}_{qg}^{(2)}}{3}
                  -n_f \frac{\hat{\tilde{\gamma}}_{qg}^{(2)}}{3}
\N\\ &&
                  +\hat{\gamma}_{qg}^{(0)}\Bigl[
                                    a_{gg,Q}^{(2)}
                                   -n_fa_{Qq}^{(2),{\sf PS}}
                                          \Bigr]
                  +a_{Qg}^{(2)}
                      \Bigl[
                              \gamma_{qq}^{(0)}
                             -\gamma_{gg}^{(0)}
                             -4\beta_0
                             -4\beta_{0,Q}
                      \Bigr]
                  +\frac{\hat{\gamma}_{qg}^{(0)}\zeta_2}{16}
                      \Bigl[
                              \gamma_{gg}^{(0)} \Bigl\{
                                                        2\gamma_{qq}^{(0)}
\N\\ &&
                                                       -\gamma_{gg}^{(0)}
                                                       -6\beta_0
                                                       +2\beta_{0,Q}
                                                \Bigr\}
                             -(n_f+1)\gamma_{gq}^{(0)}\hat{\gamma}_{qg}^{(0)}
                             +\gamma_{qq}^{(0)} \Bigl\{
                                                       -\gamma_{qq}^{(0)}
                                                       +6\beta_0
                                                \Bigr\}
                             -8\beta_0^2
\N\\ &&
                             +4\beta_{0,Q}\beta_0
                             +24\beta_{0,Q}^2
                      \Bigr]
                  + \frac{\delta m_1^{(-1)}}{2}
                      \Bigl[
                              -2\hat{\gamma}_{qg}^{(1)}
                              +3\delta m_1^{(-1)}\hat{\gamma}_{qg}^{(0)}
                              +2\delta m_1^{(0)}\hat{\gamma}_{qg}^{(0)}
                      \Bigr]
\N\\ &&
                  + \delta m_1^{(0)}\hat{\gamma}_{qg}^{(0)}
                       \Bigl[
                               \gamma_{gg}^{(0)}
                              -\gamma_{qq}^{(0)}
                              +2\beta_0
                              +4\beta_{0,Q}
                      \Bigr]
                  -\delta m_2^{(-1)}\hat{\gamma}_{qg}^{(0)}
             \Biggr)
                 +a_{Qg}^{(3)}
                  \Biggr]~. \label{AhhhQg3} \\
%%%%%%%%%%%%%%%%%%%
   \Ahathat_{qg,Q}^{(3)}&=&
                   n_f\Bigl(\frac{\hat{m}^2}{\mu^2}\Bigr)^{3\ep/2}
                     \Biggl[
           \frac{\hat{\gamma}_{qg}^{(0)}}{6\ep^3}
             \Biggl(
                    \gamma_{gq}^{(0)}\hat{\gamma}_{qg}^{(0)}
                   +2\beta_{0,Q}\Bigl[
                                  \gamma_{gg}^{(0)}
                                 -\gamma_{qq}^{(0)}
                                 +2\beta_0 
                                \Bigr]
             \Biggr)
          +\frac{1}{\ep^2}
             \Biggl(
                   \frac{\hat{\gamma}_{qg}^{(0)}}{6} \Bigl[
                                     2\hat{\gamma}_{gg}^{(1)}
\N\\ &&
                                    +\hat{\gamma}_{qq}^{(1), {\sf PS}}
                                    -2\hat{\gamma}_{qq}^{(1), {\sf NS}}
                                    +4\beta_{1,Q}
                             \Bigr]
                   +\frac{\hat{\gamma}_{qg}^{(1)}\beta_{0,Q}}{3}
             \Biggr)
          +\frac{1}{\ep}
             \Biggl(
                   \frac{\hat{\tilde{\gamma}}_{qg}^{(2)}}{3}
                  +\hat{\gamma}_{qg}^{(0)}\Bigl[
                                            a_{gg,Q}^{(2)}
                                           -a_{qq,Q}^{(2),{\sf NS}}
\N\\ &&
                                           +\beta_{1,Q}^{(1)}
                                          \Bigr]
                  -\frac{\hat{\gamma}_{qg}^{(0)}\zeta_2}{16}\Bigl[
                                    \gamma_{gq}^{(0)}\hat{\gamma}_{qg}^{(0)}
                                   +2\beta_{0,Q}\Bigl\{
                                         \gamma_{gg}^{(0)}
                                        -\gamma_{qq}^{(0)}
                                        +2\beta_0
                                                \Bigr\}
                                                           \Bigr]
            \Biggr)
         +\frac{a_{qg,Q}^{(3)}}{n_f}
                         \Biggr]~.\label{Ahhhqg3Q}
  \end{eqnarray}  
  The renormalized expressions are
  \begin{eqnarray}
   A_{Qg}^{(3), \MS}&=&
                  \frac{\hat{\gamma}_{qg}^{(0)}}{48}
                     \Biggl\{
                             (n_f+1)\gamma_{gq}^{(0)}\hat{\gamma}_{qg}^{(0)}
                            +\gamma_{gg}^{(0)}\Bigl(
                                      \gamma_{gg}^{(0)}
                                     -2\gamma_{qq}^{(0)}
                                     +6\beta_0
                                     +14\beta_{0,Q}
                                              \Bigr)
                            +\gamma_{qq}^{(0)}\Bigl(
                                      \gamma_{qq}^{(0)}
\N\\
 &&
                                     -6\beta_0
                                     -8\beta_{0,Q}
                                              \Bigr)
                            +8\beta_0^2
                            +28\beta_{0,Q}\beta_0
                            +24\beta_{0,Q}^2
                     \Biggr\}
                     \ln^3 \Bigl(\frac{m^2}{\mu^2}\Bigr)
                    +\frac{1}{8}\Biggl\{
                            \hat{\gamma}_{qg}^{(1)}
                                \Bigl(
                                        \gamma_{qq}^{(0)}
                                       -\gamma_{gg}^{(0)}
\N\\ &&
                                       -4\beta_0
                                       -6\beta_{0,Q}
                                \Bigr)
                           +\hat{\gamma}_{qg}^{(0)}
                                \Bigl(
                                        \hat{\gamma}_{gg}^{(1)}
                                       -\gamma_{gg}^{(1)}
                                       +(1-n_f) \hat{\gamma}_{qq}^{(1), {\sf PS}}
                                       +\gamma_{qq}^{(1), {\sf NS}}
                                      +\hat{\gamma}_{qq}^{(1), {\sf NS}}
                                      -2\beta_1
\N\\ &&
                                      -2\beta_{1,Q}
                                \Bigr)
                     \Biggr\}
                     \ln^2 \Bigl(\frac{m^2}{\mu^2}\Bigr)
                    +\Biggl\{
                            \frac{\hat{\gamma}_{qg}^{(2)}}{2}
                           -n_f\frac{\hat{\tilde{\gamma}}_{qg}^{(2)}}{2}
                           +\frac{a_{Qg}^{(2)}}{2}
                                \Bigl(
                                        \gamma_{qq}^{(0)}
                                       -\gamma_{gg}^{(0)}
                                       -4\beta_0
                                       -4\beta_{0,Q}
                                \Bigr)
\N\\ &&
                           +\frac{\hat{\gamma}_{qg}^{(0)}}{2}
                                \Bigl(
                                       a_{gg,Q}^{(2)}
                                      -n_fa_{Qq}^{(2), {\sf PS}}
                                \Bigr)
                           +\frac{\hat{\gamma}_{qg}^{(0)}\zeta_2}{16}
                                \Bigl(
                                       -(n_f+1)\gamma_{gq}^{(0)}
                                             \hat{\gamma}_{qg}^{(0)}
                                       +\gamma_{gg}^{(0)}\Bigl[
                                                2\gamma_{qq}^{(0)}
                                               -\gamma_{gg}^{(0)}
                                               -6\beta_0
\N\\ &&
                                               -6\beta_{0,Q}
                                                         \Bigr]
                                       -4\beta_0[2\beta_0+3\beta_{0,Q}]
                                       +\gamma_{qq}^{(0)}\Bigl[
                                               -\gamma_{qq}^{(0)}
                                               +6\beta_0
                                               +4\beta_{0,Q}
                                                         \Bigr]
                                \Bigr)
                     \Biggr\}
                     \ln \Bigl(\frac{m^2}{\mu^2}\Bigr)
                           +\overline{a}_{Qg}^{(2)}
                                \Bigl(
                                        \gamma_{gg}^{(0)}
\N\\ 
&&
                                       -\gamma_{qq}^{(0)}
                                       +4\beta_0
                                       +4\beta_{0,Q}
                                \Bigr)
                           +\hat{\gamma}_{qg}^{(0)}\Bigl(
                                        n_f\overline{a}_{Qq}^{(2), {\sf PS}}
                                       -\overline{a}_{gg,Q}^{(2)}
                                                   \Bigr)
                           +\frac{\hat{\gamma}_{qg}^{(0)}\zeta_3}{48}
                                \Bigl(
                                        (n_f+1)\gamma_{gq}^{(0)}
                                                \hat{\gamma}_{qg}^{(0)}
\N\\ &&
                                       +\gamma_{gg}^{(0)}\Bigl[
                                                 \gamma_{gg}^{(0)}
                                               -2\gamma_{qq}^{(0)}
                                               +6\beta_0
                                               -2\beta_{0,Q}
                                                         \Bigr]
                                       +\gamma_{qq}^{(0)}\Bigl[
                                                \gamma_{qq}^{(0)}
                                               -6\beta_0
                                                         \Bigr]
                                       +8\beta_0^2
                                       -4\beta_0\beta_{0,Q}
\N\\ &&
                                       -24\beta_{0,Q}^2
                                \Bigr)
                           +\frac{\hat{\gamma}_{qg}^{(1)}\beta_{0,Q}\zeta_2}{8}
                           +\frac{\hat{\gamma}_{qg}^{(0)}\zeta_2}{16}
                                \Bigl(
                                        \gamma_{gg}^{(1)}
                                       -\hat{\gamma}_{qq}^{(1), {\sf NS}}
                                       -\gamma_{qq}^{(1), {\sf NS}}
                                       -\hat{\gamma}_{qq}^{(1),{\sf PS}}
                                       +2\beta_1
\N\\ &&
                                       +2\beta_{1,Q}
                                \Bigr)
                           +\frac{\delta m_1^{(-1)}}{8}
                                \Bigl(
                                       16 a_{Qg}^{(2)}
                                  +\hat{\gamma}_{qg}^{(0)}\Bigl[
                                           -24 \delta m_1^{(0)}
                                           -8 \delta m_1^{(1)}
                                           -\zeta_2\beta_0
                                           -9\zeta_2\beta_{0,Q}
                                                          \Bigr]
                                \Bigr)
\N\\ &&
                           +\frac{\delta m_1^{(0)}}{2}
                                \Bigl(
                                       2\hat{\gamma}_{qg}^{(1)}
                                      -\delta m_1^{(0)}
                                       \hat{\gamma}_{qg}^{(0)} 
                                \Bigr)
                           +\delta m_1^{(1)}\hat{\gamma}_{qg}^{(0)}
                                \Bigl(
                                        \gamma_{qq}^{(0)}
                                       -\gamma_{gg}^{(0)}
                                       -2\beta_0
                                       -4 \beta_{0,Q}
                                \Bigr)
\N\\ &&
                           +\delta m_2^{(0)}\hat{\gamma}_{qg}^{(0)}
                           +a_{Qg}^{(3)}~. \label{AQg3MSren}  \\
%\end{eqnarray}\begin{eqnarray}
%%%%%%%%%%%%%%%%%%
    A_{qg,Q}^{(3), \MS}&=&n_f\Biggl[
         \frac{\hat{\gamma}_{qg}^{(0)}}{48}\Biggl\{
                            \gamma_{gq}^{(0)}\hat{\gamma}_{qg}^{(0)}
                           +2\beta_{0,Q}\Bigl(
                                      \gamma_{gg}^{(0)}
                                     -\gamma_{qq}^{(0)}
                                     +2\beta_0
                                        \Bigr)
                                           \Biggr\}
                         \ln^3 \Bigl(\frac{m^2}{\mu^2}\Bigr) 
        +\frac{1}{8}\Biggl\{
                     2\hat{\gamma}_{qg}^{(1)}\beta_{0,Q}
\N\\ &&
                           +\hat{\gamma}_{qg}^{(0)} 
                                      \Bigl(
                                 \hat{\gamma}_{qq}^{(1), {\sf PS}}
                                -\hat{\gamma}_{qq}^{(1), {\sf NS}}
                                +\hat{\gamma}_{gg}^{(1)}
                                +2\beta_{1,Q}
                                        \Bigr)
                   \Biggr\}
                         \ln^2 \Bigl(\frac{m^2}{\mu^2}\Bigr)
        +\frac{1}{2}\Biggl\{
                       \hat{\tilde{\gamma}}_{qg}^{(2)}
                      +\hat{\gamma}_{qg}^{(0)}  \Bigl(
                                a_{gg,Q}^{(2)}
\N\\ &&
                               -a_{qq,Q}^{(2),{\sf NS}}
                               +\beta_{1,Q}^{(1)}
                                        \Bigr)
                      -\frac{\hat{\gamma}_{qg}^{(0)}}{8}\zeta_2 \Bigl(
                                     \gamma_{gq}^{(0)}\hat{\gamma}_{qg}^{(0)}
                                    +2\beta_{0,Q}\Bigl[
                                                 \gamma_{gg}^{(0)}
                                                -\gamma_{qq}^{(0)}
                                                +2\beta_0
                                                 \Bigr]
                                        \Bigr)
                  \Biggr\}
                         \ln \Bigl(\frac{m^2}{\mu^2}\Bigr)
\N\\ &&
           +\hat{\gamma}_{qg}^{(0)}\Bigl(
                              \overline{a}_{qq,Q}^{(2),{\sf NS}}
                             -\overline{a}_{gg,Q}^{(2)}
                             -\beta_{1,Q}^{(2)}
                                   \Bigr)
           +\frac{\hat{\gamma}_{qg}^{(0)}}{48}\zeta_3\Bigl(
                            \gamma_{gq}^{(0)}\hat{\gamma}_{qg}^{(0)}
                                    +2\beta_{0,Q}\Bigl[
                                                 \gamma_{gg}^{(0)}
                                                -\gamma_{qq}^{(0)}
                                                +2\beta_0
                                                 \Bigr]
                                   \Bigr)
\N\\ &&
           -\frac{\zeta_2}{16}\Bigl(
                     \hat{\gamma}_{qg}^{(0)}\hat{\gamma}_{qq}^{(1), {\sf PS}}
                    +2\hat{\gamma}_{qg}^{(1)}\beta_{0,Q}
                                   \Bigr)
           +\frac{a_{qg,Q}^{(3)}}{n_f}
              \Biggr]~. \label{Aqg3QMSren}
   \end{eqnarray}
%%%%%%%%%%%%%%%%%%%%%%%%%%%%%%%%%%%%%%%%%%%%%%%%%%%%%%%%%%%%%%%%%%%%%%%%%
%
% Subsubsection
%
% AgqQ
%
%%%%%%%%%%%%%%%%%%%%%%%%%%%%%%%%%%%%%%%%%%%%%%%%%%%%%%%%%%%%%%%%%%%%%%%%%
 \subsubsection{$A_{gq,Q}$}
  \label{SubSec-AgqQ}
%%%%%%%%%%%%%%%%%%%%%%%%%%%%%%%%%%%%%%%%%%%%%%%%%%%%%%%%%%%%%%%%%%%%%%%%%
  The $gq$--contributions start at $O(a_s^2)$,
  \begin{eqnarray}
   A_{gq,Q}&=&
                       a_s^2A_{gq,Q}^{(2)}
                       +a_s^3A_{gq,Q}^{(3)}
                       +O(a_s^4)~. \label{AgqQpert}
  \end{eqnarray}
  The renormalization formulas in the {\sf MOM}--scheme read
  \begin{eqnarray}
    A_{gq,Q}^{(2),\MOM}&=&
                     \hat{A}_{gq,Q}^{(2),\MOM}
                    +Z_{gq}^{-1,(2)}(n_f+1)
                    -Z_{gq}^{-1,(2)}(n_f)
\N\\ 
&&
                    +\Bigl(
                           \hat{A}_{gg,Q}^{(1),\MOM}
                          +Z_{gg}^{-1,(1)}(n_f+1)
                          -Z_{gg}^{-1,(1)}(n_f)
                      \Bigr)\Gamma_{gq}^{-1,(1)}~, \\
    A_{gq,Q}^{(3),\MOM}&=& 
                        \hat{A}_{gq,Q}^{(3),\MOM}
                       +Z^{-1,(3)}_{gq}(n_f+1)
                       -Z^{-1,(3)}_{gq}(n_f)
                       +Z^{-1,(1)}_{gg}(n_f+1)\hat{A}_{gq,Q}^{(2),\MOM}
\N\\ &&
                       +Z^{-1,(1)}_{gq}(n_f+1)\hat{A}_{qq}^{(2),\MOM}
                       +\Bigl[ \hat{A}_{gg,Q}^{(1),\MOM}
                              +Z^{-1,(1)}_{gg}(n_f+1)
\N\\ &&
                              -Z^{-1,(1)}_{gg}(n_f)
                        \Bigr]
                              \Gamma^{-1,(2)}_{gq}(n_f)
                       +\Bigl[ \hat{A}_{gq,Q}^{(2),\MOM}
                              +Z^{-1,(2)}_{gq}(n_f+1)
\N\\ &&
                              -Z^{-1,(2)}_{gq}(n_f)
                        \Bigr]
                              \Gamma^{-1,(1)}_{qq}(n_f)
                       +\Bigl[ \hat{A}_{gg,Q}^{(2),\MOM}
                              +Z^{-1,(2)}_{gg}(n_f+1)
\N\\ &&
                              -Z^{-1,(2)}_{gg}(n_f)
                              +Z^{-1,(1)}_{gg}(n_f+1)\hat{A}_{gg,Q}^{(1),\MOM}
\N\\ &&
                              +Z^{-1,(1)}_{gq}(n_f+1)\hat{A}_{Qg}^{(1),\MOM}
                        \Bigr]
                              \Gamma^{-1,(1)}_{gq}(n_f)
                      \label{AgqQRen1}~,
   \end{eqnarray}
   while the unrenormalized expressions are
   \begin{eqnarray}
    \Ahathat_{gq,Q}^{(2)}&=&\Bigl(\frac{\hat{m}^2}{\mu^2}\Bigr)^{\ep}\Biggl[
                     \frac{2\beta_{0,Q}}{\ep^2}\gamma_{gq}^{(0)}
                    +\frac{\hat{\gamma}_{gq}^{(1)}}{2\ep}
                    +a_{gq,Q}^{(2)}
                    +\overline{a}_{gq,Q}^{(2)}\ep
                        \Biggr]~, \label{Ahhhgq2Q} \\
    \Ahathat_{gq,Q}^{(3)}&=&
        \Bigl(\frac{\hat{m}^2}{\mu^2}\Bigr)^{3\ep/2}\Biggl\{
               -\frac{\gamma_{gq}^{(0)}}{3\ep^3}
                   \Biggl(
                           \gamma_{gq}^{(0)}\hat{\gamma}_{qg}^{(0)}
                          +\Bigl[
                                 \gamma_{qq}^{(0)}
                                -\gamma_{gg}^{(0)}
                                +10\beta_0
                                +24\beta_{0,Q}
                                      \Bigr]\beta_{0,Q}
                   \Biggr)
\N\\ &&
               +\frac{1}{\ep^2}
                   \Biggl(
                      \gamma_{gq}^{(1)}\beta_{0,Q}
                     +\frac{\hat{\gamma}_{gq}^{(1)}}{3}\Bigl[
                          \gamma_{gg}^{(0)}
                         -\gamma_{qq}^{(0)}
                         -4\beta_0
                         -6\beta_{0,Q}
                                                        \Bigr]
                     +\frac{\gamma_{gq}^{(0)}}{3}\Bigl[
                                \hat{\gamma}_{qq}^{(1), {\sf NS}}
                               +\hat{\gamma}_{qq}^{(1), {\sf PS}}
                               -\hat{\gamma}_{gg}^{(1)}
\N\\ &&
                               +2\beta_{1,Q}
                                                        \Bigr]
                     -4\delta m_1^{(-1)}\beta_{0,Q}\gamma_{gq}^{(0)} 
                   \Biggr)
               +\frac{1}{\ep}
                   \Biggl(
                                \frac{\hat{\gamma}_{gq}^{(2)}}{3}
                               +a_{gq,Q}^{(2)}\Bigl[
                                       \gamma_{gg}^{(0)}
                                      -\gamma_{qq}^{(0)}
                                      -6\beta_{0,Q}
                                      -4\beta_0
                                              \Bigr]
\N\\ &&
                               +\gamma_{gq}^{(0)}\Bigl[
                                     a_{qq,Q}^{(2),{\sf NS}}
                                    +a_{Qq}^{(2),{\sf PS}}
                                    -a_{gg,Q}^{(2)}
                                              \Bigr]
                              +\gamma_{gq}^{(0)}\beta_{1,Q}^{(1)}
                               +\frac{\gamma_{gq}^{(0)}\zeta_2}{8}\Bigl[
                                     \gamma_{gq}^{(0)}\hat{\gamma}_{qg}^{(0)}
                                    +\beta_{0,Q} ( 
                                        \gamma_{qq}^{(0)}
\N\\&&
                                        -\gamma_{gg}^{(0)}
                                        +10\beta_0
                                                 )
                                              \Bigr]
                               -\delta m_1^{(-1)}\hat{\gamma}_{gq}^{(1)}
                               -4\delta m_1^{(0)}\beta_{0,Q}\gamma_{gq}^{(0)}
                   \Biggr)
                +a_{gq,Q}^{(3)}
                        \Biggr\}~. \label{AhhhgqQ3}
   \end{eqnarray}
   The contributions to the renormalized operator matrix element are given by
   \begin{eqnarray}
    A_{gq,Q}^{(2), \MS}&=&\frac{\beta_{0,Q}\gamma_{gq}^{(0)}}{2}
    \ln^2 \Bigl(\frac{m^2}{\mu^2}\Bigr)
    +\frac{\hat{\gamma}_{gq}^{(1)}}{2} \ln \Bigl(\frac{m^2}{\mu^2}\Bigr)
    +a_{gq,Q}^{(2)}-\frac{\beta_{0,Q}\gamma_{gq}^{(0)}}{2}\zeta_2~,
    \label{Agq2QMSren} \\
    A_{gq,Q}^{(3), \MS}&=&
                     -\frac{\gamma_{gq}^{(0)}}{24}
                      \Biggl\{
                          \gamma_{gq}^{(0)}\hat{\gamma}_{qg}^{(0)}
                         +\Bigl(
                              \gamma_{qq}^{(0)}
                             -\gamma_{gg}^{(0)}
                             +10\beta_0
                             +24\beta_{0,Q}
                                     \Bigr)\beta_{0,Q}
                      \Biggr\}
                           \ln^3 \Bigl(\frac{m^2}{\mu^2}\Bigr)
\N\\ &&
                     +\frac{1}{8}\Biggl\{
                         6\gamma_{gq}^{(1)}\beta_{0,Q}
                        +\hat{\gamma}_{gq}^{(1)}\Bigl(
                                      \gamma_{gg}^{(0)}
                                     -\gamma_{qq}^{(0)}
                                     -4\beta_0
                                     -6\beta_{0,Q}
                                                 \Bigr)
                        +\gamma_{gq}^{(0)}\Bigl(
                                       \hat{\gamma}_{qq}^{(1), {\sf NS}}
                                      +\hat{\gamma}_{qq}^{(1), {\sf PS}}
\N\\ &&
                                      -\hat{\gamma}_{gg}^{(1)}
                                      +2\beta_{1,Q}
                                                 \Bigr)
                      \Biggr\}
                           \ln^2 \Bigl(\frac{m^2}{\mu^2}\Bigr)
                     +\frac{1}{8}\Biggl\{
                              4\hat{\gamma}_{gq}^{(2)}
                            + 4a_{gq,Q}^{(2)}         \Bigl(
                                    \gamma_{gg}^{(0)}
                                   -\gamma_{qq}^{(0)}
                                   -4\beta_0
\N\\ &&
                                   -6\beta_{0,Q}
                                                       \Bigr)
                            + 4\gamma_{gq}^{(0)}       \Bigl(
                                      a_{qq,Q}^{(2),{\sf NS}}
                                     +a_{Qq}^{(2),{\sf PS}}
                                     -a_{gg,Q}^{(2)}
                                     +\beta_{1,Q}^{(1)}
                                                       \Bigr)
                            + \gamma_{gq}^{(0)}\zeta_2 \Bigl(
                               \gamma_{gq}^{(0)}\hat{\gamma}_{qg}^{(0)}
                               +\Bigl[
                                        \gamma_{qq}^{(0)}\N
\end{eqnarray}
\begin{eqnarray}
%\N\\
&&
                                       -\gamma_{gg}^{(0)}
                                       +12\beta_{0,Q}
                                       +10\beta_0
                                            \Bigr]\beta_{0,Q}
                                                       \Bigr)
                      \Biggr\}
                           \ln \Bigl(\frac{m^2}{\mu^2}\Bigr)
                  + \overline{a}_{gq,Q}^{(2)} \Bigl(
                                       \gamma_{qq}^{(0)}
                                      -\gamma_{gg}^{(0)}
                                      +4\beta_0
                                      +6\beta_{0,Q}
                                             \Bigr)
\N\\ 
&& 
                  + \gamma_{gq}^{(0)} \Bigl(
                                       \overline{a}_{gg,Q}^{(2)}
                                      -\overline{a}_{Qq}^{(2),{\sf PS}}
                                      -\overline{a}_{qq,Q}^{(2),{\sf NS}}
                                             \Bigr)
                -\gamma_{gq}^{(0)}\beta_{1,Q}^{(2)}
                -\frac{\gamma_{gq}^{(0)}\zeta_3}{24} \Bigl(
                           \gamma_{gq}^{(0)}\hat{\gamma}_{qg}^{(0)}
                          +\Bigl[
                                   \gamma_{qq}^{(0)}
                                  -\gamma_{gg}^{(0)}
\N\\
 &&
                                  +10\beta_0
                                      \Bigr]\beta_{0,Q}
                                             \Bigr)
                -\frac{3\gamma_{gq}^{(1)}\beta_{0,Q}\zeta_2}{8}
                +2 \delta m_1^{(-1)} a_{gq,Q}^{(2)}
                +\delta m_1^{(0)} \hat{\gamma}_{gq}^{(1)}
                +4 \delta m_1^{(1)} \beta_{0,Q} \gamma_{gq}^{(0)}
                +a_{gq,Q}^{(3)}~. \N \\ \label{Agq3QMSren}
   \end{eqnarray}
%%%%%%%%%%%%%%%%%%%%%%%%%%%%%%%%%%%%%%%%%%%%%%%%%%%%%%%%%%%%%%%%%%%%%%%%%
%
% Subsubsection 
%
% AggQ 
%
%%%%%%%%%%%%%%%%%%%%%%%%%%%%%%%%%%%%%%%%%%%%%%%%%%%%%%%%%%%%%%%%%%%%%%%%%
 \subsubsection{$A_{gg,Q}$}
  \label{SubSec-AggQ}
%%%%%%%%%%%%%%%%%%%%%%%%%%%%%%%%%%%%%%%%%%%%%%%%%%%%%%%%%%%%%%%%%%%%%%%%%
  The $gg$--contributions start at $O(a_s^0)$,
  \begin{eqnarray}
   A_{gg,Q}&=&1+
                        a_sA_{gg,Q}^{(1)}
                       +a_s^2A_{gg,Q}^{(2)}
                       +a_s^3A_{gg,Q}^{(3)}
                       +O(a_s^4)~. \label{AggQpert}
  \end{eqnarray}
  The corresponding renormalization formulas read in the {\sf MOM}--scheme
  \begin{eqnarray}
    A_{gg,Q}^{(1), \MOM}&=&
                    \hat{A}_{gg,Q}^{(1), \MOM}
                   +Z^{-1,(1)}_{gg}(n_f+1)
                   -Z^{-1,(1)}_{gg}(n_f)
     ~, \label{AggQ1ren1} \\ 
    A_{gg,Q}^{(2), \MOM}&=&
                    \hat{A}_{gg,Q}^{(2), \MOM}
                   +Z^{-1,(2)}_{gg}(n_f+1)
                   -Z^{-1,(2)}_{gg}(n_f)
 \N\\ &&
                   +Z^{-1,(1)}_{gg}(n_f+1)\hat{A}_{gg,Q}^{(1), \MOM}
                   +Z^{-1,(1)}_{gq}(n_f+1)\hat{A}_{Qg}^{(1), \MOM}
 \N\\ &&
                   +\Bigl[ \hat{A}_{gg,Q}^{(1), \MOM}
                          +Z_{gg}^{-1,(1)}(n_f+1)
                          -Z_{gg}^{-1,(1)}(n_f)
                    \Bigr]\Gamma^{-1,(1)}_{gg}(n_f)
    ~, \label{AggQ1ren2} \\
    A_{gg,Q}^{(3), \MOM}&=&
                     \hat{A}_{gg,Q}^{(3), \MOM}
                    +Z^{-1,(3)}_{gg}(n_f+1)
                    -Z^{-1,(3)}_{gg}(n_f)
                    +Z^{-1,(2)}_{gg}(n_f+1)\hat{A}_{gg,Q}^{(1), \MOM}
\N\\ &&
                    +Z^{-1,(1)}_{gg}(n_f+1)\hat{A}_{gg,Q}^{(2), \MOM}
                    +Z^{-1,(2)}_{gq}(n_f+1)\hat{A}_{Qg}^{(1), \MOM}
\N\\ &&
                    +Z^{-1,(1)}_{gq}(n_f+1)\hat{A}_{Qg}^{(2), \MOM}
                    +\Bigl[
                            \hat{A}_{gg,Q}^{(1), \MOM}
                           +Z^{-1,(1)}_{gg}(n_f+1)
  \N\\ &&
                           -Z^{-1,(1)}_{gg}(n_f)
                     \Bigr]\Gamma^{-1,(2)}_{gg}(n_f)
                    +\Bigl[ \hat{A}_{gg,Q}^{(2), \MOM}
                           +Z^{-1,(2)}_{gg}(n_f+1)
\N\\ &&
                           -Z^{-1,(2)}_{gg}(n_f)
                           +Z^{-1,(1)}_{gq}(n_f+1)A_{Qg}^{(1), \MOM}
\N\\ &&
                           +Z^{-1,(1)}_{gg}(n_f+1)A_{gg,Q}^{(1), \MOM}
                     \Bigr]\Gamma^{-1,(1)}_{gg}(n_f)
\N\\ &&
                    +\Bigl[ \hat{A}_{gq,Q}^{(2), \MOM}
                           +Z^{-1,(2)}_{gq}(n_f+1)
                           -Z^{-1,(2)}_{gq}(n_f)
                     \Bigr]\Gamma^{-1,(1)}_{qg}(n_f)
         ~.\label{AggQ1ren3}
  \end{eqnarray}
  The general structure of the unrenormalized $1$--loop result
  is then given by
  \begin{eqnarray}
    \Ahathat_{gg,Q}^{(1)}&=&
             \Bigl(\frac{\hat{m}^2}{\mu^2}\Bigr)^{\ep/2}\Biggl(
                          \frac{\hat{\gamma}_{gg}^{(0)}}{\ep}
                         +a_{gg,Q}^{(1)}
                         +\ep\overline{a}_{gg,Q}^{(1)}
                         +\ep^2\overline{\overline{a}}_{gg,Q}^{(1)}
                        \Biggr)
                        ~. \label{AggQ1unren1}
   \end{eqnarray}
   One obtains
   \begin{eqnarray}
    \Ahathat_{gg,Q}^{(1)}&=&
                \Bigl(\frac{\hat{m}^2}{\mu^2}\Bigr)^{\ep/2}
                \Bigl(-\frac{2\beta_{0,Q}}{\ep}\Bigr)
                \exp \Bigl(\sum_{i=2}^{\infty}\frac{\zeta_i}{i}
                       \Bigl(\frac{\ep}{2}\Bigr)^{i}\Bigr)
                        ~. \label{AggQ1unren2}
   \end{eqnarray}
   Using Eq.~(\ref{AggQ1unren2}), the $2$--loop term
   is given by
   \begin{eqnarray}
    \Ahathat_{gg,Q}^{(2)}&=&
             \Bigl(\frac{\hat{m}^2}{\mu^2}\Bigr)^{\ep}
                  \Biggl[
                          \frac{1}{2\ep^2}
                             \Bigl\{
                                \gamma_{gq}^{(0)}\hat{\gamma}_{qg}^{(0)}
                               +2\beta_{0,Q}
                                    \Bigl(
                                           \gamma_{gg}^{(0)}
                                          +2\beta_0
                                          +4\beta_{0,Q}
                                    \Bigr)
                             \Bigr\}
                         +\frac{\hat{\gamma}_{gg}^{(1)}
                                  +4\delta m_1^{(-1)}\beta_{0,Q}}{2\ep}
 \N\\ &&
                         +a_{gg,Q}^{(2)}
                         +2\delta m_1^{(0)}\beta_{0,Q}
                         +\beta_{0,Q}^2\zeta_2
                         +\ep\Bigl[\overline{a}_{gg,Q}^{(2)}
                                   +2\delta m_1^{(1)}\beta_{0,Q}
                                   +\frac{\beta_{0,Q}^2\zeta_3}{6}
                             \Bigr]
                 \Biggr]~. \label{AhhhggQ2}
  \end{eqnarray}
   Again, we have made explicit one--particle reducible contributions 
   and terms stemming from mass renormalization in order to refer to  
   the notation of Refs.~\cite{Buza:1995ie,Buza:1996wv}, cf. 
   the discussion below (\ref{AhhhQg2}).
   The $3$--loop contribution becomes
   \begin{eqnarray}
    \Ahathat_{gg,Q}^{(3)}&=&
                  \Bigl(\frac{\hat{m}^2}{\mu^2}\Bigr)^{3\ep/2}
                     \Biggl[
           \frac{1}{\ep^3}
             \Biggl(
                  -\frac{\gamma_{gq}^{(0)}\hat{\gamma}_{qg}^{(0)}}{6}
                                \Bigl[
                                        \gamma_{gg}^{(0)}
                                       -\gamma_{qq}^{(0)}
                                       +6\beta_0
                                       +4n_f\beta_{0,Q}
                                       +10\beta_{0,Q}
                                \Bigr]
\N\\ &&
                  -\frac{2\gamma_{gg}^{(0)}\beta_{0,Q}}{3}
                                \Bigl[
                                        2\beta_0
                                       +7\beta_{0,Q}
                                \Bigr]
                  -\frac{4\beta_{0,Q}}{3}
                                \Bigl[
                                        2\beta_0^2
                                       +7\beta_{0,Q}\beta_0
                                       +6\beta_{0,Q}^2
                                \Bigr]
             \Biggr)
\N\\ &&
          +\frac{1}{\ep^2}
             \Biggl(
                   \frac{\hat{\gamma}_{qg}^{(0)}}{6}
                                \Bigl[
                                        \gamma_{gq}^{(1)}
                                       -(2n_f-1)\hat{\gamma}_{gq}^{(1)}
                                \Bigr]
                  +\frac{\gamma_{gq}^{(0)}\hat{\gamma}_{qg}^{(1)}}{3}
                  -\frac{\hat{\gamma}_{gg}^{(1)}}{3}                 
                                \Bigl[
                                        4\beta_0
                                       +7\beta_{0,Q}
                                \Bigr]
\N\\ &&
                  +\frac{2\beta_{0,Q}}{3}                 
                                \Bigl[
                                        \gamma_{gg}^{(1)}
                                       +\beta_1
                                       +\beta_{1,Q}
                                \Bigr]
                  +\frac{2\gamma_{gg}^{(0)}\beta_{1,Q}}{3}
                           +\delta m_1^{(-1)}
                                \Bigl[
                                    -\hat{\gamma}_{qg}^{(0)}\gamma_{gq}^{(0)}
                                    -2\beta_{0,Q}\gamma_{gg}^{(0)}
\N\\ &&
                                    -10\beta_{0,Q}^2
                                    -6\beta_{0,Q}\beta_0
                               \Bigr]
             \Biggr)
          +\frac{1}{\ep}
             \Biggl(
                   \frac{\hat{\gamma}_{gg}^{(2)}}{3}
                  -2(2\beta_0+3\beta_{0,Q})a_{gg,Q}^{(2)}
                  -n_f\hat{\gamma}_{qg}^{(0)}a_{gq,Q}^{(2)}
\N\\ &&
                  +\gamma_{gq}^{(0)}a_{Qg}^{(2)}
                  +\beta_{1,Q}^{(1)} \gamma_{gg}^{(0)}
                  +\frac{\gamma_{gq}^{(0)}\hat{\gamma}_{qg}^{(0)}\zeta_2}{16}
                                \Bigl[
                                         \gamma_{gg}^{(0)}
                                       - \gamma_{qq}^{(0)}
                                       +2(2n_f+1)\beta_{0,Q}
                                       +6\beta_0
                                \Bigr]
\N\\ &&
                  +\frac{\beta_{0,Q}\zeta_2}{4}
                                \Bigl[
                                       \gamma_{gg}^{(0)}
                                      \{2\beta_0-\beta_{0,Q}\}
                                       +4\beta_0^2
                                       -2\beta_{0,Q}\beta_0
                                       -12\beta_{0,Q}^2
                                \Bigr]
\N\\ &&
                           +\delta m_1^{(-1)}
                                \Bigl[
                                   -3\delta m_1^{(-1)}\beta_{0,Q}
                                   -2\delta m_1^{(0)}\beta_{0,Q}
                                   -\hat{\gamma}_{gg}^{(1)} 
                                \Bigr]
                           +\delta m_1^{(0)}
                                \Bigl[
                                   -\hat{\gamma}_{qg}^{(0)}\gamma_{gq}^{(0)} 
\N\\ &&
                                   -2\gamma_{gg}^{(0)}\beta_{0,Q}
                                   -4\beta_{0,Q}\beta_0
                                   -8\beta_{0,Q}^2
                                \Bigr]
                           +2 \delta m_2^{(-1)} \beta_{0,Q}
             \Biggr)
           +a_{gg,Q}^{(3)}
                  \Biggr]~. \label{Ahhhgg3Q}
   \end{eqnarray}
   The renormalized results are
   \begin{eqnarray}
    A_{gg,Q}^{(1), \MS}&=& - \beta_{0,Q} 
\ln\left(\frac{m^2}{\mu^2}\right)~,
                \label{AggQ1MSren}\\
    A_{gg,Q}^{(2), \MS}&=&
      \frac{1}{8}\Biggl\{
                          2\beta_{0,Q}
                             \Bigl(
                                    \gamma_{gg}^{(0)}
                                   +2\beta_0 
                             \Bigr)
                         +\gamma_{gq}^{(0)}\hat{\gamma}_{qg}^{(0)}
                         +8\beta_{0,Q}^2
                 \Biggr\}
                    \ln^2 \Bigl(\frac{m^2}{\mu^2}\Bigr)
                +\frac{\hat{\gamma}_{gg}^{(1)}}{2}
                     \ln \Bigl(\frac{m^2}{\mu^2}\Bigr)
\N\\ &&
                -\frac{\zeta_2}{8}\Bigl[
                                         2\beta_{0,Q}
                                            \Bigl(
                                                   \gamma_{gg}^{(0)}
                                                  +2\beta_0
                                            \Bigr)
                                        +\gamma_{gq}^{(0)}
                                            \hat{\gamma}_{qg}^{(0)}
                                  \Bigr]
                +a_{gg,Q}^{(2)}~,   \label{AggQ2MSren} \\
  A_{gg,Q}^{(3), \MS}&=&
                    \frac{1}{48}\Biggl\{
                            \gamma_{gq}^{(0)}\hat{\gamma}_{qg}^{(0)}
                                \Bigl(
                                        \gamma_{qq}^{(0)}
                                       -\gamma_{gg}^{(0)}
                                       -6\beta_0
                                       -4n_f\beta_{0,Q}
                                       -10\beta_{0,Q}
                                \Bigr)
                           -4
                                \Bigl(
                                        \gamma_{gg}^{(0)}\Bigl[
                                            2\beta_0
                                           +7\beta_{0,Q}
                                                         \Bigr]
\N\\ &&
                                       +4\beta_0^2
                                       +14\beta_{0,Q}\beta_0
                                       +12\beta_{0,Q}^2
                                \Bigr)\beta_{0,Q}
                     \Biggr\}
                     \ln^3 \Bigl(\frac{m^2}{\mu^2}\Bigr)
                    +\frac{1}{8}\Biggl\{
                            \hat{\gamma}_{qg}^{(0)}
                                \Bigl(
                                        \gamma_{gq}^{(1)}
                                       +(1-n_f)\hat{\gamma}_{gq}^{(1)}
                                \Bigr)
\N\\ &&
                           +\gamma_{gq}^{(0)}\hat{\gamma}_{qg}^{(1)}
                           +4\gamma_{gg}^{(1)}\beta_{0,Q}
                           -4\hat{\gamma}_{gg}^{(1)}[\beta_0+2\beta_{0,Q}]
                           +4[\beta_1+\beta_{1,Q}]\beta_{0,Q}
\N\\ &&
                           +2\gamma_{gg}^{(0)}\beta_{1,Q}
                     \Biggr\}
                     \ln^2 \Bigl(\frac{m^2}{\mu^2}\Bigr)
                    +\frac{1}{16}\Biggl\{
                            8\hat{\gamma}_{gg}^{(2)}
                           -8n_fa_{gq,Q}^{(2)}\hat{\gamma}_{qg}^{(0)}
                           -16a_{gg,Q}^{(2)}(2\beta_0+3\beta_{0,Q})
\N\\ &&
                           +8\gamma_{gq}^{(0)}a_{Qg}^{(2)}
                           +8\gamma_{gg}^{(0)}\beta_{1,Q}^{(1)}
                   +\gamma_{gq}^{(0)}\hat{\gamma}_{qg}^{(0)}\zeta_2
                                \Bigl(
                                        \gamma_{gg}^{(0)}
                                       -\gamma_{qq}^{(0)}
                                       +6\beta_0
                                       +4n_f\beta_{0,Q}
                                       +6\beta_{0,Q}
                                \Bigr)
\N\\ &&
                   +4\beta_{0,Q}\zeta_2
                                \Bigl( 
                                       \gamma_{gg}^{(0)}
                                      +2\beta_0
                                \Bigr)
                                \Bigl(
                                       2\beta_0
                                      +3\beta_{0,Q}
                                \Bigr)
                     \Biggr\}
                     \ln \Bigl(\frac{m^2}{\mu^2}\Bigr)
                   +2(2\beta_0+3\beta_{0,Q})\overline{a}_{gg,Q}^{(2)}\N
\end{eqnarray}
\begin{eqnarray} 
%\N\\
 &&
                   +n_f\hat{\gamma}_{qg}^{(0)}\overline{a}_{gq,Q}^{(2)}
                   -\gamma_{gq}^{(0)}\overline{a}_{Qg}^{(2)}
                   -\beta_{1,Q}^{(2)} \gamma_{gg}^{(0)}
                   +\frac{\gamma_{gq}^{(0)}\hat{\gamma}_{qg}^{(0)}\zeta_3}{48}
                                \Bigl(
                                        \gamma_{qq}^{(0)}
                                       -\gamma_{gg}^{(0)}
                                       -2[2n_f+1]\beta_{0,Q}
\N\\ &&
                                       -6\beta_0
                                \Bigr)
                   +\frac{\beta_{0,Q}\zeta_3}{12}
                                \Bigl(
                                        [\beta_{0,Q}-2\beta_0]\gamma_{gg}^{(0)}
                                       +2[\beta_0+6\beta_{0,Q}]\beta_{0,Q}
                                       -4\beta_0^2
                                \Bigr)
\N\\ &&
                   -\frac{\hat{\gamma}_{qg}^{(0)}\zeta_2}{16}
                                \Bigl(
                                        \gamma_{gq}^{(1)}
                                       +\hat{\gamma}_{gq}^{(1)}
                                \Bigr)
                   +\frac{\beta_{0,Q}\zeta_2}{8}
                                \Bigl(
                                        \hat{\gamma}_{gg}^{(1)}
                                      -2\gamma_{gg}^{(1)}
                                      -2\beta_1
                                      -2\beta_{1,Q}
                                \Bigr)
                           +\frac{\delta m_1^{(-1)}}{4}
                                \Bigl(
                                     8 a_{gg,Q}^{(2)}
\N\\ &&
                                    +24 \delta m_1^{(0)} \beta_{0,Q}
                                    +8 \delta m_1^{(1)} \beta_{0,Q} 
                                    +\zeta_2 \beta_{0,Q} \beta_0
                                    +9 \zeta_2 \beta_{0,Q}^2
                                \Bigr)
                           +\delta m_1^{(0)}
                                \Bigl(
                                     \beta_{0,Q} \delta m_1^{(0)}
                                    +\hat{\gamma}_{gg}^{(1)}
                                \Bigr)
\N\\ &&
                           +\delta m_1^{(1)}
                                \Bigl(
                                     \hat{\gamma}_{qg}^{(0)} \gamma_{gq}^{(0)}
                                    +2 \beta_{0,Q} \gamma_{gg}^{(0)}
                                    +4 \beta_{0,Q} \beta_0
                                    +8 \beta_{0,Q}^2
                                \Bigr)
                           -2 \delta m_2^{(0)} \beta_{0,Q}
                 +a_{gg,Q}^{(3)}~. \label{Agg3QMSren}
   \end{eqnarray}
%%%%%%%%%%%%%%%%%%%%%%%%%%%%%%%%%%%%%%%%%%%%%%%%%%%%%%%%%%%%%%%%%%%%%%%%%%%%%%%
%
% Chapter 5
%
% Representation in different renormalization schemes
%
%%%%%%%%%%%%%%%%%%%%%%%%%%%%%%%%%%%%%%%%%%%%%%%%%%%%%%%%%%%%%%%%%%%%%%%%%%%%%%%
\newpage
 \section{\bf\boldmath Representation in Different Renormalization Schemes}
  \label{Sec-REP}
  \renewcommand{\theequation}{\thesection.\arabic{equation}}
  \setcounter{equation}{0}
%%%%%%%%%%%%%%%%%%%%%%%%%%%%%%%%%%%%%%%%%%%%%%%%%%%%%%%%%%%%%%%%%%%%%%%%%%%%%%%
  As outlined in Section~\ref{Sec-REN}, there are different obvious 
  possibilities to choose a scheme for the renormalization of
  the mass and the coupling constant. Concerning the coupling constant, 
  we intermediately worked in a ${\sf MOM}$--scheme, which derives from 
  the condition that the external gluon lines have to be kept on--shell.
  In the end, we transformed back to 
  the $\overline{\sf MS}$--description via. Eq.~(\ref{asmoma}), since this is
  the 
  commonly used renormalization scheme. If masses are involved,
  it is useful to renormalize them in the on--mass--shell--scheme, 
  as it was done in the previous Section. In this scheme, 
  one defines the renormalized mass $m$ as the pole of the quark propagator.
  In this Section, we present the relations required to transform the
  renormalized results from Section~\ref{SubSec-RENPred} into 
  the different, related schemes. In Section~\ref{SubSec-HQElProdWave}, 
  we show how these scheme transformations affect the ${\sf NLO}$ results. 
  Denoting the $\overline{\sf MS}$--mass by 
  $\overline{m}$, there are in addition to the $\{a^{\MS},~m\}$--scheme
  adopted in Section~\ref{SubSec-RENPred} the following schemes
  \begin{eqnarray}
    \Bigl\{a_s^{\MOM},~m\Bigr\}~,\quad~
    \Bigl\{a_s^{\MOM},~\overline{m}\Bigr\}~,\quad~
    \Bigl\{a_s^{\MS},~\overline{m}\Bigr\}~.
  \end{eqnarray}
  In case of mass renormalization in the $\overline{\sf MS}$--scheme, 
  Eq.~(\ref{mren1}) becomes
  \begin{eqnarray}
    \hat{m}=Z_m^{\MS} \overline{m}
           &=&\overline{m} \Bigl[ 1 
                       + \hat{a}_s \delta \overline{m}_1
                       + \hat{a}_s^2 \delta \overline{m}_2
                 \Bigr] + O(\hat{a}_s^3)~.
            \label{mrenms}
  \end{eqnarray}
  The corresponding coefficients read, \cite{Tarrach:1980up},
  \begin{eqnarray}
    \delta \overline{m}_1 &=&
            \frac{6}{\ep}C_F \label{delm1MSbar}
               \equiv  \frac{\delta \overline{m}_1^{(-1)}}{\ep}~, \\
    \delta \overline{m}_2 &=&
                   \frac{C_F}{\ep^2}\left(18 C_F-22 C_A+8T_F(n_f+1)
                                    \right)
                  +\frac{C_F}{\ep}\left(\frac{3}{2}C_F+\frac{97}{6}C_A
                    -\frac{10}{3}T_F(n_f+1)\right) \N\\
               &\equiv&  \frac{\delta \overline{m}_2^{(-2)}}{\ep^2}
                        +\frac{\delta \overline{m}_2^{(-1)}}{\ep}~.
                      \label{delm2MSbar}
  \end{eqnarray}
  One notices that the following relations hold between the expansion 
  coefficients in $\ep$ of the on--shell-- and $\overline{\sf MS}$--terms
  \begin{eqnarray}
   \delta \overline{m}_1^{(-1)}&=&\delta m_1^{(-1)}~, \\
   \delta \overline{m}_2^{(-2)}&=&\delta m_2^{(-2)}~, \\
   \delta \overline{m}_2^{(-1)}&=&\delta m_2^{(-1)}
                                  -\delta m_1^{(-1)}\delta m_1^{(0)}
                                  +2\delta m_1^{(0)}(\beta_0+\beta_{0,Q})~.
  \end{eqnarray}
  One has to be careful, since the choice of this scheme also affects 
  the renormalization constant of the coupling in the ${\sf MOM}$--scheme. 
  This is due to the fact that in Eq.~(\ref{GluSelfBack})
  mass renormalization had been performed in the on--shell--scheme. 
  Going through the same steps as in Eqs. (\ref{GluSelfBack})--(\ref{Zgnfp1}), 
  but using the $\overline{\sf MS}$--mass, we obtain for $Z_g$ 
  in the ${\sf MOM}$--scheme. 
  \begin{eqnarray}
   {Z_g^{\MOM}}^2(\ep,n_f+1,\mu^2,\overline{m}^2)&=&
                  1+a^{\MOM}_s(\mu^2) \Bigl[
                              \frac{2}{\ep} (\beta_0(n_f)+\beta_{0,Q}f(\ep))
                        \Bigr]
\N\\ &&\hspace{-35mm}
                  +{a^{\MOM}_s}^2(\mu^2) \Bigl[
                                \frac{\beta_1(n_f)}{\ep}
                         +\frac{4}{\ep^2} (\beta_0(n_f)+\beta_{0,Q}f(\ep))^2
                          +\frac{2\beta_{0,Q}}{\ep}\delta \overline{m}_1^{(-1)}
                            f(\ep)
\N\\ &&\hspace{-35mm}
                          +\frac{1}{\ep}\Bigl(\frac{\overline{m}^2}{\mu^2}\Bigr)^{\ep}
                           \Bigl( \overline{\beta}_{1,Q}+
                                  \ep\overline{\beta}_{1,Q}^{(1)}
                                 +\ep^2\overline{\beta}_{1,Q}^{(2)}
                           \Bigr)
                          \Bigr]+O({a^{\MOM}_s}^3)~, \label{Zgheavy2MSmass}
  \end{eqnarray}
  where in the term $f(\ep)$, cf. Eq.~(\ref{fep}), the 
  $\overline{\sf MS}$--mass has to be used. 
  The coefficients differing from the on--shell--scheme 
  in the above equation are given by, cf. Eqs.~(\ref{b1Q1},~\ref{b1Q2})
  \begin{eqnarray}
   \overline{\beta}_{1,Q}&=&\beta_{1,Q}-2\beta_{0,Q}\delta m_1^{(-1)}~,\\
   \overline{\beta}_{1,Q}^{(1)}&=&
                     \beta_{1,Q}^{(1)}-2\beta_{0,Q}\delta m_1^{(0)}~,\\
   \overline{\beta}_{1,Q}^{(2)}&=&\beta_{1,Q}^{(2)}
                                -\frac{\beta_{0,Q}}{4}\Bigl(
                                        8\delta m_1^{(1)}
                                       +\delta m_1^{(-1)}\zeta_2 
                                      \Bigr)~.
  \end{eqnarray}
  The transformation formulas between the 
  different schemes follow from the condition that 
  the unrenormalized terms are equal.

  In order to transform from the $\{a_s^{\MS},~m\}$--scheme 
  to the $\{a_s^{\MOM},~m\}$--scheme, the 
  inverse of Eq.~(\ref{asmoma})
  \begin{eqnarray}
   a_s^{\MS}(m^2)&=&
              a_s^{\MOM}\Bigl[
                        1
                       +\beta_{0,Q}
                        \ln \Bigl(\frac{m^2}{\mu^2}\Bigr)
                        a_s^{\MOM} 
\N\\ && \phantom{a_s^{\MOM}\Bigl[1}
                       +\Bigl\{
                            \beta_{0,Q}^2
                            \ln^2 \Bigl(\frac{m^2}{\mu^2}\Bigr)
                           +\beta_{1,Q}
                            \ln \Bigl(\frac{m^2}{\mu^2}\Bigr)
                           +\beta_{1,Q}^{(1)}
                        \Bigr\}{a_s^{\MOM}}^2
                        \Bigr] \label{aMSON2aMOMON}
  \end{eqnarray}
  is used.
  For the transformation to the $\{a_s^{\MS},~\overline{m}\}$--scheme 
  one obtains
  \begin{eqnarray}
   m(a_s^{\MS})&=&\overline{m}(a_s^{\MS})\Biggl(
                        1
                       +\Biggl\{
                         -\frac{\delta m^{(-1)}_1}{2}
                             \ln \Bigl(\frac{\overline{m}^2}{\mu^2}\Bigr)
                         -\delta m_1^{(0)}
                        \Biggr\}a_s^{\MS}
\N\\ &&
                       +\Biggl\{
                           \frac{\delta m_1^{(-1)}}{8}
                              \Bigl[
                                    2\beta_0
                                   +2\beta_{0,Q}
                                   +\delta m_1^{(-1)}
                              \Bigr]
                               \ln^2 \Bigl(\frac{\overline{m}^2}{\mu^2}\Bigr)
                             +\frac{1}{2}\Bigl[
                                  -\delta m_1^{(0)}
                                       \Bigl( 2\beta_0
                                        +2\beta_{0,Q}
\N\\ &&
                                        -3\delta m_1^{(-1)}
                                       \Bigr)
                                  +{\delta m_1^{(-1)}}^{2}
                                  -2\delta m_2^{(-1)}
                              \Bigr]
                               \ln \Bigl(\frac{\overline{m}^2}{\mu^2}\Bigr)
                             +\delta m_1^{(1)}
                              \Bigl[
                                    \delta m_1^{(-1)}
                                   -2\beta_0
                                   -2\beta_{0,Q}
                              \Bigr]
\N\\ &&
                             +\delta m_1^{(0)}
                               \Bigl[\delta m_1^{(-1)}+\delta m_1^{(0)}\Bigr]
                             -\delta m_2^{(0)}
                        \Biggr\}{a_s^{\MS}}^2
                                     \Biggr)~.\label{mMSON2MSMS}
  \end{eqnarray}
  Finally, the transformation to the $\{a_s^{\MOM},~\overline{m}\}$
  is achieved via
  \begin{eqnarray}
   a_s^{\MS}(m^2)&=&
              a_s^{\MOM}\Bigl[
                        1
                       +\beta_{0,Q}
                        \ln \Bigl(\frac{\overline{m}^2}{\mu^2}\Bigr)
                        a_s^{\MOM}
                       +\Bigl\{
                            \beta_{0,Q}^2\ln^2 
                               \Bigl(\frac{\overline{m}^2}{\mu^2}\Bigr)
\N\\ &&
                           +\Bigl(
                                \beta_{1,Q}
                               -\beta_{0,Q}\delta m_1^{(-1)}
                            \Bigr)
                              \ln \Bigl(\frac{\overline{m}^2}{\mu^2}\Bigr)
                           +\beta_{1,Q}^{(1)}
                           -2\delta m_1^{(0)}\beta_{0,Q}
                        \Bigr\}{a_s^{\MOM}}^2
                        \Bigr]~, \label{aMSON2aMOMMS}
  \end{eqnarray}
  and
  \begin{eqnarray}
   m(a_s^{\MS})&=&\overline{m}(a_s^{\MOM})\Biggl(
                        1
                       +\Biggl\{
                         -\frac{\delta m^{(-1)}_1}{2}
                             \ln \Bigl(\frac{\overline{m}^2}{\mu^2}\Bigr)
                         -\delta m_1^{(0)}
                        \Biggr\}a_s^{\MOM}
\N\\ &&
                       +\Biggl\{
                           \frac{\delta m_1^{(-1)}}{8}
                              \Bigl[
                                    2\beta_0
                                   -2\beta_{0,Q}
                                   +\delta m_1^{(-1)}
                              \Bigr]
                               \ln^2 \Bigl(\frac{\overline{m}^2}{\mu^2}\Bigr)
                             +\frac{1}{2}\Bigl[
                                  -\delta m_1^{(0)}
                                       \Bigl( 2\beta_0
                                        +4\beta_{0,Q}
\N\\ &&
                                        -3\delta m_1^{(-1)}
                                       \Bigr)
                                  +{\delta m_1^{(-1)}}^{2}
                                  -2\delta m_2^{(-1)}
                              \Bigr]
                               \ln \Bigl(\frac{\overline{m}^2}{\mu^2}\Bigr)
                             +\delta m_1^{(1)}
                              \Bigl[
                                    \delta m_1^{(-1)}
                                   -2\beta_0
                                   -2\beta_{0,Q}
                              \Bigr]
\N\\ &&
                             +\delta m_1^{(0)}
                               \Bigl[\delta m_1^{(-1)}+\delta m_1^{(0)}\Bigr]
                             -\delta m_2^{(0)}
                        \Biggr\}{a_s^{\MOM}}^2
                                     \Biggr)~.\label{mMSON2MOMMS}
  \end{eqnarray}
  The expressions for the OMEs in different schemes are then obtained by
  inserting the relations (\ref{aMSON2aMOMON})--(\ref{mMSON2MOMMS})
  into the general expression (\ref{PertOmeren}) and expanding 
  in the coupling constant. 
%%%%%%%%%%%%%%%%%%%%%%%%%%%%%%%%%%%%%%%%%%%%%%%%%%%%%%%%%%%%%%%%%%%%%%%%%%%%%%%
%
%	Scheme Dependence at ${\sf NLO}$
%
%%%%%%%%%%%%%%%%%%%%%%%%%%%%%%%%%%%%%%%%%%%%%%%%%%%%%%%%%%%%%%%%%%%%%%%%%%%%%%%
  \subsection{\bf\boldmath Scheme Dependence at ${\sf NLO}$}
   \label{SubSec-HQElProdWave}
%%%%%%%%%%%%%%%%%%%%%%%%%%%%%%%%%%%%%%%%%%%%%%%%%%%%%%%%%%%%%%%%%%%%%%%%%%%%%%%
   Finally, we would like to comment on how the factorization formulas 
   for the heavy flavor Wilson coefficients, (\ref{eqWIL1})--(\ref{eqWIL5}), 
   have to be applied to obtain a complete description.
   Here, the renormalization of the coupling constant has to be 
   carried out in the same way for all quantities contributing.
   The general factorization formula (\ref{CallFAC}) holds only 
   for completely inclusive quantities, including radiative corrections 
   containing heavy quark loops, \cite{Buza:1996wv}.

\begin{figure}[htb]
      \begin{center}
       \includegraphics[angle=0, width=10.0cm]{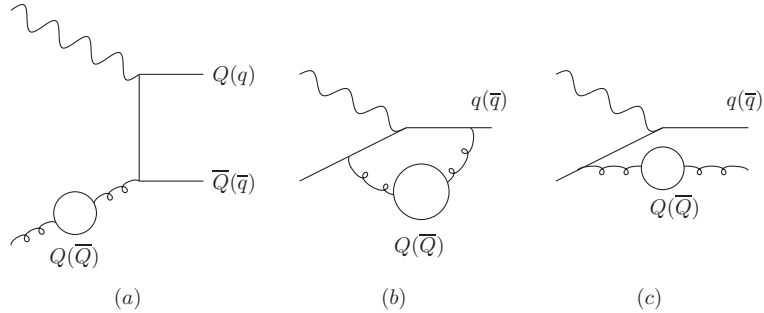}
      \end{center}
      \begin{center} 
       \caption{\sf $O(a_s^2)$ virtual heavy quark corrections.}
       \label{2LOOPIRR}
       \noindent
       \small
      \end{center}
      \normalsize
     \end{figure}

   One has to distinguish one-particle irreducible and reducible diagrams,
   which both contribute in the calculation. We would like to remind the reader
   of the background of this aspect. If one evaluates the heavy-quark Wilson
   coefficients, diagrams
   of the type shown in Figure \ref{2LOOPIRR} may appear as well. 
   Diagram (a) contains a virtual heavy quark loop correction to 
   the gluon propagator in the initial state and contributes to the terms 
   $L_{g,i}$ and $H_{g,i}$, respectively, depending on whether a light or 
   heavy quark pair is produced in the final state. Diagrams (b), (c) 
   contribute to $L_{q,i}^{\sf NS}$ and contain radiative corrections 
   to the gluon propagator due to heavy quarks as well. 
   The latter diagrams contribute to $F_{(2,L)}(x,Q^2)$ in the inclusive
   case, but are absent in the semi--inclusive $Q\overline{Q}$--production
   cross section. The same holds for diagram (a) if a $q\overline{q}$--pair
   is produced.
%%%%%%%%%%%%%%%%%%%%%%%%%%%%%%%%%%%%%%%%%%%%%%%%%%%%%%%%%%%%%%%%%%%%%%%%%%%%%%%
%%%%%%%%%%%%%%%%%%%%%%%%%%%%%%%%%%%%%%%%%%%%%%%%%%%%%%%%%%%%%%%%%%%%%%%%%%%%%%%
   In Refs.~\cite{Laenen:1992zkxLaenen:1992xs,*Riemersma:1994hv}, the coupling 
   constant was renormalized in the ${\sf MOM}$--scheme by absorbing 
   the contributions of diagram (a) into the coupling constant, as a 
   consequence of which the term $L_{g,i}$ appears for the first time 
   at $O(a_s^3)$. This can be made explicit by considering the 
   complete gluonic Wilson coefficient up to $O(a_s^2)$, including one heavy 
   quark, cf. Eqs. (\ref{eqWIL3}, \ref{eqWIL5}),
    \begin{eqnarray}
      &&C_{g,(2,L)}(n_f)+L_{g,(2,L)}(n_f+1)+H_{g,(2,L)}(n_f+1) 
      =
           a_s^{\MS} \Bigl[~A_{Qg}^{(1), \MS}~\delta_2
                     +C^{(1)}_{g,(2,L)}(n_f+1) \Bigr]
\N\\ &&\hspace{2mm}
         + {a_s^{\MS}}^2 \Bigl[~A_{Qg}^{(2), \MS}~\delta_2
     +A_{Qg}^{(1), \MS}~C^{(1), {\sf NS}}_{q,(2,L)}(n_f+1)
     +A_{gg,Q}^{(1), \MS}~C^{(1)}_{g,(2,L)}(n_f+1) 
\N\\ &&\hspace{18mm}
     +C^{(2)}_{g,(2,L)}(n_f+1) \Bigr]~.  \label{Sec5Ex1}
    \end{eqnarray}
   The above equation is given in the $\overline{\sf MS}$--scheme, and the
   structure 
   of the OMEs can be inferred from Eqs. (\ref{AQg1MSren},~\ref{AQg2MSren}).
   Here, diagram (a) gives a contribution, corresponding exactly 
   to the color factor $T_F^2$. The transformation to the
   ${\sf MOM}$--scheme for $a_s$,
   cf. Eqs.~(\ref{asmoma}, \ref{asmsa}), yields 
    \begin{eqnarray}
      &&C_{g,(2,L)}(n_f)+L_{g,(2,L)}(n_f+1)+H_{g,(2,L)}(n_f+1) 
      =
           a_s^{\MOM} \Bigl[~A_{Qg}^{(1), \MS}~\delta_2
                     +C^{(1)}_{g,(2,L)}(n_f+1) \Bigr]
\N\\ &&\hspace{2mm}
         + {a_s^{\MOM}}^2 \Bigl[~A_{Qg}^{(2), \MS}~\delta_2
     +~\beta_{0,Q}\ln\Bigl(\frac{m^2}{\mu^2}\Bigr)A_{Qg}^{(1), \MS}\delta_2
     +A_{Qg}^{(1), \MS}~C^{(1), {\sf NS}}_{q,(2,L)}(n_f+1)
\N\\ &&\hspace{2mm}
     +A_{gg,Q}^{(1), \MS}~C^{(1)}_{g,(2,L)}(n_f+1) 
     +\beta_{0,Q}\ln\Bigl(\frac{m^2}{\mu^2}\Bigr)~C^{(1)}_{g,(2,L)}(n_f+1) 
     +C^{(2)}_{g,(2,L)}(n_f+1) \Bigr]~.
     \label{Sec5Ex2}
    \end{eqnarray}
   By using the general structure of the renormalized OMEs,
   Eqs.~(\ref{AQg1MSren},~\ref{AQg2MSren},~\ref{AggQ1MSren}), one notices 
   that all contributions due to diagram (a) cancel in the 
   ${\sf MOM}$--scheme, i.e., the color factor $T_F^2$ does not occur
   at the $2$--loop level. Thus the factorization formula reads
    \begin{eqnarray}
      &&C_{g,(2,L)}(n_f)+L_{g,(2,L)}(n_f+1)+H_{g,(2,L)}(n_f+1) 
      =
\N\\ &&\hspace{5mm}
           a_s^{\MOM} \Bigl[~A_{Qg}^{(1), \MOM}~\delta_2
                     +C^{(1)}_{g,(2,L)}(n_f+1) \Bigr]
\N\\ &&\hspace{2mm}
         + {a_s^{\MOM}}^2 \Bigl[~A_{Qg}^{(2), \MOM}~\delta_2
     +A_{Qg}^{(1), \MOM}~C^{(1), {\sf NS}}_{q,(2,L)}(n_f+1)
     +C^{(2)}_{g,(2,L)}(n_f+1) \Bigr]~.  \label{Sec5Ex3}
    \end{eqnarray}
    Splitting up Eq.~(\ref{Sec5Ex3}) into $H_{g,i}$ and $L_{g,i}$, 
    one observes that $L_{g,i}$  vanishes at $O(a_s^2)$ and the term 
    $H_{g,i}$ is the one calculated in Ref.~\cite{Buza:1995ie}. This is 
    the asymptotic expression of the gluonic heavy flavor Wilson coefficient 
    as calculated
    in Refs.~\cite{Laenen:1992zkxLaenen:1992xs,*Riemersma:1994hv}.
    Note that the observed cancellation was due to the fact that the term 
    $A_{gg,Q}^{(1)}$ receives only contributions from the heavy 
    quark loops of the gluon--self energy, which also enters 
    into the definition of the ${\sf MOM}$--scheme. It is not clear whether 
    this can be achieved at the $3$--loop level as well, i.e., 
    transforming the general inclusive factorization formula (\ref{CallFAC})
    in such a way that only the contributions due to heavy flavors in the final
    state remain. Therefore one should use these asymptotic expressions only
    for completely inclusive analyzes, where heavy and light flavors are 
    treated together. This approach has also been adopted in 
    Ref.~\cite{Buza:1996wv} for the renormalization of the massive OMEs, which 
    was performed in the $\overline{\sf MS}$--scheme and not in the ${\sf MOM}$--scheme, 
    as previously in Ref.~\cite{Buza:1995ie}. The radiative corrections 
    in the ${\sf NS}$--case can be treated in the same manner.
    Here the scheme
    transformation affects only the light Wilson coefficients and not the OMEs
    at the $2$--loop level. In the ${\overline{\sf MS}}$--scheme,
    one obtains the 
    following asymptotic expression up to $O(a_s^2)$ from Eqs. 
    (\ref{LNSFAC}, \ref{eqWIL1}).
    \begin{eqnarray}
     \label{Sec5Ex4}
      &&C_{q,(2,L)}^{\sf NS}(n_f)
     +L_{q,(2,L)}^{\sf NS}(n_f+1) =
 \N\\ &&
     \hspace{2mm}
       1+a_s^{\MS} C_{q,(2,L)}^{(1), {\sf NS}}(n_f+1)
        +{a_s^{\MS}}^2
          \Bigl[A_{qq,Q}^{(2), {\sf NS}, \MS}(n_f+1)~\delta_2 +
          C^{(2), {\sf NS}}_{q,(2,L)}(n_f+1)\Bigr]~.
    \end{eqnarray}
    Transformation to the ${\sf MOM}$--scheme yields
    \begin{eqnarray}
     \label{Sec5Ex5}
     &&C_{q,(2,L)}^{\sf NS}(n_f)
     +L_{q,(2,L)}^{\sf NS}(n_f+1) =
 \N\\ &&
     \hspace{10mm}
       1+a_s^{\MOM} C_{q,(2,L)}^{(1), {\sf NS}}(n_f+1)
        +{a_s^{\MOM}}^2
            \Bigl [A_{qq,Q}^{(2), {\sf NS}, \MOM}(n_f+1)~\delta_2
\N \\ && \hspace{20mm}
                   +~\beta_{0,Q}\ln\Bigl(\frac{m^2}{\mu^2}\Bigr)
                     C_{q,(2,L)}^{(1), {\sf NS}}(n_f+1)
                  +C^{(2), {\sf NS}}_{q,(2,L)}(n_f+1)
            \Bigr]~.
    \end{eqnarray}
    Note that $A_{qq,Q}^{(2),{\sf NS}}$, Eq.~(\ref{Aqq2NSQMSren}), 
    is not affected by this scheme transformation. As is obvious from 
    Figure~\ref{2LOOPIRR}, the logarithmic term in Eq. (\ref{Sec5Ex5})
    can therefore only be attributed to the massless Wilson coefficient.
    Separating the light from the heavy part one obtains
    \begin{eqnarray}
     \label{Sec5Ex6}
      &&L_{q,(2,L)}^{(2),{\sf NS}, \MOM}(n_f+1)=
\N\\ && \hspace{5mm}
            A_{qq,Q}^{(2), {\sf NS}, \MOM}(n_f+1)~\delta_2
                   +~\beta_{0,Q}\ln\Bigl(\frac{m^2}{\mu^2}\Bigr)
                     C_{q,(2,L)}^{(1), {\sf NS}}(n_f+1)
                  +\hat{C}^{(2), {\sf NS}}_{q,(2,L)}(n_f)~. \label{LNSFAC3}
    \end{eqnarray}
    This provides the same results as Eqs.~(4.23)--(4.29) of 
    Ref.~\cite{Buza:1995ie}. These are the asymptotic expressions 
    of the ${\sf NS}$ heavy flavor Wilson coefficients from  
    Refs.~\cite{Laenen:1992zkxLaenen:1992xs,*Riemersma:1994hv}, where only 
    the case of $Q\overline{Q}$--production in the final state has been 
    considered. Hence the logarithmic term in Eq. (\ref{LNSFAC3}) just 
    cancels the contributions due to diagrams (b), (c)
    in Figure~\ref{2LOOPIRR}.
%%%%%%%%%%%%%%%%%%%%%%%%%%%%%%%%%%%%%%%%%%%%%%%%%%%%%%%%%%%%%%%%%%%%%%%%%%%%%%%
%
% Chapter 6
%
% Calculation of the Massive Operator Matrix Elements at $O(a_s^2)$ 
% up to $O(\ep)$
%
%%%%%%%%%%%%%%%%%%%%%%%%%%%%%%%%%%%%%%%%%%%%%%%%%%%%%%%%%%%%%%%%%%%%%%%%%%%%%%%
\newpage
 \section{\bf\boldmath Calculation of the Massive Operator Matrix Elements 
             up to $O(a_s^2\ep)$}
  \label{Sec-2L}
  \renewcommand{\theequation}{\thesection.\arabic{equation}}
  \setcounter{equation}{0}
%%%%%%%%%%%%%%%%%%%%%%%%%%%%%%%%%%%%%%%%%%%%%%%%%%%%%%%%%%%%%%%%%%%%%%%%%%%%%%%
   The quarkonic $2$--loop massive OMEs $A_{Qg}^{(2)}, A_{Qq}^{(2), {\sf PS}}$
   and $A_{qq,Q}^{(2)}$ have been calculated for the first time 
   in Ref.~\cite{Buza:1995ie} to construct
   asymptotic expressions for the ${\sf NLO}$ heavy flavor Wilson Coefficients
   in the limit $Q^2~\gg~m^2$, cf. Section~\ref{SubSec-HQAsym}. 
   The corresponding gluonic OMEs $A_{gg,Q}^{(2)}$
   and $A_{gq,Q}^{(2)}$ were calculated
   in Ref.~\cite{Buza:1996wv}, where they were used within 
   a VFNS description of heavy flavors in high--energy scattering processes, 
   see Section~\ref{SubSec-HQFlav}.  In these calculations, 
   the integration--by--parts
   technique, \cite{Chetyrkin:1980pr},
   has been applied to reduce the number of propagators 
   occurring in the momentum integrals. Subsequently, the integrals 
   were calculated in $z$--space, which led to a variety 
   of multiple integrals of logarithms, partially with complicated 
   arguments. The final results were given in terms 
   of polylogarithms and Nielsen--integrals, see Appendix \ref{App-SpeFunHarm}.
   The quarkonic terms have been confirmed 
   in Ref.~\cite{Bierenbaum:2007qe}, cf. also \cite{SKdiploma}, 
   where a different approach was followed. The calculation was 
   performed in Mellin--$N$ space and by avoiding the integration--by--parts 
   technique. Using representations in terms of generalized hypergeometric 
   functions, the integrals could be expressed in terms of multiple 
   finite and infinite sums with one free parameter, $N$.
   The advantage of this approach is that the evaluation of these sums 
   can be automatized
   using various techniques, simplifying the calculation. The final 
   result is then obtained in Mellin--space in terms of nested harmonic
   sums or $Z$--sums, cf. \cite{Blumlein:1998if,Vermaseren:1998uu} and Appendix
   \ref{App-SpeFunHarm}. An additional simplification was found
   since the final result, e.g., for $A_{Qg}^{(2)}$ can be expressed 
   in terms of {\sf two} basic harmonic sums only, using algebraic,
   \cite{Blumlein:2003gb}, and structural relations,
   \cite{Blumlein:2009ta,Blumlein:2009fz}, between them. This is another
   example of an observation which has been made 
   for many different single scale quantities in high--energy physics, 
   namely that the Mellin--space representation is better 
   suited to the problem than the $z$--space representation.
   
\begin{figure}[htb]
   \begin{center}
   \includegraphics[angle=0, width=3cm]{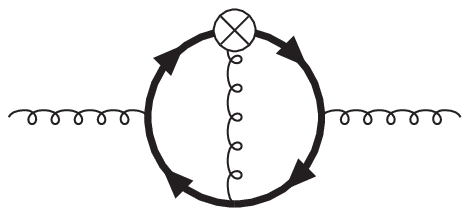}
   \includegraphics[angle=0, width=2cm]{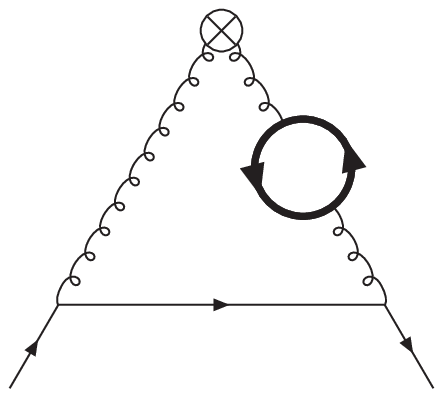}
   \includegraphics[angle=0, width=3cm]{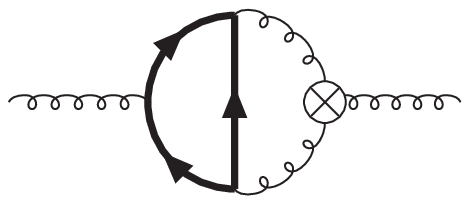}
   \includegraphics[angle=0, width=3cm]{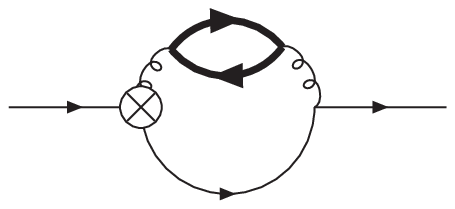}
   \includegraphics[angle=0, width=3cm]{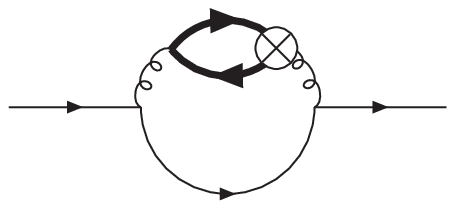}
   \end{center}
   {\small
   \hspace*{17mm}
   ($\sf Qg$) \hspace{1.5cm}
   ($\sf gq,Q$) \hspace{1.3cm}
   ($\sf gg,Q$) \hspace{1.8cm}
   ($\sf NS$) \hspace{1.7cm}
   ($\sf Qq,~PS$)
   \begin{center} 
   \caption[{\sf Examples for 2--loop diagrams contributing to the massive 
                OMEs.}]
           {\sf Examples for 2--loop diagrams contributing to the massive 
             OMEs. Thick lines: heavy quarks, 
                curly lines: gluons, full lines:  quarks.} 
   \label{diaex2L}
   \end{center}}
   \end{figure}

   As has been outlined in Section~\ref{Sec-REN}, the $O(\ep)$--terms
   of the unrenormalized $2$--loop massive OMEs are needed in the 
   renormalization of the $3$--loop contributions. In this Section, 
   we calculate these terms based on the approach 
   advocated in Ref.~\cite{Bierenbaum:2007qe},
   which is a new result, \cite{Bierenbaum:2008yu,Bierenbaum:2009zt}.
   Additionally, 
   we re--calculate the gluonic OMEs up to the constant term in $\ep$
   for the first time, cf. \cite{Bierenbaum:2009zt,Buza:1996wv}. 
   Example diagrams for each OME are shown in Figure~\ref{diaex2L}.
   
   In Section~\ref{SubSec-2LF32}, we explain how the integrals
   are obtained in terms of finite and infinite sums 
   using representations in terms of generalized hypergeometric 
   functions, cf. \cite{Slater,Bailey,*Roy:2001} 
   and Appendix~\ref{App-SpeFunFPQ}.
   For the calculation of these sums we mainly used the 
   {\sf MATHEMATICA}--based program \SigmaP, \cite{sigma1,sigma2}, 
   which is discussed in Section~\ref{SubSec-2LInfSum}. 
   The results are presented in Section~\ref{SubSec-2LRes}.
   Additionally, we make several remarks on the
   ${\sf MOM}$--scheme, which has to be adopted 
   intermediately for the renormalization
   of the coupling constant, cf. Section~\ref{SubSec-RENCo}.
   In Section~\ref{SubSec-2LChecks}, different checks of the results are
   presented.
%%%%%%%%%%%%%%%%%%%%%%%%%%%%%%%%%%%%%%%%%%%%%%%%%%%%%%%%%%%%%%%%%%%%%%%%%
%%%%%%%%%%%%%%%%%%%%%%%%%%%%%%%%%%%%%%%%%%%%%%%%%%%%%%%%%%%%%%%%%%%%%%%%%
   \subsection{\bf\boldmath Representation in Terms of Hypergeometric Functions} 
    \label{SubSec-2LF32}
%%%%%%%%%%%%%%%%%%%%%%%%%%%%%%%%%%%%%%%%%%%%%%%%%%%%%%%%%%%%%%%%%%%%%%%%%%%%%% 
    All diagrams contributing to the massive OMEs are shown 
    in Figures 1--4 in Ref.~\cite{Buza:1995ie} and in 
    Figures 3,4 in Ref.~\cite{Buza:1996wv}, respectively. They 
    represent $2$--point 
    functions with on--shell external momentum $p$, $p^2=0$.
    They are expressed in two parameters, 
    the heavy quark mass $m$ and the Mellin--parameter $N$. Since 
    the mass can be factored out of the integrals, the problem effectively 
    contains a single scale. The parameter $N$ represents the spin 
    of the composite operators, (\ref{COMP1})--(\ref{COMP3}),
    and enters the calculation 
    via the Feynman--rules for 
    these objects, cf. Appendix \ref{App-FeynRules}. 

\begin{figure}[H]
     \begin{center}
      \includegraphics[angle=0, height=2cm]{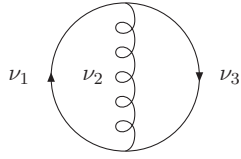}
     \end{center}
     \begin{center} 
      \caption[{\sf Basic $2$--loop massive tadpole }]
     {\sf Basic $2$--loop massive tadpole }
        \label{2LMa}
      \small
     \end{center}
     \normalsize
    \end{figure}

    Since the external momentum does not appear in the final result,
    the corresponding scalar integrals reduce to massive tadpoles
    if one sets $N=0$. 
    In order to explain our method, 
    we consider first the massive $2$--loop tadpole shown in 
    Figure \ref{2LMa}, from which all OMEs can be derived at this 
    order, by attaching $2$ outer legs and inserting the composite operator in
    all possible ways, i.e., both on the lines and on the vertices. 
%%%%%%%%%%%%%%%
%%%%%%%%%%%%%%%
    In Figure \ref{2LMa}, the wavy line is massless and the full
    lines are massive. Here $\nu_i$ labels the power of the
    propagator. We adopt the convention 
    ${\nu_{i...j}}\equiv{\nu_i}+...+{\nu_j}$ etc. The 
    corresponding dimensionless
    momentum integral reads in Minkowski--space 
    \begin{eqnarray}
     I_1&=&
            \int\int \frac{d^D k_1d^Dk_2}{(4\pi)^{4D}}
                     \frac{(4\pi)^4(-1)^{\nu_{123}-1}(m^2)^{\nu_{123}-D}}
                          {(k_1^2-m^2)^{\nu_1}(k_1^2-k_2^2)^{\nu_2}
                           (k_2^2-m^2)^{\nu_3}}~, \label{momint2L}
    \end{eqnarray}
    where we have attached a factor $(4\pi)^4(-1)^{\nu_{123}-1}$ 
    for convenience.
    Using standard Feynman--parametrization and Eq.~(\ref{Dint})
    for momentum integration, one obtains the following Feynman--parameter 
    integral
    \begin{eqnarray}
     I_1&=&
          \Gamma\Biggl[\frac[0pt]{\nu_{123}-4-\ep}{\nu_1,\nu_2,\nu_3}\Biggr]
                 \iint_0^1~dxdy~ \frac{ x^{1+\ep/2-\nu_2}
                                           (1-x)^{\nu_{23}-3-\ep/2}
                                            y^{\nu_3-1}
                                           (1-y)^{\nu_{12}-3-\ep/2}}
                                 {(4\pi)^{\ep}(1-xy)^{\nu_{123}-4-\ep}}~,\N\\
                 \label{parint2L}
    \end{eqnarray}
    which belongs to the class of the hypergeometric 
    function $\empty_3F_2$ with argument $z=1$, see Appendix 
    \ref{App-SpeFunFPQ}. Applying Eq.~(\ref{FPQint}), one obtains
    \begin{eqnarray}
     I_1&=&
           S_{\ep}^2
           \exp\Bigl( \sum_{i=2}^{\infty} \frac{\zeta_i}{i}\ep^i\Bigr)
           \Gamma\Biggl[\frac[0pt]{{\nu_{123}}-4-\ep,2+\ep/2-{\nu_2},
                                      {\nu_{23}}-2-\ep/2,{\nu_{12}}-2-\ep/2}
                                      {1-\ep,{\nu_1},{\nu_2},{\nu_3}
                                       ,{\nu_{123}}-2-\ep/2}\Biggr] 
\N \\ &&
         \times~\phantom{}_{{3}}{F}_{{2}}\Biggl[\frac[0pt]{{\nu_{123}}-4-\ep
                                           ,2+\ep/2-{\nu_2},{\nu_3}}
                                  {{\nu_3},{\nu_{123}}-2-\ep/2};{1}\Biggr]~,
                 \label{F322L}
    \end{eqnarray}
    where we have used Eq.~(\ref{SepA}). 
    The term $\nu_3$ in the argument of the $\empty_3F_2$
    cancels between nominator and denominator and thus one can use 
    Gauss's theorem, Eq.~(\ref{Gauss}), to write the result
    in terms of $\Gamma$--functions
    \begin{eqnarray}
     I_1&=&\Gamma\Biggl[\frac[0pt]{ 
                                   \nu_{123}-4-\ep,
                                   2+\ep/2-\nu_2,
                                   \nu_{12}-2-\ep/2,
                                   \nu_{23}-2-\ep/2
                                  }{
                                   1-\ep,
                                   2+\ep/2,
                                   \nu_1,
                                   \nu_3,
                                   \nu_{123}+\nu_2-4-\ep
                                  }
                 \Biggr]
           S_{\ep}^2
           \exp\Bigl( \sum_{i=2}^{\infty} \frac{\zeta_i}{i}\ep^i\Bigr)~. \N\\
                \label{F212L}
    \end{eqnarray}
    This calculation is of course trivial and Eq.~(\ref{F212L})
    can be easily checked using {\sf MATAD}, cf. Ref.~\cite{Steinhauser:2000ry}
    and Section~\ref{SubSec-3LMatad}. 
    Next, let us consider the case of arbitrary moments in presence of the
    complete numerator structure. 
    Since the final result contains the factor $(\Delta.p)^N$, one cannot 
    set $p$ to zero anymore.
    This increases the number of propagators and hence the number 
    of Feynman--parameters in Eq.~(\ref{parint2L}). 
    Additionally, the terms $(\Delta.q)^A$ in the integral lead
    to polynomials in the Feynman--parameters to a symbolic power 
    in the integral, which can not be integrated trivially.
    Hence neither Eq.~(\ref{FPQint}) nor Gauss's theorem
    can be applied anymore in the general case.
    
    However, the structure of the integral in Eq.~(\ref{parint2L})
    does not change. For {\sf any diagram} deriving from the 
    $2$--loop tadpole, a general integral of the type
    \begin{eqnarray}
     I_2&=&C_2 \iint_0^1~dxdy~ \frac{x^a(1-x)^by^c(1-y)^d}{
                      (1-xy)^e}\int_0^1dz_1...\int_0^1dz_i~ {{\sf P}}\Bigl(x,y,z_1 \ldots z_i,{N}\Bigr)~ \label{Gen2L}
    \end{eqnarray}
    is obtained.
    Here ${\sf P}$ is a rational function of $x,y$ and possibly 
    more parameters $z_1$...$z_i$. $N$ denotes the Mellin--parameter
    and occurs in some exponents. Note that operator 
    insertions with more than two legs give rise to additional 
    finite sums in ${\sf P}$, see Appendix \ref{App-FeynRules}.
    For fixed values of $N$, one can expand ${\sf P}$
    and the integral $I_2$ turns into 
    a finite sum over integrals of the type $I_1$. The terms
    $\nu_i$ in these integrals 
    might have been shifted by integers, but
    after expanding in $\ep$, the one--fold infinite 
    sum can be performed, e.g., using the ${\sf FORM}$--based 
    code ${\sf Summer}$, \cite{Vermaseren:1998uu}.
     
    To illustrate the sophistication occurring once one keeps the complete
    dependence on $N$ in an example, we consider the scalar integral
    contributing to $A_{Qg}^{(2)}$ shown in Figure \ref{diaex2L}.
    After momentum integration, it reads
    \begin{eqnarray}
     I_3 &=&\frac{(\Delta p)^{N-2}\Gamma(1-\ep)}{(4\pi)^{4+\ep}(m^2)^{1-\ep}}
        \iiiint dudzdydx 
        \frac{(1-u)^{-\ep/2}z^{-\ep/2}(1-z)^{\ep/2-1}}{(1-u+uz)^{1-\ep}(x-y)}
        \N\\
      &&\Biggl[ \Bigl(zyu+x(1-zu)\Bigr)^{N-1}
              -\Bigl((1-u)x+uy\Bigr)^{N-1}\Biggr]~, 
    \end{eqnarray}
    where we have performed the finite sum already, which stems
    from the operator insertion. 
    Here and below, the Feynman-parameter integrals are carried out
    over the respective unit-cube.
    This integral is of the type of Eq.~(\ref{Gen2L}) and 
    the term $x-y$ in the 
    denominator cancels for fixed values of $N$. 
    Due to the operator insertion on an internal vertex, it is one of the more 
    involved integrals in the $2$--loop case. For almost all other 
    integrals, all but two parameters can be integrated automatically, 
    leaving only a single infinite sum of the type of Eq.~(\ref{F322L}) 
    with $N$ appearing in the parameters of the hypergeometric 
    function, cf. e.g. \cite{Bierenbaum:2007qe,SKdiploma,Bierenbaum:2007dm}. 
    In order to render this example calculable, 
    suitable variable transformations, as, e.g., 
    given in Ref.~\cite{Hamberg:thesis}, are 
    applied, \cite{Bierenbaum:2007qe,SKdiploma}. Thus one 
    arrives at the following double sum
    \begin{eqnarray}
     I_3
     &=&\frac{S_{\ep}^2(\Delta p)^{N-2}}{(4\pi)^4(m^2)^{1-\ep}}
        \exp\Biggl\{\sum_{l=2}^{\infty}\frac{\zeta_l}{l}\ep^l\Biggr\}
        \frac{2\pi}{N\sin(\frac{\pi}{2}\ep)}
        \sum_{j=1}^{N}\Biggl\{\binom{N}{j}(-1)^j+\delta_{j,N}\Biggr\}
        \N \\
     && \times\Biggl\{
        \frac{\Gamma(j)\Gamma(j+1-\frac{\ep}{2})}
        {\Gamma(j+2-\ep)\Gamma(j+1+\frac{\ep}{2})}
       -\frac{B(1-\frac{\ep}{2},1+j)}{j}~ 
        _3F_2\Biggl[\frac[0pt]{1-\ep,\frac{\ep}{2},j+1}{1,j+2
        -\frac{\ep}{2}};1\Biggr]
        \Biggr\} \N\\
     &=&\frac{S_{\ep}^2(\Delta p)^{N-2}}{(4\pi)^4(m^2)^{1-\ep}}
          \Bigl\{I_3^{(0)}+I_3^{(1)}\ep +O(\ep^2)\Bigr\}~.\label{ffinal}
    \end{eqnarray}
    Note that in our approach no expansion in $\ep$ is needed 
    until a sum--representation of the kind of Eq.~(\ref{ffinal}) is obtained. 
    Having performed the momentum integrations, the expressions 
    of almost all diagrams were given in terms of single generalized
    hypergeometric series $_3F_2$ at $z=1$, with possibly 
    additional finite summations. 
    These infinite sums could then be safely expanded in $\ep$, leading 
    to different kinds of sums depending on the Mellin--parameter $N$. 
    The summands are typically products of harmonic sums with
    different arguments, weighted by summation parameters and contain
    hypergeometric terms~\footnote{$f(k)$ is hypergeometric in $k$ 
    iff $f(k+1)/f(k)=g(k)$ for some fixed rational function $g(k)$.}, 
    like binomials 
    or Beta--function factors $B(N,i)$, cf. Eq.~(\ref{betafun1}).
    Here $i$ is a summation--index.
    In the most difficult cases, double sums as 
    in Eq.~(\ref{ffinal}) or even triple sums were obtained, which 
    had to be treated accordingly. 
    In general, these
    sums can be expressed in terms of nested harmonic sums 
    and $\zeta$--values. Note that sums containing
    Beta--functions with different arguments, e.g. $B(i,i),~B(N+i,i)$,
    usually do not lead to harmonic sums in the final result. 
    Some of these sums can be performed by the existing packages \cite{Vermaseren:1998uu,Weinzierl:2002hv,Moch:2005uc}.
    However, there exists so far no automatic computer program to
    calculate sums which contain Beta--function factors of the 
    type $B(N,i)$ and single harmonic sums in the summand.
    These sums can be calculated applying analytic methods, 
    as integral representations, and general summation methods, as encoded 
    in the \SigmaP~package \cite{Refined,Schneider:2007,sigma1,sigma2}. In the
    next Section, we will present details on this.
    
    Before finishing this Section, we give the result in terms of 
    harmonic sums for the 
    double sum in Eq.~(\ref{ffinal}) applying these summation methods. 
    The $O(\ep^0)$ of Eq.~(\ref{ffinal}) is needed for the
    constant term $a_{Qg}^{(2)}$, cf. 
    Refs.~\cite{Bierenbaum:2007qe,Bierenbaum:2007dm}. The linear 
    term in $\ep$ reads
    \begin{eqnarray}
     I_3^{(1)}&=&\frac{1}{N} \Biggl[-2{S_{2,1}}+2{S_3} 
                                   +\frac{4N+1}{N}{S_2}
               -\frac{{S^2_1}}{N}-\frac{4}{N}{S_1}\Biggr]~, 
    \end{eqnarray}
    where we adopt the notation to take harmonic sums at argument 
    $N$, if not stated otherwise.
%%%%%%%%%%%%%%%%%%%%%%%%%%%%%%%%%%%%%%%%%%%%%%%%%%%%%%%%%%%%%%%%%%%%%%%
   \subsection{\bf\boldmath Difference Equations and Infinite Summation}
    \label{SubSec-2LInfSum}
%%%%%%%%%%%%%%%%%%%%%%%%%%%%%%%%%%%%%%%%%%%%%%%%%%%%%%%%%%%%%%%%%%%%%%%%%%%%%%%
    Single scale quantities in renormalizable quantum field theories
    are most simply represented in terms of nested harmonic sums,
    cf. \cite{Blumlein:1998if,Vermaseren:1998uu} 
    and Appendix \ref{App-SpeFunHarm}, which holds at least up to
    3--loop order for massless Yang--Mills theories and for a wide 
    class of different processes. 
    This includes the
    anomalous dimensions and massless Wilson coefficients for unpolarized 
    and polarized space- and time-like processes to 3--loop order, 
    the Wilson coefficients for the Drell-Yan process and pseudoscalar 
    and scalar Higgs--boson production in hadron scattering in the heavy 
    quark mass limit,
    as well as the soft- and virtual corrections to Bhabha scattering in 
    the on--mass--shell--scheme
    to 2--loop order, cf.~\cite{Blumlein:2004bb,Blumlein:2005im,*Blumlein:2006rr,Dittmar:2005ed,Blumlein:2007dj,Vermaseren:2005qc,Moch:2004pa,Vogt:2004mw}. 
    The corresponding Feynman--parameter integrals are such that nested
    harmonic sums appear in a
    natural way, working in Mellin space,
    \cite{Blumlein:2009ta,Blumlein:2009fz}.
    Single scale massive quantities at
    2 loops, like the unpolarized and polarized heavy-flavor Wilson 
    coefficients in the region $Q^2 \gg m^2$ as considered in this thesis, 
    belong also to this class, \cite{Buza:1995ie,Buza:1996xr,Blumlein:2006mh,Bierenbaum:2007qe,Bierenbaum:2007pn,Blumlein:pol,Bierenbaum:2006mq,Bierenbaum:2007dm,Blumlein:trans}. 
    Finite harmonic sums obey algebraic, cf. \cite{Blumlein:2003gb}, and 
    structural relations, \cite{Blumlein:2009ta}, which can be used to 
    obtain simplified expressions and both shorten the calculations
    and yield compact final results. 
    These representations have to be mapped to momentum-fraction space 
    to use the respective quantities in experimental analyzes. 
    This is obtained by an Mellin inverse transform which requires 
    the analytic continuation of the harmonic sums
    w.r.t. the Mellin index $N~{\in}~\mathbb{C}$,
    \cite{Blumlein:2000hw,*Blumlein:2005jg,Blumlein:2009ta,Blumlein:2009fz}.
    
    Calculating the massive OMEs in Mellin space, new types of 
    infinite sums occur if compared to massless calculations. In the
    latter case, summation algorithms as 
    {\sf Summer},~\cite{Vermaseren:1998uu},
    {\sf Nestedsums},~\cite{Weinzierl:2002hv}, 
    and {\sf Xsummer},~\cite{Moch:2005uc},
    may be used to calculate the respective sums. {\sf Summer} 
    and {\sf Xsummer} are based on {\sf FORM}, 
    while {\sf Nestedsums} is based on {\sf GiNaC}, \cite{Bauer:2000cp}. 
    The new sums which 
    emerge in \cite{Bierenbaum:2007qe,Bierenbaum:2007pn,Bierenbaum:2006mq,Bierenbaum:2007dm,Blumlein:pol,Bierenbaum:2008yu} 
    can be calculated in different ways. In Ref.~\cite{Bierenbaum:2007qe,Bierenbaum:2007pn}, 
    we chose analytic methods and in the former reference all sums 
    are given which are needed to calculate the constant term of the massive 
    OMEs. Few of these sums 
    can be calculated using general theorems, as 
    Gauss' theorem,~(\ref{Gauss}), Dixon's theorem, \cite{Slater},
    or summation tables in the literature,
    cf.~\cite{Vermaseren:1998uu,Devoto:1983tc,*Blumlein:2004bs,*Blumleinunp}.
    
    In order to calculate the gluonic OMEs as well as the $O(\ep)$--terms, 
    many new sums had to be evaluated. For this we adopted a more 
    systematic technique based on difference equations, which 
    are the discrete equivalent of differential equations, cf.~\cite{Norlund,*Thomson}.
    This is a promising approach, since it allowed us to obtain 
    all sums needed automatically and it may be applied to entirely 
    different single--scale processes as well.
    It is based on applying general summation algorithms in
    computer algebra. A first method is Gosper's telescoping
    algorithm,~\cite{GOSPER}, for hypergeometric terms.
    For practical applications, Zeilberger's
    extension of Gosper's algorithm to creative
    telescoping,~\cite{Zeil:91,AequalB1}, can be considered as the
    breakthrough in symbolic summation. The recent summation
    package~\SigmaP,~\cite{Refined,Schneider:2007,sigma1,sigma2}, written in
    ${\sf MATHEMATICA}$ opens up completely new
    possibilities in symbolic summation. Based on Karr's
    $\Pi\Sigma$-difference fields,~\cite{Karr:1981}, and further
    refinements,~\cite{Refined,RefinedDF}, the package contains summation
    algorithms,~\cite{SigmaAlg}, that allow to solve not only
    hypergeometric sums, like Gosper's and Zeilberger's algorithms, but also
    sums involving indefinite nested sums.
    In this algebraic setting, one can represent completely algorithmically 
    indefinite nested sums and products without introducing any algebraic 
    relations between them. Note that this  general class of expressions 
    covers as special cases the harmonic sums or
    generalized nested harmonic 
    sums,~cf.~\cite{GON,Borwein:1999js,PET,Moch:2001zr}.
    Given such an optimal representation, by introducing as less sums as 
    possible, various summation principles are available in~\SigmaP. 
    In this work, we applied the following strategy which has been 
    generalized from the hypergeometric case,~\cite{AequalB1,AequalB2}, 
    to the $\Pi\Sigma$-field setting. 
    \begin{enumerate}
      \item Given a definite sum that involves an extra parameter $N$,
            we compute a recurrence 
            relation in $N$ that is fulfilled by the input sum. 
            The underlying difference field algorithms exploit 
            Zeilberger's creative telescoping 
            principle,~\cite{AequalB1,AequalB2}.
     \item Then we solve the derived recurrence in terms of the so-called
           d'Alembertian solutions,~\cite{AequalB1,AequalB2}. 
           Since this class covers the~harmonic~sums,
           we find all solutions in terms of~harmonic~sums.
     \item Taking the initial values of the original input sum, 
           we can combine the solutions found from step~2 in order to 
           arrive at a closed representation in terms of harmonic sums.
    \end{enumerate}
    In the following, we give some examples on how \SigmaP~works. 
    A few typical sums we had to calculate are 
    listed in Appendix \ref{App-Sums} 
    and a complete set of sums needed to calculate the 2--Loop 
    OMEs up to $O(\ep)$ can be found in Appendix B of 
    Refs.~\cite{Bierenbaum:2007qe,Bierenbaum:2008yu}. 
    Note that in this calculation also more well-known 
    sums are occurring which can, e.g., be easily solved using 
    ${\sf Summer}$.
%%%%%%%%%%%%%%%%%%%%%%%%%%%%%%%%%%%%%%%%%%%%%%%%%%%%%%%%%%%%%%%%%%%%%%%
%
% The Sigma Approach 
%
%%%%%%%%%%%%%%%%%%%%%%%%%%%%%%%%%%%%%%%%%%%%%%%%%%%%%%%%%%%%%%%%%%%%%%%
    \subsubsection{The Sigma-Approach}
     \label{SubSubSec-2LSigma} 
%%%%%%%%%%%%%%%%%%%%%%%%%%%%%%%%%%%%%%%%%%%%%%%%%%%%%%%%%%%%%%%%%%%%%%%
     As a first example we consider the sum
     \begin{eqnarray}
       T_1(N)\equiv    \sum_{i=1}^{\infty}
                   \frac{B(N,i)}{i+N+2}S_1(i)S_1(N+i)~. \label{1Beta25}
     \end{eqnarray}
     We treat the upper bound of the sum as a finite integer, i.e., we
     consider the truncated version
     $$T_1(a,N)\equiv\sum_{i=1}^{a}
                \frac{B(N,i)}{i+N+2}S_1(i)S_1(N+i) ,$$
     for $a\in\mathbb{N}$. Given this sum as input, we apply \SigmaP's
     creative telescoping algorithm and find a recurrence for
     $T_1(a,N)$ of the form
     \begin{equation}
      \label{Equ:Rec}
      c_0(N)T(a,N)+\dots c_d(N)T(a,N+d)=q(a,N)
     \end{equation}
     with order $d=4$. Here, the $c_i(N)$ and $q(a,N)$ are known 
     functions of $N$ and $a$.
     Finally, we perform the limit $a\to\infty$ and we end up at the
     recurrence
     \begin{multline*}
        -N (N+1)(N+2)^2 \Bigl\{4 N^5+68 N^4+455 N^3+1494
           N^2+2402 N+1510\Bigr\} T_1(N)\\
        -(N+1)(N+2)(N+3) \Bigl\{16 N^5+260N^4+1660 N^3+5188 N^2+7912
           N+4699\Bigr\} \\ \times T_1(N+1)
        +(N+2)(N+4)(2 N+5) \Bigl\{4 N^6+74
           N^5+542 N^4+1978 N^3+3680 N^2 \\
           +3103N+767\Bigr\}  T_1(N+2)
        +(N+4)(N+5) \Bigl\{16 N^6+276
           N^5+1928 N^4+6968 N^3 \\ 
           +13716 N^2+13929N+5707\Bigr\}T_1(N+3)
        -(N+4)(N+5)^2 (N+6) \Bigl\{4 N^5+48 N^4\\
          +223 N^3
          +497 N^2+527 N
          +211\Bigr\}T_1(N+4)
        =P_1(N)+P_2(N)S_1(N)
     \end{multline*} where
     \begin{align*}
        P_1(N)&=\Big(32 N^{18}+1232 N^{17}+21512
           N^{16}+223472 N^{15}+1514464
           N^{14}+6806114 N^{13}\\
        &+18666770N^{12}+15297623 N^{11}-116877645
           N^{10}-641458913 N^9-1826931522N^8\\
        &-3507205291 N^7-4825457477 N^6-4839106893 N^5-3535231014
           N^4\\
        &-1860247616 N^3
         -684064448 N^2-160164480 N-17395200\Big) \\
        &\big/\big(N^3 (N+1)^3
           (N+2)^3 (N+3)^2 (N+4) (N+5)\big)
     \end{align*}
     and
     \begin{align*}
        P_2(N)&=-4\Big((4 N^{14}+150 N^{13}+2610
           N^{12}+27717 N^{11}+199197
           N^{10}+1017704 N^9 \\ 
        &+3786588N^8
         +10355813 N^7+20779613 N^6+30225025
           N^5+31132328 N^4\\
        &+21872237 N^3+9912442N^2
         +2672360 N+362400\Big) \\
        &\big/\big(N^2
           (N+1)^2 (N+2)^2 (N+3) (N+4) (N+5)\big)~.
     \end{align*}
     In the next step, we apply \SigmaP's recurrence solver to the
     computed recurrence and find the four linearly independent solutions
     \begin{eqnarray*}
        h_1(N)=\frac{1}{N+2},&h_2(N)&=\frac{(-1)^N}{N(N+1)(N+2)},\\
        h_3(N)=\frac{S_1(N)}{N+2},&
        h_4(N)&=(-1)^N\frac{\big(1+(N+1)S_1(N)\big)}{N(N+1)^2(N+2)}~,
     \end{eqnarray*}
     of the homogeneous version of the recurrence and the particular
     solution
     \begin{eqnarray*}
     p(N)
         &=&
            \frac{2(-1)^N}{N(N+1)(N+2)}\Biggl[
                                        2S_{-2,1}(N)
                                       -3S_{-3}(N)
                                       -2S_{-2}(N)S_1(N)
                                       -\zeta_2S_1(N)\N\\
&&                                     -\zeta_3
                                       -\frac{2S_{-2}(N)+\zeta_2}{N+1}
                                   \Biggr]
        -2\frac{S_3(N)-\zeta_3}{N+2}
        -\frac{S_2(N)-\zeta_2}{N+2}S_1(N)\N\\
&&      +\frac{2+7N+7N^2+5N^3+N^4}
              {N^3(N+1)^3(N+2)}S_1(N)
          +2\frac{2+7N+9N^2+4N^3+N^4}
              {N^4(N+1)^3(N+2)}
%        ~,\N\\ &&    \surd~(Beta25)
    \end{eqnarray*}
     of the recurrence itself. Finally, we look for constants
     $c_1,\dots,c_4$ such that
     $$T_1(N)=c_1\,h_1(N)+c_2\,h_2(N)+c_3\,h_3(N)+c_4\,h_4(N)+p(N)~.$$
     The calculation of the necessary initial values for $N=0,1,2,3$ does
     not pose a problem for \SigmaP\ and we conclude that
     $c_1=c_2=c_3=c_4=0$. Hence the final result reads
     \begin{eqnarray}
     T_1(N)
      &=&
        \frac{2(-1)^N}{N(N+1)(N+2)}\Biggl[
                                        2S_{-2,1}(N)
                                       -3S_{-3}(N)
                                       -2S_{-2}(N)S_1(N)
                                       -\zeta_2S_1(N)\N\\
&&                                     -\zeta_3
                                       -\frac{2S_{-2}(N)+\zeta_2}{N+1}
                                   \Biggr]
        -2\frac{S_3(N)-\zeta_3}{N+2}
        -\frac{S_2(N)-\zeta_2}{N+2}S_1(N)\N\\
&&      +\frac{2+7N+7N^2+5N^3+N^4}
              {N^3(N+1)^3(N+2)}S_1(N)
          +2\frac{2+7N+9N^2+4N^3+N^4}
              {N^4(N+1)^3(N+2)}~. \N\\
           \label{2Beta25}
     \end{eqnarray}
     Using more refined algorithms of \SigmaP, see e.g.
     \cite{AhlgrenPade1,*AhlgrenPade2}, even a first order difference equation 
     can be obtained
     \begin{eqnarray}
       &&(N+2)T_1(N)-(N+3)T_1(N+1)\N\\\N\\
      &=&
        2\frac{(-1)^N}{N(N+2)}\Biggl(
                   -\frac{3N+4}
                         {(N+1)(N+2)}\Bigl(\zeta_2+2S_{-2}(N)\Bigr)
                   -2\zeta_3-2S_{-3}(N)-2\zeta_2S_1(N)
\N\\ &&
                   -4S_{1,-2}(N) 
                \Biggr)
         +\frac{N^6+8N^5+31N^4+66N^3+88N^2+64N+16}{N^3(N+1)^2(N+2)^3}S_1(N)\N\\
       &&+\frac{S_2(N)-\zeta_2}{N+1}
         +2\frac{N^5+5N^4+21N^3+38N^2+28N+8}{N^4(N+1)^2(N+2)^2}~.
       \label{Beta25eq2}
     \end{eqnarray}
     However, in deriving Eq.~(\ref{Beta25eq2}), use had to be made of
     further sums of less complexity, which had to be calculated
     separately. As above, we can easily solve the recurrence and obtain
     again the result~\eqref{2Beta25}. Here and in the following we applied
     various algebraic relations
     between harmonic sums to obtain a simplification of our results,
     cf.~\cite{Blumlein:2003gb}.
%%%%%%%%%%%%%%%%%%%%%%%%%%%%%%%%%%%%%%%%%%%%%%%%%%%%%%%%%%%%%%%%%%%%%%%
%
% The Alyternative Approaches
%
%%%%%%%%%%%%%%%%%%%%%%%%%%%%%%%%%%%%%%%%%%%%%%%%%%%%%%%%%%%%%%%%%%%%%%%
    \subsubsection{Alternative Approaches}
     \label{SubSubSec-2LAlt} 
%%%%%%%%%%%%%%%%%%%%%%%%%%%%%%%%%%%%%%%%%%%%%%%%%%%%%%%%%%%%%%%%%%%%%%%
     As a second example we consider the sum
     \begin{eqnarray}
       T_2(N)\equiv\sum_{i=1}^{\infty}\frac{S^2_1(i+N)}{i^2}~,\label{1Harm8}
     \end{eqnarray}
     which does not contain a Beta--function. In a first attempt, we
     proceed as in the first example $T_1(N)$. The
     naive application of \SigmaP\ yields a fifth order difference
     equation, which is clearly too complex for this sum. However, 
     similar to the situation $T_1(N)$, \SigmaP\ can
     reduce it to a third order relation which reads
     \begin{eqnarray}
      \label{eq2}
     &&T_2(N)(N+1)^2
      -T_2(N+1)(3N^2+10N+9)\N\\
     &&+T_2(N+2)(3N^2+14N+17)
     -T_2(N+3)(N+3)^2\N\\\N\\
     &=&\frac{
             6N^5+48N^4+143N^3+186N^2+81N-12
             }
             {(N+1)^2(N+2)^3(N+3)^2}
      - 2\frac{2N^2+7N+7
             }
            {(N+1)(N+2)^2(N+3)}S_1(N)\N\\
      &&  +\frac{-2N^6-24N^5-116N^4-288N^3-386N^2-264N-72
            }
          {(N+1)^2(N+2)^3(N+3)^2}\zeta_2\label{1diffeqT3}~.
     \end{eqnarray}
     Solving this recurrence relation in terms of harmonic sums gives a
     closed form, see~\eqref{Harm8} below.
     Still  (\ref{eq2}) represents a rather involved way to
     solve the problem. It is of advantage to map the
     numerator $S_1^2(i+N)$ into a linear representation, which can be
     achieved using Euler's relation
     \begin{eqnarray}
      S_a^2(N) = 2 S_{a,a}(N) - S_{2a}(N),~~~a > 0~.
     \end{eqnarray}
     This is realized in ${\sf Summer}$ by the {\sf basis}--command 
     for general--type harmonic sums,
     \begin{eqnarray}
      T_2(N)=\sum_{i=1}^{\infty}\frac{2S_{1,1}(i+N)-S_2(i+N)}{i^2}
             ~.\label{2Harm8}
     \end{eqnarray}
     As outlined in Ref.~\cite{Vermaseren:1998uu}, sums of this type can be
     evaluated by considering the difference
     \begin{eqnarray}
      D_2(j)&=&T_2(j)-T_2(j-1)
           =2 \sum_{i=1}^{\infty}\frac{S_1(j+i)}{i}
              -\sum_{i=1}^{\infty}
                                  \frac{1}{i^2(j+i)^2}~.\label{2diffeqT2}
     \end{eqnarray}
     The solution is then obtained by summing (\ref{2diffeqT2}) 
     to
     \begin{eqnarray}
      T_2(N)=\sum_{j=1}^ND_2(j)+T_2(0)~.
     \end{eqnarray}
     The sums in Eq.~(\ref{2diffeqT2}) are now calculable trivially
     or are of less complexity than the original sum. In the case considered
     here, only the first sum on the left hand side is not trivial.
     However, after partial fractioning, one can repeat the same 
     procedure, resulting into another
     difference equation, which is now easily solved.
     Thus using this technique, the solution of Eq.~(\ref{1Harm8}) can 
     be obtained
     by summing two first order difference equations or
     solving a second order one. The above procedure is well
     known and some of the summation--algorithms of ${\sf Summer}$
     are based on it. As a consequence, infinite sums with an
     arbitrary number of harmonic sums with the same argument
     can be performed using this package. Note that sums containing harmonic
     sums with different arguments, see e.g Eq.~(\ref{1Harm27}), 
     can in principle be summed automatically using the same approach. However,
     this feature is not yet built into ${\sf Summer}$.
     A third way to obtain the sum (\ref{1Harm8}) consists
     of using integral representations for harmonic sums, 
     \cite{Blumlein:1998if}. One finds
     \begin{eqnarray}
      T_2(N)
      &=&2\sum_{i=1}^{\infty}
         \int_0^1dx\frac{x^{i+N}}{i^2}\Bigl(\frac{\ln(1-x)}{1-x}\Bigr)_+
         -\sum_{i=1}^{\infty}
          \Biggl(\int_0^1dx\frac{x^{i+N}}{i^2}\frac{\ln(x)}{1-x}
          +\frac{\zeta_2}{i^2}\Biggr)\N\\
      &=&2\M\Bigl[\Bigl(\frac{\ln(1-x)}{1-x}\Bigr)_+\Li_2(x)\Bigr](N+1)
       \N\\ &&
         -\Biggl(\M\Bigl[\frac{\ln(x)}{1-x}\Li_2(x)\Bigr](N+1)+\zeta_2^2
          \Biggr)~.
         \label{Harm8mellin}
     \end{eqnarray}
     Here the Mellin--transform is defined in Eq.~(\ref{Mellintrans}). 
     Eq.~(\ref{Harm8mellin}) can then be easily calculated since the
     corresponding
     Mellin--transforms are well--known, \cite{Blumlein:1998if}.
     Either of these three methods above lead to
     \begin{eqnarray}
      T_2(N)=
                \frac{17}{10}\zeta_2^2
                +4S_1(N)\zeta_3
                +S^2_1(N)\zeta_2
                -S_2(N)\zeta_2
                -2S_1(N)S_{2,1}(N)
                -S_{2,2}(N)
          ~.
          \label{Harm8}
     \end{eqnarray}
     As a third example we would like to evaluate the sum
     \begin{eqnarray}
      \label{1Harm27}
        T_3(N) = \sum_{i=1}^{\infty} \frac{S_1^2(i+N) S_1(i)}{i}~.
     \end{eqnarray}
     Note that (\ref{1Harm27})
     is divergent. In order to treat this divergence,
     the symbol $\sigma_1$, cf. Eq.~(\ref{sigmaval}), is used. 
     The application of \SigmaP~to this sum
     yields a fourth order difference equation
     \begin{eqnarray}
      &&(N+1)^2(N+2)T_2(N)
         -(N+2)\left(4N^2+15N+15\right)T_2(N+1) \N\\
      &&+(2N+5)\left(3N^2+15N+20\right)T_2(N+2)
      -(N+3)\left(4N^2+25N+40\right)T_2(N+3)\N\\
      &&+(N+3)(N+4)^2T_2(N+4)\N\\
      &=&
      \frac{6N^5+73N^4+329N^3+684N^2+645N+215}
           {(N+1)^2(N+2)^2(N+3)^2}
     +\frac{6N^2+19N+9}
            {(N+1)(N+2)(N+3)}S_1(N)\label{1diffeqT2}~, \N\\
     \end{eqnarray}
     which can be solved.
     As in the foregoing example the better way to calculate the sum is to
     first change $S_1^2(i+N)$ into a linear basis representation
     \begin{eqnarray}
      T_3(N)=\sum_{i=1}^{\infty}\frac{2S_{1,1}(i+N)-S_2(i+N)}{i}S_1(i)
             ~.\label{2Harm27}
     \end{eqnarray}
     One may now calculate $T_3(N)$ using telescoping for the difference
     \begin{eqnarray}
      D_3(j)&=&T_3(j)-T_3(j-1)
            =2\sum_{i=1}^{\infty}\frac{S_{1}(i+j)S_1(i)}{i(i+j)}
       -\sum_{i=1}^{\infty}\frac{S_1(i)}{i(i+j)^2}~,\label{2diffeqT3}
     \end{eqnarray}
     with
    \begin{eqnarray}
     T_3(N)=\sum_{j=1}^ND_2(j)+T_3(0)~.
    \end{eqnarray}
     One finally obtains
    \begin{eqnarray}
       T_3(N)&=&
%%%
                \frac{\sigma_1^4}{4}
                +\frac{43}{20}\zeta_2^2
                +5S_1(N)\zeta_3
                +\frac{3S^2_1(N)-S_2(N)}{2}\zeta_2
                -2S_1(N)S_{2,1}(N) \N\\
              &&+S^2_1(N)S_2(N)
                +S_1(N)S_3(N)
                -\frac{S^2_2(N)}{4}
                +\frac{S^4_1(N)}{4}
%&&         ~\surd~(Harm27)
        ~.\label{Harm27}
    \end{eqnarray}
%%%%%%%%%%%%%%%%%%%%%%%%%%%%%%%%%%%%%%%%%%%%%%%%%%%%%%%%%%%%%%%%%%%%%%%%%%%%%%%
   \subsection{\bf\boldmath Results}
    \label{SubSec-2LRes}
%%%%%%%%%%%%%%%%%%%%%%%%%%%%%%%%%%%%%%%%%%%%%%%%%%%%%%%%%%%%%%%%%%%%%%%%%%%%%%%
    For the singlet contributions, we leave out an overall factor 
    \begin{eqnarray}
     \frac{1+(-1)^N}{2}~
    \end{eqnarray}
    in the following. This factor emerges naturally in our calculation
    and is due to the fact that in the light--cone expansion, only 
    even values of $N$ contribute to $F_2$ 
    and $F_L$, cf. Section~\ref{SubSec-DISComptLCE}.
    Additionally, we do not choose a linear 
    representation in terms of harmonic sums as was done 
    in Refs.~\cite{Vermaseren:2005qc,Moch:2004pa,Vogt:2004mw}, 
    since these are non--minimal w.r.t. to the corresponding 
    quasi--shuffle algebra, \cite{Hoffman:1997,*Hoffman:2004bf}.
    Due to this a much smaller number of harmonic sums contributes. Remainder
    terms can be expressed in polynomials $P_i(N)$.
    Single harmonic sums with negative index 
    are expressed in terms of the function $\beta(N+1)$, 
    cf. Appendix \ref{App-SpeFunHarm}. 
    For completeness, 
    we also give all pole terms and the constant terms of the 
    quarkonic OMEs. The latter have been obtained before in 
    Refs.~\cite{Buza:1995ie,Bierenbaum:2007qe}.
    The pole terms can be expressed via the 
    ${\sf LO}$--, \cite{Gross:1973juxGross:1974cs,*Georgi:1951sr}, and the 
    fermionic parts of the ${\sf NLO}$,~\cite{Floratos:1977auxFloratos:1977aue1,Floratos:1978ny,GonzalezArroyo:1979df,GonzalezArroyo:1979he,*Curci:1980uw,*Furmanski:1980cm,Hamberg:1991qt},
    anomalous dimensions and the $1$--loop $\beta$--function, 
    \cite{Khriplovich:1969aa,tHooft:unpub,Politzer:1973fx,Gross:1973id}.
    
    We first consider the matrix element $A_{Qg}^{(2)}$, which 
    is the most complex of the $2$--loop OMEs. For the calculation 
    we used the projector given in Eq.~(\ref{projG1}) and therefore 
    have to include diagrams with external ghost lines as well. 
    The $1$--loop result 
    is straightforward to calculate and has already been given in 
    Eqs. (\ref{AhhhQg1},~\ref{AQg1MSren}). As explained in 
    Section~\ref{Sec-REN}, we perform the calculation accounting for 
    1--particle reducible diagrams. 
    Hence the 1--loop massive gluon self--energy term, Eq.~(\ref{GluSelf1}),
    contributes. The unrenormalized 
    $2$--loop OME is then given in terms of 1--particle irreducible 
    and reducible contributions by 
    \begin{eqnarray}
     \Ahathat_{Qg}^{(2)}&=&
            \Ahathat_{Qg}^{(2), \mbox{irr}}-
           ~\Ahathat_{Qg}^{(1)}
            \hat{\Pi}^{(1)}~\Bigl(0,\frac{\hat{m}^2}{\mu^2}\Bigr)~.
    \end{eqnarray}
    Using the techniques described in the previous Sections,
    the pole--terms predicted by renormalization in 
    Eq.~(\ref{AhhhQg2}) are obtained, which have been given in 
    Refs.~\cite{Buza:1995ie,Bierenbaum:2007qe} before. 
    Here, the contributing $1$--loop anomalous dimensions are
%%%%%%%%%%%%%%%
%
% LO Andims, gg, qq, qg
% NLO Andims, qg
%
%%%%%%%%%%%%%%%
    \begin{eqnarray}
     \gamma_{qq}^{(0)}&=&
                 4 C_F\Biggl\{2S_1-\frac{3N^2+3N+2}{2N(N+1)}\Biggr\}~, 
\\
     \hat{\gamma}_{qg}^{(0)} &=&-8T_F\frac{N^2+N+2}{N(N+1)(N+2)}~, 
 \\
     \gamma_{gg}^{(0)} &=&8C_A\Biggl\{S_1-\frac{2(N^2+N+1)}
                       {(N-1)N(N+1)(N+2)}\Biggr\}-2 \beta_0~,
    \end{eqnarray}
    and the $2$--loop contribution reads
    \begin{eqnarray}
     \hat{\gamma}_{qg}^{(1)}&=&
                 8C_FT_F
                    \Biggl\{
                           2\frac{N^2+N+2}
                                 {N(N+1)(N+2)}
                                  \left[
                                         S_2
                                        -S_1^2
                                  \right]
                          +\frac{4}{N^2}S_1
                          -\frac{P_1}
                                {N^3(N+1)^3(N+2)}
                    \Biggr\}
 \N\\ &&
                +16C_AT_F
                    \Biggl\{
                            \frac{N^2+N+2}
                                 {N(N+1)(N+2)}
                                  \left[
                                         S_2
                                        +S_1^2
                                        -2\beta'
                                        -\zeta_2
                                 \right]
 \N\\ &&
                           -\frac{4(2N+3)S_1}
                                 {(N+1)^2(N+2)^2}
                           -\frac{P_2}
                                 {(N-1)N^3(N+1)^3(N+2)^3}
                      \Biggr\}~, \\ \N\\
     P_1&=&5N^6+15N^5+36N^4+51N^3+25N^2+8N+4~, \N\\
     P_2&=&N^9+6N^8+15N^7+25N^6+36N^5+85N^4+128N^3 \N\\ &&
            +104N^2+64N+16~.
    \end{eqnarray}
    These terms agree with the literature and provide a strong 
    check on the calculation. 
    The constant term in $\ep$ in Eq.~(\ref{AhhhQg2}) is determined after mass 
    renormalization, \cite{Buza:1995ie,Bierenbaum:2007qe,SKdiploma}.
%%%%%%%%%%%%%%%%%%%%%%%%%%%%%%%%%%%%%%%%%%%%%%%%%%%%%%
    \begin{eqnarray}
      a_{Qg}^{(2)}&=&
                T_FC_F\Biggl\{
                        \frac{4(N^2+N+2)}
                             {3N(N+1)(N+2)}
                              \Bigl(
                                     4S_3
                                    -3S_2S_1
                                    -S_1^3
                                    -6S_1\zeta_2
                              \Bigr)
                       +4\frac{3N+2}
                              {N^2(N+2)}S_1^2
\N\\ && \hspace{-10mm}
                       +4\frac{N^4+17N^3+17N^2-5N-2}
                              {N^2(N+1)^2(N+2)}S_2
                       +2\frac{(3N^2+3N+2)(N^2+N+2)}
                              {N^2(N+1)^2(N+2)}\zeta_2
\N\\ &&\hspace{-10mm}
                       +4\frac{N^4-N^3-20N^2-10N-4}
                              {N^2(N+1)^2(N+2)}S_1
                       +\frac{2P_3}
                              {N^4(N+1)^4(N+2)}
\Biggr\}
\N\\ &&\hspace{-10mm}
                +T_FC_A\Biggl\{
                        \frac{2(N^2+N+2)}
                             {3N(N+1)(N+2)}
                              \Bigl(
                                    -24S_{-2,1}
                                    +6\beta''
                                    +16S_3
                                    -24\beta'S_1
                                    +18S_2S_1
                                    +2S_1^3
\N\\ &&\hspace{-10mm}
                                    -9\zeta_3
              \Bigr)
                       -16\frac{N^2-N-4}
                               {(N+1)^2(N+2)^2}\beta'
                       -4\frac{7N^5+21N^4+13N^3+21N^2+18N+16}
                              {(N-1)N^2(N+1)^2(N+2)^2}S_2
\N\\ &&\hspace{-10mm}
                       -4\frac{N^3+8N^2+11N+2}
                              {N(N+1)^2(N+2)^2}S_1^2
                       -8\frac{N^4-2N^3+5N^2+2N+2}
                              {(N-1)N^2(N+1)^2(N+2)}\zeta_2
\N\\ &&\hspace{-10mm}
                       -\frac{4P_4}
                              {N(N+1)^3(N+2)^3}S_1
                       +\frac{4P_5}
                              {(N-1)N^4(N+1)^4(N+2)^4}
                \Biggr\}~,
         \label{aQg2}
    \end{eqnarray}
    where the polynomials in Eq.~(\ref{aQg2}) are given by
    \begin{eqnarray}
     P_3&=&12N^8+52N^7+132N^6+216N^5+191N^4+54N^3-25N^2\N\\ &&
           -20N-4~, \\
     P_4&=&N^6+8N^5+23N^4+54N^3+94N^2+72N+8~, \\
     P_5&=&2N^{12}+20N^{11}+86N^{10}+192N^9+199N^8-N^7-297N^6-495N^5
           \N\\ && -514N^4-488N^3-416N^2-176N-32~.
    \end{eqnarray}
%%%%%%%%%%%%%%%%%%%%%%%%%%%%%%%%%%%%%%%%%%%%%%%%%%%%%%
    The newly calculated $O(\ep)$ contribution to $A_{Qg}^{(2)}$, 
    \cite{Bierenbaum:2008yu}, reads after mass renormalization
%%%%%%%%%%%%%%%%%%%%%%%%%%%%%%%%%%%%%%%%%%%%%%%%%%%%%%%%%%%%%%%%%%%%%%%%
%
% \overline{a}_{Qg}^{(2)}
%
%%%%%%%%%%%%%%%%%%%%%%%%%%%%%%%%%%%%%%%%%%%%%%%%%%%%%%%%%%%%%%%%%%%%%%%%
    \begin{eqnarray}
      \overline{a}_{Qg}^{(2)}&=&
         T_FC_F\Biggl\{
                      \frac{N^2+N+2}
                           {N(N+1)(N+2)}
                   \Bigl(
                     16S_{2,1,1}
                    -8S_{3,1}
                    -8S_{2,1}S_1
                    +3S_4
                    -\frac{4}{3}S_3S_1
                    -\frac{1}{2}S^2_2
\N\\&&\hspace{-10mm}
                    -S_2S^2_1
                   -\frac{1}{6}S^4_1
                    +2\zeta_2S_2
                    -2\zeta_2S^2_1
                    -\frac{8}{3}\zeta_3S_1
                   \Bigr)
                -8\frac{N^2-3N-2}
                       {N^2(N+1)(N+2)}S_{2,1}
\N\\&&\hspace{-10mm}
                +\frac{2}{3}\frac{3N+2}
                         {N^2(N+2)}S^3_1
                +\frac{2}{3}\frac{3N^4+48N^3+43N^2-22N-8}
                         {N^2(N+1)^2(N+2)}S_3
                +2\frac{3N+2}
                       {N^2(N+2)}S_2S_1
\N\\&&\hspace{-10mm}
                +4\frac{S_1}
                       {N^2}\zeta_2
                +\frac{2}{3}\frac{(N^2+N+2)(3N^2+3N+2)}
                                 {N^2(N+1)^2(N+2)}\zeta_3
                +\frac{P_{6}}
                                 {N^3(N+1)^3(N+2)}S_2
\N\\&&\hspace{-10mm}
                +\frac{N^4-5N^3-32N^2-18N-4}
                                 {N^2(N+1)^2(N+2)}S^2_1
                -2\frac{2N^5-2N^4-11N^3-19N^2-44N-12}
                                 {N^2(N+1)^3(N+2)}S_1
\N\\&&\hspace{-10mm}
                -\frac{5N^6+15N^5+36N^4+51N^3+25N^2+8N+4}
                                 {N^3(N+1)^3(N+2)}\zeta_2
                -\frac{P_{7}}
                      {N^5(N+1)^5(N+2)}
                          \Biggr\}
\N\\&&\hspace{-10mm}
        +T_F{C_A}\Biggl\{ 
                   \frac{N^2+N+2}
                         {N(N+1)(N+2)}
               \Bigl(
                        16S_{-2,1,1}
                       -4S_{2,1,1}
                       -8S_{-3,1}
                       -8S_{-2,2}
                       -4S_{3,1}
                       -\frac{2}{3}\beta'''
\N\\&&\hspace{-10mm} 
                       +9S_4
                       -16S_{-2,1}S_1
                       +\frac{40}{3}S_1S_3
                       +4\beta''S_1
                       -8\beta'S_2
                       +\frac{1}{2}S^2_2
                       -8\beta'S^2_1
                       +5S^2_1S_2
                       +\frac{1}{6}S^4_1
\N\\&&\hspace{-10mm}
                       -\frac{10}{3}S_1\zeta_3
                       -2S_2\zeta_2
                       -2S^2_1\zeta_2
                       -4\beta'\zeta_2
                       -\frac{17}{5}\zeta_2^2
              \Bigr)
                   -8\frac{N^2+N-1}
                          {(N+1)^2(N+2)^2}\zeta_2S_1
\N\\ \N \\&&\hspace{-10mm}
                  +\frac{4(N^2-N-4)}
                         {(N+1)^2(N+2)^2}
                      \Bigl(
                       -4S_{-2,1}
                       +\beta''
                       -4\beta'S_1
                     \Bigr)
                   -\frac{2}{3}\frac{N^3+8N^2+11N+2}
                            {N(N+1)^2(N+2)^2}S^3_1
\N\\ \N\\&&\hspace{-10mm}
                   -\frac{16}{3}\frac{N^5+10N^4+9N^3+3N^2+7N+6}
                             {(N-1)N^2(N+1)^2(N+2)^2}S_3
                   +8\frac{N^4+2N^3+7N^2+22N+20}
                          {(N+1)^3(N+2)^3}\beta'
\N\\ \N\\&&\hspace{-10mm}
                   +2\frac{3N^3-12N^2-27N-2}
                          {N(N+1)^2(N+2)^2}S_2S_1
                   -\frac{2}{3}\frac{9N^5-10N^4-11N^3+68N^2+24N+16}
                            {(N-1)N^2(N+1)^2(N+2)^2}\zeta_3
\N\\ \N\\ &&\hspace{-10mm}
                   -\frac{P_{8}S_2}
                         {(N-1)N^3(N+1)^3(N+2)^3}
                   -\frac{P_{10}S^2_1}
                         {N(N+1)^3(N+2)^3}
                   +\frac{2P_{11}S_1}
                          {N(N+1)^4(N+2)^4}
\N\\ \N\\ &&\hspace{-10mm}
                   -\frac{2P_{9}\zeta_2}
                          {(N-1)N^3(N+1)^3(N+2)^2}
                   -\frac{2P_{12}}
                         {(N-1)N^5(N+1)^5(N+2)^5}
                \Biggr\}~,\label{aQg2bar}
    \end{eqnarray}
%%%%%%%%%%%%%%%%%%%%%%%%%%%%%%%%%%%%%%%%%%%%%%%%%%%%%%%%%%%%%%%%%%%%%%%%
    with the polynomials
    \begin{eqnarray}
     P_{6}&=&3N^6+30N^5+15N^4-64N^3-56N^2-20N-8~,
    \end{eqnarray}
    \begin{eqnarray}
     P_{7}&=&24N^{10}+136N^9+395N^8+704N^7+739N^6 +407N^5+87N^4 \N\\  &&
             +27N^3+45N^2+24N+4~, \\
     P_{8}&=&N^9+21N^8+85N^7+105N^6+42N^5+290N^4+600N^3+456N^2\N\\ &&
             +256N+64~, \\
     P_{9}&=&(N^3+3N^2+12N+4)(N^5-N^4+5N^2+N+2)~,\\
     P_{10}&=&N^6+6N^5+7N^4+4N^3+18N^2+16N-8~,\\
     P_{11}&=&2N^8+22N^7+117N^6+386N^5+759N^4+810N^3+396N^2\N\\ &&
             +72N+32~,\\
     P_{12}&=&4N^{15}+50N^{14}+267N^{13}+765N^{12}+1183N^{11}+682N^{10}
               -826N^9 \N\\&&
               -1858N^8-1116N^7+457N^6+1500N^5+2268N^4+2400N^3 \N\\ &&
               +1392N^2+448N+64~. 
    \end{eqnarray}
%%%%
    Note that the terms $\propto~\zeta_3$ in Eq.~(\ref{aQg2})
    and $\propto~\zeta_2^2$ in Eq.~(\ref{aQg2bar}) are only 
    due to the representation using the $\beta^{(k)}$--functions and 
    are absent in representations using harmonic sums.
    The results for the individual diagrams contributing to $A_{Qg}^{(2)}$
    can be found up to $O(\ep^0)$ in Ref.~\cite{Bierenbaum:2007qe} and 
    at $O(\ep)$ in Ref.~\cite{Bierenbaum:2008yu}.
    
    Since harmonic sums appear in a wide variety of 
    applications, it is interesting to study the pattern in which they emerge. 
    In Table~\ref{table:CompRes}, we list the harmonic sums contributing 
    to each individual diagram~\footnote{Cf. Ref.~\cite{Buza:1995ie} for 
    the labeling of the diagrams.}. 
%%%%%%%%%%%%%%%%%%%%%%%%%%%%%%%%%%%%%%%%%%%%%%%%%%%%%%%%%%%%%%%%%%%%%%%%%
    {\tiny
    \begin{table}[htb]
      \caption{\sf Complexity of the results
      for the individual diagrams contributing to $A_{Qg}^{(2)}$ }
      \label{table:CompRes}
      \begin{center}
       \renewcommand{\arraystretch}{1.1}
%%%%%%%%%%%%%%%%%%%%%%
       \begin{tabular}{||l|c|c|c|c|c|c|c|c|c|c|c|c|c|r||}
        \hline \hline
  Diagram & $S_1$     & $S_2$        & $S_3$      & $S_4$       & $S_{-2}$ &
            $S_{-3}$  & $S_{-4}$     & $S_{2,1}$  & $S_{-2,1}$  & $S_{-2,2}$ &
            $S_{3,1}$ & $S_{-3,1}$   & $S_{2,1,1}$& $S_{-2,1,1}$ \\
        \hline \hline
          A             & &+&+& & & & & & & & & & &  \\
          B             &+&+&+&+& & & &+& & &+& &+&  \\
          C             & &+&+& & & & & & & & & & &  \\
          D             &+&+&+& & & & &+& & & & & &  \\
          E             &+&+&+& & & & &+& & & & & &  \\
          F             &+&+&+&+& & & &+& & & & &+&  \\
          G             &+&+&+& & & & &+& & & & & &  \\
          H             &+&+&+& & & & &+& & & & & &  \\
          I             &+&+&+&+&+&+&+&+&+&+&+&+&+& +\\
          J             & &+&+& & & & & & & & & & &  \\
          K             & &+&+& & & & & & & & & & &  \\
          L             &+&+&+&+& & & &+& & &+& &+&  \\
          M             & &+&+& & & & & & & & & & &  \\
          N             &+&+&+&+&+&+&+&+&+&+&+&+&+&+ \\
          O             &+&+&+&+& & & &+& & &+& &+&  \\
          P             &+&+&+&+& & & &+& & &+& &+&  \\
          S             & &+&+& & & & & & & & & & &  \\
          T             & &+&+& & & & & & & & & & &  \\
          \hline\hline
       \end{tabular}
%%%%%%%%%%%%%%%%%%%%%%
       \renewcommand{\arraystretch}{1.0}
      \end{center}
    \end{table} }
%%%%%%%%%%%%%%%%%%%%%%%%%%%%%%%%%%%%%%%%%%%%%%%%%%%%%%%%%%%%%%%%%%%%%%%%%
    The $\beta$--function and their derivatives can be 
    traced back to the single non--alternating harmonic sums, allowing for 
    half-integer arguments, cf. \cite{Blumlein:1998if} and Appendix 
    \ref{App-SpeFunHarm}.
    Therefore, all single harmonic sums form an equivalence class
    being represented by the sum $S_1$, from which the other single harmonic
    sums are
    easily derived through differentiation and half-integer relations
    Additionally, we have already made use of the algebraic relations, 
    \cite{Blumlein:2003gb}, between harmonic sums in deriving 
    Eqs. (\ref{aQg2},~\ref{aQg2bar}). Moreover,
    the sums $S_{-2,2}$ and $S_{3,1}$ obey
    structural relations to other harmonic sums, 
    i.e., they lie in corresponding
    equivalence classes and may be obtained by either rational argument
    relations 
    and/or differentiation w.r.t. $N$. Reference to these equivalence classes
    is 
    useful since the representation of these sums for $N~\epsilon~\mathbb{C}$
    needs 
    not to be derived newly, except of straightforward differentiations. 
    All functions involved are meromorphic, with poles at the non--negative 
    integers.
    Thus the $O(\ep^0)$--term depends on two basic functions only, 
    $S_1$ and $S_{-2,1}$~\footnote{The associated Mellin transform 
    to this sum has been discussed in 
    Ref.~\cite{GonzalezArroyo:1979df} first.}. 
    This has to be compared to the $z$--space 
    representation used in Ref.~\cite{Buza:1995ie}, in which 
    48 different functions were needed. 
    As shown in \cite{Blumlein:1998if}, various of these functions have
    Mellin transforms containing triple 
    sums, which do not occur in our approach even on the level of
    individual diagrams.
    Thus the method applied here allowed to compactify the representation of 
    the heavy flavor matrix elements and Wilson coefficients significantly.
    
    The $O(\ep)$--term consists of 6 basic functions only, which are 
    given by 
    \begin{eqnarray} 
      && \{S_1,~S_2,~S_3,~S_4,~S_{-2},~S_{-3},~S_{-4}\},~
         S_{2,1},~ S_{-2,1},~ S_{-3,1},~ S_{2,1,1},~ 
         S_{-2,1,1}~, \label{6basic} \\
      && S_{-2,2} \quad:\quad \mbox{depends on}~\quad S_{-2,1},~S_{-3,1} \N\\
      && S_{3,1}\hspace{2.4mm}\quad:\quad \mbox{depends on}~\quad S_{2,1}~. \N
    \end{eqnarray}
    The absence of harmonic sums containing $\{-1\}$ as index was noted
    before for all other classes of space-- and time--like anomalous dimensions
    and Wilson coefficients, including those for other hard processes
    having been 
    calculated so far, cf.~\cite{Blumlein:2004bb,Blumlein:2005im,*Blumlein:2006rr,Dittmar:2005ed}. This can not be seen if one applies 
    the $z$--space representation or the linear representation 
    in Mellin--space, \cite{Moch:1999eb}.
    
    Analytic continuation, e.g., for $S_{-2,1}$ proceeds
    via the equality, 
    \begin{eqnarray}
     \Mvec\left[\frac{\Li_2(x)}{1+x}\right](N+1)  - \zeta_2 \beta(N+1)
     = (-1)^{N+1} \left[S_{-2,1}(N)  + \frac{5}{8} \zeta_3\right]
     ~\label{SM21ANCONT}
    \end{eqnarray}
    with similar representations for the remaining sums, 
    \cite{Blumlein:1998if}
    ~\footnote{Note that the argument of
    the Mellin-transform in Eq.~(36), Ref.~\cite{Blumlein:2006mh}, should
    read $(N+1)$.}.
    
    As discussed in \cite{Blumlein:2006mh}, 
    the result for $a_{Qg}^{(2)}$ agrees with that in $z$--space 
    given in Ref.~\cite{Buza:1995ie}. However, there is a difference
    concerning 
    the complete renormalized expression for $A_{Qg}^{(2)}$. This 
    is due to the scheme--dependence for the renormalization of 
    the coupling constant, which has been described in 
    Sections~\ref{SubSec-RENCo},~\ref{SubSec-HQElProdWave} 
    and emerges for the first 
    time at $O(a_s^2)$. 
    Comparing Eq.~(\ref{AQg2MSren})
    for the renormalized result in the ${\sf \MS}$--scheme for the coupling
    constant
    with the transformation formula to the ${\sf MOM}$--scheme, Eq. 
    (\ref{aMSON2aMOMON}), this difference is given by 
    \begin{eqnarray}
     \label{AA1}
     A_{Qg}^{(2), \MS} &=&
        A_{Qg}^{(2),\MOM} 
          - \beta_{0,Q} \frac{\hat{\gamma}_{qg}^{(0)}}{2} 
                        \ln^2\left(\frac{m^2}{\mu^2}\right)~.
    \end{eqnarray}
    As an example, the second moment of the massive OME up to $2$--loops 
    reads in the ${\sf \MS}$--scheme for coupling constant renormalization
    \begin{eqnarray}
      A_{Qg}^{\MS}&=&
                a_s^{\MS}\Biggl\{
                     -\frac{4}{3} T_F\ln \Bigl(\frac{m^2}{\mu^2}\Bigr)
                        \Biggr\} 
              +{a_s^{\MS}}^2\Biggl\{
                      T_F\Bigl[
                         \frac{22}{9}C_A
                        -\frac{16}{9}C_F
                        -\frac{16}{9}T_F
                      \Bigr]
                        \ln^2\Bigl(\frac{m^2}{\mu^2}\Bigr)
\N\\ &+&
                      T_F\Bigl[
                        -\frac{70}{27}C_A
                        -\frac{148}{27}C_F
                      \Bigr]
                        \ln \Bigl(\frac{m^2}{\mu^2}\Bigr)
                     -\frac{7}{9}C_AT_F
                     +\frac{1352}{81}C_FT_F
                        \Biggr\}~, \label{AQg2N2MSON}
    \end{eqnarray}
    and in the ${\sf MOM}$--scheme
    \begin{eqnarray}
     A_{Qg}^{\MOM}&=&
                a_s^{\MOM}\Biggl\{
                     -\frac{4}{3} T_F\ln \Bigl(\frac{m^2}{\mu^2}\Bigr)
                        \Biggr\} 
              +{a_s^{\MOM}}^2\Biggl\{
                      T_F\Bigl[
                         \frac{22}{9}C_A
                        -\frac{16}{9}C_F
                      \Bigr]
                        \ln^2\Bigl(\frac{m^2}{\mu^2}\Bigr)
\N\\
     &+&
                      T_F\Bigl[
                        -\frac{70}{27}C_A
                        -\frac{148}{27}C_F
                      \Bigr]
                        \ln \Bigl(\frac{m^2}{\mu^2}\Bigr)
                     -\frac{7}{9}C_AT_F
                     +\frac{1352}{81}C_FT_F
                        \Biggr\} ~. \label{AQg2N2MOMON}
    \end{eqnarray}
    As one infers from the above formulas, this difference affects 
    at the $2$--loop level only the double logarithmic term 
    and stems from the treatment of the 1--particle--reducible 
    contributions. In Ref.~\cite{Buza:1995ie}, these contributions 
    were absorbed into the coupling constant, applying the ${\sf MOM}$--scheme.
    This was motivated by the need to eliminate the virtual contributions 
    due to heavier quarks (b, t) and was ${\sf also}$ extended to the 
    charm--quark, 
    thus adopting the same renormalization scheme as has been used
    in Refs.~\cite{Laenen:1992zkxLaenen:1992xs,*Riemersma:1994hv} for the exact calculation of the heavy 
    flavor contributions to the Wilson coefficients. 
    Contrary, in Ref.~\cite{Buza:1996wv}, the ${\sf \MS}$--description 
    was applied and the strong coupling constant depends on $n_f+1$
    flavors, cf. the discussion in Section~\ref{SubSec-HQElProdWave}.
    
%%%%%%%%%%%%%%%%%%%%%%%%%%
    The remaining massive OMEs are less complex than the 
    term $A_{Qg}^{(2)}$ and depend only on single harmonic sums, 
    i.e. on only one basic function, $S_1$. 
%%%%%%%%%%%%%%%%%%%%%%%%%%
%
% Results for AQq2PS
%
%%%%%%%%%%%%%%%%%%%%%%%%%%
    In the ${\sf PS}$--case, the ${\sf LO}$ and ${\sf NLO}$
    anomalous dimensions 
    \begin{eqnarray}
     \gamma_{gq}^{(0)} &=&-4C_F\frac{N^2+N+2}{(N-1) N (N+1)}~,
                         \label{ggq0} \\
     \hat{\gamma}_{qq}^{(1), {\sf PS}}&=&
                 -16C_FT_F\frac{5N^5+32N^4+49N^3+38N^2+28N+8}
                               {(N-1)N^3(N+1)^3(N+2)^2}~ \label{gqqhat1PS}
    \end{eqnarray}
    contribute. The pole--terms
    are given by Eq.~(\ref{AhhhQq2PS}) and we obtain for the 
    higher order terms in $\ep$
%%%%%%%%%%%%%%%%%%%%%%%%%%%%%%%%%%%%%%%%%%%%%%%%%%%%%%%%%%%%%%%%%%%%%%%%
%
%  a_{Qq}^{(2), {\sf PS}}, \overline{a}_{Qq}^{(2), {\sf PS}}
% 
%%%%%%%%%%%%%%%%%%%%%%%%%%%%%%%%%%%%%%%%%%%%%%%%%%%%%%%%%%%%%%%%%%%%%%%%
    \begin{eqnarray}
      a_{Qq}^{(2), {\sf PS}}&=&
              C_FT_F\Biggl\{
                     -\frac{4(N^2+N+2)^2\left(2S_2+\zeta_2\right)}
                            {(N-1)N^2(N+1)^2(N+2)}
                     +\frac{4P_{13}}
                           {(N-1)N^4(N+1)^4(N+2)^3}
                     \Biggr\},  \N\\ 
               \label{aQq2PS}\\
\N\\ 
      P_{13}&=&N^{10}+8N^9+29N^8+49N^7-11N^6-131N^5-161N^4 \N\\ &&
               -160N^3-168N^2-80N  -16~, \\
%%%%
\N\\ 
      \overline{a}_{Qq}^{(2), {\sf PS}}&=&
              C_FT_F\Biggl\{
                     -2\frac{(5N^3+7N^2+4N+4)(N^2+5N+2)}
                            {(N-1)N^3(N+1)^3(N+2)^2}\left(2S_2+\zeta_2\right)
\N\\
&&
                     -\frac{4(N^2+N+2)^2\left(3S_3+\zeta_3\right)}
                            {3(N-1)N^2(N+1)^2(N+2)}
                     +\frac{2P_{14}}
                           {(N-1)N^5(N+1)^5(N+2)^4}
                     \Biggr\}, 
           \label{aQq2PSbar} \\
\N\\
   P_{14}&=&5N^{11}+62N^{10}+252N^9+374N^8-400N^6+38N^7-473N^5\N\\&&
           -682N^4-904N^3-592N^2-208N-32~.
    \end{eqnarray}
    Since the ${\sf PS}$--OME emerges for the first time 
    at $O(a_s^2)$, there is no difference between its representation in 
    the ${\sf MOM}$-- and the 
    ${\sf \MS}$--scheme. The renormalized OME $A_{Qq}^{(2)\sf PS}$ 
    is given in Eq.~(\ref{AQq2PSMSON}) and the second moment reads
    \begin{eqnarray}
      A_{Qq}^{{\sf PS}, \MS}&=&
               {a_s^{\MS}}^2\Biggl\{
                   -\frac{16}{9}\ln^2 \Bigl(\frac{m^2}{\mu^2}\Bigr)
                   -\frac{80}{27}\ln \Bigl(\frac{m^2}{\mu^2}\Bigr)
                   -4
                        \Biggr\}C_FT_F
              +O({a_s^{\MS}}^3)~. \label{AQq2PSN2MSON}
    \end{eqnarray}
%%%%%%%%%%%%%%%%%%%%%%%%%%
%
% Results for Aqq2NSQ
%
%%%%%%%%%%%%%%%%%%%%%%%%%%
   The flavor non-singlet ${\sf NLO}$ anomalous dimension is given by
   \begin{eqnarray}
     \hat{\gamma}_{qq}^{(1), {\sf NS}}&=&
                        \frac{4C_FT_F}{3} 
                                \Biggl\{
                                   8S_2
                                 -\frac{40}{3}S_1
                                 +\frac{3N^4+6N^3+47N^2+20N-12}
                                       {3N^2(N+1)^2}
                                \Biggr\}~. \label{gqqhat1NS}
   \end{eqnarray}
   The unrenormalized OME is obtained from the 1--particle irreducible
   graphs and the contributions of heavy quark loops 
   to the quark self--energy. The latter is given at $O(\hat{a}_s^2)$
   in Eq.~(\ref{QuSelf2}). One obtains
   \begin{eqnarray}
    \Ahathat_{qq,Q}^{(2), {\sf NS}}&=&
        \Ahathat_{qq,Q}^{(2), {\sf NS}, \mbox{irred}}-
        \hat{\Sigma}^{(2)}(0,\frac{\hat{m}^2}{\mu^2})~.
   \end{eqnarray}
   Our result is of the structure given in Eq.~(\ref{Ahhhqq2NSQ})
   and the higher order terms in $\ep$ read
%%%%%%%%%%%%%%%%%%%%%%%%%%%%%%%%%%%%%%%%%%%%%%%%%%%%%%%%%%%%%%%%%%%%%%%%
%
%  a_{qq,Q}^{(2), {\sf NS}}, \overline{a}_{qq,Q}^{(2), {\sf NS}}
% 
%%%%%%%%%%%%%%%%%%%%%%%%%%%%%%%%%%%%%%%%%%%%%%%%%%%%%%%%%%%%%%%%%%%%%%%%
   \begin{eqnarray}
     a_{qq,Q}^{(2), {\sf NS}}&=&
             \frac{C_FT_F}{3}\Biggl\{
                     -8S_3
                     -8\zeta_2S_1
                     +\frac{40}{3}S_2
                     +2\frac{3N^2+3N+2}
                            {N(N+1)}\zeta_2
                     -\frac{224}{9}S_1
 \N\\ && 
                     +\frac{219N^6+657N^5+1193N^4+763N^3-40N^2-48N+72}
                           {18N^3(N+1)^3}\Biggr\} 
                     \label{aqq2NSQ},
 \\ \N\\
%%%
     \overline{a}_{qq,Q}^{(2), {\sf NS}}&=&
             \frac{C_FT_F}{3}\Biggl\{
                      4S_4
                     +4S_2\zeta_2
                     -\frac{8}{3}S_1\zeta_3
                     +\frac{112}{9}S_2
                     +\frac{3N^4+6N^3+47N^2+20N-12}
                           {6N^2(N+1)^2}\zeta_2
\N\\&& 
                     -\frac{20}{3}S_1\zeta_2
                     -\frac{20}{3}S_3
                     -\frac{656}{27}S_1
                     +2\frac{3N^2+3N+2}
                            {3N(N+1)}\zeta_3
                     +\frac{P_{15}}{216N^4(N+1)^4}
                   \Biggr\} ~,  
           \label{aqq2NSQbar}
 \\ \N\\
             P_{15}&=&1551N^8+6204N^7+15338N^6+17868N^5+8319N^4 \N\\ &&
                     +944N^3+528N^2-144N-432~. 
   \end{eqnarray}
    The anomalous dimensions in Eqs.~(\ref{ggq0},~\ref{gqqhat1PS},~\ref{gqqhat1NS}) agree with the literature. 
    Eqs.~(\ref{aQq2PS},~\ref{aqq2NSQ}), cf. Ref.~\cite{Bierenbaum:2007qe}, 
    were first given in Ref.~\cite{Buza:1995ie} and agree with the results 
    presented there. Eqs.~(\ref{aQq2PSbar},~\ref{aqq2NSQbar}),
    \cite{Bierenbaum:2008yu}, are new results of this thesis.
    As in the ${\sf PS}$ case, the ${\sf NS}$ OME emerges for the first time 
    at $O(a_s^2)$. The corresponding renormalized OME 
    $A_{qq,Q}^{(2),{\sf NS}}$ 
    is given in Eq.~(\ref{Aqq2NSQMSren}) and the second moment reads
    \begin{eqnarray}
      A_{qq,Q}^{{\sf NS}, \MS}&=&
               {a_s^{\MS}}^2\Biggl\{
                   -\frac{16}{9}\ln^2 \Bigl(\frac{m^2}{\mu^2}\Bigr)
                   -\frac{128}{27}\ln \Bigl(\frac{m^2}{\mu^2}\Bigr)
                   -\frac{128}{27}
                        \Biggr\}C_FT_F
              +O({a_s^{\MS}}^3)~.  \label{Aqq2NSQN2MSON}
    \end{eqnarray}
%%%%%%%%%%%%%%%%%%%%%%%%%%
    Note that the first moment of the ${\sf NS}$--OME vanishes, even 
    on the unrenormalized level up to $O(\ep)$. This provides 
    a check on the results in Eqs. (\ref{aqq2NSQ}, \ref{aqq2NSQbar}), because
    this is required by fermion number conservation.
    
    At this point an additional comment on the difference between the 
    ${\sf MOM}$ and the ${\sf \MS}$--scheme is in order. The ${\sf MOM}$--scheme was applied 
    in Ref.~\cite{Buza:1995ie} for two different purposes. The first one
    is described below Eq.~(\ref{AQg2N2MOMON}). It was introduced to absorb
    the contributions of one--particle reducible diagrams and heavier 
    quarks into the definition of the coupling constant. However, 
    in case of $A_{Qg}^{(2)}$, renormalization in the ${\sf MOM}$--scheme 
    and the scheme transformation from the ${\sf MOM}$--scheme 
    to the ${\sf \MS}$--scheme accidentally commute. 
    This means, that one could apply 
    Eq.~(\ref{RenAQg2MOM}) in the ${\sf \MS}$--scheme, 
    i.e., set 
    \begin{eqnarray}
     \delta a_{s,1}^{\MOM}=\delta a_{s,1}^{\MS}(n_f+1)
    \end{eqnarray}
    from the start and obtain Eq.~(\ref{AQg2MSren}) 
    for the renormalized result. This is not the case for 
    $A_{qq,Q}^{(2), {\sf NS}}$. As mentioned earlier,
    the scheme transformation 
    does not have an effect on this term at $2$--loop order. This means that 
    Eq.~(\ref{2LNSRen1}) should yield the same renormalized result 
    in the ${\sf MOM}$-- and in the ${\sf \MS}$--scheme. However, in the
    latter 
    case, the difference of $Z$--factors does not contain the mass. Thus 
    a term 
    \begin{eqnarray}
     \propto \frac{1}{\ep}\ln \Bigl(\frac{m^2}{\mu^2}\Bigr)~, 
    \end{eqnarray}
    which stems from the expansion of the unrenormalized result in 
    Eq.~(\ref{Ahhhqq2NSQ}), can not be subtracted. The reason for this 
    is the following. As pointed out 
    in Ref.~\cite{Buza:1995ie}, the term $\hat{A}_{qq,Q}^{(2), {\sf NS}}$
    is only UV--divergent.
    However, this is only the case if one 
    imposes the condition that the heavy quark contributions to the gluon 
    self--energy vanishes for on--shell momentum of the gluon. This 
    is exactly the condition we imposed for renormalization in 
    the ${\sf MOM}$--scheme, cf. Section~\ref{SubSec-RENCo}. 
    Hence in this case,
    the additional divergences absorbed into the coupling are of the 
    collinear type, contrary to the term in $A_{Qg}^{(2)}$. By applying 
    the transformation back to the ${\sf \MS}$--scheme, we treat these
    two different terms in a concise way. This is especially important 
    at the three--loop level, since in this case both effects are 
    observed for all OMEs and the renormalization would not be possible 
    if not applying the ${\sf MOM}$--scheme first. \\
    
    Let us now turn to the gluonic OMEs $A_{gg,Q}^{(2)},~A_{gq,Q}^{(2)}$, 
    which are not needed for the asymptotic 2--loop heavy flavor Wilson 
    coefficients. They contribute, however, 
    in the VFNS--description of heavy flavor parton densities,
    cf. Ref.~\cite{Buza:1996wv} and Section~\ref{SubSec-HQFlav}.
%%%%%%%%%%%%%%%%%%%%%%%%%%
%
% Results for Agg2Q
%
%%%%%%%%%%%%%%%%%%%%%%%%%%
    The $1$--loop term $A_{gg,Q}^{(1)}$ has already been given in 
    Eqs. (\ref{AggQ1unren2}, \ref{AggQ1MSren}). 
    In case of $A_{gg,Q}^{(2)}$, the part
    \begin{eqnarray}
     \hat{\gamma}_{gg}^{(1)}&=&
                         8C_FT_F
                     \frac{N^8+4N^7+8N^6+6N^5-3N^4-22N^3-10N^2-8N-8}
                          {(N-1)N^3(N+1)^3(N+2)}
\N\\ &&\hspace{-10mm}
                       +\frac{32C_AT_F}{9}
                                \Biggl\{
                    -5S_1
                    +\frac{3N^6+9N^5+22N^4+29N^3+41N^2+28N+6}
                          {(N-1)N^2(N+1)^2(N+2)}
                                \Biggr\}~
   \end{eqnarray}
   of the $2$--loop anomalous dimension is additionally needed. 
   As for $A_{Qg}^{(2)}$, the
   massive parts of the gluon self--energy contribute, 
   Eqs. (\ref{GluSelf1},~\ref{GluSelf2}).
   The unrenormalized OME at the $2$--loop level is then given in terms 
   of reducible and irreducible contributions via
   \begin{eqnarray}
    \Ahathat_{gg,Q}^{(2)}&=&
        \Ahathat_{gg,Q}^{(2), \mbox{irred}}-
       ~\Ahathat_{gg,Q}^{(1)}
        \hat{\Pi}^{(1)}\Bigl(0,\frac{\hat{m}^2}{\mu^2}\Bigr)
       -\hat{\Pi}^{(2)}\Bigl(0,\frac{\hat{m}^2}{\mu^2}\Bigr)~.
   \end{eqnarray}
   In the unrenormalized result, we observe the same pole 
   structure as predicted in Eq.~(\ref{AhhhggQ2}). 
   The constant and $O(\ep)$ contributions  
   $a_{gg,Q}^{(2)}$ and $\overline{a}_{gg,Q}^{(2)}$ are
%%%%%%%%%%%%%%%%%%%%%%%%%%
%
% agg2Q, agg2Qb
%
%%%%%%%%%%%%%%%%%%%%%%%%%%
   \begin{eqnarray}
    a_{gg,Q}^{(2)}&=&
      T_FC_A\Biggl\{
                      -\frac{8}{3}\zeta_2S_1
                      +\frac{16(N^2+N+1)\zeta_2}
                            {3(N-1)N(N+1)(N+2)}
                      -4\frac{56N+47}
                             {27(N+1)}S_1
 \N\\ &&\hspace{-10mm}
                      +\frac{2P_{16}}
                            {27(N-1)N^3(N+1)^3(N+2)}
           \Biggr\}
 \N\\ &&\hspace{-10mm}
     +T_FC_F\Biggl\{
                      \frac{4(N^2+N+2)^2\zeta_2}
                           {(N-1)N^2(N+1)^2(N+2)}
                      -\frac{P_{17}}
                            {(N-1)N^4(N+1)^4(N+2)}
         \Biggr\}~,
            \label{agg2Q} 
   \end{eqnarray}
%%%%
   \begin{eqnarray}
    \overline{a}_{gg,Q}^{(2)}&=&
     T_FC_A\Biggl\{
                     -\frac{8}{9}\zeta_3S_1
                     -\frac{20}{9}\zeta_2S_1
                     +\frac{16(N^2+N+1)}
                           {9(N-1)N(N+1)(N+2)}\zeta_3 
                     +\frac{2N+1}
                           {3(N+1)}S_2
 \N\\ &&\hspace{-10mm}
                     -\frac{S_1^2}{3(N+1)}
                     -2\frac{328N^4+256N^3-247N^2-175N+54}
                            {81(N-1)N(N+1)^2}S_1 
\N\\ && \hspace{-10mm}
                     +\frac{4P_{18}\zeta_2}
                           {9(N-1)N^2(N+1)^2(N+2)}
                     +\frac{P_{19}}
                           {81(N-1)N^4(N+1)^4(N+2)}
         \Biggr\}
\N\\ &&\hspace{-10mm}
      +T_FC_F\Biggl\{
                       \frac{4(N^2+N+2)^2\zeta_3}
                            {3(N-1)N^2(N+1)^2(N+2)}
                      +\frac{P_{20}\zeta_2}
                            {(N-1)N^3(N+1)^3(N+2)} \N\\ &&
                      +\frac{P_{21}}
                             {4(N-1)N^5(N+1)^5(N+2)}
         \Biggr\}~, 
       \label{agg2Qbar} 
   \end{eqnarray}
   \begin{eqnarray}
    P_{16}&=&15N^8+60N^7+572N^6+1470N^5+2135N^4 \N\\ &&
          +1794N^3+722N^2-24N-72~,\\
    P_{17}&=&15N^{10}+75N^9+112N^8+14N^7-61N^6+107N^5+170N^4
           +36N^3 \N\\ &&
    -36N^2-32N-16~,\\ 
    P_{18}&=&3N^6+9N^5+22N^4+29N^3+41N^2+28N+6~,\\
    P_{19}&=&3N^{10}+15N^9+3316N^8+12778N^7+22951N^6+23815N^5+14212N^4\N\\ &&
            +3556N^3-30N^2+288N+216~,\\
    P_{20}&=&N^8+4N^7+8N^6+6N^5-3N^4-22N^3-10N^2-8N-8~,\\
    P_{21}&=&31N^{12}+186N^{11}+435N^{10}+438N^9-123N^8-1170N^7-1527N^6
          \N\\ && -654N^5 +88N^4 -136N^2-96N-32~.
   \end{eqnarray}
   We agree with the result for $a_{gg,Q}^{(2)}$
   given in \cite{Buza:1996wv}, which is presented in Eq.
   (\ref{agg2Q}). The new term
   $\overline{a}_{gg,Q}^{(2)}$, Eq.~(\ref{agg2Qbar}),
   contributes to all OMEs $A_{ij}^{(3)}$ through 
   renormalization. The renormalized OME is then given by Eq. 
   (\ref{AggQ2MSren}). 
   Since this OME already emerges at ${\sf LO}$, the $O(a_s^2)$
   term changes replacing the ${\sf MOM}$-- by the ${\sf \MS}$--scheme. 
   The second moment in the ${\sf \MS}$--scheme reads 
   \begin{eqnarray}
      A_{gg,Q}^{\MS}&=&
                a_s^{\MS}\Biggl\{
                           \frac{4}{3}T_F\ln \Bigl(\frac{m^2}{\mu^2}\Bigr)
                        \Biggr\} 
              +{a_s^{\MS}}^2\Biggl\{
                      T_F\Bigl[
                          -\frac{22}{9}C_A
                          +\frac{16}{9}C_F
                          +\frac{16}{9}T_F
                      \Bigr]
                        \ln^2 \Bigl(\frac{m^2}{\mu^2}\Bigr)\N
   \end{eqnarray}
   \begin{eqnarray}
%\N\\ 
    &+&
                     T_F\Bigl[
                           \frac{70}{27}C_A
                          +\frac{148}{27}C_F
                      \Bigr]
                        \ln \Bigl(\frac{m^2}{\mu^2}\Bigr)
                     +\frac{7}{9}C_AT_F
                     -\frac{1352}{81}C_FT_F
                        \Biggr\} 
              +O({a_s^{\MS}}^3)~. \label{Agg2QN2MSON}
   \end{eqnarray} 
   In the ${\sf MOM}$--scheme it is given by
   \begin{eqnarray}
     A_{gg,Q}^{\MOM}&=&
                a_s^{\MOM}\Biggl\{
                           \frac{4}{3}T_F\ln \Bigl(\frac{m^2}{\mu^2}\Bigr)
                        \Biggr\} 
              +{a_s^{\MOM}}^2\Biggl\{
                      T_F\Bigl[
                          -\frac{22}{9}C_A
                          +\frac{16}{9}C_F
                      \Bigr]
                        \ln^2 \Bigl(\frac{m^2}{\mu^2}\Bigr)
\N\\ &+&
                      T_F\Bigl[
                           \frac{70}{27}C_A
                          +\frac{148}{27}C_F
                      \Bigr]
                        \ln \Bigl(\frac{m^2}{\mu^2}\Bigr)
                     +\frac{7}{9}C_AT_F
                     -\frac{1352}{81}C_FT_F
                        \Biggr\} 
              +O({a_s^{\MOM}}^3)~.  \label{Agg2QN2MOMON}
   \end{eqnarray}
   The difference between the schemes reads
   \begin{eqnarray}
     \label{AA2}
     A_{gg,Q}^{(2), \MS} = A_{gg,Q}^{(2), \MOM} + 
      \beta_{0,Q}^2 \ln^2\left(\frac{m^2}{\mu^2}\right)~.
   \end{eqnarray}
   The need for applying intermediately the ${\sf MOM}$--scheme for 
   renormalization becomes obvious again for the term 
   $A_{gg,Q}^{(2)}$. As in the 
   ${\sf NS}$--case, renormalization in the ${\sf \MS}$--scheme for the 
   coupling constant does not cancel all singularities.
%%%%%%%%%%%%%%%%%%%%%%%%%%
%
% Results for Agq2Q
%
%%%%%%%%%%%%%%%%%%%%%%%%%%
   The remaining term is $A_{gq,Q}^{(2)}$, which emerges for the first 
   time at $O(a_s^2)$ and the same result is obtained in the 
   ${\sf \MS}$-- and ${\sf MOM}$--schemes. The corresponding 
   ${\sf NLO}$ anomalous dimension is given by 
   \begin{eqnarray}
    \hat{\gamma}_{gq}^{(1)}&=&\frac{32C_FT_F}{3}\Biggl\{
               -\frac{(N^2+N+2)S_1}{(N-1)N(N+1)}
               +\frac{8N^3+13N^2+27N+16}{3(N-1)N(N+1)^2}
               \Biggr\}~.
   \end{eqnarray}
   Again, we obtain the pole terms as predicted in Eq.~(\ref{Ahhhgq2Q}). 
   The constant and $O(\ep)$ contributions  
   $a_{gq,Q}^{(2)}$ and $\overline{a}_{gq,Q}^{(2)}$ then read
%%%%%%%%%%%%%%%%%%%%%%%%%%
%
% agg2Q, agg2Qb
%
%%%%%%%%%%%%%%%%%%%%%%%%%%
   \begin{eqnarray}
    a_{gq,Q}^{(2)}&=&
          T_FC_F\Biggl\{
            \frac{4}{3}\frac{N^2+N+2}{(N-1)N(N+1)}
               \Bigl(2\zeta_2+S_2+S_1^2\Bigr)
 \N\\ &&
           -\frac{8}{9}\frac{8N^3+13N^2+27N+16}
                            {(N-1)N(N+1)^2}S_1
           +\frac{8}{27}\frac{P_{22}}
                            {(N-1)N(N+1)^3}
                 \Biggr\}~,  
         \label{agq2Q} 
\\
   \overline{a}_{gq,Q}^{(2)}&=&
          T_FC_F\Biggl\{
            \frac{2}{9}\frac{N^2+N+2}{(N-1)N(N+1)}
               \Bigl(-2S_3-3S_2S_1-S_1^3+4\zeta_3-6\zeta_2S_1\Bigr)\N\\ 
&&
           +\frac{2}{9}\frac{8N^3+13N^2+27N+16}
                            {(N-1)N(N+1)^2}
                \Bigl(2\zeta_2+S_2+S_1^2\Bigr)
           -\frac{4}{27}\frac{P_{22}S_1}
                             {(N-1)N(N+1)^3}\N\\ &&
           +\frac{4}{81}\frac{P_{23}}
                             {(N-1)N(N+1)^4}
                \Biggr\}~, 
           \label{agq2Qbar}
   \end{eqnarray}
   with
   \begin{eqnarray}
    P_{22}&=&43N^4+105N^3+224N^2+230N+86 \\
    P_{23}&=&248N^5+863N^4+1927N^3+2582N^2+1820N+496~. 
   \end{eqnarray}
   The second moment of the renormalized result, cf. Eq.~(\ref{Agq2QMSren}), 
   reads
   \begin{eqnarray}
      A_{gq,Q}^{\MS}&=&
              {a_s^{\MS}}^2\Biggl\{
                    \frac{32}{9}\ln^2 \Bigl(\frac{m^2}{\mu^2}\Bigr)
                   +\frac{208}{27}\ln \Bigl(\frac{m^2}{\mu^2}\Bigr)
                   +\frac{236}{27}
                        \Biggr\}C_FT_F
              +O({a_s^{\MS}}^3)~. \label{Agq2QN2MSON}
   \end{eqnarray}
   We agree with the result for $a_{gq,Q}^{(2)}$
   given in \cite{Buza:1996wv}, which is presented in 
   (\ref{agq2Q}).
   
%%%%%%%%%%%%%%%%%%%%%%%%%%%%%%%%%%%%%%%
   Let us summarize so far. 
   In this Section, we newly calculated the $O(\ep)$ terms of the 
   $2$--loop massive OMEs. We additionally recalculated for the first 
   time the terms $a_{gg,Q}^{(2)}$, Eq.~(\ref{agg2Q}), 
   and $a_{gq,Q}^{(2)}$, Eq.~(\ref{agq2Q}), which were 
   given in Ref.~\cite{Buza:1996wv} and find full agreement. For 
   completeness, we showed as well the terms $a_{qq,Q}^{(2), {\sf NS}}$,
   $a_{Qq}^{(2), {\sf PS}}$ and $a_{Qg}^{(2)}$, which have been calculated 
   for the first time in Ref.~\cite{Buza:1995ie} and were recalculated 
   in Refs.~\cite{Bierenbaum:2007qe,SKdiploma}. The latter terms 
   contribute to the heavy flavor Wilson coefficients 
   in deeply inelastic scattering to the non power-suppressed contributions
   at $O(a_s^2)$. 
   In the renormalization of the heavy flavor Wilson coefficients 
   to 3--loop order, all these terms contribute together with 
   lower order single pole terms. The $O(a_s^2 \ep)$ contributions
   form parts of the constant terms of the 3--loop 
   heavy flavor unpolarized operator matrix elements needed to describe 
   the 3--loop heavy flavor
   Wilson coefficients in the region $Q^2 \gg m^2$.
    
   The mathematical structure of our results is as follows. 
   The terms $\overline{a}_{ij}^{(2)}$ can
   be expressed in terms of polynomials of the basic nested harmonic sums 
   up to weight ${\sf w=4}$ and derivatives thereof. They belong to the
   complexity-class of the general two-loop Wilson coefficients or hard
   scattering cross sections in massless QED and QCD
   and are described by six basic functions and their derivatives 
   in Mellin space. Their analytic continuation to complex values of $N$ 
   is known in explicit form. The package {\sf Sigma}, \cite{Refined,Schneider:2007,sigma1,sigma2}, 
   proved to be a useful tool to solve the sums occurring in the 
   present problem and was extended accordingly by its author.
%%%%%%%%%%%%%%%%%%%%%%%%%%%%%%%%%%%%%%%%%%%%%%%%%%%%%%%%%%%%%%%%%%%%%%%%%%%%%%%
%
% Subsection Checks
% 
%%%%%%%%%%%%%%%%%%%%%%%%%%%%%%%%%%%%%%%%%%%%%%%%%%%%%%%%%%%%%%%%%%%%%%%%%%%%%%%
   \subsection{\bf\boldmath Checks on the Calculation}
    \label{SubSec-2LChecks}
%%%%%%%%%%%%%%%%%%%%%%%%%%%%%%%%%%%%%%%%%%%%%%%%%%%%%%%%%%%%%%%%%%%%%%%%%%%%%%%
    There are several checks which we can use for our results. 
    First of all, the terms up to $O(\ep^0)$ have been calculated in 
    Refs.~\cite{Buza:1995ie,Buza:1996wv} and we agree with all 
    unrenormalized results. As described in Sections \ref{SubSec-2LF32},
    \ref{SubSec-2LInfSum}, we keep the complete $\ep$--dependence 
    until we expand the summand of the finite or infinite sums, which 
    serves as a consistency check on the $O(\ep)$ results.
    
    Another test is provided by the sum 
    rules in Eqs. (\ref{sumrule1},~\ref{sumrule2}) for $N=2$, 
    which are fulfilled by the
    renormalized OMEs presented here and in 
    Refs.~\cite{Buza:1995ie,Buza:1996wv}.
    These rules are obeyed regardless of the renormalization scheme. 
    We observe that they hold on the unrenormalized
    level as well, even up to $O(\ep)$.
    
    For the term $A_{Qg}^{(2)}$, we evaluated fixed moments of $N$ for the 
    contributing unrenormalized diagrams using 
    the Mellin0-Barnes method, \cite{MB1a,*MB1b,*MB2,MB3,*MB4,Paris:2001}, cf. 
    also Appendix \ref{App-SpeFunMB}.
    Here, we used an extension of a method developed for massless
    propagators in Ref.~\cite{Bierenbaum:2003ud} to massive on--shell
    operator matrix elements, \cite{Bierenbaum:2006mq,Bierenbaum:2007zz,Bierenbaum:2007dm}. 
    The Mellin--Barnes integrals are then evaluated numerically using 
    the package {\sf MB}, \cite{Czakon:2005rk}. Using this method,
    we calculated the even moments $N=2,4,6,8$ and agree
    with the corresponding fixed moments of our all--$N$ result~\footnote{In Table~2 of Ref. \cite{Bierenbaum:2008yu}, the moments $N=2$ and $N=6$ for the more
difficult two--loop diagrams are presented.}. 

    For the first moment of the Abelian part of the unrenormalized 
    term $~\Ahathat_{Qg}^{(2)}$, there exists even another check. 
    After analytic continuation from the {\sf even} values of $N$ to 
    $N~\epsilon~\mathbb{C}$ is performed, one may consider the limit 
    $N \rightarrow 1$. 
    In this procedure the term $(1 + (-1)^N)/2$ equals to 1. At $O(a_s^2)$ the 
    terms $\propto T_F C_A$ contain $1/z$ contributions in momentum fraction
    space
    and their first moment diverges. For the other contributions to the 
    unrenormalized operator matrix element, after mass renormalization 
    to 2--loop order, the first moment is related to the Abelian part of the 
    transverse contribution to the gluon propagator $\Pi_V(p^2,m^2)|_{p^2=0}$,
    except the term $\propto T_F^2$ which results from wave function 
    renormalization. This  was shown in 
    \cite{Buza:1995ie} up to the 
    constant term in $\ep$. One obtains
    \begin{eqnarray}
     \hat{\Pi}_V(p^2,m^2) = \hat{a}_s T_F \hat{\Pi}_V^{(1)}(p^2,m^2) 
     + 
     \hat{a}_s^2 C_F T_F 
     \hat{\Pi}_V^{(2)}(p^2,m^2) + O(\hat{a}_s^3)~, 
    \end{eqnarray}
    with
    \begin{eqnarray}
     \label{eqPI1abel}
      \lim_{p^2 \rightarrow 0} \hat{\Pi}_V^{(1)}(p^2,m^2) &=& \frac{1}{2} 
      ~\Ahathat_{Qg}^{(1), N=1}\\
      \label{eqPI2abel}
      \lim_{p^2 \rightarrow 0} \hat{\Pi}_V^{(2)}(p^2,m^2) &=& \frac{1}{2} 
      ~\Ahathat_{Qg}^{(2), N=1}|_{C_F}~.
    \end{eqnarray}
    Here, we extend the relation to the linear terms in $\ep$.  For the first 
    moment
    the double pole contributions in $\ep$ vanish in Eq.~(\ref{eqPI2abel}).
    We compare with the corresponding QED--expression for the 
    photon--propagator, $\Pi_T^{V, (k)}$, 
    which has been obtained in Ref.~\cite{Djouadi:1993ss}. Due to the 
    transition from QED to QCD, the relative color factor at the 
    $2$--loop level has to be adjusted to 
    $1/4 = 1/(C_F C_A)$. After asymptotic expansion 
    in $m^2/p^2$, 
    the comparison can be performed up to the linear term in $\ep$. 
    One obtains
    \begin{eqnarray}
     \label{eqrPI1}
      \lim_{p^2 \rightarrow 0} \frac{1}{p^2}
      \hat{\Pi}_T^{V, (1)}(p^2,m^2) &=& \frac{1}{2 T_F} 
      ~\Ahathat_{Qg}^{(1), N=1}
      = - \left(\frac{m^2}{\mu^2}\right)^{\ep/2} 
      \left[\frac{8}{3 \ep} + \frac{\ep}{3} \zeta_2 \right] \\
      \label{eqrPI2}
      \lim_{p^2 \rightarrow 0} \frac{1}{p^2}
      \hat{\Pi}_T^{V, (2)}(p^2,m^2) &=& \frac{1}{2 T_F C_F} 
      ~\Ahathat_{Qg}^{(2), N=1}|_{C_F}
      = \left(\frac{m^2}{\mu^2}\right)^{\ep} 
       \left[ - \frac{4}{\ep} +15 - \left(\frac{31}{4} + \zeta_2\right) 
       \ep \right]~. \N\\
    \end{eqnarray}
%%%%%%%%%%%%%%%
    Additionally, we notice that the renormalized results 
    do not anymore contain $\zeta_2$--terms. The renormalized 
    terms in Eqs. (\ref{Aqq2NSQMSren}, \ref{AQq2PSMSON}, \ref{AQg2MSren},
    \ref{Agq2QMSren}, \ref{AggQ2MSren}) contain expressions
    proportional to $\zeta_2$ in the non--logarithmic contributions,
    which just cancel the corresponding $\zeta_2$--terms in
    $a_{ij}^{(2)}$, cf. Eqs. (\ref{aQg2}, \ref{aQq2PS}, \ref{aqq2NSQ}, 
    \ref{agg2Q}, \ref{agq2Q}). For explicit examples
    of this cancellation, one may compare 
    the second moments of the renormalized OMEs presented in Eqs. 
    (\ref{AQg2N2MSON}, \ref{AQg2N2MOMON}, \ref{AQq2PSN2MSON}, 
     \ref{Aqq2NSQN2MSON}, \ref{Agg2QN2MSON}, \ref{Agg2QN2MOMON}, 
     \ref{Agq2QN2MSON}). The latter provides no stringent test, 
    but is in accordance 
    with general observations made in higher loop calculations, namely 
    that even $\zeta$--values cancel for massless calculations 
    in even dimensions in the renormalized results
    if presented in the ${\sf \MS}$--scheme, \cite{Broadhurst:private1}. 
    In the present work, this observation holds for the $\zeta_2$--terms 
    in a single--scale massive calculation as well.
    
    The most powerful test is provided by the {\sf FORM}--based program 
    {\sf MATAD},~\cite{Steinhauser:2000ry}, which we used 
    to calculate fixed moments of the $2$--loop OMEs up to $O(\ep)$. 
    The setup is the same as in the $3$--loop case and is explained 
    in the next Section. At the $2$--loop level we worked in 
    general $R_{\xi}$--gauges 
    and explicitly observe the cancellation of the gauge parameter. 
    For the terms 
    $A_{Qg}^{(2)},~A_{gg}^{(2)}$ we used both projection operators given 
    in Eqs. (\ref{projG1},~\ref{projG2}), which serves as another 
    consistency check. In the singlet case,
    we calculated the even moments $N=2,4,...,12$ 
    and found full agreement with the results presented in this Section
    up to $O(\ep)$.
    The same holds in the non--singlet case, where we calculated 
    the odd moments as well, $N=1,2,3,...,12$. 
%%%%%%%%%%%%%%%%%%%%%%%%%%%%%%%%%%%%%%%%%%%%%%%%%%%%%%%%%%%%%%%%%%%%%%%%%%%%%%%
%
% Chapter 7
%
% Calculation of Moments at $O(a_s^3)$
%
%%%%%%%%%%%%%%%%%%%%%%%%%%%%%%%%%%%%%%%%%%%%%%%%%%%%%%%%%%%%%%%%%%%%%%%%%%%%%%%
\newpage
 \section{\bf\boldmath Calculation of Moments at $O(a_s^3)$}
  \label{Sec-3L}
\begin{figure}[H]
   \begin{center}
   \includegraphics[angle=0, width=1.8cm]{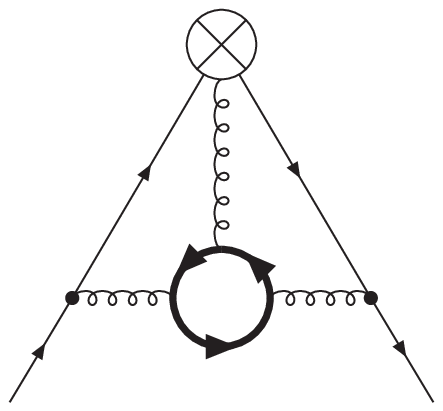}
   \includegraphics[angle=0, width=1.8cm]{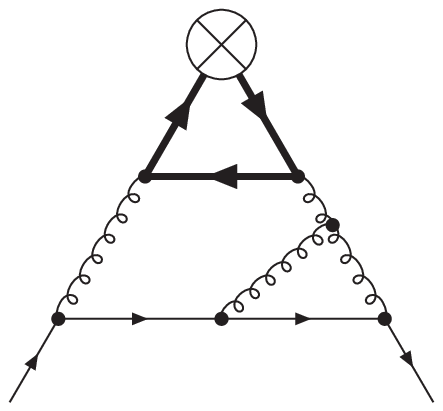}
   \includegraphics[angle=0, width=1.8cm]{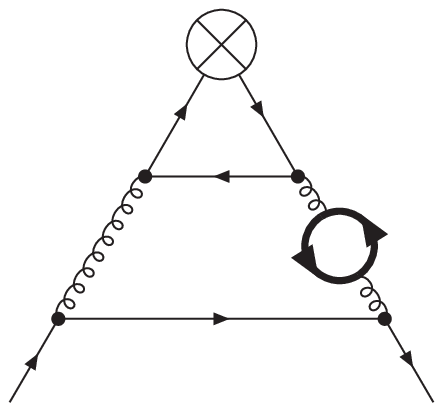}
   \includegraphics[angle=0, width=1.8cm]{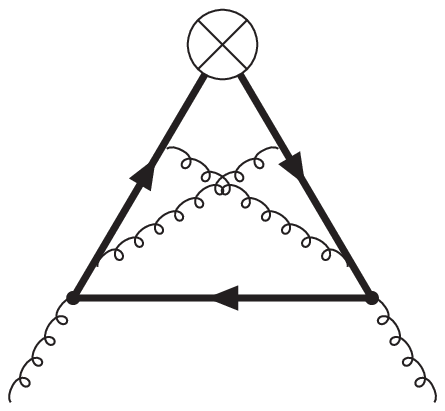}
   \includegraphics[angle=0, width=1.8cm]{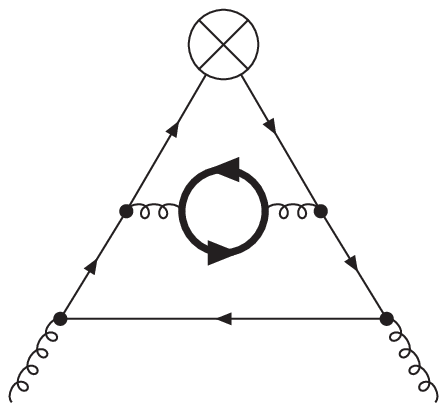}
   \includegraphics[angle=0, width=1.8cm]{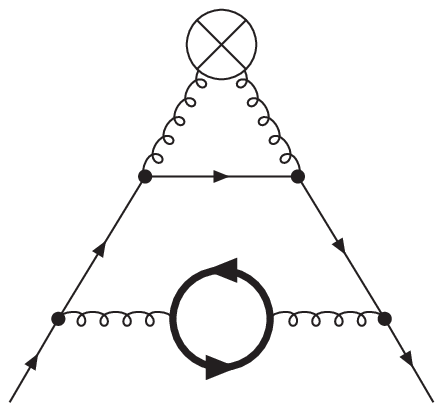}
   \includegraphics[angle=0, width=1.8cm]{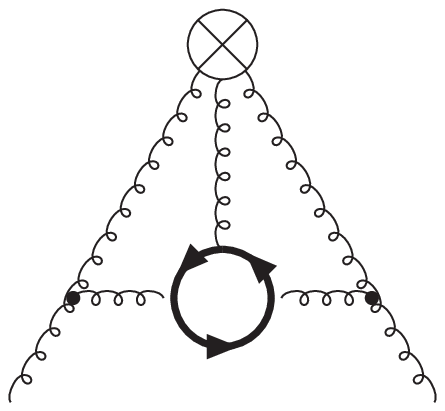}
   \includegraphics[angle=0, width=1.8cm]{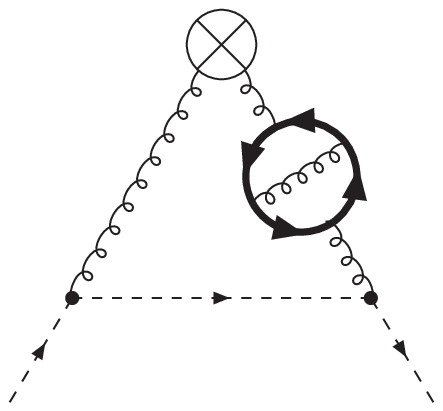}
   \end{center}
   {\small
   \hspace*{5mm}
   ($\sf NS$) \hspace{0.9cm}
   ($\sf PS_H$) \hspace{0.9cm}
   ($\sf PS_l$) \hspace{0.85cm}
   ($\sf qg_H$) \hspace{0.9cm}
   ($\sf qg_l$) \hspace{0.9cm}
   ($\sf gq$) \hspace{0.9cm}
   ($\sf gg$) \hspace{0.9cm}
   {\sf ghost}} 
   \begin{center} 
   \caption[{\sf Examples for 3--loop diagrams contributing to the massive 
                OMEs.}]
           {\sf Examples for 3--loop diagrams contributing to the massive 
                operator matrix elements: NS - non--singlet, 
                ${\sf PS_{H,l}}$ - pure--singlet, singlet ${\sf qg_{H,l}}$, 
                {\sf gq}, gg and ghost contributions. Here the coupling of the 
                gauge boson to a heavy or light fermion line is labeled by 
                {\sf H} and {\sf l}, respectively. Thick lines: heavy quarks, 
                curly lines: gluons, full lines:  quarks, dashed lines: 
                ghosts.}
   \label{diaex}
   \end{center}
   \end{figure}

  \renewcommand{\theequation}{\thesection.\arabic{equation}}
  \setcounter{equation}{0}
%%%%%%%%%%%%%%%%%%%%%%%%%%%%%%%%%%%%%%%%%%%%%%%%%%%%%%%%%%%%%%%%%%%%%%%%%
  In this Chapter, we describe the computation of the $3$--loop
  corrections to the massive operator matrix elements in detail, cf. 
  \cite{Bierenbaum:2009mv}. Typical 
  Feynman diagrams contributing for the different processes are shown in 
  Figure~\ref{diaex}, where $\otimes$ denotes the corresponding composite 
  operator insertions, cf. Appendix~\ref{App-FeynRules}. 
  The generation of these diagrams with the 
  {\sf FORTRAN}--based program {\sf QGRAF}, \cite{Nogueira:1991ex}, is 
  described in Section~\ref{SubSec-3LGen} along with the subsequent steps to
  prepare the input for the {\sf FORM}--based program {\sf MATAD}, 
  \cite{Steinhauser:2000ry}. The latter allows the calculation of massive
  tadpole integrals in $D$ dimensions up to three loops and relies on the 
  {\sf MINCER} algorithm, \cite{Gorishnii:1989gt,Larin:1991fz}. The use of 
  {\sf MATAD} and the projection onto fixed moments are explained in 
  Section~\ref{SubSec-3LMatad}. Finally, we present our results for the fixed 
  moments of the $3$--loop OMEs and the fermionic contributions to the 
  anomalous dimensions in Section~\ref{SubSec-3LResUn}. The calculation is 
  mainly performed using {\sf FORM} programs while 
  in a few cases codes have also been written in {\sf MAPLE}.
%%%%%%%%%%%%%%%%%%%%%%%%%%%%%%%%%%%%%%%%%%%%%%%%%%%%%%%%%%%%%%%%%%%%%%%%%
%%%%%%%%%%%%%%%%%%%%%%%%%%%%%%%%%%%%%%%%%%%%%%%%%%%%%%%%%%%%%%%%%%%%%%%%%
   \noindent \vspace{-18mm}
%%%%%%%%%%%%%%%%%%%%%%%%%%%%%%%%%%%%%%%%%%%%%%%%%%%%%%%%%%%%%%%%%%%%%%%%%
  \subsection{\bf\boldmath Generation of Diagrams}
   \label{SubSec-3LGen}
\begin{figure}[H]
  \begin{center}
  \includegraphics[angle=0, width=14.0cm]{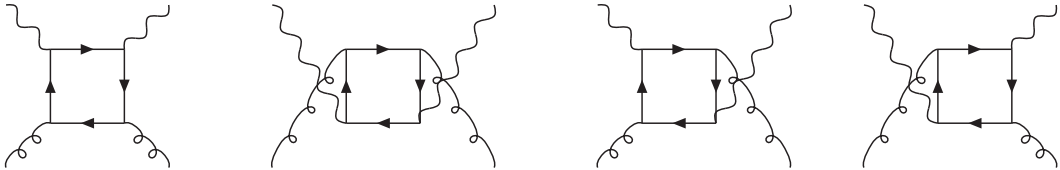}
   \end{center}
   \begin{center} 
    \caption[{\sf Diagrams contributing to $H_{g,(2,L)}^{(1)}$ via 
                 the optical theorem.}]
           {\sf Diagrams contributing to $H_{g,(2,L)}^{(1)}$ via 
                 the optical theorem. Wavy lines: photons; \\ curly lines:
                 gluons; full lines: quarks.} 
     \label{GENOPINS6}
   \end{center}
 \end{figure}

\begin{figure}[H]
  \begin{center}
    \includegraphics[angle=0, width=14.0cm]{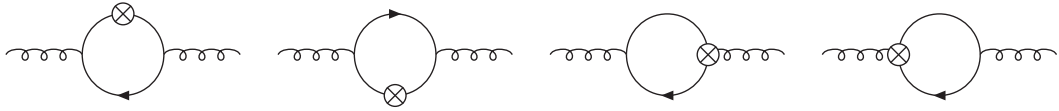}
   \end{center}
   \begin{center} 
    \caption{\sf Diagrams contributing to $A_{Qg}^{(1)}$.}
     \label{GENOPINS7}
   \end{center}
 \end{figure}

\begin{figure}[H]
  \begin{center}
   \includegraphics[angle=0, width=8.0cm]{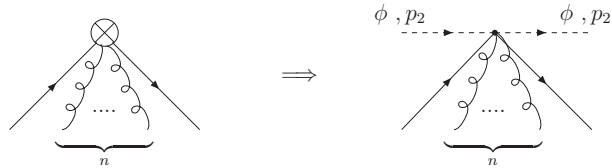}
   \end{center}
   \begin{center} 
    \caption{\sf Generation of the operator insertion.}
     \label{GENOPINS1}
   \end{center}
 \end{figure}

\begin{figure}[H]
   \begin{center}
   \includegraphics[angle=0, width=10.0cm]{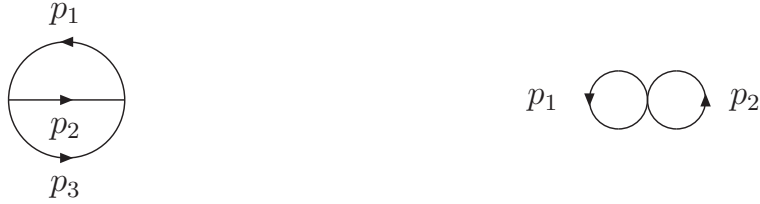}
   \end{center}
   \begin{center} 
    \caption[{\sf $2$--Loop topologies for ${\sf MATAD}$}]
            {\sf $2$--Loop topologies for ${\sf MATAD}$, indicating labeling 
                 of momenta.}
     \label{MATADTOP1}
   \end{center}
   \end{figure}

\begin{figure}[H]
     \begin{center}
   \includegraphics[angle=0, width=4.0cm]{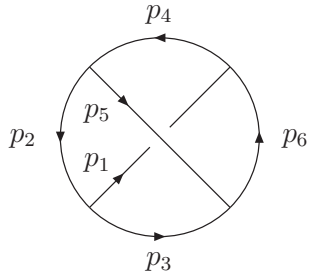}
     \end{center}
     \begin{center} 
      \caption[{\sf Master $3$--loop topology for MATAD.}]
              {\sf Master $3$--loop topology for MATAD, indicating labeling 
                   of momenta.}
        \label{MATADTOP2}
    \end{center}
  \end{figure}

\begin{figure}[htb]
     \begin{center}
    \includegraphics[angle=0, width=12.0cm]{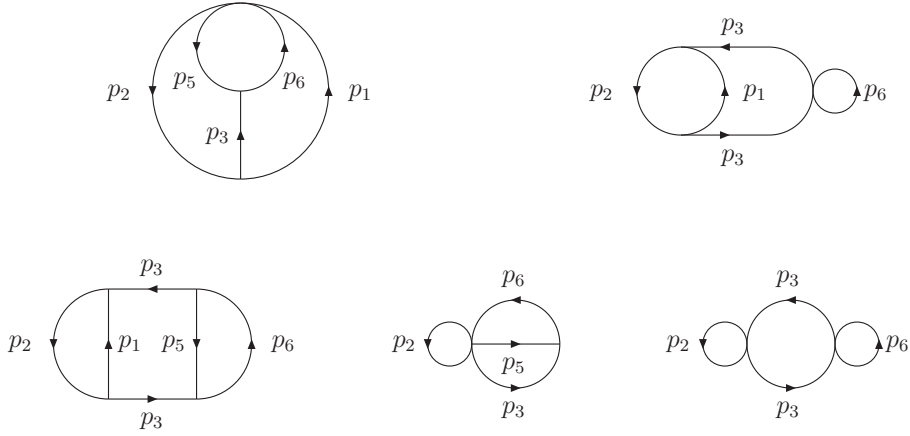}
     \end{center}
     \begin{center} 
      \caption{\sf Additional $3$--loop topologies for ${\sf MATAD}$.}
        \label{MATADTOP3}
    \end{center}
  \vspace{-8mm}
  \end{figure}

%%%%%%%%%%%%%%%%%%%%%%%%%%%%%%%%%%%%%%%%%%%%%%%%%%%%%%%%%%%%%%%%%%%%%%%%%
   {\sf QGRAF} is a quite general program to generate Feynman diagrams and  
   allows to specify various kinds of particles and interactions. 
   Our main issue is to generate diagrams which contain composite
   operator insertions, cf. (\ref{COMP1})--(\ref{COMP3}) and 
   Appendix~\ref{App-FeynRules}, as special vertices.  
   To give an example, let us consider the contributions to $A_{Qg}^{(1)}$. 
   Within the light--cone expansion, Section~\ref{SubSec-DISComptLCE}, 
   this term derives from the Born diagrams squared of the
   photon--gluon fusion process shown in Figure~\ref{GENOPINS6}, cf. 
   Section~\ref{SubSec-HQElProd} and Figure \ref{CCbarLO}.
%%%%%%%%%%%%%%%%%%%%%%%%%%%%%%%%%%%%%%%%%%%%%%%%%%%%%%%%%%%%%%%%%%
%%%%%%%%%%%%%%%%%%%%%%%%%%%%%%%%%%%%%%%%%%%%%%%%%%%%%%%%%%%%%%%%%%
  \noindent 
  After expanding these diagrams with respect to the virtuality of the photon,
  the mass effects are given by the diagrams 
  in Figure~\ref{GENOPINS7}. These are obtained 
  by contracting the lines between the external photons. 
%%%%%%%%%%%%%%%%%%%%%%%%%%%%%%%%%%%%%%%%%%%%%%%%%%%%%%%%%%%%%%%%%%
 \vspace{-8mm}
%%%%%%%%%%%%%%%%%%%%%%%%%%%%%%%%%%%%%%%%%%%%%%%%%%%%%%%%%%%%%%%%%%
  \noindent
  Thus, one may think of the operator insertion as being coupled to two
  external particles, an incoming and an outgoing one, which carry the same
  momentum. Therefore, one defines in the model file of {\sf QGRAF} vertices
  which resemble the operator insertions in this manner, using a scalar field
  $\phi$, which shall not propagate in order to ensure that there is only one
  of these vertices for each diagram. For the quarkonic operators, one defines
  the vertices
  \begin{eqnarray}
   \phi+\phi+q+\overline{q}+n~g~~, \hspace*{3mm} 0 \le n \le 3~, 
   \label{phiquark}
  \end{eqnarray}
  which is illustrated in Figure~\ref{GENOPINS1}. 
%%%%%%%%%%%%%%%%%%%%%%%%%%%%%%%%%%%%%%%%%%%%%%%%%%%%%%%%%%%%%%%%%%%%%%%%%
 \vspace{-8mm}
%%%%%%%%%%%%%%%%%%%%%%%%%%%%%%%%%%%%%%%%%%%%%%%%%%%%%%%%%%%%%%%%%%%%%%%%%
  \noindent 
  The same procedure can be used for the purely gluonic interactions 
  and one defines in this case
  \begin{eqnarray}
   \phi+\phi+n~g~~, \hspace*{3mm} 0 \le n \le 4~. \label{phigluon}
  \end{eqnarray}
  The Green's functions we have to consider and their relation to the 
  respective OMEs were given in Eqs. (\ref{omeGluOpQ})--(\ref{omelqproj}). 
  The number of diagrams we obtain contributing to each OME 
  is shown in Table \ref{table:numdiags}.
%%%%%%%%%%%%%%%%%%%%%%%%%%%%%%%%%%%%%%%%%%%%%%%%%%%%%%%%%%%%%%%%%%%%%%%%%  
  \begin{table}[H]
  \label{table:numdiags}
  \begin{center}
  \newcommand{\m}{\hphantom{ }}
  \newcommand{\cc}[1]{\multicolumn{1}{c}{#1}}
  \renewcommand{\arraystretch}{1.3} % enlarge line spacing
  \begin{tabular}{|llllllll|}
%%%%%%%%%%%%%%%%%%%%%%%%
   \hline\hline
    Term & \phantom{00}\#  & Term & \phantom{00}\# & Term & \phantom{00}\#  & Term & \phantom{00}\#  \\
   \hline\hline
%%%%%%%%%%%%%%%%%%%%%%%%
     $A_{Qg}^{(3)}$             & 1358 
   & $A_{qg,Q}^{(3)}$           & \phantom{0}140
   & $A_{Qq}^{(3),{\sf PS}}$    & \phantom{0}125 
   & $A_{qq,Q}^{(3),{\sf PS}}$  & \phantom{000}8
 \\[0.75em]
%%%%%%%%%%%%%%%%%%%%%%%%
     $A_{qq,Q}^{(3),{\sf NS}}$ & \phantom{0}129 
   & $A_{gq,Q}^{(3)}$          & \phantom{00}89
   & $A_{gg,Q}^{(3)}$          & \phantom{0}886
   &                           &               
  \\
%%%%%%%%%%%%%%%%%%%%%%%%
   [3mm]
%%%%%%%%%%%%%%%%%%%%%%%%
   \hline\hline
  \end{tabular}\\[2pt]
  \caption{\sf Number of diagrams contributing to the $3$--loop  
               massive OMEs. }
  \end{center}
  \end{table}
  \renewcommand{\arraystretch}{1.0} % enlarge line spacing
%%%%%%%%%%%%%%%%%%%%%%%%%%%%%%%%%%%%%%%%%%%%%%%%%%%%%%%%%%%%%%%%%%%%%%%%%
  \noindent
  The next step consists in rewriting the output provided by {\sf QGRAF} in
  such a way, that the Feynman rules given in Appendix~\ref{App-FeynRules} can
  be inserted. Thus, one has to introduce Lorentz and color indices and align
  the fermion lines. Additionally, the integration momenta have to be written
  in such a way that {\sf MATAD} can handle them. For the latter step, all
  information on the types of particles, the operator insertion and the
  external momentum are irrelevant, leading to only two basic topologies to be
  considered at the $2$--loop level, which are shown in Figure~\ref{MATADTOP1}.
%%%%%%%%%%%%%%%%%%%%%%%%%%%%%%%%%%%%%%
   \vspace{-8mm}
%%%%%%%%%%%%%%%%%%%%%%%%%%%%%%%%%%%%%%%%%%%%%%%%%%%%%%%%%%%%%%%%%%%%%%%%%
  \noindent
  Note, that in the case at hand the topology on the right--hand side of 
  Figure~\ref{MATADTOP1} always yields zero after integration.  
  At the $3$--loop level, the master topology is given in 
  Figure~\ref{MATADTOP2}.
%%%%%%%%%%%%%%%%%%%%%%%%%%%%%%%%%%%%%%%%%%%%%%%%%%%%%%%%%%%%%%%%%%%%%%%%%
   \vspace{-8mm}
%%%%%%%%%%%%%%%%%%%%%%%%%%%%%%%%%%%%%%%%%%%%%%%%%%%%%%%%%%%%%%%%%%%%%%%% 
  \noindent
  From this topology, five types of diagrams are derived by shrinking
  various lines. These diagrams are shown in Figure~\ref{MATADTOP3}.
%%%%%%%%%%%%%%%%%%%%%%%%%%%%%%%%%%%%%%
%%%%%%%%%%%%%%%%%%%%%%%%%%%%%%%%%%%%%%%%%%%%%%%%%%%%%%%%%%%%%%%%%%%%%%%%
  \noindent
  Finally the projectors given in Eqs. (\ref{projG1},~\ref{projQ}) are applied
  to project onto the scalar massive OMEs. We only use the physical projector
  (\ref{projG2}) as a check for lower moments, since it causes
  a significant increase of the computation time. 
  To calculate the color factor of
  each diagram, we use the program provided in Ref.~\cite{vanRitbergen:1998pn}
  and for the calculation of fermion traces we use ${\sf FORM}$.
  Up to this point, all operations have been performed for general values of 
  Mellin $N$ and the dimensional parameter $\ep$. The integrals do not contain 
  any Lorentz or color indices anymore. In order to use {\sf MATAD}, one now 
  has to assign to $N$ a specific value. Additionally, the unphysical momentum 
  $\Delta$ has to be replaced by a suitable projector, which we 
  define in the following Section.
%%%%%%%%%%%%%%%%%%%%%%%%%%%%%%%%%%%%%%%%%%%%%%%%%%%%%%%%%%%%%%%%%%%%%%%%%%%%%%%
  \subsection{\bf\boldmath Calculation of Fixed $3$--Loop Moments 
                           Using {\sf MATAD}}
   \label{SubSec-3LMatad}
%%%%%%%%%%%%%%%%%%%%%%%%%%%%%%%%%%%%%%%%%%%%%%%%%%%%%%%%%%%%%%%%%%%%%%%%%%%%%%%
   We consider integrals of the type
   \begin{eqnarray}
    I_l(p,m,n_1\ldots n_j) &\equiv&
                      \int \frac{d^Dk_1}{(2\pi)^D}\ldots
                      \int \frac{d^Dk_l}{(2\pi)^D}
                      (\Delta.q_1)^{n_1}\ldots
                      (\Delta.q_j)^{n_j} 
                      f(k_1\ldots k_l,p,m)~. \N \\ \label{ExInt1}
   \end{eqnarray}
   Here $p$ denotes the external momentum, $p^2=0$, $m$ is the heavy quark
   mass, and $\Delta$ is a light--like vector, $\Delta^2=0$.  The momenta
   $q_{i}$ are given by any linear combination of the loop momenta $k_i$ and
   external momentum $p$. The exponents $n_i$ are integers or possibly sums
   of integers, see the Feynman rules in Appendix \ref{App-FeynRules}. Their
   sum is given by
   \begin{eqnarray}
    \sum_{i=1}^j n_i = N~.
   \end{eqnarray}
   The function $f$ in Eq.~(\ref{ExInt1}) contains propagators, of which at
   least one is massive, dot-products of its arguments and powers of $m$. If
   one sets $N=0$, (\ref{ExInt1}) is given by
   \begin{eqnarray}
    I_l(p,m,0\ldots 0)=I_l(m)=
                      \int \frac{d^Dk_1}{(2\pi)^D}\ldots
                      \int \frac{d^Dk_l}{(2\pi)^D}
                      f(k_1\ldots k_l,m)~. \label{ExInt2}
   \end{eqnarray}
   From $p^2=0$ it follows, that the result can not depend on $p$ anymore.
   The above integral is a massive tadpole integral and thus of the
   type {\sf MATAD} can process. Additionally, {\sf MATAD} can calculate the
   integral up to a given order as a power series in $p^2/m^2$. Let us return
   to the general integral given in Eq.~(\ref{ExInt1}).  One notes, that for
   fixed moments of $N$, each integral of this type splits up into one or more
   integrals of the same type with the $n_i$ having fixed integer 
   values. At this
   point, it is useful to recall that the auxiliary vector $\Delta$ has only
   been introduced to get rid of the trace terms of the expectation values of
   the composite operators and has no physical significance.  By undoing the
   contraction with $\Delta$, these trace terms appear again. Consider as an
   example
   \begin{eqnarray}
    I_l(p,m,2,1) 
                    &=&
                      \int \frac{d^Dk_1}{(2\pi)^D}\ldots
                      \int \frac{d^Dk_l}{(2\pi)^D}
                      (\Delta.q_1)^2 
                      (\Delta.q_2) 
                      f(k_1\ldots k_l,p,m)  \label{ExInt5} \\
                    &=&\Delta^{\mu_1}\Delta^{\mu_2}\Delta^{\mu_3}
                      \int \frac{d^Dk_1}{(2\pi)^D}\ldots
                      \int \frac{d^Dk_l}{(2\pi)^D}
                      q_{1,\mu_1}q_{1,\mu_2}q_{2,\mu_3}
                      f(k_1\ldots k_l,p,m)~. \N \\ \label{ExInt3}
   \end{eqnarray}
   One notices that the way of distributing the indices in 
   Eq.~(\ref{ExInt3}) is somewhat arbitrary, since after the 
   contraction with the totally symmetric tensor
   $\Delta^{\mu_1}\Delta^{\mu_2}\Delta^{\mu_3}$ only the completely 
   symmetric part of the corresponding tensor integral contributes. 
   This is made explicit by distributing the indices 
   among the $q_i$ in all possible ways and dividing by the number 
   of permutations one has used. Thus Eq.~(\ref{ExInt3}) is written as
   \begin{eqnarray}
    I_l(p,m,2,1) 
                    &=&\Delta^{\mu_1}\Delta^{\mu_2}\Delta^{\mu_3}
                       \frac{1}{3}
                      \int \frac{d^Dk_1}{(2\pi)^D}\ldots
                      \int \frac{d^Dk_l}{(2\pi)^D} (
                      q_{1,\mu_2}q_{1,\mu_3}q_{2,\mu_1}
                     +q_{1,\mu_1}q_{1,\mu_3}q_{2,\mu_2} 
\N \\ &&
                     +q_{1,\mu_1}q_{1,\mu_2}q_{2,\mu_3}
                                                   )
                      f(k_1\ldots k_l,p,m)~. \label{ExInt4}
   \end{eqnarray}
   Generally speaking, the symmetrization of the tensor 
   resulting from 
   \begin{eqnarray}
                     \prod_{i=1}^j (\Delta.q_1)^{n_i}
   \end{eqnarray}
   can be achieved by shuffling indices, 
   \cite{Borwein:1999js,Blumlein:1998if,Vermaseren:1998uu,Remiddi:1999ew,
   Moch:2001zr,Blumlein:2003gb}, and dividing 
   by the number of terms. The shuffle product is given by
   \begin{eqnarray}
    C \left[\underbrace{(k_1, \ldots, k_1)}_{ \small n_1} \SHU 
        \underbrace{(k_2, \ldots, k_2)}_{ \small n_2}  \SHU \ldots 
         \SHU \underbrace{(k_I, \ldots, k_I)}_{ \small n_I} \right]~,
   \end{eqnarray}
   where $C$ is the normalization constant
   \begin{eqnarray}
    C = \binom{N}{n_1, \ldots, n_I}^{-1}~. 
   \end{eqnarray}
   As an example, the symmetrization of
   \begin{eqnarray}
    q_{1,\mu_1} q_{1,\mu_2} q_{2,\mu_3}
   \end{eqnarray}
   can be inferred from Eq.~(\ref{ExInt4}).  After undoing the contraction
   with $\Delta$ in (\ref{ExInt1}) and shuffling the indices, one may
   make the following ansatz for the result of this integral, 
   which follows from the necessity
   of complete symmetry in the Lorentz indices
   \begin{eqnarray}
    R_{\{\mu_1\ldots \mu_N\}} &\equiv&\sum_{j=1}^{[N/2]+1} A_j
                  \Bigl(\prod_{k=1}^{j-1} g_{\{\mu_{2k}\mu_{2k-1}} \Bigr)
                  \Bigl(\prod_{l=2j-1}^N p_{\mu_l\}} \Bigr)
                   ~. \label{GenResInt}
   \end{eqnarray}
   In the above equation, $[~~]$ denotes the Gauss--bracket and $\{\}$
   symmetrization with respect to the indices enclosed and dividing by the
   number of terms, as outlined above. The first few terms are then given by
   \begin{eqnarray}
    R_0 &\equiv& 1~, \\
    R_{\{\mu_1\}} 
                  &=&      A_1 p_{\mu_1}~, \\
    R_{\{\mu_1\mu_2\}} 
                  &=&      A_1 p_{\mu_1}p_{\mu_2}+A_2 g_{\mu_1\mu_2} ~, \\
    R_{\{\mu_1\mu_2\mu_3\}} 
                  &=&       
                  A_1 p_{\mu_1}p_{\mu_2}p_{\mu_3}
                 +A_2 g_{\{\mu_1\mu_2}p_{\mu_3\}} ~. 
   \end{eqnarray}
   The scalars $A_j$ have in general different mass dimensions. 
   By contracting again with $\Delta$, all trace terms vanish 
   and one obtains 
   \begin{eqnarray}
    I_l(p,m,n_1\ldots n_j) &=&\Delta^{\mu_1}\ldots \Delta^{\mu_N}
                         R_{\{\mu_1\ldots \mu_N\}}  \\
                     &=& A_1 (\Delta.p)^N
   \end{eqnarray}
   and thus the coefficient $A_1$ in Eq.~(\ref{GenResInt}) gives the desired 
   result. To obtain it, one constructs a different projector, 
   which is made up only of the external momentum $p$ and the metric tensor. 
   By making a general ansatz for this projector, applying 
   it to Eq.~(\ref{GenResInt}) and demanding that the result 
   shall be equal to $A_1$, the coefficients 
   of the different Lorentz structures can be 
   determined. The projector reads
   \begin{eqnarray}
    \Pi_{\mu_1 \ldots \mu_N}&=&F(N)
                              \sum_{i=1}^{[N/2]+1}C(i,N)
                              \Bigl(\prod_{l=1}^{[N/2]-i+1}
                                   \frac{g_{\mu_{2l-1}\mu_{2l}}}{p^2} 
                              \Bigr)
                              \Bigl(\prod_{k=2[N/2]-2i+3}^N
                                   \frac{p_{\mu_k}}{p^2}
                              \Bigr)~. \label{Proj1}
   \end{eqnarray}
   For the overall pre-factors $F(N)$ and the coefficients 
   $C(i,N)$, one has to distinguish between even and odd values of $N$, 
   \begin{eqnarray}
    C^{odd}(k,N)&=&(-1)^{N/2+k+1/2}
                 \frac{2^{2k-N/2-3/2}\Gamma(N+1)\Gamma(D/2+N/2+k-3/2)}
                      {\Gamma(N/2-k+3/2)\Gamma(2k)\Gamma(D/2+N/2-1/2)}~,\N\\ \\
    F^{odd}(N)  &=&\frac{2^{3/2-N/2}\Gamma(D/2+1/2)}
                        {(D-1)\Gamma(N/2+D/2-1)}~, \\
    C^{even}(k,N)&=&(-1)^{N/2+k+1}
                    \frac{2^{2k-N/2-2}\Gamma(N+1)\Gamma(D/2+N/2-2+k)}
                         {\Gamma(N/2-k+2)\Gamma(2k-1)\Gamma(D/2+N/2-1)}~, \\
    F^{even}(N)  &=&\frac{2^{1-N/2}\Gamma(D/2+1/2)}
                         {(D-1)\Gamma(N/2+D/2-1/2)}~.
   \end{eqnarray}
   The projector obeys the normalization 
   condition
   \begin{eqnarray}
    \Pi_{\mu_1\ldots \mu_N}R^{\mu_1\ldots \mu_N} &=&A_1 ~, 
   \end{eqnarray}
   which implies 
   \begin{eqnarray}
    \Pi_{\mu_1\ldots \mu_N}p^{\mu_1}\ldots p^{\mu_N}=1~. \\
   \end{eqnarray}
   As an example for the above procedure, we consider the case $N=3$,
   \begin{eqnarray}
    \Pi_{\mu_1\mu_2\mu_3}
                       &=&\frac{1}{D-1}
                              \Bigl(
                             -3\frac{g_{\mu_{1}\mu_{2}}p_{\mu_3}}{p^4}
                             +(D+2)
                                   \frac{p_{\mu_1}p_{\mu_2}p_{\mu3}}{p^6}
                              \Bigr)~.
   \end{eqnarray}  
   Applying this term to (\ref{ExInt4}) yields
   \begin{eqnarray}
    I_l(p,m,2,1) 
                    &=&
                       \frac{1}{(D-1)p^6}
                      \int \frac{d^Dk_1}{(2\pi)^D}\ldots
                      \int \frac{d^Dk_l}{(2\pi)^D} 
                        \Bigl(
                         -2 p^2 q_1.q_2 p.q_1  
\N\\[1em] &&
                         -p^2 q_1^2 p.q_2
                         +(D+2) (q_1.p)^2 q_2.p
                                                   \Bigr)
                      f(k_1\ldots k_l,p,m)~. \label{ExInt6}
   \end{eqnarray}
   Up to $3$--loop integrals of the type (\ref{ExInt6}) can be calculated by 
   ${\sf MATAD}$ as a Taylor series in $p^2/m^2$.
   It is important to keep $p$ artificially off--shell until the end of the
   calculation.  By construction, the overall result will not contain any term
   $\propto~1/p^2$, since the integral one starts with is free of such terms.
   Thus, at the end, these terms have to cancel. The remaining constant term in
   $p^2$ is the desired result.

   The above projectors are similar to the harmonic projectors used in the 
   ${\sf MINCER}$--program, cf. \cite{Larin:1991fz,Vermaseren:mincer}. These 
   are, however, applied to the virtual forward Compton--amplitude to 
   determine the anomalous dimensions and the moments of the massless Wilson 
   coefficients up to 3--loop order. 

   The calculation was performed in Feynman gauge in general.
   Part of the calculation was carried out keeping the gauge parameter in 
   $R_\xi$--gauges, in particular for the moments $N=2,4$ in the singlet case 
   and for $N=1,2,3,4$ in the non--singlet case, yielding agreement with the 
   results being obtained using Feynman--gauge. In  addition, for the moments 
   $N=2,4$ in the terms with external gluons, we applied the physical projector
   in Eq.~(\ref{projG2}), which serves as another verification of our results. 
   The computation of the more complicated diagrams
   was performed on various 32/64 Gb machines using {\sf FORM} and for part of
   the calculation {\sf TFORM},~\cite{Tentyukov:2007mu}, was used. 
   The complete calculation required about 250 CPU days.
%%%%%%%%%%%%%%%%%%%%%%%%%%%%%%%%%%%%%%%%%%%%%%%%%%%%%%%%%%%%%%%%%%%%%%%%%%%%%%%
  \subsection{\bf\boldmath Results}
   \label{SubSec-3LResUn}
%%%%%%%%%%%%%%%%%%%%%%%%%%%%%%%%%%%%%%%%%%%%%%%%%%%%%%%%%%%%%%%%%%%%%%%%%%%%%%%
   We calculated the unrenormalized operator matrix elements treating the 
   1PI-contributions explicitly. They contribute to $A_{Qg}^{(3)}, 
   A_{gg,Q}^{(3)}$ and $A_{qq,Q}^{(3), {\sf NS}}$. One obtains the following 
   representations
   \begin{eqnarray}
     \Ahathat_{Qg}^{(3)}&=&
            \Ahathat_{Qg}^{(3), \mbox{\small \sf irr}}
           -~\Ahathat_{Qg}^{(2), \mbox{\small \sf irr}}
            \hat{\Pi}^{(1)}\Bigl(0,\frac{\hat{m}^2}{\mu^2}\Bigr)
           -~\Ahathat_{Qg}^{(1)}
            \hat{\Pi}^{(2)}\Bigl(0,\frac{\hat{m}^2}{\mu^2}\Bigr)
\N\\ &&
           +~\Ahathat_{Qg}^{(1)}
            \hat{\Pi}^{(1)}\Bigl(0,\frac{\hat{m}^2}{\mu^2}\Bigr)
            \hat{\Pi}^{(1)}\Bigl(0,\frac{\hat{m}^2}{\mu^2}\Bigr)~,\\
     \Ahathat_{gg,Q}^{(3)}&=&
            \Ahathat_{gg,Q}^{(3), \mbox{\small \sf irr}}
           -\hat{\Pi}^{(3)}\Bigl(0,\frac{\hat{m}^2}{\mu^2}\Bigr)
           -~\Ahathat_{gg,Q}^{(2), \mbox{\small \sf irr}}
            \hat{\Pi}^{(1)}\Bigl(0,\frac{\hat{m}^2}{\mu^2}\Bigr)
\N\\ &&
           -2~\Ahathat_{gg,Q}^{(1)}
            \hat{\Pi}^{(2)}\Bigl(0,\frac{\hat{m}^2}{\mu^2}\Bigr)
           +~\Ahathat_{gg,Q}^{(1)}
            \hat{\Pi}^{(1)}\Bigl(0,\frac{\hat{m}^2}{\mu^2}\Bigr)
            \hat{\Pi}^{(1)}\Bigl(0,\frac{\hat{m}^2}{\mu^2}\Bigr)~,
\\
     \Ahathat_{qq,Q}^{(3), {\sf NS}}&=&
            \Ahathat_{qq,Q}^{(3), {\sf NS}, \mbox{\small \sf irr}}
           -~\hat{\Sigma}^{(3)}\Bigl(0,\frac{\hat{m}^2}{\mu^2}\Bigr)~.
   \end{eqnarray}
   The self-energies are given in
   Eqs.~(\ref{GluSelf1}, \ref{GluSelf2}, \ref{GluSelf3}, \ref{QuSelf3}). 
   The calculation of 
   the one-particle irreducible 3--loop contributions is performed 
   as described in the previous Section~\footnote{Partial results of the 
   calculation were presented in \cite{Bierenbaum:2008dk,Bierenbaum:2008tt}.}.
   The amount of moments, which could be calculated, depended on the available 
   computer resources w.r.t. memory and computational time, as well as the 
   possible
   parallelization using {\sf TFORM}.  Increasing the Mellin moment 
   from $N~\rightarrow~N+2$ 
   demands both a factor of 6--8 larger memory and CPU time. We have 
   calculated the even
   moments $N = 2, \ldots, 10$ for $A_{Qg}^{(3)}$, $A_{gg,Q}^{(3)}$, and
   $A_{qg,Q}^{(3)}$, for $A_{Qq}^{(3), \rm PS}$ up to $ N = 12$, and for 
   $A_{qq,Q}^{(3), \rm NS}, A_{qq,Q}^{(3),\rm PS}, A_{gq,Q}^{(3)}$ 
   up to $N=14$. In the ${\sf NS}$--case, we also calculated the odd 
   moments $N=1,\ldots, 13$, which correspond to the ${\sf NS}^-$--terms.

   \vspace*{7mm}\noindent
   \underline {\large \sf $(i)$ Anomalous Dimensions :}
   \vspace*{2mm}\noindent 

   The pole terms of the unrenormalized
   OMEs emerging in the calculation agree with the general structure we 
   presented in Eqs.~(\ref{Ahhhqq3NSQ}, \ref{AhhhQq3PS}, \ref{Ahhhqq3PSQ}, \ref{AhhhQg3}, \ref{Ahhhqg3Q}, \ref{AhhhgqQ3}, \ref{Ahhhgg3Q}). 
   Using lower order renormalization coefficients and the constant terms of the
   $2$--loop results, \cite{Buza:1995ie,Buza:1996wv,Bierenbaum:2007qe,Bierenbaum:2009zt}, allows to determine the fixed moments of the 2--loop anomalous 
   dimensions and the contributions $\propto T_F$ of the $3$--loop anomalous 
   dimensions, cf. Appendix~\ref{App-AnDim}.  All our results agree with the 
   results of Refs.~\cite{Gracey:1993nn,Larin:1996wd,Retey:2000nq,Moch:2002sn,Moch:2004pa,Vogt:2004mw}. 
   The anomalous dimensions 
   $\gamma_{qg}^{(2)}$ and $\gamma_{qq}^{(2), {\sf PS}}$ are obtained 
   completely. The present calculation is fully independent both in the 
   algorithms and codes compared to Refs.~\cite{Larin:1996wd,Retey:2000nq,Moch:2002sn,Moch:2004pa,Vogt:2004mw}
   and thus provides a stringent check on these results.

   \vspace*{7mm}\noindent
   \underline {\large \sf $(ii)$ The constant terms $a_{ij}^{(3)}(N)$:}
   \vspace*{2mm}\noindent 

   The constant terms at $O(a_s^3)$, cf. Eqs. (\ref{Ahhhqq3NSQ}, \ref{AhhhQq3PS}, \ref{Ahhhqq3PSQ}, \ref{AhhhQg3}, \ref{Ahhhqg3Q}, \ref{AhhhgqQ3}, \ref{Ahhhgg3Q}), 
   are the new contributions to the non--logarithmic part of the 3--loop 
   massive operator matrix elements, which can not be constructed by other
   renormalization constants calculated previously. They are given in
   Appendix~\ref{App-OMEs}. All other contributions to the heavy 
   flavor Wilson coefficients in the region $Q^2 \gg m^2$ are known for general
   values of $N$, cf. Sections~\ref{SubSec-RENPred} and \ref{Sec-2L}. 
   The functions $a_{ij}^{(3)}(N)$ still contain coefficients 
   $\propto \zeta_2$ and we will see below, under which circumstances these 
   terms will contribute to the heavy flavor contributions to the 
   deep--inelastic structure functions. The constant ${\sf B_4}$, (\ref{B4}), 
   emerges as in other massive single--scale calculations, \cite{Broadhurst:1991fi,Avdeev:1994db,*Laporta:1996mq,Broadhurst:1998rz,Boughezal:2004ef}.

   \vspace*{7mm}\noindent
   \underline {\large \sf $(iii)$ Moments of the Constant Terms of the 
                                  $3$--loop Massive OMEs}

   \vspace*{2mm}\noindent

   The logarithmic terms of the renormalized $3$--loop massive OMEs 
   are determined by known renormalization constants 
   and lower order contributions to the massive OMEs. 
   They can be inferred from
   Eqs. (\ref{Aqq3NSQMSren}, \ref{AQq3PSMSren}, \ref{Aqq3PSQMSren}, 
   \ref{AQg3MSren}, \ref{Aqg3QMSren}, \ref{Agq3QMSren}, \ref{Agg3QMSren}).
   In the following, we consider as examples the non--logarithmic 
   contributions to the second moments of the renormalized massive OMEs. 
   We refer to coupling constant renormalization in the 
   $\overline{\sf MS}$--scheme and
   compare the results performing the mass renormalization in the 
   on--shell--scheme $(m)$ and the $\overline{\sf MS}$--scheme 
   $(\overline{m})$, cf. Section~\ref{Sec-REP}.
   For the matrix elements with external gluons, we obtain~:
\begin{eqnarray}
A_{Qg}^{(3), \MS}(\mu^2=m^2,2) &=&
T_FC_A^2
      \Biggl( 
                 \frac{174055}{4374}
                -\frac{88}{9}{\sf B_4}+72\zeta_4
                -\frac{29431}{324}\zeta_3
      \Biggr)
\N \\ \N \\ && \hspace{-35mm}
+T_FC_FC_A
      \Biggl( 
                -\frac{18002}{729}
                +\frac{208}{9}{\sf B_4}-104\zeta_4
                +\frac{2186}{9}\zeta_3
                -\frac{64}{3}\zeta_2+64\zeta_2\ln(2)
      \Biggr) \N
\end{eqnarray}
\begin{eqnarray}
%\N \\ \N \\
&& \hspace{-35mm}
+T_FC_F^2
      \Biggl( 
                -\frac{8879}{729}
                -\frac{64}{9}{\sf B_4}+32\zeta_4
                -\frac{701}{81}\zeta_3+80\zeta_2-128\zeta_2\ln(2)
      \Biggr)
\N \\ \N \\ && \hspace{-35mm}
+T_F^2C_A
      \Biggl( 
                -\frac{21586}{2187}
                +\frac{3605}{162}\zeta_3
      \Biggr)
+T_F^2C_F
      \Biggl( 
                -\frac{55672}{729}
                +\frac{889}{81}\zeta_3
                +\frac{128}{3}\zeta_2
      \Biggr)
\N \\ \N \\ && \hspace{-35mm}
+n_fT_F^2C_A
      \Biggl( 
                -\frac{7054}{2187}
                -\frac{704}{81}\zeta_3
      \Biggr)
+n_fT_F^2C_F
      \Biggl( 
                -\frac{22526}{729}
                +\frac{1024}{81}\zeta_3
                -\frac{64}{3}\zeta_2
      \Biggr)~.\label{AQg3N2ONMS} \\
%%%%%%%%%%%%%%%%%%%%%%%%%%%%%%%%%%%%%%%%%%%%%%%%
A_{Qg}^{(3), \MS}(\mu^2=\overline{m}^2,2) &=&
T_FC_A^2
      \Biggl( 
                 \frac{174055}{4374}
                -\frac{88}{9}{\sf B_4}+72\zeta_4
                -\frac{29431}{324}\zeta_3
      \Biggr)
\N \\ \N \\ && \hspace{-35mm}
+T_FC_FC_A
      \Biggl( 
                -\frac{123113}{729}
                +\frac{208}{9}{\sf B_4}-104\zeta_4
                +\frac{2330}{9}\zeta_3
      \Biggr)
+T_FC_F^2
      \Biggl( 
                -\frac{8042}{729}
                -\frac{64}{9}{\sf B_4}
\N \\ \N\\ && \hspace{-35mm}
                +32\zeta_4-\frac{3293}{81}\zeta_3
      \Biggr)
+T_F^2C_A
      \Biggl( 
                -\frac{21586}{2187}
                +\frac{3605}{162}\zeta_3
      \Biggr)
+T_F^2C_F
      \Biggl( 
                -\frac{9340}{729}
                +\frac{889}{81}\zeta_3
      \Biggr)
\N \\ \N \\ 
&& \hspace{-35mm}
+n_fT_F^2C_A
      \Biggl( 
                -\frac{7054}{2187}
                -\frac{704}{81}\zeta_3
      \Biggr)
+n_fT_F^2C_F
      \Biggl( 
                 \frac{478}{729}
                +\frac{1024}{81}\zeta_3
      \Biggr)
~. \label{AQg3N2MSMS} \\
%%%%%%%%%%%%%%%%%%%%%%%%%%%%%%%%5
A_{qg,Q}^{(3), \MS}(\mu^2=m^2,2) &=&
n_fT_F^2C_A
      \Biggl( 
                 \frac{64280}{2187}
                -\frac{704}{81}\zeta_3
      \Biggr)
+n_fT_F^2C_F
      \Biggl( 
                -\frac{7382}{729}
                +\frac{1024}{81}\zeta_3
      \Biggr)
~. \N \\ \label{Aqg3QN2ONMS} \\
%%%%%%%%%%%%%%%%%%%%%%%%%%%%%%%%%%
A_{gg,Q}^{(3), \MS}(\mu^2=m^2,2) &=&
T_FC_A^2
      \Biggl( 
                -\frac{174055}{4374}
                +\frac{88}{9}{\sf B_4}-72\zeta_4
                +\frac{29431}{324}\zeta_3
      \Biggr)
\N \\ \N\\ 
&& \hspace{-35mm}
+T_FC_FC_A
      \Biggl( 
                 \frac{18002}{729}
                -\frac{208}{9}{\sf B_4}+104\zeta_4
                -\frac{2186}{9}\zeta_3
                +\frac{64}{3}\zeta_2-64\zeta_2\ln(2)
      \Biggr)
\N \\ \N \\ 
&& \hspace{-35mm}
+T_FC_F^2
      \Biggl( 
                 \frac{8879}{729}
                +\frac{64}{9}{\sf B_4}-32\zeta_4
                +\frac{701}{81}\zeta_3-80\zeta_2+128\zeta_2\ln(2)
      \Biggr)
\N \\ \N \\ 
&& \hspace{-35mm}
+T_F^2C_A
      \Biggl( 
                 \frac{21586}{2187}
                -\frac{3605}{162}\zeta_3
      \Biggr)
+T_F^2C_F
      \Biggl( 
                 \frac{55672}{729}
                -\frac{889}{81}\zeta_3
                -\frac{128}{3}\zeta_2
      \Biggr)
\N \\ \N \\ && \hspace{-35mm}
+n_fT_F^2C_A
      \Biggl( 
                -\frac{57226}{2187}
                +\frac{1408}{81}\zeta_3
      \Biggr)
+n_fT_F^2C_F
      \Biggl( 
                 \frac{29908}{729}
                -\frac{2048}{81}\zeta_3
                +\frac{64}{3}\zeta_2
      \Biggr)~. \label{Agg3QN2ONMS} \\
%%%%%%%%%%%%%%%%%%%%%%%%%%%%%%%%
A_{gg,Q}^{(3), \MS}(\mu^2=\overline{m}^2,2) &=&
T_FC_A^2
      \Biggl( 
                -\frac{174055}{4374}
                +\frac{88}{9}{\sf B_4}-72\zeta_4
                +\frac{29431}{324}\zeta_3
      \Biggr)
\N \\ \N \\ && \hspace{-35mm}
+T_FC_FC_A
      \Biggl( 
                 \frac{123113}{729}
                -\frac{208}{9}{\sf B_4}+104\zeta_4
                -\frac{2330}{9}\zeta_3
      \Biggr)
+T_FC_F^2
      \Biggl( 
                 \frac{8042}{729}
                +\frac{64}{9}{\sf B_4}
\N \\ \N \\ && \hspace{-35mm}
                -32\zeta_4
                +\frac{3293}{81}\zeta_3
      \Biggr)
+T_F^2C_A
      \Biggl( 
                 \frac{21586}{2187}
                -\frac{3605}{162}\zeta_3
      \Biggr)
+T_F^2C_F
      \Biggl( 
                 \frac{9340}{729}
                -\frac{889}{81}\zeta_3
      \Biggr)\N
\end{eqnarray} 
\begin{eqnarray} 
%\N \\ \N \\ && \hspace{-35mm}
&& \hspace{-35mm}
+n_fT_F^2C_A
      \Biggl( 
                -\frac{57226}{2187}
                +\frac{1408}{81}\zeta_3
      \Biggr)
+n_fT_F^2C_F
      \Biggl( 
                 \frac{6904}{729}
                -\frac{2048}{81}\zeta_3
      \Biggr)~. \label{Agg3QN2MSMS}
\end{eqnarray} 

   Comparing the operator matrix elements in case of the on--shell--scheme and
   $\overline{\sf MS}$--scheme, one notices that the terms 
   $\ln(2) \zeta_2$ and $\zeta_2$ are
   absent in the latter. The $\zeta_2$ terms, which contribute to
   $a_{ij}^{(3)}$, are canceled by other contributions through
   renormalization. Although the present process is massive, this observation
   resembles the known result that $\zeta_2$--terms do not contribute in
   space--like massless higher order calculations in even 
   dimensions,~\cite{Broadhurst:private1}. 
   This behavior is found for all calculated 
   moments. The occurring $\zeta_4$--terms may partly cancel
   with those in the $3$--loop light Wilson coefficients, 
   \cite{Vermaseren:2005qc}. 
   Note, that Eq.~(\ref{Aqg3QN2ONMS}) is not sensitive to mass renormalization 
   due to the structure of the contributing diagrams.

   An additional check is provided by the sum rule (\ref{sumrule2}),
   which is fulfilled in all renormalization schemes and also on the 
   unrenormalized level.

   Unlike the operator matrix elements with external gluons, 
   the second moments of the quarkonic OMEs emerge for the first time at
   $O(a_s^2)$. 
   To 3--loop order, the renormalized quarkonic
   OMEs do not contain terms $\propto \zeta_2$. Due to their simpler structure,
   mass renormalization in the on--shell--scheme does not give rise to terms
   $\propto \zeta_2, \ln(2) \zeta_2$. Only the rational contribution in the
   color factor $\propto T_F C_F^2$ turns out to be different compared 
   to the on--mass--shell--scheme and 
   $A_{qq,Q}^{\sf PS, (3)}$, (\ref{eqqqQ3}), is not affected at all. This 
   holds again for all moments
   we calculated. The non--logarithmic contributions are given by
\begin{eqnarray}
A_{Qq}^{(3), {\sf PS}, \MS}(\mu^2=m^2,2) &=&
T_FC_FC_A
      \Biggl( 
                 \frac{830}{2187}
                +\frac{64}{9}{\sf B_4}-64\zeta_4
                +\frac{1280}{27}\zeta_3
      \Biggr)
\N \\ \N \\ && \hspace{-35mm}
+T_FC_F^2
      \Biggl( 
                 \frac{95638}{729}
                -\frac{128}{9}{\sf B_4}+64\zeta_4
                -\frac{9536}{81}\zeta_3
      \Biggr)
+T_F^2C_F
      \Biggl( 
                 \frac{53144}{2187}
                -\frac{3584}{81}\zeta_3
      \Biggr)
\N \\ \N \\ 
&& \hspace{-35mm}
+n_fT_F^2C_F
      \Biggl( 
                -\frac{34312}{2187}
                +\frac{1024}{81}\zeta_3
      \Biggr)
~. \\
%%%%%%%%%%%%%%%%%%%%%%%%%5
A_{Qq}^{(3),{\sf PS}, \MS}(\mu^2=\overline{m}^2,2) &=&
T_FC_FC_A
      \Biggl( 
                 \frac{830}{2187}
                +\frac{64}{9}{\sf B_4}-64\zeta_4
                +\frac{1280}{27}\zeta_3
      \Biggr)
\N \\ \N \\ && \hspace{-35mm}
+T_FC_F^2
      \Biggl( 
                 \frac{78358}{729}
                -\frac{128}{9}{\sf B_4}+64\zeta_4
                -\frac{9536}{81}\zeta_3
      \Biggr)
+T_F^2C_F
      \Biggl( 
                 \frac{53144}{2187}
                -\frac{3584}{81}\zeta_3
      \Biggr)
\N \\ \N \\ && \hspace{-35mm}
+n_fT_F^2C_F
      \Biggl( 
                -\frac{34312}{2187}
                +\frac{1024}{81}\zeta_3
      \Biggr)~. \\
%%%%%%%%%%%%%%%%%%%%%%%
A_{qq,Q}^{(3),{\sf PS}, \MS}(\mu^2=m^2,2) &=&
n_fT_F^2C_F
      \Biggl( 
                -\frac{52168}{2187}
                +\frac{1024}{81}\zeta_3
      \Biggr)~. \label{eqqqQ3} \\
%%%%%%%%%%%%%%%%%%%%
A_{qq,Q}^{(3),{\sf NS}, \MS}(\mu^2=m^2,2) &=&
T_FC_FC_A
      \Biggl( 
                -\frac{101944}{2187}
                +\frac{64}{9}{\sf B_4}-64\zeta_4
                +\frac{4456}{81}\zeta_3
      \Biggr) \N
\N \\ \N \\ &&
+T_FC_F^2
      \Biggl( 
                 \frac{283964}{2187}
                -\frac{128}{9}{\sf B_4}+64\zeta_4
                -\frac{848}{9}\zeta_3
      \Biggr)\N
\end{eqnarray}
\begin{eqnarray}
&&\hspace{-35mm}
%\N \\ \N \\ &&\hspace{-35mm}
+T_F^2C_F
      \Biggl( 
                 \frac{25024}{2187}
                -\frac{1792}{81}\zeta_3
      \Biggr)
+n_fT_F^2C_F
      \Biggl( 
                -\frac{46336}{2187}
                +\frac{1024}{81}\zeta_3
      \Biggr)
~. \label{Aqq3NSQN2MSON} \\
%%%%%%%%%%%%%%%%%%%%%%
A_{qq,Q}^{(3),{\sf NS}, \MS}(\mu^2=\overline{m}^2,2) &=&
T_FC_FC_A
      \Biggl( 
                -\frac{101944}{2187}
                +\frac{64}{9}{\sf B_4}-64\zeta_4
                +\frac{4456}{81}\zeta_3
      \Biggr)
\N \\ \N \\ &&\hspace{-35mm}
+T_FC_F^2
      \Biggl( 
                 \frac{201020}{2187}
                -\frac{128}{9}{\sf B_4}+64\zeta_4
                -\frac{848}{9}\zeta_3
      \Biggr)
\N \\ \N \\ &&\hspace{-35mm}
+T_F^2C_F
      \Biggl( 
                 \frac{25024}{2187}
                -\frac{1792}{81}\zeta_3
      \Biggr)
+n_fT_F^2C_F
      \Biggl( 
                -\frac{46336}{2187}
                +\frac{1024}{81}\zeta_3
      \Biggr)
~. \\
%%%%%%%%%%%%%%%%%%%%
A_{gq,Q}^{(3), \MS}(\mu^2=m^2,2) &=&
T_FC_FC_A
      \Biggl( 
                 \frac{101114}{2187}
                -\frac{128}{9}{\sf B_4}+128\zeta_4
                -\frac{8296}{81}\zeta_3
      \Biggr)
\N \\ \N \\ 
&&\hspace{-35mm}
+T_FC_F^2
      \Biggl( 
                -\frac{570878}{2187}
                +\frac{256}{9}{\sf B_4}-128\zeta_4
                +\frac{17168}{81}\zeta_3
      \Biggr)
\N \\ \N \\ &&\hspace{-35mm}
+T_F^2C_F
      \Biggl( 
                -\frac{26056}{729}
                +\frac{1792}{27}\zeta_3
      \Biggr)
+n_fT_F^2C_F
      \Biggl( 
                 \frac{44272}{729}
                -\frac{1024}{27}\zeta_3
      \Biggr)
~.  \\
A_{gq,Q}^{(3), \MS}(\mu^2=\overline{m}^2,2) &=&
T_FC_FC_A
      \Biggl( 
                 \frac{101114}{2187}
                -\frac{128}{9}{\sf B_4}+128\zeta_4
                -\frac{8296}{81}\zeta_3
      \Biggr)
\N \\ \N \\ 
&&\hspace{-35mm}
+T_FC_F^2
      \Biggl( 
                -\frac{436094}{2187}
                +\frac{256}{9}{\sf B_4}-128\zeta_4
                +\frac{17168}{81}\zeta_3
      \Biggr)
\N \\ \N \\ &&\hspace{-35mm}
+T_F^2C_F
      \Biggl( 
                -\frac{26056}{729}
                +\frac{1792}{27}\zeta_3
      \Biggr)
+n_fT_F^2C_F
      \Biggl( 
                 \frac{44272}{729}
                -\frac{1024}{27}\zeta_3
      \Biggr)
~. 
\end{eqnarray}
   Finally, the sum rule (\ref{sumrule2})
   holds on the unrenormalized level, as well as for the renormalized
   expressions in all schemes considered.

   {\sf FORM}--codes for the constant terms $a_{ij}^{(3)}$, 
   Appendix~\ref{App-OMEs}, and the corresponding  moments
   of the renormalized massive operator matrix elements, both for the mass 
   renormalization carried out in the on--shell-- and $\overline{\sf MS}$--scheme, are 
   attached to Ref.~\cite{Bierenbaum:2009mv} and can be obtained upon request. 
   Phenomenological 
   studies of the 3--loop heavy flavor Wilson coefficients 
   in the region $Q^2 \gg m^2$ will be given elsewhere, \cite{Blumlein:prep1}.
%%%%%%%%%%%%%%%%%%%%%%%%%%%%%%%%%%%%%%%%%%%%%%%%%%%%%%%%%%%%%%%%%%%%%%%%%%%%%%%
%%%%%%%%%%%%%%%%%%%%%%%%%%%%%%%%%%%%%%%%%%%%%%%%%%%%%%%%%%%%%%%%%%%%%%%%%%%%%%%
%
% Chapter 8
%
% Heavy Flavor Corrections to Polarized Deep-Inelastic Scattering
%
%%%%%%%%%%%%%%%%%%%%%%%%%%%%%%%%%%%%%%%%%%%%%%%%%%%%%%%%%%%%%%%%%%%%%%%%%%%%%%%
%%%%%%%%%%%%%%%%%%%%%%%%%%%%%%%%%%%%%%%%%%%%%%%%%%%%%%%%%%%%%%%%%%%%%%%%%%%%%%%
\newpage
\section{\bf\boldmath Heavy Flavor Corrections to Polarized Deep-Inelastic 
                      Scattering}
 \label{Sec-POL}
 \renewcommand{\theequation}{\thesection.\arabic{equation}}
 \setcounter{equation}{0}
%%%%%%%%%%%%%%%%%%%%%%%%%%%%%%%%%%%%%%%%%%%%%%%%%%%%%%%%%%%%%%%%%%%%%%%%%%%%%%
  The composition of the proton spin in terms of partonic degrees of freedom
  has attracted much interest after the initial experimental finding,
  \cite{Alguard:1976bmxAlguard:1978gf,*Baum:1983ha,*Ashman:1987hvxAshman:1989ig},
  that the polarization of the three light quarks alone does not add to the
  required value 1/2. Subsequently, the polarized proton structure functions
  have been measured in great detail by various experiments, \cite{Adeva:1993km,*Anthony:1996mw,*Ackerstaff:1997ws,*Abe:1997cx,*Adeva:1998vv,*Abe:1998wq,*Airapetian:1998wi,*Anthony:1999rm,*Anthony:2000fn,*Zheng:2003un,*Airapetian:2004zf,*Ageev:2005gh,*Ageev:2007du,*Airapetian:2007mh}~\footnote{For theoretical surveys see \cite{Reya:1992ye,Lampe:1998eu,Burkardt:2008jw}.}.
  To determine the different contributions to the nucleon spin, both the flavor
  dependence as well as the contributions due to gluons and angular
  excitations at virtualities $Q^2$ in the perturbative region have to be
  studied in more detail in the future. As the nucleon spin contributions are
  related to the first moments of the respective distribution functions, it is
  desirable to measure to very small values of $x$ at high energies, cf.
  \cite{Bluemlein:1995uc,Blumlein:1997ch,Blumlein:1995gh}. 

  A detailed treatment of the flavor structure requires the inclusion of heavy
  flavor. As in the unpolarized case, this contribution is driven by the gluon
  and sea--quark densities. Exclusive data on charm--quark pair production in 
  polarized deep--inelastic scattering are available only in the region of very
  low photon virtualities at present, \cite{Kurek:2006fw,*Brona:2007ug}.
  However, the inclusive measurement of the structure functions $g_1(x,Q^2)$ 
  and $g_2(x,Q^2)$ contains the heavy flavor contributions for hadronic masses
  $W^2 \geq (2 m + M)^2$. 

  The polarized heavy flavor Wilson coefficients are known to first
  order in the whole kinematic range, 
\cite{Watson:1981ce,Gluck:1990in,Vogelsang:1990ug}. In these references,
  numerical illustrations for the ${\sf LO}$ contributions were given as well,
  cf. also \cite{Blumlein:2003wk}. 
  The polarized parton densities have been extracted from deep-inelastic 
scattering data in 
\cite{Altarelli:1998nb,Gluck:2000dy,Bluemlein:2002be,*Hirai:2006sr,*Leader:2006xc,deFlorian:2009vb}.
  Unlike the case for photo-production, \cite{Bojak:1998zm}, the ${\sf NLO}$
  Wilson coefficients have not been calculated for the whole kinematic domain,
  but only in the region $Q^2 \gg m^2$, \cite{Buza:1996xr}, applying the same 
  technique as described in Section~\ref{SubSec-HQAsym}. 
  As outlined in the same Section, the 
  heavy flavor contributions to the structure function $F_2(x,Q^2)$ are very 
  well described by the asymptotic representation for 
  $Q^2/m^2 \gsim 10$, i.e.,
  $Q^2 \gsim 22.5\GeV^2$, in case of charm. A similar 
  approximation should hold in case of the polarized structure function 
  $g_1(x,Q^2)$.

  In this chapter, we re-calculate for the first time the heavy flavor 
  contributions to the longitudinally polarized structure function $g_1(x,Q^2)$
  to $O(a_s^2)$ in the asymptotic region $Q^2 \gg m^2$, \cite{Buza:1996xr}.
  The corresponding contributions to the structure function $g_2(x,Q^2)$ can be
  obtained by using the Wandzura--Wilczek relation, \cite{Wandzura:1977qf}, at 
  the level of twist--2 operators, as has been shown in
  Refs.~\cite{Jackson:1989ph,*Roberts:1996ub,Blumlein:1996vs,Blumlein:1996tp,Blumlein:2003wk} within the covariant parton model.

  In the polarized case, the twist-$2$ heavy flavor Wilson coefficients 
  factorize in the limit $Q^2\gg~m^2$ in the same way as in the unpolarized 
  case, cf. Section~\ref{SubSec-HQAsym} and \cite{Buza:1996xr}. The 
  corresponding
  light flavor Wilson coefficients were obtained in Ref.~\cite{Zijlstra:1993shxZijlstra:1993she1xZijlstra:1993she2}. We proceed by calculating the $2$--loop
  polarized massive quarkonic OMEs, as has been done in 
  Ref.~\cite{Buza:1996xr}. Additionally, we newly calculate
  the $O(\ep)$ terms of these objects, which will be needed to evaluate the 
  $O(a_s^3)$ corrections, cf. Section~\ref{Sec-REN}. 

  The calculation is performed in the same way as described in 
  Section~\ref{Sec-2L} and we therefore only discuss aspects that are
  specific to the 
  polarized case. The notation for the heavy flavor Wilson coefficients 
  is the same as in Eq.~(\ref{Calldef}) and below, except that the index 
  $(2,L)$ has to be replaced by $(g_1,g_2)$. The polarized massive operator 
  matrix elements are denoted by $\Delta A_{ij}$ and obey the same relations
  as in 
  Sections \ref{Sec-HQDIS} and \ref{Sec-REN}, if one replaces the anomalous
  dimensions, cf. Eq.~(\ref{gammazetNS}, \ref{gammazetS}), by their 
  polarized counterparts, $\Delta \gamma_{ij}$. 

  The asymptotic heavy flavor corrections for polarized deeply inelastic 
  scattering to $O(a_s^2)$, \cite{Buza:1996xr}, were calculated in a  
  specific scheme for the treatment of $\gamma_5$ in dimensional 
  regularization. This was done in order to use the same scheme as has been 
  applied in the calculation of the massless Wilson coefficients in 
  \cite{Zijlstra:1993shxZijlstra:1993she1xZijlstra:1993she2}. Here, we refer to the version prior to an Erratum 
  submitted in 2007, which connected the calculation to the 
  $\overline{\sf MS}$--scheme.
  In this chapter we would like to compare to the results given in 
  Ref.~\cite{Buza:1996xr}, which requires to apply the conventions used there.

  In Section~\ref{sec-P2}, we summarize main relations such as the 
  differential cross
  sections for polarized deeply inelastic scattering and the leading order
  heavy 
  flavor corrections. We give a brief outline on the representation of the 
  asymptotic heavy flavor corrections at ${\sf NLO}$. 
  In Sections (\ref{sec-P4})--(\ref{sec-P5}),
  the contributions to the operator 
  matrix elements $\Delta A_{qq,Q}^{(2), {\sf NS}}$, $\Delta A_{Qg}^{(2)}$
  and $\Delta A_{Qq}^{{\sf PS},(2)}$ are calculated up to the linear terms in 
  $\ep$.
%%%%%%%%%%%%%%%%%%%%%%%%%%%%%%%%%%%%%%%%%%%%%%%%%%%%%%%%%%%%%%%%%%%%%%%%
  \subsection{\bf \boldmath Polarized Scattering Cross Sections}
   \label{sec-P2}
%%%%%%%%%%%%%%%%%%%%%%%%%%%%%%%%%%%%%%%%%%%%%%%%%%%%%%%%%%%%%%%%%%%%%%%%
    We consider the process of deeply inelastic longitudinally polarized 
    charged lepton scattering off longitudinally (L) or transversely (T) 
    polarized nucleons in case of single photon exchange~\footnote{For the basic kinematics of DIS, see Section~\ref{SubSec-DISKin}.}. 
    The differential scattering cross section is given by 
    \begin{eqnarray}
     \frac{d^3 \sigma}{dx dy d \theta} = 
         \frac{y \alpha^2}{Q^4} L^{\mu\nu} W_{\mu\nu}~,
    \end{eqnarray}
    cf.~\cite{Lampe:1998eu,Blumlein:1996tp}. Here, $\theta$ is the azimuthal
    angle of the final state lepton. One may define an asymmetry between the 
    differential cross sections for opposite nucleon polarization 
    \begin{equation}
     A(x,y,\theta)_{L,T} = 
      \frac{d^3 \sigma_{L,T}^{\rightarrow}}{dx dy d \theta} 
     -\frac{d^3 \sigma_{L,T}^{\leftarrow}}{dx dy d \theta}~, 
    \end{equation}
    which projects onto the asymmetric part of both the leptonic and hadronic 
    tensors, $L^A_{\mu\nu}$ and $W^A_{\mu\nu}$. The hadronic tensor is then 
    expressed by two nucleon structure functions
    \begin{equation}
     W^A_{\mu\nu} = i \ep_{\mu\nu\lambda\sigma} 
           \left[\frac{q^\lambda S^\sigma}{P.q} g_1(x,Q^2)
         + \frac{q^\lambda (P.q S^\sigma - S.q P^\sigma)}{(P.q)^2} g_2(x,Q^2) 
           \right]~. 
    \end{equation}
    Here $S$ denotes the nucleon's spin vector 
    \begin{eqnarray}
     S_L &=& (0,0,0,M) \N\\
     S_T &=& M(0,\cos(\bar{\theta}),\sin(\bar{\theta}),0)~,
    \end{eqnarray}
    with $\bar{\theta}$ a fixed angle in the plane transverse to the nucleon
    beam. $\ep_{\mu\nu\lambda\sigma}$ is the Levi--Civita symbol.
    The asymmetries $A(x,y,\theta)_{L,T}$ read 
    \begin{eqnarray}
     A(x,y)_{L} &=& 4 \lambda \frac{\alpha^2}{Q^2} \left[  
             \left(2 - y - \frac{2xy M^2}{s} \right)
              g_1(x,Q^2) + 4 \frac{yx  M^2}{s} g_2(x,Q^2) \right]~, \\
     A(x,y,\bar{\theta},\theta)_{T} &=& -8 \lambda \frac{\alpha^2}{Q^2} 
                  \sqrt{\frac{M^2}{s}} \sqrt{\frac{x}{y} 
               \left[1-y-\frac{xy M^2}{S}\right]}
               \cos(\bar{\theta} - \theta) [ y g_1(x,Q^2)
\N\\ &&
               + 2 g_2(x,Q^2)]~,
    \end{eqnarray}
    where $\lambda$ is the degree of polarization. In case of $A(x,y)_{L}$, 
    the 
    azimuthal angle was integrated out, since the differential cross section 
    depends on it only through phase space.

    The twist--2 heavy flavor contributions to the structure function 
    $g_1(x,Q^2)$ are calculated using the collinear parton model. This is not 
    possible in case of the structure function $g_2(x,Q^2)$. As has been shown 
    in Ref.~\cite{Blumlein:2003wk}, the Wandzura--Wilczek relation holds for
    the gluonic heavy flavor contributions as well
    \begin{eqnarray}
     g_2^{\tau = 2}(x,Q^2) = - g_1^{\tau = 2}(x,Q^2) 
     +\int_x^1 \frac{dz}{z} g_1^{\tau = 2}(z,Q^2)~, 
    \end{eqnarray}
    from which $g_2(x,Q^2)$ can be calculated for twist $\tau=2$.
    At leading order the
    heavy flavor corrections are known for the whole kinematic region, 
    \cite{Watson:1981ce,Gluck:1990in,Vogelsang:1990ug}, 
    \begin{eqnarray}
     g_1^{Q\overline{Q}}(x,Q^2,m^2) = 4 e_Q^2  a_s \int_{ax}^1 \frac{dz}{z} 
                       H_{g,g_1}^{(1)}\left(\frac{x}{z},\frac{m^2}{Q^2}\right)
                       \Delta G(z,n_f,Q^2)~
    \end{eqnarray}
    and are of the same structure as in the unpolarized case, cf. Eq. 
    (\ref{FcLO}). Here, $\Delta G$ is the polarized gluon density. The 
    ${\sf LO}$ heavy flavor Wilson coefficient then reads 
    \begin{eqnarray}
     H_{g,g_1}^{(1)}\left(\tau,\frac{m^2}{Q^2}\right) = 
        4 T_F \left[v (3- 4\tau) + (1- 2\tau) 
        \ln \Biggl( \frac{1 - v}{1 + v} \Biggr)\right]~. 
    \end{eqnarray}
    The support of $H_{g,g_1}^{(1)}\left(\tau,{m^2}/{Q^2}\right)$ is 
    $\tau~\epsilon~[0,1/a]$. As is well known, its first moment vanishes
    \begin{eqnarray}
     \label{eq2new}
     \int_0^{1/a} d\tau H_{g_1}^{(1)}\left(\tau,\frac{m^2}{Q^2}\right) =  0~,
    \end{eqnarray}
    which has a phenomenological implication on the heavy flavor contributions 
    to polarized structure functions, resulting in an oscillatory profile, 
    \cite{Blumlein:2003wk}. The unpolarized heavy flavor Wilson coefficients,
    \cite{Laenen:1992zkxLaenen:1992xs,*Riemersma:1994hv,Buza:1995ie,Bierenbaum:2007qe,Bierenbaum:2007dm,Bierenbaum:2006mq,Bierenbaum:2007rg,Bierenbaum:2008tm}, 
    do not obey a relation like (\ref{eq2new}) but exhibit a rising behavior 
    towards smaller values of $x$. 

    At asymptotic values $Q^2 \gg m^2$ one obtains
    \begin{eqnarray}
      \label{eq1new}
      H_{g,g_1}^{(1),{\rm as}}\left(\tau,\frac{m^2}{Q^2}\right)
       = 4 T_F \left[(3- 4 \tau) - (1- 2\tau) 
          \ln \left(\frac{Q^2}{m^2} \frac{1-\tau}{\tau}\right)\right]~.
    \end{eqnarray}
    The factor in front of the logarithmic term $\ln(Q^2/m^2)$ in 
    (\ref{eq1new}) is the leading order splitting function $\Delta P_{qg}(\tau)$,~\cite{Altarelli:1977zs,Ito:1975pf,*Sasaki:1975hk,*Ahmed:1976ee}~\footnote{Early calculations of the leading order polarized singlet splitting functions in Refs.~\cite{Ito:1975pf,*Sasaki:1975hk,*Ahmed:1976ee} still contained some errors.},
    \begin{eqnarray}
    \Delta P_{qg}(\tau) =8T_F\left[\tau^2-(1-\tau)^2\right]=8T_F\left[2\tau- 
                                                            1\right]~. 
    \end{eqnarray}
    The sum--rule (\ref{eq2new}) also holds in the asymptotic case extending
    the range of integration to $\tau~\epsilon~[0,1]$,
    \begin{eqnarray}
     \label{eq2as}
     \int_0^{1}d\tau H_{g,g_1}^{(1), {\rm as}}\left(\tau,\frac{m^2}{Q^2}\right)
                =0~.
    \end{eqnarray}
%%%%%%%%%%%%%%%%%%%%%%%%%%%%%%%%%%%%%%%%%%%%%%%%%%%%%%%%%%%%%%%%%%%%%%%%
  \subsection{\bf \boldmath Polarized Massive Operator Matrix Elements}
   \label{sec-P2a}
%%%%%%%%%%%%%%%%%%%%%%%%%%%%%%%%%%%%%%%%%%%%%%%%%%%%%%%%%%%%%%%%%%%%%%%%
    The asymptotic heavy flavor Wilson coefficients obey the same factorization
    relations in the limit $Q^2\gg~m^2$ as in the unpolarized case, Eqs. 
    (\ref{LNSFAC})--(\ref{HgFAC}), if one replaces all quantities by their 
    polarized counterparts. 
 
    The corresponding polarized twist--$2$ composite operators, cf. Eqs. 
    (\ref{COMP1})--(\ref{COMP3}), are given by 
    \begin{eqnarray}
         \label{COMP1pol}     % \label{eq3}
          O^{\sf NS}_{q,r;\mu_1, \ldots, \mu_N} &=& i^{N-1} {\bf S} 
                  [\overline{\psi}\gamma_5\gamma_{\mu_1} 
                                  D_{\mu_2} \ldots D_{\mu_N} 
                   \frac{\lambda_r}{2}\psi] - {\rm trace~terms}~, \\
         \label{COMP2pol}     %\label{eq4}
          O^{\sf S}_{q;\mu_1, \ldots, \mu_N} &=& i^{N-1} {\bf S} 
                  [\overline{\psi}\gamma_5
                                  \gamma_{\mu_1} D_{\mu_2} \ldots D_{\mu_N}
                   \psi] - {\rm trace~terms}~, \\
         \label{COMP3pol}
          O^{\sf S}_{g;\mu_1, \ldots, \mu_N} &=& 2 i^{N-2} {\bf S} 
            {\rm \bf Sp}[\frac{1}{2} 
               \ep^{\mu_1 \alpha \beta \gamma} F_{\beta\gamma}^a
               D^{\mu_2} \ldots D^{\mu_{N-1}}  F_{\alpha, a}^{\mu_N}]
                         - {\rm trace~terms}~.
    \end{eqnarray}
    The Feynman rules needed are given in Appendix~\ref{App-FeynRules}.
    The polarized anomalous dimensions of these operators are defined in the 
    same way as in Eqs. (\ref{gammazetNS}, \ref{gammazetS}), as is the case 
    for the polarized massive OMEs, cf. Eq.~(\ref{pertomeren}) and below. 
 
    In the subsequent investigation,
    we will follow Ref.~\cite{Buza:1996xr} and 
    calculate the quarkonic heavy quark contributions to $O(a_s^2)$. The 
    diagrams contributing to the corresponding massive OMEs are the same as in 
    the unpolarized case and are shown in Figures 1--4 in 
    Ref.~\cite{Buza:1995ie}.
    The formal factorization relations for the heavy flavor Wilson 
    coefficients can be inferred from Eqs. 
    (\ref{eqWIL1}, \ref{eqWIL4}, \ref{eqWIL5}). Here, we perform the 
    calculation in the ${\sf MOM}$--scheme,
    cf. Section~\ref{SubSec-HQElProdWave}, to account for heavy quarks in the
    final state only.
    The same scheme has been adopted in Ref.~\cite{Buza:1996xr}.
    Identifying $\mu^2=Q^2$, the heavy flavor Wilson coefficients in the 
    limit $Q^2\gg~m^2$ become, \cite{Buza:1996xr}, 
    \begin{eqnarray}
     H_{g,g_1}^{(1)}\left(\frac{Q^2}{m^2}, N\right)&=&
                  -\frac{1}{2} \Delta \hat{\gamma}_{qg}^{(0)}
                               \ln\left(\frac{Q^2}{m^2}\right) 
                  +\hat{c}^{(1)}_{g,g_1}, \label{POLHgLO} \\
     H_{g,g_1}^{(2)}\left(\frac{Q^2}{m^2}, N\right)&=&
         \Biggl\{ \frac{1}{8}\Delta\hat{\gamma}^{(0)}_{qg}
                   \left[ 
                          \Delta \gamma_{qq}^{(0)}
                         -\Delta \gamma_{gg}^{(0)}
                         -2\beta_0
                   \right] \ln^2\left(\frac{Q^2}{m^2}\right)
\N\\ &&
                 -\frac{1}{2}\left[
                                    \Delta\hat{\gamma}_{qg}^{(1)}
                                 +\Delta\hat{\gamma}_{qg}^{(0)}c_{q,g_1}^{(1)}
                            \right] \ln\left(\frac{Q^2}{m^2}\right)
\N\\ &&
                 +\left[
                         \Delta \gamma_{gg}^{(0)}
                        -\Delta \gamma_{qq}^{(0)}
                        +2\beta_0
                  \right] \frac{\Delta\hat{\gamma}_{qg}^{(0)}\zeta_2}{8}
                 +\hat{c}_{g,g_1}^{(2)}
                 +\Delta a_{Qg}^{(2)}
         \Biggr\} ~, \label{POLHgNLO} \\
     H_{q,g_1}^{(2), {\sf PS}}\left(\frac{Q^2}{m^2}, N\right)&=& 
         \Biggl\{
                 -\frac{1}{8}\Delta\hat{\gamma}_{qg}^{(0)}
                             \Delta\gamma_{gq}^{(0)} 
                              \ln^2\left(\frac{Q^2}{m^2}\right)
                -\frac{1}{2}\Delta\hat{\gamma}_{qq}^{(1), {\sf PS}}
                            \ln\left(\frac{Q^2}{m^2}\right) 
\N\\ && 
                +\frac{\Delta\hat{\gamma}_{qg}^{(0)}\Delta\gamma_{gq}^{(0)}}{8}
                  \zeta_2
                +\hat{c}_{q,g_1}^{(2), {\sf PS}}
                +\Delta a_{Qq}^{(2), {\sf PS}}
        \Biggr\}~, \label{POLHPSNLO} \\
     L_{q,g_1}^{(2), {\sf NS}}\left(\frac{Q^2}{m^2},N\right)&=&
        \Biggl\{
                 \frac{1}{4}\beta_{0,Q}\Delta\gamma^{(0)}_{qq}
                            \ln^2\left(\frac{Q^2}{m^2}\right) 
                -\left[
                        \frac{1}{2}\Delta\hat{\gamma}_{qq}^{(1), {\sf NS}}
                       +\beta_{0,Q} c_{q,g_1}^{(1)}
                 \right] \ln\left(\frac{Q^2}{m^2}\right)
\N\\ && 
                -\frac{1}{4}\beta_{0,Q}\zeta_2\Delta\gamma_{qq}^{(0)}
                +\hat{c}_{q,g_1,Q}^{(2), {\sf NS}}
                +\Delta a_{qq,Q}^{(2), {\sf NS}}
       \Biggr\}~.\label{POLLNSNLO}
    \end{eqnarray} 
    $c_{i,g_1}^{(k)}$ are the $k$th order non--logarithmic terms of the 
    polarized coefficient functions. As has been described in 
    \cite{Buza:1996xr}, the relations (\ref{POLHgNLO})--(\ref{POLLNSNLO}) 
    hold if one uses the same scheme for the description of $\gamma_5$
    in dimensional regularization for the massive OMEs and the light flavor 
    Wilson coefficients. This is the case for the massive OMEs as calculated 
    in \cite{Buza:1996xr}, to which we refer, and the light flavor Wilson 
    coefficients as calculated in Ref.~\cite{Zijlstra:1993shxZijlstra:1993she1xZijlstra:1993she2}.
%%%%%%%%%%%%%%%%%%%%%%%%%%%%%%%%%%%%%%%%%%%%%%%%%%%%%%%%%%%%%%%%%%%%%%%%
  \subsubsection{$\Delta A_{qq,Q}^{(2), {\sf NS}}$}
   \label{sec-P4}
%%%%%%%%%%%%%%%%%%%%%%%%%%%%%%%%%%%%%%%%%%%%%%%%%%%%%%%%%%%%%%%%%%%%%%%%
    The non--singlet operator matrix element 
    $\Delta A_{qq,Q}^{(2), {\sf NS}}$ has to be the same as in the unpolarized 
    case due to the Ward--Takahashi identity,
    \cite{Ward:1950xp,*Takahashi:1957xn}. Since it is obtained as
    zero--momentum
    insertion on a graph for the transition 
    $\langle p| \rightarrow |p \rangle$,
    one may write it equivalently in terms 
    of the momentum derivative of the self--energy. The latter is independent 
    of the operator insertion and yields therefore the same in case of 
    $\adag \Delta (\Delta.p)^{N-1}$ and 
    $\adag \Delta \gamma_5 (\Delta.p)^{N-1}$. Hence, 
    $\Delta A_{qq,Q}^{(2), {\sf NS}}$ reads, cf. Eq.~(\ref{Aqq2NSQMSren}), 
    \begin{eqnarray}
      \Delta A_{qq,Q}^{(2), {\sf NS}}\left(N,\frac{m^2}{\mu^2}\right)
             &&\hspace{-5mm}=
             A_{qq,Q}^{(2), {\sf NS}}\left(N,\frac{m^2}{\mu^2}\right)
\N \\
      &&\hspace{-5mm}=
          \frac{\beta_{0,Q} \gamma_{qq}^{(0)}}{4}
          \ln^2\left(\frac{m^2}{\mu^2}\right)
         +\frac{\hat{\gamma}_{qq}^{(1), {\sf NS}}}{2} 
          \ln\left(\frac{m^2}{\mu^2}\right)
         + a_{qq,Q}^{(2), {\sf NS}}
         - \frac{\gamma_{qq}^{(0)}}{4} \beta_{0,Q}\zeta_2~, \N \\ 
    \end{eqnarray}
    where the constant term in $\ep$ of the unrenormalized result, 
    Eq.~(\ref{Ahhhqq2NSQ}), is given in Eq.~(\ref{aqq2NSQ}) and the 
    $O(\ep)$--term in Eq.~(\ref{aqq2NSQbar}).
%%%%%%%%%%%%%%%%%%%%%%%%%%%%%%%%%%%%%%%%%%%%%%%%%%%%%%%%%%%%%%%%%%%%%%%%
  \subsubsection{$\Delta A_{Qg}^{(2)}$}
   \label{sec-P3}
%%%%%%%%%%%%%%%%%%%%%%%%%%%%%%%%%%%%%%%%%%%%%%%%%%%%%%%%%%%%%%%%%%%%%%%%
    To calculate the OME $\Delta A_{Qg}$ up to $O(a_s^2)$, the Dirac-matrix 
    $\gamma_5$ is represented in $D = 4 + \ep$ dimensions via,
    \cite{Buza:1996xr,'tHooft:1972fi,Akyeampong:1973xixAkyeampong:1973vkxAkyeampong:1973vj,*Breitenlohner:1976te},
    \begin{eqnarray}
     \adag \Delta \gamma^5 &=& \frac{i}{6}\ep_{\mu\nu\rho\sigma}\Delta^{\mu}
              \gamma^{\nu}\gamma^{\rho}\gamma^{\sigma}~. \label{gamma5}
    \end{eqnarray}
    The Levi--Civita symbol will be contracted later with a second Levi--Civita
    symbol emerging in the general expression for the Green's function, 
    cf. Eq.~(\ref{omeGluOpQ}), 
    \begin{eqnarray}
    \Delta \hat{G}^{ab}_{Q,\mu\nu}&=&
                        \Delta~\Ahathat_{Qg}
                           \Bigl(\frac{\hat{m}^2}{\mu^2},\ep,N\Bigr)
                           \delta^{ab}(\Delta \cdot p)^{N-1}
                        \ep_{\mu\nu\alpha\beta}\Delta^{\alpha}p^{\beta}~,
                        \label{greensingpol}
    \end{eqnarray}
    by using the following relation in $D$--dimensions, \cite{Itzykson:1980rh},
    \begin{eqnarray}
    \ep_{\mu\nu\rho\sigma}\ep^{\alpha\lambda\tau\gamma}&=&
              -{\sf Det} \left[g^{\beta}_{\omega}\right]~,
                                                    \label{levidet} \\
                &&\beta=\alpha,\lambda,\tau,\gamma~,\quad
                 \omega=\mu,\nu,\rho,\sigma~. \N
    \end{eqnarray}
    In particular, anti--symmetry relations of the Levi-Civita tensor or the 
    relation $\gamma_5^2 = {\bf 1}$,
    holding in four dimensions, are not used. The 
    projector for the gluonic OME then reads
    \begin{eqnarray}
     \Delta~\Ahathat_{Qg}&=&
                    \frac{\delta^{ab}}{N_c^2-1}
                    \frac{1}{(D-2)(D-3)}
                    (\Delta.p)^{-N-1}\ep^{\mu\nu\rho\sigma}
                    \Delta \hat{G}^{ab}_{Q,\mu\nu}
                    \Delta_{\rho}p_{\sigma}\label{projecsing}~.
    \end{eqnarray}
    In the following, we will present the results for the operator matrix
    element using the above prescription for $\gamma_5$. This  
    representation allows a direct comparison to Ref.~\cite{Buza:1996xr} 
    despite the fact that in this scheme even some of the anomalous dimensions 
    are not those of the $\overline{\rm MS}$--scheme. We will
    discuss operator matrix elements for which only mass 
    renormalization was carried out, cf. Section~\ref{SubSec-RENMa}. Due to 
    the crossing relations of the forward Compton amplitude corresponding to 
    the polarized case, only odd moments contribute. Therefore the overall 
    factor
    \begin{eqnarray}
     \frac{1}{2} \left[1 - (-1)^N\right],~~N~\in~{\mathbb{N}},
    \end{eqnarray}
    is implied in the following. To obtain the results in $x$--space the 
    analytic continuation to complex values of $N$ can be performed 
    starting from the odd integers. 
    The $O(a_s)$ calculation is straightforward
    \begin{eqnarray}
     \Delta~\Ahathat^{(1)}_{Qg}
         &=&    \left(\frac{m^2}{\mu^2}\right)^{\ep/2}
                \left[\frac{1}{\ep}+\frac{\zeta_2}{8}\ep^2
                                            +\frac{\zeta_3}{24}\ep
                \right]
                ~ \Delta \hat{\gamma}_{qg}^{(0)} + O(\ep^3) \\
         &=&
                 \left(\frac{m^2}{\mu^2}\right)^{\ep/2}
                \left[\frac{1}{\ep} \Delta \hat{\gamma}_{qg}^{(0)} 
                     +\Delta a_{Qg}^{(1)} 
                     +\ep\Delta \overline{a}_{Qg}^{(1)} 
                     +\ep^2\Delta\overline{\overline{a}}_{Qg}^{(1)} 
               \right] + O(\ep^3)~.
    \end{eqnarray}
    The matrix element contains the leading order anomalous dimension
    $\Delta \hat{\gamma}_{qg}^{(0)}$,
    \begin{eqnarray}
     \Delta A_{Qg}^{(1)}=\frac{1}{2} \Delta \hat{\gamma}_{qg}^{(0)}
                         \ln\left(\frac{m^2}{\mu^2}\right)~,
    \end{eqnarray}
    where
    \begin{eqnarray}
     \label{eq11}
      \Delta \hat{\gamma}_{qg}^{(0)}&=&-8 T_F \frac{N-1}{N(N+1)}~.
    \end{eqnarray}
    The leading order polarized Wilson coefficient $c_{g,g_1}^{(1)}$ 
    reads,~\cite{Bodwin:1989nz,Vogelsang:1990ug,Zijlstra:1993shxZijlstra:1993she1xZijlstra:1993she2},
    \begin{eqnarray}
     \label{eq11a}
     c_{g,g_1}^{(1)}
              &=& -4 T_F\frac{N-1}{N(N+1)}
                         \left[S_1+\frac{N-1}{N}\right]~.
    \end{eqnarray}
    The Mellin transform of Eq.~(\ref{eq1new}) then yields the same expression
    as one 
    obtains from Eq.~(\ref{POLHgLO})
    \begin{eqnarray}
     H_{g,g_1}^{(1),\rm as}\left(N,\frac{m^2}{Q^2}\right)=
      \left[ -\frac{1}{2} \Delta \hat{\gamma}_{qg}^{(0)}
                                  \ln\left(\frac{Q^2}{m^2}\right)
             +c_{g,g_1}^{(1)}\right]~, 
    \end{eqnarray}
    for which the proportionality
    \begin{eqnarray}
     H_{g, g_1}^{(1),\rm as}\left(N,\frac{m^2}{Q^2}\right) \propto (N-1)~
    \end{eqnarray}
    holds, leading to a vanishing first moment.
   
    At the $2$--loop level, we express the operator matrix 
    element $\Delta~\Ahathat_{Qg}^{(2)}$, after mass renormalization, 
    in terms of anomalous dimensions,
    cf. \cite{Buza:1995ie,Bierenbaum:2007qe,Bierenbaum:2007dm,Bierenbaum:2006mq,Bierenbaum:2007rg,Bierenbaum:2008tm}, by 
    \begin{eqnarray}
     \label{eq30}
      \Delta~\Ahathat_{Qg}^{(2)}&=&
         \Bigl(\frac{m^2}{\mu^2}\Bigr)^\ep
           \Biggl[
                   \frac{\Delta \hat{\gamma}^{(0)}_{qg}}{2\ep^2}
                     \Bigl\{
                                  \Delta \gamma^{(0)}_{qq}
                                 -\Delta \gamma^{(0)}_{gg}
                                 -2\beta_0
                     \Bigr\}
%\N \\&&
                  +\frac{\Delta \hat{\gamma}^{\prime (1)}_{qg}}{\ep}
                  +\Delta a^{\prime (2)}_{Qg}
                  +\Delta \overline{a}^{\prime (2)}_{Qg}\ep
          \Bigg]
\N\\ && 
        -\frac{2}{\ep}\beta_{0,Q}\Bigl(\frac{m^2}{\mu^2}\Bigr)^{\ep/2}
                 \Biggl(1+\frac{\ep^2}{8}\zeta_2+\frac{\ep^3}{24}\zeta_3
                 \Biggr)\Delta~\Ahathat^{(1)}_{Qg}
        +O(\ep^2)~, \label{AhhhQg2polD}
    \end{eqnarray}
    The remaining ${\sf LO}$ anomalous dimensions are
    \begin{eqnarray}
      \Delta \gamma_{qq}^{(0)} \!&=&\!
                -C_F\Biggl(-8S_1+2\frac{3N^2+3N+2}{N(N+1)}\Biggr)~,
       \label{eq42}\\
      \Delta \gamma_{gg}^{(0)}\!&=&\!-C_A\Biggl(
                   -8S_1
                   +2\frac{11N^2+11N+24}{3N(N+1)}
                  \Biggr)
                  +\frac{8}{3}T_Fn_f~. \label{eq38}
    \end{eqnarray}
    The renormalized expression in the ${\sf MOM}$--scheme is given by
    \begin{eqnarray}
     \Delta A_{Qg}^{\prime (2), \MOM} &=& 
       \frac{\Delta \hat{\gamma}^{(0)}_{qg}}{8} \Bigl[
                            \Delta \gamma^{(0)}_{qq}
                           -\Delta \gamma^{(0)}_{gg}
                           -2\beta_0
                  \Bigr] \ln^2\left(\frac{m^2}{\mu^2}\right)
      +\frac{\hat{\gamma}^{\prime (1)}_{qg}}{2}
                 \ln\left(\frac{m^2}{\mu^2}\right)
\N\\ &&                             
      +\left(
              \Delta \gamma^{(0)}_{gg}
             -\Delta \gamma^{(0)}_{qq}
             +2\beta_0\right
            ) \frac{\hat{\gamma}_{qg}^{(0)}}{8}\zeta_2
      +\Delta a^{\prime (2)}_{Qg}~.  \label{AQg2polD}
    \end{eqnarray}
    The ${\sf LO}$ anomalous dimensions which enter the double pole term in 
    Eq.~(\ref{AhhhQg2polD}) and the $\ln^2(m^2/\mu^2)$ 
    term in Eq.~(\ref{AQg2polD}),
    respectively, are scheme--independent. This is not the case for the 
    remaining terms, which depend on the particular scheme we adopted in 
    Eqs.~(\ref{levidet},~\ref{gamma5}) and are therefore denoted by a prime. 
    The 
    ${\sf NLO}$ anomalous dimension we obtain is given by
    \begin{eqnarray}
     \label{eq39}
      \Delta \hat{\gamma}_{qg}^{\prime (1)}\!&=&\!
                 -T_FC_F\Biggl(
                         -16\frac{N-1}
                                {N(N+1)}S_2
                        +16\frac{N-1}
                               {N(N+1)}S_1^2
                       -32\frac{N-1}
                               {N^2(N+1)}S_1 \N\\
&&\!                    +8\frac{(N-1)(5N^4+10N^3+8N^2+7N+2)}
                              {N^3(N+1)^3}
                      \Biggr) \N\\
&&\!              +T_FC_A\Biggl(
                        32\frac{N-1}
                               {N(N+1)}\beta'
                       +16\frac{N-1}
                               {N(N+1)}S_2
                       +16\frac{N-1}
                               {N(N+1)}S_1^2 \N\\
&&\!                     -16\frac{N-1}
                               {N(N+1)}\zeta_2
                       -\frac{64S_1}
                               {N(N+1)^2}
                       -16\frac{N^5+N^4-4N^3+3N^2-7N-2}
                               {N^3(N+1)^3}
                       \Biggr)~.\N\\
    \end{eqnarray}
    It differs from the result in the $\overline{\sf MS}$--scheme,
    \cite{Mertig:1995ny,Vogelsang:1995vh,*Vogelsang:1996im}, by a
    finite renormalization. This is due to the fact that we contracted the 
    Levi--Civita symbols in $D$ dimensions. The correct ${\sf NLO}$ splitting 
    function is obtained by
    \begin{eqnarray}
      \Delta \hat{\gamma}_{qg}^{(1)}=\Delta \hat{\gamma}_{qg}^{\prime (1)}
                                      +64T_FC_F\frac{N-1}{N^2(N+1)^2}~.
     \label{Pqg1r}
    \end{eqnarray}
    In an earlier version of Ref.~\cite{Zijlstra:1993shxZijlstra:1993she1xZijlstra:1993she2},
    $\Delta \hat{\gamma}_{qg}^{\prime (1)}$ was used as the 
    anomalous dimension 
    departing from the $\overline{\sf MS}$ scheme. Therefore, in Ref.~\cite{Buza:1996xr} the 
    finite renormalization (\ref{Pqg1r}), as the corresponding one for 
    $c_{g,g_1}^{(2)}$, \cite{Zijlstra:1993shxZijlstra:1993she1xZijlstra:1993she2}, was not used for the 
    calculation of
    $\Delta A_{Qg}^{(2)}$. For the higher order terms in $\ep$ in Eq. 
    (\ref{AhhhQg2polD}) we obtain
    \begin{eqnarray}
    \Delta a_{Qg}^{\prime (2)}&=&
           -T_F C_F\Biggl\{
                        \frac{4(N-1)}{3N(N+1)}\Bigl(-4S_3
                                                  +S^3_1
                                                  +3S_1S_2
                                                  +6S_1\zeta_2 
                                            \Bigr)
                         -\frac{4(3N^2+3N-2)S^2_1}
                                {N^2(N+1)(N+2)}
\N\\ &&\hspace{-15mm}
                         -4\frac{N^4+17N^3+43N^2+33N+2}
                                {N^2(N+1)^2(N+2)}S_2
                         -2\frac{(N-1)(3N^2+3N+2)}
                                {N^2(N+1)^2}\zeta_2
\N\\ &&\hspace{-15mm}
                         -4\frac{N^3-2N^2-22N-36}
                                {N^2(N+1)(N+2)}S_1
                         +\frac{2P_1}
                                {N^4(N+1)^4(N+2)}
                  \Biggr\} 
\N\\ &&\hspace{-15mm}
          -T_FC_A\Biggl\{
                         4\frac{N-1}{3N(N+1)}\Bigl(
                             12\M\left[\frac{\Li_2(x)}{1+x}\right](N+1)
                                    +3\beta''
                                    -8S_3
                                    -S^3_1 
                                    -9S_1S_2
\N\\ &&\hspace{-15mm}
                                    -12S_1\beta'
                                    -12\beta\zeta_2
                                    -3\zeta_3
                                            \Bigr)
                         -16\frac{N-1}
                                 {N(N+1)^2}\beta' 
                      +4\frac{N^2+4N+5}
                                {N(N+1)^2(N+2)}S^2_1
\N\\ &&\hspace{-15mm} 
                         +4\frac{7N^3+24N^2+15N-16}
                                {N^2(N+1)^2(N+2)}S_2 
                         +8\frac{(N-1)(N+2)}
                                {N^2(N+1)^2}\zeta_2
\N\\ &&\hspace{-15mm} 
                      +4\frac{N^4+4N^3-N^2-10N+2}
                                {N(N+1)^3(N+2)}S_1
                         -\frac{4P_2}
                                {N^4(N+1)^4(N+2)}
                  \Biggr\}~, \label{aQg2t}  \\
     \Delta \overline{a}_{Qg}^{\prime (2)}&=&
       T_F C_F   \Biggl\{
                            \frac{N-1}
                                 {N(N+1)}
                             \Bigl(
                               16S_{2,1,1}
                              -8S_{3,1}
                              -8S_{2,1}S_1
                              +3S_4
                              -\frac{4}{3}S_3S_1
                              -\frac{1}{2}S^2_2
                              -\frac{1}{6}S^4_1
\N\\&&\hspace{-15mm}
                              -\frac{8}{3}S_1\zeta_3 
                              -S_2S^2_1
                              +2S_2\zeta_2
                              -2S^2_1\zeta_2
                             \Bigr)
                           -8\frac{S_{2,1}}{N^2}
                           +\frac{3N^2+3N-2}
                                 {N^2(N+1)(N+2)}
                             \Bigl(
                              2S_2S_1
                              +\frac{2}{3}S^3_1
                             \Bigl)
\N\\&&\hspace{-15mm}
                           +2\frac{3N^4+48N^3+123N^2+98N+8}
                                 {3N^2(N+1)^2(2+N)}S_3
                           +\frac{4(N-1)}
                                  {N^2(N+1)}S_1\zeta_2
\N\\&&\hspace{-15mm}                         
                           +\frac{2}{3}
                            \frac{(N-1)(3N^2+3N+2)}
                                 {N^2(N+1)^2}\zeta_3
                           +\frac{P_3S_2}
                                 {N^3(N+1)^3(N+2)}
                           +\frac{N^3-6N^2-22N-36}
                                 {N^2(N+1)(N+2)}S^2_1
\N\\&&\hspace{-15mm}
                           +\frac{P_4\zeta_2}
                                 {N^3(N+1)^3} 
                           -2\frac{2N^4-4N^3-3N^2+20N+12}
                                  {N^2(N+1)^2(N+2)}S_1
                           +\frac{P_5}
                                 {N^5(N+1)^5(N+2)}
                   \Biggr\}
\N\\&&\hspace{-15mm}
     + T_F C_A   \Biggl\{
                            \frac{N-1}
                                 {N(N+1)}
                             \Bigl(
                               16S_{-2,1,1}
                              -4S_{2,1,1}
                              -8S_{-3,1}
                              -8S_{-2,2}
                               -4S_{3,1}
                              +\frac{2}{3}\beta'''
\N\\&&\hspace{-15mm}
                              -16S_{-2,1} S_1
                              -4\beta'' S_1
                              +8\beta' S_2
                              +8 \beta' S^2_1
                              +9S_4 
                              +\frac{40}{3} S_3 S_1
                              +\frac{1}{2}S^2_2
                              +5 S_2 S^2_1
                              +\frac{1}{6} S^4_1
\N\\&&\hspace{-15mm} 
                              +4\zeta_2 \beta'
                              -2\zeta_2 S_2
                              -2\zeta_2  S^2_1
                              -\frac{10}{3} S_1\zeta_3
                              -\frac{17}{5}\zeta_2^2
                             \Bigr)
                           -\frac{N-1}
                                 {N(N+1)^2}
                             \Bigl(
                              16 S_{-2,1}
                              +4\beta''
                              -16 \beta' S_1
                             \Bigr)
\N\\&&\hspace{-15mm}
                           -\frac{16}{3}\frac{N^3+7N^2+8N-6}
                                             {N^2(N+1)^2(N+2)}S_3
                           +\frac{2(3N^2-13)S_2S_1}
                                  {N(N+1)^2(N+2)}
                           -\frac{2(N^2+4N+5)}
                                 {3N(N+1)^2(N+2)}S^3_1
\N\\&&\hspace{-15mm}
                           -\frac{8\zeta_2S_1}
                                  {(N+1)^2}
                           -\frac{2}{3} 
                            \frac{(N-1)(9N+8)}
                                 {N^2(N+1)^2}\zeta_3
                           -\frac{8(N^2+3)}
                                  {N(N+1)^3}\beta'
                           -\frac{P_6S_2}
                                 {N^3(N+1)^3(N+2)}
\N\\&&\hspace{-15mm}
                           -\frac{N^4+2N^3-5N^2-12N+2}
                                 {N(N+1)^3(N+2)}S^2_1
                           -\frac{2P_7\zeta_2}
                                  {N^3(N+1)^3}
                           +\frac{2P_8S_1}
                                  {N(N+1)^4(N+2)}
\N\\&&\hspace{-15mm}
                           -\frac{2P_9}
                                  {N^5(N+1)^5(N+2)}
                   \Biggr\}~, \label{aQg2tbar} 
    \end{eqnarray}
    with the polynomials
    \begin{eqnarray}
     P_1 &=& 4N^8+12N^7+4N^6-32N^5-55N^4-30N^3-3N^2-8N-4~, \\
     P_2 &=& 2N^8+10N^7+22N^6+36N^5+29N^4+4N^3+33N^2+12N+4~,\\
     P_3 &=&3N^6+30N^5+107N^4+124N^3+48N^2+20N+8~, \\
     P_4 &=&(N-1)(7N^4+14N^3+4N^2-7N-2)~, \\
     P_5 &=&8N^{10}+24N^9-11N^8-160N^7-311N^6-275N^5
            -111N^4-7N^3 \N\\ &&
            +11N^2+12N+4 ~,  \\
     P_6 &=&N^6+18N^5+63N^4+84N^3+30N^2-64N-16~, \\
     P_7 &=&N^5-N^4-4N^3-3N^2-7N-2~, \\
     P_8 &=&2N^5+10N^4+29N^3+64N^2+67N+8~, \\
     P_9 &=&4N^{10}+22N^9+45N^8+36N^7-11N^6
            -15N^5+25N^4-41N^3 \N\\ &&
            -21N^2-16N-4~. 
    \end{eqnarray}
    The Mellin--transform in Eq.~(\ref{aQg2t}) is given in 
    Eq. (\ref{SM21ANCONT}) in terms of harmonic sums.
    As a check, we calculated several lower moments ($N=1\ldots 9$) of each 
    individual diagram contributing to $A_{Qg}^{(2)}$~\footnote{These are shown in Figure 2 of Ref.~\cite{Buza:1995ie}.} using the
    Mellin--Barnes method, \cite{Czakon:2005rk,Bierenbaum:2007dm}.
    In Table \ref{table:MBcheckpol}, we present the numerical results we obtain
    for the moments $N = 3,7$ of the individual diagrams.
    We agree with the 
    results obtained for general values of $N$.
    The contributions from the individual diagrams are 
    given in \cite{Bierenbaum:prep1}. Our results up to $O(\ep^0)$, 
    Eqs. (\ref{AhhhQg2polD}, \ref{aQg2t}), agree with the results presented 
    in \cite{Buza:1996xr}, which we thereby confirm for the first time.     
    Eq.~(\ref{aQg2tbar}) is a new result.

    In this calculation extensive, use was made
    of the representation of the Feynman-parameter integrals in terms of 
    generalized hypergeometric functions, cf. Section~\ref{Sec-2L}. The 
    infinite sums, which occur in the polarized calculation, are widely the 
    same as in the 
    unpolarized case, \cite{Bierenbaum:2007qe,Bierenbaum:2007dm,Bierenbaum:2006mq,Bierenbaum:2007rg,Bierenbaum:2008tm}.
    The structure of the result for the higher order 
    terms in $\ep$ is completely the same as in the unpolarized case as well,
    see Eq.~(\ref{aQg2}) and the following discussion. 
    Especially, the structural relations between the finite harmonic sums,
    \cite{Blumlein:2004bb,Blumlein:2007dj,Blumlein:2009ta,Blumlein:2009fz},
    allow to express $\Delta a_{Qg}^{\prime (2)}$ by only two basic Mellin 
    transforms, $S_1$ and $S_{-2,1}$. This has to be compared to the $24$ 
    functions needed in Ref.~\cite{Buza:1996xr} to express the constant term in
    $z$--space. 
    Thus we reached a more compact
    representation. $\Delta \overline{a}_{Qg}^{\prime (2)}$ depends on the 
    six sums $S_1(N), S_{\pm 2,1}(N), S_{-3,1}(N), S_{\pm 2,1,1}(N)$, after 
    applying the structural relations. The $O(\ep^0)$ term has the same 
    complexity as the 2--loop anomalous dimensions, whereas the 
    complexity of the $O(\ep)$ term corresponds to the level observed 
    for 2--loop Wilson 
    coefficients and other hard scattering processes which depend on a single 
    scale, cf.~\cite{Blumlein:2005im,*Blumlein:2006rr,Dittmar:2005ed}.
%%%%%%%%%%%%%%%%%%%%%%%%%%%%%%%%%%%%%%%%%%%%%%%%%%%%%%%%%%%%%%%%%%%%%%%%%
  \subsubsection{$\Delta A_{Qq}^{(2), {\sf PS}}$}
   \label{sec-P5}
%%%%%%%%%%%%%%%%%%%%%%%%%%%%%%%%%%%%%%%%%%%%%%%%%%%%%%%%%%%%%%%%%%%%%%%%%
    The operator matrix element $\Delta A_{Qq}^{(2), {\sf PS}}$ is obtained 
    from the 
    diagrams shown in Figure~3 of Ref.~\cite{Buza:1995ie}. 
    In this calculation, we did not adopt any specific scheme for $\gamma_5$,
    but calculated the corresponding integrals without performing any 
    traces or (anti)commuting $\gamma_5$. 
%%%%%%%%%%%%%%%%%%%%%%%%%%%%%%%%%%%%%%%%%%%%%%%%%%%%%%%%%%%%%%%%%%%%%%%%%
    \begin{table}[H]
      \begin{center}
       \includegraphics[angle=0, width=10.0cm]{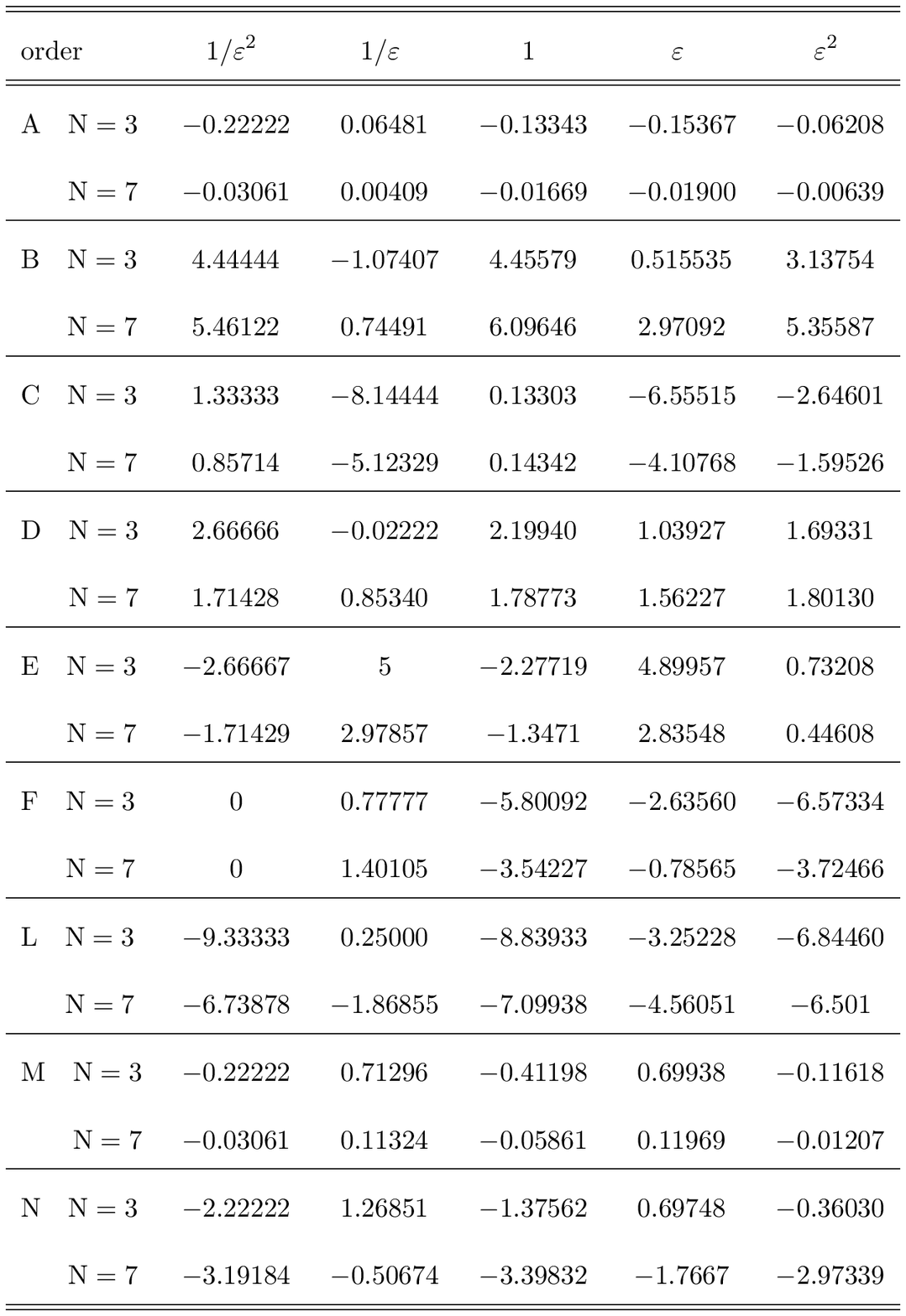}
      \end{center}
      \begin{center}
       \caption{\sf Numerical values for moments of individual diagrams of 
                     $\Delta~\Ahathat_{Qg}^{(2)}$.}
      \label{table:MBcheckpol}
      \small
     \end{center}
     \normalsize
    \end{table}
%%%%%%%%%%%%%%%%%%%%%%%%%%%%%%%%%%%%%%%%%%%%%%%%%%%%%%%%%%%%%%%%%%%%%%%%%
    \noindent
    The result can then be represented in
    terms of three bi--spinor structures
    \begin{eqnarray}
     \label{eq12}
      C_1(\ep)&=&\frac{1}{\Delta.p}
           Tr\Bigl\{\adag\Delta~\adag p\gamma^{\mu}\gamma^{\nu}\gamma_5\Bigr\}
             \adag\Delta\gamma_{\mu}\gamma_{\nu}
               =\frac{24}{3+\ep}\adag\Delta \gamma_5 \\
     \label{eq13}
      C_2(\ep)&=&
       Tr\Bigl\{\adag\Delta\gamma^{\mu}\gamma^{\nu}\gamma^{\rho}\gamma_5\Bigr\}
            \gamma_{\mu}\gamma_{\nu}\gamma_{\rho}
             =24 \adag\Delta \gamma_5 \\
      C_3(\ep)&=&\frac{1}{m^2}
           Tr\Bigl\{\adag\Delta~\adag p\gamma^{\mu}\gamma^{\nu}\gamma_5\Bigr\}
             \adag p \gamma_{\mu}\gamma_{\nu}~.\label{factors}
    \end{eqnarray}
    These are placed between the states $\langle p| \ldots |p \rangle$, with
    \begin{eqnarray}
     \adag p |p \rangle &=&  m_0 |p \rangle 
    \end{eqnarray}
    and $m_0$ the light quark mass. Therefore, the contribution due to 
    $C_3(\ep)$ vanishes in the limit $m_0 \rightarrow 0$. The results for 
    $C_{1,2}(\ep)$ in the r.h.s. of Eqs.~(\ref{eq12},~\ref{eq13}) can be 
    obtained by applying the projector
    \begin{eqnarray}
     \frac{-3}{2(D-1)(D-2)(D-3)}{\sf Tr}[~\adag p \gamma_5~C_i]
    \end{eqnarray}
    and performing the trace in $D$--dimensions using relations
    (\ref{gamma5}, \ref{levidet}). Note that the result in $4$--dimensions 
    is recovered by setting $\ep=0$.
    One obtains from the truncated $2$--loop Green's function 
    $\Delta~G_{Qq}^{ij,(2)}$
    \begin{eqnarray}
     \Delta~G_{Qq}^{ij,(2)} &= &  
     \Delta~\Ahathat^{(2), {\sf PS}}_{Qq}~\adag \Delta \gamma_5
     \delta^{ij}(\Delta.p)^{N-1}~, \label{GreenfPS}
     \label{eq14}
    \end{eqnarray}
    the following result for the massive OME
    \begin{eqnarray}
     \label{eq32}
     \Delta~\Ahathat^{(2),{\sf PS}}_{Qq}\adag \Delta \gamma_5&=&
       \Bigl(\frac{m^2}{\mu^2}\Bigr)^{\ep} C(\ep) 8(N+2)
            \Biggl\{
                    -\frac{1}{\ep^2}\frac{2(N-1)}
                                         {N^2(N+1)^2}
                    +\frac{1}{\ep}\frac{N^3+2N+1}
                                       {N^3(N+1)^3}
\N\\&&\hspace{-15mm}
                    -\frac{(N-1)(\zeta_2+2S_2)}
                          {2N^2(N+1)^2}
                    -\frac{4N^3-4N^2-3N-1}
                          {2N^4(N+1)^4} 
                    +\ep\Bigl[
                               \frac{(N^3+2N+1) (\zeta_2+2S_2)}
                                    {4N^3(N+1)^3}
\N\\ &&\hspace{-15mm}
                              -\frac{(N-1)(\zeta_3+3S_3)}
                                    {6N^2(N+1)^2}
                              +\frac{N^5-7N^4+6N^3+7N^2+4N+1}
                                    {4N^5(N+1)^5}
                        \Bigr]
                 \Biggr\}+O(\ep^2)~, \label{AhhhQq2PSpol}
    \end{eqnarray}
    where
    \begin{eqnarray}
     C(\ep)&=&\frac{C_1(\ep)\cdot(N-1)+C_2(\ep)}{8(N+2)}
            = \adag \Delta \gamma_5 \frac{3(N+2+\ep)}{(N+2)(3+\ep)} 
\N\\   
           &=&1+\frac{N-1}{3(N+2)}\Bigl(
                                        -\ep
                                        +\frac{\ep^2}{3}
                                        -\frac{\ep^3}{9}
                               \Bigr)+O(\ep^4)~.
    \end{eqnarray}
    Comparing our result, Eq.~(\ref{AhhhQq2PSpol}), to the result obtained in 
    \cite{Buza:1996xr}, one notices that there the factor $C(\ep)$ was 
    calculated in $4$--dimensions, i.e. $C(\ep)=1$. Therefore, we do the same 
    and obtain 
    \begin{eqnarray}
     \Delta~\Ahathat^{(2), {\sf PS}}_{Qq}&=&
       S_\ep^2 \Bigl(\frac{m^2}{\mu^2}\Bigr)^{\ep}
           \Biggl[
                  -\frac{\Delta\hat{\gamma}^{(0)}_{qg}
                          \Delta\gamma^{(0)}_{gq}}{2\ep^2} 
                  +\frac{\Delta \hat{\gamma}^{(1), {\sf PS}}_{qq}}{2\ep}
                  + \Delta a^{(2), {\sf PS}}_{Qq}
                  + \Delta \overline{a}^{(2), {\sf PS}}_{Qq}\ep
           \Biggr]~. \N \\ \label{AQq2f}
    \end{eqnarray}
    with 
    \begin{eqnarray}
     \label{eq41}
      \Delta \gamma_{gq}^{(0)}&=&-4C_F\frac{N+2}{N(N+1)}~,\\
      \Delta \hat{\gamma}_{qq}^{(1), {\sf PS}}&=&
                  16T_FC_F\frac{(N+2)(N^3+2N+1)}{N^3(N+1)^3}
                           \label{splitpolPS}~,  \\
      \Delta a^{(2), {\sf PS}}_{Qq}&=&
                    -\frac{(N-1)(\zeta_2+2S_2)}
                          {2N^2(N+1)^2}
                    -\frac{4N^3-4N^2-3N-1}
                          {2N^4(N+1)^4} ~, \\
      \Delta\overline{a}^{(2), {\sf PS}}_{Qq}&=&
                              -\frac{(N-1)(\zeta_3+3S_3)}
                                    {6N^2(N+1)^2}
                               \frac{(N^3+2N+1) (\zeta_2+2S_2)}
                                    {4N^3(N+1)^3}
\N\\ &&
                              +\frac{N^5-7N^4+6N^3+7N^2+4N+1}
                                    {4N^5(N+1)^5}~. \label{aQq2PStbar}
    \end{eqnarray}
    Here, we agree up to $O(\ep^0)$ with Ref.~\cite{Buza:1996xr} and 
    Eq.~(\ref{aQq2PStbar}) is a new result.
    Note, that Eq.~(\ref{splitpolPS}) is already the
    $\overline{\sf MS}$ anomalous
    dimension as obtained in 
    Refs.~\cite{Mertig:1995ny,Vogelsang:1995vh,*Vogelsang:1996im}. 
    Therefore any additional scheme dependence due to $\gamma_5$ can 
    only be contained in the higher order terms in $\ep$. As a comparison
    the anomalous dimension
    $\Delta \hat{\gamma}_{qq}^{\prime (1), {\sf PS}}$
    which is obtained by calculating $C(\ep)$ in $D$ dimensions, is related 
    to the $\overline{\sf MS}$ one by
    \begin{eqnarray}
      \label{eq43}
      \Delta\hat{\gamma}_{qq}^{(1), {\sf PS}} &=&
      \Delta\hat{\gamma}_{qq}^{\prime (1), {\sf PS}}
        -T_FC_F\frac{16(N-1)^2}{3N^2(N+1)^2}~.
    \end{eqnarray}
    The renormalized result becomes
    \begin{eqnarray}
     \Delta A_{Qq}^{(2), {\sf PS}}&=&
        -\frac{\Delta\hat{\gamma}_{qg}^{(0)}\Delta\gamma_{gq}^{(0)}}{8}
         \ln^2\left(\frac{m^2}{\mu^2}\right)
       +\frac{\Delta\hat{\gamma}_{qq}^{(1), {\sf PS}}}{2}
             \ln\left(\frac{m^2}{\mu^2}\right)
       +\Delta a_{Qq}^{(2), {\sf PS}}
       +\frac{\Delta \hat{\gamma}_{qg}^{(0)} \Delta \gamma_{gq}^{(0)}}{8}
        \zeta_2 ~. \N\\
       \label{eqA2}
    \end{eqnarray}
    The results in this Section constitute 
    a partial step towards the calculation of the
    asymptotic heavy flavor contributions at $O(a_s^2)$ in the 
    ${\sf \overline{MS}}$--scheme, thereby going beyond the results of 
    Ref.~\cite{Buza:1996xr}. The same holds for the $O(a_s^2\ep)$--terms, 
    which we calculated for the first time, using the same description for 
    $\gamma_5$ as has been done in \cite{Buza:1996xr}.
    The correct finite renormalization to transform to the
    ${\sf \overline{MS}}$--scheme remains to be worked out and will be presented
    elsewhere, \cite{Bierenbaum:prep1}.
%%%%%%%%%%%%%%%%%%%%%%%%%%%%%%%%%%%%%%%%%%%%%%%%%%%%%%%%%%%%%%%%%%%%%%%%
%%%%%%%%%%%%%%%%%%%%%%%%%%%%%%%%%%%%%%%%%%%%%%%%%%%%%%%%%%%%%%%%%%%%%%%%
%
% Chapter 9
%
% Heavy Flavor Contributions to Transversity
%
%%%%%%%%%%%%%%%%%%%%%%%%%%%%%%%%%%%%%%%%%%%%%%%%%%%%%%%%%%%%%%%%%%%%%%%%
%%%%%%%%%%%%%%%%%%%%%%%%%%%%%%%%%%%%%%%%%%%%%%%%%%%%%%%%%%%%%%%%%%%%%%%%
\newpage
 \section{Heavy Flavor Contributions to Transversity}
  \label{sec-1}
  \renewcommand{\theequation}{\thesection.\arabic{equation}}
  \setcounter{equation}{0}
%%%%%%%%%%%%%%%%%%%%%%%%%%%%%%%%%%%%%%%%%%%%%%%%%%%%%%%%%%%%%%%%%%%%%%%%
   The transversity distribution $\Delta_T f(x,Q^2)$ is one of the three 
   possible quarkonic twist-2 parton distributions besides the unpolarized 
   and the longitudinally polarized quark distribution, 
   $f(x,Q^2)$ and $\Delta f(x,Q^2)$, respectively. Unlike the latter 
   distribution functions, it cannot be measured in inclusive deeply inelastic 
   scattering in case of massless partons since it is chirally odd. However, it
   can be extracted from semi--inclusive deep-inelastic scattering (SIDIS) 
   studying isolated meson production, \cite{Ralston:1979ys,*Jaffe:1991kpxJaffe:1991ra,Cortes:1991ja}, 
   and in the Drell-Yan process,~\cite{Artru:1989zv,Cortes:1991ja,Collins:1992kk,*Jaffe:1993xb,*Tangerman:1994bb,*Boer:1997nt}, off transversely polarized 
   targets~\footnote{For a review see Ref.~\cite{Barone:2001sp}.}. Measurements 
   of the transversity distribution in different polarized hard scattering 
   processes are currently performed or in preparation, \cite{Airapetian:2004twxAirapetian:2008sk,*Alexakhin:2005iw,*Afanasev:2007qh,*:2008dn,*Lutz:2009ff}. 
   In the past, phenomenological models for the transversity distribution were 
   developed based on bag-like models, chiral models, light--cone models, 
   spectator models, and non-perturbative QCD calculations, cf. Section~8 of Ref.~\cite{Barone:2001sp}.
   The main characteristics of the transversity distributions are that they 
   vanish by some power law both at small and large values of Bjorken--$x$
   and exhibit a
   shifted bell-like shape. Recent attempts to extract the distribution out of 
   data were made in Refs.~\cite{Anselmino:2007fs,*Anselmino:2008jk}. The 
   moments of the transversity distribution can be measured in lattice 
   simulations, which help to constrain it ab initio, where first results were 
   given in Refs.~\cite{Aoki:1996pi,*Gockeler:1996es,*Khan:2004vw,*Diehl:2005ev,*Gockeler:2006zu,*Renner:private1,Dolgov:2002zm}. From these investigations 
   there is evidence, that the up-quark distribution is positive while the 
   down-quark distribution is negative, with first moments between 
   $\{0.85 \ldots 1.0\}$ and $\{-0.20 \ldots -0.24\}$, respectively. This is in
   qualitative agreement with phenomenological fits. 

   Some of the processes which have been proposed to measure transversity
   contain $k_\perp-$ and higher twist effects, cf.~\cite{Barone:2001sp}. We 
   will limit our considerations
   to the class of purely twist--2 contributions, 
   for which the formalism to calculate the heavy flavor corrections is 
   established, cf. Section~\ref{Sec-HQDIS}. As for the unpolarized flavor
   non--singlet contributions, we apply the factorization relation of
   the heavy flavor Wilson coefficient (\ref{LNSFAC}) in the region 
   $Q^2 \gg m^2$. 

   As physical processes one may consider the SIDIS process  
   $l N \rightarrow l' h + X$ off transversely polarized targets in which the 
   transverse momentum of the produced final state hadron $h$ is integrated. 
   The differential scattering cross section in case of single photon 
   exchange reads
   \begin{eqnarray}
    \label{sidis1}
      \frac{d^3\sigma}{dxdydz}&=&
         \frac{4\pi\alpha^2}{xyQ^2} \sum_{i=q,\overline{q}}e_i^2 x 
           \Biggl\{
                    \frac{1}{2}\Bigl[1+(1-y)^2\Bigr] F_i(x,Q^2) 
                    \tilde{D}_i(z,Q^2)
\N\\ &&
                   -(1-y)|{\bf S}_\perp||{\bf S}_{h\perp}| 
                    \cos\left(\phi_S + \phi_{S_h}\right)\Delta_T 
                    F_i(x,Q^2) \Delta_T \tilde{D}_i(z,Q^2)\Biggr\}~.
   \end{eqnarray}
   Here, in addition to the Bjorken variables $x$ and $y$, the
   fragmentation variable $z$ occurs.
    ${\bf S}_\perp$ and ${\bf S}_{h\perp}$ are the 
   transverse spin vectors of the incoming nucleon $N$ and the measured hadron 
   $h$. The angles $\phi_{S,S_h}$ are measured in the plane perpendicular to
   the $\gamma^* N$ (z--) axis between the $x$-axis and the respective vector.
   The transversity distribution can be obtained from Eq.~(\ref{sidis1}) for a 
   {\sf transversely} polarized hadron $h$ by measuring its polarization. 
   The functions $F_i, \tilde{D}_i, \Delta_T F_i, \Delta_T \tilde{D}_i$ are
   given by
   \begin{eqnarray}
    F_i(x,Q^2) &=& {\cal C}_i(x,Q^2) \otimes f_i(x,Q^2) \\
    \tilde{D}_i(z,Q^2)  &=& \tilde{{\cal C}}_i(z,Q^2) \otimes {D}_i(z,Q^2) \\
    \Delta_T F_i(x,Q^2) &=& \Delta_T {\cal C}_i(x,Q^2) \otimes 
                            \Delta_T f_i(x,Q^2)\\
    \Delta_T \tilde{D}_i(z,Q^2)  &=& \Delta_T \tilde{{\cal C}}_i(z,Q^2) 
                                     \otimes \Delta_T {D}_i(z,Q^2)~.
   \end{eqnarray}
   Here, $D_i, \Delta_T {D}_i$ are the fragmentation functions and 
   $\tilde{{\cal C}}_i,~\Delta_T {\cal C}_i,~\Delta_T \tilde{{\cal C}}_i$ 
   are the corresponding
   space- and time-like Wilson coefficients. 
   The functions
   ${\cal C}_i$ are the Wilson 
   coefficients as have been considered in the unpolarized case, cf. 
   Sections \ref{Sec-DIS} and \ref{Sec-HQDIS}. The 
   Wilson coefficient for transversity, $\Delta_T {\cal C}_i(x,Q^2)$,
   contains light-- and heavy flavor contributions,
   cf. Eq.~(\ref{Callsplit}),
   \begin{eqnarray}
    \Delta_T {\cal C}_i(x,Q^2) = \Delta_T C_i(x,Q^2) 
                    + \Delta_T H_i(x,Q^2)~.
   \end{eqnarray}
   $\Delta_T C_i$ denotes the light flavor transversity Wilson 
   coefficient and $\Delta_T H_i(x,Q^2)$ the heavy flavor part. We dropped
   arguments of the type $n_f,~m^2,~\mu^2$ for brevity, since they
   can all be inferred from the discussion in Section~\ref{Sec-HQDIS}.

   Eq.~(\ref{sidis1}) holds for spin--$1/2$ hadrons in the final state, but
   the transversity distribution may also be measured in the leptoproduction
   process of spin--1 hadrons, \cite{Ji:1993vw}. In this case, the 
   ${\bf P}_{h \perp}$-integrated Born cross section reads 
   \begin{eqnarray}
     \frac{d^3\sigma}{dxdydz}&=&\frac{4\pi\alpha^2}{xyQ^2} 
         \sin\left(\phi_S + \phi_{S_{LT}}\right)
             |{\bf S}_\perp||{S}_{LT}| (1-y)
            \sum_{i =q,\overline{q}} e_i^2 
                x \Delta_T F_i(x,Q^2) \widehat{H}_{i,1,LT}(z,Q^2)~.\N\\
    \label{sidis2}
   \end{eqnarray}
   Here, the polarization state of a spin--1 particle is described by a tensor 
   with five independent components, \cite{Bacchetta:2000jk}. $\phi_{LT}$ 
   denotes the azimuthal angle of $\vec{S}_{LT}$, with 
   \begin{eqnarray}
    |S_{LT}| = \sqrt{\left(S_{LT}^x\right)^2 +\left(S_{LT}^y\right)^2}~. 
   \end{eqnarray}
   $\widehat{H}_{a,1,LT}(z,Q^2)$ is a $T$- and chirally odd twist-2 
   fragmentation function at vanishing $k_\perp$. The process (\ref{sidis2}) 
   has the advantage that the transverse polarization of the produced hadron 
   can be measured from its decay products. 

   The transversity distribution can also be measured in the transversely 
   polarized Drell--Yan process, see Refs.~\cite{Vogelsang:1992jn,Vogelsang:1997ak,Shimizu:2005fp}. However, the SIDIS processes have the advantage that in 
   high luminosity experiments the heavy flavor contributions can be tagged 
   like in deep-inelastic scattering. This is 
   not the case for the Drell-Yan process, where the heavy flavor effects
   appear as inclusive radiative corrections in the Wilson coefficients only.

   The same argument as in Section~\ref{SubSec-HQAsym} can be applied 
   to obtain the heavy flavor Wilson coefficients for transversity in 
   the asymptotic limit $Q^2\gg~m^2$. Since transversity is a ${\sf NS}$
   quantity,
   the relation is the same as in the unpolarized ${\sf NS}$ case and
   reads up to $O(a_s^3)$, cf. Eq.~(\ref{eqWIL1}),
   \begin{eqnarray}
     \Delta_T H^{\mbox{\small Asym}}_q(n_f+1)=&&
                 a_s^2\Bigl[ \Delta_T A_{qq,Q}^{(2),\sf NS}(n_f+1) 
                             +\Delta_T \hat{C}_q^{(2)}(n_f)
                      \Bigr]
\N \\
                &+&a_s^3\Bigl[ 
                              \Delta_T A_{qq,Q}^{(3),\sf NS}(n_f+1)
                             +\Delta_T A_{qq,Q}^{(2),\sf NS}(n_f+1)
                              \Delta_T C_q^{(1)}(n_f+1)
\N\\ && \hspace{5mm}
                             +\Delta_T\hat{C}_q^{(3)}(n_f)\Bigr]~.
    \label{HwcofTR}
   \end{eqnarray}
   The operator matrix elements $\Delta_T A_{qq,Q}^{(2,3),\sf NS}$ are --
   as in the unpolarized case -- universal and account for all mass
   contributions but power corrections. 
   The respective asymptotic heavy flavor Wilson 
   coefficients are obtained in combination with the light flavor 
   process--dependent Wilson coefficients~\footnote{Apparently, the light flavor Wilson coefficients for SIDIS were not yet calculated even at $O(a_s)$, although this calculation and the corresponding soft-exponentiation should be straightforward. For the transversely polarized Drell-Yan process the $O(a_s)$ light flavor Wilson coefficients were given in \cite{Vogelsang:1997ak} and higher order terms due to soft resummation were derived in \cite{Shimizu:2005fp}.}.
  In the following, we will concentrate on the calculation of the massive
  operator matrix 
  elements. The twist--$2$ local operator in case of transversity has a 
  different Lorentz--structure compared to Eqs. (\ref{COMP1})--(\ref{COMP3}) and 
  is given by 
  \begin{eqnarray}
   \label{op2} 
    O_{q,r;\mu, \mu_1, \ldots, \mu_N}^{\sf TR,NS} &=& 
                i^{N-1}  {\bf S} 
                         [\overline{\psi} \sigma^{\mu \mu_1} D^{\mu_2} 
                                 \ldots D^{\mu_N} \frac{\lambda_r}{2}
                                 \psi] - {\rm trace~terms}~,
  \end{eqnarray}
  with $\sigma^{\nu\mu} = (i/2)\left[\gamma_\nu \gamma_\mu - \gamma_\mu \gamma_\nu \right]$ and the definition of the massive operator matrix element 
  is the same as in Section~\ref{SubSec-HQAsym}. 
  Since (\ref{op2}) denotes a twist--2 flavor non--singlet operator, it does 
  not mix with other operators.
  After multiplying with the external source $J_N$, cf. Eq.~(\ref{Jsource}) and
  below, the Green's function in momentum space corresponding to the 
  transversity operator between quarkonic states is given by
  \begin{eqnarray}
  \overline{u}(p,s) G^{ij, {\sf TR,NS}}_{\mu,q,Q} \lambda_r u(p,s)&=&
   J_N\bra{\overline{\Psi}_i(p)}O_{q,r;\mu, \mu_1, \ldots, \mu_N}^{{\sf TR,NS}}
        \ket{\Psi^j(p)}_Q~\label{GijTRNS}~. 
  \end{eqnarray}
  It relates to the unrenormalized transversity OME via
  \begin{eqnarray}
   \hat{G}^{ij, {\sf TR,NS}}_{\mu,q,Q}&=&
            \delta_{ij}(\Delta \cdot p)^{N-1} 
             \Bigl(
                   \Delta_{\rho}\sigma^{\mu\rho}
                   \Delta_T~\Ahathat_{qq,Q}^{\sf NS}
                        \Bigl(\frac{\hat{m}^2}{\mu^2},\ep,N\Bigr) 
                  +c_1 \Delta^\mu + c_2 p^\mu 
                  +c_3 \gamma^\mu p \hspace*{-2mm} / 
\N\\ && \hspace{30mm}
                  +c_4 \Delta \hspace*{-3mm}/~p \hspace*{-2mm}/ \Delta^\mu
                  +c_5 \Delta \hspace*{-3mm}/~p \hspace*{-2mm}/ p^\mu
             \Bigr)~.
  \end{eqnarray}
  The Feynman rules for the operators multiplied with the external source are 
  given in Appendix \ref{App-FeynRules}. The projection onto the massive 
  OME is found to be
  \begin{eqnarray}
    \label{eqc3}
     \Delta_T~\Ahathat_{qq,Q}^{\sf NS}\Bigl(\frac{\hat{m}^2}{\mu^2},\ep,N\Bigr)
     &=& 
         - \frac{i\delta^{ij}}{4N_c(\Delta.p)^2 (D-2)}
           \Bigl\{
     {\sf Tr}[ \Delta\hspace*{-3mm}/~p\hspace*{-2mm}/~
               p^{\mu}\hat{G}^{ij, {\sf TR,NS}}_{\mu,q,Q}]
    -\Delta.p {\sf Tr}[p^{\mu}\hat{G}^{ij, {\sf TR,NS}}_{\mu,q,Q}] 
\N\\ && \hspace{40mm}
    +i\Delta.p {\sf Tr}[\sigma_{\mu \rho} p^\rho 
                        \hat{G}^{ij, {\sf TR,NS}}_{\mu,q,Q}]
          \Bigr\}~.
  \end{eqnarray}
  Renormalization for transversity proceeds in the same manner as in the 
  ${\sf NS}$--case. The structure of the unrenormalized expressions at the
  $2$-- and $3$--loop level are the same as shown in 
  Eqs.~(\ref{Ahhhqq2NSQ},~\ref{Ahhhqq3NSQ}), if one inserts the respective 
  transversity anomalous dimensions. The expansion coefficients of the 
  renormalized OME then read up to $O(a_s^3)$ in the
  $\overline{\sf MS}$--scheme, 
  cf. Eqs.~(\ref{Aqq2NSQMSren},~\ref{Aqq3NSQMSren}), 
  \begin{eqnarray}
     \Delta_T A_{qq,Q}^{(2),{\sf NS, \MS}}&=&
                  \frac{\beta_{0,Q}\gamma_{qq}^{(0),\sf TR}}{4}
                    \ln^2 \Bigl(\frac{m^2}{\mu^2}\Bigr)
                 +\frac{\hat{\gamma}_{qq}^{(1), {\sf TR}}}{2}
                    \ln \Bigl(\frac{m^2}{\mu^2}\Bigr)
                 +a_{qq,Q}^{(2),{\sf TR}}
                 -\frac{\beta_{0,Q}\gamma_{qq}^{(0),\sf TR}}{4}\zeta_2~, \N \\
                  \label{Aqq2NSTRQMSren} \\
%%%
    \Delta_T    A_{qq,Q}^{(3),{\sf NS}, \MS}&=&
     -\frac{\gamma_{qq}^{(0),{\sf TR}}\beta_{0,Q}}{6}
          \Bigl(
                 \beta_0
                +2\beta_{0,Q}
          \Bigr)
             \ln^3 \Bigl(\frac{m^2}{\mu^2}\Bigr)
         +\frac{1}{4}
          \Biggl\{
                   2\gamma_{qq}^{(1),{\sf TR}}\beta_{0,Q}\N
\end{eqnarray}
\begin{eqnarray}
%\N\\ 
&&
                  -2\hat{\gamma}_{qq}^{(1),{\sf TR}}
                             \Bigl(
                                    \beta_0
                                   +\beta_{0,Q}
                             \Bigr)
                  +\beta_{1,Q}\gamma_{qq}^{(0),{\sf TR}}
          \Biggr\}
             \ln^2 \Bigl(\frac{m^2}{\mu^2}\Bigr)
         +\frac{1}{2}
          \Biggl\{
                   \hat{\gamma}_{qq}^{(2),{\sf TR}}
\N\\ &&
                  -\Bigl(
                           4a_{qq,Q}^{(2),{\sf TR}}
                          -\zeta_2\beta_{0,Q}\gamma_{qq}^{(0),{\sf TR}}
                                    \Bigr)(\beta_0+\beta_{0,Q})
                  +\gamma_{qq}^{(0),{\sf TR}}\beta_{1,Q}^{(1)}
          \Biggr\}
             \ln \Bigl(\frac{m^2}{\mu^2}\Bigr)
\N\\&&
         +4\overline{a}_{qq,Q}^{(2),{\sf TR}}(\beta_0+\beta_{0,Q})
         -\gamma_{qq}^{(0)}\beta_{1,Q}^{(2)}
         -\frac{\gamma_{qq}^{(0),{\sf TR}}\beta_0\beta_{0,Q}\zeta_3}{6}
         -\frac{\gamma_{qq}^{(1),{\sf TR}}\beta_{0,Q}\zeta_2}{4}
\N\\ \N \\&&
         +2 \delta m_1^{(1)} \beta_{0,Q} \gamma_{qq}^{(0),{\sf TR}}
         +\delta m_1^{(0)} \hat{\gamma}_{qq}^{(1),{\sf TR}}
         +2 \delta m_1^{(-1)} a_{qq,Q}^{(2),{\sf TR}}
         +a_{qq,Q}^{(3),{\sf TR}}~. \label{Aqq3NSTRQMSren}
  \end{eqnarray}
  Here, $\gamma_{qq}^{(k),{\sf TR}},~\{k = 0, 1, 2\}$, denote the 
  transversity quark anomalous dimensions at $O(a_s^{k+1})$ and 
  $a_{qq,Q}^{(2,3),{\sf TR}}, \overline{a}_{qq,Q}^{(2),{\sf TR}}$ are the 
  constant and $O(\ep)$ terms of the massive operator matrix element
  at 2-- and 3--loop order, respectively, cf. the discussion in 
  Section~\ref{Sec-REN}. 
  At ${\sf LO}$ the transversity anomalous dimension was 
  calculated in~\cite{Baldracchini:1980uq,*Shifman:1980dk,*Bukhvostov:1985rn,*Mukherjee:2001zx,Artru:1989zv,Blumlein:2001ca}~\footnote{The small $x$ limit of the 
${\sf LO}$ anomalous dimension was calculated in \cite{Kirschner:1996jj}.}, and
  at ${\sf NLO}$ in~\cite{Hayashigaki:1997dn,*Kumano:1997qp,Vogelsang:1997ak}~\footnote{For calculations in the non-forward case, see 
\cite{Belitsky:1997rh,*Hoodbhoy:1998vm,*Belitsky:2000yn,Blumlein:2001ca}.}. 
  At three-loop order the moments $N=1 \ldots 8$ are known, \cite{Gracey:2003yrxGracey:2003mrxGracey:2006zrxGracey:2006ah}.

  The 2--loop calculation for all $N$ proceeds in the same way as described in 
  Section~\ref{Sec-2L}. We also calculated the unprojected Green's 
  function to check the projector (\ref{eqc3}). 
  Fixed moments at the $2$-- and $3$--loop level were calculated using 
  ${\sf MATAD}$ as described in Section~\ref{Sec-3L}.
  From the pole terms of the unrenormalized $2$--loop OMEs, the leading and 
  next-to-leading order anomalous dimensions $\gamma_{qq}^{(0),{\sf TR}}$
  and $\hat{\gamma}_{qq}^{(1),{\sf TR}}$ can be determined.
  We obtain 
  \begin{eqnarray}
    \gamma_{qq}^{(0),{\sf TR}} = 2 C_F \left[ -3 + 4 S_1\right],
   \label{gqqTR0}
  \end{eqnarray}
  and
  \begin{eqnarray}
   \hat{\gamma}_{qq}^{(1), {\sf TR}} = \frac{32}{9} C_F T_F \left[ 3 S_2
                                       - 5 S_1 + \frac{3}{8} \right]~,
   \label{gqqTR1}
  \end{eqnarray}
  confirming earlier results, 
  \cite{Hayashigaki:1997dn,*Kumano:1997qp,Vogelsang:1997ak}.
  The finite and $O(\ep)$ contributions are given by
  \begin{eqnarray}
   a_{qq,Q}^{(2), {\sf TR}} &=& C_F T_F \Biggl\{ 
    -\frac{8}{3}    S_3
    +\frac{40}{9}   S_2
    -\left[ \frac{224}{27} 
           +\frac{8}{3}\zeta_2 \right] S_1
    +2 \zeta_2
    +{\frac{ \left( 24+73\,N+73\,{N}^{2} \right)}{18 N \left( N+1 \right) }}
    \Biggr\}~,  \N\\  \label{aqqTR2} \\
   \overline{a}_{qq,Q}^{(2), {\sf TR}} &=& C_F T_F \Biggl\{
      - \left[
               {\frac {656}{81}}\, 
              +{\frac {20}{9}}\, \zeta_2
              +{\frac {8}{9}}\, \zeta_3 \right] S_1
        +\left[{\frac {112}{27}}\, +\frac{4}{3}\, \zeta_2 \right] S_2
        -{\frac {20}{9}}\, S_3
         \N\\ &&
         +\frac{4}{3}\, S_4
         +\frac{1}{6}\, \zeta_2
         +\frac{2}{3}\, \zeta_3
      +\frac{ 
        \left( -144-48\,N+757\,{N}^{2}+1034\,{N}^{3}+517\,{N}^{4} \right) }
              {216 {N}^{2} \left( N+1 \right) ^{2}}
        \Biggr\}~. \N\\
            \label{aqqTR2bar}
  \end{eqnarray}
  The renormalized $2$--loop massive OME (\ref{Aqq2NSTRQMSren}) reads
  \begin{eqnarray}
   \Delta_T A_{qq,Q}^{(2),{\sf NS, \MS}} &=& C_F T_F \Biggl\{\left[
    -\frac{8}{3} S_1 + 2 \right] \ln^2\left(\frac{m^2}{\mu^2}\right)
    + \left[-{\frac {80}{9}} S_1 + \frac{2}{3} +\frac{16}{3} S_2
    \right] \ln\left(\frac{m^2}{\mu^2}\right)
    \N\\ && \hspace*{1cm}
    - \frac{8}{3}    S_3
    + \frac{40}{9}   S_2 
    - \frac{224}{27} S_1
    + \frac {24+73N+73{N}^{2}}{18 N \left( N+1\right) }    
     \Biggr\}~.  
  \end{eqnarray}
  Using ${\sf MATAD}$, we calculated the moments $N=1\ldots 13$ at $O(a_s^2)$ 
  and $O(a_s^3)$. At the $2$--loop level, we find complete agreement 
  with the results presented in Eqs. (\ref{gqqTR0})--(\ref{aqqTR2bar}).
  At $O(a_s^3)$, we also obtain $\hat{\gamma}_{qq}^{(2),{\sf TR}}$, 
  which can be compared to the $T_F$-terms in the calculation
  \cite{Gracey:2003yrxGracey:2003mrxGracey:2006zrxGracey:2006ah} 
  for $N = 1 ...8$. This is the first re-calculation of these terms and we find
  agreement. For the moments $N = 9...13$ this contribution
  to the transversity anomalous dimension 
  is calculated for 
  the first time. We list the anomalous dimensions in 
  Appendix~\ref{App-Trans}. There, also the constant contributions 
  $a_{qq,Q}^{(3),{\sf TR}}$ are given for $N=1\ldots 13$, which is a new 
  result.
  Furthermore, we obtain in the 
  $3$--loop calculation the moments $N = 1...13$ of the complete 2--loop
  anomalous dimensions. These are in accordance with Refs.~\cite{Vogelsang:1997ak,Hayashigaki:1997dn,*Kumano:1997qp}. 

  Finally, we show as examples the first moments of the
  $\overline{\sf MS}$--renormalized 
  $O(a_s^3)$ massive transversity OME. Unlike the case for
  the vector current, the first moment does not vanish, since there is no
  conservation law to enforce this.
   \begin{eqnarray}
    \Delta_T~A_{qq,Q}^{(3),{\sf NS},\MS}(1)&=&C_FT_F\Biggl\{
        \Bigl(
            \frac{16}{27}T_F(1-n_f)
           +\frac{44}{27}C_A
        \Bigr)\ln^3\Bigl(\frac{m^2}{\mu^2}\Bigr)
       +\Bigl(
            \frac{104}{27}T_F
\N\\&&\hspace{-25mm}
           -\frac{106}{9}C_A
           +\frac{32}{3}C_F
        \Bigr)\ln^2\Bigl(\frac{m^2}{\mu^2}\Bigr)
       +\Biggl[
           -\frac{604}{81}n_fT_F
           -\frac{4}{3}T_F
           +\Bigl(
                  -\frac{2233}{81}
                  -16\zeta_3
            \Bigr)C_A
\N\\&&\hspace{-25mm}
           +\Bigl(
                   16\zeta_3
                  +\frac{233}{9}
            \Bigr)C_F
        \Biggr]\ln\Bigl(\frac{m^2}{\mu^2}\Bigr)
       +\Bigl(
              -\frac{6556}{729}
              +\frac{128}{27}\zeta_3
        \Bigr)T_Fn_f
\N\\&&\hspace{-25mm}
       +\Bigl(
               \frac{2746}{729}
              -\frac{224}{27}\zeta_3
        \Bigr)T_F
        +\Bigl(
                \frac{8}{3}B_4
               +\frac{437}{27}\zeta_3
               -24\zeta_4
               -\frac{34135}{729}
         \Bigr)C_A
\N\\&&\hspace{-25mm}
        +\Bigl(
               -\frac{16}{3}B_4
               +24\zeta_4
               -\frac{278}{9}\zeta_3
               +\frac{7511}{81}
         \Bigr)C_F
             \Biggr\}~,\\
%%%%%
    \Delta_T~A_{qq,Q}^{(3),{\sf NS}, \MS}(2)&=&C_FT_F\Biggl\{
        \Bigl(
               \frac{16}{9}T_F(1-n_f)
              +\frac{44}{9}C_A
        \Bigr)\ln^3\Bigl(\frac{m^2}{\mu^2}\Bigr)
       +\Bigl(
              -\frac{34}{3}C_A
\N\\&&\hspace{-25mm}
              +8T_F
        \Bigr)\ln^2\Bigl(\frac{m^2}{\mu^2}\Bigr)
       +\Bigl[
              -\frac{196}{9}n_fT_F
              -\frac{92}{27}T_F
            +\Bigl(
                   -48\zeta_3
                   -\frac{73}{9}
             \Bigr)C_A
            +\Bigl(
                    48\zeta_3
\N\\&&\hspace{-25mm}
                   +15
             \Bigr)C_F
        \Bigr]\ln\Bigl(\frac{m^2}{\mu^2}\Bigr)
      +\Bigl(
              \frac{128}{9}\zeta_3
             -\frac{1988}{81}
       \Bigr)T_Fn_f
      +\Bigl(
              \frac{338}{27}
             -\frac{224}{9}\zeta_3
       \Bigr)T_F
      +\Bigl(
             -56
\N\\&&\hspace{-25mm}
             -72\zeta_4
             +8B_4
             +\frac{533}{9}\zeta_3
       \Bigr)C_A
      +\Bigl(
             -16B_4
             +\frac{4133}{27}
             +72\zeta_4
             -\frac{310}{3}\zeta_3
       \Bigr)C_F
                    \Biggr\}~.
\end{eqnarray}
  The structure of the result and the contributing numbers are the same as in 
  the unpolarized case, cf. Eq.~(\ref{Aqq3NSQN2MSON}). We checked the moments 
  $N=1\ldots 4$ keeping the complete dependence on the gauge--parameter $\xi$
  and find that it cancels in the final result. Again, we observe that 
  the massive OMEs do not depend on $\zeta_2$, cf. 
  Section~\ref{SubSec-3LResUn}.

  Since the light flavor Wilson coefficients for the processes from which the 
  transversity distribution can be extracted are not known 
  to 2-- and 3--loop order, phenomenological studies on  the effect of the 
  heavy flavor contributions cannot yet be performed. However, the
  results of this Section can be used in comparisons with 
  upcoming lattice simulations with (2+1+1)-dynamical fermions including the 
  charm quark. More details on this calculation are given in~\cite{Blumlein:trans}.
%%%%%%%%%%%%%%%%%%%%%%%%%%%%%%%%%%%%%%%%%%%%%%%%%%%%%%%%%%%%%%%%%%%%%%%%%%%%%%%
%%%%%%%%%%%%%%%%%%%%%%%%%%%%%%%%%%%%%%%%%%%%%%%%%%%%%%%%%%%%%%%%%%%%%%%%%%%%%%%
%
% Chapter 10
%
% Towards a full 3--Loop Calculation
%
%%%%%%%%%%%%%%%%%%%%%%%%%%%%%%%%%%%%%%%%%%%%%%%%%%%%%%%%%%%%%%%%%%%%%%%%%%%%%%%
%%%%%%%%%%%%%%%%%%%%%%%%%%%%%%%%%%%%%%%%%%%%%%%%%%%%%%%%%%%%%%%%%%%%%%%%%%%%%%%
\newpage
 \section{\bf\boldmath First Steps Towards a Calculation of
                       $A_{ij}^{(3)}$ for all Moments.}
  \label{Sec-FULL3L}
  \renewcommand{\theequation}{\thesection.\arabic{equation}}
  \setcounter{equation}{0}
%%%%%%%%%%%%%%%%%%%%%%%%%%%%%%%%%%%%%%%%%%%%%%%%%%%%%%%%%%%%%%%%%%%%%%%%%%%%%%%
   In Section \ref{Sec-3L}, we described how the various massive OMEs
   are calculated for fixed integer values of the Mellin variable $N$ 
   at $3$--loop order using ${\sf MATAD}$. The ultimate goal is to 
   calculate these quantities for general values of $N$. So far no massive 
   {\sf single scale calculation} at $O(a_s^3)$ has been performed. In the 
   following we would like to present some first results and a general method,
   which may be of use in later work calculating the general $N$--dependence 
   of the massive OMEs $A_{ij}^{(3)}$. 

   In Section~\ref{Sec-FULL3LF1}, we solve, as an example, a $3$--loop ladder 
   graph contributing to $A_{Qg}^{(3)}$ for general values of $N$ by direct 
   integration, avoiding the integration--by--parts method.
   In Section~\ref{SubSec-FULL3LGuess}, 
   Ref.~\cite{Blumlein:2009tm,Blumlein:2009tj}, 
   we discuss a general algorithm which allows to determine from a sufficiently
   large but ${\sf finite}$ number of moments for a recurrent quantity its 
   general $N$--dependence. This algorithm has been successfully applied 
   in \cite{Blumlein:2009tj} to reconstruct the $3$--loop anomalous 
   dimensions, 
   \cite{Moch:2004pa,Vogt:2004mw}, and massless $3$--loop Wilson coefficients, 
   \cite{Vermaseren:2005qc}, from their moments.
   These are the largest single scale quantities known at the moment and are 
   well suited to demonstrate the power of this formalism. Similarly, one may
   apply this method to new problems of smaller size which emerge in course of 
   the calculation of the OMEs $A_{ij}^{(3)}$ for general values of $N$.
%%%%%%%%%%%%%%%%%%%%%%%%%%%%%%%%%%%%%%%%%%%%%%%%%%%%%%%%%%%%%%%%%%%%%%%%%%%%%%%
  \subsection{\bf\boldmath Results for all--$N$ Using Generalized
                       Hypergeometric Functions}
   \label{Sec-FULL3LF1}
%%%%%%%%%%%%%%%%%%%%%%%%%%%%%%%%%%%%%%%%%%%%%%%%%%%%%%%%%%%%%%%%%%%%%%%%%%%%%%%
    In Section \ref{SubSec-2LF32}, we showed that there is only one basic 
    $2$--loop massive tadpole which needs to be considered. From it, all
    diagrams contributing to the massive $2$--loop OMEs can be derived
    by attaching external quark--, gluon-- and ghost--lines, respectively,
    and including one operator insertion according to the
    Feynman rules given in Appendix~\ref{App-FeynRules}. 
    The corresponding parameter--integrals are then all of the same structure,
    Eq. (\ref{Gen2L}). If one knows
    a method to calculate the basic topology for arbitrary
    integer powers of the propagators, the calculation of the $2$--loop 
    OMEs is straightforward for fixed values of $N$. In the general 
    case, we arrived at infinite sums containing the 
    parameter $N$. To calculate these sums, additional tools
    are needed, e.g. the program \SigmaP, cf. Section~\ref{SubSec-2LInfSum}.
%%%%%%%%%%%%%%%
    \begin{figure}[H]
     \begin{center}
      \includegraphics[angle=0, height=3.7cm]{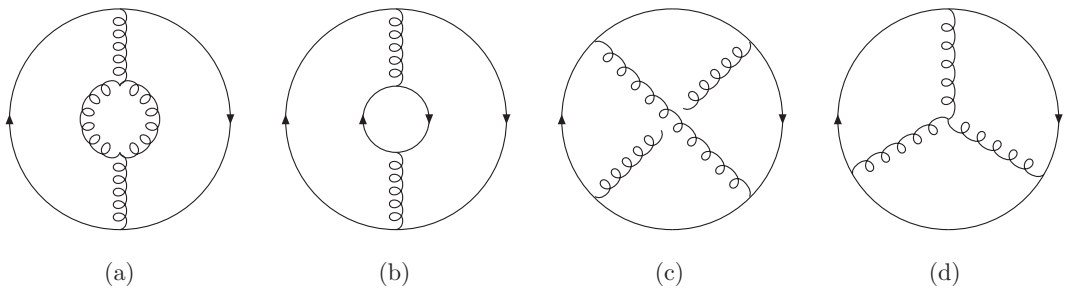}
     \end{center}
     \begin{center} 
      \caption[{\sf Basic $3$--loop topologies}]
     {\sf Basic $3$--loop topologies. Straight lines: quarks,
          curly lines: gluons. \\
          \phantom{Figurex17:y}The gluon loop in (a) can also be replaced 
          by a ghost loop.}
        \label{3La}
      \small
     \end{center}
     \normalsize
    \end{figure} 
%%%%%%%%%%%%%%%
    \noindent
    We would like to follow the same approach in the $3$--loop case.
    Here, five basic topologies need to considered, which are shown in 
    Figures~\ref{3La},~\ref{3Lb}.
    Diagram (a) and (b) --
    if one of the quark loops corresponds to a massless quark -- can 
    be reduced to $2$--loop integrals, 
    because the massless loop can be integrated trivially.
    For the remaining terms, this is not the case. 
    Diagrams (c) and (d) are the most complex topologies, the former giving
    rise to the number $B_4$, Eq. (\ref{B4}), whereas the latter
    yields single $\zeta$--values up to 
    weight $4$, cf. e.g. \cite{Broadhurst:1991fi}. Diagram (b) -- if 
    both quarks are massive -- and (e) are ladder topologies and of
    less complexity. Let us, as an example, consider diagram (e).
%%%%%%%%%%%%%%%
    \begin{figure}[H]
     \begin{center}
      \includegraphics[angle=0, height=3.3cm]{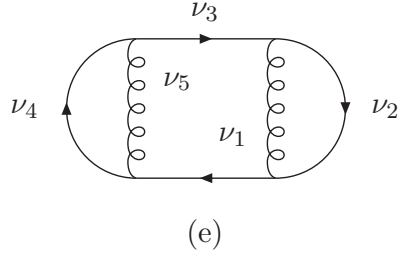}
     \end{center}
     \begin{center} 
      \caption[{\sf $3$--loop ladder graph }]
     {\sf $3$--loop ladder graph}
        \label{3Lb}
      \small
     \end{center}
     \normalsize
    \end{figure} 
%%%%%%%%%%%%%%%
    \noindent
    Our notation is the same as in Section \ref{SubSec-2LF32}. The scalar
    $D$--dimensional integral corresponding to diagram (e) reads for 
    arbitrary exponents of the propagators
    \begin{eqnarray}
     T_e&=&\iiint \frac{d^Dqd^Dkd^Dl}{(2\pi)^{3D}} 
                  \frac{i (-1)^{\nu_{12345}} (m^2)^{\nu_{12345}-3D/2}
                        (4\pi)^{3D/2}}
                        {
                         (k^2)^{\nu_1}((k-l)^2-m^2)^{\nu_2}
                         (l^2-m^2)^{\nu_3}
                         ((q-l)^2-m^2)^{\nu_4}(q^2)^{\nu_5}
                        }~.  \N\\
    \end{eqnarray}
    Again, we have attached suitable normalization factors for convenience. 
    After loop--by--loop integration
    of the momenta $k,q,l$ (in this order)
    using Feynman--parameters, one obtains after a few steps 
    the following parameter integral
    \begin{eqnarray}
     T_e&=&
           \Gamma\Biggl[\frac[0pt]{\nu_{12345}-6-3\ep/2}{\nu_1,\nu_2,\nu_3,\nu_4,\nu_5}\Biggr] 
\N\\ &&
           \int_0^1~dw_1\ldots \int_0^1~dw_4~\theta(1-w_1-w_2)
           \frac{w_1^{-3-\ep/2+\nu_{12}}w_2^{-3-\ep/2+\nu_{45}}(1-w_1-w_2)^{\nu_3-1}}
                {\D{(1+w_1\frac{w_3}{1-w_3}+w_2\frac{w_4}{1-w_4})^{\nu_{12345}-6-3\ep/2}}}
\N\\ \N\\&&
       \times
                  w_3^{1+\ep/2-\nu_1}(1-w_3)^{1+\ep/2-\nu_{2}}
                  w_4^{1+\ep/2-\nu_5}(1-w_4)^{1+\ep/2-\nu_{4}}~.
                  \label{bubble3}
    \end{eqnarray}
    The $\theta$--function enforces $w_1+w_2\le 1$.
    In order to perform the $\{w_1,~w_2\}$ integration, one considers
    \begin{eqnarray}
     I&=&\int_0^1~dw_1\int_0^1~dw_2~\theta(1-w_1-w_2)w_1^{b-1}w_2^{b'-1}
                          (1-w_1-w_2)^{c-b-b'-1}(1-w_1x-w_2y)^{-a}~.  \N\\
    \end{eqnarray}
    The parameters $a,b,b',c$ shall be such that this integral is convergent. 
    It can be expressed in terms of the Appell function $F_1$
    via, \cite{Slater}~\footnote{Note that Eq.~(8.2.2) of Ref.~\cite{Slater} contains typos.},
    \begin{eqnarray}
     I&=&\Gamma \Biggl[\frac[0pt]{b,b',c-b-b'}{c}\Biggr]
                    \sum_{m,n=0}^{\infty}
                     \frac{(a)_{m+n}(b)_n(b')_m}
                       {(1)_m(1)_n(c)_{m+n}}
                     x^ny^m~ \\
      &=& \Gamma
        \Biggl[\frac[0pt]{b,b',c-b-b'}{c}\Biggr]
            F_1\Bigl[a;b,b';c;x,y\Bigr]~.   \label{F1def}
    \end{eqnarray}
    The parameters $x,~y$ correspond to $w_3/(1-w_3)$ and $w_4/(1-w_4)$ 
    in Eq.~(\ref{bubble3}), respectively. Hence the integral over 
    these variables would yield a divergent sum. Therefore one uses the 
    following analytic continuation relation for $F_1$, \cite{Slater},
    \begin{eqnarray}
     F_1[a;b,b';c;\frac{x}{x-1},\frac{y}{y-1}] =
         (1-x)^b(1-y)^{b'}F_1[c-a;b,b';c;x,y]~.  \label{F1ancont}
    \end{eqnarray}
    Finally one arrives at an infinite double sum
    \begin{eqnarray}
     T_e&=&
           \Gamma\Biggl[\frac[0pt]{-2-\ep/2+\nu_{12},-2-\ep/2+\nu_{45},-6-3\ep/2+\nu_{12345}}{\nu_2,\nu_4,-4-\ep+\nu_{12345}}\Biggr] 
\N\\ && \times
           \sum_{m,n=0}^{\infty}
           \Gamma\Biggl[\frac[0pt]{2+m+\ep/2-\nu_1,2+n+\ep/2-\nu_5}{1+m,1+n,2+m+\ep/2,2+n+\ep/2}\Biggr] 
\N\\ && \times
           \frac{(2+\ep/2)_{n+m}(-2-\ep/2+\nu_{12})_m(-2-\ep/2+\nu_{45})_n}{(-4-\ep+\nu_{12345})_{n+m}}~. \label{bubble3Res}
    \end{eqnarray} 
    Here, we have adopted the notation for the $\Gamma$--function defined in 
    (\ref{gammashort}) and $(a)_b$ is Pochhammer's symbol,
    Eq. (\ref{pochhammer}).
    As one expects, Eq.~(\ref{bubble3Res}) is symmetric w.r.t. exchanges of
    the indices $\{\nu_1,~\nu_2\}~\leftrightarrow~\{\nu_4,\nu_5\}$. 
    For any values of $\nu_i$ of the type $\nu_i=a_i+b_i\ep$, with 
    $a_i\in~{\mathbb N},~b_i\in~{\mathbb C}$, the Laurent--series in $\ep$
    of Eq.~(\ref{bubble3Res}) can be calculated using e.g. ${\sf Summer}$, 
    \cite{Vermaseren:1998uu}. We have checked Eq.~(\ref{bubble3Res}) for 
    various values of the $\nu_i$ using ${\sf MATAD}$, cf. Section 
    \ref{SubSec-3LMatad}.

\begin{figure}[H]
     \begin{center}
      \includegraphics[angle=0, height=3.3cm]{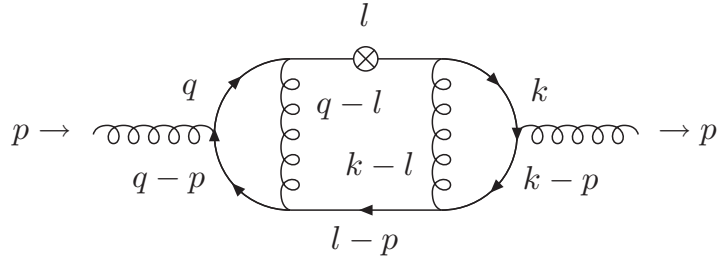}
     \end{center}
     \begin{center} 
      \caption[{\sf Example $3$--loop graph }]
     {\sf Example $3$--loop graph }
        \label{3Lc}
      \small
     \end{center}
     \normalsize
    \end{figure}

    Next, let us consider the diagram shown in Figure \ref{3Lc}, 
    which contributes to $A_{Qg}^{(3)}$ and derives from diagram (e).
    We treat the case where all exponents of the propagators are 
    equal to one.
%%%%%%%%%%%%%%%
%%%%%%%%%%%%%%%
    \noindent
    Including the factor $i(m^2)^{2-3\ep/2}(4\pi)^{3D/2}$ and 
    integrating $q,k,l$ (in this order), we obtain
    \begin{eqnarray}
     I_{ex}&=&\Gamma(2-3\ep/2)\int_0^1 dw_i~
              \theta(1-w_1-w_2)
              \frac{w_1^{-\ep/2} w_2^{-\ep/2} (1-w_1-w_2)}
                   {\D{(1+w_1\frac{1-w_3}{w_3}+w_2\frac{1-w_4}{w_4})^{2-3\ep/2}}} 
\N \\ \N \\&& \hspace{35mm}
          \times
              w_3^{-1+\ep/2}(1-w_3)^{\ep/2}
              w_4^{-1+\ep/2} (1-w_4)^{\ep/2}
\N \\ && \hspace{35mm}
          \times
              (1-w_5w_1-w_6w_2-(1-w_1-w_2)w_7)^N~, \label{IL1}
    \end{eqnarray}
    where all parameters $w_1\ldots w_7$ have to be integrated from 
    $0\ldots 1$.
    As in the $2$--loop case, (\ref{Gen2L}), one observes
    that the integral--kernel given by the corresponding 
    massive tadpole integral (\ref{bubble3}) is multiplied with a polynomial
    containing various integration parameters to the power $N$. The same 
    holds true for the remaining $3$--loop diagrams. Hence, if a general 
    sum representation for the corresponding tadpoles integrals is known and
    one knows how to evaluate the corresponding sums, at least fixed 
    moments of the $3$--loop massive OMEs can be calculated right away. 
    The presence of the polynomial to the power $N$ (which may also 
    involve a finite sum, cf. the Feynman--rules in 
    Appendix~\ref{App-FeynRules},) complicates the calculation further. One 
    has to find a suitable way to deal with this situation, which depends on 
    the integral
    considered. For $I_{ex}$, we split it up into several finite sums, 
    rendering the integrals calculable in the same way as for $T_e$.
    We obtain
    \begin{eqnarray}
     I_{ex}&=& \frac{-\Gamma(2-3\ep/2)}{(N+1)(N+2)(N+3)}\sum_{m,n=0}^{\infty}
         \Biggl\{
\N\\ && \hspace{5mm}
              \sum_{t=1}^{N+2}
              \binom{3+N}{t}
              \frac{(t-\ep/2)_m(2+N+\ep/2)_{n+m}(3-t+N-\ep/2)_n}
                   {(4+N-\ep)_{n+m}}
\N \\ && \hspace{10mm} \times
              \Gamma\Biggl[\frac[0pt]{t,t-\ep/2,1+m+\ep/2,1+n+\ep/2,3-t+N,3-t+N-\ep/2}{4+N-\ep,1+m,1+n,1+t+m+\ep/2,4-t+n+N+\ep/2}\Biggr]
\N\\ && \hspace{1mm}
             -\sum_{s=1}^{N+3}\sum_{r=1}^{s-1}
              \binom{s}{r}\binom{3+N}{s}(-1)^s
              \frac{(r-\ep/2)_m(-1+s+\ep/2)_{n+m}(s-r-\ep/2)_n}
                   {(1+s-\ep)_{n+m}}
\N\\ && \hspace{10mm} \times
              \Gamma\Biggl[\frac[0pt]{r,r-\ep/2,s-r,1+m+\ep/2,1+n+\ep/2,s-r-\ep/2}{1+m,1+n,1+r+m+\ep/2,1+s-r+n+\ep/2,1+s-\ep}\Biggr]
         \Biggr\}~. \N \\\label{IL2}
    \end{eqnarray}
    After expanding in $\ep$, the summation can be performed using \SigmaP~and 
    the summation techniques explained in 
    Section~\ref{SubSec-2LInfSum}. The result reads
    \begin{eqnarray}
      I_{ex}&=&
       -\frac{4(N+1){S_1}+4}{(N+1)^2(N+2)}{\zeta_3}
       +\frac{2{S_{2,1,1}}}{(N+2)(N+3)}
       +\frac{1}{(N+1)(N+2)(N+3)}\Biggl\{                        
\N\\ &&
                        -2(3N+5){S_{3,1}}
                        -\frac{{S_1}^4}{4}
                        +\frac{4(N+1){S_1}-4N}{N+1}{S_{2,1}}
                        +2\Bigl(
                           (2N+3){S_1}
                          +\frac{5N+6}{N+1}
                          \Bigr){S_3}
\N\\ &&
                        +\frac{9+4N}{4}{S_2}^2
                        +\Bigl(
                           2\frac{7N+11}{(N+1)(N+2)}
                          +\frac{5N}{N+1}{S_1}
                          -\frac{5}{2}{S_1}^2
                         \Bigr){S_2}
                        +\frac{2(3N+5){S_1}^2}{(N+1)(N+2)}
\N\\ &&
                        +\frac{N}{N+1}{S_1}^3
                        +\frac{4(2N+3){S_1}}{(N+1)^2(N+2)}
                        -\frac{(2N+3){S_4}}{2}
                        +8\frac{2N+3}{(N+1)^3(N+2)}
                               \Biggr\}
\N\\ &&                +O(\ep)~,
    \end{eqnarray}
    which agrees with the fixed moments $N=1\ldots 10$ obtained 
    using ${\sf MATAD}$, cf. Section \ref{SubSec-3LMatad}.

    We have shown that in principle one can be apply similar techniques as 
    on the $2$--loop level, Section \ref{SubSec-2LF32}, to calculate the 
    massive $3$--loop OMEs considering only the five basic topologies. In this 
    approach the integration-by-parts method is not used.
    We have given the necessary formulas for one non--trivial 
    topology $(e)$ and showed for on of the cases there how 
    the calculation proceeds
    keeping the all--$N$ dependence. In order to obtain complete results 
    for the massive OMEs, suitable integral representations for 
    diagrams (b), (c) and (d) of Figure~\ref{3La} have to be derived first.
    This will allow for a calculation of fixed moments not relying on 
    ${\sf MATAD}$. Next, an automatization of the step from 
    (\ref{IL1}) to (\ref{IL2}) has to be found in order to obtain 
    sums which can be handled e.g. by \SigmaP. The latter step is not trivial, 
    since it depends on the respective diagram and the flow of the 
    outer momentum $p$ through it.
%%%%%%%%%%%%%%%%%%%%%%%%%%%%%%%%%%%%%%%%%%%%%%%%%%%%%%%%%%%%%%%%%%%%%%%%%%%%%%%
  \subsection{\bf\boldmath Reconstructing General--$N$ Relations \\
                           from a Finite
                           Number of Mellin--Moments}
   \label{SubSec-FULL3LGuess}
%%%%%%%%%%%%%%%%%%%%%%%%%%%%%%%%%%%%%%%%%%%%%%%%%%%%%%%%%%%%%%%%%%%%%%%%%%%%%%%
    Higher order calculations in Quantum Field Theories easily become tedious 
    due to the large number of terms emerging and the sophisticated form of 
    the contributing Feynman parameter integrals. This applies already to 
    zero scale and single scale quantities. Even more this is the case for
    problems containing at least two scales. While in the latter case the
    mathematical structure of the solution of the Feynman integrals is widely 
    unknown, it is explored to a certain extent for zero-  and single scale
    quantities. Zero scale quantities emerge as the expansion coefficients of
    the running couplings and masses, as fixed moments of splitting functions,
    etc.. They can be expressed by rational numbers and certain special numbers
    as multiple zeta-values (MZVs), \cite{Borwein:1999js,Blumlein:2009Zet} and
    related quantities.

    Single scale quantities depend on a scale $z$ which may be given as a ratio
    of Lorentz invariants $s'/s$ in the respective physical problem. One may 
    perform a Mellin transform over $z$,  Eq.~(\ref{Mellintrans}). All
    subsequent calculations are then carried out in Mellin space and one
    assumes $N \in {\mathbb N},~N > 0$. By this transformation, the problem at
    hand becomes discrete and one may seek a description in terms of difference
    equations, \cite{Norlund,*Thomson}. 
    Zero scale problems are obtained from single scale problems 
    treating $N$ as a fixed integer or considering the limit 
    $N \rightarrow \infty$.

    A main question concerning zero scale quantities is: Do the corresponding
    Feynman integrals always lead to MZVs? In the lower orders this is the 
    case. However, starting at some order, even for single-mass problems, other
    special numbers will occur,
    \cite{Broadhurst:1998rz,Andre:2008,*Brown:2008um}. 
    Since one has to known the respective basis completely, 
    this makes it difficult
    to use methods like ${\sf PSLQ}$, \cite{Bailey:1991}, to determine the 
    analytic structure of the corresponding terms even if one may calculate 
    them numerically at high enough precision.
    Zero scale problems are much easier to calculate than single scale 
    problems. In some analogy to the determination of the analytic structure in
    zero scale problems through integer relations over a known basis 
    ({\sf PSLQ}) one may think of an automated reconstruction of the all--$N$ 
    relation out of a finite number of Mellin moments given in analytic form. 
    This is possible for recurrent quantities. At least up to 3-loop order, 
    presumably even to higher orders, single scale quantities belong to this 
    class. Here we report on a {\sf general algorithm}
    for this purpose, which we 
    applied to the problem being currently the most sophisticated one: 
    the anomalous dimensions and massless Wilson coefficients to 
    $3$--loop order for unpolarized DIS,
    \cite{Moch:2004pa,Vogt:2004mw,Vermaseren:2005qc}. Details of our 
    calculation are given in Refs.~\cite{Blumlein:2009tj,Blumlein:2009tm}.
%%%%%%%%%%%%%%%%%%%%%%%%%%%%%%%%%%%%%%%%%%%%%%%%%%%%%%%%%%%%%%%%%%%%%%%%%
    \subsubsection{Single Scale Feynman Integrals as Recurrent Quantities}
     \label{SubSec-FULL3LGuessFeyn}
%%%%%%%%%%%%%%%%%%%%%%%%%%%%%%%%%%%%%%%%%%%%%%%%%%%%%%%%%%%%%%%%%%%%%%%%%
      For a large variety of massless problems single scale Feynman integrals
      can be represented as polynomials in the ring formed by the nested 
      harmonic sums, cf. Appendix \ref{App-SpeFunHarm}, and the MZVs
      $\zeta_{a_1, \dots, a_l}$, which we set equal to the $\sigma$--values
      defined in Eq. (\ref{sigmaval}). Rational functions in $N$ and harmonic 
      sums obey recurrence relations. Thus, due to closure properties, 
      \cite{Salvy:1994,Mallinger:1996}, also any polynomial expression in such
      terms is a solution of a recurrence. Consider as an example the recursion
      \begin{eqnarray}
       F(N+1) - F(N) = \frac{{\rm sign}(a)^{N+1}}{(N+1)^{|a|}}~. 
      \end{eqnarray}
      It is solved by the harmonic sum $S_a(N)$.
      Corresponding difference equations hold for
      harmonic sums of deeper nestedness. 
      Feynman integrals can often be 
      decomposed into a combination containing terms of the  form
      \begin{eqnarray}
       \int_0^1 dz \frac{z^{N-1}-1}{1-z} H_{\vec{a}}(z),~~~~\int_0^1 dz
        \frac{(-z)^{N-1}-1}{1+z} H_{\vec{a}}(z)~, 
      \end{eqnarray}
      with $H_{\vec{a}}(z)$ being a harmonic polylogarithm, 
      \cite{Remiddi:1999ew}. This structure also leads to recurrences, 
      \cite{Blumlein:2008pa}. Therefore, it is
      very likely that single scale Feynman 
      diagrams do always obey difference equations.
%%%%%%%%%%%%%%%%%%%%%%%%%%%%%%%%%%%%%%%%%%%%%%%%%%%%%%%%%%%%%%%%%%%%%%%%%%%%%
    \subsubsection{Establishing and Solving Recurrences}
     \label{SubSec-FULL3LGuessRec}
%%%%%%%%%%%%%%%%%%%%%%%%%%%%%%%%%%%%%%%%%%%%%%%%%%%%%%%%%%%%%%%%%%%%%%%%%%%%%
      Suppose we are given a finite array of rational numbers,
      \begin{alignat*}1
       &q_1,\ q_2,\ \dots,\ q_K~,
      \end{alignat*}
      which are the first terms of an infinite sequence~$F(N)$,~i.e., 
      $F(1)=q_1$, $F(2)=q_2$, etc. Let us assume that $F(N)$ represents a
      physical quantity and satisfies a recurrence of type
      \begin{eqnarray}
       \label{DEQ}
       \sum_{k=0}^l\Bigl(\sum_{i=0}^d c_{i,k} N^i\Bigr)F(N+k)=0~,
      \end{eqnarray}
      which we would like to deduce from the given numbers 
      $q_m$. In a strict sense, this is not possible without 
      knowing how the sequence continues for $N>K$. One thing we can do is to
      determine the recurrence equations satisfied by the data we are given.
      Any recurrence for $F(N)$ must certainly be among those. 

      To find the recurrence equations of $F(N)$ valid for the first terms, the
      simplest way to proceed is by making an ansatz with undetermined
      coefficients. Let us fix an order~$l\in\mathbb{N}$ and a degree 
      $d\in\mathbb{N}$ and consider the generic recurrence (\ref{DEQ}), 
      where the $c_{i,k}$ are unknown. For each specific choice 
      $N=1,2,\dots,K-l$, we can evaluate the ansatz, because we know all the
      values of $F(N+k)$ in this range, and we obtain a system of $K-l$
      homogeneous linear equations for $(l+1)(d+1)$ unknowns~$c_{i,j}$.

      If $K-l>(l+1)(d+1)$, this system is under-determined and is thus
      guaranteed to have nontrivial solutions. All these solutions will be
      valid recurrences for $F(N)$ for $N=1,\dots,K-l$, but they will most
      typically fail to hold beyond. If, on the other hand, 
      $K-l\leq(l+1)(d+1)$, then the system is overdetermined and nontrivial
      solutions are not to be expected. But at least recurrence equations valid
      for all~$N$, if there are any, must appear among the solutions. We
      therefore expect in this case that the solution set will precisely
      consist of the recurrences of~$F(N)$ of order~$l$ and degree~$d$ valid
      for all~$N$.

      As an example, let us consider the contribution to the gluon splitting
      function $\propto C_A$ at leading order, $P_{gg}^{(0)}(N)$. The first 20
      terms, starting with $N=3$, of the sequence $F(N)$ are
      \begin{alignat*}1
       &\tfrac{14}{5},\ \tfrac{21}{5},\ \tfrac{181}{35},\ \tfrac{83}{14},\
       \tfrac{4129}{630},\ \tfrac{319}{45},\ \tfrac{26186}{3465},\
       \tfrac{18421}{2310},\ \tfrac{752327}{90090},\ \tfrac{71203}{8190},\
       \tfrac{811637}{90090},\ \tfrac{128911}{13860},\
       \tfrac{29321129}{3063060},\\ &
       \tfrac{2508266}{255255},\ \tfrac{292886261}{29099070},\
       \tfrac{7045513}{684684},\ \tfrac{611259269}{58198140},\
       \tfrac{1561447}{145860},\ \tfrac{4862237357}{446185740},\
       \tfrac{988808455}{89237148}~.
      \end{alignat*}
      Making an ansatz for a recurrence of order~3 with polynomial coefficients
      of degree~3 leads to an overdetermined homogeneous linear system with 16
      unknowns and 17 equations. Despite of being overdetermined and dense, 
      this system has two linearly independent solutions. Using bounds for the 
      absolute value of determinants depending on the size of a matrix and the
      bit size of its coefficients, one can very roughly estimate the
      probability for this to happen ``by coincidence'' to about~$10^{-65}$. 
      And in fact, it did not happen by coincidence. The solutions to the
      system correspond to the two recurrence equations
      \begin{alignat}1
       &(7 N^3+113 N^2+494 N+592) F(N)-(12 N^3+233 N^2+1289 N+2156) F(N+1)
        \notag\\
       &{}+(3 N^3+118 N^2+1021 N+2476) F(N+2)+(2 N^3+2 N^2-226 N-912) F(N+3)
        \notag\\
       &{}=0 
       \label{eq1}
      \end{alignat}
      and
      \begin{alignat}1
       &(4 N^3+64 N^2+278 N+332) F(N)-(7 N^3+134 N^2+735N+1222) F(N+1)\notag\\
       &{}+(2 N^3+71 N^2+595 N+1418) F(N+2)+(N^3-N^2-138 N-528) F(N+3)
         \notag\\
       &{}=0,
      \end{alignat}
      which both are valid for all~$N\geq1$. If we had found that the linear
      system did not have a nontrivial solution, then we could have concluded
      that the sequence $F(N)$ would \emph{definitely} (i.e.\ without any
      uncertainty) not satisfy a recurrence of order~3 and degree~3. It might
      then still have satisfied recurrences with larger order or degree, but
      more terms of the sequence had to be known for detecting those. 

      The method of determining (potential) recurrence equations for sequences
      as just described is not new. It is known to the experimental mathematics
      community as automated guessing and is frequently applied in the study of
      combinatorial sequences. Standard software packages for generating
      functions such as \textsf{gfun}~\cite{Salvy:1994} for ${\sf MAPLE}$ or
      \textsf{GeneratingFunctions.m}~\cite{Mallinger:1996} for 
      ${\sf MATHEMATICA}$ provide functions which take as input a finite array
      of numbers, thought of as the first terms of some infinite sequence, and
      produce as output recurrence equations that are, with high probability,
      satisfied by the infinite sequence.

      These packages apply the method described above more or less literally,
      and this is perfectly sufficient for small examples. But if thousands of
      terms of a sequence are needed, there is no way to solve the linear
      systems using rational number arithmetic. Even worse, already for
      medium sized problems from our collection, the size of the linear system
      exceeds by far typical memory capacities of~16--64Gb.
      Let us consider as an example the difference 
      equation associated to the contribution of the color factor $C_F^3$
      for the 3-loop Wilson coefficient  $C_{2,q}^{(3)}$ in unpolarized deeply
      inelastic scattering. {11~Tb} of memory would be required to 
      establish~(\ref{DEQ}) in a naive way. Therefore refined methods have to 
      be applied. We use arithmetic in finite fields together with Chinese 
      remaindering, \cite{Geddes:1992,*Gathen:1999,*Kauers:2008zz}, which
      reduces the storage requirements to a few Gb of memory. The linear system
      approximately minimizes for {$l \approx d$}. If one finds more than one 
      recurrence the different recurrences are joined to reduce {$l$} to a 
      minimal value. It seems to be a general phenomenon that the recurrence of
      minimal order is that with the smallest integer coefficients, cf. also 
      \cite{bostan:08}. For even larger problems than those dealt with in the 
      present analysis, a series of further technical improvements may be
      carried out, \cite{Beckermann:1992,*Beckermann:2000}.

      For the solution of the recurrence low orders are clearly preferred. It 
      is solved in depth-optimal {$\Pi\Sigma$ fields}, \cite{Karr:1981,Karr:1985,sigma1,Schneider:2007,Schneider:2001,*Schneider:2007a,*RISC3389,*Schneider:2008};
      here we apply advanced symbolic summation methods as: efficient 
      recurrence solvers and refined telescoping algorithms. They are available
      in the summation package \SigmaP, \cite{sigma1,sigma2}. 

      The solutions are found as linear combinations of rational terms in $N$ 
      combined with functions, which cannot be further reduced in the 
      $\Pi\Sigma$ fields. In the present application they turn out to be 
      nested harmonic sums, cf. Appendix \ref{App-SpeFunHarm}. Other or higher 
      order applications may lead to sums of different type as well, which are 
      uniquely found by the present algorithm.
%%%%%%%%%%%%%%%%%%%%%%%%%%%%%%%%%%%%%%%%%%%%%%%%%%%%%%%%%%%%%%%%%%%%%%%%%
    \subsubsection{Determination of the 3-Loop Anomalous Dimensions \\
                   and Wilson Coefficients}
     \label{SubSec-FULL3LGuessWil}
%%%%%%%%%%%%%%%%%%%%%%%%%%%%%%%%%%%%%%%%%%%%%%%%%%%%%%%%%%%%%%%%%%%%%%%%%
      We apply the method to determine the unpolarized anomalous dimensions and
      massless Wilson coefficients to $3$--loop order. 
      Here we apply the above method to the contributions 
      stemming from a single color/$\zeta_i$-factor. These are 186 
      terms. As input we use the respective Mellin moments, 
      which were calculated by a {\sf MAPLE}--code based on the harmonic sum 
      representation calculated in Refs.~\cite{Moch:2004pa,Vogt:2004mw,Vermaseren:2005qc}.
      We need very high moments and calculate the input recursively. As an 
      example, let us illustrate the size of the moments for the 
      $C_F^3$-contribution to the Wilson coefficient $C^{(3)}_{2,q}$. 
      The highest 
      moment required is $N = 5114$. It cannot be calculated simply using 
      {\sf Summer}, \cite{Vermaseren:1998uu}, and we used a recursive 
algorithm in {\sf MAPLE} for it. 

      The corresponding difference equations (\ref{DEQ}) are determined by a 
      recurrence finder. Furthermore, the order of the difference equation is 
      reduced to the smallest value possible. The difference equations are then
      solved order by order using the summation package \SigmaP.
      For the $C_F^3$-term in $C^{(3)}_{q,2}$, the recurrence was established after
      20.7 days of CPU time. Here 4h were required for the modular prediction
      of the dimension of the system, 5.8 days were spent on solving modular
      linear systems, and 11 days for the modular operator GCDs. The Chinese 
      remainder method and rational reconstruction took 3.8 days. 140 word size
      primes were needed. As output one obtains a recurrence of 31 Mb, which is
      of order 35 and degree 938, with a largest integer of 1227 digits. The 
      recurrence was solved by \SigmaP~after 5.9 days. We reached a 
      compactification from 289 harmonic sums needed in
      \cite{Moch:2004pa,Vogt:2004mw,Vermaseren:2005qc} to 58 harmonic sums.
      The determination of the $3$--loop anomalous dimensions is a much smaller
      problem. Here the computation takes only about 18~h for the complete
      result.

      For the three most complicated cases, establishing and solving of the 
      difference equations took $3+1$ weeks each, requiring $\leq 10$Gb on a 
      2~GHz processor. This led to an overall computation time of about sixteen
      weeks.

      In the final representation, we account for algebraic reduction,
      \cite{Blumlein:2003gb}. For this task we used the package 
      {\sf HarmonicSums}, \cite{Ablinger:09}, which complements the 
      functionality of~\SigmaP. One observes that different color factor
      contributions lead to the same, or nearly the same, amount of sums for a
      given quantity. This points to the fact that the amount of sums
      contributing, after the algebraic reduction has been carried out,
      is governed by topology rather than the field- and color structures 
      involved. The linear harmonic sum representations used in 
      \cite{Moch:2004pa,Vogt:2004mw,Vermaseren:2005qc} require many more sums
      than in the representation reached by the present analysis. A further
      reduction can be obtained using the structural relations, which leads to 
      maximally 35 different sums up to the level of the $3$--loop Wilson
      coefficients, \cite{Blumlein:2008pa,Blumlein:2009ta,Blumlein:2009fz}. It 
     is not unlikely that the present
      method can be applied to single scale problems in even higher orders. As 
      has been found before in \cite{Blumlein:2008pa,Blumlein:2005im,*Blumlein:2006rr,Blumlein:2007dj,Blumlein:prep2,Blumlein:2006mh,Buza:1995ie,Bierenbaum:2008yu,Bierenbaum:2007qe,Berends:1987abxBerends:1987abe1,Blumlein:2007tr},
      representing a large number of 2- and 3-loop processes in terms of 
      harmonic sums, the basis elements emerging are always the same.

      In practice no method does yet exist to calculate such a high number of
      moments ab initio as required for the determination of the all--$N$ 
      formulas in the 3--loop case. On the other hand, a proof of existence has
      been delivered of a quite general and powerful automatic 
      difference-equation solver, standing rather demanding tests.
      It opens up good prospects for the development of 
      even more powerful methods, which can be applied in establishing and 
      solving difference equations for single scale quantities such as the
      classes of Feynman--parameter integrals contributing to the massive 
      operator matrix elements for general values of $N$.
%%%%%%%%%%%%%%%%%%%%%%%%%%%%%%%%%%%%%%%%%%%%%%%%%%%%%%%%%%%%%%%%%%%%%%%%%%%%%%%
%
% Chapter 11
%
% Conclusions
%
%%%%%%%%%%%%%%%%%%%%%%%%%%%%%%%%%%%%%%%%%%%%%%%%%%%%%%%%%%%%%%%%%%%%%%%%%%%%%%%
\newpage
 \section{\bf\boldmath Conclusions}
  \label{Sec-CONC}
  \renewcommand{\theequation}{\thesection.\arabic{equation}}
  \setcounter{equation}{0}
%%%%%%%%%%%%%%%%%%%%%%%%%%%%%%%%%%%%%%%%%%%%%%%%%%%%%%%%%%%%%%%%%%%%%%%%%%%%%%
   In this thesis, we extended the description of the contributions of a 
   single heavy
   quark to the unpolarized Wilson coefficients 
   ${\cal C}_{(q,g),2}^{\sf S,PS,NS}$ to $O(a_s^3)$.
   In upcoming precision analyzes
   of deep--inelastic data, this will allow more precise determinations of 
   parton distribution functions and of the strong coupling constant. 
   We applied a factorization relation for the complete inclusive heavy flavor
   Wilson coefficients, which holds in the limit $Q^2\gg 10m^2$ in case of
   $F_2(x,Q^2)$, \cite{Buza:1995ie}, at the level of twist--$2$. 
   It relates the asymptotic heavy flavor Wilson
   coefficients to a convolution of the corresponding light flavor Wilson
   coefficients, which
   are known up to $O(a_s^3)$, \cite{Vermaseren:2005qc},
   and describe all process
   dependence, with the massive operator matrix elements. The latter are
   process independent quantities and describe all mass--dependent
   contributions but the
   power--suppressed terms ($(m^2/Q^2)^k,~k\ge~1$).
   They are obtained from the unpolarized twist--$2$ local composite operators
   stemming from the light--cone expansion of the electromagnetic current 
   between on--shell partonic states, including virtual heavy quark
   lines.
   The first calculation of fixed moments of all $3$--loop massive OMEs is 
   the main result of this thesis. \\
 
   In Section~\ref{SubSec-HQAsym}, we applied the 
   factorization formula at the $O(a_s^3)$--level. It holds
   for the inclusive heavy flavor Wilson coefficients, including radiative
   corrections due to heavy quark loops.
   In order to describe the production of heavy
   quarks in the final states only, further assumptions have to be made.
   This description succeeded at the $2$--loop level in 
   Ref.~\cite{Buza:1995ie} because of
   the possible comparison with the exact calculation in 
   Refs.~\cite{Laenen:1992zkxLaenen:1992xs,*Riemersma:1994hv} and since the
   contributing virtual heavy flavor corrections are easily identified,
   cf. Section~\ref{SubSec-HQElProdWave}. 
   At $O(a_s^3)$ this is not possible 
   anymore and only the inclusive description should be used, as has been done
   in Ref.~\cite{Buza:1996wv} in order to derive heavy flavor 
   parton densities. These are obtained as convolutions of the light flavor 
   densities with the massive OMEs, cf. Section~\ref{SubSec-HQFlav}. \\

   In Section~\ref{Sec-REN}, we derived and presented in detail the
   renormalization of the massive operator matrix elements up to $O(a_s^3)$. 
   This led to an intermediary representation in a defined 
   ${\sf MOM}$--scheme to maintain the partonic description required for the
   factorization of the heavy flavor Wilson coefficients into OMEs and the
   light flavor Wilson coefficients. Finally, we applied the 
   $\overline{\sf MS}$--scheme for coupling constant renormalization 
   in order to refer to the inclusive heavy flavor Wilson coefficients 
   and to be able to combine 
   our results with the light flavor Wilson coefficients, which have been 
   calculated in the same scheme. For mass renormalization we chose 
   the on--mass--shell--scheme and provided in Section~\ref{Sec-REP} 
   all necessary formulas to transform between the 
   ${\sf MOM}$-- and the on--shell--scheme, respectively,
   and the ${\sf \overline{MS}}$--scheme. \\

   For renormalization at $O(a_s^3)$, all $O(a_s^2)$ massive OMEs 
   $A_{Qg},~A_{Qq}^{\sf PS},~A_{qq,Q}^{\sf NS},~A_{gg,Q},~A_{gq,Q}$
   are needed up to $O(\ep)$ in dimensional regularization. 
   In Section~\ref{Sec-2L}, we newly calculated all the corresponding $O(\ep)$
   contributions in Mellin space for general values of $N$. 
   This involved a first re--calculation of the 
   complete terms $A_{gg,Q}^{(2)}$ and $A_{gq,Q}^{(2)}$, in which we agree with
   the literature, \cite{Buza:1996wv}.
   We made use of the representation of the Feynman--parameter
   integrals in terms of generalized hypergeometric functions. 
   The $O(\ep)$--expansion led
   to new infinite sums which had to be solved by analytic and
   algebraic methods. The results can be
   expressed in terms of polynomials of the basic nested harmonic sums up to
   weight ${\sf w=4}$ and derivatives thereof. They belong to the
   complexity-class of the general two-loop Wilson coefficients or hard
   scattering cross sections in massless QED and QCD and are described by six 
   basic functions and their derivatives in Mellin space. The package 
   \SigmaP, \cite{sigma1,sigma2,Refined,Schneider:2007},
   proved to be a useful tool to solve the sums occurring in the
   present problem, leading to extensions of this code by the author.\\

   The main part of the thesis was the calculation of fixed moments of all
   3--loop massive operator matrix elements
   $A_{Qg},~A_{qg,Q},~A_{Qq}^{\sf PS},~A_{qq,Q}^{\sf PS},~A_{qq,Q}^{\sf NS},~A_{gq,Q},~A_{gg,Q}$, cf. Section~\ref{Sec-3L}. These 
   are needed to describe the asymptotic heavy flavor Wilson
   coefficients at $O(a_s^3)$
   and to derive massive quark--distributions at the same level,
   \cite{Buza:1996wv}. 
   We developed computer algebra codes which allow based on ${\sf QGRAF}$,
   \cite{Nogueira:1991ex}, the automatic generation of $3$--loop Feynman
   diagrams with local operator insertions. These were then projected onto
   massive tadpole diagrams for fixed values of the Mellin variable $N$.
   For the final calculation of the diagrams,
   use was made of the ${\sf FORM}$--code
   ${\sf MATAD}$, \cite{Steinhauser:2000ry}.
   The representation of the massive OMEs is available for general values of 
   $N$ in analytic form, apart from the constant terms $a_{ij}^{(3)}$ of the 
   unrenormalized 3--loop OMEs. This is achieved by combining our general
   expressions for the renormalized results, 
   the all--$N$ results up to $O(a_s^2\ep)$
   and results given in the literature.
   A number of fixed Mellin moments of
   the terms $a_{ij}^{(3)}$ were calculated,
   reaching up to $N = 10, 12, 14$, depending on the complexity of the
   corresponding operator matrix element.  The computation required about 
   $250$ CPU days on $32/64~Gb$--machines. 

   Through the renormalization of the massive OMEs, the
   corresponding moments of the complete 2-loop anomalous dimensions and the
   $T_F$--terms of the 3--loop anomalous dimensions were obtained, as were the
   moments of the complete anomalous dimensions $\gamma_{qq}^{(2), \sf PS}$
   and $\gamma_{qg}^{(2)}$, in which we agree with the literature.
   This provides
   a first independent check of the moments of the fermionic contributions to
   the $3$--loop anomalous dimensions, which have been obtained in 
   Refs.~\cite{Larin:1996wd,Retey:2000nq}.  \\
   
   In Section~\ref{Sec-POL}, we presented results on the effects of heavy
   quarks in polarized deep--inelastic scattering, using essentially 
   the same description as in the unpolarized case. We worked in the 
   scheme for $\gamma_5$ in dimensional regularization used in 
   Ref.~\cite{Buza:1996xr} and could confirm the results given there 
   for the $2$--loop 
   massive OMEs $\Delta A_{Qq}^{\sf PS}$ and $\Delta A_{Qg}$.
   Additionally, we newly presented the $O(\ep)$ contributions of
   these terms.\\

   We calculated the $2$--loop massive OMEs of transversity for all--$N$
   and the $3$--loop terms for the moments $N=1,\ldots,13$ 
   in Section~\ref{sec-1}.
   This calculation is not yet of phenomenological use,
   since the corresponding light flavor Wilson coefficients have not 
   been calculated so far. However, these results could be obtained by making
   only minor changes to the computer programs written for the 
   unpolarized case. We confirmed for the first time 
   the moments $N=1,\ldots,8$ of the fermionic contributions to the $3$--loop
   transversity anomalous dimension obtained in 
   Refs.~\cite{Gracey:2003yrxGracey:2003mrxGracey:2006zrxGracey:2006ah}. Our 
   results can, however, be used in comparison with lattice calculations. \\

   Several steps were undertaken towards an all--$N$ calculation of the 
   massive OMEs. Four non--trivial $3$--loop massive topologies 
   contribute. We presented in an example
   a first all--$N$ result for a ladder--topology in
   Section~\ref{Sec-FULL3LF1}.

   In Section~\ref{SubSec-FULL3LGuess}, we
   described a general algorithm to calculate the exact expression for 
   single scale quantities from a finite (suitably large) number of moments, 
   which are zero scale quantities. The latter are much more easily
   calculable than single scale quantities. We applied the method to the
   anomalous dimensions and massless Wilson coefficients up to $3$--loop order,
   \cite{Moch:2004pa,Vogt:2004mw,Vermaseren:2005qc}.
   Solving $3$--loop problems in this way directly is not possible at present,
   since the
   number of required moments is too large for the methods available.
   Yet this method constitutes a proof of principle and may find application
   in medium--sized problems in the future.
%%%%%%%%%%%%%%%%%%%%%%%%%%%%%%%%%%%%%%%%%%%%%%%%%%%%%%%%%%%%%%%%%%%%%%%%
%%%%%%%%%%%%%%%%%%%%%%%%%%%%%%%%%%%%%%%%%%%%%%%%%%%%%%%%%%%%%%%%%%%%%%%%
%
% End of the main part
%
%%%%%%%%%%%%%%%%%%%%%%%%%%%%%%%%%%%%%%%%%%%%%%%%%%%%%%%%%%%%%%%%%%%%%%%%
%%%%%%%%%%%%%%%%%%%%%%%%%%%%%%%%%%%%%%%%%%%%%%%%%%%%%%%%%%%%%%%%%%%%%%%%
 \newpage
 \thispagestyle{empty}
 \begin{flushleft}
 \end{flushleft}
 \vspace{70mm}
 \begin{center}
%%%%%%%%%%%%%%%%%%%%%%%%%%%%%%%%%%%%%%%%%%%%%%%%%%%%%%%%%%%%%%%%%%%%%%%%%%%%%%%
%  \section{\bf\boldmath Appendix}
%%%%%%%%%%%%%%%%%%%%%%%%%%%%%%%%%%%%%%%%%%%%%%%%%%%%%%%%%%%%%%%%%%%%%%%%%%%%%%%
 \end{center}
 \newpage
 \thispagestyle{empty}
 \newpage
 \begin{flushleft}
 \end{flushleft}
 \newpage
%
%%%%%%%%%%%%%%%%%%%%%%%%%%%%%%%%%%%%%%%%%%%%%%%%%%%%%%%%%%%%%%%%%%%%%%%%%%%%%%%
% 
%   Appendix
%
%%%%%%%%%%%%%%%%%%%%%%%%%%%%%%%%%%%%%%%%%%%%%%%%%%%%%%%%%%%%%%%%%%%%%%%%%%%%%%%
%%%%%%%%%%%%%%%%%%%%%%%%%%%%%%%%%%%%%%%%%%%%%%%%%%%%%%%%%%%%%%%%%%%%%%%%%%%%%%%
   \begin{appendix}
%%%%%%%%%%%%%%%%%%%%%%%%%%%%%%%%%%%%%%%%%%%%%%%%%%%%%%%%%%%%%%%%%%%%%%%%%%%%%
%
% Appendix 1 
%
% Conventions
%
%%%%%%%%%%%%%%%%%%%%%%%%%%%%%%%%%%%%%%%%%%%%%%%%%%%%%%%%%%%%%%%%%%%%%%%%%%%%%
    \section{\bf \boldmath Conventions}
%%%%%%%%%%%%%%%%%%%%%%%%%%%%%%%%%%%%%%%%%%%%%%%%%%%%%%%%%%%%%%%%%%%%%%%%%%%%%
     \label{App-Con}
     \renewcommand{\theequation}{\thesection.\arabic{equation}}
     \setcounter{equation}{0}
%%%%%%%%%%%%%%%%%%%%%%%%%%%%%%%%%%%%%%%%%%%%%%%%%%%%%%%%%%%%%%%%%%%%%%%%%%%%%
     We use natural units
     \begin{eqnarray}
      \hbar=1~,\quad c=1~,\quad \ep_0=1~,
     \end{eqnarray}
     where $\hbar$ denotes Planck's constant, $c$ the vacuum speed 
     of light and $\ep_0$ the permittivity of vacuum.
     The electromagnetic fine--structure constant $\alpha$~is
     given by 
     \begin{eqnarray} 
      \alpha=\alpha'(\mu^2=0)=\frac{e^2}{4\pi\ep_0\hbar c}
       =\frac{e^2}{4\pi}\approx 
      \frac{1}{137.03599911(46)}~.
     \end{eqnarray}
     In this convention, energies and momenta are given in the 
     same units, electron volt ($\eV$). 

     The space--time dimension is taken to be $D=4+\ep$ and 
     the metric tensor $g_{\mu\nu}$ in Minkowski--space 
     is defined as
     \begin{eqnarray}
      g_{00}=1~,\quad g_{ii}=-1~,i=1\ldots D-1~,\quad g_{ij}=0~,i\neq j~. 
      \label{metricDdim}
     \end{eqnarray}
     Einstein's summation convention is used, i.e.
     \begin{eqnarray}
      x_{\mu}y^{\mu}:=\sum^{D-1}_{\mu=0}x_{\mu}y^{\mu}~.
     \end{eqnarray}
     Bold--faced symbols represent $(D-1)$--dimensional spatial vectors:
     \begin{eqnarray}
      x=(x_0,{\bf x})~.
     \end{eqnarray}
     If not stated otherwise, Greek indices refer to the $D$--component 
     space--time vector and Latin ones to the $D-1$ spatial components 
     only. The dot product of two vectors is defined 
     by 
     \begin{eqnarray}
        p.q=p_0q_0-\sum_{i=1}^{D-1}p_iq_i~. 
     \end{eqnarray}
     The $\gamma$--matrices $\gamma_{\mu}$ are taken to be of dimension
     $D$ and fulfill the anti--commutation relation
     \begin{eqnarray}
      \{\gamma_{\mu},\gamma_{\nu}\}=2g_{\mu\nu} \label{gammaanticom}~.
     \end{eqnarray}
     It follows that
     \begin{eqnarray}
      \gamma_{\mu}\gamma^{\mu}&=&D \\
   Tr \left(\gamma_{\mu}\gamma_{\nu}\right)&=&4g_{\mu\nu} \\
   Tr \left(\gamma_{\mu}\gamma_{\nu}\gamma_{\alpha}\gamma_{\beta}\right)
       &=&4[g_{\mu\nu}g_{\alpha\beta}+
            g_{\mu\beta}g_{\nu\alpha}-
            g_{\mu\alpha}g_{\nu\beta}] \label{gammarelations}~.
     \end{eqnarray}
      The slash--symbol for a $D$-momentum $p$ is defined by
      \begin{eqnarray}
       \adag p:=\gamma_{\mu}p^{\mu} \label{dagger}~.
      \end{eqnarray}
      The conjugate of a bi--spinor $u$ of a particle is given by 
      \begin{eqnarray}
       \overline{u}=u^{\dagger}\gamma_0~,
      \end{eqnarray}
      where $\dagger$ denotes Hermitian and $*$ complex 
      conjugation, respectively. The bi--spinors 
      $u$ and $v$ fulfill the free
      Dirac--equation,
      \begin{eqnarray}
       (\adag p-m )u(p)&=&0~,~\quad \overline{u}(p)(\adag p-m )=0 \\
       (\adag p+m )v(p)&=&0~,~\quad \overline{v}(p)(\adag p+m )=0~. 
      \end{eqnarray}
      Bi--spinors and polarization vectors are normalized to 
      \begin{eqnarray}
       \sum_{\sigma}u(p,\sigma)\overline{u}(p,\sigma)&=&\adag p+m \\   
       \sum_{\sigma}v(p,\sigma)\overline{v}(p,\sigma)&=&\adag p-m \\
       \sum_{\lambda}\epsilon^{\mu}(k,\lambda )\epsilon^{\nu}(k,\lambda)
       &=&-g^{\mu \nu}~,
      \end{eqnarray}
      where $\lambda$ and $\sigma$ represent the spin.

      The commonly used caret~``~$\hat{\empty}~$''~to signify an operator, 
      e.g. $\hat{O}$, is omitted if confusion is not to be expected. 

      The gauge symmetry group of QCD is the Lie--Group $SU(3)_c$. We
      consider the general case of $SU(N_c)$. The 
      non--commutative generators are denoted by $t^a$, where 
      $a$ runs from $1$ to $N_c^2-1$. The generators can 
      be represented by Hermitian, traceless matrices, \cite{Muta:1998vi}.
      The structure constants $f^{abc}$
      and $d^{abc}$ of $SU(N_c)$ are defined via the commutation and 
      anti--commutation relations of its generators, \cite{Yndurain:1999ui},
      \begin{eqnarray}
       [t^a,t^b]&=&if^{abc}t^c \label{structconstf} \\
      \{t^a,t^b\}&=& d^{abc}t^c+\frac{1}{N_c}\delta_{ab} \label{structconstd}~.
      \end{eqnarray}
      The indices of the color matrices, in a certain representation, 
      are denoted by $i,j,k,l,..$. The color invariants
      most commonly encountered are
      \begin{eqnarray}
       \delta_{ab} C_A&=&f^{acd}f^{bcd}   \label{CA} \\
       \delta_{ij} C_F&=&t^a_{il}t^a_{lj} \label{CF} \\
       \delta_{ab} T_F&=&t^a_{ik}t^b_{ki} \label{TR}~.
      \end{eqnarray}
      These constants evaluate to 
      \begin{eqnarray}
       C_A=N_c~,\quad~C_F=\frac{N_c^2-1}{2N_c}~,\quad~T_F=\frac{1}{2}~.\label{Cval}
      \end{eqnarray}
      At higher loops, more color--invariants emerge.
      At $3$--loop order, one additionally obtains
      \begin{eqnarray}
        d^{abd}d_{abc}=(N_c^2-1)(N_c^2-4)/N_c~. \label{dabc2}
      \end{eqnarray}
      In case of $SU(3)_c$, $C_A=3~,~C_F=4/3~,~d^{abc}d_{abc}=40/3$ holds.
%%%%%%%%%%%%%%%%%%%%%%%%%%%%%%%%%%%%%%%%%%%%%%%%%%%%%%%%%%%%%%%%%%%%%%%%%%%%%
%
% Appendix 2
%
% Feynman Rules
%
%%%%%%%%%%%%%%%%%%%%%%%%%%%%%%%%%%%%%%%%%%%%%%%%%%%%%%%%%%%%%%%%%%%%%%%%%%%%%
\newpage
  \section{\bf \boldmath Feynman Rules}
\begin{figure}[H]
     \begin{center}
      \includegraphics[angle=0, height=17cm]{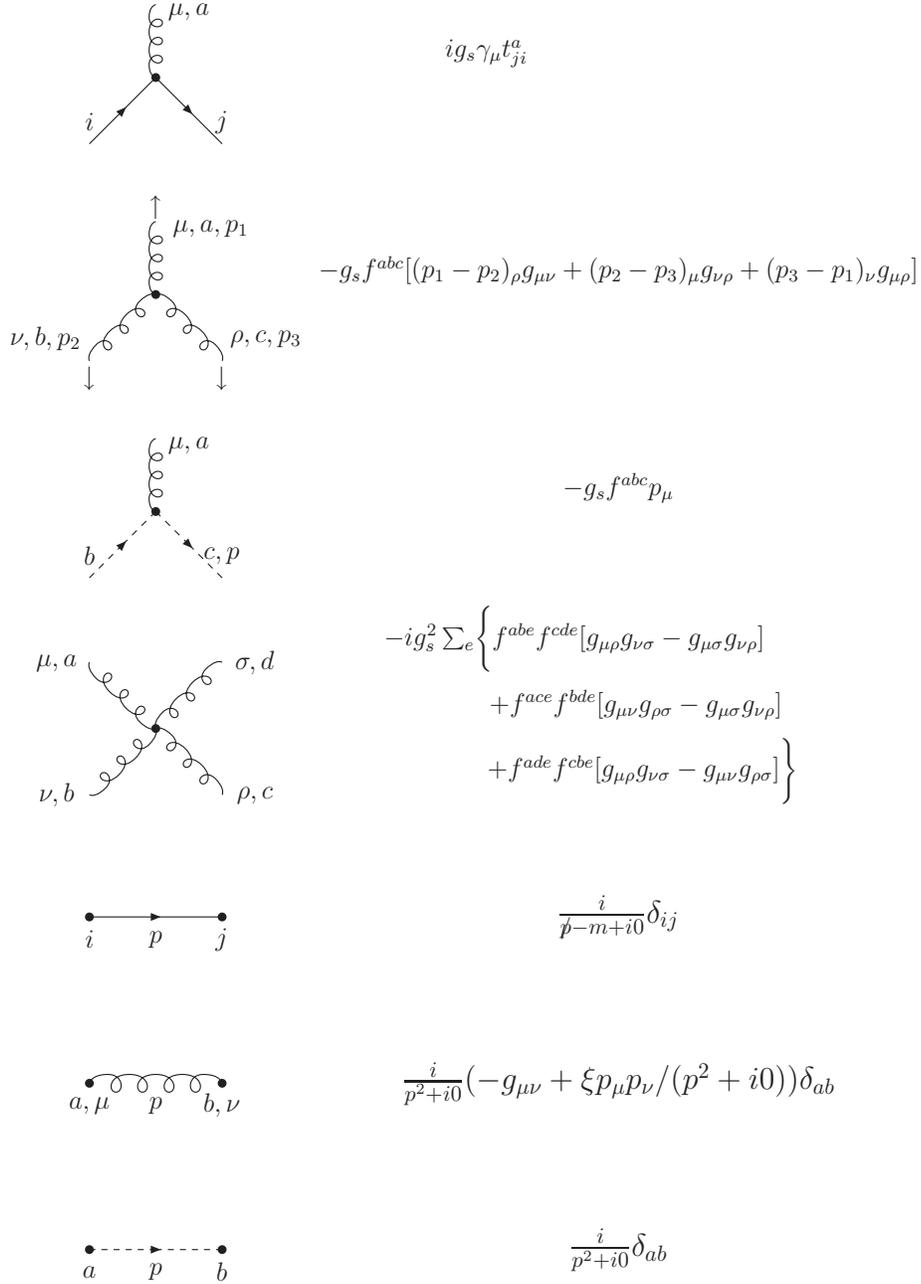}
     \end{center}
     \begin{center} 
      \caption{\sf Feynman rules of QCD.}
        \label{feynrulesqcd}
      \noindent
      \small
     \end{center}
     \normalsize
    \end{figure}

\begin{figure}[H]
     \begin{center}
      \includegraphics[angle=0, height=16.5cm]{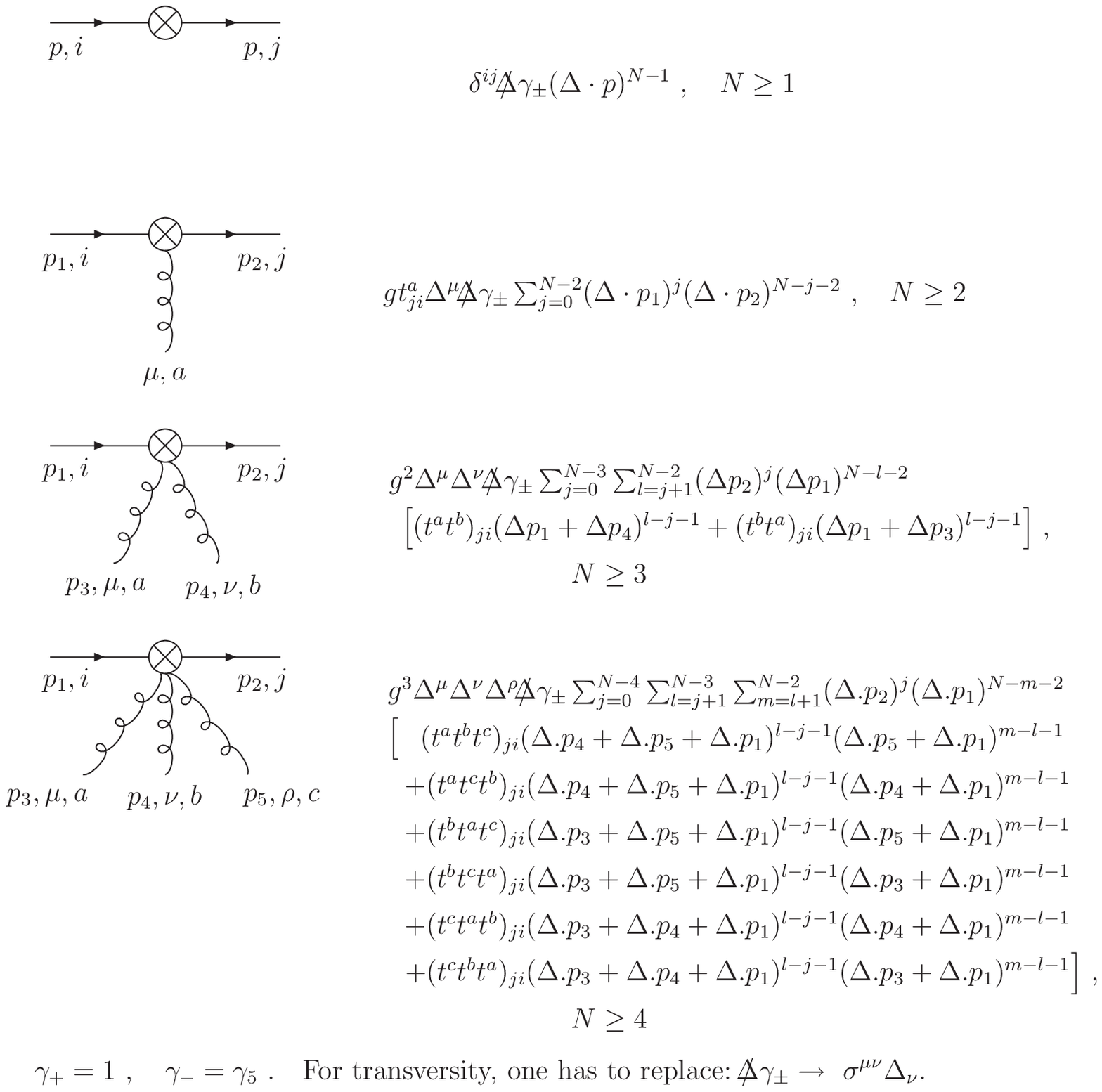}
     \end{center}
     \begin{center} 
      \caption[{\sf Feynman rules for quarkonic composite operators.}]
     {\sf Feynman rules for quarkonic composite operators. $\Delta$
     denotes  a light-like $4$-vector,\\ 
              \phantom{abcdefghijk} $\Delta^2=0$;
              $N$ is a suitably large positive integer.}
        \label{feynrulescompqua}
      \noindent
      \small
     \end{center}
     \normalsize
    \end{figure}

\begin{figure}[H]
     \begin{center}
      \includegraphics[angle=0, height=17cm]{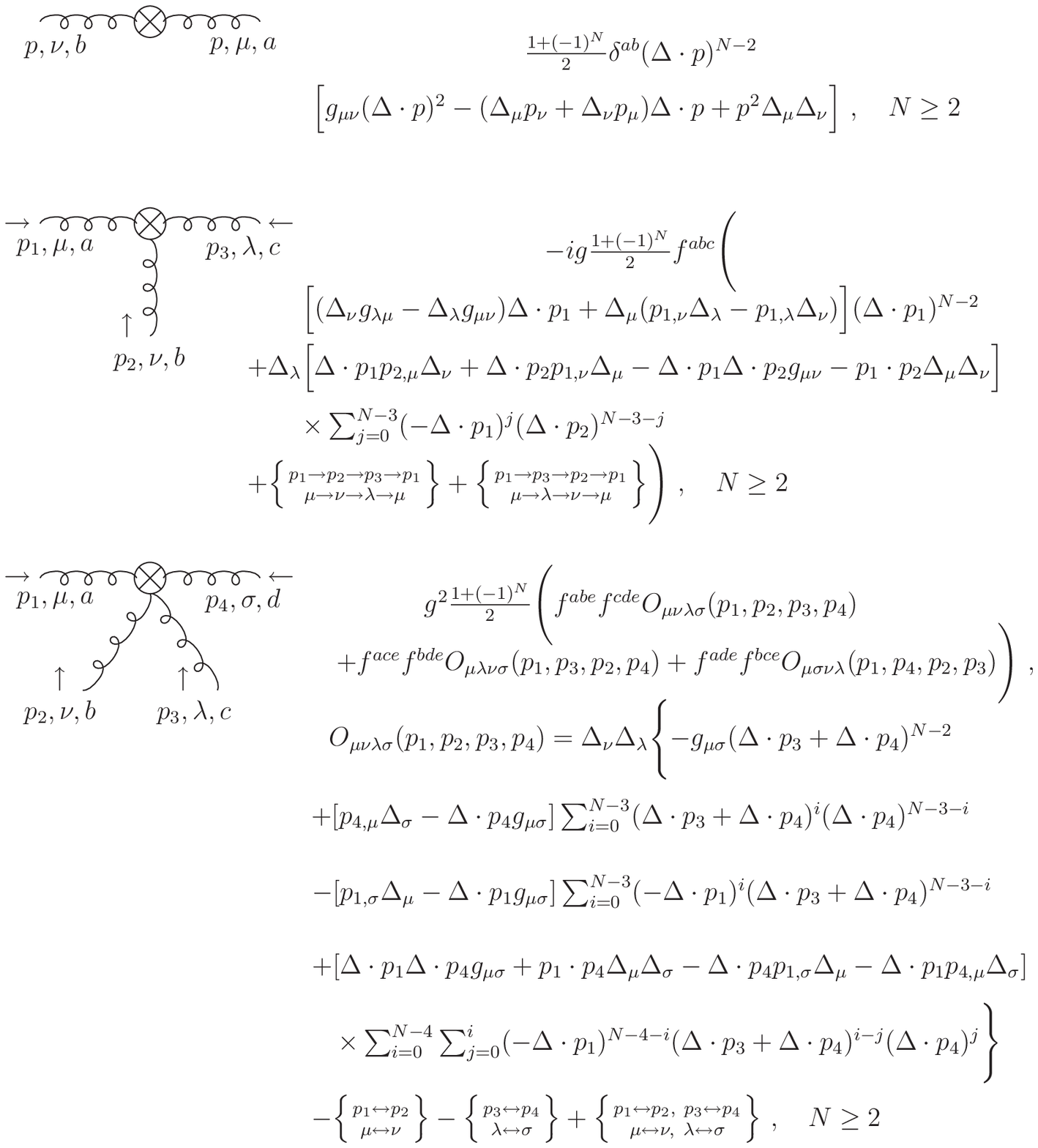}
     \end{center}
     \begin{center} 
      \caption[{\sf Feynman rules for gluonic composite operators.}]
     {\sf Feynman rules for gluonic composite operators. $\Delta$
     denotes  a light-like $4$-vector,\\ \phantom{abcdefghijk}
        $\Delta^2=0$; $N$ is an integer.}
        \label{feynrulescompglu}
      \noindent
      \small
     \end{center}
     \normalsize
    \end{figure}

%%%%%%%%%%%%%%%%%%%%%%%%%%%%%%%%%%%%%%%%%%%%%%%%%%%%%%%%%%%%%%%%%%%%%%%%%%%%%
   \label{App-FeynRules}
   \renewcommand{\theequation}{\thesection.\arabic{equation}}
   \setcounter{equation}{0}
%%%%%%%%%%%%%%%%%%%%%%%%%%%%%%%%%%%%%%%%%%%%%%%%%%%%%%%%%%%%%%%%%%%%%%%%%%%%%
    For the QCD Feynman rules, Figure \ref{feynrulesqcd}, we follow Ref.
    \cite{Yndurain:1999ui}, cf. also 
    Refs.~\cite{Veltman:1994wz,*'tHooft:1973pz}.
    $D$--dimensional momenta are denoted by $p_i$ and Lorentz-indices by Greek
    letters.
    Color indices are $a,b,...$ and $i,j$ are indices of the color matrices.
    Solid lines represent fermions, wavy lines gluons and dashed lines ghosts.
    Arrows denote the direction of the momenta. A factor $(-1)$ has to be
    included for each closed fermion-- or ghost loop.
%%%%%%%%%%%%%%%
%%%%%%%%%%%%%%%
   \noindent
   The Feynman rules for the quarkonic composite operators are given 
   in Figure \ref{feynrulescompqua}. Up to $O(g^2)$ they can be found 
   in Ref. \cite{Floratos:1977auxFloratos:1977aue1} 
   and also in \cite{Mertig:1995ny}. Note that the 
   $O(g)$ term in the former reference contains a typographical error.
   We have checked these terms and agree up to normalization 
   factors, which may be due to other conventions being applied there. 
   We newly derived the rule with three external gluons.
   The terms $\gamma_{\pm}$ refer to the unpolarized ($+$) and polarized 
   ($-$) case, respectively.
   Gluon momenta are taken to be incoming. 
%%%%%%%%%%%%%%%
%%%%%%%%%%%%%%%
   \newpage
   \noindent 
   The Feynman rules for the unpolarized gluonic composite operators are given 
   in Figure \ref{feynrulescompglu}. Up to $O(g^2)$, they can be found 
   in Refs. \cite{Floratos:1978ny} and \cite{Hamberg:1991qt}. 
   We have checked these terms and agree up to $O(g^0)$. At $O(g)$, 
   we agree with \cite{Floratos:1978ny}, but not with \cite{Hamberg:1991qt}.
   At $O(g^2)$, we do not agree with either of these results, 
   which even differ from each other\footnote{We would like to thank J. Smith for the possibility to compare with their {\sf FORM}--code used in Refs. \cite{Buza:1995ie,Buza:1996xr,Matiounine:1998re,Matiounine:1998ky}, to which we agree.}.
%%%%%%%%%%%%%%%   
%%%%%%%%%%%%%%%%%%%%%%%%%%%%%%%%%%%%%%%%%%%%%%%%%%%%%%%%%%%%%%%%%%%%%%%%%%%%%
%
% Appendix 3
%
% Special Functions
%
%%%%%%%%%%%%%%%%%%%%%%%%%%%%%%%%%%%%%%%%%%%%%%%%%%%%%%%%%%%%%%%%%%%%%%%%%%%%%
\newpage
  \section{\bf \boldmath Special Functions}
%%%%%%%%%%%%%%%%%%%%%%%%%%%%%%%%%%%%%%%%%%%%%%%%%%%%%%%%%%%%%%%%%%%%%%%%%%%%%
   \label{App-SpeFun}
   \renewcommand{\theequation}{\thesection.\arabic{equation}}
   \setcounter{equation}{0}
%%%%%%%%%%%%%%%%%%%%%%%%%%%%%%%%%%%%%%%%%%%%%%%%%%%%%%%%%%%%%%%%%%%%%%%%%%%%%
    In the following we summarize for convenience some relations for special
    functions 
    which occur repeatedly in quantum field theory and are used in this thesis.
%%%%%%%%%%%%%%%%%%%%%%%%%%%%%%%%%%%%%%%%%%%%%%%%%%%%%%%%%%%%%%%%%%%%%%%%%%%%%
   \subsection{The $\Gamma$--function}
   \label{App-SpeFunGA}
%%%%%%%%%%%%%%%%%%%%%%%%%%%%%%%%%%%%%%%%%%%%%%%%%%%%%%%%%%%%%%%%%%%%%%%%%%%%%
    The $\Gamma$-function, cf. \cite{stegun,Nielsen:1906}, is analytic 
    in the whole complex plane except at single poles at the non-positive
    integers. Its inverse is given
    by Euler's infinite product
    \begin{eqnarray}
     \frac{1}{\Gamma(z)}=z\exp(\gamma_Ez)
               \prod_{i=1}^{\infty} 
               \Biggl[\Bigl(1+\frac{z}{i}\Bigr)\exp(-z/i)\Biggr]~. 
               \label{eulerprod}
    \end{eqnarray}
   The residues of the $\Gamma$-function at its poles are given 
   by 
   \begin{eqnarray}
   {\sf Res}[\Gamma(z)]_{z=-N}=\frac{(-1)^N}{N!}~,\quad N \in {\mathbb{N}}\cup 0~.
       \label{gammares} 
   \end{eqnarray}
   In case of ${\sf Re}(z) > 0$, it can be expressed by Euler's
   integral
   \begin{eqnarray}
    \Gamma(z)=\int_0^{\infty} \exp(-t)t^{z-1}dt~,
   \end{eqnarray}
   from which one infers the well known
   functional equation of the $\Gamma$-function
   \begin{eqnarray}
    \Gamma(z+1)=z\Gamma(z)~,\label{funcrelgam}
   \end{eqnarray}
   which is used for analytic continuation.
   Around $z=1$, the following series expansion is obtained
   \begin{eqnarray}
    \Gamma(1-\ep)
      &=&\exp(\ep \gamma_E)
         \exp\Biggl\{\sum_{i=2}^{\infty}\zeta_i\frac{\ep^i}{i}\Biggr\}~,
          \label{gammaser}\\
      |\ep|&<&1~.
   \end{eqnarray}
   Here and in (\ref{eulerprod}), $\gamma_E$ denotes the Euler-Mascheroni 
   constant, see Eq. (\ref{gammaesum}). 
   In (\ref{gammaser}) Riemann's $\zeta$--function is given by
   \begin{eqnarray}
    \zeta_k=\sum_{i=1}^{\infty}\frac{1}{i^k}~,\quad 2\le k \in \mathbb{N}~.
    \label{zeta}
   \end{eqnarray}
   A shorthand notation for rational functions of $\Gamma$--functions is
   \begin{eqnarray}
    \Gamma\Biggl[\frac[0pt]{a_1,...,a_i}{b_1,...,b_j}\Biggr]:=
    \frac{\Gamma(a_1)...\Gamma(a_i)}{\Gamma(b_1)...\Gamma(b_j)}~.
    \label{gammashort}
   \end{eqnarray}
   Functions closely related to the $\Gamma$-function 
   are the function $\psi(x)$, the Beta-function $B(A,C)$
   and the function $\beta(x)$.

   The Beta-function can be defined by Eq.~(\ref{gammashort})
   \begin{eqnarray}
    B(A,C)=\Gamma\Biggl[\frac[0pt]{A,C}{A+C}\Biggr]~. \label{betafun1}
   \end{eqnarray}
   If ${\sf Re}(A),{\sf Re}(C) > 0 $, the following 
   integral representation is valid 
   \begin{eqnarray}
    B(A,C)=\int_0^1 dx~x^{A-1}(1-x)^{C-1}~. \label{betafun2}
   \end{eqnarray}
   For arbitrary values of $A$ and $C$, (\ref{betafun2})
   can be continued analytically using Eqs. (\ref{eulerprod},~\ref{betafun1}).
   Its expansion around singularities can be performed 
   via Eqs. (\ref{gammares},~\ref{gammaser}).
   The $\psi$-function and $\beta(x)$ are defined as derivatives of the 
   $\Gamma$-function via
   \begin{eqnarray}
    \psi(x)  &=& \frac{1}{\Gamma(x)} \frac{d}{dx} \Gamma(x)~. \label{psifun}\\
    \beta(x) &=& \frac{1}{2} \left[
                  \psi\Bigl(\frac{x+1}{2}\Bigr)
                -  \psi\Bigl(\frac{x}{2}\Bigr)\right]~. \label{smallbeta}
   \end{eqnarray}
%%%%%%%%%%%%%%%%%%%%%%%%%%%%%%%%%%%%%%%%%%%%%%%%%%%%%%%%%%%%%%%%%%%%%%%%%%%%%
   \subsection{The Generalized Hypergeometric Function}
   \label{App-SpeFunFPQ}
%%%%%%%%%%%%%%%%%%%%%%%%%%%%%%%%%%%%%%%%%%%%%%%%%%%%%%%%%%%%%%%%%%%%%%%%%%%%%
   The generalized hypergeometric function $\empty_{P}F_Q$ is defined by,
   cf. \cite{Slater,Bailey,*Roy:2001}, 
   \begin{eqnarray}
    \empty_{P}F_Q\Biggl[\frac[0pt]{a_1,...,a_P}
                                  {b_1,...,b_Q}
                                  ;z\Biggr]
    =\sum_{i=0}^{\infty}
     \frac{(a_1)_i...(a_P)_i}
          {(b_1)_i...(b_Q)_i}
          \frac{z^i}{\Gamma(i+1)}~.
     \label{fpq}
   \end{eqnarray}
   Here $(c)_n$ is Pochhammer's symbol
   \begin{eqnarray}
    (c)_n=\frac{\Gamma(c+n)}{\Gamma(c)} \label{pochhammer}~, 
   \end{eqnarray} 
   for which the following relation holds
   \begin{eqnarray}
    (N+1)_{-i}&=&\frac{(-1)^i}{(-N)_i}~,~N\in~\mathbb{N}~.  \label{reflect}
   \end{eqnarray}
   In (\ref{fpq}), there are $P$ numerator parameters $a_1...a_P$, 
   $Q$ denominator parameters $b_1...b_Q$ and one variable 
   $z$, all of which may be real or complex. Additionally, the 
   denominator parameters must not be negative integers, since in 
   that case (\ref{fpq}) is not defined. The generalized 
   hypergeometric series $\empty_{P}F_Q$ are evaluated 
   at a certain value of $z$, which in this thesis is always $z=1$ for the
   final integrals.  \\
   Gauss was the first to study this 
   kind of functions, introducing the Gauss function $\empty_2F_1$, and proved 
   the theorem, cf. \cite{Slater},
   \begin{eqnarray} 
    \empty_{2}F_1[a,b;c;1]=
    \Gamma\Biggl[\frac[0pt]{c,c-a-b}{c-a,c-b}\Biggr]
                 \label{Gauss}~, \quad {\sf Re}(c-a-b)>0 
   \end{eqnarray}
  which is called Gauss' theorem.
  An integral representation for the Gauss function 
  is given by the integral, cf. \cite{Slater},
  \begin{eqnarray}
   \empty_2F_1\Biggl[\frac[0pt]{a,b+1}{c+b+2};z\Biggl]=
   \Gamma\Biggl[\frac[0pt]{c+b+2}{c+1,b+1}\Biggr]
   \int_0^1 dx~x^{b}(1-x)^c(1-zx)^{-a}~,\label{pochint}
  \end{eqnarray}
  provided that the conditions 
  \begin{eqnarray}
    |z|< 1~,\quad  {\sf Re}(c+1),~{\sf Re}(b+1)~> 0~,\label{condpoch}
  \end{eqnarray}
  are obeyed. 
  Applying Eq. (\ref{pochint}) recursively, one obtains the following 
  integral representation for a general $\empty_{P+1}F_P$--function
  \begin{eqnarray}
   &&\empty_{P+1}F_P\Biggl[\frac[0pt]{a_0,a_1,\ldots ,a_P}
                                  {b_1,\ldots ,b_P}
                                  ;z\Biggr]
    =
        \Gamma\Biggl[\frac[0pt]{b_1,\ldots ,b_P}
                             {a_1,\ldots ,a_P,b_1-a_1,\ldots ,b_P-a_P}\Biggr] 
        \times
\N\\ &&~
                \int_0^1dx_1\ldots \int_0^1dx_P
              x_1^{a_1-1}(1-x_1)^{b_1-a_1-1}\ldots x_P^{a_P-1}(1-x_P)^{b_P-a_P-1} 
                 (1-zx_1\ldots x_P)^{-a_0}~,\N \\ \label{FPQint}
  \end{eqnarray}
  under similar conditions as in Eq. (\ref{condpoch}). 
%%%%%%%%%%%%%%%%%%%%%%%%%%%%%%%%%%%%%%%%%%%%%%%%%%%%%%%%%%%%%%%%%%%%%%%%%%%%%
   \subsection{Mellin--Barnes Integrals}
   \label{App-SpeFunMB}
%%%%%%%%%%%%%%%%%%%%%%%%%%%%%%%%%%%%%%%%%%%%%%%%%%%%%%%%%%%%%%%%%%%%%%%%%%%%%
  For the Gauss function, there exists a representation in terms 
  of a complex contour integral over $\Gamma$-functions. It is given by, 
  cf. \cite{Slater},
  \begin{eqnarray}
   _2F_1\Biggl[\frac[0pt]{a,b}{c};z\Biggr]=
       \frac{\Gamma(c)}{2\pi i \Gamma(a)\Gamma(b)} 
       \int_{-i\infty +\alpha}^{i\infty +\alpha}
       \frac{\Gamma(a+s)\Gamma(b+s)\Gamma(-s)}{\Gamma(c+s)}(-z)^s ds~,
       \label{intrepgauss}
  \end{eqnarray}
  under the conditions 
  \begin{eqnarray}
   |z| < 1~,\quad |\arg(-z)| < \pi~. \label{comcon}
  \end{eqnarray}
  (\ref{intrepgauss}) only holds if one chooses the integration contour 
  in the complex plane and the positive constant $\alpha$ in such a way 
  that the poles of the $\Gamma$-functions containing $(+s)$ are separated 
  from those arising from the $\Gamma$-functions containing $(-s)$ and closes
  the contour to the right. \\
  Setting $b=1,~c=1$ in (\ref{intrepgauss}) one obtains 
  \begin{eqnarray}
   _1F_0[a;z]=\frac{1}{(1-z)^a}~,
  \end{eqnarray}
  which yields the Mellin-Barnes transformation, 
  cf. \cite{MB1a,*MB1b,*MB2,Paris:2001,Smirnov:2004ym}, 
  \begin{eqnarray}
   \frac{1}{(X+Y)^{\lambda}}=\frac{1}{2\pi i\Gamma(\lambda)}
   \int_{-i\infty+\alpha}^{+i\infty+\alpha} ds \Gamma(\lambda + s) \Gamma(-s)
   \frac{Y^s}{X^{\lambda +s}}~. \label{mbtrafo}
  \end{eqnarray}  
  The contour has to be chosen as in (\ref{intrepgauss}) and the 
  conditions $0 < \alpha < {\sf Re}(\lambda)$~, $|\arg(Y/X)|< \pi$ have to 
  be fulfilled. 
%%%%%%%%%%%%%%%%%%%%%%%%%%%%%%%%%%%%%%%%%%%%%%%%%%%%%%%%%%%%%%%%%%%%%%%%%%%%%
   \subsection{Harmonic Sums and Nielsen--Integrals}
   \label{App-SpeFunHarm}
%%%%%%%%%%%%%%%%%%%%%%%%%%%%%%%%%%%%%%%%%%%%%%%%%%%%%%%%%%%%%%%%%%%%%%%%%%%%%
    Expanding the $\Gamma$--function in $\ep$, 
    its logarithmic derivatives, the $\psi^{(k)}$-functions, emerge. 
    In many applications of perturbative QCD and 
    QED, harmonic sums occur, cf. \cite{Blumlein:1998if,Vermaseren:1998uu},
    which can be considered as generalization 
    of the $\psi$-function and the $\beta$-function. 
    These are defined by
    \begin{eqnarray}
      S_{a_1, \ldots, a_m}(N)&=&
        \sum_{n_1=1}^N  \sum_{n_2=1}^{n_1} \ldots \sum_{n_m=1}^{n_{m-1}} 
        \frac{({\rm sign}(a_1))^{n_1}}{n_1^{|a_1|}}
        \frac{({\rm sign}(a_2))^{n_2}}{n_2^{|a_2|}} \ldots
        \frac{({\rm sign}(a_m))^{n_m}}{n_m^{|a_m|}}~, \N\\
        && N~\in~{\mathbb{N}},~\forall~l~a_l~\in {\mathbb{Z}}\setminus 0~, 
      \label{harmdef} \\
     S_{\emptyset}&=&1~.
    \end{eqnarray}
    We adopt the convention
    \begin{eqnarray}
     S_{a_1, \ldots ,a_m} \equiv S_{a_1, \ldots ,a_m}(N)~, 
    \end{eqnarray}
    i.e. harmonic sums are taken at argument $(N)$, if no argument is 
    indicated.
    Related quantities are the $Z$--sums defined by 
    \begin{eqnarray}
        \label{SZSums}
     Z_{m_1, \ldots, m_k}(N) &= \sum_{N \geq i_1 > i_2 \ldots > i_k > 0}
     \D{\frac {\prod_{l=1}^k [{\rm sign}(m_l)]^{i_l}}{i_l^{|m_l|}}}~.
    \end{eqnarray}
    The depth $d$ and the weight $w$ of a harmonic sum are
    given by
    \begin{eqnarray}
     d&:=&m~,\label{depth} \\
     w&:=&\sum_{i=1}^m |a_i| ~.\label{weight}
    \end{eqnarray}
    Harmonic sums of depth $d=1$ are referred to as single harmonic sums. 
    The complete set of algebraic relations connecting harmonic 
    sums to other harmonic sums of the same or lower weight 
    is known~\cite{Blumlein:2003gb}, see also \cite{Vermaseren:1998uu}
    for an implementation in ${\sf FORM}$. Thus the number of 
    independent harmonic sums can be reduced significantly,
    e.g., for $w=3$ the $18$ possible harmonic 
    sums can be expressed algebraically in terms of $8$ basic harmonic 
    sums only.
    One introduces a product for the harmonic sums, 
    the shuffle product \SH, cf.~\cite{Blumlein:2003gb}. 
    For the product of a single and
    a general finite harmonic sum it is given by
    \begin{eqnarray}
      S_{a_1}(N) \SH S_{b_1, \ldots, b_m}(N)
       = S_{a_1, b_1, \ldots, b_m}(N) 
        + S_{b_1, a_1, b_2, \ldots, b_m}(N)
         + \ldots + S_{b_1, b_2,  \ldots, b_m, a_1}(N)~. \N\\
         \label{shuffle}  
    \end{eqnarray}
    For sums $S_{a_1, \ldots, a_n}(N)$ and 
    $S_{b_1, \ldots,b_m}(N)$ of arbitrary depth, the shuffle 
    product is then
     the sum of all harmonic sums of depth $m+n$ in the index set of 
     which $a_i$ occurs left of $a_j$ for $i < j$, likewise for
     $b_k$ and $b_l$ for $k < l$. Note that the shuffle 
     product is symmetric. \\
     One can show that the following relation holds, cf.~\cite{Blumlein:2003gb}, 
     \begin{eqnarray}
      S_{a_1}(N) \cdot S_{b_1, \ldots, b_m}(N)
      &=& S_{a_1}(N) \SH S_{b_1, \ldots, b_m}(N) \N\\
      & & -S_{a_1 \wedge b_1, b_2, \ldots, b_m}(N) - \ldots -
          S_{b_1, b_2, \ldots, a_1 \wedge b_m}(N)~,
          \label{genshuff}
     \end{eqnarray}
     where the $\wedge$ symbol is defined as
     \begin{eqnarray}
      a \wedge b = \si(a) \si(b) \left(|a| + |b|\right)~.
      \label{wedge}
     \end{eqnarray}
     Due to the additional terms containing wedges ($\wedge$) between 
     indices, harmonic sums form a quasi--shuffle 
     algebra, \cite{Hoffman:1997,*Hoffman:2004bf}.
     By summing (\ref{genshuff}) over permutations, one 
     obtains the symmetric algebraic relations between harmonic sums.  
     At depth $2$ and $3$ these read,~\cite{Blumlein:1998if},
     \begin{eqnarray}
      S_{m,n} + S_{n,m} &=&  S_m S_n + S_{m \wedge n}~, \label{algrel1}\\
      \sum_{{\rm perm}\{l,m,n\}} S_{l,m,n} &=& S_l S_m S_n 
      + \sum_{{\rm inv~perm}\{l,m,n\}} S_{l}
      S_{m \wedge n}
      + 2~S_{l \wedge m \wedge n}~, \label{algrel2}
     \end{eqnarray}
     \normalsize
     which we used extensively to simplify our expressions. 
     In (\ref{algrel1},~\ref{algrel2}), 
     ``{\sf perm}'' denotes all permutations and ``{\sf inv perm}''
     invariant ones.

     The limit $N \rightarrow \infty$ of finite harmonic sums exists
     only if $a_1\neq 1$ in (\ref{harmdef}). Additionally, one defines
     all $\sigma$-values symbolically as
     \begin{eqnarray}
      \sigma_{k_l, \ldots, k_1} = \lim_{N \rightarrow \infty}
      S_{a_1, \ldots, a_l}(N)~. \label{sigmaval}
     \end{eqnarray}
     The finite 
     $\sigma$-values are related to multiple $\zeta$-values,~\cite{Blumlein:1998if,Vermaseren:1998uu,Euler:1775,Zagier:1994,Borwein:1999js},
     Eq.~(\ref{zeta}). Further we define the symbol
     \begin{eqnarray}
      \sigma_0:=\sum_{i=1}^{\infty}1~.
     \end{eqnarray}
     It is useful to include these $\sigma$-values into the algebra, since they
     allow to treat parts of sums individually, accounting 
     for the respective divergences, cf. also \cite{Vermaseren:1998uu}.
     These divergent pieces
     cancel in the end if the overall sum is finite. 

     The relation of single harmonic sums with positive 
     or negative indices to the $\psi^{(k)}$--functions is then given by
     \begin{eqnarray}
      S_1(N)  &=&\psi(N+1)+\gamma_E~,\label{s1psi}    \\
      S_a(N)  &=&\frac{(-1)^{a-1}}{\Gamma(a)}\psi^{(a-1)}(N+1)
                 +\zeta_a~, k \ge 2~,\label{sapsi}    \\
      S_{-1}(N)&=&(-1)^N\beta(N+1)-\ln(2)~,\label{sm1beta}\\
      S_{-a}(N)&=&-\frac{(-1)^{N+a}}{\Gamma(a)}\beta^{(a-1)}(N+1)-
                  \left(1-2^{1-a}\right)\zeta_a~, 
                 \label{smabeta}k \ge 2~.
     \end{eqnarray}
     Thus single harmonic sums can be analytically continued 
     to complex values of $N$ by these relations. 
     At higher depths, harmonic sums can be expressed in terms 
     of Mellin--transforms of polylogarithms and the more general 
     Nielsen-integrals,~\cite{Nielsen:1909,Kolbig:1983qt,Devoto:1983tc}. 
     The latter are defined by
     \begin{eqnarray}
      {\rm S}_{n,p}(z) = \frac{(-1)^{n+p-1}}{(n-1)! p!} \int_0^1 \frac{dx}{x}
      \log^{n-1}(x) \log^p(1-zx)~ \label{nielsenint}
     \end{eqnarray}
     and fulfill the relation
     \begin{eqnarray}
      \frac{d \SN_{n,p}(x)}{d \log(x)} = \SN_{n-1,p}(x)~.
     \end{eqnarray}
     If $p=1$, one obtains the polylogarithms
     \begin{eqnarray}
      \Li_n(x) =  \SN_{n-1,1}(x)~,
     \end{eqnarray}
     where 
     \begin{eqnarray}
      \Li_0(x) =  \frac{x}{1-x}~.
     \end{eqnarray}
     These functions
     do not suffice for arbitrary harmonic sums, in which case 
     the harmonic polylogarithms have to be considered, \cite{Remiddi:1999ew}.
     The latter functions obey a direct shuffle algebra, cf. 
     \cite{Borwein:1999js,Blumlein:2003gb}.
     The representation in terms of Mellin--transforms then allows 
     an analytic continuation of arbitrary harmonic sums to complex $N$, 
     cf. \cite{Carlson:thesis,*Titchmarsh:1939,Blumlein:2000hw,*Blumlein:2005jg}.
     Equivalently, one may express harmonic sums by factorial 
     series, \cite{Nielsen:1906,Knopp:1947,*Landau:1906}, up to polynomials 
     of $S_1(N)$ and harmonic sums of lower degree, and use this 
     representation for the analytic continuation to $N \in \mathbb{C}$, cf. 
     \cite{GonzalezArroyo:1979df,Blumlein:2009ta}.
%%%%%%%%%%%%%%%%%%%%%%%%%%%%%%%%%%%%%%%%%%%%%%%%%%%%%%%%%%%%%%%%%%%%%%%%%%%%%
%
% Appendix 4
%
% Finite and Infinite Sums
%
%%%%%%%%%%%%%%%%%%%%%%%%%%%%%%%%%%%%%%%%%%%%%%%%%%%%%%%%%%%%%%%%%%%%%%%%%%%%%
\newpage
    \section{\bf \boldmath Finite and Infinite Sums}
%%%%%%%%%%%%%%%%%%%%%%%%%%%%%%%%%%%%%%%%%%%%%%%%%%%%%%%%%%%%%%%%%%%%%%%%%%%%%
     \label{App-Sums}
     \renewcommand{\theequation}{\thesection.\arabic{equation}}
     \setcounter{equation}{0}
%%%%%%%%%%%%%%%%%%%%%%%%%%%%%%%%%%%%%%%%%%%%%%%%%%%%%%%%%%%%%%%%%%%%%%%%%%%%%
\vspace{1mm}\noindent
In this appendix, we list some examples for infinite sums which were needed in 
the present analysis and are newly calculated. The calculation was done
using the ${\sf Sigma}$--package as explained in Section \ref{SubSec-2LInfSum}.
A complete set of sums contributing to the calculation of the 2--loop massive
OMEs can be found in Appendix B of Refs. \cite{Bierenbaum:2007qe,Bierenbaum:2008yu}. 
%%%%%%%%%%%%%%%%%%%%%%%%%%%%%%%%%%%%%%%%%%%%%%%%%%%%%%%%%%%%%%%%%%%%%%%%%%%%%%%
   \begin{eqnarray}
    \sum_{i=1}^{\infty} \frac{B(N-2,i)}{(i+N)^3}
     &=& 
         (-1)^N 
           \frac{ 4S_{1,-2}
                 +2S_{-3}
                 +2\zeta_2S_1
                 +2\zeta_3
                 -6S_{-2}
                 -3\zeta_2
                }
                {(N-2)(N-1)N}
\N\\ &&
          +\frac{1}{(N-2)(N-1)N^2}~.
           \label{Beta2}
    \end{eqnarray}
%%%%%%%%%%%%%%%%%%%%%%%%%%%%%%
    \begin{eqnarray}
    \sum_{i=1}^{\infty} \frac{B(N-2,i)}{(i+N)^2}S_1(i+N-2)
     &=&
        \frac{(-1)^{N+1}}{(N-2)(N-1)N}
             \Bigl(
                    8S_{1,-2}
                   -4S_{-3}
                   -4S_1S_{-2}
                   -2\zeta_3
\N\\ &&
                   +2\zeta_2S_1
                   -10S_{-2}
                   -5\zeta_2
                                \Bigr) 
       +\frac{N^2-3N+3}{(N-2)(N-1)^2N^2}S_1
\N\\ &&
       -\frac{N^3-5N+3}{(N-2)(N-1)^3N^3}~.
    \end{eqnarray}
%%%%%%%%%%%%%%%%%%%%%%%%%%%%%%
    \begin{eqnarray}
    \sum_{i=1}^{\infty}\frac{B(N,i)}{i+N+2}S_1(i)S_1(N+i)
     &=&
        \frac{(-1)^N}{N(N+1)(N+2)}
             \Bigl(
                    4S_{-2,1}
                   -6S_{-3}
                   -4S_{-2}S_1
                   -2\zeta_3
\N\\ &&
                   -2\zeta_2S_1
                   -2\frac{\zeta_2}{(N+1)}
                   -4\frac{S_{-2}}{(N+1)}
             \Bigr)
\N\\ &&  
        +\frac{
               -2S_3
               -S_1 S_2
               +\zeta_2 S_1
               +2 \zeta_3
              }{N+2}
\N\\ &&
        +\frac{2+7N+7N^2+5N^3+N^4}
              {N^3(N+1)^3(N+2)}S_1
\N\\  && 
        +2\frac{2+7N+9N^2+4N^3+N^4}
              {N^4(N+1)^3(N+2)}
        ~.
           \label{Beta25}
    \end{eqnarray}
%%%%%%%%%%%%%%%%%%%%%%%%%%%%%%%%%%%%%
    \begin{eqnarray}
     \sum_{i=1}^{\infty}\frac{S_{1}(i+N)S^2_1(i)}{i+N}
      &=&
         \frac{\sigma^4_1}{4}
        -\frac{3\zeta^2_2}{4}
        +\Bigl(\frac{2}{N}
        -2S_1\Bigr)\zeta_3
        +\Bigl(\frac{S_1}{N}
        -\frac{S^2_1}{2}
        -\frac{S_2}{2}\Bigr)\zeta_2
        +\frac{S^3_1}{N}
 \N\\ &&
        -\frac{S^4_1}{4}
        +S^2_1\Bigl(
                     -\frac{1}{N^2}
                     -\frac{3S_2}{2}
              \Bigr)
        -\frac{S_2}{N^2}
        -\frac{S^2_2}{4}
        -\frac{S_{2,1}}{N}
\N\\ &&
        +S_1\Bigl(
                   3\frac{S_2}{N}
                  +S_{2,1}
                  -2S_3
            \Bigr)
        +2\frac{S_3}{N}
        +S_{3,1}
        -S_4
         ~.
         \label{Harm58}
    \end{eqnarray}
%%%%%%%%%%%%%%%%%%%%%%%%%%%%
    \begin{eqnarray}
     \sum_{i=1}^{\infty}\Bigl(S_1(i+N)-S_1(i)\Bigr)^3
      &=&
        -\frac{3}{2}S^2_1
        -S^3_1
        -\frac{1}{2}S_2
        +3NS_{2,1}
        -NS_3
        +N\zeta_3
        ~.
        \label{Harm37}
    \end{eqnarray} 
%%%%%%%%%%%%%%%%%%%%%%%%
    \begin{eqnarray}
     \sum_{k=1}^{\infty}\frac{B(k+\ep/2,N+1)}{N+k}
       &=&
          (-1)^N\Bigl[2S_{-2}+\zeta_2\Bigr] 
\N\\
       &+&\frac{\ep}{2}(-1)^N\Bigl[
          -\zeta_3+\zeta_2S_1+2S_{1,-2}-2S_{-2,1}
          \Bigr] 
\N\\
       &+&\frac{\ep^2}{4}(-1)^N\Biggl[
           \frac{2}{5}\zeta_2^2-\zeta_3S_1+\zeta_2S_{1,1}
\N\\   && 
          +2\Bigl\{S_{1,1,-2}+S_{-2,1,1}-S_{1,-2,1}\Bigr\}
          \Biggr]
\N\\
       &+& \ep^3(-1)^N\Biggl[
                     -\frac{\zeta_5}{8}+\frac{S_1}{20}\zeta_2^2
                     -\frac{S_{1,1}}{8}\zeta_3
                     +\frac{S_{1,1,1}}{8}\zeta_2\N\\
                   &&+\frac{S_{1,-2,1,1}+S_{1,1,1,-2}-S_{-2,1,1,1}
                            -S_{1,1,-2,1}}{4}
                    \Biggr]\N\\
       &&+O(\ep^4)~. \label{specialsum1}
    \end{eqnarray}
%%%%%%%%%%%%%%%%%%%%%%%%%%%%%%%%%%%%%%%%%%%%%%%%%%%%%%%%%%%%%%%%%%%%%%%%%%%%%%%
   An example for a double infinite sum we encountered is given by 
   \begin{eqnarray}
      N~\sum_{i,j=1}^{\infty}\frac{S_1(i)S_1(i+j+N)}{i(i+j)(j+N)}
       &=& 4S_{2,1,1} -2S_{3,1} +S_1\Bigl( -3S_{2,1}+\frac{4S_3}{3} \Bigr)
           -\frac{S_4}{2} 
\N\\ &&    
           -S^2_2 +S^2_1S_2 +\frac{S^4_1}{6} +6S_1\zeta_3
           +\zeta_2\Bigl( 2S^2_1+S_2  \Bigr)~. \label{DoubleSum1}
   \end{eqnarray} 
   A detailed description of the method to calculate this sum 
   can be found in Appendix B of Ref. \cite{Bierenbaum:2008yu}. 
%%%%%%%%%%%%%%%%%%%%%%%%%%%%%%%%%%%%%%%%%%%%%%%%%%%%%%%%%%%%%%%%%%%%%%%%%%%%%
%
% Appendix 5
%
% Moments of the Fermionic Contributions to the $3$--Loop Anomalous Dimensions
%
%%%%%%%%%%%%%%%%%%%%%%%%%%%%%%%%%%%%%%%%%%%%%%%%%%%%%%%%%%%%%%%%%%%%%%%%%%%%%
\newpage
  \section{\bf \boldmath Moments of the Fermionic Contributions to the \\
                         $3$--Loop Anomalous Dimensions}
%%%%%%%%%%%%%%%%%%%%%%%%%%%%%%%%%%%%%%%%%%%%%%%%%%%%%%%%%%%%%%%%%%%%%%%%%%%%%
   \label{App-AnDim}
   \renewcommand{\theequation}{\thesection.\arabic{equation}}
   \setcounter{equation}{0}
%%%%%%%%%%%%%%%%%%%%%%%%%%%%%%%%%%%%%%%%%%%%%%%%%%%%%%%%%%%%%%%%%%%%%%%%%%%%%
   The pole terms of the unrenormalized OMEs in our calculation
   agree with the general structure we presented in Eqs. (\ref{Ahhhqq3NSQ},
   \ref{AhhhQq3PS}, \ref{Ahhhqq3PSQ}, \ref{AhhhQg3}, \ref{Ahhhqg3Q},
   \ref{AhhhgqQ3}, \ref{Ahhhgg3Q}). Using the lower order renormalization
   coefficients and the constant terms of the $2$--loop results,
   Eqs. (\ref{aQg2}, \ref{aQq2PS}, \ref{aqq2NSQ}, \ref{agg2Q}, \ref{agq2Q}), 
   allows to determine the $T_F$--terms 
   of the $3$--loop anomalous dimensions for fixed values of $N$.
   All our results 
   agree with the results of Refs. \cite{Gracey:1993nn,Larin:1996wd,Retey:2000nq,Moch:2002sn,Moch:2004pa,Vogt:2004mw}. Note that 
   in this way we obtain the complete expressions for the 
   terms $\gamma_{qg}^{(2)}$ and $\gamma_{qq}^{(2), {\sf PS}}$, since 
   they always involve an overall factor $T_F$. For them we obtain

\vspace*{2mm}\noindent
\underline{$(i)$~~~\large $\hat{\gamma}_{qg}^{(2)}$}~:
%%%%%%%%%%%%%%%%%%%%%%%%%%%%%%%%%%
%
% gqg2hat
%
%%%%%%%%%%%%%%%%%%%%%%%%%%%%%%%%%%
\begin{eqnarray}
  \hat{\gamma}_{qg}^{(2)}(2)&=&T_F\Biggl[
                          (1+2n_f)T_F \Bigl(
                             \frac{8464}{243}C_A
                            -\frac{1384}{243}C_F
                                        \Bigr)
                          +\frac{\zeta_3}{3} \Bigl(
                           -416{C_AC_F}
                           +288{C_A^2}
\N\\ \N\\ && \hspace{-15mm}
                           +128{C_F^2}
                          \Bigr)
                         -\frac{7178}{81}{C_A^2}
                         +\frac{556}{9}{C_AC_F}
                         -\frac{8620}{243}{C_F^2}~\Biggr]~,  \\ \N \\
%%%%%%%%%%%%%%%%%%%%%%%%%%%%
  \hat{\gamma}_{qg}^{(2)}(4)&=&T_F\Biggl[
                          (1+2n_f)T_F \Bigl(
                             \frac{4481539}{303750}C_A
                            +\frac{9613841}{3037500}C_F
                                        \Bigr)
                          +\frac{\zeta_3}{25} \Bigl(
                             2832{C_A^2}
                            -3876{C_AC_F}
\N\\ \N\\ && \hspace{-15mm}
                            +1044{C_F^2}
                         \Bigr)
                         -\frac{295110931}{3037500}{C_A^2}
                         +\frac{278546497}{2025000}{C_AC_F}
                         -\frac{757117001}{12150000}{C_F^2}~\Biggr]~, \\ \N \\
%%%%%%%%%%%%%%%%%%%%%%%%%%%%
  \hat{\gamma}_{qg}^{(2)}(6)&=&T_F\Biggl[
                          (1+2n_f)T_F \Bigl(
                             \frac{86617163}{11668860}C_A
                            +\frac{1539874183}{340341750}C_F
                                        \Bigr)
                          +\frac{\zeta_3}{735} \Bigl(
                              69864{C_A^2}
\N\\ \N\\ && \hspace{-15mm}
                             -94664{C_AC_F}
                             +24800{C_F^2}
                          \Bigr)
                         -\frac{58595443051}{653456160}{C_A^2}
                         +\frac{1199181909343}{8168202000}{C_AC_F}
\N\\ \N\\ && \hspace{-15mm}
                         -\frac{2933980223981}{40841010000}{C_F^2}~\Biggr]~, 
                   \\  \N \\
%%%%%%%%%%%%%%%%%%%%%%%%%%%%
  \hat{\gamma}_{qg}^{(2)}(8)&=&T_F\Biggl[
                          (1+2n_f)T_F \Bigl(
                          \frac{10379424541}{2755620000}C_A
                         +\frac{7903297846481}{1620304560000}C_F
                                        \Bigr)
\N\\ \N\\ && \hspace{-15mm}
                          +\zeta_3 \Bigl(
                          \frac{128042}{1575}{C_A^2}
                         -\frac{515201}{4725}{C_AC_F}
                         +\frac{749}{27}{C_F^2}
                          \Bigr)
                         -\frac{24648658224523}{289340100000}{C_A^2}
\N\\ \N\\ && \hspace{-15mm}
                         +\frac{4896295442015177}{32406091200000}{C_AC_F}
                     -\frac{4374484944665803}{56710659600000}{C_F^2}~\Biggr]~,
             \\ \N \\
%%%%%%%%%%%%%%%%%%%%%%%%%%%%
  \hat{\gamma}_{qg}^{(2)}(10)&=&T_F\Biggl[
                          (1+2n_f)T_F \Bigl(
                              \frac{1669885489}{988267500}C_A
                             +\frac{1584713325754369}{323600780868750}C_F
               \Bigr)
\N \\ \N \\
          && \hspace{-15mm}
                          +\zeta_3 \Bigl(
                          \frac{1935952}{27225}{C_A^2}
                         -\frac{2573584}{27225}{C_AC_F}
                         +\frac{70848}{3025}{C_F^2}
                          \Bigr)
                         -\frac{21025430857658971}{255684567600000}{C_A^2} \N
\end{eqnarray}
\begin{eqnarray}
           && \hspace{-15mm}
                       +\frac{926990216580622991}{6040547909550000}{C_AC_F}
                       -\frac{1091980048536213833}{13591232796487500}{C_F^2}~\Biggr]~.
\end{eqnarray}
%%%%%%%%%%%%%%%%%%%%%%%%%%%%

%
\vspace*{2mm}\noindent
\underline{$(ii)$~~~\large $\hat{\gamma}_{qq}^{(2), {\sf PS}}$}~:
%%%%%%%%%%%%%%%%%%%%%%%%%%%%%%%%%%
%
% gqq2PS
%
%%%%%%%%%%%%%%%%%%%%%%%%%%%%%%%%%%
\begin{eqnarray}
 \hat{\gamma}_{qq}^{(2),{\sf PS}}(2)&=&T_FC_F\Biggl[
                         -(1+2n_f)T_F\frac{5024}{243}
                         +\frac{256}{3}\Bigl(C_F-C_A\Bigl)\zeta_3
                         +\frac{10136}{243}C_A
\N\\ \N\\ && \hspace{-15mm}
                         -\frac{14728}{243}C_F~\Biggr]~,  \\ \N \\
%%%%%%%%%%%%%%%%%%%%%%%%%%%%
 \hat{\gamma}_{qq}^{(2),{\sf PS}}(4)&=&T_FC_F\Biggl[
                         -(1+2n_f)T_F\frac{618673}{151875}
                         +\frac{968}{75}\Bigl(C_F-C_A\Bigl)\zeta_3
                         +\frac{2485097}{506250}C_A
\N\\ \N\\ && \hspace{-15mm}
                         -\frac{2217031}{675000}C_F~\Biggr]~, \\ \N \\
%%%%%%%%%%%%%%%%%%%%%%%%%%%%
 \hat{\gamma}_{qq}^{(2),{\sf PS}}(6)&=&T_FC_F\Biggl[
                         -(1+2n_f)T_F\frac{126223052}{72930375}
                         +\frac{3872}{735}\Bigl(C_F-C_A\Bigl)\zeta_3
\N\\ \N\\ && \hspace{-15mm}
                         +\frac{1988624681}{4084101000}C_A
                         +\frac{11602048711}{10210252500}C_F~\Biggr], \\ \N \\
%%%%%%%%%%%%%%%%%%%%%%%%%%%%
 \hat{\gamma}_{qq}^{(2),{\sf PS}}(8)&=&T_FC_F\Biggl[
                         -(1+2n_f)T_F\frac{13131081443}{13502538000}
                         +\frac{2738}{945}\Bigl(C_F-C_A\Bigl)\zeta_3
\N\\ \N\\ && \hspace{-15mm}
                         -\frac{343248329803}{648121824000}C_A
                         +\frac{39929737384469}{22684263840000}C_F~\Biggr]~,
                        \\ \N \\
%%%%%%%%%%%%%%%%%%%%%%%%%%%%
 \hat{\gamma}_{qq}^{(2),{\sf PS}}(10)&=&T_FC_F\Biggl[
                          -(1+2n_f)T_F\frac{265847305072}{420260754375}
                         +\frac{50176}{27225}\Bigl(C_F-C_A\Bigl)\zeta_3
\N\\ \N\\ && \hspace{-15mm}
                         -\frac{1028766412107043}{1294403123475000}C_A
                         +\frac{839864254987192}{485401171303125}C_F~\Biggr]~,
                       \\ \N \\
%%%%%%%%%%%%%%%%%%%%%%%%%%%%
 \hat{\gamma}_{qq}^{(2),{\sf PS}}(12)&=&T_FC_F\Biggl[
                          -(1+2n_f)T_F\frac{2566080055386457}{5703275664286200}
                         +\frac{49928}{39039}\Bigl(C_F-C_A\Bigl)\zeta_3
\N\\ \N\\ && \hspace{-15mm}
                         -\frac{69697489543846494691}{83039693672007072000}C_A
                         +\frac{86033255402443256197}{54806197823524667520}C_F~\Biggr]~.
\end{eqnarray}
%%%%%%%%%%%%%%%%%%%%%%%%%%%%

  For the remaining terms, only the projection onto the color factor $T_F$
  can be obtained~:

\vspace*{2mm}\noindent
\underline{$(iii)$~~~\large $\hat{\gamma}_{qq}^{(2), {\sf NS,+}}$}~:
%%%%%%%%%%%%%%%%%%%%%%%%%%%%%%%%%%
%
% gqq2NSplus
%
%%%%%%%%%%%%%%%%%%%%%%%%%%%%%%%%%%
\begin{eqnarray}
 \hat{\gamma}_{qq}^{(2),{\sf NS},+}(2)&=&T_FC_F\Biggl[
                         -(1+2n_f)T_F\frac{1792}{243}
                         +\frac{256}{3}\Bigl(C_F-C_A\Bigl)\zeta_3
                         -\frac{12512}{243}C_A \N
\end{eqnarray}
\begin{eqnarray}
%\N\\ \N\\
&& \hspace{-15mm}
                         -\frac{13648}{243}C_F~\Biggr]~, \\ \N \\
%%%%%%%%%%%%%%%%%%%%%%%%%%%%
 \hat{\gamma}_{qq}^{(2),{\sf NS},+}(4)&=&T_FC_F\Biggl[
                         -(1+2n_f)T_F\frac{384277}{30375}
                         +\frac{2512}{15}\Bigl(C_F-C_A\Bigl)\zeta_3
\N\\ \N\\ && \hspace{-15mm}
                         -\frac{8802581}{121500}C_A
                         -\frac{165237563}{1215000}C_F~\Biggr]~,\\ \N \\
%%%%%%%%%%%%%%%%%%%%%%%%%%%%
 \hat{\gamma}_{qq}^{(2),{\sf NS},+}(6)&=&T_FC_F\Biggl[
                         -(1+2n_f)T_F\frac{160695142}{10418625}
                         +\frac{22688}{105}\Bigl(C_F-C_A\Bigl)\zeta_3
\N\\ \N\\ && \hspace{-15mm}
                         -\frac{13978373}{171500}C_A
                         -\frac{44644018231}{243101250}C_F~\Biggr]~,\\ \N \\
%%%%%%%%%%%%%%%%%%%%%%%%%%%%
 \hat{\gamma}_{qq}^{(2),{\sf NS},+}(8)&=&T_FC_F\Biggl[
                         -(1+2n_f)T_F\frac{38920977797}{2250423000}
                         +\frac{79064}{315}\Bigl(C_F-C_A\Bigl)\zeta_3
\N\\ \N\\ && \hspace{-15mm}
                         -\frac{1578915745223}{18003384000}C_A
                         -\frac{91675209372043}{420078960000}C_F~\Biggr]~,
\\ \N \\
%%%%%%%%%%%%%%%%%%%%%%%%%%%%
 \hat{\gamma}_{qq}^{(2),{\sf NS},+}(10)&=&T_FC_F\Biggl[
                         -(1+2n_f)T_F\frac{27995901056887}{1497656506500}
                         +\frac{192880}{693}\Bigl(C_F-C_A\Bigl)\zeta_3
\N\\ \N\\ && \hspace{-15mm}
                         -\frac{9007773127403}{97250422500}C_A
                         -\frac{75522073210471127}{307518802668000}C_F~\Biggr]~,
\\ \N \\
%%%%%%%%%%%%%%%%%%%%%%%%%%%%
 \hat{\gamma}_{qq}^{(2),{\sf NS},+}(12)&=&T_FC_F\Biggl[
                         -(1+2n_f)T_F\frac{65155853387858071}{3290351344780500}
                         +\frac{13549568}{45045}\Bigl(C_F-C_A\Bigl)\zeta_3
\N\\ \N\\ && \hspace{-15mm}
                         -\frac{25478252190337435009}{263228107582440000}C_A
                         -\frac{35346062280941906036867}{131745667845011220000}C_F~\Biggr]~,
\N \\ \\
%%%%%%%%%%%%%%%%%%%%%%%%%%%%
 \hat{\gamma}_{qq}^{(2),{\sf NS},+}(14)&=&T_FC_F\Biggl[
                       -(1+2n_f)T_F\frac{68167166257767019}{3290351344780500}
                       +\frac{2881936}{9009}\Bigl(C_F-C_A\Bigr)\zeta_3
\N\\ && \hspace{-15mm}
                       -\frac{92531316363319241549}{921298376538540000}C_A
                     -\frac{37908544797975614512733}{131745667845011220000}C_F
                                       \Biggr]~.\N\\ 
\end{eqnarray}
%%%%%%%%%%%%%%%%%%%%%%%%%%%%

%
\vspace*{2mm}\noindent
\underline{$(iv)$~~~\large $\hat{\gamma}_{qq}^{(2), {\sf NS,-}}$}~:
%%%%%%%%%%%%%%%%%%%%%%%%%%%%%%%%%%
%
% gqq2NSminus
%
%%%%%%%%%%%%%%%%%%%%%%%%%%%%%%%%%%
\begin{eqnarray}
 {\hat{\gamma}_{qq}^{(2),{\sf NS},-}(1)}&=&  0~, \\
%%%%%%%%%%%%%%%%%%%%%%%%%%%%
 \hat{\gamma}_{qq}^{(2),{\sf NS},-}(3)&=&T_FC_F\Biggl[
                         -(1+2n_f)T_F\frac{2569}{243}
                         +\frac{400}{3}\Bigl(C_F-C_A\Bigl)\zeta_3
                         -\frac{62249}{972}C_A \N
\end{eqnarray}
\begin{eqnarray}
%\N\\ \N\\
 && \hspace{-15mm}
                         -\frac{203627}{1944}C_F~\Biggr]~, \\ \N \\
%%%%%%%%%%%%%%%%%%%%%%%%%%%%
 \hat{\gamma}_{qq}^{(2),{\sf NS},-}(5)&=&T_FC_F\Biggl[
                         -(1+2n_f)T_F\frac{431242}{30375}
                         +\frac{2912}{15}\Bigl(C_F-C_A\Bigl)\zeta_3
\N\\ \N\\ && \hspace{-15mm}
                         -\frac{38587}{500}C_A
                         -\frac{5494973}{33750}C_F~\Biggr]~, \\ \N \\
%%%%%%%%%%%%%%%%%%%%%%%%%%%%
 \hat{\gamma}_{qq}^{(2),{\sf NS},-}(7)&=&T_FC_F\Biggl[
                         -(1+2n_f)T_F\frac{1369936511}{83349000}
                         +\frac{8216}{35}\Bigl(C_F-C_A\Bigl)\zeta_3
\N\\ \N\\ && \hspace{-15mm}
                         -\frac{2257057261}{26671680}C_A
                         -\frac{3150205788689}{15558480000}C_F~\Biggr]~,\\ \N \\
%%%%%%%%%%%%%%%%%%%%%%%%%%%%
 \hat{\gamma}_{qq}^{(2),{\sf NS},-}(9)&=&T_FC_F\Biggl[
                         -(1+2n_f)T_F\frac{20297329837}{1125211500}
                         +\frac{16720}{63}\Bigl(C_F-C_A\Bigl)\zeta_3
\N\\ \N\\ && \hspace{-15mm}
                         -\frac{126810403414}{1406514375}C_A
                         -\frac{1630263834317}{7001316000}C_F~\Biggr]~,\\ \N \\
%%%%%%%%%%%%%%%%%%%%%%%%%%%%
 \hat{\gamma}_{qq}^{(2),{\sf NS},-}(11)&=&T_FC_F\Biggl[
                         -(1+2n_f)T_F\frac{28869611542843}{1497656506500}
                         +\frac{1005056}{3465}\Bigl(C_F-C_A\Bigl)\zeta_3
\N\\ \N\\ && \hspace{-15mm}
                         -\frac{1031510572686647}{10892047320000}C_A
                         -\frac{1188145134622636787}{4612782040020000}C_F~\Biggr]~,
 \\ \N \\
%%%%%%%%%%%%%%%%%%%%%%%%%%%%
 \hat{\gamma}_{qq}^{(2),{\sf NS},-}(13)&=&T_FC_F\Biggl[
                         -(1+2n_f)T_F\frac{66727681292862571}{3290351344780500}
                         +\frac{13995728}{45045}\Bigl(C_F-C_A\Bigl)\zeta_3
\N\\ \N\\ && \hspace{-15mm}
                         -\frac{90849626920977361109}{921298376538540000}C_A
                         -\frac{36688336888519925613757}{131745667845011220000}C_F~\Biggr]~. \N \\
\end{eqnarray}
%%%%%%%%%%%%%%%%%%%%%%%%%%%%

%
\vspace*{2mm}\noindent
\underline{$(v)$~~~\large $\hat{\gamma}_{gg}^{(2)}$}~:
%%%%%%%%%%%%%%%%%%%%%%%%%%%%%%%%%%
%
% ggg2hat
%                        
%%%%%%%%%%%%%%%%%%%%%%%%%%%%%%%%%%
\begin{eqnarray}
  \hat{\gamma}_{gg}^{(2)}(2)&=&T_F\Biggl[
                         (1+2n_f)T_F \Bigl(
                             -\frac{8464}{243}C_A
                             +\frac{1384}{243}C_F
                                       \Bigr)
                         +\frac{\zeta_3}{3} \Bigl(
                               -288{C_A^2}
                               +416C_AC_F 
\N \\ \N \\
     && \hspace{-15mm}
                               -128{C_F^2}
                                                  \Bigr)
                        +\frac{7178}{81}{C_A^2}
                        -\frac{556}{9}C_AC_F
                        +\frac{8620}{243}{C_F^2}~\Biggr]~,   \\ \N \\
%%%%%%%%%%%%%%%%%%%%%%%%%%%%
  \hat{\gamma}_{gg}^{(2)}(4)&=&T_F\Biggl[
                         (1+2n_f)T_F \Bigl(
                             -\frac{757861}{30375}C_A
                             -\frac{979774}{151875}C_F
                                       \Bigr)
                         +\frac{\zeta_3}{25} \Bigl(
                             -6264{C_A^2}
                             +6528C_AC_F \N
\end{eqnarray}
\begin{eqnarray}
%\N \\ \N \\
&& \hspace{-15mm}
                             -264{C_F^2}
                          \Bigr)
                         +\frac{53797499}{607500}{C_A^2}
                         -\frac{235535117}{1012500}C_AC_F
                         +\frac{2557151}{759375}{C_F^2}~\Biggr]~,\\ \N \\
%%%%%%%%%%%%%%%%%%%%%%%%%%%%
  \hat{\gamma}_{gg}^{(2)}(6)&=&T_F\Biggl[
                         (1+2n_f)T_F \Bigl(
                             -\frac{52781896}{2083725}C_A
                             -\frac{560828662}{72930375}C_F
                                       \Bigr)
                         +\zeta_3 \Bigl(
                        -\frac{75168}{245}{C_A^2}
\N \\ \N \\ && \hspace{-15mm}
                        +\frac{229024}{735}C_AC_F
                        -\frac{704}{147}{C_F^2}
                                                  \Bigr)
                         +\frac{9763460989}{116688600}{C_A^2}
                         -\frac{9691228129}{32672808}C_AC_F
\N \\ \N \\ && \hspace{-15mm}
                         -\frac{11024749151}{10210252500}{C_F^2}~\Biggr]~, 
\\ \N \\
%%%%%%%%%%%%%%%%%%%%%%%%%%%%
  \hat{\gamma}_{gg}^{(2)}(8)&=&T_F\Biggl[
                         (1+2n_f)T_F \Bigl(
                           -\frac{420970849}{16074450}C_A
                           -\frac{6990254812}{843908625}C_F
                                       \Bigr)
\N \\ \N \\ && \hspace{-15mm}
                         +\zeta_3 \Bigl(
                             -\frac{325174}{945}{C_A^2}
                             +\frac{327764}{945}C_AC_F
                             -\frac{74}{27}{C_F^2}
                          \Bigr)
                         +\frac{2080130771161}{25719120000}{C_A^2}
\N \\ \N \\ && \hspace{-15mm}
                         -\frac{220111823810087}{648121824000}C_AC_F
                         -\frac{14058417959723}{5671065960000}{C_F^2}~\Biggr]~, 
\\ \N \\
%%%%%%%%%%%%%%%%%%%%%%%%%%%%
  \hat{\gamma}_{gg}^{(2)}(10)&=&T_F\Biggl[
                         (1+2n_f)T_F \Bigl(
                            -\frac{2752314359}{101881395}C_A
                            -\frac{3631303571944}{420260754375}C_F
                                       \Bigr)
\N \\ \N \\ && \hspace{-15mm}
                         +\zeta_3 \Bigl(
                            -\frac{70985968}{190575}{C_A^2}
                            +\frac{71324656}{190575}C_AC_F
                            -\frac{5376}{3025}{C_F^2}
                                                  \Bigr)
\N \\ \N \\ && \hspace{-15mm}
                         +\frac{43228502203851731}{549140719050000}{C_A^2}
                         -\frac{3374081335517123191}{9060821864325000}C_FC_A
\N \\ \N \\ && \hspace{-15mm}
                         -\frac{3009386129483453}{970802342606250}{C_F^2}~\Biggr]~. 
\end{eqnarray}
%%%%%%%%%%%%%%%%%%%%%%%%%%%%

%
\vspace*{2mm}\noindent
\underline{$(vi)$~~~\large $\hat{\gamma}_{gq}^{(2)}$}~:
%%%%%%%%%%%%%%%%%%%%%%%%%%%%%%%%%%
%
% ggq2hat
%
%%%%%%%%%%%%%%%%%%%%%%%%%%%%%%%%%%
\begin{eqnarray}
  \hat{\gamma}_{gq}^{(2)}(2)&=&T_FC_F\Biggl[
                          (1+2n_f)T_F\frac{2272}{81}
                         +\frac{512}{3}\Bigl(C_A-C_F\Bigl)\zeta_3
                         +\frac{88}{9}C_A
                         +\frac{28376}{243}C_F~\Biggr]~, \N \\  \\
%%%%%%%%%%%%%%%%%%%%%%%%%%%%
  \hat{\gamma}_{gq}^{(2)}(4)&=&T_FC_F\Biggl[
                          (1+2n_f)T_F\frac{109462}{10125}
                         +\frac{704}{15}\Bigl(C_A-C_F\Bigl)\zeta_3
                         -\frac{799}{12150}C_A
\N\\ \N\\ && \hspace{-15mm}
                         +\frac{14606684}{759375}C_F~\Biggr]~,\\ \N \\
%%%%%%%%%%%%%%%%%%%%%%%%%%%%
  \hat{\gamma}_{gq}^{(2)}(6)&=&T_FC_F\Biggl[
                          (1+2n_f)T_F\frac{22667672}{3472875}
                         +\frac{2816}{105}\Bigl(C_A-C_F\Bigl)\zeta_3
                         -\frac{253841107}{145860750}C_A \N
\end{eqnarray}
\begin{eqnarray}
%\N\\ \N\\ 
&& \hspace{-15mm}
                         +\frac{20157323311}{2552563125}C_F~\Biggr]~,\\ \N \\
%%%%%%%%%%%%%%%%%%%%%%%%%%%%
  \hat{\gamma}_{gq}^{(2)}(8)&=&T_FC_F\Biggl[
                          (1+2n_f)T_F\frac{339184373}{75014100}
                         +\frac{1184}{63}\Bigl(C_A-C_F\Bigl)\zeta_3
\N\\ \N\\ && \hspace{-15mm}
                         -\frac{3105820553}{1687817250}C_A
                         +\frac{8498139408671}{2268426384000}C_F~\Biggr]~, 
\\ \N \\
%%%%%%%%%%%%%%%%%%%%%%%%%%%%
  \hat{\gamma}_{gq}^{(2)}(10)&=&T_FC_F\Biggl[
                          (1+2n_f)T_F\frac{1218139408}{363862125}
                         +\frac{7168}{495}\Bigl(C_A-C_F\Bigl)\zeta_3
\N\\ \N\\ && \hspace{-15mm}
                         -\frac{18846629176433}{11767301122500}C_A
                         +\frac{529979902254031}{323600780868750}C_F~\Biggr]~,
\\ \N \\
%%%%%%%%%%%%%%%%%%%%%%%%%%%%
  \hat{\gamma}_{gq}^{(2)}(12)&=&T_FC_F\Biggl[
                          (1+2n_f)T_F\frac{13454024393417}{5222779912350}
                         +\frac{5056}{429}\Bigl(C_A-C_F\Bigl)\zeta_3
\N\\ \N\\ && \hspace{-15mm}
                         -\frac{64190493078139789}{48885219979596000}C_A
                         +\frac{1401404001326440151}{3495293228541114000}C_F~\Biggr]~,
\\ \N \\
%%%%%%%%%%%%%%%%%%%%%%%%%%%%
  \hat{\gamma}_{gq}^{(2)}(14)&=&T_FC_F\Biggl[
                        (1+2n_f)T_F\frac{19285002274}{9495963477}
                       +\frac{13568}{1365}\Bigl(C_A-C_F\Bigr) \zeta_3 
\N \\ \N \\ && \hspace{-15mm}
                       -\frac{37115284124613269}{35434552943790000}C_A
                       -\frac{40163401444446690479}{104797690331258925000}C_F
                                       \Biggr]~.
\end{eqnarray}
%%%%%%%%%%%%%%%%%%%%%%%%%%%%%%%%%%
%%%%%%%%%%%%%%%%%%%%%%%%%%%%%%%%%%%%%%%%%%%%%%%%%%%%%%%%%%%%%%%%%%%%%%%%%%%%%
%
% Appendix 6
%
% Moments of the constants Terms of the $3$--Loop Massive OMEs
%
%%%%%%%%%%%%%%%%%%%%%%%%%%%%%%%%%%%%%%%%%%%%%%%%%%%%%%%%%%%%%%%%%%%%%%%%%%%%%
\newpage
  \section{\bf \boldmath 
            The $O(\ep^0)$ Contributions to $\Ahathat_{ij}^{(3)}$}
%%%%%%%%%%%%%%%%%%%%%%%%%%%%%%%%%%%%%%%%%%%%%%%%%%%%%%%%%%%%%%%%%%%%%%%%%%%%%
   \label{App-OMEs}
   \renewcommand{\theequation}{\thesection.\arabic{equation}}
   \setcounter{equation}{0}
%%%%%%%%%%%%%%%%%%%%%%%%%%%%%%%%%%%%%%%%%%%%%%%%%%%%%%%%%%%%%%%%%%%%%%%%%%%%%
   Finally, we present all moments we calculated. We only give the constant 
   term in $\ep$ of the unrenormalized result, cf. Eqs. (\ref{Ahhhqq3NSQ},
   \ref{AhhhQq3PS}, \ref{Ahhhqq3PSQ}, \ref{AhhhQg3}, \ref{Ahhhqg3Q}, 
   \ref{AhhhgqQ3}, \ref{Ahhhgg3Q}). These terms have to be inserted 
   into the general results on the renormalized level, cf. 
   Eqs. (\ref{Aqq3NSQMSren}, \ref{AQq3PSMSren}, \ref{Aqq3PSQMSren}, 
   \ref{AQg3MSren}, \ref{Aqg3QMSren}, \ref{Agq3QMSren}, \ref{Agg3QMSren}). 
   We obtain 

\vspace*{2mm}\noindent
\underline{$(i)$~\large $a_{Qq}^{(3), \sf PS}$}~:
%%%%%%%%%%%%%%%
\begin{eqnarray}
a_{Qq}^{(3), {\sf PS}}(2)&=&
T_FC_FC_A
      \Biggl( 
                 \frac{117290}{2187}
                +\frac{64}{9}{\sf B_4}-64\zeta_4
                +\frac{1456}{27}\zeta_3
                +\frac{224}{81}\zeta_2
      \Biggr)
\N \\ \N \\ && \hspace{-15mm}
+T_FC_F^2
      \Biggl( 
                 \frac{42458}{243}
                -\frac{128}{9}{\sf B_4}+64\zeta_4
                -\frac{9664}{81}\zeta_3
                +\frac{704}{27}\zeta_2
      \Biggr)
\N \\ \N \\ && \hspace{-15mm}
+T_F^2C_F
      \Biggl( 
                -\frac{36880}{2187}
                -\frac{4096}{81}\zeta_3
                -\frac{736}{81}\zeta_2
      \Biggr)
\N \\ \N \\ && \hspace{-15mm}
+n_fT_F^2C_F
      \Biggl( 
                -\frac{76408}{2187}
                +\frac{896}{81}\zeta_3
                -\frac{112}{81}\zeta_2
      \Biggr)~,
\\
a_{Qq}^{(3), {\sf PS}}(4)&=&
T_FC_FC_A
      \Biggl( 
                 \frac{23115644813}{1458000000}
                +\frac{242}{225}{\sf B_4}
                -\frac{242}{25}\zeta_4
                +\frac{1403}{180}\zeta_3
                +\frac{283481}{270000}\zeta_2
      \Biggr)
\N \\ \N \\ && \hspace{-15mm}
+T_FC_F^2
      \Biggl( 
                -\frac{181635821459}{8748000000}
                -\frac{484}{225}{\sf B_4}
                +\frac{242}{25}\zeta_4
                +\frac{577729}{40500}\zeta_3
\N \\ \N \\ && \hspace{-15mm}
                +\frac{4587077}{1620000}\zeta_2
      \Biggr)
+T_F^2C_F
      \Biggl( 
                -\frac{2879939}{5467500}
                -\frac{15488}{2025}\zeta_3
                -\frac{1118}{2025}\zeta_2
      \Biggr)
\N \\ \N \\ && \hspace{-15mm}
+n_fT_F^2C_F
      \Biggl( 
                -\frac{474827503}{109350000}
                +\frac{3388}{2025}\zeta_3
                -\frac{851}{20250}\zeta_2
      \Biggr)~,
\\
a_{Qq}^{(3), {\sf PS}}(6)&=&
T_FC_FC_A
      \Biggl( 
                 \frac{111932846538053}{10291934520000}
                +\frac{968}{2205}{\sf B_4}
                -\frac{968}{245}\zeta_4
                +\frac{2451517}{1852200}\zeta_3
\N \\ \N \\ && \hspace{-15mm}
                +\frac{5638039}{7779240}\zeta_2
      \Biggr)
+T_FC_F^2
      \Biggl( 
                -\frac{238736626635539}{5145967260000}
                -\frac{1936}{2205}{\sf B_4}
                +\frac{968}{245}\zeta_4
\N \\ \N \\ && \hspace{-15mm}
                +\frac{19628197}{555660}\zeta_3
                +\frac{8325229}{10804500}\zeta_2
      \Biggr)
+T_F^2C_F
      \Biggl( 
                 \frac{146092097}{1093955625}
                -\frac{61952}{19845}\zeta_3
\N \\ \N \\ && \hspace{-15mm}
                -\frac{7592}{99225}\zeta_2
      \Biggr)
+n_fT_F^2C_F
      \Biggl( 
                -\frac{82616977}{45378900}
                +\frac{1936}{2835}\zeta_3
                -\frac{16778}{694575}\zeta_2
      \Biggr)~,
\\
a_{Qq}^{(3), {\sf PS}}(8)&=&
T_FC_FC_A
      \Biggl( 
                 \frac{314805694173451777}{32665339929600000}
                +\frac{1369}{5670}{\sf B_4}
                -\frac{1369}{630}\zeta_4\N
\end{eqnarray}
\begin{eqnarray}
%\N \\ \N \\
&& \hspace{-15mm}
                -\frac{202221853}{137168640}\zeta_3
                +\frac{1888099001}{3429216000}\zeta_2
      \Biggr)
+T_FC_F^2
      \Biggl( 
                -\frac{25652839216168097959}{457314759014400000}
\N \\ \N \\ && \hspace{-15mm}
                -\frac{1369}{2835}{\sf B_4}
                +\frac{1369}{630}\zeta_4
                +\frac{2154827491}{48988800}\zeta_3
                +\frac{12144008761}{48009024000}\zeta_2
      \Biggr)
\N \\ \N \\ && \hspace{-15mm}
+T_F^2C_F
      \Biggl( 
                 \frac{48402207241}{272211166080}
                -\frac{43808}{25515}\zeta_3
                +\frac{1229}{142884}\zeta_2
      \Biggr)
\N \\ \N \\ && \hspace{-15mm}
+n_fT_F^2C_F
      \Biggl( 
                -\frac{16194572439593}{15122842560000}
                +\frac{1369}{3645}\zeta_3
                -\frac{343781}{14288400}\zeta_2
      \Biggr)~,
\\
%\end{eqnarray}
%\begin{eqnarray}
a_{Qq}^{(3), {\sf PS}}(10)&=&
T_FC_FC_A
      \Biggl( 
                 \frac{989015303211567766373}{107642563748181000000}
                +\frac{12544}{81675}{\sf B_4}
                -\frac{12544}{9075}\zeta_4
\N \\ \N \\ && \hspace{-15mm}
                -\frac{1305489421}{431244000}\zeta_3
                +\frac{2903694979}{6670805625}\zeta_2
      \Biggr)
+T_FC_F^2
      \Biggl( 
                -\frac{4936013830140976263563}{80731922811135750000}
\N \\ \N \\ && \hspace{-15mm}
                -\frac{25088}{81675}{\sf B_4}
                +\frac{12544}{9075}\zeta_4
                +\frac{94499430133}{1940598000}\zeta_3
                +\frac{282148432}{4002483375}\zeta_2
      \Biggr)
\N \\ \N \\ && \hspace{-15mm}
+T_F^2C_F
      \Biggl( 
                 \frac{430570223624411}{2780024890190625}
                -\frac{802816}{735075}\zeta_3
                +\frac{319072}{11026125}\zeta_2
      \Biggr)
\N \\ \N \\ && \hspace{-15mm}
+n_fT_F^2C_F
      \Biggl( 
                -\frac{454721266324013}{624087220246875}
                +\frac{175616}{735075}\zeta_3
                -\frac{547424}{24257475}\zeta_2
      \Biggr)~,
\\
a_{Qq}^{(3), {\sf PS}}(12)&=&
T_FC_FC_A
      \Biggl( 
                 \frac{968307050156826905398206547}{107727062441920086477312000}
                +\frac{12482}{117117}{\sf B_4}
                -\frac{12482}{13013}\zeta_4
\N \\ \N \\ && \hspace{-15mm}
                -\frac{64839185833913}{16206444334080}\zeta_3
                +\frac{489403711559293}{1382612282251200}\zeta_2
      \Biggr)
\N \\ \N \\ && \hspace{-15mm}
+T_FC_F^2
      \Biggl( 
                -\frac{190211298439834685159055148289}{2962494217152802378126080000}
                -\frac{24964}{117117}{\sf B_4}
                +\frac{12482}{13013}\zeta_4
\N \\ \N \\ && \hspace{-15mm}
                +\frac{418408135384633}{8103222167040}\zeta_3
                -\frac{72904483229177}{15208735104763200}\zeta_2
      \Biggr)
\N \\ \N \\ && \hspace{-15mm}
+T_F^2C_F
      \Biggl( 
                 \frac{1727596215111011341}{13550982978344011200}
                -\frac{798848}{1054053}\zeta_3
                +\frac{11471393}{347837490}\zeta_2
      \Biggr)
\N \\ \N \\ && \hspace{-15mm}
+n_fT_F^2C_F
      \Biggl( 
                -\frac{6621557709293056160177}{12331394510293050192000}
                +\frac{24964}{150579}\zeta_3
\N \\ \N \\ && \hspace{-15mm}
                -\frac{1291174013}{63306423180}\zeta_2
      \Biggr)~.
\end{eqnarray} 
%%%%%%%%%%%%%%%%%

\vspace*{2mm}\noindent
\underline{$(ii)$~\large $a_{qq,Q}^{(3), \sf PS}$}~: 
%%%%%%%%%%%%%%%
\begin{eqnarray}
a_{qq,Q}^{(3), {\sf PS}}(2)&=&
n_fT_F^2C_F
      \Biggl( 
                -\frac{100096}{2187}
                +\frac{896}{81}\zeta_3
                -\frac{256}{81}\zeta_2
      \Biggr)~, \\
a_{qq,Q}^{(3), {\sf PS}}(4)&=&
n_fT_F^2C_F
      \Biggl( 
                -\frac{118992563}{21870000}
                +\frac{3388}{2025}\zeta_3
                -\frac{4739}{20250}\zeta_2
      \Biggr)~, \\
a_{qq,Q}^{(3), {\sf PS}}(6)&=&
n_fT_F^2C_F
      \Biggl( 
                -\frac{17732294117}{10210252500}
                +\frac{1936}{2835}\zeta_3
                -\frac{9794}{694575}\zeta_2
      \Biggr)~,\\
a_{qq,Q}^{(3), {\sf PS}}(8)&=&
n_fT_F^2C_F
      \Biggl( 
                -\frac{20110404913057}{27221116608000}
                +\frac{1369}{3645}\zeta_3
                +\frac{135077}{4762800}\zeta_2
      \Biggr)~,\\
a_{qq,Q}^{(3), {\sf PS}}(10)&=&
n_fT_F^2C_F
      \Biggl( 
                -\frac{308802524517334}{873722108345625}
                +\frac{175616}{735075}\zeta_3
                +\frac{4492016}{121287375}\zeta_2
      \Biggr)~,\\
a_{qq,Q}^{(3), {\sf PS}}(12)&=&
n_fT_F^2C_F
      \Biggl( 
                -\frac{6724380801633998071}{38535607844665781850}
                +\frac{24964}{150579}\zeta_3
                +\frac{583767694}{15826605795}\zeta_2
      \Biggr)~, \N \\ \\
a_{qq,Q}^{(3), {\sf PS}}(14)&=&
n_fT_F^2C_F
      \Biggl( 
                -\frac{616164615443256347333}{7545433703850642600000}
                +\frac{22472}{184275}\zeta_3
\N\\ \N \\ && \hspace{-15mm}
                +\frac{189601441}{5533778250}\zeta_2
      \Biggr)~.
\N\\
\end{eqnarray}
%%%%%%%%%%%%%%%

\vspace*{2mm}\noindent
\underline{$(iii)$~\large $a_{Qg}^{\rm (3)}$}~: 
%%%%%%%%%%%%%%%
\begin{eqnarray}
a_{Qg}^{(3)}(2)&=&
T_FC_A^2
      \Biggl( 
                 \frac{170227}{4374}
                -\frac{88}{9}{\sf B_4}
                +72\zeta_4
                -\frac{31367}{324}\zeta_3
                +\frac{1076}{81}\zeta_2
      \Biggr)
\N \\ \N \\ && \hspace{-15mm}
+T_FC_FC_A
      \Biggl( 
                -\frac{154643}{729}
                +\frac{208}{9}{\sf B_4}
                -104\zeta_4
                +\frac{7166}{27}\zeta_3
                -54\zeta_2
      \Biggr)
\N \\ \N \\ && \hspace{-15mm}
+T_FC_F^2
      \Biggl( 
                -\frac{15574}{243}
                -\frac{64}{9}{\sf B_4}+32\zeta_4
                -\frac{3421}{81}\zeta_3
                +\frac{704}{27}\zeta_2
      \Biggr)
+T_F^2C_A
      \Biggl( 
                -\frac{20542}{2187}
\N \\ \N \\ && \hspace{-15mm}
                +\frac{4837}{162}\zeta_3
                -\frac{670}{81}\zeta_2
      \Biggr)
+T_F^2C_F
      \Biggl( 
                 \frac{11696}{729}
                +\frac{569}{81}\zeta_3
                +\frac{256}{9}\zeta_2
      \Biggr)
                -\frac{64}{27}T_F^3\zeta_3
\N \\ \N \\ && \hspace{-15mm}
+n_fT_F^2C_A
      \Biggl( 
                -\frac{6706}{2187}
                -\frac{616}{81}\zeta_3
                -\frac{250}{81}\zeta_2
      \Biggr)
+n_fT_F^2C_F
      \Biggl( 
                 \frac{158}{243}
                +\frac{896}{81}\zeta_3
                +\frac{40}{9}\zeta_2
      \Biggr)~, \N\\
\end{eqnarray}
\begin{eqnarray}
a_{Qg}^{(3)}(4)&=&
T_FC_A^2
      \Biggl( 
                -\frac{425013969083}{2916000000}
                -\frac{559}{50}{\sf B_4}
                +\frac{2124}{25}\zeta_4
                -\frac{352717109}{5184000}\zeta_3
\N \\ \N \\ && \hspace{-15mm}
                -\frac{4403923}{270000}\zeta_2
      \Biggr)
+T_FC_FC_A
      \Biggl( 
                -\frac{95898493099}{874800000}
                +\frac{646}{25}{\sf B_4}
                -\frac{2907}{25}\zeta_4
\N \\ \N \\ && \hspace{-15mm}
                +\frac{172472027}{864000}\zeta_3
                -\frac{923197}{40500}\zeta_2
      \Biggr)
+T_FC_F^2
      \Biggl( 
                -\frac{87901205453}{699840000}
                -\frac{174}{25}{\sf B_4}
\N \\ \N \\ && \hspace{-15mm}
                +\frac{783}{25}\zeta_4
                +\frac{937829}{12960}\zeta_3
                +\frac{62019319}{3240000}\zeta_2
      \Biggr)
+T_F^2C_A
      \Biggl( 
                 \frac{960227179}{29160000}
                +\frac{1873781}{51840}\zeta_3
\N \\ \N \\&& \hspace{-15mm}
                +\frac{120721}{13500}\zeta_2
      \Biggr)
+T_F^2C_F
      \Biggl( 
                -\frac{1337115617}{874800000}
                +\frac{73861}{324000}\zeta_3
                +\frac{8879111}{810000}\zeta_2
      \Biggr)
\N \\ \N \\ && \hspace{-15mm}
                -\frac{176}{135}T_F^3\zeta_3
+n_fT_F^2C_A
      \Biggl( 
                 \frac{947836283}{72900000}
                -\frac{18172}{2025}\zeta_3
                -\frac{11369}{13500}\zeta_2
      \Biggr)
\N \\ \N \\ && \hspace{-15mm}
+n_fT_F^2C_F
      \Biggl( 
                 \frac{8164734347}{4374000000}
                +\frac{130207}{20250}\zeta_3
                +\frac{1694939}{810000}\zeta_2
      \Biggr)~, \\
a_{Qg}^{(3)}(6)&=&
T_FC_A^2
      \Biggl( 
                -\frac{48989733311629681}{263473523712000}
                -\frac{2938}{315}{\sf B_4}
                +\frac{17466}{245}\zeta_4
                -\frac{748603616077}{11379916800}\zeta_3
\N \\ \N \\ && \hspace{-15mm}
                -\frac{93013721}{3457440}\zeta_2
      \Biggr)
+T_FC_FC_A
      \Biggl( 
                 \frac{712876107019}{55319040000}
                +\frac{47332}{2205}{\sf B_4}
                -\frac{23666}{245}\zeta_4
\N \\ \N \\ && \hspace{-15mm}
                +\frac{276158927731}{1896652800}\zeta_3
                +\frac{4846249}{11113200}\zeta_2
      \Biggr)
+T_FC_F^2
      \Biggl( 
                -\frac{38739867811364113}{137225793600000}
\N \\ \N \\ && \hspace{-15mm}
                -\frac{2480}{441}{\sf B_4}
                +\frac{1240}{49}\zeta_4
                +\frac{148514798653}{711244800}\zeta_3
                +\frac{4298936309}{388962000}\zeta_2
      \Biggr)
\N \\ \N \\ && \hspace{-15mm}
+T_F^2C_A
      \Biggl( 
                 \frac{706058069789557}{18819537408000}
                +\frac{3393002903}{116121600}\zeta_3
                +\frac{6117389}{555660}\zeta_2
      \Biggr)
\N \\ \N \\ && \hspace{-15mm}
+T_F^2C_F
      \Biggl( 
                -\frac{447496496568703}{54890317440000}
                -\frac{666922481}{284497920}\zeta_3
                +\frac{49571129}{9724050}\zeta_2
      \Biggr)
\N \\ \N \\ && \hspace{-15mm}
                -\frac{176}{189}T_F^3\zeta_3
+n_fT_F^2C_A
      \Biggl( 
                 \frac{12648331693}{735138180}
                -\frac{4433}{567}\zeta_3
                +\frac{23311}{111132}\zeta_2
      \Biggr)
\N \\ \N \\ && \hspace{-15mm}
+n_fT_F^2C_F
      \Biggl( 
                -\frac{8963002169173}{1715322420000}
                +\frac{111848}{19845}\zeta_3
                +\frac{11873563}{19448100}\zeta_2
      \Biggr)~, \\
a_{Qg}^{(3)}(8)&=&
T_FC_A^2
      \Biggl( 
                -\frac{358497428780844484961}{2389236291993600000}
                -\frac{899327}{113400}{\sf B_4}
                +\frac{64021}{1050}\zeta_4\N
\end{eqnarray}\begin{eqnarray}
\N \\ \N \\ && \hspace{-15mm}
                -\frac{12321174818444641}{112368549888000}\zeta_3
                -\frac{19581298057}{612360000}\zeta_2
      \Biggr)
\N \\ \N \\ && \hspace{-15mm}
+T_FC_FC_A
      \Biggl( 
                 \frac{941315502886297276939}{8362327021977600000}
                +\frac{515201}{28350}{\sf B_4}
                -\frac{515201}{6300}\zeta_4
\N \\ \N \\ && \hspace{-15mm}
                +\frac{5580970944338269}{56184274944000}\zeta_3
                +\frac{495290785657}{34292160000}\zeta_2
      \Biggr)
\N \\ \N \\ && \hspace{-15mm}
+T_FC_F^2
      \Biggl( 
                -\frac{23928053971795796451443}{36585180721152000000}
                -\frac{749}{162}{\sf B_4}
                +\frac{749}{36}\zeta_4
\N \\ \N \\ && \hspace{-15mm}
                +\frac{719875828314061}{1404606873600}\zeta_3
                +\frac{2484799653079}{480090240000}\zeta_2
      \Biggr)
+T_F^2C_A
      \Biggl( 
                 \frac{156313300657148129}{4147979673600000}
\N \\ \N \\ && \hspace{-15mm}
                +\frac{58802880439}{2388787200}\zeta_3
                +\frac{46224083}{4082400}\zeta_2
      \Biggr)
+T_F^2C_F
      \Biggl( 
                -\frac{986505627362913047}{87107573145600000}
\N \\ \N \\ && \hspace{-15mm}
                -\frac{185046016777}{50164531200}\zeta_3
                +\frac{7527074663}{3429216000}\zeta_2
      \Biggr)
                -\frac{296}{405}T_F^3\zeta_3
\N \\ \N \\ && \hspace{-15mm}
+n_fT_F^2C_A
      \Biggl( 
                 \frac{24718362393463}{1322697600000}
                -\frac{125356}{18225}\zeta_3
                +\frac{2118187}{2916000}\zeta_2
      \Biggr)
\N \\ \N \\ && \hspace{-15mm}
+n_fT_F^2C_F
      \Biggl( 
                -\frac{291376419801571603}{32665339929600000}
                +\frac{887741}{174960}\zeta_3
                -\frac{139731073}{1143072000}\zeta_2
      \Biggr)~, \\
a_{Qg}^{(3)}(10)&=&
T_FC_A^2
      \Biggl( 
                 \frac{6830363463566924692253659}{685850575063965696000000}
                -\frac{563692}{81675}{\sf B_4}
\N \\ \N \\ && \hspace{-15mm}
                +\frac{483988}{9075}\zeta_4
                -\frac{103652031822049723}{415451499724800}\zeta_3
                -\frac{20114890664357}{581101290000}\zeta_2
      \Biggr)
\N \\ \N \\ && \hspace{-15mm}
+T_FC_FC_A
      \Biggl( 
                 \frac{872201479486471797889957487}{2992802509370032128000000}
                +\frac{1286792}{81675}{\sf B_4}
\N \\ \N \\ && \hspace{-15mm}
                -\frac{643396}{9075}\zeta_4
                -\frac{761897167477437907}{33236119977984000}\zeta_3
                +\frac{15455008277}{660342375}\zeta_2
      \Biggr)
\N \\ \N \\ && \hspace{-15mm}
+T_FC_F^2
      \Biggl( 
                -\frac{247930147349635960148869654541}{148143724213816590336000000}
                -\frac{11808}{3025}{\sf B_4}
\N \\ \N \\ && \hspace{-15mm}
                +\frac{53136}{3025}\zeta_4
                +\frac{9636017147214304991}{7122025709568000}\zeta_3
                +\frac{14699237127551}{15689734830000}\zeta_2
      \Biggr)\N
\end{eqnarray}\begin{eqnarray}
%\N \\ \N \\ 
&& \hspace{-15mm}
+T_F^2C_A
      \Biggl( 
                 \frac{23231189758106199645229}{633397356480430080000}
                +\frac{123553074914173}{5755172290560}\zeta_3
                +\frac{4206955789}{377338500}\zeta_2
      \Biggr)
\N \\ \N \\ && \hspace{-15mm}
+T_F^2C_F
      \Biggl( 
                -\frac{18319931182630444611912149}{1410892611560158003200000}
                -\frac{502987059528463}{113048027136000}\zeta_3
\N \\ \N \\ && \hspace{-15mm}
                +\frac{24683221051}{46695639375}\zeta_2
      \Biggr)
                -\frac{896}{1485}T_F^3\zeta_3
+n_fT_F^2C_A
      \Biggl( 
                 \frac{297277185134077151}{15532837481700000}
\N \\ \N \\ && \hspace{-15mm}
                -\frac{1505896}{245025}\zeta_3
                +\frac{189965849}{188669250}\zeta_2
      \Biggr)
+n_fT_F^2C_F
      \Biggl( 
                -\frac{1178560772273339822317}{107642563748181000000}
\N \\ \N \\ && \hspace{-15mm}
                +\frac{62292104}{13476375}\zeta_3
                -\frac{49652772817}{93391278750}\zeta_2
      \Biggr)~.
\end{eqnarray}
%%%%%%%%%%%%%%%

\vspace*{2mm}\noindent
\underline{$(iv)$~\large $a_{qg,Q}^{\rm (3)}$}~: 
%%%%%%%%%%%%%%%
\begin{eqnarray}\hspace{-6mm}
a_{qg,Q}^{(3)}(2)&=&
n_fT_F^2C_A
      \Biggl( 
                 \frac{83204}{2187}
                -\frac{616}{81}\zeta_3
                +\frac{290}{81}\zeta_2
      \Biggr)
\N \\ \N \\ && \hspace{-15mm}
+n_fT_F^2C_F
      \Biggl( 
                -\frac{5000}{243}
                +\frac{896}{81}\zeta_3
                -\frac{4}{3}\zeta_2
      \Biggr)~, \\
a_{qg,Q}^{(3)}(4)&=&
n_fT_F^2C_A
      \Biggl( 
                 \frac{835586311}{14580000}
                -\frac{18172}{2025}\zeta_3
                +\frac{71899}{13500}\zeta_2
      \Biggr)
\N \\ \N \\ && \hspace{-15mm}
+n_fT_F^2C_F
      \Biggl( 
                -\frac{21270478523}{874800000}
                +\frac{130207}{20250}\zeta_3
                -\frac{1401259}{810000}\zeta_2
      \Biggr)~,\\
a_{qg,Q}^{(3)}(6)&=&
n_fT_F^2C_A
      \Biggl( 
                 \frac{277835781053}{5881105440}
                -\frac{4433}{567}\zeta_3
                +\frac{2368823}{555660}\zeta_2
      \Biggr)
\N \\ \N \\ && \hspace{-15mm}
+n_fT_F^2C_F
      \Biggl( 
                -\frac{36123762156197}{1715322420000}
                +\frac{111848}{19845}\zeta_3
                -\frac{26095211}{19448100}\zeta_2
      \Biggr)~,\\
a_{qg,Q}^{(3)}(8)&=&
n_fT_F^2C_A
      \Biggl( 
                 \frac{157327027056457}{3968092800000}
                -\frac{125356}{18225}\zeta_3
                +\frac{7917377}{2268000}\zeta_2
      \Biggr)
\N \\ \N \\ && \hspace{-15mm}
+n_fT_F^2C_F
      \Biggl( 
                -\frac{201046808090490443}{10888446643200000}
                +\frac{887741}{174960}\zeta_3
\N \\ \N \\ && \hspace{-15mm}
                -\frac{3712611349}{3429216000}\zeta_2
      \Biggr)~,
\end{eqnarray}\begin{eqnarray}
a_{qg,Q}^{(3)}(10)&=&
n_fT_F^2C_A
      \Biggl( 
                 \frac{6542127929072987}{191763425700000}
                -\frac{1505896}{245025}\zeta_3
                +\frac{1109186999}{377338500}\zeta_2
      \Biggr) \N
\end{eqnarray}\begin{eqnarray}
%\N \\ \N \\ 
&& \hspace{-15mm}
+n_fT_F^2C_F
      \Biggl( 
                -\frac{353813854966442889041}{21528512749636200000}
                +\frac{62292104}{13476375}\zeta_3
%\N \\ \N \\ && \hspace{-15mm}
                -\frac{83961181063}{93391278750}\zeta_2
      \Biggr)~. \N \\
\end{eqnarray}
%%%%%%%%%%%%%%%

\vspace*{2mm}\noindent
\underline{$(v)$~\large $a_{gq,Q}^{\rm (3)}$}~: 
%%%%%%%%%%%%%%%
\begin{eqnarray}
a_{gq,Q}^{(3)}(2)&=&
T_FC_FC_A
      \Biggl( 
                -\frac{126034}{2187}
                -\frac{128}{9}{\sf B_4}+128\zeta_4
                -\frac{9176}{81}\zeta_3
                -\frac{160}{81}\zeta_2
      \Biggr)
\N \\ \N \\ && \hspace{-15mm}
+T_FC_F^2
      \Biggl( 
                -\frac{741578}{2187}
                +\frac{256}{9}{\sf B_4}-128\zeta_4
                +\frac{17296}{81}\zeta_3
                -\frac{4496}{81}\zeta_2
      \Biggr)
\N \\ \N \\ && \hspace{-15mm}
+T_F^2C_F
      \Biggl( 
                 \frac{21872}{729}
                +\frac{2048}{27}\zeta_3
                +\frac{416}{27}\zeta_2
      \Biggr)
+n_fT_F^2C_F
      \Biggl( 
                 \frac{92200}{729}
                -\frac{896}{27}\zeta_3
\N \\ \N \\ && \hspace{-15mm}
                +\frac{208}{27}\zeta_2
      \Biggr)~,\\
a_{gq,Q}^{(3)}(4)&=&
T_FC_FC_A
      \Biggl( 
                -\frac{5501493631}{218700000}
                -\frac{176}{45}{\sf B_4}
                +\frac{176}{5}\zeta_4
                -\frac{8258}{405}\zeta_3
\N \\ \N \\ && \hspace{-15mm}
                +\frac{13229}{8100}\zeta_2
      \Biggr)
+T_FC_F^2
      \Biggl( 
                -\frac{12907539571}{145800000}
                +\frac{352}{45}{\sf B_4}
                -\frac{176}{5}\zeta_4
\N \\ \N \\ && \hspace{-15mm}
                +\frac{132232}{2025}\zeta_3
                -\frac{398243}{27000}\zeta_2
      \Biggr)
\N \\ \N \\ && \hspace{-15mm}
+T_F^2C_F
      \Biggl( 
                 \frac{1914197}{911250}
                +\frac{2816}{135}\zeta_3
                +\frac{1252}{675}\zeta_2
      \Biggr)
\N \\ \N \\ && \hspace{-15mm}
+n_fT_F^2C_F
      \Biggl( 
                 \frac{50305997}{1822500}
                -\frac{1232}{135}\zeta_3
                +\frac{626}{675}\zeta_2
      \Biggr)~,\\
a_{gq,Q}^{(3)}(6)&=&
T_FC_FC_A
      \Biggl( 
                -\frac{384762916141}{24504606000}
                -\frac{704}{315}{\sf B_4}
                +\frac{704}{35}\zeta_4
                -\frac{240092}{19845}\zeta_3
\N \\ \N \\ && \hspace{-15mm}
                +\frac{403931}{463050}\zeta_2
      \Biggr)
+T_FC_F^2
      \Biggl( 
                -\frac{40601579774533}{918922725000}
                +\frac{1408}{315}{\sf B_4}
                -\frac{704}{35}\zeta_4
\N \\ \N \\ && \hspace{-15mm}
                +\frac{27512264}{694575}\zeta_3
                -\frac{24558841}{3472875}\zeta_2
      \Biggr)
+T_F^2C_F
      \Biggl( 
                -\frac{279734446}{364651875}
\N \\ \N \\ && \hspace{-15mm}
                +\frac{11264}{945}\zeta_3
                +\frac{8816}{33075}\zeta_2
      \Biggr)
+n_fT_F^2C_F
      \Biggl( 
                 \frac{4894696577}{364651875}
                -\frac{704}{135}\zeta_3\N
\end{eqnarray}\begin{eqnarray}
%\N \\ \N \\ && \hspace{-15mm}
&& \hspace{-15mm}
                +\frac{4408}{33075}\zeta_2
      \Biggr)~,\\
a_{gq,Q}^{(3)}(8)&=&
T_FC_FC_A
      \Biggl( 
                -\frac{10318865954633473}{816633498240000}
                -\frac{296}{189}{\sf B_4}
                +\frac{296}{21}\zeta_4
                -\frac{1561762}{178605}\zeta_3
\N \\ \N \\ && \hspace{-15mm}
                +\frac{30677543}{85730400}\zeta_2
      \Biggr)
+T_FC_F^2
      \Biggl( 
                -\frac{305405135103422947}{11432868975360000}
                +\frac{592}{189}{\sf B_4}
                -\frac{296}{21}\zeta_4
\N \\ \N \\ && \hspace{-15mm}
                +\frac{124296743}{4286520}\zeta_3
                -\frac{4826251837}{1200225600}\zeta_2
      \Biggr)
+T_F^2C_F
      \Biggl( 
                -\frac{864658160833}{567106596000}
\N \\ \N \\ && \hspace{-15mm}
                +\frac{4736}{567}\zeta_3
                -\frac{12613}{59535}\zeta_2
      \Biggr)
+n_fT_F^2C_F
      \Biggl( 
                 \frac{9330164983967}{1134213192000}
                -\frac{296}{81}\zeta_3
\N \\ \N \\ && \hspace{-15mm}
                -\frac{12613}{119070}\zeta_2
      \Biggr)~,\\
a_{gq,Q}^{(3)}(10)&=&
T_FC_FC_A
      \Biggl( 
                -\frac{1453920909405842897}{130475834846280000}
                -\frac{1792}{1485}{\sf B_4}
                +\frac{1792}{165}\zeta_4
                -\frac{1016096}{147015}\zeta_3
\N \\ \N \\ && \hspace{-15mm}
                +\frac{871711}{26952750}\zeta_2
      \Biggr)
+T_FC_F^2
      \Biggl( 
                -\frac{11703382372448370173}{667205973645750000}
                +\frac{3584}{1485}{\sf B_4}
\N \\ \N \\ && \hspace{-15mm} 
               -\frac{1792}{165}\zeta_4
                +\frac{62282416}{2695275}\zeta_3
                -\frac{6202346032}{2547034875}\zeta_2
      \Biggr)
+T_F^2C_F
      \Biggl( 
                -\frac{1346754066466}{756469357875}
\N \\ \N \\ && \hspace{-15mm}
                +\frac{28672}{4455}\zeta_3
                -\frac{297472}{735075}\zeta_2
      \Biggr)
+n_fT_F^2C_F
      \Biggl( 
                 \frac{4251185859247}{756469357875}
                -\frac{12544}{4455}\zeta_3
\N \\ \N \\ && \hspace{-15mm}
                -\frac{148736}{735075}\zeta_2
      \Biggr)~,\\
a_{gq,Q}^{(3)}(12)&=&
T_FC_FC_A
      \Biggl( 
                -\frac{1515875996003174876943331}{147976734123516602304000}
                -\frac{1264}{1287}{\sf B_4}
                +\frac{1264}{143}\zeta_4
\N \\ \N \\ && \hspace{-15mm}
                -\frac{999900989}{173918745}\zeta_3
                -\frac{693594486209}{3798385390800}\zeta_2
      \Biggr)
\N \\ \N \\ && \hspace{-15mm}
+T_FC_F^2
      \Biggl( 
                -\frac{48679935129017185612582919}{4069360188396706563360000}
                +\frac{2528}{1287}{\sf B_4}
                -\frac{1264}{143}\zeta_4
\N \\ \N \\ && \hspace{-15mm}
                +\frac{43693776149}{2260943685}\zeta_3
                -\frac{2486481253717}{1671289571952}\zeta_2
      \Biggr)
\N \\ \N \\ && \hspace{-15mm}
+T_F^2C_F
      \Biggl( 
                -\frac{2105210836073143063}{1129248581528667600}
                +\frac{20224}{3861}\zeta_3
                -\frac{28514494}{57972915}\zeta_2
      \Biggr)
\N \\ \N \\ && \hspace{-15mm}
+n_fT_F^2C_F
      \Biggl( 
                 \frac{9228836319135394697}{2258497163057335200}
                -\frac{8848}{3861}\zeta_3
                -\frac{14257247}{57972915}\zeta_2
      \Biggr)~,
\end{eqnarray}\begin{eqnarray}
a_{gq,Q}^{(3)}(14)&=&
T_FC_FC_A
      \Biggl( 
                -\frac{1918253569538142572718209}{199199449781656964640000}
                -\frac{3392}{4095}{\sf B_4}
\N \\ \N \\ && \hspace{-15mm}
                +\frac{3392}{455}\zeta_4
                -\frac{2735193382}{553377825}\zeta_3
                -\frac{1689839813797}{5113211103000}\zeta_2
      \Biggr)
\N \\ \N \\ && \hspace{-15mm}
+T_FC_F^2
      \Biggl( 
                -\frac{143797180510035170802620917}{17429951855894984406000000}
                +\frac{6784}{4095}{\sf B_4}
\N \\ \N \\ && \hspace{-15mm}
                -\frac{3392}{455}\zeta_4
                +\frac{12917466836}{774728955}\zeta_3
                -\frac{4139063104013}{4747981738500}\zeta_2
      \Biggr)
\N \\ \N \\ && \hspace{-15mm}
+T_F^2C_F
      \Biggl( 
                -\frac{337392441268078561}{179653183425015300}
                +\frac{54272}{12285}\zeta_3
                -\frac{98112488}{184459275}\zeta_2
      \Biggr)
\N \\ \N \\ && \hspace{-15mm}
+n_fT_F^2C_F
      \Biggl( 
                 \frac{222188365726202803}{71861273370006120}
                -\frac{3392}{1755}\zeta_3
                -\frac{49056244}{184459275}\zeta_2
      \Biggr)~.
\end{eqnarray}
%%%%%%%%%%%%%%%

\vspace*{2mm}\noindent
\underline{$(vi)$~\large $a_{gg,Q}^{\rm (3)}$}~: 
%%%%%%%%%%%%%%%
\begin{eqnarray}
a_{gg,Q}^{(3)}(2)&=&
T_FC_A^2
      \Biggl( 
                -\frac{170227}{4374}
                +\frac{88}{9}{\sf B_4}-72\zeta_4
                +\frac{31367}{324}\zeta_3
                -\frac{1076}{81}\zeta_2
      \Biggr)
\N \\ \N \\ && \hspace{-15mm}
+T_FC_FC_A
      \Biggl( 
                 \frac{154643}{729}
                -\frac{208}{9}{\sf B_4}
                +104\zeta_4
                -\frac{7166}{27}\zeta_3+54\zeta_2
      \Biggr)
\N \\ \N \\ && \hspace{-15mm}
+T_FC_F^2
      \Biggl( 
                 \frac{15574}{243}
                +\frac{64}{9}{\sf B_4}-32\zeta_4
                +\frac{3421}{81}\zeta_3
                -\frac{704}{27}\zeta_2
      \Biggr)
\N \\ \N \\ && \hspace{-15mm}
+T_F^2C_A
      \Biggl( 
                 \frac{20542}{2187}
                -\frac{4837}{162}\zeta_3
                +\frac{670}{81}\zeta_2
      \Biggr)
+T_F^2C_F
      \Biggl( 
                -\frac{11696}{729}
                -\frac{569}{81}\zeta_3
\N \\ \N \\ && \hspace{-15mm}
                -\frac{256}{9}\zeta_2
      \Biggr)
                +\frac{64}{27}T_F^3\zeta_3
+n_fT_F^2C_A
      \Biggl( 
                -\frac{76498}{2187}
                +\frac{1232}{81}\zeta_3
                -\frac{40}{81}\zeta_2
      \Biggr)
\N \\ \N \\ && \hspace{-15mm}
+n_fT_F^2C_F
      \Biggl( 
                 \frac{538}{27}
                -\frac{1792}{81}\zeta_3
                -\frac{28}{9}\zeta_2
      \Biggr)~,\\
a_{gg,Q}^{(3)}(4)&=&
T_FC_A^2
      \Biggl( 
                 \frac{29043652079}{291600000}
                +\frac{533}{25}{\sf B_4}
                -\frac{4698}{25}\zeta_4
                +\frac{610035727}{2592000}\zeta_3
\N \\ \N \\ && \hspace{-15mm}
                +\frac{92341}{6750}\zeta_2
      \Biggr)
+T_FC_FC_A
      \Biggl( 
                 \frac{272542528639}{874800000}
                -\frac{1088}{25}{\sf B_4}
                +\frac{4896}{25}\zeta_4
\N \\ \N \\ && \hspace{-15mm}
                -\frac{3642403}{17280}\zeta_3
                +\frac{73274237}{810000}\zeta_2
      \Biggr)
+T_FC_F^2
      \Biggl( 
                 \frac{41753961371}{1749600000}
\N
\end{eqnarray}\begin{eqnarray}
%\N \\ \N \\
&& \hspace{-15mm}
                +\frac{44}{25}{\sf B_4}
                -\frac{198}{25}\zeta_4
                +\frac{2676077}{64800}\zeta_3
                -\frac{4587077}{1620000}\zeta_2
      \Biggr)
+T_F^2C_A
      \Biggl( 
                -\frac{1192238291}{14580000}
\N \\ \N \\&& \hspace{-15mm}
                -\frac{2134741}{25920}\zeta_3
                -\frac{16091}{675}\zeta_2
      \Biggr)
+T_F^2C_F
      \Biggl( 
                -\frac{785934527}{43740000}
                -\frac{32071}{8100}\zeta_3
\N \\ \N \\ && \hspace{-15mm}
                -\frac{226583}{8100}\zeta_2
      \Biggr)
                +\frac{64}{27}T_F^3\zeta_3
+n_fT_F^2C_A
      \Biggl( 
                -\frac{271955197}{1822500}
                +\frac{13216}{405}\zeta_3
\N \\ \N \\ && \hspace{-15mm}
                -\frac{6526}{675}\zeta_2
      \Biggr)
+n_fT_F^2C_F
      \Biggl( 
                -\frac{465904519}{27337500}
                -\frac{6776}{2025}\zeta_3
                -\frac{61352}{10125}\zeta_2
      \Biggr)~,\\
a_{gg,Q}^{(3)}(6)&=&
T_FC_A^2
      \Biggl( 
                 \frac{37541473421359}{448084224000}
                +\frac{56816}{2205}{\sf B_4}
                -\frac{56376}{245}\zeta_4
                +\frac{926445489353}{2844979200}\zeta_3
\N \\ \N \\ && \hspace{-15mm}
                +\frac{11108521}{555660}\zeta_2
      \Biggr)
+T_FC_FC_A
      \Biggl( 
                 \frac{18181142251969309}{54890317440000}
                -\frac{114512}{2205}{\sf B_4}
                +\frac{57256}{245}\zeta_4
\N \\ \N \\ && \hspace{-15mm}
                -\frac{12335744909}{67737600}\zeta_3
                +\frac{94031857}{864360}\zeta_2
      \Biggr)
+T_FC_F^2
      \Biggl( 
                 \frac{16053159907363}{635304600000}
                +\frac{352}{441}{\sf B_4}
\N \\ \N \\ && \hspace{-15mm}
                -\frac{176}{49}\zeta_4
                +\frac{3378458681}{88905600}\zeta_3
                -\frac{8325229}{10804500}\zeta_2
      \Biggr)
+T_F^2C_A
      \Biggl( 
                -\frac{670098465769}{6001128000}
\N \\ \N \\ && \hspace{-15mm}
                -\frac{25725061}{259200}\zeta_3
                -\frac{96697}{2835}\zeta_2
      \Biggr)
+T_F^2C_F
      \Biggl( 
                -\frac{8892517283287}{490092120000}
                -\frac{12688649}{2540160}\zeta_3
\N \\ \N \\ && \hspace{-15mm}
                -\frac{2205188}{77175}\zeta_2
      \Biggr)
                +\frac{64}{27}T_F^3\zeta_3
+n_fT_F^2C_A
      \Biggl( 
                -\frac{245918019913}{1312746750}
                +\frac{3224}{81}\zeta_3
\N \\ \N \\ && \hspace{-15mm}
                -\frac{250094}{19845}\zeta_2
      \Biggr)
+n_fT_F^2C_F
      \Biggl( 
                -\frac{71886272797}{3403417500}
                -\frac{3872}{2835}\zeta_3
                -\frac{496022}{77175}\zeta_2
      \Biggr)~,\\
a_{gg,Q}^{(3)}(8)&=&
T_FC_A^2
      \Biggl( 
                 \frac{512903304712347607}{18665908531200000}
                +\frac{108823}{3780}{\sf B_4}
                -\frac{162587}{630}\zeta_4
\N \\ \N \\ && \hspace{-15mm}
                +\frac{2735007975361}{6502809600}\zeta_3
                +\frac{180224911}{7654500}\zeta_2
      \Biggr)
\N \\ \N \\ && \hspace{-15mm}
+T_FC_FC_A
      \Biggl( 
                 \frac{13489584043443319991}{43553786572800000}
                -\frac{163882}{2835}{\sf B_4}
                +\frac{81941}{315}\zeta_4
\N \\ \N \\ && \hspace{-15mm}
                -\frac{3504113623243}{25082265600}\zeta_3
                +\frac{414844703639}{3429216000}\zeta_2
      \Biggr)
\N \\ \N \\ && \hspace{-15mm}
+T_FC_F^2
      \Biggl( 
                 \frac{5990127272073225467}{228657379507200000}\N
                +\frac{37}{81}{\sf B_4}
                -\frac{37}{18}\zeta_4
                +\frac{3222019505879}{87787929600}\zeta_3 \N
\N
\end{eqnarray}\begin{eqnarray}
%\N \\ \N \\ 
&& \hspace{-15mm}
                -\frac{12144008761}{48009024000}\zeta_2
      \Biggr)
+T_F^2C_A
      \Biggl( 
                -\frac{16278325750483243}{124439390208000}\N
\N \\ \N \\ && \hspace{-15mm}
                -\frac{871607413}{7962624}\zeta_3
                -\frac{591287}{14580}\zeta_2
      \Biggr)
+T_F^2C_F
      \Biggl( 
                -\frac{7458367007740639}{408316749120000}
\N \\ \N \\ && \hspace{-15mm}
                -\frac{291343229}{52254720}\zeta_3
                -\frac{2473768763}{85730400}\zeta_2
      \Biggr)
                +\frac{64}{27}T_F^3\zeta_3
\N \\ \N \\ && \hspace{-15mm}
+n_fT_F^2C_A
      \Biggl( 
                -\frac{102747532985051}{486091368000}
                +\frac{54208}{1215}\zeta_3
                -\frac{737087}{51030}\zeta_2
      \Biggr)
\N \\ \N \\ && \hspace{-15mm}
+n_fT_F^2C_F
      \Biggl( 
                -\frac{1145917332616927}{51039593640000}
                -\frac{2738}{3645}\zeta_3
                -\frac{70128089}{10716300}\zeta_2
      \Biggr)~,\\
a_{gg,Q}^{(3)}(10)&=&
T_FC_A^2
      \Biggl( 
                -\frac{15434483462331661005275759}{327337774462347264000000}
                +\frac{17788828}{571725}{\sf B_4}
\N \\ \N \\ && \hspace{-15mm}
                -\frac{17746492}{63525}\zeta_4
                +\frac{269094476549521109}{519314374656000}\zeta_3
                +\frac{1444408720649}{55468759500}\zeta_2
      \Biggr)
\N \\ \N \\ && \hspace{-15mm}
+T_FC_FC_A
      \Biggl( 
                 \frac{207095356146239371087405921}{771581896946961408000000}
                -\frac{35662328}{571725}{\sf B_4}
\N \\ \N \\ && \hspace{-15mm}
                +\frac{17831164}{63525}\zeta_4
                -\frac{3288460968359099}{37093883904000}\zeta_3
                +\frac{6078270984602}{46695639375}\zeta_2
      \Biggr)
\N \\ \N \\ && \hspace{-15mm}
+T_FC_F^2
      \Biggl( 
                 \frac{553777925867720521493231}{20667372239650752000000}
                +\frac{896}{3025}{\sf B_4}
                -\frac{4032}{3025}\zeta_4
\N \\ \N \\ && \hspace{-15mm}
                +\frac{7140954579599}{198717235200}\zeta_3
                -\frac{282148432}{4002483375}\zeta_2
      \Biggr)
\N \\ \N \\ && \hspace{-15mm}
+T_F^2C_A
      \Biggl( 
                -\frac{63059843481895502807}{433789788579840000}
                -\frac{85188238297}{729907200}\zeta_3
                -\frac{33330316}{735075}\zeta_2
      \Biggr)
\N \\ \N \\ && \hspace{-15mm}
+T_F^2C_F
      \Biggl( 
                -\frac{655690580559958774157}{35787657557836800000}
                -\frac{71350574183}{12043468800}\zeta_3
                -\frac{3517889264}{121287375}\zeta_2
      \Biggr)
\N \\ \N \\ && \hspace{-15mm}
                +\frac{64}{27}T_F^3\zeta_3
+n_fT_F^2C_A
      \Biggl( 
                -\frac{6069333056458984}{26476427525625}
                +\frac{215128}{4455}\zeta_3
                -\frac{81362132}{5145525}\zeta_2
      \Biggr)
\N \\ \N \\ && \hspace{-15mm}
+n_fT_F^2C_F
      \Biggl( 
                -\frac{100698363899844296}{4368610541728125}
                -\frac{351232}{735075}\zeta_3
                -\frac{799867252}{121287375}\zeta_2
      \Biggr)~.
\end{eqnarray}
%%%%%%%%%%%%%%%

\vspace*{2mm}\noindent
\underline{$(vii)$~\large $a_{qq,Q}^{(3), \sf NS}$}~: 
%%%%%%%%%%%%%%%
\begin{eqnarray}
a_{qq,Q}^{(3), {\sf NS}}(1)&=& 0~,\\
a_{qq,Q}^{(3), {\sf NS}}(2)&=&
T_FC_FC_A
      \Biggl( 
                 \frac{8744}{2187}
                +\frac{64}{9}{\sf B_4}
                -64\zeta_4
                +\frac{4808}{81}\zeta_3
                -\frac{64}{81}\zeta_2
      \Biggr)
\N \\ \N \\ && \hspace{-15mm}
+T_FC_F^2
      \Biggl( 
                 \frac{359456}{2187}
                -\frac{128}{9}{\sf B_4}
                +64\zeta_4
                -\frac{848}{9}\zeta_3
                +\frac{2384}{81}\zeta_2
      \Biggr)
\N \\ \N \\ && \hspace{-15mm}
+T_F^2C_F
      \Biggl( 
                -\frac{28736}{2187}
                -\frac{2048}{81}\zeta_3
                -\frac{512}{81}\zeta_2
      \Biggr)
+n_fT_F^2C_F
      \Biggl( 
                -\frac{100096}{2187}
\N \\ \N \\ && \hspace{-15mm}
                +\frac{896}{81}\zeta_3
                -\frac{256}{81}\zeta_2
      \Biggr)~,\\
a_{qq,Q}^{(3), {\sf NS}}(3)&=&
T_FC_FC_A
      \Biggl( 
                 \frac{522443}{34992}
                +\frac{100}{9}{\sf B_4}-100\zeta_4
                +\frac{15637}{162}\zeta_3
                +\frac{175}{162}\zeta_2
      \Biggr)
\N \\ \N \\ && \hspace{-15mm}
+T_FC_F^2
      \Biggl( 
                 \frac{35091701}{139968}
                -\frac{200}{9}{\sf B_4}+100\zeta_4
                -\frac{1315}{9}\zeta_3
                +\frac{29035}{648}\zeta_2
      \Biggr)
\N \\ \N \\ 
&& \hspace{-15mm}
+T_F^2C_F
      \Biggl( 
                -\frac{188747}{8748}
                -\frac{3200}{81}\zeta_3
                -\frac{830}{81}\zeta_2
      \Biggr)
\N \\ \N \\ && \hspace{-15mm}
+n_fT_F^2C_F
      \Biggl( 
                -\frac{1271507}{17496}
                +\frac{1400}{81}\zeta_3
                -\frac{415}{81}\zeta_2
      \Biggr)~,\\
a_{qq,Q}^{(3), {\sf NS}}(4)&=&
T_FC_FC_A
      \Biggl( 
                 \frac{419369407}{21870000}
                +\frac{628}{45}{\sf B_4}
                -\frac{628}{5}\zeta_4
                +\frac{515597}{4050}\zeta_3
                +\frac{10703}{4050}\zeta_2
      \Biggr)
\N \\ \N \\ && \hspace{-15mm}
+T_FC_F^2
      \Biggl( 
                 \frac{137067007129}{437400000}
                -\frac{1256}{45}{\sf B_4}
                +\frac{628}{5}\zeta_4
                -\frac{41131}{225}\zeta_3
\N \\ \N \\ && \hspace{-15mm}
                +\frac{4526303}{81000}\zeta_2
      \Biggr)
+T_F^2C_F
      \Biggl( 
                -\frac{151928299}{5467500}
                -\frac{20096}{405}\zeta_3
                -\frac{26542}{2025}\zeta_2
      \Biggr)
\N \\ \N \\ && \hspace{-15mm}
+n_fT_F^2C_F
      \Biggl( 
                -\frac{1006358899}{10935000}
                +\frac{8792}{405}\zeta_3
                -\frac{13271}{2025}\zeta_2
      \Biggr)~,\\
a_{qq,Q}^{(3), {\sf NS}}(5)&=&
T_FC_FC_A
      \Biggl( 
                 \frac{816716669}{43740000}
                +\frac{728}{45}{\sf B_4}
                -\frac{728}{5}\zeta_4
                +\frac{12569}{81}\zeta_3
                +\frac{16103}{4050}\zeta_2
      \Biggr)
\N \\ \N \\ && \hspace{-15mm}
+T_FC_F^2
      \Biggl( 
                 \frac{13213297537}{36450000}
                -\frac{1456}{45}{\sf B_4}
                +\frac{728}{5}\zeta_4
                -\frac{142678}{675}\zeta_3
\N \\ \N \\ && \hspace{-15mm}
                +\frac{48391}{750}\zeta_2
      \Biggr)
+T_F^2C_F
      \Biggl( 
                -\frac{9943403}{303750}
                -\frac{23296}{405}\zeta_3
                -\frac{31132}{2025}\zeta_2
      \Biggr)
\end{eqnarray}\begin{eqnarray}
%\N \\ \N \\ 
&& \hspace{-15mm}
+n_fT_F^2C_F
      \Biggl( 
                -\frac{195474809}{1822500}
                +\frac{10192}{405}\zeta_3
                -\frac{15566}{2025}\zeta_2
      \Biggr)~, \\
a_{qq,Q}^{(3), {\sf NS}}(6)&=&
T_FC_FC_A
      \Biggl( 
                 \frac{1541550898907}{105019740000}
                +\frac{5672}{315}{\sf B_4}
                -\frac{5672}{35}\zeta_4
\N \\ \N \\ && \hspace{-15mm}
                +\frac{720065}{3969}\zeta_3
                +\frac{1016543}{198450}\zeta_2
      \Biggr)
+T_FC_F^2
      \Biggl( 
                 \frac{186569400917}{463050000}
\N \\ \N \\ && \hspace{-15mm}
                -\frac{11344}{315}{\sf B_4}
                +\frac{5672}{35}\zeta_4
                -\frac{7766854}{33075}\zeta_3
                +\frac{55284811}{771750}\zeta_2
      \Biggr)
\N \\ \N \\ && \hspace{-15mm}
+T_F^2C_F
      \Biggl( 
                -\frac{26884517771}{729303750}
                -\frac{181504}{2835}\zeta_3
                -\frac{1712476}{99225}\zeta_2
      \Biggr)
\N \\ \N \\ && \hspace{-15mm}
+n_fT_F^2C_F
      \Biggl( 
                -\frac{524427335513}{4375822500}
                +\frac{11344}{405}\zeta_3
                -\frac{856238}{99225}\zeta_2
      \Biggr)~,\\
a_{qq,Q}^{(3), {\sf NS}}(7)&=&
T_FC_FC_A
      \Biggl( 
                 \frac{5307760084631}{672126336000}
                +\frac{2054}{105}{\sf B_4}
                -\frac{6162}{35}\zeta_4
\N \\ \N \\ && \hspace{-15mm}
                +\frac{781237}{3780}\zeta_3
                +\frac{19460531}{3175200}\zeta_2
      \Biggr)
+T_FC_F^2
      \Biggl( 
                 \frac{4900454072126579}{11202105600000}
\N \\ \N \\ && \hspace{-15mm}
                -\frac{4108}{105}{\sf B_4}
                +\frac{6162}{35}\zeta_4
                -\frac{8425379}{33075}\zeta_3
                +\frac{1918429937}{24696000}\zeta_2
      \Biggr)
\N \\ \N \\ && \hspace{-15mm}
+T_F^2C_F
      \Biggl( 
                -\frac{8488157192423}{210039480000}
                -\frac{65728}{945}\zeta_3
                -\frac{3745727}{198450}\zeta_2
      \Biggr)
\N \\ \N \\ && \hspace{-15mm}
+n_fT_F^2C_F
      \Biggl( 
                -\frac{54861581223623}{420078960000}
                +\frac{4108}{135}\zeta_3
                -\frac{3745727}{396900}\zeta_2
      \Biggr)~,\\
a_{qq,Q}^{(3), {\sf NS}}(8)&=&
T_FC_FC_A
      \Biggl( 
                -\frac{37259291367883}{38887309440000}
                +\frac{19766}{945}{\sf B_4}
                -\frac{19766}{105}\zeta_4
\N \\ \N \\ && \hspace{-15mm}
                +\frac{1573589}{6804}\zeta_3
                +\frac{200739467}{28576800}\zeta_2
      \Biggr)
+T_FC_F^2
      \Biggl( 
                 \frac{3817101976847353531}{8166334982400000}
\N \\ \N \\ && \hspace{-15mm}
                -\frac{39532}{945}{\sf B_4}
                +\frac{19766}{105}\zeta_4
                -\frac{80980811}{297675}\zeta_3
                +\frac{497748102211}{6001128000}\zeta_2
      \Biggr)
\N \\ \N \\ && \hspace{-15mm}
+T_F^2C_F
      \Biggl( 
                -\frac{740566685766263}{17013197880000}
                -\frac{632512}{8505}\zeta_3
                -\frac{36241943}{1786050}\zeta_2
      \Biggr)
\N \\ \N \\ && \hspace{-15mm}
+n_fT_F^2C_F
      \Biggl( 
                -\frac{4763338626853463}{34026395760000}
                +\frac{39532}{1215}\zeta_3
                -\frac{36241943}{3572100}\zeta_2
      \Biggr)~,
\end{eqnarray}\begin{eqnarray}
a_{qq,Q}^{(3), {\sf NS}}(9)&=&
T_FC_FC_A
      \Biggl( 
                -\frac{3952556872585211}{340263957600000}
                +\frac{4180}{189}{\sf B_4}
                -\frac{4180}{21}\zeta_4
\N \\ \N \\ && \hspace{-15mm}
                +\frac{21723277}{85050}\zeta_3
                +\frac{559512437}{71442000}\zeta_2
      \Biggr)
+T_FC_F^2
      \Biggl( 
                 \frac{1008729211999128667}{2041583745600000}
\N \\ \N \\ && \hspace{-15mm}
                -\frac{8360}{189}{\sf B_4}
                +\frac{4180}{21}\zeta_4
                -\frac{85539428}{297675}\zeta_3
                +\frac{131421660271}{1500282000}\zeta_2
      \Biggr)
\N \\ \N \\ && \hspace{-15mm}
+T_F^2C_F
      \Biggl( 
                -\frac{393938732805271}{8506598940000}
                -\frac{133760}{1701}\zeta_3
                -\frac{19247947}{893025}\zeta_2
      \Biggr)
\N \\ \N \\ && \hspace{-15mm}
+n_fT_F^2C_F
      \Biggl( 
                -\frac{2523586499054071}{17013197880000}
                +\frac{8360}{243}\zeta_3
                -\frac{19247947}{1786050}\zeta_2
      \Biggr)~,\\
a_{qq,Q}^{(3), {\sf NS}}(10)&=&
T_FC_FC_A
      \Biggl( 
                -\frac{10710275715721975271}{452891327565600000}
                +\frac{48220}{2079}{\sf B_4}
\N \\ \N \\ && \hspace{-15mm}
                -\frac{48220}{231}\zeta_4
                +\frac{2873636069}{10291050}\zeta_3
                +\frac{961673201}{112266000}\zeta_2
      \Biggr)
\N \\ \N \\ && \hspace{-15mm}
+T_FC_F^2
      \Biggl( 
                 \frac{170291990048723954490137}{328799103812625600000}
                -\frac{96440}{2079}{\sf B_4}
\N \\ \N \\ && \hspace{-15mm}
                +\frac{48220}{231}\zeta_4
                -\frac{10844970868}{36018675}\zeta_3
                +\frac{183261101886701}{1996875342000}\zeta_2
      \Biggr)
\N \\ \N \\ && \hspace{-15mm}
+T_F^2C_F
      \Biggl( 
                -\frac{6080478350275977191}{124545115080540000}
                -\frac{1543040}{18711}\zeta_3
\N \\ \N \\ && \hspace{-15mm}
                -\frac{2451995507}{108056025}\zeta_2
      \Biggr)
+n_fT_F^2C_F
      \Biggl( 
                -\frac{38817494524177585991}{249090230161080000}
\N \\ \N \\ && \hspace{-15mm}
                +\frac{96440}{2673}\zeta_3
                -\frac{2451995507}{216112050}\zeta_2
      \Biggr)~, \\
a_{qq,Q}^{(3), {\sf NS}}(11)&=&
T_FC_FC_A
      \Biggl( 
                -\frac{22309979286641292041}{603855103420800000}
                +\frac{251264}{10395}{\sf B_4}
\N \\ \N \\ && \hspace{-15mm}
                -\frac{251264}{1155}\zeta_4
                +\frac{283300123}{935550}\zeta_3
                +\frac{1210188619}{130977000}\zeta_2
      \Biggr)
\N \\ \N \\ && \hspace{-15mm}
+T_FC_F^2
      \Biggl( 
                 \frac{177435748292579058982241}{328799103812625600000}
                -\frac{502528}{10395}{\sf B_4}
\N \\ \N \\ && \hspace{-15mm}
                +\frac{251264}{1155}\zeta_4
                -\frac{451739191}{1440747}\zeta_3
                +\frac{47705202493793}{499218835500}\zeta_2
      \Biggr) \N
\end{eqnarray}\begin{eqnarray}
%\N \\ \N \\ 
&& \hspace{-15mm}
+T_F^2C_F
      \Biggl( 
                -\frac{6365809346912279423}{124545115080540000}
                -\frac{8040448}{93555}\zeta_3
\N \\ \N \\ && \hspace{-15mm}
                -\frac{512808781}{21611205}\zeta_2
      \Biggr)
+n_fT_F^2C_F
      \Biggl( 
                -\frac{40517373495580091423}{249090230161080000}
\N \\ \N \\ && \hspace{-15mm}
                +\frac{502528}{13365}\zeta_3
                -\frac{512808781}{43222410}\zeta_2
      \Biggr)~,\\
a_{qq,Q}^{(3), {\sf NS}}(12)&=&
T_FC_FC_A
      \Biggl( 
                -\frac{126207343604156227942043}{2463815086971638400000}
                +\frac{3387392}{135135}{\sf B_4}
\N \\ \N \\ && \hspace{-15mm}
                -\frac{3387392}{15015}\zeta_4
                +\frac{51577729507}{158107950}\zeta_3
                +\frac{2401246832561}{243486243000}\zeta_2
      \Biggr)
\N \\ \N \\ && \hspace{-15mm}
+T_FC_F^2
      \Biggl( 
                 \frac{68296027149155250557867961293}{122080805651901196900800000}
                -\frac{6774784}{135135}{\sf B_4}
\N \\ \N \\ && \hspace{-15mm}
                +\frac{3387392}{15015}\zeta_4
                -\frac{79117185295}{243486243}\zeta_3
                +\frac{108605787257580461}{1096783781593500}\zeta_2
      \Biggr)
\N \\ \N \\ && \hspace{-15mm}
+T_F^2C_F
      \Biggl( 
                -\frac{189306988923316881320303}{3557133031815302940000}
                -\frac{108396544}{1216215}\zeta_3
\N \\ \N \\ && \hspace{-15mm}
                -\frac{90143221429}{3652293645}\zeta_2
      \Biggr)
+n_fT_F^2C_F
      \Biggl( 
                -\frac{1201733391177720469772303}{7114266063630605880000}
\N \\ \N \\ && \hspace{-15mm}
                +\frac{6774784}{173745}\zeta_3
                -\frac{90143221429}{7304587290}\zeta_2
      \Biggr)~,\\
a_{qq,Q}^{(3), {\sf NS}}(13)&=&
T_FC_FC_A
      \Biggl( 
                -\frac{12032123246389873565503373}{181090408892415422400000}
                +\frac{3498932}{135135}{\sf B_4}
\N \\ \N \\ && \hspace{-15mm}
                -\frac{3498932}{15015}\zeta_4
                +\frac{2288723461}{6548850}\zeta_3
                +\frac{106764723181157}{10226422206000}\zeta_2
      \Biggr)
\N \\ \N \\ && \hspace{-15mm}
+T_FC_F^2
      \Biggl( 
              \frac{10076195142551036234891679659}{17440115093128742414400000}
                -\frac{6997864}{135135}{\sf B_4}
\N \\ \N \\ && \hspace{-15mm}
                +\frac{3498932}{15015}\zeta_4
                -\frac{81672622894}{243486243}\zeta_3
                +\frac{448416864235277759}{4387135126374000}\zeta_2
      \Biggr)
\N \\ \N \\ && \hspace{-15mm}
+T_F^2C_F
      \Biggl( 
                -\frac{196243066652040382535303}{3557133031815302940000}
                -\frac{111965824}{1216215}\zeta_3
\N \\ \N \\ && \hspace{-15mm}
                -\frac{93360116539}{3652293645}\zeta_2
      \Biggr)
+n_fT_F^2C_F
      \Biggl( 
                -\frac{1242840812874342588467303}{7114266063630605880000}\N
\end{eqnarray}
\begin{eqnarray}
%\N \\ \N \\ 
&& \hspace{-15mm}
                +\frac{6997864}{173745}\zeta_3
                -\frac{93360116539}{7304587290}\zeta_2
      \Biggr)~,\\
a_{qq,Q}^{(3), {\sf NS}}(14)&=&
T_FC_FC_A
      \Biggl( 
                -\frac{994774587614536873023863}{12072693926161028160000}
                +\frac{720484}{27027}{\sf B_4}
\N \\ \N \\ && \hspace{-15mm}
                -\frac{720484}{3003}\zeta_4
                +\frac{6345068237}{17027010}\zeta_3
                +\frac{37428569944327}{3408807402000}\zeta_2
      \Biggr)
\N \\ \N \\ && \hspace{-15mm}
+T_FC_F^2
      \Biggl( 
                 \frac{72598193631729215117875463981}{122080805651901196900800000}
                -\frac{1440968}{27027}{\sf B_4}
\N \\ \N \\ && \hspace{-15mm}
                +\frac{720484}{3003}\zeta_4
                -\frac{2101051892878}{6087156075}\zeta_3
                +\frac{461388998135343407}{4387135126374000}\zeta_2
      \Biggr)
\N \\ \N \\ && \hspace{-15mm}
+T_F^2C_F
      \Biggl( 
                -\frac{40540032063650894708251}{711426606363060588000}
                -\frac{23055488}{243243}\zeta_3
\N \\ \N \\ && \hspace{-15mm}
                -\frac{481761665447}{18261468225}\zeta_2
      \Biggr)
+n_fT_F^2C_F
      \Biggl( 
                -\frac{256205552272074402170491}{1422853212726121176000}
\N \\ \N \\ && \hspace{-15mm}
                +\frac{1440968}{34749}\zeta_3
                -\frac{481761665447}{36522936450}\zeta_2
      \Biggr)~.
\end{eqnarray}
%%%%%%%%%%%%%%%%%%%%%%%%%%%%%%%%%%%%%%%%%%%%%%%%%%%%%%%%%%%%%%%%%%%%%%%%%%%%%
%%%%%%%%%%%%%%%%%%%%%%%%%%%%%%%%%%%%%%%%%%%%%%%%%%%%%%%%%%%%%%%%%%%%%%%%%%%%%
%
% Appendix 7
%
% $3$--loop Moments for Transversity
%
%%%%%%%%%%%%%%%%%%%%%%%%%%%%%%%%%%%%%%%%%%%%%%%%%%%%%%%%%%%%%%%%%%%%%%%%%%%%%
\newpage
  \section{\bf \boldmath $3$--loop Moments for Transversity}
%%%%%%%%%%%%%%%%%%%%%%%%%%%%%%%%%%%%%%%%%%%%%%%%%%%%%%%%%%%%%%%%%%%%%%%%%%%%%
   \label{App-Trans}
   \renewcommand{\theequation}{\thesection.\arabic{equation}}
   \setcounter{equation}{0}
%%%%%%%%%%%%%%%%%%%%%%%%%%%%%%%%%%%%%%%%%%%%%%%%%%%%%%%%%%%%%%%%%%%%%%%%%%%%%
   We obtain the following fixed moments of the fermionic contributions 
   to the $3$--loop 
   transversity anomalous dimension $\gamma_{qq}^{(2),{\sf TR}}(N)$
%%%%%%%%%%%%%%%%%%%%%%
\begin{eqnarray}
\hat{\gamma}_{qq}^{(2),{\sf TR}}(1)&=&C_FT_F\Biggl[
          -\frac{8}{3}T_F(1+2n_f)
          -\frac{2008}{27}C_A
\N\\ && \hspace{-15mm}
          +\frac{196}{9}C_F
          +32(C_F-C_A)\zeta_3
\Biggr]~,\\
%%%%%%%%%%%%%%%%%%%%%%
\hat{\gamma}_{qq}^{(2),{\sf TR}}(2)&=&C_FT_F\Biggl[
          -\frac{184}{27}T_F(1+2n_f)
          -\frac{2084}{27}C_A
\N\\ && \hspace{-15mm}
          -60C_F
          +96(C_F-C_A)\zeta_3
\Biggr]~,\\
%%%%%%%%%%%%%%%%%%%%%%
\hat{\gamma}_{qq}^{(2),{\sf TR}}(3)&=&C_FT_F\Biggl[
          -\frac{2408}{243}T_F(1+2n_f)
          -\frac{19450}{243}C_A
\N\\ && \hspace{-15mm}
          -\frac{25276}{243}C_F
          +\frac{416}{3}(C_F-C_A)\zeta_3
\Biggr]~,\\
%%%%%%%%%%%%%%%%%%%%%%
\hat{\gamma}_{qq}^{(2),{\sf TR}}(4)&=&C_FT_F\Biggl[
          -\frac{14722}{1215}T_F(1+2n_f)
          -\frac{199723}{2430}C_A
\N\\ && \hspace{-15mm}
          -\frac{66443}{486}C_F
          +\frac{512}{3}(C_F-C_A)\zeta_3
\Biggr]~,\N\\ \\
%%%%%%%%%%%%%%%%%%%%%%
\hat{\gamma}_{qq}^{(2),{\sf TR}}(5)&=&C_FT_F\Biggl[
          -\frac{418594}{30375}T_F(1+2n_f)
          -\frac{5113951}{60750}C_A
\N\\ && \hspace{-15mm}
          -\frac{49495163}{303750}C_F
          +\frac{2944}{15}(C_F-C_A)\zeta_3
\Biggr]~,\\
%%%%%%%%%%%%%%%%%%%%%%
\hat{\gamma}_{qq}^{(2),{\sf TR}}(6)&=&C_FT_F\Biggl[
          -\frac{3209758}{212625}T_F(1+2n_f)
          -\frac{3682664}{42525}C_A
\N\\ && \hspace{-15mm}
          -\frac{18622301}{101250}C_F
          +\frac{1088}{5}(C_F-C_A)\zeta_3
\Biggr]~,\\
%%%%%%%%%%%%%%%%%%%%%%
\hat{\gamma}_{qq}^{(2),{\sf TR}}(7)&=&C_FT_F\Biggl[
          -\frac{168501142}{10418625}T_F (1+2n_f)
          -\frac{1844723441}{20837250}C_A
\N\\ && \hspace{-15mm}
          -\frac{49282560541}{243101250}C_F
          +\frac{8256}{35}(C_F-C_A)\zeta_3
\Biggr]
\\
%%%%%%%%%%%%%%%%%%%%%%
\hat{\gamma}_{qq}^{(2),{\sf TR}}(8)&=&C_FT_F\Biggl[
          -\frac{711801943}{41674500} T_F(1+2n_f)
          -\frac{6056338297}{66679200}C_A \N
\end{eqnarray}
\begin{eqnarray}
\N\\ && \hspace{-15mm}
          -\frac{849420853541}{3889620000}C_F
          +\frac{8816}{35}(C_F-C_A)\zeta_3
\Biggr]
\end{eqnarray}
%%%%%%%%%%%%%%%%%%%%%%
These moments $(N=1..8)$ agree with the corresponding terms obtained in 
\cite{Gracey:2003yrxGracey:2003mrxGracey:2006zrxGracey:2006ah}.
The newly calculated moments read
%%%%%%%%%%%%%%%%%%%%%%
\begin{eqnarray}
\hat{\gamma}_{qq}^{(2),{\sf TR}}(9)&=&C_FT_F\Biggl[
          -\frac{20096458061}{1125211500}T_F(1+2n_f)
          -\frac{119131812533}{1285956000}C_A
\N\\ && \hspace{-15mm}
          -\frac{24479706761047}{105019740000}C_F
          +\frac{83824}{315}(C_F-C_A)\zeta_3
\Biggr]
\\
%%%%%%%%%%%%%%%%%%%%%%
\hat{\gamma}_{qq}^{(2),{\sf TR}}(10)&=&C_FT_F\Biggl[
          -\frac{229508848783}{12377326500}T_F(1+2n_f)
          -\frac{4264058299021}{45008460000}C_A
\N\\ && \hspace{-15mm}
          -\frac{25800817445759}{105019740000}C_F
          +\frac{87856}{315}(C_F-C_A)\zeta_3
\Biggr]
\\
%%%%%%%%%%%%%%%%%%%%%%
\hat{\gamma}_{qq}^{(2),{\sf TR}}(11)&=&C_FT_F\Biggl[
          -\frac{28677274464343}{1497656506500}T_F(1+2n_f)
          -\frac{75010870835743}{778003380000}C_A
\N \\
&& \hspace{-15mm}
          -\frac{396383896707569599}{1537594013340000}C_F
          +\frac{1006736}{3465}(C_F-C_A)\zeta_3
\Biggr]
\\
%%%%%%%%%%%%%%%%%%%%%%
\hat{\gamma}_{qq}^{(2),{\sf TR}}(12)&=&C_FT_F\Biggl[
          -\frac{383379490933459}{19469534584500}T_F(1+2n_f)
\N\\ && \hspace{-15mm}
          -\frac{38283693844132279}{389390691690000}C_A
\N\\ && \hspace{-15mm}
          -\frac{1237841854306528417}{4612782040020000}C_F
          +\frac{1043696}{3465}(C_F-C_A)\zeta_3
\Biggr]
\\
%%%%%%%%%%%%%%%%%%%%%%
\hat{\gamma}_{qq}^{(2),{\sf TR}}(13)&=&C_FT_F\Biggl[
          -\frac{66409807459266571}{3290351344780500}T_F(1+2n_f)
\N\\ && \hspace{-15mm}
          -\frac{6571493644375020121}{65807026895610000}C_A
\N\\ && \hspace{-15mm}
          -\frac{36713319015407141570017}{131745667845011220000}C_F
          +\frac{14011568}{45045}(C_F-C_A)\zeta_3
\Biggr]~.
\end{eqnarray}
%%%%%%%%%%%%%%%%%%%%%%%%%%%%%%%%%%%%%%%%%%%%%%%%%%%%%%%%%%%%%%%%%%%%%%%%%%%%%
%
%  Constant terms of transversity
%
%%%%%%%%%%%%%%%%%%%%%%%%%%%%%%%%%%%%%%%%%%%%%%%%%%%%%%%%%%%%%%%%%%%%%%%%%%%%%
  The fixed moments of the constant terms $a_{qq,Q}^{(3), \rm TR}(N)$
  of the unrenormalized OME, see Eq. (\ref{Aqq3NSTRQMSren}), are given by
%%%%%%%%%%%%%%%
\begin{eqnarray}
a_{qq,Q}^{(3), {\rm TR}}(1)&=&
T_FC_FC_A
      \Biggl( 
                -\frac{26441}{1458}
                +\frac{8}{3}{\sf B_4}
                -24\zeta_4
                +\frac{481}{27}\zeta_3
                -\frac{61}{27}\zeta_2
      \Biggr)
\N \\ \N \\ && \hspace{-15mm}
+T_FC_F^2
      \Biggl( 
                 \frac{15715}{162}
                -\frac{16}{3}{\sf B_4}
                +24\zeta_4
                -\frac{278}{9}\zeta_3
                +\frac{49}{3}\zeta_2
      \Biggr)\N
\end{eqnarray}\begin{eqnarray}
%\N \\ \N \\ 
&& \hspace{-15mm}
+T_F^2C_F
      \Biggl( 
                -\frac{6548}{729}
                -\frac{256}{27}\zeta_3
                -\frac{104}{27}\zeta_2
      \Biggr)
\N \\ \N \\ && \hspace{-15mm}
+n_fT_F^2C_F
      \Biggl( 
                -\frac{15850}{729}
                +\frac{112}{27}\zeta_3
                -\frac{52}{27}\zeta_2
      \Biggr)
~, \\
%%%%%%%%%%%%%%%
a_{qq,Q}^{(3), {\rm TR}}(2)&=&
T_FC_FC_A
      \Biggl( 
                 \frac{1043}{162}+8{\sf B_4}-72\zeta_4
                +\frac{577}{9}\zeta_3+\frac{\zeta_2}{3}
      \Biggr)
\N \\ \N \\ && \hspace{-15mm}
+T_FC_F^2
      \Biggl( 
                 \frac{10255}{54}-16{\sf B_4}+72\zeta_4
                -\frac{310}{3}\zeta_3+33\zeta_2
      \Biggr)
\N \\ \N \\ && \hspace{-15mm}
+T_F^2C_F
      \Biggl( 
                -\frac{1388}{81}
                -\frac{256}{9}\zeta_3-8\zeta_2
      \Biggr)
\N \\ \N \\ && \hspace{-15mm}
+n_fT_F^2C_F
      \Biggl( 
                -\frac{4390}{81}
                +\frac{112}{9}\zeta_3-4\zeta_2
      \Biggr)
~, \\
%%%%%%%%%%%%%%%
a_{qq,Q}^{(3), {\rm TR}}(3)&=&
T_FC_FC_A
      \Biggl( 
                 \frac{327967}{21870}
                +\frac{104}{9}{\sf B_4}-104\zeta_4
                +\frac{40001}{405}\zeta_3
                +\frac{121}{81}\zeta_2
      \Biggr)
\N \\ \N \\ && \hspace{-15mm}
+T_FC_F^2
      \Biggl( 
                 \frac{1170943}{4374}
                -\frac{208}{9}{\sf B_4}+104\zeta_4
                -\frac{1354}{9}\zeta_3
                +\frac{3821}{81}\zeta_2
      \Biggr)
\N \\ \N \\ && \hspace{-15mm}
+T_F^2C_F
      \Biggl( 
                -\frac{52096}{2187}
                -\frac{3328}{81}\zeta_3
                -\frac{904}{81}\zeta_2
      \Biggr)
\N \\ \N \\ && \hspace{-15mm}
+n_fT_F^2C_F
      \Biggl( 
                -\frac{168704}{2187}
                +\frac{1456}{81}\zeta_3
                -\frac{452}{81}\zeta_2
      \Biggr)
~, \\
%%%%%%%%%%%%%%%
a_{qq,Q}^{(3), {\rm TR}}(4)&=&
T_FC_FC_A
      \Biggl( 
                 \frac{4400353}{218700}
                +\frac{128}{9}{\sf B_4}-128\zeta_4
                +\frac{52112}{405}\zeta_3
                +\frac{250}{81}\zeta_2
      \Biggr)
\N \\ \N \\ && \hspace{-15mm}
+T_FC_F^2
      \Biggl( 
                 \frac{56375659}{174960}
                -\frac{256}{9}{\sf B_4}+128\zeta_4
                -\frac{556}{3}\zeta_3
                +\frac{4616}{81}\zeta_2
      \Biggr)
\N \\ \N \\ && \hspace{-15mm}
+T_F^2C_F
      \Biggl( 
                -\frac{3195707}{109350}
                -\frac{4096}{81}\zeta_3
                -\frac{1108}{81}\zeta_2
      \Biggr)
\N \\ \N \\ && \hspace{-15mm}
+n_fT_F^2C_F
      \Biggl( 
                -\frac{20731907}{218700}
                +\frac{1792}{81}\zeta_3
                -\frac{554}{81}\zeta_2
      \Biggr)
~, \\
%%%%%%%%%%%%%%%
a_{qq,Q}^{(3), {\rm TR}}(5)&=&
T_FC_FC_A
      \Biggl( 
                 \frac{1436867309}{76545000}
                +\frac{736}{45}{\sf B_4}
                -\frac{736}{5}\zeta_4
                +\frac{442628}{2835}\zeta_3
                +\frac{8488}{2025}\zeta_2
      \Biggr)
\N \\ \N \\ && \hspace{-15mm}
+T_FC_F^2
      \Biggl( 
                 \frac{40410914719}{109350000}
                -\frac{1472}{45}{\sf B_4}
                +\frac{736}{5}\zeta_4
                -\frac{47932}{225}\zeta_3
                +\frac{662674}{10125}\zeta_2
      \Biggr)\N
\end{eqnarray}\begin{eqnarray}
%\N \\ \N \\
&& \hspace{-15mm}
+T_F^2C_F
      \Biggl( 
                -\frac{92220539}{2733750}
                -\frac{23552}{405}\zeta_3
                -\frac{31924}{2025}\zeta_2
      \Biggr)
\N \\ \N \\ && \hspace{-15mm}
+n_fT_F^2C_F
      \Biggl( 
                -\frac{596707139}{5467500}
                +\frac{10304}{405}\zeta_3
                -\frac{15962}{2025}\zeta_2
      \Biggr)
~, \\
%%%%%%%%%%%%%%%
a_{qq,Q}^{(3), {\rm TR}}(6)&=&
T_FC_FC_A
      \Biggl( 
                 \frac{807041747}{53581500}
                +\frac{272}{15}{\sf B_4}
                -\frac{816}{5}\zeta_4
                +\frac{172138}{945}\zeta_3
\N \\ \N \\ && \hspace{-15mm}
                +\frac{10837}{2025}\zeta_2
      \Biggr)
+T_FC_F^2
      \Biggl( 
                 \frac{14845987993}{36450000}
                -\frac{544}{15}{\sf B_4}
                +\frac{816}{5}\zeta_4
                -\frac{159296}{675}\zeta_3
\N \\ \N \\ && \hspace{-15mm}
                +\frac{81181}{1125}\zeta_2
      \Biggr)
+T_F^2C_F
      \Biggl( 
                -\frac{5036315611}{133953750}
                -\frac{8704}{135}\zeta_3
                -\frac{35524}{2025}\zeta_2
      \Biggr)
\N \\ \N \\ && \hspace{-15mm}
+n_fT_F^2C_F
      \Biggl( 
                -\frac{32472719011}{267907500}
                +\frac{3808}{135}\zeta_3
                -\frac{17762}{2025}\zeta_2
      \Biggr)
~, \\
%%%%%%%%%%%%%%%
a_{qq,Q}^{(3), {\rm TR}}(7)&=&
T_FC_FC_A
      \Biggl( 
                 \frac{413587780793}{52509870000}
                +\frac{688}{35}{\sf B_4}
                -\frac{6192}{35}\zeta_4
                +\frac{27982}{135}\zeta_3
\N \\ \N \\ && \hspace{-15mm}
                +\frac{620686}{99225}\zeta_2
      \Biggr)
+T_FC_F^2
      \Biggl( 
                 \frac{12873570421651}{29172150000}
                -\frac{1376}{35}{\sf B_4}
                +\frac{6192}{35}\zeta_4
\N \\ \N \\ && \hspace{-15mm}
                -\frac{8454104}{33075}\zeta_3
                +\frac{90495089}{1157625}\zeta_2
      \Biggr)
+T_F^2C_F
      \Biggl( 
                -\frac{268946573689}{6563733750}
                -\frac{22016}{315}\zeta_3
\N \\ \N \\ && \hspace{-15mm}
                -\frac{1894276}{99225}\zeta_2
      \Biggr)
+n_fT_F^2C_F
      \Biggl( 
                -\frac{1727972700289}{13127467500}
                +\frac{1376}{45}\zeta_3
                -\frac{947138}{99225}\zeta_2
      \Biggr)
~, \\
%%%%%%%%%%%%%%%
a_{qq,Q}^{(3), {\rm TR}}(8)&=&
T_FC_FC_A
      \Biggl( 
                -\frac{91321974347}{112021056000}
                +\frac{2204}{105}{\sf B_4}
                -\frac{6612}{35}\zeta_4
                +\frac{87613}{378}\zeta_3
\N \\ \N \\ && \hspace{-15mm}
                +\frac{11372923}{1587600}\zeta_2
      \Biggr)
+T_FC_F^2
      \Biggl( 
                 \frac{1316283829306051}{2800526400000}
                -\frac{4408}{105}{\sf B_4}
                +\frac{6612}{35}\zeta_4
\N \\ \N \\ && \hspace{-15mm}
                -\frac{9020054}{33075}\zeta_3
                +\frac{171321401}{2058000}\zeta_2
      \Biggr)
+T_F^2C_F
      \Biggl( 
                -\frac{4618094363399}{105019740000}
                -\frac{70528}{945}\zeta_3
\N \\ \N \\ && \hspace{-15mm}
                -\frac{2030251}{99225}\zeta_2
      \Biggr)
+n_fT_F^2C_F
      \Biggl( 
                -\frac{29573247248999}{210039480000}
                +\frac{4408}{135}\zeta_3
                -\frac{2030251}{198450}\zeta_2
      \Biggr)
~, \N \\ \\
%%%%%%%%%%%%%%%
a_{qq,Q}^{(3), {\rm TR}}(9)&=&
T_FC_FC_A
      \Biggl( 
                -\frac{17524721583739067}{1497161413440000}
                +\frac{20956}{945}{\sf B_4}
                -\frac{20956}{105}\zeta_4
\N \\ \N \\ && \hspace{-15mm}
                +\frac{9574759}{37422}\zeta_3
                +\frac{16154189}{2041200}\zeta_2
      \Biggr)
+T_FC_F^2
      \Biggl( 
                 \frac{1013649109952401819}{2041583745600000}
                -\frac{41912}{945}{\sf B_4}\N
\end{eqnarray}\begin{eqnarray}
%\N \\ \N \\ 
&& \hspace{-15mm}
                +\frac{20956}{105}\zeta_4
                -\frac{85698286}{297675}\zeta_3
                +\frac{131876277049}{1500282000}\zeta_2
      \Biggr)
\N \\ \N \\ && \hspace{-15mm}
+T_F^2C_F
      \Biggl( 
                -\frac{397003835114519}{8506598940000}
                -\frac{670592}{8505}\zeta_3
                -\frac{19369859}{893025}\zeta_2
      \Biggr)
\N \\ \N \\ && \hspace{-15mm}
+n_fT_F^2C_F
      \Biggl( 
                -\frac{2534665670688119}{17013197880000}
                +\frac{41912}{1215}\zeta_3
                -\frac{19369859}{1786050}\zeta_2
      \Biggr)
~, \\
%%%%%%%%%%%%%%%
a_{qq,Q}^{(3), {\rm TR}}(10)&=&
T_FC_FC_A
      \Biggl( 
                -\frac{176834434840947469}{7485807067200000}
                +\frac{21964}{945}{\sf B_4}
                -\frac{21964}{105}\zeta_4
\N \\ \N \\ && \hspace{-15mm}
                +\frac{261607183}{935550}\zeta_3
                +\frac{618627019}{71442000}\zeta_2
      \Biggr)
+T_FC_F^2
      \Biggl( 
                 \frac{11669499797141374121}{22457421201600000}
\N \\ \N \\ && \hspace{-15mm}
                -\frac{43928}{945}{\sf B_4}
                +\frac{21964}{105}\zeta_4
                -\frac{3590290}{11907}\zeta_3
                +\frac{137983320397}{1500282000}\zeta_2
      \Biggr)
\N \\ \N \\ && \hspace{-15mm}
+T_F^2C_F
      \Biggl( 
                -\frac{50558522757917663}{1029298471740000}
                -\frac{702848}{8505}\zeta_3
                -\frac{4072951}{178605}\zeta_2
      \Biggr)
\N \\ \N \\ && \hspace{-15mm}
+n_fT_F^2C_F
      \Biggl( 
                -\frac{321908083399769663}{2058596943480000}
                +\frac{43928}{1215}\zeta_3
                -\frac{4072951}{357210}\zeta_2
      \Biggr)
~, \\
%%%%%%%%%%%%%%%
a_{qq,Q}^{(3), {\rm TR}}(11)&=&
T_FC_FC_A
      \Biggl( 
                -\frac{436508000489627050837}{11775174516705600000}
                +\frac{251684}{10395}{\sf B_4}
                -\frac{251684}{1155}\zeta_4
\N \\ \N \\ && \hspace{-15mm}
                +\frac{3687221539}{12162150}\zeta_3
                +\frac{149112401}{16038000}\zeta_2
      \Biggr)
+T_FC_F^2
      \Biggl( 
                 \frac{177979311179110818909401}{328799103812625600000}
\N \\ \N \\ && \hspace{-15mm}
                -\frac{503368}{10395}{\sf B_4}
                +\frac{251684}{1155}\zeta_4
                -\frac{452259130}{1440747}\zeta_3
                +\frac{191230589104127}{1996875342000}\zeta_2
      \Biggr)
\N \\ \N \\ && \hspace{-15mm}
+T_F^2C_F
      \Biggl( 
                -\frac{6396997235105384423}{124545115080540000}
                -\frac{8053888}{93555}\zeta_3
                -\frac{514841791}{21611205}\zeta_2
      \Biggr)
\N \\ \N \\ && \hspace{-15mm}
+n_fT_F^2C_F
      \Biggl( 
                -\frac{40628987857774916423}{249090230161080000}
                +\frac{503368}{13365}\zeta_3
                -\frac{514841791}{43222410}\zeta_2
      \Biggr)
~, \\
%%%%%%%%%%%%%%%
a_{qq,Q}^{(3), {\rm TR}}(12)&=&
T_FC_FC_A
      \Biggl( 
                -\frac{245210883820358086333}{4783664647411650000}
                +\frac{260924}{10395}{\sf B_4}
                -\frac{260924}{1155}\zeta_4 
\N \\ \N \\ && \hspace{-15mm}
                +\frac{3971470819}{12162150}\zeta_3
                +\frac{85827712409}{8644482000}\zeta_2
      \Biggr)
+T_FC_F^2
      \Biggl( 
                 \frac{2396383721714622551610173}{4274388349564132800000}
\N \\ \N \\ && \hspace{-15mm}
                -\frac{521848}{10395}{\sf B_4}
                +\frac{260924}{1155}\zeta_4
                -\frac{468587596}{1440747}\zeta_3
                +\frac{198011292882437}{1996875342000}\zeta_2
      \Biggr) \N
\end{eqnarray}\begin{eqnarray}
%\N \\ \N \\ 
&& \hspace{-15mm}
+T_F^2C_F
      \Biggl( 
                -\frac{1124652164258976877487}{21048124448611260000}
                -\frac{8349568}{93555}\zeta_3
\N \\ \N \\ && \hspace{-15mm}
                -\frac{535118971}{21611205}\zeta_2
      \Biggr)
+n_fT_F^2C_F
      \Biggl( 
                -\frac{7126865031281296825487}{42096248897222520000}
\N \\ \N \\ && \hspace{-15mm}
                +\frac{521848}{13365}\zeta_3
                -\frac{535118971}{43222410}\zeta_2
      \Biggr)
~,\\
%%%%%%%%%%%%%%%
a_{qq,Q}^{(3), {\rm TR}}(13)&=&
T_FC_FC_A
      \Biggl( 
                -\frac{430633219615523278883051}{6467514603300550800000}
                +\frac{3502892}{135135}{\sf B_4}
\N \\ \N \\ && \hspace{-15mm}
                -\frac{3502892}{15015}\zeta_4
                +\frac{327241423}{935550}\zeta_3
                +\frac{15314434459241}{1460917458000}\zeta_2
      \Biggr)
\N \\ \N \\ && \hspace{-15mm}
+T_FC_F^2
      \Biggl( 
                 \frac{70680445585608577308861582893}{122080805651901196900800000}
                -\frac{7005784}{135135}{\sf B_4}
\N \\ \N \\ && \hspace{-15mm}
                +\frac{3502892}{15015}\zeta_4
                -\frac{81735983092}{243486243}\zeta_3
                +\frac{449066258795623169}{4387135126374000}\zeta_2
      \Biggr)
\N \\ \N \\ && \hspace{-15mm}
+T_F^2C_F
      \Biggl( 
                -\frac{196897887865971730295303}{3557133031815302940000}
                -\frac{112092544}{1216215}\zeta_3
\N \\ \N \\ && \hspace{-15mm}
                -\frac{93611152819}{3652293645}\zeta_2
      \Biggr)
+n_fT_F^2C_F
      \Biggl( 
                -\frac{1245167831299024242467303}{7114266063630605880000}
\N \\ \N \\ && \hspace{-15mm}
                +\frac{7005784}{173745}\zeta_3
                -\frac{93611152819}{7304587290}\zeta_2
      \Biggr)
~.
\end{eqnarray}
%%%%%%%%%%%%%%%
%%%%%%%%%%%%%%%%%%%%%%%%%%%%%%%%%%%%%%%%%%%%%%%%%%%%%%%%%%%%%%%%
%
%end appendix
%
%%%%%%%%%%%%%%%%%%%%%%%%%%%%%%%%%%%%%%%%%%%%%%%%%%%%%%%%%%%%%%%
\end{appendix}
%%%%%%%%%%%%%%%%%%%%%%%%%%%%%%%%%%%%%%%%%%%%%%%%%%%%%%%%%%%%%%%%%%%%%%%%%%%%%%%
%   \end{appendix}
%%%%%%%%%%%%%%%%%%%%%%%%%%%%%%%%%%%%%%%%%%%%%%%%%%%%%%%%%%%%%%%%%%%%%%%%%
 \newpage
\thispagestyle{empty}
 \begin{flushleft}
 \end{flushleft}
 \newpage
\thispagestyle{empty}
 \begin{flushleft}
 \end{flushleft}
 \newpage
%%%%%%%%%%%%%%%%%%%%%%%%%%%%%%%%%%%%%%%%%%%%%%%%%%%%%%%%%%%%%%%%%%%%%%%%%%%%%%
%       Bibliography
%%%%%%%%%%%%%%%%%%%%%%%%%%%%%%%%%%%%%%%%%%%%%%%%%%%%%%%%%%%%%%%%%%%%%%%%%%%%%%
\bibliographystyle{h-physrev3}
{
\bibliography{MYBIB}

\begin{thebibliography}{100}

\bibitem{GellMann:1964nj}
M.~Gell-Mann,
\newblock Phys. Lett. {\bf 8}, 214 (1964).
%%CITATION = PHLTA,8,214;%%

\bibitem{Zweig:1964jf}
G.~Zweig,
\newblock {\sf An $SU(3)$ model for the strong interaction symmetry and its
  breaking},\\ CERN-TH-401,~412~(1964).

\bibitem{GellMann:1964xy}
M.~Gell-Mann and Y.~Neemam,
\newblock {\sf The eightfold way: a review with a collection of reprints},
  (Benjamin Press, New York, 1964), 317~p.

\bibitem{Kokkedee:1969}
J.~J.~J. Kokkedee,
\newblock {\sf The Quark Model}, (Benjamin Press, New York, 1969), 239~p.

\bibitem{Close:1979bt}
F.~E. Close,
\newblock {\sf An Introduction to Quarks and Partons}, (Academic Press, London,
  1979),~481~p.

\bibitem{Barnes:1964pd}
V.~E. Barnes {\em et~al.},
\newblock Phys. Rev. Lett. {\bf 12}, 204 (1964).
%%CITATION = PRLTA,12,204;%%

\bibitem{Gursey:1992dc}
F.~G{\"u}rsey and L.~A. Radicati,
\newblock Phys. Rev. Lett. {\bf 13}, 173 (1964).
%%CITATION = PRLTA,13,173;%%

\bibitem{Beg:1964nm}
M.~A.~B. Beg, B.~W. Lee, and A.~Pais,
\newblock Phys. Rev. Lett. {\bf 13}, 514 (1964).
%%CITATION = PRLTA,13,514;%%

\bibitem{Sakita:1964qr}
B.~Sakita,
\newblock Phys. Rev. Lett. {\bf 13}, 643 (1964).
%%CITATION = PRLTA,13,643;%%

\bibitem{Pauli:1940zz}
W.~Pauli,
\newblock Phys. Rev. {\bf 58}, 716 (1940).
%%CITATION = PHRVA,58,716;%%

\bibitem{Greenberg:1964pe}
O.~W. Greenberg,
\newblock Phys. Rev. Lett. {\bf 13}, 598 (1964).
%%CITATION = PRLTA,13,598;%%

\bibitem{Nambu:1966}
Y.~Nambu,
\newblock {\sf A Systematics Of Hadrons In Subnuclear Physics}, in: Preludes in
  Theoretical Physics, eds. A. De-Shalit, H.~Fehsbach and L. van Hove
  (North-Holland, Amsterdam, 1966), pp. 133.

\bibitem{Han:1965pf}
M.~Y. Han and Y.~Nambu,
\newblock Phys. Rev. {\bf 139}, B1006 (1965).
%%CITATION = PHRVA,139,B1006;%%

\bibitem{Fritzsch:1972jv}
H.~Fritzsch and M.~Gell-Mann,
\newblock {\sf Current algebra: Quarks and what else?}, Proceedings of 16th
  International Conference on High-Energy Physics, Batavia, Illinois, 6-13 Sep
  Vol. {\bf 2}, J.D.~Jackson, A.~Roberts, R.~Donaldson, eds., pp. 135 (1972),
  hep-ph/0208010.
%%CITATION = HEP-PH/0208010;%%

\bibitem{Mo:1965dv}
L.~W. Mo and C.~Peck,
\newblock {\sf 8-$\GeV$/c Spectrometer}, SLAC-TN-65-029 (1965).

\bibitem{Taylor:1967qv}
R.~E. Taylor,
\newblock {\sf Nucleon form--factors above 6-$\GeV$}, in: Proc. of the Int.
  Symp. on Electron and Photon Interactions at High Energies, SLAC, September
  5--9, 1967, (SLAC, Stanford CA, 1967), SLAC-PUB-0372, pp. 78.

\bibitem{Drell:1963ej}
S.~D. Drell and J.~D. Walecka,
\newblock Ann. Phys. {\bf 28}, 18 (1964).
%%CITATION = APNYA,28,18;%%

\bibitem{Derman:1978iz}
E.~Derman,
\newblock Phys. Rev. {\bf D19}, 133 (1979).
%%CITATION = PHRVA,D19,133;%%

\bibitem{Bjorken:1968dy}
J.~D. Bjorken,
\newblock Phys. Rev. {\bf 179}, 1547 (1969).
%%CITATION = PHRVA,179,1547;%%

\bibitem{Coward:1967au}
D.~H. Coward {\em et~al.},
\newblock Phys. Rev. Lett. {\bf 20}, 292 (1968).
%%CITATION = PRLTA,20,292;%%

\bibitem{Panofsky:1968pb}
W.~K.~H. Panofsky,
\newblock {\sf Low $q^2$ electrodynamics, elastic and inelastic electron (and
  muon) scattering}, Proc. 14th International Conference on High-Energy
  Physics, Vienna, 1968, J.~Prentki and J.~Steinberger, eds., (CERN, Geneva,
  1968), pp.~23.

\bibitem{Bloom:1969kc}
E.~D. Bloom {\em et~al.},
\newblock Phys. Rev. Lett. {\bf 23}, 930 (1969).
%%CITATION = PRLTA,23,930;%%

\bibitem{Breidenbach:1969kd}
M.~Breidenbach {\em et~al.},
\newblock Phys. Rev. Lett. {\bf 23}, 935 (1969).
%%CITATION = PRLTA,23,935;%%

\bibitem{Kendall:1991np}
H.~W. Kendall,
\newblock Rev. Mod. Phys. {\bf 63}, 597 (1991).
%%CITATION = RMPHA,63,597;%%

\bibitem{Taylor:1991ew}
R.~E. Taylor,
\newblock Rev. Mod. Phys. {\bf 63}, 573 (1991).
%%CITATION = RMPHA,63,573;%%

\bibitem{Friedman:1991nq}
J.~I. Friedman,
\newblock Rev. Mod. Phys. {\bf 63}, 615 (1991).
%%CITATION = RMPHA,63,615;%%

\bibitem{Feynman:1969wa}
R.~P. Feynman,
\newblock {\sf The behavior of hadron collisions at extreme energies}, Proc. of
  3rd International Conference on High Energy Collisions, Stony Brook, 1969,
  C.N.~Yang, J.A.~Cole, M.~Good, R.~Hwa, and J.~Lee-Franzini, eds., (Gordon and
  Breach, New York, 1970), pp.~237.
%%CITATION = CONFP,C690905,237;%%

\bibitem{Feynman:1969ej}
R.~P. Feynman,
\newblock Phys. Rev. Lett. {\bf 23}, 1415 (1969).
%%CITATION = PRLTA,23,1415;%%

\bibitem{Feynman:1973xc}
R.~P. Feynman,
\newblock {\sf Photon-hadron interactions}, (Benjamin Press, Reading, 1972),
  282~p.

\bibitem{Callan:1969uq}
C.~G. Callan and D.~J. Gross,
\newblock Phys. Rev. Lett. {\bf 22}, 156 (1969).
%%CITATION = PRLTA,22,156;%%

\bibitem{Lee:1967iu}
T.~D. Lee, S.~Weinberg, and B.~Zumino,
\newblock Phys. Rev. Lett. {\bf 18}, 1029 (1967).
%%CITATION = PRLTA,18,1029;%%

\bibitem{Sakurai:1969}
J.~Sakurai,
\newblock {\sf Vector-Meson dominance - present status and future prospects},
  Proc. 4th International Symposium on Electron and Photon Interactions at High
  Energies, Liverpool, 1969, (Daresbury Laboratory, 1969), eds. D.W. Braben and
  R.E. Rand, pp. 91.

\bibitem{Sakurai:1969ss}
J.~J. Sakurai,
\newblock Phys. Rev. Lett. {\bf 22}, 981 (1969).
%%CITATION = PRLTA,22,981;%%

\bibitem{Tsai:1969yk}
Y.-S. Tsai,
\newblock SLAC-PUB-0600 (1969).

\bibitem{Fraas:1970vj}
H.~Fraas and D.~Schildknecht,
\newblock Nucl. Phys. {\bf B14}, 543 (1969).
%%CITATION = NUPHA,B14,543;%%

\bibitem{Bjorken:1969ja}
J.~D. Bjorken and E.~A. Paschos,
\newblock Phys. Rev. {\bf 185}, 1975 (1969).
%%CITATION = PHRVA,185,1975;%%

\bibitem{Weinberg:1967tq}
S.~Weinberg,
\newblock Phys. Rev. Lett. {\bf 19}, 1264 (1967).
%%CITATION = PRLTA,19,1264;%%

\bibitem{Glashow:1961tr}
S.~L. Glashow,
\newblock Nucl. Phys. {\bf 22}, 579 (1961).
%%CITATION = NUPHA,22,579;%%

\bibitem{Salam:1964ry}
A.~Salam and J.~C. Ward,
\newblock Phys. Lett. {\bf 13}, 168 (1964).
%%CITATION = PHLTA,13,168;%%

\bibitem{Salam:1968rm}
A.~Salam,
\newblock {\sf Weak and Electromagnetic Interactions}, Proc. of the 8th Nobel
  Symposium, G{\"oteborg}, Sweden, 19--25 May 1968, ed. N.~Svartholm, (Almqvist
  and Wiskell, Stockholm, 1968), pp. 367.

\bibitem{'tHooft:1972ue}
G.~'t~Hooft and M.~J.~G. Veltman,
\newblock Nucl. Phys. {\bf B50}, 318 (1972).
%%CITATION = NUPHA,B50,318;%%

\bibitem{Taylor:1971ff}
J.~C. Taylor,
\newblock Nucl. Phys. {\bf B33}, 436 (1971).
%%CITATION = NUPHA,B33,436;%%

\bibitem{Slavnov:1972fg}
A.~A. Slavnov,
\newblock Theor. Math. Phys. {\bf 10}, 99 (1972).
%%CITATION = TMPHA,10,99;%%

\bibitem{Lee:1972fjxLee:1973fn}
B.~W. Lee and J.~Zinn-Justin,
\newblock Phys. Rev.{\bf D5}, 3121 (1972); Phys. Rev.{\bf D7}, 1049 (1973).

\bibitem{Bell:1969ts}
J.~S. Bell and R.~Jackiw,
\newblock Nuovo Cim. {\bf A60}, 47 (1969).
%%CITATION = NUCIA,A60,47;%%

\bibitem{Adler:1969gk}
S.~L. Adler,
\newblock Phys. Rev. {\bf 177}, 2426 (1969).
%%CITATION = PHRVA,177,2426;%%

\bibitem{Bertlmann:1996xk}
R.~A. Bertlmann,
\newblock {\sf Anomalies in quantum field theory}, (Clarendon, Oxford, 1996),
  566~p.

\bibitem{'tHooft:1971fh}
G.~'t~Hooft,
\newblock Nucl. Phys. {\bf B33}, 173 (1971).
%%CITATION = NUPHA,B33,173;%%

\bibitem{Yang:1954ek}
C.-N. Yang and R.~L. Mills,
\newblock Phys. Rev. {\bf 96}, 191 (1954).
%%CITATION = PHRVA,96,191;%%

\bibitem{Fritzsch:1973pi}
H.~Fritzsch, M.~Gell-Mann, and H.~Leutwyler,
\newblock Phys. Lett. {\bf B47}, 365 (1973).
%%CITATION = PHLTA,B47,365;%%

\bibitem{Reya:1979zk}
E.~Reya,
\newblock Phys. Rept. {\bf 69}, 195 (1981).
%%CITATION = PRPLC,69,195;%%

\bibitem{Muta:1998vi}
T.~Muta,
\newblock {\sf Foundations of Quantum Chromodynamics}, World Sci. Lect. Notes
  Phys. {\bf 57}, (World Scientific, Singapore, 1998), 2nd edition.
%%CITATION = 00327,57,1;%%

\bibitem{Gross:1973id}
D.~J. Gross and F.~Wilczek,
\newblock Phys. Rev. Lett. {\bf 30}, 1343 (1973).
%%CITATION = PRLTA,30,1343;%%

\bibitem{Politzer:1973fx}
H.~D. Politzer,
\newblock Phys. Rev. Lett. {\bf 30}, 1346 (1973).
%%CITATION = PRLTA,30,1346;%%

\bibitem{tHooft:unpub}
G.~t'Hooft,
\newblock (1972), {\rm unpublished}.

\bibitem{Wilson:1969zs}
K.~G. Wilson,
\newblock Phys. Rev. {\bf 179}, 1499 (1969).
%%CITATION = PHRVA,179,1499;%%

\bibitem{Zimmermann:1970}
W.~Zimmermann,
\newblock {\sf Lect. on Elementary Particle Physics and Quantum Field Theory},
  Brandeis Summer Inst., Vol.~{\bf 1}, (MIT Press, Cambridge, 1970),~pp. 395.

\bibitem{Frishman:1971qn}
Y.~Frishman,
\newblock Annals Phys. {\bf 66}, 373 (1971).
%%CITATION = APNYA,66,373;%%

\bibitem{Brandt:1970kg}
R.~A. Brandt and G.~Preparata,
\newblock Nucl. Phys. {\bf B27}, 541 (1972).
%%CITATION = NUPHA,B27,541;%%

\bibitem{Gross:1971wn}
D.~J. Gross and S.~B. Treiman,
\newblock Phys. Rev. {\bf D4}, 1059 (1971).
%%CITATION = PHRVA,D4,1059;%%

\bibitem{Chang:1975sv}
C.~Chang {\em et~al.},
\newblock Phys. Rev. Lett. {\bf 35}, 901 (1975).
%%CITATION = PRLTA,35,901;%%

\bibitem{Watanabe:1975su}
Y.~Watanabe {\em et~al.},
\newblock Phys. Rev. Lett. {\bf 35}, 898 (1975).
%%CITATION = PRLTA,35,898;%%

\bibitem{Ferrara:1973eg}
S.~Ferrara, R.~Gatto, and A.~F. Grillo,
\newblock Springer Tracts Mod. Phys. {\bf 67}, 1 (1973),
\newblock and references therein.
%%CITATION = STPHB,67,1;%%

\bibitem{Gross:1973juxGross:1974cs}
D.~J. Gross and F.~Wilczek,
\newblock Phys. Rev. {\bf D8}, 3633 (1973); {\bf D9}, 980 (1974).

\bibitem{Georgi:1951sr}
H.~Georgi and H.~D. Politzer,
\newblock Phys. Rev. {\bf D9}, 416 (1974).
%%CITATION = PHRVA,D9,416;%%

\bibitem{PHOHAD:1971}
{\sf Photon-Hadron Interactions I}, International Summer Institute in
  Theoretical Physics, DESY, July 12--14, 1971, Springer Tracts in Modern
  Physics {\bf 62}, 147~p.

\bibitem{Politzer:1974fr}
H.~D. Politzer,
\newblock Phys. Rept. {\bf 14}, 129 (1974).
%%CITATION = PRPLC,14,129;%%

\bibitem{Marciano:1977su}
W.~J. Marciano and H.~Pagels,
\newblock Phys. Rept. {\bf 36}, 137 (1978).
%%CITATION = PRPLC,36,137;%%

\bibitem{Ellis:1979kt}
J.~R. Ellis and C.~T. Sachrajda,
\newblock NATO Adv. Study Inst. Ser. B Phys. {\bf 59}, 285 (1980).
%%CITATION = NASBD,59,285;%%

\bibitem{Buras:1979yt}
A.~J. Buras,
\newblock Rev. Mod. Phys. {\bf 52}, 199 (1980).
%%CITATION = RMPHA,52,199;%%

\bibitem{Altarelli:1981ax}
G.~Altarelli,
\newblock Phys. Rept. {\bf 81}, 1 (1982).
%%CITATION = PRPLC,81,1;%%

\bibitem{Wilczek:1982yx}
F.~Wilczek,
\newblock Ann. Rev. Nucl. Part. Sci. {\bf 32}, 177 (1982).
%%CITATION = ARNUA,32,177;%%

\bibitem{Jaffe:1985je}
R.~L. Jaffe,
\newblock {\sf Deep Inelastic Scattering with Application to nuclear targets},
  Lectures presented at the Los Alamos School on Quark Nuclear Physics, Los
  Alamos, NM, June 10-14, 1985, M.B. Johnson and A. Picklesimer, eds., (Wiley,
  New York, 1986), pp.~82.

\bibitem{Collins:1987pm}
J.~C. Collins and D.~E. Soper,
\newblock Ann. Rev. Nucl. Part. Sci. {\bf 37}, 383 (1987).
%%CITATION = ARNUA,37,383;%%

\bibitem{Ellis:1988vi}
R.~K. Ellis,
\newblock {\sf An Introduction to the QCD Parton Model}, Lectures given at 1987
  Theoretical Advanced Study Inst. in Elementary Particle Physics, Santa Fe,
  NM, Jul 5 - Aug 1, 1987, R. Slansky and Geoffrey West, eds., (World
  Scientific, Singapore, 1988), pp.~214.

\bibitem{Mueller:1989hs}
A.~H. Mueller,
\newblock ed., {\sf Perturbative Quantum Chromodynamics}, (World Scientific,
  Singapore, 1989), 614~p.

\bibitem{Roberts:1990ww}
R.~G. Roberts,
\newblock {\sf The Structure of the proton: Deep inelastic scattering},
  (Cambridge University Press, Cambridge, 1990), 182~p.

\bibitem{Sterman:1994ce}
G.~Sterman,
\newblock {\sf An Introduction to quantum field theory}, (Cambridge University
  Press, Cambridge, 1993), 572~p.

\bibitem{Ellis:1991qj}
R.~K. Ellis, W.~J. Stirling, and B.~R. Webber,
\newblock {\sf QCD and collider physics}, Camb. Monogr. Part. Phys. Nucl. Phys.
  Cosmol. {\bf 8}, (Cambridge University Press, Cambridge, 1996), 435~p.
%%CITATION = CMPCE,8,1;%%

\bibitem{Brock:1993sz}
CTEQ collaboration, R.~Brock {\em et~al.},
\newblock Rev. Mod. Phys. {\bf 67}, 157 (1995).
%%CITATION = RMPHA,67,157;%%

\bibitem{Blumlein:1993ar}
J.~Bl{\"umlein},
\newblock Surveys High Energ. Phys. {\bf 7}, 181 (1994).
%%CITATION = SHEPD,7,181;%%

\bibitem{Mulders:1996}
P.~Mulders,
\newblock {\sf Quantum Chromodynamics and Hard Scattering Processes}, Lectures,
  Dutch Research School for Theoretical Physics, Dalfsen, January 1996, and
  Dutch Research School for Subatomic Physics, Beekbergen, February, 1996,
  (NIKHEF, Amsterdam, The Netherlands).

\bibitem{Augustin:1974xw}
SLAC-SP-017 collaboration, J.~E. Augustin {\em et~al.},
\newblock Phys. Rev. Lett. {\bf 33}, 1406 (1974).
%%CITATION = PRLTA,33,1406;%%

\bibitem{Abrams:1974yy}
G.~S. Abrams {\em et~al.},
\newblock Phys. Rev. Lett. {\bf 33}, 1453 (1974).
%%CITATION = PRLTA,33,1453;%%

\bibitem{Aubert:1974js}
E598 collaboration, J.~J. Aubert {\em et~al.},
\newblock Phys. Rev. Lett. {\bf 33}, 1404 (1974).
%%CITATION = PRLTA,33,1404;%%

\bibitem{Maki:1964ux}
Z.~Maki and M.~Nakagawa,
\newblock Prog. Theor. Phys. {\bf 31}, 115 (1964).
%%CITATION = PTPKA,31,115;%%

\bibitem{Hara:1963gw}
Y.~Hara,
\newblock Phys. Rev. {\bf 134}, B701 (1964).
%%CITATION = PHRVA,134,B701;%%

\bibitem{Bjorken:1964gz}
J.~D. Bjorken and S.~L. Glashow,
\newblock Phys. Lett. {\bf 11}, 255 (1964).
%%CITATION = PHLTA,11,255;%%

\bibitem{Glashow:1970gm}
S.~L. Glashow, J.~Iliopoulos, and L.~Maiani,
\newblock Phys. Rev. {\bf D2}, 1285 (1970).
%%CITATION = PHRVA,D2,1285;%%

\bibitem{Amsler:2008zzb}
Particle Data Group, C.~Amsler {\em et~al.},
\newblock Phys. Lett. {\bf B667}, 1 (2008).
%%CITATION = PHLTA,B667,1;%%

\bibitem{Herb:1977ek}
S.~W. Herb {\em et~al.},
\newblock Phys. Rev. Lett. {\bf 39}, 252 (1977).
%%CITATION = PRLTA,39,252;%%

\bibitem{Abe:1994xtxAbe:1994stxAbe:1995hr}
CDF collaboration, F.~Abe {\em et~al.},
\newblock Phys. Rev. Lett. {\bf 73}, 225 (1994), hep-ex/9405005; Phys. Rev.
  {\bf D50}, 2966 (1994); Phys. Rev. Lett. {\bf 74}, 2626 (1995),
  hep-ex/9503002.

\bibitem{Abachi:1995iq}
D0 collaboration, S.~Abachi {\em et~al.},
\newblock Phys. Rev. Lett. {\bf 74}, 2632 (1995), hep-ex/9503003.
%%CITATION = HEP-EX/9503003;%%

\bibitem{Stein:1975yy}
S.~Stein {\em et~al.},
\newblock Phys. Rev. {\bf D12}, 1884 (1975).
%%CITATION = PHRVA,D12,1884;%%

\bibitem{Atwood:1976ys}
W.~B. Atwood {\em et~al.},
\newblock Phys. Lett. {\bf B64}, 479 (1976).
%%CITATION = PHLTA,B64,479;%%

\bibitem{Bodek:1979rx}
A.~Bodek {\em et~al.},
\newblock Phys. Rev. {\bf D20}, 1471 (1979).
%%CITATION = PHRVA,D20,1471;%%

\bibitem{Mestayer:1982ba}
M.~D. Mestayer {\em et~al.},
\newblock Phys. Rev. {\bf D27}, 285 (1983).
%%CITATION = PHRVA,D27,285;%%

\bibitem{Allkover:1981}
EMC collaboration, O.~Allkofer {\em et~al.},
\newblock Nucl. Instr. Meth. {\bf 179}, 445 (1981).

\bibitem{Aubert:1985fx}
EMC collaboration, J.~J. Aubert {\em et~al.},
\newblock Nucl. Phys. {\bf B259}, 189 (1985).
%%CITATION = NUPHA,B259,189;%%

\bibitem{Bollini:1982ac}
D.~Bollini {\em et~al.},
\newblock Nucl. Instr. Meth. {\bf 204}, 333 (1983).
%%CITATION = NUIMA,204,333;%%

\bibitem{Benvenuti:1984duxBenvenuti:1987zjxBenvenuti:1989rhxBenvenuti:1989fm}
BCDMS collaboration, A.~C. Benvenuti {\em et~al.},
\newblock Nucl. Instr. Meth. {\bf A226}, 330 (1984); Phys. Lett. {\bf B195}, 91
  (1987); Phys. Lett. {\bf B223}, 485 (1989); Phys. Lett. {\bf B237}, 592
  (1990).

\bibitem{Amaudruz:1991nwxAmaudruz:1992bf}
NMC collaboration, P.~Amaudruz {\em et~al.},
\newblock Nucl. Phys. {\bf B371}, 3 (1992); Phys. Lett. {\bf B295}, 159 (1992).

\bibitem{Arneodo:1995cq}
NMC collaboration, M.~Arneodo {\em et~al.},
\newblock Phys. Lett. {\bf B364}, 107 (1995), hep-ph/9509406.
%%CITATION = HEP-PH/9509406;%%

\bibitem{Anderson:1979mt}
H.~L. Anderson {\em et~al.},
\newblock Phys. Rev. {\bf D20}, 2645 (1979).
%%CITATION = PHRVA,D20,2645;%%

\bibitem{Adams:1989emxAdams:1996gu}
E665 collaboration, M.~R. Adams {\em et~al.},
\newblock Nucl. Instrum. Meth {\bf A291}, 533 (1990); Phys. Rev. {\bf D54},
  3006 (1996).

\bibitem{Jonker:1981dc}
CHARM collaboration, M.~Jonker {\em et~al.},
\newblock Phys. Lett. {\bf B109}, 133 (1982).
%%CITATION = PHLTA,B109,133;%%

\bibitem{Bergsma:1982ckxBergsma:1984ny}
CHARM collaboration, F.~Bergsma {\em et~al.},
\newblock Phys. Lett. {\bf B123}, 269 (1983); Phys. Lett. {\bf B153}, 111
  (1985).

\bibitem{Berge:1989hr}
J.~P. Berge {\em et~al.},
\newblock Z. Phys. {\bf C49}, 187 (1991).
%%CITATION = ZEPYA,C49,187;%%

\bibitem{Jones:1994pw}
Birmingham-CERN-Imperial College-M{\"u}nchen(MPI)-Oxford collaboration, G.~T.
  Jones {\em et~al.},
\newblock Z. Phys. {\bf C62}, 575 (1994).
%%CITATION = ZEPYA,C62,575;%%

\bibitem{Shaevitz:1995yc}
CCFR collaboration, M.~H. Shaevitz {\em et~al.},
\newblock Nucl. Phys. Proc. Suppl. {\bf 38}, 188 (1995).
%%CITATION = NUPHZ,38,188;%%

\bibitem{Bosetti:1978kz}
Aachen-Bonn-CERN-London-Oxford-Saclay collaboration, P.~C. Bosetti {\em
  et~al.},
\newblock Nucl. Phys. {\bf B142}, 1 (1978).
%%CITATION = NUPHA,B142,1;%%

\bibitem{deGroot:1978hr}
J.~G.~H. de~Groot {\em et~al.},
\newblock Z. Phys. {\bf C1}, 143 (1979).
%%CITATION = ZEPYA,C1,143;%%

\bibitem{Heagy:1980wj}
S.~M. Heagy {\em et~al.},
\newblock Phys. Rev. {\bf D23}, 1045 (1981).
%%CITATION = PHRVA,D23,1045;%%

\bibitem{Morfin:1981kg}
Gargamelle SPS collaboration, J.~G. Morfin {\em et~al.},
\newblock Phys. Lett. {\bf B104}, 235 (1981).
%%CITATION = PHLTA,B104,235;%%

\bibitem{Bosetti:1982yy}
Aachen-Bonn-CERN-Democritos-London-Oxford-Saclay collaboration, P.~C. Bosetti
  {\em et~al.},
\newblock Nucl. Phys. {\bf B203}, 362 (1982).
%%CITATION = NUPHA,B203,362;%%

\bibitem{Abramowicz:1982re}
H.~Abramowicz {\em et~al.},
\newblock Z. Phys. {\bf C17}, 283 (1983).
%%CITATION = ZEPYA,C17,283;%%

\bibitem{MacFarlane:1983ax}
D.~MacFarlane {\em et~al.},
\newblock Z. Phys. {\bf C26}, 1 (1984).
%%CITATION = ZEPYA,C26,1;%%

\bibitem{Allasia:1985hw}
D.~Allasia {\em et~al.},
\newblock Z. Phys. {\bf C28}, 321 (1985).
%%CITATION = ZEPYA,C28,321;%%

\bibitem{:1981uka}
{\sf HERA - a proposal for a large electron proton colliding beam facility at
  DESY}, (Hamburg, DESY, 1981), DESY HERA 81-10,~292~p.

\bibitem{Abt:1993wz}
H1 collaboration, I.~Abt {\em et~al.},
\newblock {\sf The H1 detector at HERA}, DESY-93-103 (1993), 194~p.

\bibitem{Derrick:1992nw}
ZEUS collaboration, M.~Derrick {\em et~al.},
\newblock Phys. Lett. {\bf B303}, 183 (1993).
%%CITATION = PHLTA,B303,183;%%

\bibitem{Ackerstaff:1998av}
HERMES collaboration, K.~Ackerstaff {\em et~al.},
\newblock Nucl. Instrum. Meth. {\bf A417}, 230 (1998), hep-ex/9806008.
%%CITATION = HEP-EX/9806008;%%

\bibitem{Hartouni:1995cf}
E.~Hartouni {\em et~al.},
\newblock {\sf HERA-B: An experiment to study CP violation in the B system
  using an internal target at the HERA proton ring. Design report},
  DESY-PRC-95-01 (1995), 491~p.

\bibitem{:2008tx}
H1 collaboration, F.~D. Aaron {\em et~al.},
\newblock Phys. Lett. {\bf B665}, 139 (2008), hep-ex/0805.2809.
%%CITATION = 0805.2809;%%

\bibitem{Collaboration:2009na}
ZEUS collaboration,
\newblock {\sf Measurement of the Longitudinal Proton Structure Function at
  HERA}, (2009), hep-ex/0904.1092.

\bibitem{Whitlow:1990gk}
L.~W. Whitlow, S.~Rock, A.~Bodek, E.~M. Riordan, and S.~Dasu,
\newblock Phys. Lett. {\bf B250}, 193 (1990).
%%CITATION = PHLTA,B250,193;%%

\bibitem{Dasu:1993vk}
S.~Dasu {\em et~al.},
\newblock Phys. Rev. {\bf D49}, 5641 (1994).
%%CITATION = PHRVA,D49,5641;%%

\bibitem{Tao:1995uh}
E140X collaboration, L.~H. Tao {\em et~al.},
\newblock Z. Phys. {\bf C70}, 387 (1996).
%%CITATION = ZEPYA,C70,387;%%

\bibitem{Arneodo:1996qexArneodo:1996kd}
NMC collaboration, M.~Arneodo {\em et~al.},
\newblock Nucl. Phys. {\bf B483}, 3 (1997), hep-ph/9610231; Nucl. Phys. {\bf
  B487}, 3 (1997), hep-ex/9611022.

\bibitem{Liang:2004tk}
Y.~Liang, M.~E. Christy, R.~Ent, and C.~E. Keppel,
\newblock Phys. Rev. {\bf C73}, 065201 (2006), nucl-ex/0410028.
%%CITATION = NUCL-EX/0410028;%%

\bibitem{Adloff:1996yz}
H1 collaboration, C.~Adloff {\em et~al.},
\newblock Phys. Lett. {\bf B393}, 452 (1997), hep-ex/9611017.
%%CITATION = HEP-EX/9611017;%%

\bibitem{Klein:2004DISProc}
M.~Klein,
\newblock {\sf On the future measurement of the longitudinal structure function
  at low $x$ at HERA}, in: Proc. of the 12th Int. Workshop on Deep Inelastic
  Scattering, DIS 2004, Strebske Pleso, Slovakia, 14--18 April 2004, D.
  Bruncko, J. Ferencei and P. Stri{\v z}enec, eds., (Academic Electronic Press,
  Bratislava, 2004), pp.~309.

\bibitem{Feltesse:2005aa}
J.~Feltesse,
\newblock {\sf Measurement of the longitudinal proton structure function at low
  $x$ at HERA}, in: Proc. of Ringberg Workshop on New Trends in HERA Physics
  2005, Ringberg Castle, Tegernsee, Germany, 2-7 Oct. 2005, G.~Grindhammer,
  B.A.~Kniehl, G.~Kramer and W.~Ochs, eds., (World Scientific, Singapore,
  2005), pp. 370.

\bibitem{Lipka:2006ny}
H1 and ZEUS collaboration, K.~Lipka,
\newblock Nucl. Phys. Proc. Suppl. {\bf 152}, 128 (2006).
%%CITATION = NUPHZ,152,128;%%

\bibitem{Thompson:2007mx}
P.~D. Thompson,
\newblock J. Phys. {\bf G34}, N177 (2007), hep-ph/0703103.
%%CITATION = HEP-PH/0703103;%%

\bibitem{Jung:2009eq}
H.~Jung and A.~De~Roeck,
\newblock eds., {\sf Proceedings of the workshop: HERA and the LHC workshop
  series on the implications of HERA for LHC physics}, (2006--2008, Hamburg,
  Geneve), DESY-PROC-2009-02, March 2009, 794~p. hep-ph/0903.3861.

\bibitem{Chekanov:2008yd}
ZEUS collaboration, S.~Chekanov,
\newblock {\sf Measurement of $D^{\pm}$ and $D^0$ production in deep inelastic
  scattering using a lifetime tag at HERA}, (2008), hep-ex/0812.3775.

\bibitem{Blumlein:1996sc}
J.~Bl{\"umlein} and S.~Riemersma,
\newblock {\sf QCD corrections to $F_L(x,Q^2)$}, (1996), hep-ph/9609394.

\bibitem{Brodsky:1980pb}
S.~J. Brodsky, P.~Hoyer, C.~Peterson, and N.~Sakai,
\newblock Phys. Lett. {\bf B93}, 451 (1980).
%%CITATION = PHLTA,B93,451;%%

\bibitem{Hoffmann:1983ah}
E.~Hoffmann and R.~Moore,
\newblock Z. Phys. {\bf C20}, 71 (1983).
%%CITATION = ZEPYA,C20,71;%%

\bibitem{Derrick:1995sc}
ZEUS collaboration, M.~Derrick {\em et~al.},
\newblock Phys. Lett. {\bf B349}, 225 (1995), hep-ex/9502002.
%%CITATION = HEP-EX/9502002;%%

\bibitem{Harris:1995jx}
B.~W. Harris, J.~Smith, and R.~Vogt,
\newblock Nucl. Phys. {\bf B461}, 181 (1996), hep-ph/9508403.
%%CITATION = HEP-PH/9508403;%%

\bibitem{Adloff:1996xq}
H1 collaboration, C.~Adloff {\em et~al.},
\newblock Z. Phys. {\bf C72}, 593 (1996), hep-ex/9607012.
%%CITATION = HEP-EX/9607012;%%

\bibitem{Bethke:2000ai}
S.~Bethke,
\newblock J. Phys. {\bf G26}, R27 (2000), hep-ex/0004021.
%%CITATION = HEP-EX/0004021;%%

\bibitem{Bethke:2004uy}
S.~Bethke,
\newblock Nucl. Phys. Proc. Suppl. {\bf 135}, 345 (2004), hep-ex/0407021.
%%CITATION = HEP-EX/0407021;%%

\bibitem{Blumlein:2004ip}
J.~Bl{\"umlein}, H.~B{\"o}ttcher, and A.~Guffanti,
\newblock Nucl. Phys. Proc. Suppl. {\bf 135}, 152 (2004), hep-ph/0407089.
%%CITATION = HEP-PH/0407089;%%

\bibitem{Alekhin:2005dxxAlekhin:2005dy}
S.~Alekhin {\em et~al.},
\newblock {\sf HERA and the LHC - A workshop on the implications of HERA for
  LHC physics: Proceedings Part A, B}, (2005), hep-ph/0601012, hep-ph/0601013.

\bibitem{Dittmar:2005ed}
M.~Dittmar {\em et~al.},
\newblock {\sf Parton distributions: Summary report for the HERA - LHC
  workshop}, (2005), hep-ph/0511119.

\bibitem{Gluck:2006yz}
M.~Gl{\"u}ck, E.~Reya, and C.~Schuck,
\newblock Nucl. Phys. {\bf B754}, 178 (2006), hep-ph/0604116.
%%CITATION = HEP-PH/0604116;%%

\bibitem{Alekhin:2006zm}
S.~Alekhin, K.~Melnikov, and F.~Petriello,
\newblock Phys. Rev. {\bf D74}, 054033 (2006), hep-ph/0606237.
%%CITATION = HEP-PH/0606237;%%

\bibitem{Blumlein:2006be}
J.~Bl{\"umlein}, H.~B{\"o}ttcher, and A.~Guffanti,
\newblock Nucl. Phys. {\bf B774}, 182 (2007), hep-ph/0607200.
%%CITATION = HEP-PH/0607200;%%

\bibitem{Blumlein:2007dk}
J.~Bl{\"umlein},
\newblock {\sf $\Lambda_{\rm QCD}$ and $\alpha_s(M_Z^2)$ from DIS Structure
  Functions}, (2007), hep-ph/0706.2430.

\bibitem{Jung:2008tq}
H.~Jung {\em et~al.},
\newblock {\sf What HERA may provide?}, (2008), hep-ph/0809.0549.

\bibitem{Witten:1975bh}
E.~Witten,
\newblock Nucl. Phys. {\bf B104}, 445 (1976).
%%CITATION = NUPHA,B104,445;%%

\bibitem{Babcock:1977fi}
J.~Babcock, D.~W. Sivers, and S.~Wolfram,
\newblock Phys. Rev. {\bf D18}, 162 (1978).
%%CITATION = PHRVA,D18,162;%%

\bibitem{Shifman:1977yb}
M.~A. Shifman, A.~I. Vainshtein, and V.~I. Zakharov,
\newblock Nucl. Phys. {\bf B136}, 157 (1978).
%%CITATION = NUPHA,B136,157;%%

\bibitem{Leveille:1978px}
J.~P. Leveille and T.~J. Weiler,
\newblock Nucl. Phys. {\bf B147}, 147 (1979).
%%CITATION = NUPHA,B147,147;%%

\bibitem{Gluck:1980cp}
M.~Gl{\"uck}, E.~Hoffmann, and E.~Reya,
\newblock Z. Phys. {\bf C13}, 119 (1982).
%%CITATION = ZEPYA,C13,119;%%

\bibitem{Laenen:1992zkxLaenen:1992xs}
E.~Laenen, S.~Riemersma, J.~Smith, and W.~L. van Neerven,
\newblock Nucl. Phys. {\bf B392}, 162 (1993); Nucl. Phys. {\bf B392}, 229
  (1993) .

\bibitem{Riemersma:1994hv}
S.~Riemersma, J.~Smith, and W.~L. van Neerven,
\newblock Phys. Lett. {\bf B347}, 143 (1995), hep-ph/9411431.
%%CITATION = HEP-PH/9411431;%%

\bibitem{Taylor:1976rk}
R.~E. Taylor,
\newblock {\sf Inelastic electron-Nucleon Scattering Experiments}, Invited
  paper presented at Int. Symposium on Lepton and Photon Interactions, Stanford
  Univ., Calif., Aug 21-27, 1975, W.T. Kirk, ed., (SLAC, Stanford CA, 1976),
  SLAC--PUB--1729, 29pp.

\bibitem{Zee:1974du}
A.~Zee, F.~Wilczek, and S.~B. Treiman,
\newblock Phys. Rev. {\bf D10}, 2881 (1974).
%%CITATION = PHRVA,D10,2881;%%

\bibitem{Bardeen:1978yd}
W.~A. Bardeen, A.~J. Buras, D.~W. Duke, and T.~Muta,
\newblock Phys. Rev. {\bf D18}, 3998 (1978).
%%CITATION = PHRVA,D18,3998;%%

\bibitem{Furmanski:1981cw}
W.~Furmanski and R.~Petronzio,
\newblock Z. Phys. {\bf C11}, 293 (1982),
\newblock and references therein.
%%CITATION = ZEPYA,C11,293;%%

\bibitem{Duke:1981ga}
D.~W. Duke, J.~D. Kimel, and G.~A. Sowell,
\newblock Phys. Rev. {\bf D25}, 71 (1982).
%%CITATION = PHRVA,D25,71;%%

\bibitem{Devoto:1984wu}
A.~Devoto, D.~W. Duke, J.~D. Kimel, and G.~A. Sowell,
\newblock Phys. Rev. {\bf D30}, 541 (1984).
%%CITATION = PHRVA,D30,541;%%

\bibitem{Kazakov:1987jk}
D.~I. Kazakov and A.~V. Kotikov,
\newblock Nucl. Phys. {\bf B307}, 721 (1988).
%%CITATION = NUPHA,B307,721;%%

\bibitem{Kazakov:1990fu}
D.~I. Kazakov, A.~V. Kotikov, G.~Parente, O.~A. Sampayo, and
  J.~Sanchez~Guillen,
\newblock Phys. Rev. Lett. {\bf 65}, 1535 (1990).
%%CITATION = PRLTA,65,1535;%%

\bibitem{SanchezGuillen:1990iq}
J.~Sanchez~Guillen, J.~Miramontes, M.~Miramontes, G.~Parente, and O.~A.
  Sampayo,
\newblock Nucl. Phys. {\bf B353}, 337 (1991).
%%CITATION = NUPHA,B353,337;%%

\bibitem{vanNeerven:1991nnxZijlstra:1991qcxZijlstra:1992qd}
W.~L. van Neerven and E.~B. Zijlstra,
\newblock Phys. Lett. {\bf B272}, 127 (1991); Phys. Lett. {\bf B273}, 476
  (1991); Nucl. Phys. {\bf B383}, 525 (1992).

\bibitem{Kazakov:1992xj}
D.~I. Kazakov and A.~V. Kotikov,
\newblock Phys. Lett. {\bf B291}, 171 (1992).
%%CITATION = PHLTA,B291,171;%%

\bibitem{Larin:1991fv}
S.~A. Larin and J.~A.~M. Vermaseren,
\newblock Z. Phys. {\bf C57}, 93 (1993).
%%CITATION = ZEPYA,C57,93;%%

\bibitem{Moch:1999eb}
S.~Moch and J.~A.~M. Vermaseren,
\newblock Nucl. Phys. {\bf B573}, 853 (2000), hep-ph/9912355.
%%CITATION = HEP-PH/9912355;%%

\bibitem{Larin:1993vu}
S.~A. Larin, T.~van Ritbergen, and J.~A.~M. Vermaseren,
\newblock Nucl. Phys. {\bf B427}, 41 (1994).
%%CITATION = NUPHA,B427,41;%%

\bibitem{Larin:1996wd}
S.~A. Larin, P.~Nogueira, T.~van Ritbergen, and J.~A.~M. Vermaseren,
\newblock Nucl. Phys. {\bf B492}, 338 (1997), hep-ph/9605317.
%%CITATION = HEP-PH/9605317;%%

\bibitem{Retey:2000nq}
A.~Retey and J.~A.~M. Vermaseren,
\newblock Nucl. Phys. {\bf B604}, 281 (2001), hep-ph/0007294.
%%CITATION = HEP-PH/0007294;%%

\bibitem{Moch:2004xu}
S.~Moch, J.~A.~M. Vermaseren, and A.~Vogt,
\newblock Phys. Lett. {\bf B606}, 123 (2005), hep-ph/0411112.
%%CITATION = HEP-PH/0411112;%%

\bibitem{Blumlein:2004xt}
J.~Bl{\"umlein} and J.~A.~M. Vermaseren,
\newblock Phys. Lett. {\bf B606}, 130 (2005), hep-ph/0411111.
%%CITATION = HEP-PH/0411111;%%

\bibitem{Vermaseren:2005qc}
J.~A.~M. Vermaseren, A.~Vogt, and S.~Moch,
\newblock Nucl. Phys. {\bf B724}, 3 (2005), hep-ph/0504242.
%%CITATION = HEP-PH/0504242;%%

\bibitem{Alekhin:2003ev}
S.~I. Alekhin and J.~Bl{\"umlein},
\newblock Phys. Lett. {\bf B594}, 299 (2004), hep-ph/0404034.
%%CITATION = HEP-PH/0404034;%%

\bibitem{Altarelli:1977zs}
G.~Altarelli and G.~Parisi,
\newblock Nucl. Phys. {\bf B126}, 298 (1977).
%%CITATION = NUPHA,B126,298;%%

\bibitem{Baikov:2006ai}
P.~A. Baikov and K.~G. Chetyrkin,
\newblock Nucl. Phys. Proc. Suppl. {\bf 160}, 76 (2006).
%%CITATION = NUPHZ,160,76;%%

\bibitem{Floratos:1977auxFloratos:1977aue1}
E.~G. Floratos, D.~A. Ross, and C.~T. Sachrajda,
\newblock Nucl. Phys. {\bf B129}, 66 (1977); [Erratum-ibid.] {\bf B139}, 545
  (1978).

\bibitem{Floratos:1978ny}
E.~G. Floratos, D.~A. Ross, and C.~T. Sachrajda,
\newblock Nucl. Phys. {\bf B152}, 493 (1979).
%%CITATION = NUPHA,B152,493;%%

\bibitem{GonzalezArroyo:1979df}
A.~Gonzalez-Arroyo, C.~Lopez, and F.~J. Yndurain,
\newblock Nucl. Phys. {\bf B153}, 161 (1979).
%%CITATION = NUPHA,B153,161;%%

\bibitem{GonzalezArroyo:1979he}
A.~Gonzalez-Arroyo and C.~Lopez,
\newblock Nucl. Phys. {\bf B166}, 429 (1980).
%%CITATION = NUPHA,B166,429;%%

\bibitem{Curci:1980uw}
G.~Curci, W.~Furmanski, and R.~Petronzio,
\newblock Nucl. Phys. {\bf B175}, 27 (1980).
%%CITATION = NUPHA,B175,27;%%

\bibitem{Furmanski:1980cm}
W.~Furmanski and R.~Petronzio,
\newblock Phys. Lett. {\bf B97}, 437 (1980).
%%CITATION = PHLTA,B97,437;%%

\bibitem{Hamberg:1991qt}
R.~Hamberg and W.~L. van Neerven,
\newblock Nucl. Phys. {\bf B379}, 143 (1992).
%%CITATION = NUPHA,B379,143;%%

\bibitem{Moch:2004pa}
S.~Moch, J.~A.~M. Vermaseren, and A.~Vogt,
\newblock Nucl. Phys. {\bf B688}, 101 (2004), hep-ph/0403192.
%%CITATION = HEP-PH/0403192;%%

\bibitem{Vogt:2004mw}
A.~Vogt, S.~Moch, and J.~A.~M. Vermaseren,
\newblock Nucl. Phys. {\bf B691}, 129 (2004), hep-ph/0404111.
%%CITATION = HEP-PH/0404111;%%

\bibitem{Buza:1995ie}
M.~Buza, Y.~Matiounine, J.~Smith, R.~Migneron, and W.~L. van Neerven,
\newblock Nucl. Phys. {\bf B472}, 611 (1996), hep-ph/9601302.
%%CITATION = HEP-PH/9601302;%%

\bibitem{Blumlein:2006mh}
J.~Bl{\"umlein}, A.~De~Freitas, W.~L. van Neerven, and S.~Klein,
\newblock Nucl. Phys. {\bf B755}, 272 (2006), hep-ph/0608024.
%%CITATION = HEP-PH/0608024;%%

\bibitem{Bierenbaum:2007qe}
I.~Bierenbaum, J.~Bl{\"umlein}, and S.~Klein,
\newblock Nucl. Phys. {\bf B780}, 40 (2007), hep-ph/0703285.
%%CITATION = HEP-PH/0703285;%%

\bibitem{Buza:1996wv}
M.~Buza, Y.~Matiounine, J.~Smith, and W.~L. van Neerven,
\newblock Eur. Phys. J. {\bf C1}, 301 (1998), hep-ph/9612398.
%%CITATION = HEP-PH/9612398;%%

\bibitem{Bierenbaum:2009zt}
I.~Bierenbaum, J.~Bl{\"u}mlein, and S.~Klein,
\newblock Phys. Lett. {\bf B672}, 401 (2009), hep-ph/0901.0669.
%%CITATION = 0901.0669;%%

\bibitem{Bierenbaum:2008dk}
I.~Bierenbaum, J.~Bl{\"umlein}, and S.~Klein,
\newblock Nucl. Phys. Proc. Suppl. {\bf 183}, 162 (2008), hep-ph/0806.4613.
%%CITATION = 0806.4613;%%

\bibitem{Bierenbaum:2008tt}
I.~Bierenbaum, J.~Bl{\"umlein}, and S.~Klein,
\newblock PoS {\bf {\sf Confinement8}}, 185 (2008), hep-ph/0812.2427.
%%CITATION = 0812.2427;%%

\bibitem{Bierenbaum:2009HERA}
I.~Bierenbaum, J.~Bl{\"u}mlein, and S.~Klein,
\newblock {\sf 2-- and 3--loop heavy flavor contributions to $F_2(x,Q^2),
  F_L(x,Q^2)$ and $g_{1,2}(x,Q^2)$} in \cite{Jung:2009eq}, pp~363.

\bibitem{Bierenbaum:2009mv}
I.~Bierenbaum, J.~Bl{\"u}mlein, and S.~Klein,
\newblock {\sf Mellin Moments of the {$O(\alpha_s^3$)} Heavy Flavor
  Contributions to unpolarized Deep-Inelastic Scattering at $Q^2 \gg m^2$ and
  Anomalous Dimensions}, Nucl. Phys. {\bf B} (in print), (2009),
  hep-ph/0904.3563.

\bibitem{Bierenbaum:2007rg}
I.~Bierenbaum, J.~Bl{\"umlein}, and S.~Klein,
\newblock Acta Phys. Polon. {\bf B38}, 3543 (2007); Pos {\sf RADCOR} 2007, 034
  (2007), hep-ph/0710.3348.
%%CITATION = 0710.3348;%%

\bibitem{Bierenbaum:2008tm}
I.~Bierenbaum, J.~Bl{\"umlein}, and S.~Klein,
\newblock {Acta Phys. Polon.} {\bf B39}, 1531 (2008), hep-ph/0806.0451.
%%CITATION = 0806.0451;%%

\bibitem{Bierenbaum:2008yu}
I.~Bierenbaum, J.~Bl{\"umlein}, S.~Klein, and C.~Schneider,
\newblock Nucl. Phys. {\bf B803}, 1 (2008), hep-ph/0803.0273.
%%CITATION = 0803.0273;%%

\bibitem{Blumlein:2007dj}
J.~Bl{\"umlein} and S.~Klein,
\newblock PoS {\bf {\sf ACAT 2007}}, 084 (2007), hep-ph/0706.2426.
%%CITATION = 0706.2426;%%

\bibitem{Bierenbaum:2007zu}
I.~Bierenbaum, J.~Bl{\"umlein}, S.~Klein, and C.~Schneider,
\newblock PoS {\bf {\sf ACAT 2007}}, 082 (2007), math-ph/0707.4659.
%%CITATION = 0707.4659;%%

\bibitem{Blumlein:2009tm}
J.~Bl{\"u}mlein, M.~Kauers, S.~Klein, and C.~Schneider,
\newblock PoS {\bf {\sf ACAT 2008}}, 106 (2008), hep-ph/0902.4095.
%%CITATION = 0902.4095;%%

\bibitem{Blumlein:2009tj}
J.~Bl{\"u}mlein, M.~Kauers, S.~Klein, and C.~Schneider,
\newblock {\sf Determining the closed forms of the $O(a_s^3)$ anomalous
  dimensions and Wilson coefficients from Mellin moments by means of computer
  algebra}, Comp. Phys. Commun. (in print), (2009), hep-ph/0902.4091.

\bibitem{Blumlein:1998if}
J.~Bl{\"umlein} and S.~Kurth,
\newblock Phys. Rev. {\bf D60}, 014018 (1999), hep-ph/9810241.
%%CITATION = HEP-PH/9810241;%%

\bibitem{Vermaseren:1998uu}
J.~A.~M. Vermaseren,
\newblock Int. J. Mod. Phys. {\bf A14}, 2037 (1999), hep-ph/9806280.
%%CITATION = HEP-PH/9806280;%%

\bibitem{Blumlein:2004bb}
J.~Bl{\"umlein},
\newblock Nucl. Phys. Proc. Suppl. {\bf 135}, 225 (2004), hep-ph/0407044.
%%CITATION = HEP-PH/0407044;%%

\bibitem{Blumlein:2005im}
J.~Bl{\"umlein} and V.~Ravindran,
\newblock Nucl. Phys. {\bf B716}, 128 (2005), hep-ph/0501178.
%%CITATION = HEP-PH/0501178;%%

\bibitem{Blumlein:2006rr}
J.~Bl{\"umlein} and V.~Ravindran,
\newblock Nucl. Phys. {\bf B749}, 1 (2006), hep-ph/0604019.
%%CITATION = HEP-PH/0604019;%%

\bibitem{Blumlein:2003gb}
J.~Bl{\"umlein},
\newblock Comput. Phys. Commun. {\bf 159}, 19 (2004), hep-ph/0311046.
%%CITATION = HEP-PH/0311046;%%

\bibitem{Blumlein:2009ta}
J.~Bl{\"u}mlein,
\newblock {\sf Structural Relations of Harmonic Sums and Mellin Transforms up
  to Weight w=5}, (2009), hep-ph/0901.3106.

\bibitem{Blumlein:2009fz}
J.~Bl{\"u}mlein,
\newblock {\sf Structural Relations of Harmonic Sums and Mellin Transforms at
  Weight w=6}, in Proc. of the Workshop ``Motives, Quantum Field Theory, and
  Pseudodifferential Operators, June (2008)'', (Clay Institute, Boston
  University, 2009), in print, math-ph/0901.0837.
%%CITATION = 0901.0837;%%

\bibitem{Weinzierl:2002hv}
S.~Weinzierl,
\newblock Comput. Phys. Commun. {\bf 145}, 357 (2002), math-ph/0201011.
%%CITATION = MATH-PH/0201011;%%

\bibitem{Moch:2005uc}
S.~Moch and P.~Uwer,
\newblock Comput. Phys. Commun. {\bf 174}, 759 (2006), math-ph/0508008.
%%CITATION = MATH-PH/0508008;%%

\bibitem{Refined}
C.~Schneider,
\newblock Ann.Comb., {\bf 9} (1) (2005) 75; Proc. ISSAC'05, (2005) pp. 285 (ACM
  Press); Proc. FPSAC'07, (2007) 1.

\bibitem{Schneider:2007}
C.~Schneider,
\newblock J. Algebra Appl. {\bf 6 (3)}, 415 (2007).

\bibitem{sigma1}
C.~Schneider,
\newblock J. Diffr. Equations Appl., {\bf 11} (9) (2005) 799.

\bibitem{sigma2}
C.~Schneider,
\newblock S\'{e}m. Lothar. Combin. {\bf 56} (2007) Article B56b and
  Habilitation Thesis, JKU Linz, (2007).

\bibitem{Borwein:1999js}
J.~M. Borwein, D.~M. Bradley, D.~J. Broadhurst, and P.~Lisonek,
\newblock Trans. Am. Math. Soc. {\bf 353}, 907 (2001), math/9910045.
%%CITATION = MATH/9910045;%%

\bibitem{Blumlein:2009Zet}
J.~Bl{\"umlein}, D.~Broadhurst, and J.~Vermaseren,
\newblock {\sf The multiple zeta value data mine}, {\rm DESY 09--003}.

\bibitem{Bierenbaum:2007pn}
I.~Bierenbaum, J.~Bl{\"umlein}, and S.~Klein,
\newblock {\sf Two-Loop Massive Operator Matrix Elements for Polarized and
  Unpolarized Deep-Inelastic Scattering}, in: Proc. of {\sf 15th International
  Workshop On Deep-Inelastic Scattering And Related Subjects (DIS2007)}, G.
  Grindhammer, K. Sachs, eds., (16--20 April 2007, Munich), Vol. {\bf 2}, pp.
  821, hep-ph/0706.2738.
%%CITATION = 0706.2738;%%

\bibitem{Bierenbaum:2007zz}
I.~Bierenbaum, J.~Bl{\"umlein}, and S.~Klein,
\newblock {\sf Two-loop massive operator matrix elements for polarized and
  unpolarized deep-inelastic scattering}, PoS ACAT {\bf 2007}, 070 (2007).
%%CITATION = POSCI,ACAT,070;%%

\bibitem{Bierenbaum:prep1}
I.~Bierenbaum, J.~Bl{\"u}mlein, and S.~Klein,
\newblock {\sf in preparation}.

\bibitem{Blumlein:trans}
J.~Bl{\"umlein}, S.~Klein, and B.~T{\"o}dtli,
\newblock {\sf $O(\alpha_s^2)$ and $O(\alpha_s^3$) Heavy Flavor Contributions
  to Transversity at $Q^2 \gg m^2$}, DESY 09--60 (2009).

\bibitem{Nogueira:1991ex}
P.~Nogueira,
\newblock J. Comput. Phys. {\bf 105}, 279 (1993).
%%CITATION = JCTPA,105,279;%%

\bibitem{Vermaseren:2000nd}
J.~A.~M. Vermaseren,
\newblock {\sf New features of FORM}, (2000), math-ph/0010025.

\bibitem{vanRitbergen:1998pn}
T.~van Ritbergen, A.~N. Schellekens, and J.~A.~M. Vermaseren,
\newblock Int. J. Mod. Phys. {\bf A14}, 41 (1999), hep-ph/9802376.
%%CITATION = HEP-PH/9802376;%%

\bibitem{Steinhauser:2000ry}
M.~Steinhauser,
\newblock Comput. Phys. Commun. {\bf 134}, 335 (2001), hep-ph/0009029.
%%CITATION = HEP-PH/0009029;%%

\bibitem{Buza:1996xr}
M.~Buza, Y.~Matiounine, J.~Smith, and W.~L. van Neerven,
\newblock Nucl. Phys. {\bf B485}, 420 (1997), hep-ph/9608342.
%%CITATION = HEP-PH/9608342;%%

\bibitem{Barone:2001sp}
V.~Barone, A.~Drago, and P.~G. Ratcliffe,
\newblock Phys. Rept. {\bf 359}, 1 (2002), hep-ph/0104283.
%%CITATION = HEP-PH/0104283;%%

\bibitem{Vermaseren:1994je}
J.~A.~M. Vermaseren,
\newblock Comput. Phys. Commun. {\bf 83}, 45 (1994).
%%CITATION = CPHCB,83,45;%%

\bibitem{Albrecht:1969zyxAlbrecht:1969qmxAlbrecht:1969zb}
W.~Albrecht {\em et~al.},
\newblock Phys. Lett. {\bf B28}, 225 (1968); Nucl. Phys. {\bf B13}, 1 (1969);
  {\sf Separation of sigma-L and sigma-t in the region of deep inelastic
  electron - proton scattering}, DESY-69-046 (1969).

\bibitem{Clifft:1974zt}
R.~Clifft and N.~Doble,
\newblock {\sf Proposed Design of a High-Energy, High Intensity Muon Beam for
  the SPS North Experimental Area}, CERN/LAB. II/EA/74-2 (1974).

\bibitem{Fox:1974ry}
D.~J. Fox {\em et~al.},
\newblock Phys. Rev. Lett. {\bf 33}, 1504 (1974).
%%CITATION = PRLTA,33,1504;%%

\bibitem{Sloan:1988qj}
T.~Sloan, R.~Voss, and G.~Smadja,
\newblock Phys. Rept. {\bf 162}, 45 (1988).
%%CITATION = PRPLC,162,45;%%

\bibitem{Holder:1977gn}
M.~Holder {\em et~al.},
\newblock Nucl. Instrum. Meth. {\bf 151}, 69 (1978).
%%CITATION = NUIMA,151,69;%%

\bibitem{VonRuden:1982fp}
CDHSW collaboration, W.~Von~Ruden,
\newblock IEEE Trans. Nucl. Sci. {\bf 29}, 360 (1982).
%%CITATION = IETNA,29,360;%%

\bibitem{Harigel:1977}
G.~Harigel,
\newblock {\sf BEBC user's handbook}, (CERN, Geneva, 1977).

\bibitem{Sakumoto:1990py}
W.~K. Sakumoto {\em et~al.},
\newblock Nucl. Instrum. Meth. {\bf A294}, 179 (1990).
%%CITATION = NUIMA,A294,179;%%

\bibitem{King:1991gs}
B.~J. King {\em et~al.},
\newblock Nucl. Instrum. Meth. {\bf A302}, 254 (1991).
%%CITATION = NUIMA,A302,254;%%

\bibitem{Diemoz:1986kt}
M.~Diemoz, F.~Ferroni, and E.~Longo,
\newblock Phys. Rept. {\bf 130}, 293 (1986).
%%CITATION = PRPLC,130,293;%%

\bibitem{Eisele:1986uz}
F.~Eisele,
\newblock Rept. Prog. Phys. {\bf 49}, 233 (1986).
%%CITATION = RPPHA,49,233;%%

\bibitem{Mishra:1989jc}
S.~R. Mishra and F.~Sciulli,
\newblock Ann. Rev. Nucl. Part. Sci. {\bf 39}, 259 (1989).
%%CITATION = ARNUA,39,259;%%

\bibitem{Winter:1991ua}
K.~Winter,
\newblock ed., {\sf Neutrino physics}, Camb. Monogr. Part. Phys. Nucl. Phys.
  Cosmol. Vol. {\bf 1}, (Cambridge University Press, Cambridge, 1991), 670~p.
%%CITATION = CMPCE,1,1;%%

\bibitem{Schmitz:1997}
N.~Schmitz,
\newblock {\sf Neutrinophysik}, (Teubner, Stuttgart, 1997), 478~p.

\bibitem{Peccei:1988pa}
R.~D. Peccei,
\newblock ed., {\sf Proceedings, HERA Workshop}, Hamburg, F.R. Germany, October
  12-14, 1987. Vol. {\bf 1,2}, 937~p.

\bibitem{Buchmuller:1992rq}
W.~Buchm{\"u}ller and G.~Ingelman,
\newblock eds., {\sf Proceedings, Physics at HERA Workshop, Hamburg, F.R.
  Germany, October 29-30, 1991. Vol. {\bf 1-3}, 1566 p}.

\bibitem{Blumlein:1992qi}
J.~Bl{\"u}mlein and T.~Riemann,
\newblock eds., {\sf Deep inelastic scattering. Proceedings, Zeuthen Workshop
  on Elementary Particle Theory, Teupitz, Germany, April 6-10, 1992}, Nucl.
  Phys. Proc. Suppl. {\bf B29A}, 295 p. (1992).

\bibitem{Faessler:1993ku}
{\sf HERA - The new frontier for QCD. Proceedings, Workshop, Durham, UK, March
  21-26, 1993}, J. Phys. {\bf G19}, (1993), pp. 1427.
%%CITATION = JPHGB,G19,1427;%%

\bibitem{Mathiot:1995ir}
J.~F. Mathiot and J.~Tran Thanh~Van,
\newblock eds., Prepared for 6th Rencontres de Blois: {\sf The Heart of the
  Matter: from Nuclear Interactions to Quark - Gluon Dynamics}, Blois, France,
  20-25 Jun 1994, (Ed. Frontieres, Gif--sur--Yvette, 1995), 556~p.

\bibitem{Bluemlein:1995uc}
J.~Bl{\"u}mlein and W.~D. Nowak,
\newblock eds., {\sf Prospects of spin physics at HERA. Proc., Workshop,
  Zeuthen, Germany, August 28-31, 1995, DESY-95-200}, (DESY, Hamburg, 1995),
  387~p.

\bibitem{Ingelman:1996ge}
G.~Ingelman, A.~De~Roeck, and R.~Klanner,
\newblock eds., {\sf Future physics at HERA. Proceedings, Workshop, Hamburg,
  Germany, September 25, 1995-May 31, 1996. Vol. {\bf 1,2}, DESY-96-235},
  1231~p.

\bibitem{Blumlein:1997ch}
J.~Bl{\"u}mlein, A.~De~Roeck, T.~Gehrmann, and W.~D. Nowak,
\newblock eds., {\sf Deep inelastic scattering off polarized targets: Theory
  meets experiment. Physics with polarized protons at HERA}. Proceedings,
  Workshops, SPIN'97, Zeuthen, Germany, September 1-5, 1997 and Hamburg,
  Germany, March-September 1997, DESY-97-200.

\bibitem{Blumlein:2001je}
J.~Bl{\"u}mlein, W.~D. Nowak, and G.~Schnell,
\newblock eds., {\sf Transverse spin physics}, Proceedings, Topical Workshop,
  Zeuthen, Germany, July 9-11, 2001, DESY-Zeuthen-01-01, Aug 2001. 374~p.

\bibitem{Kogut:1972di}
J.~B. Kogut and L.~Susskind,
\newblock Phys. Rept. {\bf 8}, 75 (1973).
%%CITATION = PRPLC,8,75;%%

\bibitem{Yan:1976np}
T.-M. Yan,
\newblock Ann. Rev. Nucl. Part. Sci. {\bf 26}, 199 (1976).
%%CITATION = ARNUA,26,199;%%

\bibitem{Blumlein:1992we}
J.~Bl{\"u}mlein and M.~Klein,
\newblock Nucl. Instrum. Meth. {\bf A329}, 112 (1993).
%%CITATION = NUIMA,A329,112;%%

\bibitem{Blumlein:1994ii}
J.~Bl{\"umlein},
\newblock Z. Phys. {\bf C65}, 293 (1995), hep-ph/9403342.
%%CITATION = HEP-PH/9403342;%%

\bibitem{Arbuzov:1995id}
A.~Arbuzov, D.~Y. Bardin, J.~Bl{\"umlein}, L.~Kalinovskaya, and T.~Riemann,
\newblock Comput. Phys. Commun. {\bf 94}, 128 (1996), hep-ph/9511434,
\newblock and references therein.
%%CITATION = HEP-PH/9511434;%%

\bibitem{Bjorken:1969mm}
J.~D. Bjorken,
\newblock Phys. Rev. {\bf D1}, 1376 (1970).
%%CITATION = PHRVA,D1,1376;%%

\bibitem{Blumlein:1987xk}
J.~Bl{\"umlein}, M.~Klein, T.~Naumann, and T.~Riemann,
\newblock {\sf Structure functions, quark distributions and $\Lambda_{\rm QCD}$
  at HERA} in Proc. of DESY Theory Workshop on Physics at HERA (ed. R.D.
  Peccei), Hamburg, F.R. Germany, Oct 12-14, 1987, Vol {\bf 1}, 67pp.

\bibitem{Kwiatkowski:1990es}
A.~Kwiatkowski, H.~Spiesberger, and H.~J. M{\"o}hring,
\newblock Comp. Phys. Commun. {\bf 69}, 155 (1992),
\newblock and references therein.
%%CITATION = CPHCB,69,155;%%

\bibitem{Engelen:1998rf}
J.~Engelen and P.~Kooijman,
\newblock Prog. Part. Nucl. Phys. {\bf 41}, 1 (1998).
%%CITATION = PPNPD,41,1;%%

\bibitem{Abramowicz:1998ii}
H.~Abramowicz and A.~Caldwell,
\newblock Rev. Mod. Phys. {\bf 71}, 1275 (1999), hep-ex/9903037.
%%CITATION = HEP-EX/9903037;%%

\bibitem{Blumlein:1999sc}
J.~Bl{\"umlein}, B.~Geyer, and D.~Robaschik,
\newblock Nucl. Phys. {\bf B560}, 283 (1999), hep-ph/9903520.
%%CITATION = HEP-PH/9903520;%%

\bibitem{Yndurain:1999ui}
F.~J. Yndurain,
\newblock {\sf The theory of quark and gluon interactions}, (Springer, Berlin,
  2006), 474 p, 4th edition.

\bibitem{Field:1989uq}
R.~Field,
\newblock {\sf Applications of perturbative QCD}, (Addison-Wesley, Redwood
  City, 1989), 366~p.

\bibitem{Dolgov:2002zm}
LHPC collaboration, D.~Dolgov {\em et~al.},
\newblock Phys. Rev. {\bf D66}, 034506 (2002), hep-lat/0201021.
%%CITATION = HEP-LAT/0201021;%%

\bibitem{Gockeler:2007qs}
QCDSF collaboration, M.~G{\"o}ckeler {\em et~al.},
\newblock PoS {\bf {\sf LAT} 2007}, 147 (2007), hep-lat/0710.2489.
%%CITATION = 0710.2489;%%

\bibitem{Baron:2007ti}
ETM collaboration, R.~Baron {\em et~al.},
\newblock PoS {\bf {\sf LAT} 2007}, 153 (2007), hep-lat/0710.1580.
%%CITATION = 0710.1580;%%

\bibitem{Bietenholz:2008fe}
W.~Bietenholz {\em et~al.},
\newblock PoS {\bf {\sf LAT} 2008}, 149 (2008), hep-lat/0808.3637.
%%CITATION = 0808.3637;%%

\bibitem{Syritsyn:2009np}
S.~N. Syritsyn {\em et~al.},
\newblock PoS {\bf {\sf LAT} 2008}, 169 (2008), hep-lat/0903.3063.
%%CITATION = 0903.3063;%%

\bibitem{Itzykson:1980rh}
C.~Itzykson and J.~Zuber,
\newblock {\sf Quantum Field Theory}, (McGraw-Hill, New York, 1980), 705~p.

\bibitem{Blumlein:1996vs}
J.~Bl{\"umlein} and N.~Kochelev,
\newblock Nucl. Phys. {\bf B498}, 285 (1997), hep-ph/9612318.
%%CITATION = HEP-PH/9612318;%%

\bibitem{Blumlein:1998nv}
J.~Bl{\"umlein} and A.~Tkabladze,
\newblock Nucl. Phys. {\bf B553}, 427 (1999), hep-ph/9812478.
%%CITATION = HEP-PH/9812478;%%

\bibitem{Rutherford:1911zz}
E.~Rutherford,
\newblock Phil. Mag. {\bf 21}, 669 (1911).
%%CITATION = PHMAA,21,669;%%

\bibitem{Mcallister:1956ng}
R.~W. Mcallister and R.~Hofstadter,
\newblock Phys. Rev. {\bf 102}, 851 (1956).
%%CITATION = PHRVA,102,851;%%

\bibitem{Schopper:1961}
D.~N. Olson, H.~F. Schopper, and R.~R. Wilson,
\newblock Phys. Rev. Lett. {\bf 6}, 286 (1961).

\bibitem{Hofstadter:1963}
R.~Hofstadter,
\newblock {\sf Electron scattering and nuclear and nucleon structure. A
  collection of reprints with an introduction}, (New York, Benjamin, 1963),
  690~p.

\bibitem{Nash:1971aw}
C.~Nash,
\newblock Nucl. Phys. {\bf B31}, 419 (1971).
%%CITATION = NUPHA,B31,419;%%

\bibitem{Landshoff:1971xb}
P.~V. Landshoff and J.~C. Polkinghorne,
\newblock Phys. Rept. {\bf 5}, 1 (1972).
%%CITATION = PRPLC,5,1;%%

\bibitem{Jackson:1989ph}
J.~D. Jackson, G.~G. Ross, and R.~G. Roberts,
\newblock Phys. Lett. {\bf B226}, 159 (1989).
%%CITATION = PHLTA,B226,159;%%

\bibitem{Roberts:1996ub}
R.~G. Roberts and G.~G. Ross,
\newblock Phys. Lett. {\bf B373}, 235 (1996), hep-ph/9601235.
%%CITATION = HEP-PH/9601235;%%

\bibitem{Blumlein:1996tp}
J.~Bl{\"umlein} and N.~Kochelev,
\newblock Phys. Lett. {\bf B381}, 296 (1996), hep-ph/9603397.
%%CITATION = HEP-PH/9603397;%%

\bibitem{Blumlein:2003wk}
J.~Bl{\"umlein}, V.~Ravindran, and W.~L. van Neerven,
\newblock Phys. Rev. {\bf D68}, 114004 (2003), hep-ph/0304292.
%%CITATION = HEP-PH/0304292;%%

\bibitem{Amati:1978wx}
D.~Amati, R.~Petronzio, and G.~Veneziano,
\newblock Nucl. Phys. {\bf B140}, 54 (1978).
%%CITATION = NUPHA,B140,54;%%

\bibitem{Libby:1978qf}
S.~B. Libby and G.~Sterman,
\newblock Phys. Rev. {\bf D18}, 3252 (1978).
%%CITATION = PHRVA,D18,3252;%%

\bibitem{Libby:1978bx}
S.~B. Libby and G.~Sterman,
\newblock Phys. Rev. {\bf D18}, 4737 (1978).
%%CITATION = PHRVA,D18,4737;%%

\bibitem{Mueller:1978xu}
A.~H. Mueller,
\newblock Phys. Rev. {\bf D18}, 3705 (1978).
%%CITATION = PHRVA,D18,3705;%%

\bibitem{Collins:1981ta}
J.~C. Collins and G.~Sterman,
\newblock Nucl. Phys. {\bf B185}, 172 (1981).
%%CITATION = NUPHA,B185,172;%%

\bibitem{Bodwin:1984hc}
G.~T. Bodwin,
\newblock Phys. Rev. {\bf D31}, 2616 (1985); [Erratum-ibid.] {\bf D34}, 3932
  (1986).
%%CITATION = PHRVA,D31,2616;%%

\bibitem{Collins:1985ue}
J.~C. Collins, D.~E. Soper, and G.~Sterman,
\newblock Nucl. Phys. {\bf B261}, 104 (1985).
%%CITATION = NUPHA,B261,104;%%

\bibitem{Drell:1970yt}
S.~D. Drell and T.-M. Yan,
\newblock Ann. Phys. {\bf 66}, 578 (1971).
%%CITATION = APNYA,66,578;%%

\bibitem{Blumlein:1997}
J.~Bl{\"umlein},
\newblock {\sf Introduction into QCD}, Lecture Notes  (1997).

\bibitem{Jackiw:1972ee}
R.~Jackiw,
\newblock {\sf Canonical light-cone commutators and their applications} in
  \cite{PHOHAD:1971}, pp 1.

\bibitem{Frishman:1973pp}
Y.~Frishman,
\newblock Phys. Rept. {\bf 13}, 1 (1974).
%%CITATION = PRPLC,13,1;%%

\bibitem{Geyer:1977gv}
B.~Geyer, D.~Robaschik, and E.~Wieczorek,
\newblock Fortschr. Phys. {\bf 27}, 75 (1979).
%%CITATION = FPYKA,27,75;%%

\bibitem{Mueller:1981sg}
A.~H. Mueller,
\newblock Phys. Rept. {\bf 73}, 237 (1981).
%%CITATION = PRPLC,73,237;%%

\bibitem{Blumlein:2000hw}
J.~Bl{\"umlein},
\newblock Comput. Phys. Commun. {\bf 133}, 76 (2000), hep-ph/0003100.
%%CITATION = HEP-PH/0003100;%%

\bibitem{Blumlein:2005jg}
J.~Bl{\"umlein} and S.-O. Moch,
\newblock Phys. Lett. {\bf B614}, 53 (2005), hep-ph/0503188.
%%CITATION = HEP-PH/0503188;%%

\bibitem{Carlson:thesis}
E.~Carlson,
\newblock {\sf Sur une classe de s{\'e}ries de Taylor}, PhD Thesis, Uppsala,
  1914.

\bibitem{Titchmarsh:1939}
E.~Titchmarsh,
\newblock {\sf Theory of Functions}, (Oxford University Press, Oxford, 1939),
  Chapt. 9.5.

\bibitem{Landau:1959fi}
L.~D. Landau,
\newblock Nucl. Phys. {\bf 13}, 181 (1959).
%%CITATION = NUPHA,13,181;%%

\bibitem{Bjorken:1959fd}
J.~D. Bjorken,
\newblock {\sf Experimental tests of quantum electrodynamics and spectral
  representations of Green's functions in perturbation theory}, PhD Thesis,
  RX--1037 (1959).

\bibitem{Bassetto:1984ik}
A.~Bassetto, M.~Ciafaloni, and G.~Marchesini,
\newblock Phys. Rept. {\bf 100}, 201 (1983).
%%CITATION = PRPLC,100,201;%%

\bibitem{Stuckelberg:1951gg}
E.~C.~G. St{\"u}ckelberg and A.~Petermann,
\newblock Helv. Phys. Acta {\bf 24}, 317 (1951).
%%CITATION = HPACA,24,317;%%

\bibitem{GellMann:1954fq}
M.~Gell-Mann and F.~E. Low,
\newblock Phys. Rev. {\bf 95}, 1300 (1954).
%%CITATION = PHRVA,95,1300;%%

\bibitem{Bogolyubov:1980nc}
N.~N. Bogolyubov and D.~V. Shirkov,
\newblock {\sf Introduction to the theory of quantized fields}, (New York,
  Interscience, 1959),~720~p.
%%CITATION = IMTPA,3,1;%%

\bibitem{Symanzik:1970rt}
K.~Symanzik,
\newblock Commun. Math. Phys. {\bf 18}, 227 (1970).
%%CITATION = CMPHA,18,227;%%

\bibitem{Callan:1970yg}
C.~G. Callan,
\newblock Phys. Rev. {\bf D2}, 1541 (1970).
%%CITATION = PHRVA,D2,1541;%%

\bibitem{Altarelli:1989ue}
G.~Altarelli,
\newblock Ann. Rev. Nucl. Part. Sci. {\bf 39}, 357 (1989).
%%CITATION = ARNUA,39,357;%%

\bibitem{Owens:1992hd}
J.~F. Owens and W.-K. Tung,
\newblock Ann. Rev. Nucl. Part. Sci. {\bf 42}, 291 (1992).
%%CITATION = ARNUA,42,291;%%

\bibitem{Blumlein:1995cm}
J.~Bl{\"umlein},
\newblock {\sf On the Theoretical Status of Deep Inelastic Scattering}, (1995),
  hep-ph/9512272.
%%CITATION = HEP-PH/9512272;%%

\bibitem{Duke:1984ge}
D.~W. Duke and R.~G. Roberts,
\newblock Phys. Rept. {\bf 120}, 275 (1985).
%%CITATION = PRPLC,120,275;%%

\bibitem{Bethke:1992gh}
S.~Bethke and J.~E. Pilcher,
\newblock Ann. Rev. Nucl. Part. Sci. {\bf 42}, 251 (1992).
%%CITATION = ARNUA,42,251;%%

\bibitem{Moch:2008fj}
S.~Moch, J.~A.~M. Vermaseren, and A.~Vogt,
\newblock {\sf Third-order QCD corrections to the charged-current structure
  function $F_3$}, Nucl. Phys. {\bf B} (in print), (2009), hep-ph/0812.4168.
%%CITATION = 0812.4168;%%

\bibitem{Bloch:1937pw}
F.~Bloch and A.~Nordsieck,
\newblock Phys. Rev. {\bf 52}, 54 (1937).
%%CITATION = PHRVA,52,54;%%

\bibitem{Yennie:1961ad}
D.~R. Yennie, S.~C. Frautschi, and H.~Suura,
\newblock Ann. Phys. {\bf 13}, 379 (1961).
%%CITATION = APNYA,13,379;%%

\bibitem{Kinoshita:1962ur}
T.~Kinoshita,
\newblock J. Math. Phys. {\bf 3}, 650 (1962).
%%CITATION = JMAPA,3,650;%%

\bibitem{Lee:1964is}
T.~D. Lee and M.~Nauenberg,
\newblock Phys. Rev. {\bf 133}, B1549 (1964).
%%CITATION = PHRVA,133,B1549;%%

\bibitem{Peterman:1978tb}
A.~Peterman,
\newblock Phys. Rept. {\bf 53}, 157 (1979).
%%CITATION = PRPLC,53,157;%%

\bibitem{Collins:1984xc}
J.~C. Collins,
\newblock {\sf Renormalization}, (Cambridge University Press, Cambridge, 1984),
  380~p.

\bibitem{Gluck:1988xx}
M.~Gl{\"u}ck, R.~M. Godbole, and E.~Reya,
\newblock Z. Phys. {\bf C41}, 667 (1989).
%%CITATION = ZEPYA,C41,667;%%

\bibitem{Gluck:1989ze}
M.~Gl{\"u}ck, E.~Reya, and A.~Vogt,
\newblock Z. Phys. {\bf C48}, 471 (1990).
%%CITATION = ZEPYA,C48,471;%%

\bibitem{Gluck:1991ng}
M.~Gl{\"u}ck, E.~Reya, and A.~Vogt,
\newblock Z. Phys. {\bf C53}, 127 (1992).
%%CITATION = ZEPYA,C53,127;%%

\bibitem{Gluck:1994uf}
M.~Gl{\"u}ck, E.~Reya, and A.~Vogt,
\newblock Z. Phys. {\bf C67}, 433 (1995).
%%CITATION = ZEPYA,C67,433;%%

\bibitem{Gluck:1998xa}
M.~Gl{\"u}ck, E.~Reya, and A.~Vogt,
\newblock Eur. Phys. J. {\bf C5}, 461 (1998), hep-ph/9806404.
%%CITATION = HEP-PH/9806404;%%

\bibitem{Gluck:2007ck}
M.~Gl{\"u}ck, P.~Jimenez-Delgado, and E.~Reya,
\newblock Eur. Phys. J. {\bf C53}, 355 (2008), hep-ph/0709.0614.
%%CITATION = 0709.0614;%%

\bibitem{JimenezDelgado:2008hf}
P.~Jimenez-Delgado and E.~Reya,
\newblock {\sf Dynamical NNLO parton distributions}, (2008), hep-ph/0810.4274.
%%CITATION = 0810.4274;%%

\bibitem{Alekhin:2005gq}
S.~Alekhin,
\newblock JETP Lett. {\bf 82}, 628 (2005), hep-ph/0508248.
%%CITATION = HEP-PH/0508248;%%

\bibitem{Martin:2009iq}
A.~D. Martin, W.~J. Stirling, R.~S. Thorne, and G.~Watt,
\newblock {\sf Parton distributions for the LHC}, (2009), hep-ph/0901.0002.
%%CITATION = 0901.0002;%%

\bibitem{Lai:1999wy}
CTEQ collaboration, H.~L. Lai {\em et~al.},
\newblock Eur. Phys. J. {\bf C12}, 375 (2000), hep-ph/9903282.
%%CITATION = HEP-PH/9903282;%%

\bibitem{Ball:2008by}
NNPDF collaboration, R.~D. Ball {\em et~al.},
\newblock Nucl. Phys. {\bf B809}, 1 (2009), hep-ph/0808.1231.
%%CITATION = 0808.1231;%%

\bibitem{Gribov:1981ac}
L.~V. Gribov, E.~M. Levin, and M.~G. Ryskin,
\newblock Nucl. Phys. {\bf B188}, 555 (1981).
%%CITATION = NUPHA,B188,555;%%

\bibitem{Mueller:1985wy}
A.~H. Mueller and J.-W. Qiu,
\newblock Nucl. Phys. {\bf B268}, 427 (1986).
%%CITATION = NUPHA,B268,427;%%

\bibitem{Collins:1990cw}
J.~C. Collins and J.~Kwiecinski,
\newblock Nucl. Phys. {\bf B335}, 89 (1990).
%%CITATION = NUPHA,B335,89;%%

\bibitem{Bartels:1990zk}
J.~Bartels, G.~A. Schuler, and J.~Bl{\"u}mlein,
\newblock Z. Phys. {\bf C50}, 91 (1991).
%%CITATION = ZEPYA,C50,91;%%

\bibitem{Altmann:1992vm}
M.~Altmann, M.~Gl{\"u}ck, and E.~Reya,
\newblock Phys. Lett. {\bf B285}, 359 (1992).
%%CITATION = PHLTA,B285,359;%%

\bibitem{DelDuca:1995hf}
V.~Del~Duca,
\newblock {\sf An introduction to the perturbative QCD pomeron and to jet
  physics at large rapidities}, (1995), hep-ph/9503226.
%%CITATION = HEP-PH/9503226;%%

\bibitem{Lipatov:1996ts}
L.~N. Lipatov,
\newblock Phys. Rept. {\bf 286}, 131 (1997), hep-ph/9610276.
%%CITATION = HEP-PH/9610276;%%

\bibitem{Fadin:1975cb}
V.~S. Fadin, E.~A. Kuraev, and L.~N. Lipatov,
\newblock Phys. Lett. {\bf B60}, 50 (1975).
%%CITATION = PHLTA,B60,50;%%

\bibitem{Balitsky:1978ic}
I.~I. Balitsky and L.~N. Lipatov,
\newblock Sov. J. Nucl. Phys. {\bf 28}, 822 (1978).
%%CITATION = SJNCA,28,822;%%

\bibitem{Kirschner:1983di}
R.~Kirschner and L.~N. Lipatov,
\newblock Nucl. Phys. {\bf B213}, 122 (1983).
%%CITATION = NUPHA,B213,122;%%

\bibitem{Bartels:1996wc}
J.~Bartels, B.~I. Ermolaev, and M.~G. Ryskin,
\newblock Z. Phys. {\bf C72}, 627 (1996), hep-ph/9603204.
%%CITATION = HEP-PH/9603204;%%

\bibitem{Fadin:1998py}
V.~S. Fadin and L.~N. Lipatov,
\newblock Phys. Lett. {\bf B429}, 127 (1998), hep-ph/9802290.
%%CITATION = HEP-PH/9802290;%%

\bibitem{Blumlein:1995jpxBlumlein:1996ddxBlumlein:1996hbxBlumlein:1997em}
J.~Bl{\"u}mlein and A.~Vogt,
\newblock Phys. Lett. {\bf B370}, 149 (1996), hep-ph/9510410; Acta Phys. Polon.
  {\bf B27}, 1309 (1996), hep-ph/9603450; Phys. Lett. {\bf B386}, 350 (1996),
  hep-ph/9606254; Phys. Rev. {\bf D58}, 014020 (1998), hep-ph/9712546.

\bibitem{Blumlein:1998pp}
J.~Bl{\"u}mlein, V.~Ravindran, W.~L. van Neerven, and A.~Vogt,
\newblock (1998), hep-ph/9806368.
%%CITATION = HEP-PH/9806368;%%

\bibitem{Blumlein:1998mg}
J.~Bl{\"u}mlein and W.~L. van Neerven,
\newblock Phys. Lett. {\bf B450}, 412 (1999), hep-ph/9811519.
%%CITATION = HEP-PH/9811519;%%

\bibitem{Altarelli:2008aj}
G.~Altarelli, R.~D. Ball, and S.~Forte,
\newblock Nucl. Phys. {\bf B799}, 199 (2008), 0802.0032.
%%CITATION = 0802.0032;%%

\bibitem{Ciafaloni:2007gf}
M.~Ciafaloni, D.~Colferai, G.~P. Salam, and A.~M. Stasto,
\newblock JHEP {\bf 08}, 046 (2007), 0707.1453.
%%CITATION = 0707.1453;%%

\bibitem{Alekhin:2002fv}
S.~Alekhin,
\newblock Phys. Rev. {\bf D68}, 014002 (2003), hep-ph/0211096.
%%CITATION = HEP-PH/0211096;%%

\bibitem{Chuvakin:2000jm}
A.~Chuvakin, J.~Smith, and W.~L. van Neerven,
\newblock Phys. Rev. {\bf D62}, 036004 (2000), hep-ph/0002011.
%%CITATION = HEP-PH/0002011;%%

\bibitem{Schuler:1987wj}
G.~A. Schuler,
\newblock Nucl. Phys. {\bf B299}, 21 (1988).
%%CITATION = NUPHA,B299,21;%%

\bibitem{Baur:1987ai}
U.~Baur and J.~J. van~der Bij,
\newblock Nucl. Phys. {\bf B304}, 451 (1988).
%%CITATION = NUPHA,B304,451;%%

\bibitem{Gluck:1987ukxGluck:1987uke1}
M.~Gl{\"u}ck, R.~M. Godbole, and E.~Reya,
\newblock Z. Phys. {\bf C38}, 441 (1988); [Erratum-ibid.] {\bf C39}, 590
  (1988).

\bibitem{Breitweg:1999ad}
ZEUS collaboration, J.~Breitweg {\em et~al.},
\newblock Eur. Phys. J. {\bf C12}, 35 (2000), hep-ex/9908012.
%%CITATION = HEP-EX/9908012;%%

\bibitem{Adloff:2001zj}
H1 collaboration, C.~Adloff {\em et~al.},
\newblock Phys. Lett. {\bf B528}, 199 (2002), hep-ex/0108039.
%%CITATION = HEP-EX/0108039;%%

\bibitem{Chekanov:2003rb}
ZEUS collaboration, S.~Chekanov {\em et~al.},
\newblock Phys. Rev. {\bf D69}, 012004 (2004), hep-ex/0308068.
%%CITATION = HEP-EX/0308068;%%

\bibitem{Aktas:2004az}
H1 collaboration, A.~Aktas {\em et~al.},
\newblock Eur. Phys. J. {\bf C40}, 349 (2005), hep-ex/0411046.
%%CITATION = HEP-EX/0411046;%%

\bibitem{Aktas:2005iw}
H1 collaboration, A.~Aktas {\em et~al.},
\newblock Eur. Phys. J. {\bf C45}, 23 (2006), hep-ex/0507081.
%%CITATION = HEP-EX/0507081;%%

\bibitem{Eichten:1984eu}
E.~Eichten, I.~Hinchliffe, K.~D. Lane, and C.~Quigg,
\newblock Rev. Mod. Phys. {\bf 56}, 579 (1984).
%%CITATION = RMPHA,56,579;%%

\bibitem{Gluck:1993dpa}
M.~Gl{\"uck}, E.~Reya, and M.~Stratmann,
\newblock Nucl. Phys. {\bf B422}, 37 (1994).
%%CITATION = NUPHA,B422,37;%%

\bibitem{Georgi:1974sy}
H.~Georgi and S.~L. Glashow,
\newblock Phys. Rev. Lett. {\bf 32}, 438 (1974).
%%CITATION = PRLTA,32,438;%%

\bibitem{Fritzsch:1974nn}
H.~Fritzsch and P.~Minkowski,
\newblock Ann. Phys. {\bf 93}, 193 (1975).
%%CITATION = APNYA,93,193;%%

\bibitem{Djouadi:2005gixDjouadi:2005gj}
A.~Djouadi,
\newblock Phys. Rept. {\bf 457}, 1 (2008), hep-ph/0503172; Phys. Rept. {\bf
  459}, 1 (2008), hep-ph/0503173; and references therein.

\bibitem{Gluck:1979aw}
M.~Gl{\"u}ck and E.~Reya,
\newblock Phys. Lett. {\bf B83}, 98 (1979).
%%CITATION = PHLTA,B83,98;%%

\bibitem{Gluck:1978bf}
M.~Gl{\"u}ck and E.~Reya,
\newblock Phys. Lett. {\bf B79}, 453 (1978).
%%CITATION = PHLTA,B79,453;%%

\bibitem{Berger:1980ni}
E.~L. Berger and D.~L. Jones,
\newblock Phys. Rev. {\bf D23}, 1521 (1981).
%%CITATION = PHRVA,D23,1521;%%

\bibitem{Ingelman:1988qn}
G.~Ingelman and G.~A. Schuler,
\newblock Z. Phys. {\bf C40}, 299 (1988).
%%CITATION = ZEPYA,C40,299;%%

\bibitem{Berends:1987abxBerends:1987abe1}
F.~A. Berends, W.~L. van Neerven, and G.~J.~H. Burgers,
\newblock Nucl. Phys. {\bf B297}, 429 (1988); [Erratum-ibid.] {\bf B304}, 921
  (1988).

\bibitem{vanNeerven:1997gf}
W.~L. van Neerven,
\newblock Acta Phys. Polon. {\bf B28}, 2715 (1997), hep-ph/9708452.
%%CITATION = HEP-PH/9708452;%%

\bibitem{Aivazis:1993pi}
M.~A.~G. Aivazis, J.~C. Collins, F.~I. Olness, and W.-K. Tung,
\newblock Phys. Rev. {\bf D50}, 3102 (1994), hep-ph/9312319.
%%CITATION = HEP-PH/9312319;%%

\bibitem{Thorne:2008xf}
R.~S. Thorne and W.~K. Tung,
\newblock {\sf PQCD Formulations with Heavy Quark Masses and Global Analysis},
  (2008), hep-ph/0809.0714.
%%CITATION = 0809.0714;%%

\bibitem{Blumlein:1998sh}
J.~Bl{\"umlein} and W.~L. van Neerven,
\newblock Phys. Lett. {\bf B450}, 417 (1999), hep-ph/9811351.
%%CITATION = HEP-PH/9811351;%%

\bibitem{Collins:1994ee}
J.~C. Collins and R.~J. Scalise,
\newblock Phys. Rev. {\bf D50}, 4117 (1994), hep-ph/9403231.
%%CITATION = HEP-PH/9403231;%%

\bibitem{Harris:1994tp}
B.~W. Harris and J.~Smith,
\newblock Phys. Rev. {\bf D51}, 4550 (1995), hep-ph/9409405.
%%CITATION = HEP-PH/9409405;%%

\bibitem{Chetyrkin:2008jk}
K.~G. Chetyrkin, B.~A. Kniehl, and M.~Steinhauser,
\newblock Nucl. Phys. {\bf B814}, 231 (2009), hep-ph/0812.1337.
%%CITATION = 0812.1337;%%

\bibitem{'tHooft:1972fi}
G.~'t~Hooft and M.~J.~G. Veltman,
\newblock Nucl. Phys. {\bf B44}, 189 (1972).
%%CITATION = NUPHA,B44,189;%%

\bibitem{Ashmore:1972uj}
J.~F. Ashmore,
\newblock Lett. Nuovo Cim. {\bf 4}, 289 (1972).
%%CITATION = NCLTA,4,289;%%

\bibitem{Cicuta:1972jf}
G.~M. Cicuta and E.~Montaldi,
\newblock Nuovo Cim. Lett. {\bf 4}, 329 (1972).
%%CITATION = NCLTA,4,329;%%

\bibitem{Bollini:1972ui}
C.~G. Bollini and J.~J. Giambiagi,
\newblock Nuovo Cim. {\bf B12}, 20 (1972).
%%CITATION = NUCIA,B12,20;%%

\bibitem{Pauli:1949zm}
W.~Pauli and F.~Villars,
\newblock Rev. Mod. Phys. {\bf 21}, 434 (1949).
%%CITATION = RMPHA,21,434;%%

\bibitem{Speer:1974cz}
E.~R. Speer,
\newblock J. Math. Phys. {\bf 15}, 1 (1974).
%%CITATION = JMAPA,15,1;%%

\bibitem{'tHooft:1973mm}
G.~'t~Hooft,
\newblock Nucl. Phys. {\bf B61}, 455 (1973).
%%CITATION = NUPHA,B61,455;%%

\bibitem{Matiounine:1998ky}
Y.~Matiounine, J.~Smith, and W.~L. van Neerven,
\newblock Phys. Rev. {\bf D57}, 6701 (1998), hep-ph/9801224.
%%CITATION = HEP-PH/9801224;%%

\bibitem{Hamberg:thesis}
R.~Hamberg,
\newblock {\sf Second order gluonic contributions to physical quantities}, PhD
  Thesis, Leiden, 1991.

\bibitem{Tarrach:1980up}
R.~Tarrach,
\newblock Nucl. Phys. {\bf B183}, 384 (1981).
%%CITATION = NUPHA,B183,384;%%

\bibitem{Nachtmann:1981zg}
O.~Nachtmann and W.~Wetzel,
\newblock Nucl. Phys. {\bf B187}, 333 (1981).
%%CITATION = NUPHA,B187,333;%%

\bibitem{Gray:1990yh}
N.~Gray, D.~J. Broadhurst, W.~Gr{\"a}fe, and K.~Schilcher,
\newblock Z. Phys. {\bf C48}, 673 (1990).
%%CITATION = ZEPYA,C48,673;%%

\bibitem{Broadhurst:1991fy}
D.~J. Broadhurst, N.~Gray, and K.~Schilcher,
\newblock Z. Phys. {\bf C52}, 111 (1991).
%%CITATION = ZEPYA,C52,111;%%

\bibitem{Fleischer:1998dw}
J.~Fleischer, F.~Jegerlehner, O.~V. Tarasov, and O.~L. Veretin,
\newblock Nucl. Phys. {\bf B539}, 671 (1999), hep-ph/9803493.
%%CITATION = HEP-PH/9803493;%%

\bibitem{Khriplovich:1969aa}
I.~B. Khriplovich,
\newblock Yad. Fiz. {\bf 10}, 409 (1969).
%%CITATION = YAFIA,10,409;%%

\bibitem{Caswell:1974gg}
W.~E. Caswell,
\newblock Phys. Rev. Lett. {\bf 33}, 244 (1974).
%%CITATION = PRLTA,33,244;%%

\bibitem{Jones:1974mm}
D.~R.~T. Jones,
\newblock Nucl. Phys. {\bf B75}, 531 (1974).
%%CITATION = NUPHA,B75,531;%%

\bibitem{Abbott:1980hw}
L.~F. Abbott,
\newblock Nucl. Phys. {\bf B185}, 189 (1981).
%%CITATION = NUPHA,B185,189;%%

\bibitem{Rebhan:1985yf}
A.~Rebhan,
\newblock Z. Phys. {\bf C30}, 309 (1986).
%%CITATION = ZEPYA,C30,309;%%

\bibitem{Jegerlehner:1998zg}
F.~Jegerlehner and O.~V. Tarasov,
\newblock Nucl. Phys. {\bf B549}, 481 (1999), hep-ph/9809485.
%%CITATION = HEP-PH/9809485;%%

\bibitem{Chetyrkin:1999ysxChetyrkin:1999qi}
K.~G. Chetyrkin and M.~Steinhauser,
\newblock Phys. Rev. Lett. {\bf 83}, 4001 (1999), hep-ph/9907509; Nucl. Phys.
  {\bf B573}, 617 (2000), hep-ph/9911434.

\bibitem{Broadhurst:1991fi}
D.~J. Broadhurst,
\newblock Z. Phys. {\bf C54}, 599 (1992).
%%CITATION = ZEPYA,C54,599;%%

\bibitem{Avdeev:1994db}
L.~Avdeev, J.~Fleischer, S.~Mikhailov, and O.~Tarasov,
\newblock Phys. Lett. {\bf B336}, 560 (1994), hep-ph/9406363.
%%CITATION = HEP-PH/9406363;%%

\bibitem{Laporta:1996mq}
S.~Laporta and E.~Remiddi,
\newblock Phys. Lett. {\bf B379}, 283 (1996), hep-ph/9602417.
%%CITATION = HEP-PH/9602417;%%

\bibitem{Broadhurst:1998rz}
D.~J. Broadhurst,
\newblock Eur. Phys. J. {\bf C8}, 311 (1999), hep-th/9803091.
%%CITATION = HEP-TH/9803091;%%

\bibitem{Boughezal:2004ef}
R.~Boughezal, J.~B. Tausk, and J.~J. van~der Bij,
\newblock Nucl. Phys. {\bf B713}, 278 (2005), hep-ph/0410216.
%%CITATION = HEP-PH/0410216;%%

\bibitem{Chetyrkin:1980pr}
K.~G. Chetyrkin, A.~L. Kataev, and F.~V. Tkachov,
\newblock Nucl. Phys. {\bf B174}, 345 (1980).
%%CITATION = NUPHA,B174,345;%%

\bibitem{SKdiploma}
S.~Klein,
\newblock Diploma Thesis, University of Potsdam (2006).

\bibitem{Slater}
L.~Slater,
\newblock {\sf Generalized Hypergeometric Functions}, (Cambridge University
  Press, Cambridge, 1966), 273~p.

\bibitem{Bailey}
W.~Bailey,
\newblock {\sf Generalized Hypergeometric Series}, (Cambridge University Press,
  Cambridge, 1935), 108~p.

\bibitem{Roy:2001}
G.~Andrews, R.~Askey, and R.~Roy,
\newblock {\sf Special Functions}, Encyclopedia of Mathematics and its
  Applications {\bf 71}, (Cambridge University Press, Cambridge, 2001), 663~p.

\bibitem{Bierenbaum:2007dm}
I.~Bierenbaum, J.~Bl{\"umlein}, and S.~Klein,
\newblock Phys. Lett. {\bf B648}, 195 (2007), hep-ph/0702265.
%%CITATION = HEP-PH/0702265;%%

\bibitem{Blumlein:pol}
I.~Bierenbaum, J.~Bl{\"umlein}, and S.~Klein,
\newblock to appear.

\bibitem{Bierenbaum:2006mq}
I.~Bierenbaum, J.~Bl{\"umlein}, and S.~Klein,
\newblock Nucl. Phys. Proc. Suppl. {\bf 160}, 85 (2006), hep-ph/0607300.
%%CITATION = HEP-PH/0607300;%%

\bibitem{Bauer:2000cp}
C.~W. Bauer, A.~Frink, and R.~Kreckel,
\newblock {\sf Introduction to the GiNaC Framework for Symbolic Computation
  within the C++ Programming Language}, (2000), arxiv:~0004015.
%%CITATION = CS/0004015;%%

\bibitem{Devoto:1983tc}
A.~Devoto and D.~W. Duke,
\newblock Riv. Nuovo Cim. {\bf 7N6}, 1 (1984).
%%CITATION = RNCIB,7N6,1;%%

\bibitem{Blumlein:2004bs}
J.~Bl{\"u}mlein and H.~Kawamura,
\newblock Nucl. Phys. {\bf B708}, 467 (2005), hep-ph/0409289.
%%CITATION = HEP-PH/0409289;%%

\bibitem{Blumleinunp}
J.~Bl{\"umlein},
\newblock {\sf Collection of Polylog-Integrals}, unpublished.

\bibitem{Norlund}
N.~N{\"o}rlund,
\newblock {\sf Vorlesungen {\"u}ber Differenzenrechnung}, (Springer, Berlin,
  1924), 551~p.

\bibitem{Thomson}
L.~Milne-Thomson,
\newblock {\sf The Calculus of finite Differences}, (MacMillan, London, 1951),
  273~p.

\bibitem{GOSPER}
R.~Gosper,
\newblock Proc. Nat. Acad. Sci. USA {\bf 75}, 40 (1978).

\bibitem{Zeil:91}
D.~Zeilberger,
\newblock J.~Symbolic Comput. {\bf 11}, 195 (1991).

\bibitem{AequalB1}
M.~Petkov{\v s}ek, H.~S. Wilf, and D.~Zeilberger,
\newblock {${\sf A=B}$}, (A. K. Peters, Wellesley, MA, 1996).

\bibitem{Karr:1981}
M.~Karr,
\newblock J.~ACM, {\bf 28} (1981) 305. .

\bibitem{RefinedDF}
C.~Schneider,
\newblock J. Symbolic Comput. (2008), doi:10.1016/j.jsc.2008.01.001 .

\bibitem{SigmaAlg}
C.~Schneider,
\newblock {Proc. ISSAC'04}, (2004) pp. 282 (ACM Press).

\bibitem{GON}
A.~Goncharov,
\newblock Math. Res. Lett. {\bf 5} (1998) 497 .

\bibitem{PET}
M.~P. Hoang Ngoc~Minh and J.~van~der Hoeven,
\newblock Discr. Math. {\bf 225} (2000) 217 .

\bibitem{Moch:2001zr}
S.~Moch, P.~Uwer, and S.~Weinzierl,
\newblock J. Math. Phys. {\bf 43}, 3363 (2002), hep-ph/0110083.
%%CITATION = HEP-PH/0110083;%%

\bibitem{AequalB2}
D.~Zeilberger,
\newblock J.~Symbolic Comput. \textbf{11} (1991), 195 .

\bibitem{AhlgrenPade1}
P.~Paule and C.~Schneider,
\newblock Adv. in Appl. Math. {\bf 31 (2)}, 359 (2003).

\bibitem{AhlgrenPade2}
K.~Driver, H.~Prodinger, C.~Schneider, and A.~Weideman,
\newblock Ramanujan Journal {\bf 12 (3)}, 299 (2006).

\bibitem{Hoffman:1997}
M.~E. Hoffman,
\newblock J. Algebra {\bf 194}, 477 (1997).

\bibitem{Hoffman:2004bf}
M.~E. Hoffman,
\newblock Nucl. Phys. Proc. Suppl. {\bf 135}, 215 (2004), math/0406589.
%%CITATION = MATH/0406589;%%

\bibitem{MB1a}
E.~Barnes,
\newblock Proc. Lond. Math. Soc. (2) {\bf 6} (1908) 141 .

\bibitem{MB1b}
E.~Barnes,
\newblock Quart. J. Math. {\bf 41} (1910) 136 .

\bibitem{MB2}
H.~Mellin,
\newblock Math. Ann. {\bf 68} (1910) 305 .

\bibitem{MB3}
E.~Whittaker and G.~Watson,
\newblock {\sf A Course of Modern Analysis}, (Cambridge University Press,
  Cambridge, 1927; reprinted 1996) 616~p .

\bibitem{MB4}
E.~Titchmarsh,
\newblock {\sf Introduction to the Theory of Fourier Integrals}, (Calendron
  Press, Oxford, 1937; 2nd Edition 1948) .

\bibitem{Paris:2001}
R.~Paris and D.~D.~Kaminski,
\newblock {\sf Asymptotics and Mellin-Barnes Integrals}, (Cambridge University
  Press, Cambridge, 2001), 438~p.

\bibitem{Bierenbaum:2003ud}
I.~Bierenbaum and S.~Weinzierl,
\newblock Eur. Phys. J. {\bf C32}, 67 (2003), hep-ph/0308311.
%%CITATION = HEP-PH/0308311;%%

\bibitem{Czakon:2005rk}
M.~Czakon,
\newblock Comput. Phys. Commun. {\bf 175}, 559 (2006), hep-ph/0511200.
%%CITATION = HEP-PH/0511200;%%

\bibitem{Djouadi:1993ss}
A.~Djouadi and P.~Gambino,
\newblock Phys. Rev. {\bf D49}, 3499 (1994), hep-ph/9309298.
%%CITATION = HEP-PH/9309298;%%

\bibitem{Broadhurst:private1}
D.~Broadhurst,
\newblock private communication, 2009.

\bibitem{Gorishnii:1989gt}
S.~G. Gorishnii, S.~A. Larin, L.~R. Surguladze, and F.~V. Tkachov,
\newblock Comput. Phys. Commun. {\bf 55}, 381 (1989).
%%CITATION = CPHCB,55,381;%%

\bibitem{Larin:1991fz}
S.~A. Larin, F.~V. Tkachov, and J.~A.~M. Vermaseren,
\newblock {\sf The FORM version of MINCER}, NIKHEF-H-91-18 (1991).

\bibitem{Remiddi:1999ew}
E.~Remiddi and J.~A.~M. Vermaseren,
\newblock Int. J. Mod. Phys. {\bf A15}, 725 (2000), hep-ph/9905237.
%%CITATION = HEP-PH/9905237;%%

\bibitem{Vermaseren:mincer}
J.~A.~M. Vermaseren,
\newblock {\sf The Form version of MINCER}, unpublished.

\bibitem{Tentyukov:2007mu}
M.~Tentyukov and J.~A.~M. Vermaseren,
\newblock {\sf The multithreaded version of FORM}, (2007), hep-ph/0702279.

\bibitem{Gracey:1993nn}
J.~A. Gracey,
\newblock Phys. Lett. {\bf B322}, 141 (1994), hep-ph/9401214.
%%CITATION = HEP-PH/9401214;%%

\bibitem{Moch:2002sn}
S.~Moch, J.~A.~M. Vermaseren, and A.~Vogt,
\newblock Nucl. Phys. {\bf B646}, 181 (2002), hep-ph/0209100.
%%CITATION = HEP-PH/0209100;%%

\bibitem{Blumlein:prep1}
J.~Bl{\"u}mlein and S.~Klein,
\newblock {\sf in preparation}.

\bibitem{Alguard:1976bmxAlguard:1978gf}
M.~J. Alguard {\em et~al.},
\newblock Phys. Rev. Lett. {\bf 37}, 1261 (1976); Phys. Rev. Lett. {\bf 41}, 70
  (1978).

\bibitem{Baum:1983ha}
G.~Baum {\em et~al.},
\newblock Phys. Rev. Lett. {\bf 51}, 1135 (1983).
%%CITATION = PRLTA,51,1135;%%

\bibitem{Ashman:1987hvxAshman:1989ig}
EMC collaboration, J.~Ashman {\em et~al.},
\newblock Phys. Lett. {\bf B206}, 364 (1988); Nucl. Phys. {\bf B328}, 1 (1989).

\bibitem{Adeva:1993km}
SMC collaboration, B.~Adeva {\em et~al.},
\newblock Phys. Lett. {\bf B302}, 533 (1993).
%%CITATION = PHLTA,B302,533;%%

\bibitem{Anthony:1996mw}
E142 collaboration, P.~L. Anthony {\em et~al.},
\newblock Phys. Rev. {\bf D54}, 6620 (1996), hep-ex/9610007.
%%CITATION = HEP-EX/9610007;%%

\bibitem{Ackerstaff:1997ws}
HERMES collaboration, K.~Ackerstaff {\em et~al.},
\newblock Phys. Lett. {\bf B404}, 383 (1997), hep-ex/9703005.
%%CITATION = HEP-EX/9703005;%%

\bibitem{Abe:1997cx}
E154 collaboration, K.~Abe {\em et~al.},
\newblock Phys. Rev. Lett. {\bf 79}, 26 (1997), hep-ex/9705012.
%%CITATION = HEP-EX/9705012;%%

\bibitem{Adeva:1998vv}
SMC collaboration, B.~Adeva {\em et~al.},
\newblock Phys. Rev. {\bf D58}, 112001 (1998).
%%CITATION = PHRVA,D58,112001;%%

\bibitem{Abe:1998wq}
E143 collaboration, K.~Abe {\em et~al.},
\newblock Phys. Rev. {\bf D58}, 112003 (1998), hep-ph/9802357.
%%CITATION = HEP-PH/9802357;%%

\bibitem{Airapetian:1998wi}
HERMES collaboration, A.~Airapetian {\em et~al.},
\newblock Phys. Lett. {\bf B442}, 484 (1998), hep-ex/9807015.
%%CITATION = HEP-EX/9807015;%%

\bibitem{Anthony:1999rm}
E155 collaboration, P.~L. Anthony {\em et~al.},
\newblock Phys. Lett. {\bf B463}, 339 (1999), hep-ex/9904002.
%%CITATION = HEP-EX/9904002;%%

\bibitem{Anthony:2000fn}
E155 collaboration, P.~L. Anthony {\em et~al.},
\newblock Phys. Lett. {\bf B493}, 19 (2000), hep-ph/0007248.
%%CITATION = HEP-PH/0007248;%%

\bibitem{Zheng:2003un}
Jefferson Lab Hall A collaboration, X.~Zheng {\em et~al.},
\newblock Phys. Rev. Lett. {\bf 92}, 012004 (2004), nucl-ex/0308011.
%%CITATION = NUCL-EX/0308011;%%

\bibitem{Airapetian:2004zf}
HERMES collaboration, A.~Airapetian {\em et~al.},
\newblock Phys. Rev. {\bf D71}, 012003 (2005), hep-ex/0407032.
%%CITATION = HEP-EX/0407032;%%

\bibitem{Ageev:2005gh}
COMPASS collaboration, E.~S. Ageev {\em et~al.},
\newblock Phys. Lett. {\bf B612}, 154 (2005), hep-ex/0501073.
%%CITATION = HEP-EX/0501073;%%

\bibitem{Ageev:2007du}
COMPASS collaboration, E.~S. Ageev {\em et~al.},
\newblock Phys. Lett. {\bf B647}, 330 (2007), hep-ex/0701014.
%%CITATION = HEP-EX/0701014;%%

\bibitem{Airapetian:2007mh}
HERMES collaboration, A.~Airapetian {\em et~al.},
\newblock Phys. Rev. {\bf D75}, 012007 (2007), hep-ex/0609039.
%%CITATION = HEP-EX/0609039;%%

\bibitem{Reya:1992ye}
E.~Reya,
\newblock {\sf The Spin structure of the nucleon}, in: {\sf QCD - 20 years
  later}. Proceedings, Workshop, Aachen, P.M. Zerwas and H.A. Kastrup, eds.,
  Germany, June 9-13, 1992. (World Scientific, Singapore, 1993), Vol. {\bf 1},
  pp.272.

\bibitem{Lampe:1998eu}
B.~Lampe and E.~Reya,
\newblock Phys. Rept. {\bf 332}, 1 (2000), hep-ph/9810270.
%%CITATION = HEP-PH/9810270;%%

\bibitem{Burkardt:2008jw}
M.~Burkardt, A.~Miller, and W.~D. Nowak,
\newblock {\sf Spin-polarized high-energy scattering of charged leptons on
  nucleons}, (2008), hep-ph/0812.2208.
%%CITATION = 0812.2208;%%

\bibitem{Blumlein:1995gh}
J.~Bl{\"u}mlein,
\newblock {\sf On the measurability of the structure function $g_1(x,Q^2)$ in
  ep collisions at HERA}, (1995), hep-ph/9508387.
%%CITATION = HEP-PH/9508387;%%

\bibitem{Kurek:2006fw}
K.~Kurek,
\newblock {\sf $\Delta G$ from COMPASS}, (2006), hep-ex/0607061.
%%CITATION = HEP-EX/0607061;%%

\bibitem{Brona:2007ug}
COMPASS collaboration, G.~Brona,
\newblock {\sf Measurement of the gluon polarisation at COMPASS}, (2007),
  hep-ex/0705.2372.
%%CITATION = 0705.2372;%%

\bibitem{Watson:1981ce}
A.~D. Watson,
\newblock Z. Phys. {\bf C12}, 123 (1982).
%%CITATION = ZEPYA,C12,123;%%

\bibitem{Gluck:1990in}
M.~Gl{\"u}ck, E.~Reya, and W.~Vogelsang,
\newblock Nucl. Phys. {\bf B351}, 579 (1991).
%%CITATION = NUPHA,B351,579;%%

\bibitem{Vogelsang:1990ug}
W.~Vogelsang,
\newblock Z. Phys. {\bf C50}, 275 (1991).
%%CITATION = ZEPYA,C50,275;%%

\bibitem{Altarelli:1998nb}
G.~Altarelli, R.~D. Ball, S.~Forte, and G.~Ridolfi,
\newblock Acta Phys. Polon. {\bf B29}, 1145 (1998), hep-ph/9803237.
%%CITATION = HEP-PH/9803237;%%

\bibitem{Gluck:2000dy}
M.~Gl{\"u}ck, E.~Reya, M.~Stratmann, and W.~Vogelsang,
\newblock Phys. Rev. {\bf D63}, 094005 (2001), hep-ph/0011215.
%%CITATION = HEP-PH/0011215;%%

\bibitem{Bluemlein:2002be}
J.~Bl{\"u}mlein and H.~B{\"o}ttcher,
\newblock Nucl. Phys. {\bf B636}, 225 (2002), hep-ph/0203155.
%%CITATION = HEP-PH/0203155;%%

\bibitem{Hirai:2006sr}
M.~Hirai, S.~Kumano, and N.~Saito,
\newblock Phys. Rev. {\bf D74}, 014015 (2006), hep-ph/0603213.
%%CITATION = HEP-PH/0603213;%%

\bibitem{Leader:2006xc}
E.~Leader, A.~V. Sidorov, and D.~B. Stamenov,
\newblock Phys. Rev. {\bf D75}, 074027 (2007), hep-ph/0612360.
%%CITATION = HEP-PH/0612360;%%

\bibitem{deFlorian:2009vb}
D.~de~Florian, R.~Sassot, M.~Stratmann, and W.~Vogelsang,
\newblock (2009), 0904.3821.
%%CITATION = 0904.3821;%%

\bibitem{Bojak:1998zm}
I.~Bojak and M.~Stratmann,
\newblock Nucl. Phys. {\bf B540}, 345 (1999), hep-ph/9807405.
%%CITATION = HEP-PH/9807405;%%

\bibitem{Wandzura:1977qf}
S.~Wandzura and F.~Wilczek,
\newblock Phys. Lett. {\bf B72}, 195 (1977).
%%CITATION = PHLTA,B72,195;%%

\bibitem{Zijlstra:1993shxZijlstra:1993she1xZijlstra:1993she2}
E.~B. Zijlstra and W.~L. van Neerven,
\newblock Nucl. Phys. {\bf B417}, 61 (1994); [Erratum-ibid.] {\bf B426}, 245
  (1994); [Erratum-ibid.] {\bf B773}, 105 (2007).

\bibitem{Ito:1975pf}
H.~Ito,
\newblock Prog. Theor. Phys. {\bf 54}, 555 (1975).
%%CITATION = PTPKA,54,555;%%

\bibitem{Sasaki:1975hk}
K.~Sasaki,
\newblock Prog. Theor. Phys. {\bf 54}, 1816 (1975).
%%CITATION = PTPKA,54,1816;%%

\bibitem{Ahmed:1976ee}
M.~A. Ahmed and G.~G. Ross,
\newblock Nucl. Phys. {\bf B111}, 441 (1976).
%%CITATION = NUPHA,B111,441;%%

\bibitem{Ward:1950xp}
J.~C. Ward,
\newblock Phys. Rev. {\bf 78}, 182 (1950).
%%CITATION = PHRVA,78,182;%%

\bibitem{Takahashi:1957xn}
Y.~Takahashi,
\newblock Nuovo Cim. {\bf 6}, 371 (1957).
%%CITATION = NUCIA,6,371;%%

\bibitem{Akyeampong:1973xixAkyeampong:1973vkxAkyeampong:1973vj}
D.~A. Akyeampong and R.~Delbourgo,
\newblock Nuovo Cim.~{\bf A17},~578~(1973); Nuovo Cim.~{\bf A18},~94~(1973);
  Nuovo Cim.~{\bf A19},~219~(1974).

\bibitem{Breitenlohner:1976te}
P.~Breitenlohner and D.~Maison,
\newblock Commun. Math. Phys. {\bf 52}, 55 (1977).
%%CITATION = CMPHA,52,55;%%

\bibitem{Bodwin:1989nz}
G.~T. Bodwin and J.-W. Qiu,
\newblock Phys. Rev. {\bf D41}, 2755 (1990).
%%CITATION = PHRVA,D41,2755;%%

\bibitem{Mertig:1995ny}
R.~Mertig and W.~L. van Neerven,
\newblock Z. Phys. {\bf C70}, 637 (1996), hep-ph/9506451.
%%CITATION = HEP-PH/9506451;%%

\bibitem{Vogelsang:1995vh}
W.~Vogelsang,
\newblock Phys. Rev. {\bf D54}, 2023 (1996), hep-ph/9512218.
%%CITATION = HEP-PH/9512218;%%

\bibitem{Vogelsang:1996im}
W.~Vogelsang,
\newblock Nucl. Phys. {\bf B475}, 47 (1996), hep-ph/9603366.
%%CITATION = HEP-PH/9603366;%%

\bibitem{Ralston:1979ys}
J.~P. Ralston and D.~E. Soper,
\newblock Nucl. Phys. {\bf B152}, 109 (1979).
%%CITATION = NUPHA,B152,109;%%

\bibitem{Jaffe:1991kpxJaffe:1991ra}
R.~L. Jaffe and X.-D. Ji,
\newblock Phys. Rev. Lett. {\bf 67}, 552 (1991); Nucl. Phys. {\bf B375}, 527
  (1992).

\bibitem{Cortes:1991ja}
J.~L. Cortes, B.~Pire, and J.~P. Ralston,
\newblock Z. Phys. {\bf C55}, 409 (1992).
%%CITATION = ZEPYA,C55,409;%%

\bibitem{Artru:1989zv}
X.~Artru and M.~Mekhfi,
\newblock Z. Phys. {\bf C45}, 669 (1990).
%%CITATION = ZEPYA,C45,669;%%

\bibitem{Collins:1992kk}
J.~C. Collins,
\newblock Nucl. Phys. {\bf B396}, 161 (1993), hep-ph/9208213.
%%CITATION = HEP-PH/9208213;%%

\bibitem{Jaffe:1993xb}
R.~L. Jaffe and X.-D. Ji,
\newblock Phys. Rev. Lett. {\bf 71}, 2547 (1993), hep-ph/9307329.
%%CITATION = HEP-PH/9307329;%%

\bibitem{Tangerman:1994bb}
R.~D. Tangerman and P.~J. Mulders,
\newblock {\sf Polarized twist - three distributions $g_T$ and $h_L$ and the
  role of intrinsic transverse momentum}, (1994), hep-ph/9408305.
%%CITATION = HEP-PH/9408305;%%

\bibitem{Boer:1997nt}
D.~Boer and P.~J. Mulders,
\newblock Phys. Rev. {\bf D57}, 5780 (1998), hep-ph/9711485.
%%CITATION = HEP-PH/9711485;%%

\bibitem{Airapetian:2004twxAirapetian:2008sk}
HERMES collaboration, A.~Airapetian {\em et~al.},
\newblock Phys. Rev. Lett. {\bf 94}, 012002 (2005), hep-ex/0408013; JHEP {\bf
  06}, 017 (2008), hep-ex/0803.2367.

\bibitem{Alexakhin:2005iw}
COMPASS collaboration, V.~Y. Alexakhin {\em et~al.},
\newblock Phys. Rev. Lett. {\bf 94}, 202002 (2005), hep-ex/0503002.
%%CITATION = HEP-EX/0503002;%%

\bibitem{Afanasev:2007qh}
A.~Afanasev {\em et~al.},
\newblock {\sf Transversity and transverse spin in nucleon structure through
  SIDIS at Jefferson Lab}, (2007), hep-ph/0703288.
%%CITATION = HEP-PH/0703288;%%

\bibitem{:2008dn}
COMPASS collaboration, M.~Alekseev {\em et~al.},
\newblock Phys. Lett. {\bf B673}, 127 (2009), hep-ex/0802.2160.
%%CITATION = 0802.2160;%%

\bibitem{Lutz:2009ff}
The PANDA collaboration, M.~F.~M. Lutz, B.~Pire, O.~Scholten, and
  R.~Timmermans,
\newblock (2009), hep-ex/0903.3905.
%%CITATION = 0903.3905;%%

\bibitem{Anselmino:2007fs}
M.~Anselmino {\em et~al.},
\newblock Phys. Rev. {\bf D75}, 054032 (2007), hep-ph/0701006.
%%CITATION = HEP-PH/0701006;%%

\bibitem{Anselmino:2008jk}
M.~Anselmino {\em et~al.},
\newblock {\sf Update on transversity and Collins functions from SIDIS and
  $e^+$ $e^-$ data}, (2008), hep-ph/0812.4366.
%%CITATION = 0812.4366;%%

\bibitem{Aoki:1996pi}
S.~Aoki, M.~Doui, T.~Hatsuda, and Y.~Kuramashi,
\newblock Phys. Rev. {\bf D56}, 433 (1997), hep-lat/9608115.
%%CITATION = HEP-LAT/9608115;%%

\bibitem{Gockeler:1996es}
M.~G{\"o}ckeler {\em et~al.},
\newblock Nucl. Phys. Proc. Suppl. {\bf 53}, 315 (1997), hep-lat/9609039.
%%CITATION = HEP-LAT/9609039;%%

\bibitem{Khan:2004vw}
A.~A. Khan {\em et~al.},
\newblock Nucl. Phys. Proc. Suppl. {\bf 140}, 408 (2005), hep-lat/0409161.
%%CITATION = HEP-LAT/0409161;%%

\bibitem{Diehl:2005ev}
QCDSF collaboration, M.~Diehl {\em et~al.},
\newblock (2005), hep-ph/0511032.
%%CITATION = HEP-PH/0511032;%%

\bibitem{Gockeler:2006zu}
QCDSF collaboration, M.~G{\"o}ckeler {\em et~al.},
\newblock Phys. Rev. Lett. {\bf 98}, 222001 (2007), hep-lat/0612032.
%%CITATION = HEP-LAT/0612032;%%

\bibitem{Renner:private1}
D.~Renner,
\newblock private communication, 2009.

\bibitem{Ji:1993vw}
X.-D. Ji,
\newblock Phys. Rev. {\bf D49}, 114 (1994), hep-ph/9307235.
%%CITATION = HEP-PH/9307235;%%

\bibitem{Bacchetta:2000jk}
A.~Bacchetta and P.~J. Mulders,
\newblock Phys. Rev. {\bf D62}, 114004 (2000), hep-ph/0007120.
%%CITATION = HEP-PH/0007120;%%

\bibitem{Vogelsang:1992jn}
W.~Vogelsang and A.~Weber,
\newblock Phys. Rev. {\bf D48}, 2073 (1993).
%%CITATION = PHRVA,D48,2073;%%

\bibitem{Vogelsang:1997ak}
W.~Vogelsang,
\newblock Phys. Rev. {\bf D57}, 1886 (1998), hep-ph/9706511.
%%CITATION = HEP-PH/9706511;%%

\bibitem{Shimizu:2005fp}
H.~Shimizu, G.~Sterman, W.~Vogelsang, and H.~Yokoya,
\newblock Phys. Rev. {\bf D71}, 114007 (2005), hep-ph/0503270.
%%CITATION = HEP-PH/0503270;%%

\bibitem{Baldracchini:1980uq}
F.~Baldracchini, N.~S. Craigie, V.~Roberto, and M.~Socolovsky,
\newblock Fortschr. Phys. {\bf 30}, 505 (1981).
%%CITATION = FPYKA,30,505;%%

\bibitem{Shifman:1980dk}
M.~A. Shifman and M.~I. Vysotsky,
\newblock Nucl. Phys. {\bf B186}, 475 (1981).
%%CITATION = NUPHA,B186,475;%%

\bibitem{Bukhvostov:1985rn}
A.~P. Bukhvostov, G.~V. Frolov, L.~N. Lipatov, and E.~A. Kuraev,
\newblock Nucl. Phys. {\bf B258}, 601 (1985).
%%CITATION = NUPHA,B258,601;%%

\bibitem{Mukherjee:2001zx}
A.~Mukherjee and D.~Chakrabarti,
\newblock Phys. Lett. {\bf B506}, 283 (2001), hep-ph/0102003.
%%CITATION = HEP-PH/0102003;%%

\bibitem{Blumlein:2001ca}
J.~Bl{\"umlein},
\newblock Eur. Phys. J. {\bf C20}, 683 (2001), hep-ph/0104099.
%%CITATION = HEP-PH/0104099;%%

\bibitem{Kirschner:1996jj}
R.~Kirschner, L.~Mankiewicz, A.~Schafer, and L.~Szymanowski,
\newblock Z. Phys. {\bf C74}, 501 (1997), hep-ph/9606267.
%%CITATION = HEP-PH/9606267;%%

\bibitem{Hayashigaki:1997dn}
A.~Hayashigaki, Y.~Kanazawa, and Y.~Koike,
\newblock Phys. Rev. {\bf D56}, 7350 (1997), hep-ph/9707208.
%%CITATION = HEP-PH/9707208;%%

\bibitem{Kumano:1997qp}
S.~Kumano and M.~Miyama,
\newblock Phys. Rev. {\bf D56}, 2504 (1997), hep-ph/9706420.
%%CITATION = HEP-PH/9706420;%%

\bibitem{Belitsky:1997rh}
A.~V. Belitsky and D.~M{\"u}ller,
\newblock Phys. Lett. {\bf B417}, 129 (1998), hep-ph/9709379.
%%CITATION = HEP-PH/9709379;%%

\bibitem{Hoodbhoy:1998vm}
P.~Hoodbhoy and X.-D. Ji,
\newblock Phys. Rev. {\bf D58}, 054006 (1998), hep-ph/9801369.
%%CITATION = HEP-PH/9801369;%%

\bibitem{Belitsky:2000yn}
A.~V. Belitsky, A.~Freund, and D.~M{\"u}ller,
\newblock Phys. Lett. {\bf B493}, 341 (2000), hep-ph/0008005.
%%CITATION = HEP-PH/0008005;%%

\bibitem{Gracey:2003yrxGracey:2003mrxGracey:2006zrxGracey:2006ah}
J.~A. Gracey,
\newblock Nucl. Phys. {\bf B662}, 247 (2003), hep-ph/0304113; Nucl. Phys. {\bf
  B667}, 242 (2003), hep-ph/0306163; JHEP {\bf 10}, 040 (2006), hep-ph/0609231;
  Phys. Lett. {\bf B643}, 374 (2006), hep-ph/0611071.

\bibitem{Andre:2008}
Y.~Andr{\'e},
\newblock Proc. of the Int. Conf. ``{\sf Motives, Quantum Field Theory, an
  Pseudo Differential Operators}'', Clay Mathematical Institute, Boston, June,
  2008.

\bibitem{Brown:2008um}
F.~Brown,
\newblock Commun. Math. Phys. {\bf 287}, 925 (2009), arxiv:~0804.1660.
%%CITATION = 0804.1660;%%

\bibitem{Bailey:1991}
H.~Ferguson and D.~H. Bailey,
\newblock {\sf A Polynomial Time, Numerically Stable Integer Relation
  Algorithm}, RNR Techn. Rept. {\sf RNR-91-032}, 199 (1991).

\bibitem{Salvy:1994}
B.~Salvy and P.~Zimmermann,
\newblock ACM Transactions on Mathematical Software (2) {\bf 20}, 163 (1994).

\bibitem{Mallinger:1996}
C.~Mallinger,
\newblock {\sf Algorithmic Manipulations and Transformations of Univariate
  Holonomic Functions and Sequences}, Master Thesis, J. Kepler University,
  Linz, 1996.

\bibitem{Blumlein:2008pa}
J.~Bl{\"umlein},
\newblock Nucl. Phys. Proc. Suppl. {\bf 183}, 232 (2008), math-ph/0807.0700.
%%CITATION = 0807.0700;%%

\bibitem{Geddes:1992}
K.~Geddes, S.~Czapor, and G.~Labahn,
\newblock {\sf Algorithms for Computer Algebra}, (Kluwer Academic Publishers,
  Boston(USA), 1992, 585 p).

\bibitem{Gathen:1999}
J.~von~zur Gathen and J.~Gerhard,
\newblock {\sf Modern Computer Algebra}, (Cambridge University Press,
  Cambridge, 1999), 754~p.

\bibitem{Kauers:2008zz}
M.~Kauers,
\newblock Nucl. Phys. Proc. Suppl. {\bf 183}, 245 (2008).
%%CITATION = NUPHZ,183,245;%%

\bibitem{bostan:08}
A.~Bostan and M.~Kauers,
\newblock {\sf The full counting function for Gessel walks is algebraic},
  INRIA-Rocquencourt report, 2009, in preparation.

\bibitem{Beckermann:1992}
B.~Beckermann and G.~Labahn,
\newblock Numerical Algorithms {\bf 3}, 45 (1992).

\bibitem{Beckermann:2000}
B.~Beckermann and G.~Labahn,
\newblock SIAM Journal of Matrix Analysis and Applications {\bf 22}, 114
  (2000).

\bibitem{Karr:1985}
M.~Karr,
\newblock J.~Symbolic Comput., {\bf 1} (1985) 303 .

\bibitem{Schneider:2001}
C.~Schneider,
\newblock {\sf Small Symbolic Summation in Difference Fields}, PhD Thesis,
  RISC--Linz, J.~Kepler~University, Linz, 2001.

\bibitem{Schneider:2007a}
C.~Schneider,
\newblock {\sf Symbolic summation finds optimal nested sum representations},
  SFB-Report 2007-26, SFB F013, J. Kepler University, Linz, 2007.

\bibitem{RISC3389}
C.~Schneider,
\newblock J. Symbolic Comput. {\bf 43}, 611 (2008).

\bibitem{Schneider:2008}
C.~Schneider,
\newblock {\sf Parameterized telescoping proves algebraic independence of
  sums}, {Ann. Comb.}, to appear, 2009.

\bibitem{Ablinger:09}
J.~Ablinger,
\newblock {\sf A Computer Algebra Toolbox for Harmonic Sums Related to Particle
  Physics}, Diploma Thesis, J. Kepler University, Linz, 2009.

\bibitem{Blumlein:prep2}
J.~Bl{\"u}mlein and S.~Moch,
\newblock {\sf in preparation}.

\bibitem{Blumlein:2007tr}
J.~Bl{\"u}mlein, A.~de~Freitas, and W.~van Neerven,
\newblock PoS {\bf {\sf RADCOR} 2007}, 005 (2007), 0812.1588.
%%CITATION = 0812.1588;%%

\bibitem{Veltman:1994wz}
M.~J.~G. Veltman,
\newblock {\sf Diagrammatica: The Path to Feynman rules}, (Cambridge University
  Press, Cambridge, 1994), 284~p.

\bibitem{'tHooft:1973pz}
G.~'t~Hooft and M.~J.~G. Veltman,
\newblock {\sf Diagrammar}, CERN Yellow Report 73--9 (1973).
%%CITATION = NASBD,4,177;%%

\bibitem{Matiounine:1998re}
Y.~Matiounine, J.~Smith, and W.~L. van Neerven,
\newblock Phys. Rev. {\bf D58}, 076002 (1998), hep-ph/9803439.
%%CITATION = HEP-PH/9803439;%%

\bibitem{stegun}
M.~Abramowitz and I.~A. Stegun,
\newblock {\sf Handbook of Mathematical Functions}, (Dover Publications Inc.,
  New York, 1972), 1046~p.

\bibitem{Nielsen:1906}
N.~Nielsen,
\newblock {\sf Handbuch der Theorie der Gammafunktion}, (Chelsea Publishing
  Company, New York, 1965), 328 p; first published: (Teubner, Leipzig, 1906),
  326~p.

\bibitem{Smirnov:2004ym}
V.~A. Smirnov,
\newblock {\sf Evaluating Feynman integrals}, Springer Tracts Mod. Phys. {\bf
  211}, 1 (2004).
%%CITATION = STPHB,211,1;%%

\bibitem{Euler:1775}
L.~Euler,
\newblock Novi Comm. Acad. Sci Petropolitanae {\bf 1}, 140 (1775).

\bibitem{Zagier:1994}
D.~Zagier,
\newblock Proc. First European Congress Math. (Paris) {\bf II}, 497 (1994).

\bibitem{Nielsen:1909}
N.~Nielsen,
\newblock Nova Acta Leopoldina {\bf 90}, 123 (1909).

\bibitem{Kolbig:1983qt}
K.~S. K{\"o}lbig,
\newblock SIAM J. Math. Anal. {\bf 17}, 1232 (1986).
%%CITATION = SJMAA,17,1232;%%

\bibitem{Knopp:1947}
K.~Knopp,
\newblock {\sf Theorie und Anwendung der unendlichen Reihen}, (Springer,
  Berlin, 1947), 583~p.

\bibitem{Landau:1906}
E.~Landau,
\newblock S.-Ber. K{\"o}nigl. Bayerische Akad. Wiss. M{\"u}nchen, math.-naturw.
  Kl. {\bf 36}, 151 (1906).

\end{thebibliography}
}
%%%%%%%%%%%%%%%%%%%%%%%%%%%%%%%%%%%%%%%%%%%%%%%%%%%%%%%%%%%%%%%%%%%%%%%%%%%%
 \newpage
 \begin{flushleft}
 \end{flushleft}
 \thispagestyle{empty}
 \newpage
 \thispagestyle{empty}
 \setcounter{page}{0}
\begin{center}
{\bf Acknowledgement}
\end{center}
 Foremost, I would like to thank Johannes Bl{\"umlein} for
 his constant support during the last years, putting very much
 time and effort into supervising and teaching me. \\
 Further I would like to thank Prof. Reya. for giving me the 
 opportunity of getting my Ph.D. at the University of Dortmund. \\
 I am particularly grateful to I. Bierenbaum for her advice and friendship 
 during the last years and for reading the manuscript. \\
 I would like to thank
 D.~Broadhurst, K.~Chetyrkin, J.~Kallarackal, M.~Kauers, D.~Renner, 
 C.~Schneider, J.~Smith,
 F.~Stan, M.~Steinhauser and J.~Vermaseren for useful discussions. 
 Additionally, I would like to thank M.~Steinhauser for help with the use 
 of ${\sf MATAD}$, J.~Vermaseren for help with the use of ${\sf FORM}$, 
 and C.~Schneider 
 for help with the use of \SigmaP. 
 Further thanks go to B. T{\"o}dtli and K. Litten for reading
 parts of the manuscript.\\
 Finally, I also would like to thank my family, Frank, Katharina, Max and Ute
 for moral support. \\
 This work was supported in part by DFG Sonderforschungsbereich 
 Transregio 9, Computergest\"utzte Theoretische Teilchenphysik, 
 Studienstiftung des Deutschen Volkes, the European Commission MRTN
 HEPTOOLS under Contract No. MRTN-CT-2006-035505, and DESY. I thank
 both IT groups of DESY providing me access to special facilities to 
 perform the calculations involved in this thesis. \\
%%%%%%%%%%%%%%%%%%%%%%%%%%%%%%%%%%%%%%%%%%%%%%%%%%%%%%%%%%%%%%%%%%%%%%%%%%%%
 \end{document}